\documentclass{dune}
\pdfoutput=1

\usepackage[pdftex,bookmarks,hidelinks]{hyperref}
\usepackage{cite}
\graphicspath{ {graphics/} }

\newif\ifdp
\newif\ifsp

\def\expshort{DUNE\xspace}
\def\dune{\expshort}
\def\explong{The Deep Underground Neutrino Experiment\xspace}

\def\thedocsubtitle{Deep Underground Neutrino Experiment (DUNE)} 
\def\tdrtitle{Near Detector Conceptual Design Report}

\newcommand{\refsec}[2]{Volume~\csname volnumber#1\endcsname \xspace Section~#2}
\newcommand{\refch}[2]{Volume~\csname volnumber#1\endcsname \xspace Chapter~#2}
\newcommand{\refinch}[2]{#2 in Volume~\csname volnumber#1\endcsname \xspace}

\newcommand{\numu}{\ensuremath{\nu_\mu}\xspace}
\newcommand{\nue}{\ensuremath{\nu_e}\xspace}

\newcommand{\anumu}{\ensuremath{\bar\nu_\mu}\xspace}
\newcommand{\anue}{\ensuremath{\bar\nu_e}\xspace}

\newcommand{\dm}[1]{\ensuremath{\Delta m^2_{#1}}\xspace} 

\newcommand{\sinst}[1]{\ensuremath{\sin^2\theta_{#1}}\xspace} 

\newcommand{\deltacp}{\ensuremath{\delta_{\rm CP}}\xspace}   

\newcommand{\numubartonumubar}{
\ensuremath{\overline{\numu}\rightarrow\overline{\numu}}\xspace
}

\def\argon40{${}^{40}$Ar}       
\def\Ar39{$^{39}$Ar}
\def\Cl40{$^{40}$Cl}
\def\K40{$^{40}$K}
\def\B8{$^{8}$B}

\def\fdfiducialmass{\SI{40}{\kt}\xspace}

\def\larmass{\SI{17.5}{\kt}\xspace} 
\def\nominalmodsize{\SI{10}{kt}\xspace} 
\newcommand{\efield}{E field\xspace}

\newcommand{\threed}{3D\xspace}
\newcommand{\twod}{2D\xspace}
\newcommand{\phel}{photoelectron\xspace} 
\newcommand{\frfour}{FR-4\xspace} 

\newcommand{\lsim}{{\;\raise0.3ex\hbox{$<$\kern-0.75em\raise-1.1ex\hbox{$\sim$}}\;}}
\newcommand{\gsim}{{\;\raise0.3ex\hbox{$>$\kern-0.75em\raise-1.1ex\hbox{$\sim$}}\;}}
\newcommand{\beq}{\begin{equation}}
\newcommand{\eeq}{\end{equation}}
\newcommand{\bea}{\begin{eqnarray}}
\newcommand{\eea}{\end{eqnarray}}

\mathchardef\minus="002D

\newcommand{\rrt}[1]{\ifthenelse{\equal{#1}{}}{[RT:TBD]}{[RT:#1]}}

\newcommand{\microboone}{MicroBooNE\xspace} 

\newcommand{\lartpc}{LArTPC\xspace}

\newcommand{\nch}{\ensuremath{\left\langle n_{ch} \right\rangle}}

\DeclareUnicodeCharacter{223C}{~} 

\DeclareUnicodeCharacter{03C0}{$\pi$}

\DeclareSIUnit \s {\second}
\DeclareSIUnit \MB {\mega\byte}
\DeclareSIUnit \GB {\giga\byte}
\DeclareSIUnit \TB {\tera\byte}
\DeclareSIUnit \PB {\peta\byte}
\DeclareSIUnit \Mbps {\mega\bit/\s}
\DeclareSIUnit \Gbps {\giga\bit/\s}
\DeclareSIUnit \Tbps {\tera\bit/\s}
\DeclareSIUnit \Pbps {\peta\bit/\s}
\DeclareSIUnit \kton {\kilo\tonne} 
\DeclareSIUnit \kt {\kilo\tonne}
\DeclareSIUnit \Mt {\mega\tonne}
\DeclareSIUnit \eV {\electronvolt}
\DeclareSIUnit \keV {\kilo\electronvolt}
\DeclareSIUnit \MeV {\mega\electronvolt}
\DeclareSIUnit \GeV {\giga\electronvolt}
\DeclareSIUnit \m {\meter}
\DeclareSIUnit \cm {\centi\meter}
\DeclareSIUnit \in {in}
\DeclareSIUnit \km {\kilo\meter}
\DeclareSIUnit \kV {\kilo\volt}
\DeclareSIUnit \kW {\kilo\watt}
\DeclareSIUnit \MW {\mega\watt}
\DeclareSIUnit \MHz {\mega\hertz}
\DeclareSIUnit \mrad {\milli\radian}
\DeclareSIUnit \year {year}
\DeclareSIUnit \POT {POT}
\DeclareSIUnit \sig {$\sigma$}
\DeclareSIUnit\parsec{pc}
\DeclareSIUnit\lightyear{ly}
\DeclareSIUnit\foot{ft}
\DeclareSIUnit\ft{ft}
\DeclareSIUnit \ppb{ppb}
\DeclareSIUnit \ppt{ppt}
\DeclareSIUnit \samples{S}
\DeclareSIUnit \MJ{\mega\joule}
\DeclareSIUnit \T{\tesla}

\DeclareSIUnit \metricton {metric\thinspace ton}

\usepackage[toc]{glossaries}
\makeglossaries

\newcommand{\dshort}[1]{\glsentrytext{#1}}  
\newcommand{\dshorts}[1]{\glsentryshortpl{#1}}  

\newcommand{\dfirst}[1]{\glsfirst{#1}\glsunset{#1}}

\newcommand{\dword}[1]{\gls{#1}}
\newcommand{\dwords}[1]{\glspl{#1}}

\newcommand{\newduneword}[3]{
    \newglossaryentry{#1}{
        text={#2},
        long={#2},
        name={\glsentrylong{#1}},
        first={\glsentryname{#1}},
        firstplural={\glsentrylong{#1}\glspluralsuffix},
        description={#3},
        sort={#2}
    }
}

\newcommand{\newduneabbrev}[4]{
  \newglossaryentry{#1}{
    text={#2},
    long={#3},
    shortplural={{#2}\glspluralsuffix},
    longplural={{#3}\glspluralsuffix{}},
    name={\glsentrylong{#1}{} (\glsentrytext{#1}{})},
    first={#3 (#2)},
    firstplural={#3\glspluralsuffix{} (\glsentrytext{#1}\glspluralsuffix{})},
    description={#4},
    sort={#2}
  }
}

\newcommand{\newduneabbrevs}[5]{
  \newglossaryentry{#1}{
    text={#2},
    long={#3},
    plural={#4},
    shortplural={{#2}\glspluralsuffix},
    longplural={#4},
    name={\glsentrylong{#1}{} (\glsentrytext{#1}{})},
    first={#3 (#2)},
    firstplural={#4 (\glsentrytext{#1}\glspluralsuffix{})},
    description={#5},
    sort={#2}    
  }
}

\newduneword{dword}{DUNE Word}{A term in the DUNE lexicon}

\newduneword{nasa}{NASA}{U.S. National Aereonautics and Space Administration}

\newduneabbrev{nd}{ND}{near detector}{Refers to the detector(s) 
 installed close to the neutrino source at \gls{fnal} }

\newduneabbrev{fd}{FD}{far detector}{The \SI{70}{kt} total (\fdfiducialmass fiducial) mass \gls{lartpc} DUNE detector, composed of four \larmass total (\nominalmodsize fiducial) mass modules,  
  to be installed at the far site at \gls{surf} in
  Lead, SD, USA}

\newduneabbrev{sp}{SP}{single-phase}{Distinguishes one of the DUNE far detector technologies by the fact that it operates using argon in its liquid phase only}

\newduneabbrev{dp}{DP}{dual-phase}{Distinguishes one of the DUNE far detector technologies by the fact that it operates using argon 
 in both gas and liquid phases; sometimes called double-phase} 

\newduneabbrev{pds}{PDS}{photon detection system}{The detector 
  subsystem sensitive to light produced in the \gls{lar} }

\newduneabbrev{hvs}{HVS}{high voltage system}{The detector 
  subsystem that provides the \gls{tpc} drift field}

\newduneabbrev{tpc}{TPC}{time projection chamber}{A type of particle detector that uses an \efield together with a sensitive volume of gas or liquid, e.g., \gls{lar}, to perform a \threed reconstruction of a particle trajectory or interaction. The activity is recorded by digitizing the waveforms of current
  induced on the anode as the distribution of ionization charge passes by
  or is collected on the electrode (TPC is also used for ``total project cost'')} 

\newduneabbrev{lartpc}{LArTPC}{liquid argon time-projection chamber}{A \gls{tpc} filled with liquid argon; 
the basis for the \gls{dune} \gls{fd} modules} 

\newduneabbrevs{apa}{APA}{anode plane assembly}{anode plane assemblies}{A unit of the \gls{sphd}
  detector module containing the elements sensitive to ionization in the \gls{lar}. 
  It contains two faces each of three planes of wires, and interfaces to the cold
  electronics and photon detection system} 

\newduneabbrev{awg}{AWG}{American wire gauge} {U.S. standard set of non-ferrous wire conductor sizes}

\newduneabbrev{ufer}{Ufer}{concrete encased electrode} {U.S. National Electrical Code grounding method refered to as Concrete Encased Electrode}

\newduneabbrev{cro}{CRO}{charge readout}{The system for detecting
  ionization charge distributions in a \gls{dp} detector module}

\newduneabbrev{lro}{LRO}{light readout}{The system for detecting
  scintillation photons in a \gls{dp}  detector module}

\newduneabbrev{shv}{SHV}{safe high voltage}{Type of bayonet mount
connector used on coaxial cables that has additional insulation 
compared to standard BNC and MHV connectors that makes it safer
for handling \gls{hv} by preventing accidental contact with the
live wire connector in an unmated connector or plug}

\newduneabbrev{fe}{FE}{front-end}{The front-end refers to a point that is
  ``upstream'' of the data flow for a particular subsystem. 
 For example the \gls{sphd} front-end electronics is where the cold electronics
  meet the sense wires of the TPC and the front-end \gls{daq} is where the \gls{daq} meets the output of the electronics}

\newduneabbrev{daqrou}{DAQ RU}{DAQ readout unit}{The first element in the data flow of the \gls{daq}}

\newduneabbrev{cots}{COTS}{commercial off-the-shelf}{Items, typically hardware such as 
computers, that may be purchased whole, without any custom design or fabrication and 
thus at normal consumer prices and availability}

\newduneabbrev{i2c}{I2C}{Inter-Integrated Circuit}{I$^2$C or I2C is a synchronous, 
multi-master, multi-slave, packet switched, single-ended, serial computer bus widely used 
for attaching lower-speed peripheral ICs to processors and microcontrollers in short-distance, 
intra-board communication} 

\newduneabbrev{spi}{SPI}{Serial Peripheral Interface}{The Serial Peripheral Interface is a 
synchronous serial communication interface specification used for short distance 
communication, primarily in embedded systems}

\newduneabbrev{miso}{MISO}{master in slave out}{The Master In Slave Out is a logic
signal on the \gls{spi} bus on which the data from the slave are transmitted once
a request from the master is received} 

\newduneabbrev{mosi}{MOSI}{master out slave in}{The Master Out Slave In is a logic
signal on the \gls{spi} bus on which the data from the master is transmitted} 

\newduneabbrev{uart}{UART}{Universal Asynchrous Receiver/Transmitter}{A universal 
asynchronous receiver-transmitter is a computer hardware device for asynchronous 
serial communication in which the data format and transmission speeds are configurable}

\newduneword{cr}{CR}{Capacitance-Resistance} 

\newduneword{dc}{DC}{direct coupling} 

\newduneword{ac}{AC}{Alternating Current; when used in the phrase ``AC coupling'' refers to a circuit element that filters out low-frequency components, such as constant offsets, leaving higher frequency signal components. The frequency filtering is determined both by a resistor and a capacitor}

\newduneabbrev{pll}{PLL}{Phase-Locked Loop}{A control system that generates an
output signal whose phase is related to the phase of an input signal}  

\newduneword{fifo}{FIFO}{First-In-First-Out} 

\newduneword{tsmc}{TSMC}{Taiwan Semiconductor Manufacturing Company}

\newduneword{saci}{SACI}{\gls{slac} \gls{asic} Control Interface}

\newduneword{om3}{OM3}{Type of multi-mode fiber optic cable, typically capable of \SI{10}{Gbps} data transmission at lengths up to \SI{300}{m}}

\newduneword{om4}{OM4}{Type of multi-mode fiber optic cable, typically capable of \SI{10}{Gbps} data transmission at lengths up to \SI{550}{m}}

\newduneword{qfp}{QFP}{Quad Flat Package} 

\newduneabbrev{ams}{AMS}{analog and mixed signal}{Verilog-AMS is a derivative of the Verilog hardware description language that includes analog and mixed-signal extensions (AMS) in order to define the behavior of analog and mixed-signal systems}

\newduneabbrev{hepa}{HEPA}{High Efficiency Particulate Air}{The High Efficiency Particulate Air filters are a type of air filter that remove 99.97\% of particles that have a size greater than or equal to \SI{0.3}{\micro\meter}}  

\newduneabbrev{uvm}{UVM}{universal verification methodology}{The Universal Verification Methodology is a standardized methodology for verifying integrated circuit designs}   

\newduneword{lhc}{LHC}{Large Hadron Collider}

\newduneabbrev{lsb}{LSB}{least significant bit}{The bit with the lowest numerical value in a binary number}

\newduneabbrev{ldo}{LDO}{low-dropout regulator}{A low-dropout or LDO regulator is a \gls{dc} linear voltage regulator that can regulate the output voltage even when the supply voltage is very close to the output voltage}

\newduneabbrev{adc}{ADC}{analog-to-digital converter}{A sampling of a voltage
  resulting in a discrete integer count corresponding in some way to
  the input}

\newduneabbrev{inl}{INL}{integral non-linearity}{A commonly used measure of performance in \glspl{adc}. It is the deviation between the ideal input threshold value and the measured threshold level of a certain output code}

\newduneabbrev{dnl}{DNL}{differential non-linearity}{A commonly used measure of performance in \glspl{adc}. The DNL error is defined as the difference between an actual step width and the ideal value of one \gls{lsb}}

\newduneword{pnp}{PNP}{Type of bipolar junction transistor consistning of a
layer of N-doped semiconductor sandwiched between two layers of P-doped material}

\newduneword{spice}{SPICE}{SPICE
(``Simulation Program with Integrated Circuit Emphasis'') is a general-purpose, 
open-source analog electronic circuit simulator. It is a program used in integrated 
circuit and board-level design to check the integrity of circuit designs and to 
predict circuit behavior}

\newduneabbrev{daq}{DAQ}{data acquisition}{The data acquisition system
  accepts data from the detector \gls{fe} electronics, buffers
  the data, performs a \gls{trigdecision}, builds events from the selected
  data and delivers the result to the offline \gls{diskbuffer}}

\newduneabbrev{iov}{IOV}{interval of validity}{Interval over which something is valid}

\newduneword{calci}{CALCI}{Calibration and Cryogenic Instrumentation}

\newduneword{detmodule}{far detector module}{The entire DUNE far detector is
  segmented into four modules, each with a nominal \SI{10}{\kton}
  fiducial mass}
  
\newduneword{module}{module}{Many aspects of the DUNE far and near detectors are modular, so ``module'' must be understood in context. It may refer to one of the four far detector modules, distinct portions of a subdetector as in a ``field cage module,'' a software or electronics module, e.g., a separate framework plug-in, and so on}

\newduneword{detunit}{detector unit}{A portion of a \gls{detmodule} may be further partitioned into a number of similar parts.   For example, the \gls{sphd} \gls{tpc} is made up of \gls{apa}  units (and other elements)}

\newduneword{diskbuffer}{secondary DAQ buffer}{A secondary
  \gls{daq} buffer holds a small subset of the full rate as
  selected by a \gls{trigcommand}. 
  This buffer also marks the interface with the DUNE Offline}

\newduneabbrev{om}{OM}{online monitoring}{Processes that run inside
  the \gls{daq} on data ``in flight,'' specifically before landing on the
  offline disk buffer, and that provide feedback on the operation of
  the \gls{daq} itself and the general health of the data it is marshalling}

\newduneabbrev{dqm}{DQM}{data quality monitoring}{Analysis of the raw
  data to monitor the integrity of the data and the performance of the
  detectors and their electronics. This type of monitoring may be
  performed in real time, within the \gls{daq} system, or in later
  stages of processing, using disk files as input}

\newduneword{dumpbuffer}{DAQ dump buffer}{This \gls{daq} buffer
  accepts a high-rate data stream, in aggregate, from an associated
  portion of a \gls{detmodule} sufficient to collect all data likely relevant to
  a potential \gls{snb}}
\newduneabbrev{etf}{ETF}{Experiment Test Framework}{\gls{wlcg} testing middleware running grid jobs that actively test distributed sites services and capabilities, and report back to monitoring services}

\newduneabbrev{etl}{ETL}{external trigger logic}{Trigger processing
  that consumes \gls{detmodule} level \gls{trignote} information
  and other global sources of trigger input and emits
  \gls{trigcommand} information back to the \glspl{mtl}}
\newduneabbrev{daqeti}{ETI}{external trigger interface}{Interface between \glspl{mtl} and external source and sinks of relevant trigger information}

\newduneword{trignote}{trigger notification}{Information provided by
  \gls{mtl} to \gls{etl} about \gls{trigdecision} 
  processing}

\newduneword{trigprimitive}{trigger primitive}{Information derived by
  the \gls{daq} \gls{fe} hardware that describes a region of space (e.g.,
  one or several neighboring channels) and time (e.g., a contiguous set
  of \gls{adc} sample ticks) associated with some activity}

\newduneword{externtrigger}{external trigger candidate}{Information
  provided to the \gls{mtl} about events external to a
  \gls{detmodule} so that it may be considered in forming
  \glspl{trigcommand}}

\newduneabbrev{daqoob}{OOB dispatcher}{out-of-band trigger command
  dispatcher}{This component is responsible for dispatching a \gls{snb} dump
  command to all \glspl{daqfer} in the \gls{detmodule}}

\newduneabbrev{mtl}{MTL}{module trigger logic}{Trigger processing
  that consumes \gls{detunit} level \gls{trigcommand} information
  and emits \glspl{trigcommand}. 
  It provides the \gls{etl} with \glspl{trignote} and receives back any
  \glspl{externtrigger}}

\newduneword{octant}{octant}{Any of the eight parts into which 4$\pi$
  is divided by three mutually perpendicular axes. 
  In particular in referencing the value for the mixing angle
  $\theta_{23}$}

\newduneword{trigcandidate}{trigger candidate}{Summary information derived
  from the full data stream and representing a contribution toward
  forming a \gls{trigdecision}}

\newduneword{trigcommand}{trigger command}{Information derived from
  one or more \glspl{trigcandidate}  that directs elements of the
  \gls{detmodule} to read out a portion of the data stream}

\newduneabbrev{tcm}{TCM}{trigger command message}{A message flowing
  down the trigger hierarchy from global to local context.  Also see \gls{tpm}}

\newduneabbrev{mlt}{MLT}{module level trigger}{The \gls{daq} component responsible for producing a \gls{trigdecision} that will be used to command the readout of a detector module}

\newduneword{trigdecision}{trigger decision}{The process by which
  \glspl{trigcandidate} are converted into \glspl{trigcommand}}

\newduneabbrev{tpm}{TPM}{trigger primitive message}{A message flowing
  up the trigger hierarchy from local to global context.  Also see \gls{tcm}}

\newduneabbrev{ipc}{IPC}{inter-process communication}{A system for software elements to exchange information between threads, local processes or across a data network.  An IPC system is typically specified in terms of protocols  composed of message types and their associated data schema}

\newduneword{daqdispre}{discovery and presence}{As used in the context of the \gls{ipc}, a system that provides mechanisms for a node on a communication network to learn of the existence of peers and their identity (discovery) as well as determine if they are currently operational or have become unresponsive (presence)}

\newduneabbrev{pubsub}{PUB/SUB}{publish-subscribe communication pattern}{An \gls{ipc} communication pattern where one element, the publisher, sends data to all connected elements, the subscribers.  Each subscriber may connect to multiple publishers.  A variant is PUB/SUB with topics where a subscriber may register an identifier, the topic, to limit the information received to just an associated subset}

\newduneabbrev{eb}{EB}{event builder}{A software agent that executes \glspl{trigcommand}  for one  \gls{detmodule} by reading out the requested data}

\newduneabbrev{daqdfo}{DFO}{data flow orchestrator}{The process by which trigger commands are executed in parallel and asynchronous manner by the back-end output subsystem of the \gls{daq}}

\newduneabbrev{daqubi}{UBI}{upstream DAQ buffer interface}{The process which provides read-only access to data residing in the upstream \gls{daq} buffers to processes on the network}

\newduneabbrev{cob}{COB}{cluster on board}{An ATCA motherboard housing four RCEs}

\newduneabbrev{rce}{RCE}{reconfigurable computing element}{Data processor located outside of the cryostat on a \gls{cob} that contains \gls{fpga}, RAM and \gls{ssd} resources, responsible for buffering data, producing trigger primitives, responding to triggered requests for data and synching \gls{snb} dumps}

\newduneabbrev{bow}{BOW}{Bump On Wire}{A working name for the front-end readout computing elements used in the nominal \gls{daq} design to interface the \gls{dp}  crates to the \gls{daq} front-end computers}

\newduneabbrev{atca}{ATCA}{Advanced Telecommunications Computing
  Architecture}{An advanced computer architecture specification developed for the telecommunications, military, and aerospace industries that incorporates the latest trends in  high-speed interconnect technologies, next-generation processors, and improved reliability, availability and serviceability} 

\newduneabbrev{utca}{$\mu$TCA}{Micro Telecommunications Computing Architecture}{The computer architecture specification followed by the crates that house charge and light readout electronics in the \gls{dpmod}} 

\newduneabbrev{udp}{UDP}{user datagram protocol}{A simple,
  connectionless Internet protocol that supports data integrity
  checksums, requires no handshaking, and does not guarantee packet delivery}

\newduneabbrev{amc}{AMC}{advanced mezzanine card}{Holds digitizing
  electronics and lives in \gls{utca} crates}

\newduneabbrev{rf}{RF}{radio frequency}{Electromagnetic emissions
  that are within the (radio) frequency band of sensitivity of the detector
  electronics}

\newduneabbrev{fpga}{FPGA}{field programmable gate array}{An
integrated circuit technology that allows the hardware to be reconfigured to
execute different algorithms after its manufacture and deployment}

\newduneabbrev{fmc}{FMC}{FPGA mezzanine card}{Boards holding \glspl{fpga} and other integrated circuitry that attach to a motherboard}

\newduneabbrev{felix}{FELIX}{Front-End Link eXchange}{A
  high-throughput interface between \gls{fe} and trigger electronics
  and the standard PCIe computer bus}

\newduneword{daqpart}{DAQ partition}{A cohesive and
 coherent collection of \gls{daq} hardware and software working together to trigger and read out some portion of one detector module; it consists of an integral number of \glspl{daqfrag}. 
 Multiple \gls{daq} partitions may operate simultaneously, but each instance operates independently}

\newduneabbrev{fec}{DAQ FEC}{DAQ front-end computer}{The portion of one
  \gls{daqpart} that hosts the \gls{daqdr}, \gls{daqbuf} and
  \gls{daqds}.  It hosts the \gls{daqfer} and corresponding portion of the \gls{daqbuf}}

\newduneword{daqfrag}{DAQ front-end fragment}{The portion of one
  \gls{daqpart} relating to a single \gls{fec} and corresponding to an
  integral number of \glspl{detunit}.  See also \gls{datafrag}}

\newduneword{datafrag}{data fragment}{A block of data read out from a single \gls{daqfrag} that
span a contiguous period of time as requested by a \gls{trigcommand}}

\newduneabbrev{daqfer}{FER}{DAQ front-end readout}{The portion of a
  \gls{daqfrag} that accepts data from the detector electronics and
  provides it to the \gls{fec}}

\newduneabbrev{daqdr}{DDR}{DAQ data receiver}{The portion of the
  \gls{daqfrag} that accepts data from the \gls{daqfer}, emits
  trigger candidates produced from the input trigger primitives, and
  forwards the full data stream to the \gls{daqbuf}}

\newduneword{daqbuf}{DAQ primary buffer}{The portion
  of the \gls{daqfrag} that accepts full data stream from the
  corresponding \gls{detunit} and retains it sufficiently long for it
  to be available to produce a \gls{datafrag}}

\newduneword{daqds}{data selector}{The portion of the \gls{daqfrag}
  that accepts \glspl{trigcommand} and returns the corresponding
  \gls{datafrag}.  Not to be confused with \gls{daqdsn}}

\newduneword{daqdsn}{data selection}{The process of forming a trigger decision for selecting a subset of detector data for output by the \gls{daq} from the content of the detector data itself.  Not to be confused with \gls{daqds}}

\newduneabbrev{daqros}{DAQ RO}{DAQ readout subsystem}{The subsystem of the \gls{daq} for accepting and buffering data input from detector electronics}

\newduneabbrev{daqdss}{DAQ DS}{DAQ data selection subsystem}{The subsystem of the \gls{daq} responsible for forming a trigger decision based on a portion of the input data stream.  The majority subset of the \gls{daqtrs}}

\newduneabbrev{daqtrs}{DAQ TS}{DAQ trigger subsystem}{The subsystem of the \gls{daq} responsible for forming a trigger decision}

\newduneabbrev{daqbes}{DAQ BE}{DAQ back-end subsystem}{The portion of the \gls{daq} that is generally toward its output end.  It is responsible for accepting and executing trigger commands and marshaling the data they address to output storage buffers}

\newduneabbrev{daqtss}{DAQ TSS}{DAQ timing and synchronization subsystem}{The portion of the \gls{daq} that provides for timing and synchronization to various components}

\newduneabbrev{femb}{FEMB}{front-end mother board}{Refers a unit of
  the \gls{sp} \gls{ce} that contains the \gls{fe} amplifier
  and \gls{adc} \glspl{asic} covering 128 channels}

\newduneword{asic}{ASIC}{application-specific integrated circuit}

\newduneword{lv}{LV}{low voltage}

\newduneabbrev{iceberg}{ICEBERG}{ICEBERG R\&D cryostat and electronics}{Integrated Cryostat and Electronics Built for Experimental Research Goals: a new double-walled cryostat built and installed at \gls{fnal}  
for liquid argon detector R\&D and for testing of DUNE detector components}

\newduneword{coldadc}{ColdADC}{A newly developed 16-channels \gls{asic} providing analog to digital conversion}

\newduneword{coldata}{COLDATA}{A 64-channel control and communications \gls{asic}}

\newduneword{cryo}{CRYO}{(1) Integrated ASIC including \gls{fe} circuitry providing signal amplification and pulse shaping, analog to digital conversion, and control and communication functionalities for 64 channels; (2) acryonym for cryogenic systems and cryostat work scopes in \gls{lbnf}}

\newduneword{larasic}{LArASIC}{A 16-channel \gls{fe} \gls{asic} that provides signal amplification and pulse shaping}

\newduneword{cmos}{CMOS}{Complementary metal-oxide-semiconductor}

\newduneabbrev{enc}{ENC}{equivalent noise charge}{The equivalent noise charge is the input charge that corresponds to a \gls{snr}$=1$}

\newduneword{sar}{SAR}{successive approximation register}

\newduneword{protodune}{ProtoDUNE}{Either of the two DUNE prototype detectors constructed at \gls{cern}. 
  One prototype implements \gls{sp} technology and the other \gls{dp}}
  
\newduneword{protodune2}{ProtoDUNE-II}{The second run of a \gls{protodune} detector}  

\newduneword{pdsp}{ProtoDUNE-SP}{The \gls{sphd} \gls{protodune} detector at \gls{cern}}

\newduneword{pddp}{ProtoDUNE-DP}{The \gls{dp} \gls{protodune} detector at \gls{cern}}

\newduneword{wa105}{WA105 DP demonstrator}{The \SI[product-units=power]{3x1x1}{m} WA105 \gls{dp} prototype detector at \gls{cern}}

\newduneword{rawevent}{DAQ event block}{The unit of data output by the
  \gls{daq}.  
  It contains trigger and detector data spanning a unique, contiguous
  time period and a subset of the detector channels}

\newduneabbrev{ssd}{SSD}{solid-state disk}{Any storage device that
  may provide sufficient write throughput to receive, both collectively and
  distributed, the sustained full rate of data from a \gls{detmodule}
  for many seconds}
\newduneabbrev{nvme}{NVMe}{Non-volatile memory express}{A specification for an interface to storage media attached via PCIe}

\newduneabbrev{hlt}{HLT}{high-level trigger}{This is actually a filter applied to data that has been triggered and aggregated in order to further reduce or characterize it}

\newduneabbrev{pid}{PID}{particle ID}{Particle identification}

\newduneword{readout window}{readout window}{A fixed, atomic and
  continuous period of time over which data from a \gls{detmodule}, in
  whole or in part, is recorded. 
  This period may differ based on the trigger that initiated the
  readout}

\newduneabbrev{zs}{ZS}{zero-suppression}{Used to delete some portion of a
  data stream that does not significantly deviate from zero or
  intrinsic noise levels. 
  It may be applied at different granularity from per-channel to per
  \gls{detunit}}

\newduneword{rc}{RC}{Depending on context, one of (1) resistive-capacitive (circuit), (2) run control, the system for configuring, starting and terminating the \gls{daq}, or (3) resource coordinator, a member of the \gls{dune} management team responsible for coordinating the financial resources of the project} 

\newduneabbrev{daqccm}{CCM}{DAQ control, configuration and monitoring subsystem}{A system for controlling, configuring and monitoring other systems in particular those that make up the \gls{daq} where the CCM encompasses \gls{rc}}

\newduneword{daqrun}{DAQ run}{A period of time over which relevant data taking conditions and \gls{daq} configuration are asserted to be unchanged. 
  Multiple \gls{daq} runs may occur simultaneously when multiple \glspl{daqpart} are active. 
  This term should not be confused with DUNE experiment or beam ``runs'' that typically span many \gls{daq} runs}
\newduneword{daqrunnum}{DAQ run number}{A monotonically increasing count that uniquely and globally identifies a \gls{daqrun}}

\newduneabbrev{snb}{SNB}{supernova neutrino burst}{A prompt 
  increase in the flux of low-energy neutrinos emitted in the first few seconds of a core-collapse supernova.  It can also refer to a trigger command type that may be due to this phenomenon,
  or detector conditions that mimic its interaction signature}

\newduneabbrev{snble}{SNB/LE}{supernova neutrino burst and low
  energy}{Supernova neutrino burst and low-energy physics program}

\newduneabbrev{snews}{SNEWS}{SuperNova Early Warning System}{A global
  supernova neutrino burst trigger formed by a coincidence of \gls{snb} 
  triggers collected from participating experiments}

\newduneabbrev{pps}{1PPS signal}{one-pulse-per-second signal}{An
  electrical signal with a fast rise time and that arrives in real
  time with a precise period of one second}

\newduneabbrev{sls}{SLS}{spill location system}{A system residing at
  the DUNE far detector site that provides information, possibly
  predictive, indicating periods of time when neutrinos are being
  produced by the \gls{fnal} Main Injector beam spills}

\newduneabbrev{wib}{WIB}{warm interface board}{Digital electronics
  situated just outside a FD cryostat that receives digital data
  from the \glspl{femb} (part of \gls{ce}) over cold copper connections and sends it to the \gls{rce}
  \gls{fe} readout hardware}

\newduneabbrev{gps}{GPS}{Global Positioning System}{A satellite-based system that provides a highly accurate \gls{pps} that may be used to synchronize clocks and determine location}

\newduneabbrev{ntp}{NTP}{Network Time Protocol}{A networking protocol that allows synchronizing of clocks to within a few \si{\milli\second} of a time standard on a local network and within a few tens of \si{\milli\second} over the Internet} 

\newduneword{ptp}{PTP}{Depending on context, either p-terphenyl, a \gls{wls} material, or Precision Time Protocol, a networking protocol that allows synchronizing of clocks to within a few \si{\micro\second} of a time standard on a local network}  

\newduneabbrev{irig}{IRIG}{inter-range instrumentation group}{A standards body that defined a time-code standard for transferring timing information}

\newduneabbrev{nic}{NIC}{network interface controller}{Hardware for controlling the interface to a communication network.  Typically, one that obeys the Ethernet protocol}

\newduneabbrev{wiec}{WIEC}{warm interface electronics crate}{Crates mounted on the signal flanges that contain the \glspl{wib}}

\newduneabbrev{ptc}{PTC}{power and timing card}{Cards that provide further processing and distribution of the signals entering and exiting the \gls{sp} cryostat}

\newduneabbrev{ptb}{PTB}{power and timing backplane}{Backplane used to connect the \glspl{wib} and the \glspl{ptc} on the \gls{wiec}. Also connects the \gls{ce} flange on the cryostat penetration}

\newduneabbrev{sipm}{SiPM}{silicon photomultiplier}{A solid-state
  avalanche photodiode sensitive to single \phel signals}

\newduneabbrev{cisc}{CISC}{cryogenic instrumentation and slow controls}{Includes equipment to monitor all detector  components and  \gls{lar} quality and behavior, and provides a control system for many of the detector components}

\newduneword{fte}{FTE}{full-time equivalent. A unit of labor
  for the project. One year of work from one person}

\newduneword{art}{art}{A software framework implementing an
  event-based execution paradigm} 

\newduneabbrev{sam}{SAM}{sequential
  access via metadata}{A data-handling system to store and retrieve
  files and associated metadata, including a complete record of the
  processing that has used the files}

\newduneword{artdaq}{artdaq}{A data acquisition toolkit for data transfer, aggregation and processing}

\newduneword{beamline}{beamline}{A sequence of control and monitoring devices used for the formation of a directed collection of particles; a subproject within \gls{lbnf}} 

\newduneword{cdr}{CDR}{Depending on context, either ``conceptual design report,'' a formal project  document  that describes the experiment at a conceptual level, or ``conceptual design review,'' a formal review of the conceptual design of the experiment or of a component}  

\newduneabbrev{cf}{CF}{conventional facilities}{Pertaining to
  construction and operation of buildings and conventional infrastructure, and for the \gls{lbnf-dune}, CF includes the excavation caverns}

\newduneabbrev{cp}{CP}{charge conjugation and parity}{Product of charge conjugation and parity
  transformations} 

\newduneabbrev{cpt}{CPT}{charge, parity, and time reversal symmetry}{product of charge, parity
  and time-reversal transformations}

\newduneabbrev{cpv}{CPV}{charge-parity symmetry violation}{Lack of
  symmetry in a system before and after charge and parity
  transformations are applied. 
  For CP symmetry to hold,  a particle turns into its
 corresponding antiparticle under a charge transformation, and a parity
transformation inverts its space coordinates, i.e. produces the mirror image}

\newduneword{doe}{DOE}{U.S. Department of Energy}

\newduneabbrev{fra}{FRA}{Fermi Research Alliance}{A joint partnership of the University of Chicago and the Universities Research Association (URA) that manages and operates Fermilab on behalf of the \gls{doe}}

\newduneabbrev{dune}{DUNE}{Deep Underground Neutrino Experiment}{A leading-edge, international experiment for neutrino science and proton decay studies}

\newduneabbrev{esh}{ES\&H}{environment, safety and health}{A discipline and specialty that studies and implements practical aspects of environmental protection and safety at work} 

\newduneabbrev{ppe}{PPE}{personnel protective equipment}{Equipment worn to minimize exposure to hazards that cause serious workplace injuries and illnesses}

\newduneabbrev{odh}{ODH}{oxygen deficiency hazard}{a hazard that occurs when inert gases such as nitrogen, helium, or argon displace room air and thus reduce the percentage of oxygen below the level required for human life}

\newduneabbrev{feshm}{FESHM}{Fermilab Environment, Safety and Health Manual}{The document that contains Fermilab's policies and procedures designed to manage environment, safety, and health in all its programs}

\newduneabbrev{fscf}{FSCF}{far site conventional facilities}{The
  \gls{cf} at the DUNE far detector site, \gls{surf}}
  
\newduneabbrev{nscf}{NSCF}{near site conventional facilities}{The
  \gls{cf} at the DUNE near detector site, \gls{fnal}}

\newduneabbrevs{gut}{GUT}{grand unified theory}{grand unified theories}{A class of theories that unifies the electroweak and strong forces}

\newduneabbrev{lar}{LAr}{liquid argon}{Argon in its liquid phase; it is a cryogenic liquid with a boiling point of \SI{87}{K} and density of \SI{1.4}{g/ml}}

\newduneabbrev{lbl}{LBL}{long-baseline}{Refers to the distance between the 
  neutrino source  and the \gls{fd}.  It can also refer to the distance between the near and far detectors. 
  The ``long'' designation is an approximate and relative distinction. For DUNE, this distance  (between \gls{fnal} and \gls{surf}) is approximately \SI{1300}{km}}

\newduneabbrev{lbnf}{LBNF}{Long-Baseline Neutrino Facility}{The project scope and  
  organizational entity responsible for developing the neutrino beam, the cryostats
  and cryogenics systems, and the conventional facilities for DUNE} 
  
\newduneabbrev{lbnf-dune}{LBNF/DUNE}{LBNF and DUNE enterprise}{The overall enterprise including the LBNF/DUNE-US \gls{doe} project, other international contributing projects, and the \gls{dune} collaboration and experiment}

\newduneabbrev{lbnc}{LBNC}{Long-Baseline Neutrino Committee}{The committee, composed of internationally prominent scientists with relevant expertise, charged by the \gls{fnal} director to review the scientific, technical, and managerial progress, plans and decisions associated with \gls{dune}}

\newduneabbrev{ncg}{NCG}{Neutrino Cost Group}{A group of internationally prominent scientists with relevant experience that is charged by the \gls{fnal} director to review the cost, schedule, and associated risks for the \gls{dune} experiment}

\newduneabbrev{mh}{MH}{mass hierarchy}{Describes the separation
  between the mass squared differences related to the solar and
  atmospheric neutrino problems (also written as \gls{mo})}

\newduneabbrev{mo}{MO}{mass ordering}{See \gls{mh}}

\newduneabbrev{mi}{MI}{Fermilab Main Injector}{An accelerator at
  \gls{fnal} that provides a beam of high-energy protons that upon
  striking a target produce secondaries that decay to provide the
  neutrinos directed toward the DUNE far detector}

\newduneabbrev{pot}{POT}{protons on target}{Typically used as a unit
  of normalization for the number of protons striking the neutrino
  production target}

\newduneabbrev{qa}{QA}{quality assurance}{The process of ensuring that 
the quality of each element meets requirements during design and development, and to detect and correct poor results prior to production} 

\newduneabbrev{qc}{QC}{quality control}{The process (e.g., inspection, testing, measurements) 
of ensuring that each manufactured element meets its quality requirements prior to assembly or installation} 

\newduneabbrev{sm}{SM}{Standard Model}{Refers to a theory describing
  the interaction of elementary particles}

\newduneword{tdr}{TDR}{Depending on context, either ``technical design report,'' a formal project  document  that describes the experiment at a technical level, or ``technical design review,'' a formal review of the technical design of the experiment or of a component}  

\newduneabbrev{tp}{IDR}{interim design report}{An intermediate
milestone on the path to a full \gls{tdr}} 

\newduneabbrev{ckm}{CKM matrix}{Cabibbo-Kobayashi-Maskawa
  matrix}{Refers to the matrix describing the mixing between mass and
  weak eigenstates of quarks}

\newduneabbrev{cl}{CL}{confidence level}{Refers to a probability
  used to determine the value of a random variable given its
  distribution}

\newduneabbrev{pmns}{PMNS}{Pontecorvo-Maki-Nakagawa-Sakata}{A type of matrix that describes the mixing between mass and weak eigenstates of
  the neutrino}

\newduneabbrevs{cpa}{CPA}{cathode plane assembly}{cathode plane assemblies}{The component of the \gls{sp} detector module that provides the drift HV cathode}

\newduneabbrev{fc}{FC}{field cage}{The component of a \gls{lartpc} that contains and shapes the applied \efield}

\newduneword{cpafc}{CPA/FC}{A pair of \gls{cpa} panels and the top and bottom \gls{fc} portions that attach to the pair; an intermediate assembly for installation into the \gls{spmod} }

\newduneabbrev{topfc}{top FC}{top field cage}{The horizontal portions of the \gls{sphd} \gls{fc}   on the top of the \gls{tpc}}

\newduneabbrev{botfc}{bottom FC}{bottom field cage}{The horizontal portions of the \gls{sphd} \gls{fc} on the bottom of the \gls{tpc}}

\newduneabbrev{ewfc}{endwall FC}{endwall field cage}{The vertical portions of the \gls{fc} near the end walls}

\newduneabbrev{gp}{GP}{ground plane}{An electrode held electrically neutral relative to Earth ground voltage; it is mounted on the \gls{fc} in a \gls{spmod} to protect the cryostat wall}

\newduneword{gg}{ground grid}{An electrode held electrically neutral relative to Earth ground voltage; it is installed between the cathode and the \glspl{pd} in a \gls{dpmod} to protect the \glspl{pmt}, maintaining high transparency to light}

\newduneabbrev{alara}{ALARA}{as low as reasonably
  achievable}{Typically used with regard management of radiation
  exposure but may be used more generally. It means making every
  reasonable effort to maintain e.g., exposures, to as far below the
  limits as practical, consistent with the purpose for that the
  activity is undertaken}

\newduneabbrev{ecal}{ECAL}{electromagnetic calorimeter}{A detector
  component that measures energy deposition of traversing particles (in the near detector conceptual design)}

\newduneabbrev{hv}{HV}{high voltage}{Generally describes a voltage
  applied to drive the motion of free electrons through some media, e.g., LAr}

\newduneword{spmod}{SP module}{single-phase DUNE \gls{fd} module}
\newduneword{dpmod}{DP module}{dual-phase DUNE \gls{fd} module}
\newduneword{dsp}{DUNE-SP}{a single-phase DUNE far detector module} 
\newduneword{ddp}{DUNE-DP}{a dual-phase DUNE far detector module}

\newduneabbrev{tcoord}{TC}{technical coordinator}{A member of the \gls{dune} management team responsible for organizing the technical aspects of the project effort; is head of \gls{tc}}

\newduneabbrev{tc}{TCN}{technical coordination}{The DUNE organization responsible for overall integration of the detector elements and successful execution of the detector construction project; areas of responsibility include general project oversight, systems engineering, \gls{qa} and safety}  

\newduneabbrev{exb}{EB}{executive board}{The highest level DUNE
  decision-making body for the collaboration}

\newduneabbrev{tb}{TB}{technical board}{The DUNE organization responsible for
  evaluating technical decisions}

\newduneabbrev{rrb}{RRB}{Resources Review Board}{A part of \gls{dune}'s international project governance structure, composed of representatives of all funding agencies that sponsor the project, and of  \gls{fnal} management, established to provide coordination among funding partners and oversight of \gls{dune}}

\newduneabbrev{inc}{INC}{International Neutrino Council}{A highest-level international advisory body to the U.S. \gls{doe} and the  \gls{fnal} directorate on matters related to the  \gls{lbnf} and the  \gls{pip2} projects. This council is composed of representatives from the international funding agencies and  \gls{cern} that make major contributions the infrastructure}

\newduneabbrev{cc}{CC}{charged current}{Refers to an interaction
  between elementary particles where a charged weak force carrier
  ($W^+$ or $W^-$) is exchanged}

\newduneabbrev{dis}{DIS}{deep inelastic scattering}{Refers to interaction between
  elementary particles and a nucleus in an energy range where the
  interaction can be modeled as occurring between constituent quarks
  of one nucleon and resulting in no bulk recoil of the resulting
  nucleus} 

\newduneabbrev{fsi}{FSI}{final-state interactions}{Refers to
  interactions between elementary or composite particles subsequent to
  the initial, fundamental particle interaction, such as may occur as
  the products exit a nucleus}
  
\newduneabbrev{fsint}{FSI}{far site integration}{The scope of work at the \gls{fs} for the \gls{integoff}} 

\newduneword{geant4}{Geant4}{A
  software toolkit for the simulation of the passage of particles
  through matter using \gls{mc} methods}

\newduneabbrev{genie}{GENIE}{Generates Events for Neutrino Interaction
  Experiments}{Software providing an object-oriented neutrino
  interaction simulation resulting in kinematics of the products of
  the interaction}

\newduneabbrev{mc}{MC}{Monte Carlo}{Refers to a method of numerical
  integration that entails the statistical sampling of the integrand
  function. 
  Forms the basis for some types of detector and physics simulations}

\newduneabbrev{qe}{QE}{quasi-elastic}{Refers to the 
  interaction of an elementary charged particle with a nucleus in an
  energy range where the interaction can be modeled as taking place with
  individual nucleons} 

\newduneabbrev{mou}{MoU}{memorandum of understanding}{A project management methodology that documents an agreement between \gls{fnal} and the LBNF/DUNE-US Project for how Fermilab will support the project. More generally, a document
  summarizing an agreement between two or more parties} 

\newduneabbrev{pip2}{PIP-II}{Proton Improvement Plan II}{A \gls{fnal} project for
  improving the protons on target delivered delivered by the \gls{lbnf} neutrino production beam. 
  This is version two of this plan and it is planned to be followed by a PIP-III}
  
\newduneabbrev{sdsta}{SDSTA}{South Dakota Science and Technology
  Authority}{The legal entity that manages \gls{surf}, in Lead, S.D}
  
\newduneabbrev{sdsd}{SDSD}{Fermilab South Dakota Services Division}{A Fermilab division responsible providing host laboratory functions at SURF in South Dakota}

\newduneabbrev{firus}{FIRUS}{Facility Information Reporting Utility System}
 {Facility incident reporting systems, one at \gls{fnal} and at \gls{surf}, that monitors and reports the status of various fire, security and utility sensors} 

\newduneabbrev{bsi}{BSI}{building and site infrastructure}
 {The work package for outfitting of the \gls{lbnf} underground infrastructure}

\newduneabbrev{wbs}{WBS}{work breakdown structure}{An organizational
  project management tool by which the tasks to be performed are
  partitioned in a hierarchical manner}

\newduneabbrev{br}{BR}{branching ratio}{A fractional probability for a
  decay of a composite particle to occur into some specified set or
  sets of products}
\newduneword{bsm}{BSM}{beyond the Standard Model}

\newduneabbrev{dm}{DM}{dark matter}{The term given to the unknown
  matter or force that explains measurements of galaxy motion 
  that are otherwise inconsistent with the amount of mass associated
  with the observed amount of photon production}
  
\newduneabbrev{bdm}{BDM}{boosted dark matter}{A new model that describes a relativistic dark matter particle boosted by the annihilation of heavier dark matter particles in the galactic center or the sun}

\newduneabbrev{cern}{CERN}{European Laboratory for Particle Physics}{The leading particle physics laboratory in Europe and home to the ProtoDUNEs} 

\newduneabbrev{dsnb}{DSNB}{diffuse supernova neutrino background}{The
  term describing the pervasive, constant flux of neutrinos due to all
  past supernova neutrino bursts}

\newduneabbrev{espp}{ESPP}{European Strategy for Particle Physics}{The
cornerstone of Europe's
decision-making process for the long-term future of the
field. Mandated by the \gls{cern} Council, it is formed through a broad
consultation of the grass-roots particle physics community, it
actively solicits the opinions of physicists from around the world,
and it is developed in close coordination with similar processes in
the USA and Japan in order to ensure coordination between regions and
optimal use of resources globally}

\newduneabbrev{gar}{GAr}{gaseous argon}{argon in its gas phase}
\newduneabbrev{gartpc}{GArTPC}{gaseous argon time-projection chamber}{A \gls{tpc} filled with gaseous argon} 

\newduneabbrev{globes}{GLoBES}{General Long-Baseline Experiment
  Simulator}{A software package for simulating energy spectra of
  neutrino flux, interactions, and energy spectra measured after application of some
  model of a detector response)}

\newduneabbrev{snowglobes}{SNOwGLoBES}{SuperNova
Observatories with GLoBES} {From the official description: SNOwGLoBES is public software for computing interaction rates and distributions of observed quantities for \gls{snb} neutrinos in common detector materials} 

\newduneword{l/e}{L/E}{length-to-energy ratio}
\newduneword{lri}{LRI}{long-range interactions}

\newduneabbrev{nc}{NC}{neutral current}{Refers to an interaction
  between elementary particles where a neutrally charged weak force carrier
  ($Z^0$) is exchanged}

\newduneabbrev{nh}{NH}{normal hierarchy}{Refers to the neutrino mass
  eigenstate ordering whereby the sign of the mass squared difference
  associated with the atmospheric neutrino problem is positive}

\newduneabbrev{ih}{IH}{inverted hierarchy}{Refers to the neutrino mass
  eigenstate ordering whereby the sign of the mass squared difference
  associated with the atmospheric neutrino problem is negative}

\newduneabbrev{no}{NO}{normal ordering}{Refers to the neutrino mass
  eigenstate ordering whereby the sign of the mass squared difference
  associated with the atmospheric neutrino problem is positive}

\newduneabbrev{io}{IO}{inverted ordering}{Refers to the neutrino mass
  eigenstate ordering whereby the sign of the mass squared difference
  associated with the atmospheric neutrino problem is negative}

\newduneabbrev{msw}{MSW}{Mikheyev-Smirnov-Wolfenstein effect}{Explains
  the oscillatory behavior of neutrinos produced inside the sun as
  they traverse the solar matter}

\newduneabbrev{nsi}{NSI}{nonstandard interaction}{A general class of
  theory of elementary particles other than the Standard Model}

\newduneabbrev{pfive}{P5}{Particle Physics Project Prioritization
Panel}{The Particle Physics Project Prioritization Panel (P5) was a
subpanel of the High Energy Physics Advisory Panel (HEPAP). It completed
its Report, a ten-year strategic plan for high energy physics in the
U.S., in 2014. This report included a recommendation that ``host a world-leading neutrino
program that will have an optimized set of short- and long-baseline neutrino oscillation experiments, and its long-term focus
is a reformulated venture referred to here as the Long Baseline
Neutrino Facility (LBNF)''}

\newduneabbrev{sme}{SME}{standard-model extension}{an effective field theory that contains the \gls{sm}, general relativity, and all possible operators that break Lorentz symmetry (Wikipedia)}

\newduneabbrev{susy}{SUSY}{supersymmetry}{Theoretical symmetry between a fermion and a boson}

\newduneabbrev{wimp}{WIMP}{weakly-interacting massive particle}{A
  hypothesized particle that may be a component of dark matter}

\newduneabbrev{ce}{CE}{cold electronics}{Analog and digital readout electronics that operate at cryogenic temperatures}

\newduneabbrev{crp}{CRP}{charge-readout plane}{In the \gls{dp} technology, a  collection of
  electrodes in a planar arrangement placed at a particular voltage
  relative to some applied \efield such that drifting electrons
  may be collected and their number and time may be measured}

\newduneabbrev{dram}{DRAM}{dynamic random access memory}{A computer memory technology}

\newduneabbrev{fnal}{Fermilab}
{Fermi National Accelerator Laboratory}{U.S. national laboratory in Batavia, IL. It is the laboratory that hosts \gls{dune} and serves as its near site}

\newduneabbrev{bnl}{BNL}{Brookhaven National Laboratory}{US national laboratory in Upton, NY}

\newduneabbrev{slac}{SLAC}{SLAC National Accelerator Laboratory}{US national laboratory in Menlo Park, CA}

\newduneabbrev{lbnl}{LBNL}{Lawrence Berkeley National Laboratory}{US national laboratory in Berkeley, CA}

\newduneabbrev{anl}{ANL}{Argonne National Laboratory}{US national laboratory in Lemont, IL}

\newduneabbrev{lanl}{LANL}{Los Alamos National Laboratory}{US national laboratory in Los Alamos, NM}

\newduneword{fs}{FS}{(1) The far site, \gls{surf}, where the DUNE far detector is located; (2) ``Full stream'' relates to a data stream that has not undergone selection, compression or other form of reduction} 

\newduneabbrev{lem}{LEM}{large electron multiplier}{A micro-pattern detector suitable for use in ultra-pure argon vapor; LEMs consist of copper-clad PCB boards with sub-millimeter-size holes through which electrons undergo amplification}

\newduneabbrev{lng}{LNG}{liquefied natural gas}{Pertaining to natural gas in its liquid phase}

\newduneabbrev{mip}{MIP}{minimum ionizing particle}{Refers to a
  particle traversing some medium such that the particle's mean energy loss is  
  near the minimum}

\newduneabbrev{pd}{PD}{photon detector}{The detector
  elements involved in measurement of the number and arrival times of
  optical photons produced in a detector module} 

\newduneabbrev{pmt}{PMT}{photomultiplier tube}{A device that makes use
  of the photoelectric effect to produce an electrical signal from the
  arrival of optical photons}

\newduneabbrev{ppm}{ppm}{parts per million}{A concentration equal to one part in $10^{-6}$}
\newduneabbrev{ppb}{ppb}{parts per billion}{A concentration equal to one part in $10^{-9}$}
\newduneabbrev{ppt}{ppt}{parts per trillion}{A concentration equal to one part in $10^{-12}$}

\newduneword{rio}{RIO}{reconfigurable input output}

\newduneabbrev{s/n}{S/N}{signal-to-noise}{signal-to-noise ratio}
\newduneword{snr}{\mbox{S/N}}{signal-to-noise ratio}

\newduneword{ssp}{SSP}{\gls{sipm} signal processor}

\newduneabbrev{sbn}{SBN}{Short-Baseline Neutrino}{A \gls{fnal} program consisting of three collaborations, \gls{microboone}, \gls{sbnd}, and \gls{icarus}, to perform sensitive searches for $\nue$ appearance and $\numu$ disappearance in the Booster Neutrino Beam}

\newduneword{stt}{STT}{straw tube tracker}

\newduneword{wire board}{wire board}{At the head end of the APA in the \gls{sphd} \gls{tpc}, stacks of electronics boards referred to as ``wire boards'' are arrayed to anchor the wires.  They also provide the connection between the wires and the cold electronics} 

\newduneabbrev{wls}{WLS}{wavelength-shifting}{A material or process by
  which incident photons are absorbed by a material and photons are
  emitted at a different, typically longer, wavelength}
  
\newduneabbrev{tpb}{TPB}{tetra-phenyl butadiene}{A \gls{wls} material}

\newduneabbrev{sft}{SFT}{signal feedthrough}{A cryostat penetration allowing for the passage of cables or other extended parts}

\newduneabbrev{sftchimney}{SFT chimney}{signal feedthrough chimney}{In the \gls{dp} technology, a volume above the cryostat penetration used for a signal feedthrough}

\newduneabbrev{catiroc}{CATIROC}{charge and time integrated readout chip}{A complete read-out chip manufactured in AustriaMicroSystem designed to read arrays of 16 photomultipliers}

\newduneabbrev{wr}{WR}{White Rabbit}{A component of the timing system that forwards clock signal and time-of-day reference data to the master timing unit}

\newduneabbrev{mch}{MCH}{MicroTCA Carrier Hub}{An network switching device}

\newduneabbrev{wrmch}{WR-MCH}{White Rabbit \gls{utca} Carrier Hub}{A card mounted in \gls{utca} crate that recieves time syncronization information and trigger data packets over \gls{wr} network and disributes them to the \gls{amc} over \gls{utca} backplane} 

\newduneabbrev{wrtsn}{WR-TSN}{White Rabbit TimeStamping Node}{A unit on the \gls{wr} network that timestamps the trigger signals and sends out trigger data packets to \gls{wrmch}}

\newduneword{qap}{QAP}{quality assurance plan} 
\newduneword{ieshp}{IESHP}{integrated environmental, safety and health plan}
\newduneword{dmp}{DMP}{data management plan} 
\newduneword{qam}{QAM}{quality assurance manager} 

\newduneabbrev{dss}{DSS}{detector support system}{The system of rails suspended from the cryostat ceiling in a \gls{spmod} used to support the \gls{apa}s, \gls{cpa}s, and the \glspl{ewfc}} 

\newduneabbrev{ddss}{DDSS}{DUNE detector safety system}{The hardware system responsible for the safety of the detector, implemented either via a \gls{plc} or via custom hardware protections} 

\newduneabbrev{tco}{TCO}{temporary construction opening}{An opening in the side of a cryostat through which detector elements are brought into the cryostat; utilized during construction and installation}

\newduneabbrev{surf}{SURF}{Sanford Underground Research Facility}{The laboratory in South Dakota where the \gls{dune} \gls{fd} will be installed and operated; also where the \gls{lbnf} \gls{fscf} and the \gls{fs} cryostat and cryogenic systems will be constructed} 

\newduneabbrev{uit}{UIT}{underground installation team}{An organizational unit responsible for installation in the underground area at the \gls{surf} site}

\newduneabbrev{cmgc}{CMGC}{construction manager/general contractor}{The contracted company hired to manage overall construction, used by \gls{lbnf} at the \gls{surf} site for the \gls{fscf} construction} 

\newduneword{pdr}{PDR}{Depending on context, either ``preliminary design report,'' a formal project document  that describes the experiment at a preliminary level, or ``preliminary design review,'' a formal review of the preliminary design of the experiment or of a component} 

\newduneword{fdr}{FDR}{Depending on context, either ``final design report,'' a formal project document  that describes the experiment at a final level, or ``final design review,'' a formal review of the final design of the experiment or of a component} 

\newduneabbrev{prr}{PRR}{production readiness review}{A project management device by which the production readiness is reviewed}  

\newduneabbrev{irr}{IRR}{installation readiness review}{A project management device by which the plan for installation is reviewed}  
\newduneabbrev{orr}{ORR}{operational readiness review}{A project management device by which the operational readiness is reviewed}  

\newduneabbrev{ppr}{PPR}{production progress review}{A project management device by which the progress of production is reviewed}  

\newduneabbrev{edms}{EDMS}{engineering document management system}{A computerized document management system developed and supported at \gls{cern} in which some LBNF/DUNE documents, drawings and engineering models are managed} 
\newduneabbrev{docdb}{DocDB}{Document DataBase}{A computerized document management system developed and supported at \gls{fnal} in which virtually all LBNF and most DUNE documents are managed (docs.dunescience.org)}

\newduneword{wrgm}{WR grandmaster}{White Rabbit grandmaster}

\newduneabbrev{larsoft}{LArSoft}{Liquid Argon Software}{A shared base of physics software across \gls{lartpc} experiments}
\newduneword{nova}{NOvA}{The \gls{nova} off-axis neutrino oscillation experiment at \gls{fnal}}
\newduneword{minerva}{MINERvA}{Neutrino cross sections experiment at \gls{fnal}}
\newduneword{microboone}{MicroBooNE}{A \gls{lartpc} neutrino oscillation experiment at \gls{fnal}}
\newduneword{sbnd}{SBND}{The Short-Baseline Near Detector experiment at  \gls{fnal}}
\newduneabbrev{nexo}{nEXO}{Enriched Xenon Observatory}{Experiment at Lawrence Livermore National Laboratory (U.S. national lab in Livermore, CA) searching for new physics with neutrinoless double-beta decay}
\newduneword{argoneut}{ArgoNeuT}{The ArgoNeuT test-beam experiment and \gls{lartpc} prototype at  \gls{fnal}}
\newduneword{icarus}{ICARUS}{A neutrino experiment that was located at the Laboratori Nazionali del Gran Sasso (LNGS) in Italy, then refurbished at \gls{cern} for re-use in the same neutrino beam from \gls{fnal} used by the \gls{miniboone} , \gls{microboone} and \gls{sbnd} experiments. The ICARUS detector is being reassembled at \gls{fnal}}
\newduneword{atlas}{ATLAS}{One of two general-purpose detectors at the \gls{lhc}. It investigates a wide range of physics, from the measurements of the Higgs boson properties to searches for extra dimensions and particles that could make up \gls{dm}}

\newduneword{lbne}{LBNE}{Long Baseline Neutrino Experiment; (1) a terminated U.S. experiment that was reformulated in 2014 under the auspices of the new \gls{dune} collaboration, an internationally coordinated and internationally funded program, with \gls{fnal} as host; and (2) the former name of the DOE LBNF/DUNE Project} 

\newduneabbrev{lbno}{LBNO}{Long Baseline Neutrino Observatory} {A terminated European project that, during its six-year duration, assessed the feasibility of a next-generation deep underground neutrino observatory in Europe)}

\newduneword{wirecell}{Wire-Cell}{A tomographic automated \threed neutrino event reconstruction method for \glspl{lartpc}}
\newduneabbrev{wct}{WCT}{Wire-Cell Toolkit}{A software toolkit with data flow processing components for \gls{lartpc} noise and signal simulation, noise filtering, signal processing, and tomographic \threed ionization activity imaging}
\newduneword{ftslite}{F-FTS-lite}{Light-weight version of the \gls{fnal} File Transfer system used for rapid data transfers out of the online systems}
\newduneabbrev{fts}{FTS}{File Transfer System}{A file transfer system developed at \gls{fnal} to catalog and move data to permanent storage}

\newduneword{35t}{35 ton prototype}{A prototype cryostat and \gls{sp} detector built at \gls{fnal} before the \gls{protodune} detectors}

\newduneabbrev{mcr}{MCR}{main communications room}{Space at the \gls{fd} site for cyber infrastructure}

\newduneabbrev{cuc}{CUC}{central utility cavern}{The utility cavern at the 4850L of \gls{surf} located between the two detector caverns. It contains utilities such as central cryogenics and other systems, and the underground data center and control room}

\newduneabbrev{cfd}{CFD}{computational fluid dynamics}{High performance computer-assisted modeling of fluid dynamical systems}
\newduneword{vuv}{VUV}{vacuum ultra-violet}
\newduneword{tallbo}{TallBo}{A cylindrical cryostat at \gls{fnal} primarily used for developing scintillation light collection technologies for \gls{lartpc} detectors}

\newduneword{root}{ROOT}{A modular scientific software toolkit. It provides all the functionalities needed to deal with big data processing, statistical analysis, visualisation and storage. It is mainly written in C++ but integrated with other languages such as Python and R}

\newduneabbrev{eos}{EOS}{EOS}{The XRootD-based distributed file system developed by CERN}
\newduneabbrev{ehn1}{EHN1}{Experiment Hall North One}{Location at CERN of the ProtoDUNE experiments}
\newduneword{led}{LED}{Light-emitting diode}
\newduneabbrev{rtd}{RTD}{resistance temperature detector}{A temperature sensor consisting of a material with an accurate and reproducible resistance/temperature relationship}
\newduneword{swc}{SWC}{Software \& Computing}
\newduneabbrev{las}{LAS}{LEM-anode Sandwich}{In the \gls{dp} technology, a \gls{lem} and its corresponding anode are mounted together in a module called a LEM-anode sandwich}

\newduneword{roi}{ROI}{region of interest}
\newduneabbrev{hpc}{HPC}{high-performance computing}{high-performance computing facilities; generally computing facilities emphasizing parallel computing with aggregate power of more than a teraflop}

\newduneword{comfund}{common fund}{The shared resources of the collaboration}
\newduneabbrev{ims}{IMS}{integrated master schedule}{A project management device consisting of linked tasks and milestones}

\newduneword{hvdb}{HVDB}{HV divider board}

\newduneword{hvft}{HVFT}{HV feedthrough}  

\newduneword{sas}{SAS}{Another term for the materials airlock; a pass-through chamber used to ensure safe transfer of materials into a clean room, avoiding contamination in both directions}

\newduneabbrev{fea}{FEA}{finite element analysis}{Simulation of a physical phenomenon using the numerical technique called Finite Element Method (FEM), a numerical method for solving problems of engineering and mathematical physics}

\newduneword{fss}{FSS}{field shaping strips}
\newduneword{lvds}{LVDS}{low-voltage differential signaling}

\newduneword{esd}{ESD}{electrostatic discharge}

\newduneabbrev{rp}{RP}{resistive panel}{Resistive panels form the constant potential surfaces for a \gls{spmod} \gls{cpa}; they are composed of a thin layer of carbon-impregnated Kapton and laminated to both sides of a \frfour sheet}

\newduneword{uhmwpe}{UHMWPE}{ultra-high molecular weight polyethylene}

\newduneword{cts}{CTS}{Cryogenic Test System}

\newduneabbrev{plc}{PLC}{programmable logic controller}{An industrial digital computer that has been ruggedized and adapted for the control of manufacturing or other processes that require high reliability, ease of programming, and process fault diagnosis} 

\newduneword{mppc}{MPPC}{\SI{6}{mm}$\times$\SI{6}{mm} Multi-Pixel Photon Counters produced by Hamamatsu\texttrademark{} Photonics K.K}

\newduneabbrev{sfp}{SFP}{small form-factor pluggable}{a particular standard for optical transceivers}

\newduneabbrev{minipod}{MiniPOD}{miniature parallel optical device}{a family of types of multi-channel optical transceivers}

\newduneword{ccc}{CCC}{configuration change command}
\newduneword{act}{ACT}{activation time stamp}
\newduneword{lcm}{LCM}{light calibration module}
\newduneword{lpm}{LPM}{light pulser module}
\newduneword{dac}{DAC}{digital-to-analog converter}
\newduneword{arapuca}{ARAPUCA}{A \gls{pds} design that consists of a light trap that captures wavelength-shifted photons inside boxes with highly reflective internal surfaces until they are eventually detected by \gls{sipm} detectors or are lost}
\newduneword{sarapu}{S-ARAPUCA}{Standard \gls{arapuca} design with different \gls{wls} coatings on both faces of the dichroic filter window(s) of the cell}
\newduneword{xarapu}{X-ARAPUCA}{Extended \gls{arapuca} design with \gls{wls} coating on only the external face of the dichroic filter window(s) but with a \gls{wls} doped plate inside the cell}
\newduneword{feb}{FEB}{front-end board}

\newduneabbrev{lsnd}{LSND}{Liquid Scintilator Neutrino Detector}{A scintillation detector and associated experiment located at Los Alamos National Laboratory}

\newduneabbrev{cvn}{CVN}{convolutional visual network}{An algorithm for identifying neutrino interactions based on their topology and without the need for detailed reconstruction algorithms}

\newduneword{pandora}{Pandora}{The Pandora multi-algorithm approach to pattern recognition} 

\newduneabbrev{pma}{PMA}{Projection Matching Algorithm}{A reconstruction algorithm that combines \twod reconstructed objects to form a \threed representation}
\newduneabbrev{bdt}{BDT}{boosted decision tree}{A method of multivariate analysis}
\newduneabbrev{cnn}{CNN}{convolutional neural network}{A deep learning technique most commonly applied to analyzing visual imagery}
\newduneword{pdg}{PDG}{Particle Data Group}

\newduneword{pci}{PCI}{Peripheral Component Interconnect}

\newduneword{labview}{LabVIEW}{Laboratory Virtual Instrument Engineering Workbench is a system-design platform and development environment for a visual programming language from National Instruments}

\newduneword{pcb}{PCB}{printed circuit board}

\newduneword{crio}{cRIO}{Compact Reconfigurable Input Output}

\newduneword{dcs}{DCS}{Distributed Communications System}

\newduneword{opc-ua}{OPC-UA}{OPC  Unified Architecture is a machine to machine communication protocol for industrial automation developed by the OPC Foundation. OPC stands for Object Linking and Embedding for Process Control}

\newduneword{cabangle}{Cabibbo angle}{A quark mixing parameter that governs the coupling of up quarks to strange quarks}
\newduneword{valor}{VALOR}{A neutrino oscillation fitting framework that is used by \gls{t2k}; the name stands for VALencia-Oxford-Rutherford, the original three institutions that developed it}
\newduneword{cafana}{CAFAna}{Common Analysis File Analysis}
\newduneabbrev{pca}{PCA}{principal component analysis}{A statistical procedure that uses an orthogonal transformation to convert a set of observations of possibly correlated variables into a set of values of linearly uncorrelated variables called principal components (Wikipedia)}
\newduneword{numi}{NuMI}{a set of facilities at \gls{fnal}, collectively called ``Neutrinos at the Main Injector.''  The NuMI neutrino beamline target system converts an intense proton beam into a focused neutrino beam}
\newduneword{gibuu}{GiBUU}{Giessen Boltzmann-Uehling-Uhlenback Project; a unified theory and transport framework in the MeV and GeV energy regimes for elementary reactions on nuclei }
\newduneabbrev{rpa}{RPA}{random phase approximation} {an approximation method commonly used for describing the dynamic linear electronic response of electron systems (Wikipedia)}
\newduneword{t2k}{T2K}{T2K (Tokai to Kamioka) is a long-baseline neutrino experiment in Japan studying neutrino oscillations}
\newduneword{mptdet}{MPT detector}{multipurpose tracking detector}

\newduneword{lariat}{LArIAT}{The repurposed ArgoNeuT \gls{lartpc}, modified for use in a charged particle beam, dedicated to the calibration and precise characterization of the output response of these detectors}

\newduneword{captain}{CAPTAIN}{Experimental program sited at \gls{lanl} that is designed to make measurements of scientific importance to \gls{lbl} neutrino physics and physics topics that will be explored by large underground detectors}

\newduneword{dayabay}{Daya Bay}{a neutrino-oscillation experiment in Daya Bay, China, designed to measure the mixing angle $\Theta_{13}$  using antineutrinos produced by the reactors of the Daya Bay and Ling Ao nuclear power plants}

\newduneword{nuwro}{NuWro}{neutrino interaction generator}

\newduneabbrev{neut}{NEUT}{neutrino interaction generator}{A neutrino interaction simulation program library for the studies of atmospheric and accelerator neutrinos}

\newduneword{minos}{MINOS}{A long-baseline neutrino experiment, with a near detector at \gls{fnal} and a far detector in the Soudan mine in Minnesota, designed to observe the phenomena of neutrino oscillations (ended data runs in 2012)}

\newduneabbrev{efig}{EFIG}{Experimental Facilities Interface Group}{The body responsible for the required high-level coordination between the \gls{lbnf} and \gls{dune} projects}
\newduneword{ashriver}{Ash River}{The Ash River, Minnesota, USA \gls{nova} experiment far site, used as an assembly test site for \gls{dune}} 

\newduneword{ipd}{project integration director}{Responsible for integration and installation of DUNE detector deliverables. Manages the integration project} 

\newduneabbrev{jpo}{JPO}{Joint Project Office}{The framework through which team members from the LBNF project office, \gls{integoff}, and DUNE \gls{tc} work together to provide coherence in project support functions across the global enterprise. 
JPO functions include systems engineering, procurement, \gls{esh}, \gls{qa}, finance, project controls, risk management, compliance, internal review, partner agreement management, document management, and administrative support} 

\newduneword{ifbeam}{IFbeam}{Database that stores beamline information indexed by timestamp}

\newduneabbrev{marley}{MARLEY}{Model of Argon Reaction Low Energy Yields}{Developed at UC Davis, MARLEY is the first realistic model of neutrino electron interactions on argon for enegies less than \SI{50}{MeV}. This includes the energy range important for \gls{snb} neutrinos and also solar 8--boron neutrinos}

\newduneabbrev{es}{ES}{elastic scattering}{Events in which a neutrino
elastically scatters off of another particle}

\newduneabbrev{cno}{CNO}{carbon nitrogen oxygen}{The CNO cycle (for carbon-nitrogen-oxygen) is one of the two known sets of fusion reactions by which stars convert hydrogen to helium, the other being the proton-proton chain reaction (pp-chain reaction). In the CNO cycle, four protons fuse, using carbon, nitrogen, and oxygen isotopes as catalysts, to produce one alpha particle, two positrons and two electron neutrinos}

\newduneabbrev{sdwf}{SDWF}{South Dakota Warehouse Facility}{Warehousing operations in South Dakota responsible for receiving LBNF and DUNE goods and coordinating shipments to the access shaft (Ross Shaft) at \gls{surf}}

\newduneabbrev{wms}{WMS}{warehouse management system}{Commercial software package used to track shipments and interface to freight forwarders. This includes a database for shipping}

\newduneabbrev{dcdb}{DCDB}{DUNE construction database}{Database used by DUNE to track the history and testing of all parts of each \gls{detmodule}}

\newduneabbrev{aup}{AUP}{acceptance for use and possession}{Required for beneficial occupancy of the underground areas at \gls{surf} for \gls{lbnf} and \gls{dune}}

\newduneabbrev{bms}{BMS}{building management system}{A system provided by the \gls{cf} to manage the utility (cooling, ventilation, power, etc.) and fire/life safety systems. Separate systems are provided at \gls{surf} and at \gls{fnal}} 

\newduneabbrev{fls}{FLS}{fire and life safety system}{Fire and life safety; systems designed with \gls{cf} to meet building/safety code compliance for safe facilities at \gls{surf} and at \gls{fnal}} 

\newduneabbrev{sno}{SNO}{Sudbury Neutrino Observatory}{The Sudbury Neutrino Observatory was a detector built 6800 feet under ground, in INCO's Creighton mine near Sudbury, Ontario, Canada. SNO was a heavy-water Cherenkov detector designed to detect neutrinos produced by fusion reactions in the sun}

\newduneword{sk}{Super-Kamiokande}{Experiment sited in the Kamioka-mine, Hida-city, Gifu, Japan that uses a large water Cherenkov detector to study neutrino properties through the observation of solar neutrinos, atmospheric neutrinos and man-made neutrinos}

\newduneabbrev{id}{ID}{inner diameter}{Inner diameter of a tube}

\newduneabbrev{od}{OD}{outer diameter}{Outer diameter of a tube}

\newduneabbrev{rms}{RMS}{root mean square}{The square root of the arithmetic mean of the squares of a set of values, used as a measure of the typical magnitude of a set of numbers, regardless of their sign}

\newduneabbrev{orc}{ORC}{operational readiness clearance}{Final safety approval prior to the start of operation}

\newduneabbrev{gsc}{GSC group}{global safety coordination group}{DUNE group that evaluates applicable codes and standards, including international code equivalency, for the design, assembly, and installation of the \gls{fd}}

\newduneabbrev{ha}{HA}{hazard analysis}{A first step in a process to assess risk; the result of hazard analysis is the identification of the hazards present for a task or process}
\newduneword{har}{HAR}{hazard analysis report}

\newduneabbrev{tap}{TAP}{trip action plan}{A document required for any trip by a worker to the underground area at \gls{surf}, per that site's access control program; it describes the work to be accomplished during the trip} 

\newduneword{em}{EM}{emergency management}
\newduneword{ert}{ERT}{emergency response team}

\newduneabbrev{ndk}{NDK}{nucleon decay}{The hypothetical, baryon number violating decay of a proton or a bound neutron into lighter particles}

\newduneabbrev{emi}{EMI}{electromagnetic interference}{Disturbance generated by an external source that affects an electrical circuit by electromagnetic induction, electrostatic coupling, or conduction}

\newduneabbrev{pe}{PE}{photoelectron}{An electron ejected from the surface of a material by the photoelectric effect}

\newduneabbrev{spe}{SPE}{single photoelectron}{A single photoelectron}

\newduneabbrev{fwhm}{FWHM}{full width at half maximum}{Width of a distribution measured between those points at which the distribution is equal to half of its maximum amplitude}

\newduneabbrev{gdml}{GDML}{geometry description markup language}{An application-indepedent, geometry-description format based on XML}

\newduneabbrev{xml}{XML}{extensible markup language}{A markup language that defines a set of rules for encoding documents in a format that is both human-readable and machine-readable}

\newduneabbrev{crt}{CRT}{cosmic-ray tagger}{Detector external to the TPC designed to tag TPC-traversing cosmic ray particles}

\newduneabbrev{sn}{SN}{supernova}{Event that occurs upon the death of certain types of stars}

\newduneabbrev{wg}{WG}{working group}{A group of persons working together to achieve specified goals}

\newduneabbrev{ctsf}{CTSF}{coating, testing and storage facility}{A facility where the the \gls{dp} photon detectors will be coated, tested, and stored}

\newduneword{rucio}{Rucio}{Data management system originally developed
by \gls{atlas} but now open-source and shared across HEP}
\newduneabbrev{doma}{DOMA}{data organization, management, and access}{data organization, management, and access efforts through the HEP Software Foundation}

\newduneabbrev{hsf}{HSC}{HEP Software Foundation Collaboration}{A foundation that facilitates cooperation and common efforts in high energy physics software and computing internationally}

\newduneabbrev{wlcg}{WLCG}{Worldwide LHC Computing Grid}{Worldwide LHC
Computing Grid}
\newduneabbrev{osg}{OSG}{Open Science Grid}{Open Science Grid}
\newduneabbrev{sci}{SCI}{Scientific Computing Infrastructure}{Proposed
extension of the infrastructure component of \gls{wlcg} to other
experiments}
\newduneabbrev{csc}{CSC}{computing and software consortium}{DUNE
computing and software consortium}

\newduneword{dirac}{DIRAC}{Computing workflow management designed for LHCb and now used by many HEP experiments}

\newduneword{frp}{FRP}{fiber-reinforced plastic}
\newduneabbrev{hdpe}{HDPE}{high-density polyethylene}{A thermoplastic polymer made from petroleum commonly used to make plastic bottles}
\newduneword{hvps}{HVPS}{\gls{hv} power supply}
\newduneword{aisi}{AISI}{American Iron and Steel Institute}
\newduneword{ific}{IFIC}{Instituto de Fisica Corpuscular (in Valencia, Spain)}
\newduneabbrev{rsds}{RSDS}{radioactive source deployment system}{Proposed calibration system based on the deployment of
radioactive sources inside the \gls{dune} cryostat}
\newduneword{2p2h}{2p2h}{two particle, two hole}
\newduneabbrev{duneprism}{DUNE-PRISM}{\gls{dune} Precision Reaction-Independent Spectrum Measurement}{a mobile near detector that can perform measurements over a range of angles off-axis from the neutrino beam direction in order to sample many different neutrino energy distributions}
\newduneword{arcube}{ArgonCube}{The name of the core part of the \gls{dune} \gls{nd}, a \gls{lartpc}}

\newduneabbrev{citf}{CITF}{cryogenic instrumentation test facility}{A facility at \gls{fnal} with small ($<$\SI{1}{ton}) to intermediate ($\sim$\SI{1}{ton}) volumes of instrumented, purified TPC-grade \gls{lar}, used for testing devices intended for use in \gls{dune}}

\newduneabbrev{3dst}{3DST}{3D scintillator tracker}{The core part of the \threed projection scintillator tracker spectrometer in the near detector conceptual design}
\newduneabbrev{3dsts}{3DST-S}{3D scintillator tracker spectrometer}{The \threed projection scintillator tracker spectrometer  in the near detector conceptual design}
\newduneabbrev{mpd}{MPD}{multi-purpose detector}{A component of the near detector conceptual design; it is a magnetized system consisting of a \gls{hpgtpc} and a surrounding \gls{ecal}}
\newduneabbrev{hpg}{HPG}{high-pressure gas}{gas at high pressure to be used in a \gls{hpgtpc}} 
\newduneabbrev{hpgtpc}{HPgTPC}{high-pressure gaseous argon TPC}{A \gls{tpc} filled with gaseous argon; a possible component of the \gls{dune} \gls{nd}}

\newduneword{src}{SRC}{short-range correlated nucleon-nucleon interactions}
\newduneword{larpix}{LArPix}{ \gls{asic} pixelated charge readout for a \gls{tpc} }
\newduneword{arclt}{ArCLight}{a light detector for the \gls{arcube} effort}
\newduneword{fhc}{FHC}{forward horn current ($\numu$ mode)}
\newduneword{rhc}{RHC}{reverse horn current (\numubartonumubar mode)}
\newduneword{mwpc}{MWPC}{multi-wire proportional chamber}
\newduneword{na61}{NA61}{CERN hadron production experiment}
\newduneword{pdnd}{ProtoDUNE-ND}{a prototype \gls{dune} \gls{nd}}
\newduneabbrev{roc}{ROC}{readout chamber}{readout chamber for gaseous argon \gls{tpc}}
\newduneabbrev{iroc}{IROC}{inner readout chamber}{inner (radial) readout chamber for gaseous argon \gls{tpc}}
\newduneabbrev{oroc}{OROC}{outer readout chamber}{outer (radial) readout chamber for gaseous argon \gls{tpc}}

\newduneword{lux}{LUX}{Large Underground Xenon (LUX) dark matter detector at \gls{surf} }

\newduneword{mjdemo}{Majorana Demonstrator}{Experiment sited at \gls{surf} that  seeks to determine whether neutrinos are their own antiparticles}

\newduneword{lz}{LZ}{Experiment sited at \gls{surf} that  seeks to detect faint interactions between galactic dark matter and regular matter}

\newduneword{mu2e}{Mu2e}{An experiment sited at \gls{fnal} that searches for charged-lepton flavor violation and seeks to discover physics beyond the \gls{sm}}

\newduneword{pdsp2}{ProtoDUNE-SP-II}{A second test run in the singe-phase
ProtoDUNE test stand at CERN, acting as a validation of the final
single-phase detector design}  

\newduneword{osha}{OSHA}{Occupational Safety and Health Administration (USA Department of Labor) formed by the Occupational Safety and Health Act of 1970}
\newduneabbrev{pns}{PNS}{pulsed neutron source}{Calibration system based
on neutron capture gamma showers spread out in the whole detector}

\newduneabbrev{fv}{FV}{fiducial volume}{The detector volume within the \gls{tpc} that is selected for physics analysis through cuts on reconstructed event position}

\newduneword{p6}{P6}{framework used to plan and status the resource-loaded schedule of activities associated with the USA contributions to \gls{lbnf} and \gls{dune} }
\newduneabbrev{evms}{EVMS}{earned value management system}{Earned Value Management is a systematic approach to the integration and measurement of cost, schedule, and technical (scope) accomplishments on a project or task. It provides both the government and contractors the ability to examine detailed schedule information, critical program and technical milestones, and cost data (text from the US DOE); the EVMS is a system that implements this approach}

\newduneword{core}{CORE}{CORE contributions are in either monetary units or labor hours. They can be technical components for the facility or experiment and the effort of the staff needed to produce, install, and test them;  major facilities for the experiment; or other products and services relevant for the completion of the facility or experiment} 

\newduneabbrev{ahj}{AHJ}{Authority Having Jurisdiction}{An organization, office, or individual responsible for enforcing the requirements of a code or standard, or for approving equipment, materials, an installation, or a procedure (\gls{osha})}
\newduneword{cte}{CTE}{coefficient of thermal expansion}

\newduneabbrev{opc}{OPC}{open platform communications}{Open platform communications is a series of standards and specifications for industrial telecommunication} 
\newduneword{scada}{SCADA}{supervisory control and data acquisition}
\newduneword{ln}{LN$_2$}{liquid nitrogen}
\newduneabbrev{lapd}{LAPD}{Liquid Argon Purity Demonstrator}{Cryostat at Fermilab for long-term studies requiring a large volume of argon}

\newduneabbrev{pab}{PAB}{Proton Assembly Building}{Home of several \gls{lar} facilities at Fermilab}
\newduneword{hep}{HEP}{high energy physics}
\newduneword{cms}{CMS}{Compact Muon Solenoid experiment at CERN}
\newduneword{alice}{ALICE}{A Large Ion Collider Experiment, at CERN}
\newduneword{gpib}{GPIB}{general purpose interface bus}

\newduneabbrev{pfparticle}{PFParticle}{particle flow particle}{Each of the individual reconstructed particles in the hierarchy (or particle flow) describing the reconstructed event interaction}

\newduneabbrev{mcparticle}{MCParticle}{Monte Carlo Particle}{Individual true simulated particle}
\newduneword{au}{AU}{astronomical unit}
\newduneword{nufit}{NuFIT 4.0}{The NuFIT 4.0 global fit to neutrino oscillation data}

\newduneword{uhv}{UHV}{ultra high vacuum}
\newduneword{lps}{LPS}{laser positioning system}

\newduneword{unicamp}{UNICAMP}{University of Campinas, Sao Paulo, Brazil}
 
\newduneabbrev{fbk}{FBK}{Fondazione Bruno Kessler}{FBK is a research non-profit entity in Trento, Italy that partners in the development of technology with applications in various fields including High Energy Physics}

\newduneword{fft}{FFT}{fast Fourier transform}
\newduneabbrev{enob}{ENOB}{effective number of bits}{The effective number of bits is a measure of the dynamic range of an \gls{adc} and its associated circuitry. The resolution of an \gls{adc} is specified by the number of bits used to represent the analog value, in principle giving 2N signal levels for an N-bit signal. However, all real \gls{adc} circuits introduce noise and distortion. ENOB specifies the resolution of an ideal \gls{adc} circuit that would have the same resolution as the circuit under consideration}
\newduneabbrev{sndr}{SNDR}{signal to noise and distortion ratio}{Also known as SINAD. Ratio of the \gls{rms} signal amplitude to the mean value of the root-sum-square of all other spectral components, including harmonics, but excluding \gls{dc} levels. It is a good indication of the overall dynamic performance of an \gls{adc} because it includes all components which make up noise and distortion}
\newduneabbrev{sfdr}{SFDR}{spurious free dynamic range}{Spurious free dynamic range is the ratio of the \gls{rms} value of the signal to the \gls{rms} value of the worst spurious signal regardless of where it falls in the frequency spectrum. The worst spur may or may not be a harmonic of the original signal}
\newduneabbrev{thd}{THD}{total harmonic distortion}{Total harmonic distortion is the ratio of the \gls{rms} value of the fundamental signal to the mean value of the root-sum-square of its harmonics} 
\newduneword{tvs}{TVS}{transient voltage suppression}

\newduneword{riskprob}{risk probabilities}{The risk probability, after taking into account the planned mitigation activities, is ranked as 
 L (low $<\,$\SI{10}{\%}), 
M (medium \SIrange{10}{25}{\%}), or 
H (high $>\,$\SI{25}{\%}). 
The cost and schedule impacts are ranked as 
L (cost increase $<\,$\SI{5}{\%}, schedule delay $<\,$2 months), 
M (\SIrange{5}{25}{\%} and 2--6 months, respectively) and 
H ($>\,$\SI{20}{\%} and $>\,$2 months, respectively)}

\newduneabbrev{lbls}{LBLS}{laser beam location system}
{Auxiliary calibration system providing an independent location measurement of the ionization laser beams direction}

\newduneabbrev{lsst}{LSST}{Large Synoptic Survey Telescope}{8.4 m telescope with 3.2G-pixel camera that will start taking data in 2023}
\newduneabbrev{ska}{SKA}{Square Kilometer Array}{International radio telescope array planned to start data-taking in 2027}
\newduneabbrev{hyperk}{HyperK}{Hyper Kamiokande}{260 kt water Cerenkov neutrino detector to begin construction at Kamiokande in 2020}
\newduneword{lhcb}{LHCb}{LHC experiment dedicated to forward physics}
\newduneword{belleii}{Belle II}{B-factory experiment now running at KEK}

\newduneabbrev{ldm}{LDM}{light-mass dark matter}{Refers to dark matter particles with mass values much lower than the electroweak scale, specifically below the 1~GeV level}
 
\newduneabbrev{bnv}{BNV}{baryon-number violating}{Describing an interaction where \gls{baryonnumber} is not conserved}

\newduneword{bugey}{Bugey}{Neutrino experiment that operated at the Bugey nuclear power plant in France}

\newduneword{minosplus}{MINOS$+$}{The successor to the \gls{minos} experiment, utilizing the same detectors and beam line, but operating at higher beam energy tune than \gls{minos}, parasitic with \gls{nova}}

\newduneword{baryonnumber}{baryon number}{A quantity expressing the total number of baryons in a system minus the number of antibaryons}

\newduneword{np04}{NP04}{Experiment in the CERN North Area \gls{h4} hadron beamline;  \gls{pdsp}}  

\newduneword{np02}{NP02}{Experiment in the CERN North Area \gls{h2} hadron beamline;  \gls{pddp}}  

\newduneword{h4}{H4}{CERN North Area hadron beamline used for the \gls{pdsp} test beam run}  

\newduneword{h2}{H2}{CERN North Area hadron beamline used for the \gls{pddp} test beam run}  

\newduneword{ua1}{UA1}{UA1 (Underground Area 1) was a particle detector at \gls{cern}'s  Super Proton Synchrotron (SPS). It ran from 1981 until 1990, when the SPS was used as a proton-antiproton collider, searching for traces of W and Z particles in collisions. (CERN) The UA1 dipole magnet was reused in the NOMAD experiment and currently provides the magnetic field for the \gls{t2k} ND280 detector}

\newduneword{ssc}{SSC}{The Superconducting Super Collider was to be a huge underground ring complex beneath the area near Waxahachie, Texas, USA, that would have been the world’s most energetic particle accelerator. It was begun in 1990, but canceled by the U.S. Congress in 1993 (scientificamerican.com Oct 2013)}

\newduneword{daphne}{DAPHNE}{Detector electronics for Acquiring PHotons from NEutrinos is a custom-developed warm front-end waveform digitizing electronics module derived from the readout system developed at Fermilab for the Mu2e experiment}
 
\newduneword{nersc}{NERSC}{National Energy Research Computing Facility at \gls{lbnl}}

\newduneword{integoff}{integration project}{The \gls{doe} project element that organizes the onsite teams responsible for coordinating far detector installation and detector-facility integration activities at \gls{surf} as well as near detector installation activities at \gls{fnal}.  The integration project office includes overall LBNF/DUNE systems engineering, compliance and review offices} 

\newduneabbrev{sma}{SMA}{SubMiniature version A}{Connector interface for coaxial cables
with a screw-type coupling mechanism}

\newduneword{kloe}{KLOE}{KLOE is a $e^+ e^-$ collider detector spectrometer operated at DAFNE, the $\phi$-meson factory at Frascati, Rome. In DUNE it will consist of a \SI{26}{cm} Pb+scintillating fiber ECAL surrounding a cylindrical open detector region that is  \SI{4.00}{m} in diameter and \SI{4.30}{m} long. The ECAL and detector region are embedded in a \SI{0.6}{T} magnetic field created by a \SI{4.86}{m} diameter superconducting coil and a \SI{475}{tonne} iron yoke}

\newduneabbrev{ro}{RO}{review office}{An office within the \gls{integoff} that organizes reviews} 

\newduneabbrev{doecd}{CD}{critical decision}{The U.S. DOE's Order 413.3B outlines a series of staged project approvals, each of which is referred to as a critical decision (CD)}

\newduneabbrev{lbnfspac}{LBNF/DUNE-US SPAC}{LBNF/DUNE-US Strategic Project Advisory Committee}{A committee charged by the host laboratory director to provide expert, independent advice on significant issues and strategies related to LBNF/DUNE-US project organization, management, and risks} 

\newduneabbrev{sand}{SAND}{System for on-Axis Neutrino Detection}{The beam monitor component of the near detector that remains on-axis at all times and serves as a dedicated neutrino spectrum monitor}

\newduneword{4850l}{4850L}{The depth in feet (1480 m) of the access level for the DUNE underground area at SURF; called the ``4850 level''} 

\newduneword{apb}{APB}{authorship and publications board}
\newduneword{crb}{CRB}{collaboration resources board}
\newduneword{cube}{CuBe}{beryllium copper, used to make \gls{sphd} \gls{apa} readout planes}
\newduneword{drift}{drift}{(1) refers to electron drift under the influence of an electric field in a \gls{tpc}; (2) an excavated horizontal corridors (tunnels) in the underground areas at \gls{surf}}
\newduneword{exposure}{exposure}{The integrated detector fiducial mass times beam intensity; it is proportional to the number of interactions and is used to normalize cross sections in a data sample}
\newduneword{fr4}{FR-4}{Flame-retardant fiberglass-reinforced epoxy resin laminate used in making PCBs and other detector components}
\newduneword{g10}{G-10}{Non-flame-retardant fiberglass-reinforced epoxy resin laminate used in making PCBs}
\newduneword{kapton}{Kapton}{A polyimide plastic film that is stable over a broad range of temperatures and is resistant to radiation damage}
\newduneword{shaft}{shaft}{A vertical excavation at \gls{surf} connecting with the surface}
\newduneword{winze}{winze}{A vertical excavation at \gls{surf} connecting two drifts, not connecting to the surface}
\newduneword{ib}{IB}{institutional board}
\newduneword{irb}{IBR}{institutional board representative}
\newduneword{sc}{SC}{depending on context, either speakers committee or scientific computing} 
\newduneword{sac}{SAC}{spokespersons advisory committee}
\newduneword{ccondc}{CCC}{code of conduct committee}
\newduneword{eoc}{EOC}{education and outreach committee}
\newduneword{indico}{Indico}{Web-based meeting organization tool}
\newduneabbrev{htc}{HTC}{High Throughput Computing}{Computing facilities typically consisting of large numbers of commodity servers as opposed to a single large machine. Best suited for running large numbers of independent jobs in parallel, these facilities are what is usually meant by ``grid computing''}
\newduneword{dcache}{dCache}{A distributed, highly scalable (multi-PB) storage system, usable as both a standalone system and as a high-speed frontend to a tape storage system (such as /pnfs at Fermilab)}
\newduneabbrev{ifdhc}{IFDHC}{Intensity Frontier Data Handling Client}{A multi-protocol tool for data transfer and file delivery in jobs. It is able to automatically select transfer protocols based on source and destination characteristics}
\newduneabbrev{ifdh}{ifdh}{Intensity Frontier Data Handling}{The actual command invoked when using \gls{ifdhc}, on the command line, e.g. ifdh cp source\_file dest\_file}
\newduneabbrev{pnfs}{PNFS}{Pseudo Network File System}{A file system often used in large storage systems. Typically interaction is very similar to a regular NFS volume, but there can be some subtle and important differences}

\newduneword{nde}{NDE}{non-destructive evaluation} 
\newduneword{psv}{PSV}{pressure safety valve}
\newduneword{pickling}{pickling}{steel pickling and oiling is a metal surface treatment finishing process used to remove surface impurities such as rust and carbon scale from hot rolled carbon steel}

\newduneabbrev{tof}{ToF}{time of flight}{The time a particle takes to fly between two visible interactions observed in the detector. If combined with the distance traveled by the particle, for example a neutron, it can be used for energy reconstruction}

\newduneword{pep4}{PEP-4}{TPC for the Positron Electron Project 4 Collider at Stanford}

\newduneword{ndlar}{ND-LAr}{\gls{lartpc} component of the near detector based on \gls{arcube} technology}

\newduneword{ndgar}{ND-GAr}{component of the near detector with a core gaseous argon \gls{tpc} surrounded by an \gls{ecal} and a magnet}

\newduneword{sfgd}{SuperFGD}{Super Fine-Grained Detector (SuperFGD) is a 3D granular plastic scintillator detector that adopts the same technology as \dword{3dst}. It will be installed in the \dword{t2k} \dword{nd280} system \cite{Abe:2019whr}. The \dword{3dst} design will inherit in large part from the SuperFGD detector}

\newduneword{nd280}{ND280}{Near Detector 280, is the \dword{t2k} magnetized near detector \cite{Abe:2011ks}}

\newduneword{fee}{FEE}{front-end electronics}

\newduneword{mpgd}{MPGD}{MicroPattern Gas Detectors}

\newduneword{rmm}{RMM}{Resistive MicroMegas}

\newduneabbrev{stv}{STV}{Single Transverse Variables}{Kinematical variables obtained by projecting the neutrino interaction onto the transverse plane}

\newduneabbrev{ccqe}{CCQE}{charged current quasielastic interaction} {An interaction where a neutrino scatters from a nucleon, producing a charged lepton and converting a neutron to a proton or vice versa}

\newduneword{sis}{SIS}{shallow inelastic scattering}

\newduneabbrev{res}{RES}{resonant scattering}{The mode of scattering where the target nucleon is excited to a resonant state and decays, typically producing one or more pions}

\newduneabbrev{hadw}{W}{invariant mass of the hadronic system}{Refers to the invariant mass of the hadronic system formed during the neutrino scatter}

\newduneabbrev{agky}{AGKY}{Andreopoulos-Gallagher-Kehayias-Yang}{A model for hadronization of non-resonant inelastic neutrino reactions used in \gls{genie}. At low invariant hadronic masses, typically less than 2.3\,GeV/c$^2$, it is a KNO-inspired empirical model anchored on several bubble chamber measurements of neutrino-induced shower characteristics. For invariant hadronic masses between 2.3 and 3.0\,GeV/c$^2$, the model transitions linearly to a \gls{genie}-tuned version of PYTHIA, which is also used for the simulation of events at higher invariant masses} 

\newduneabbrev{clas}{CLAS}{CEBAF Large Acceptance Spectrometer}{A nuclear and particle physics detector located in the experimental Hall B at Jefferson Laboratory in Newport News, Virginia, United States. It is used to study the properties of the nuclear matter by the collaboration of over 200 physicists. Of particular relevance is its study of electron interactions with nuclei, including argon} 

\newduneabbrev{e4nu}{e4nu}{Electrons for Neutrinos}{A collaboration dedicated to using JLab's electron-scattering data to deliver improved neutrino-nucleus cross sections} 

\newduneabbrev{ldmx}{LDMX}{Light Dark Matter eXperiment}{The LDMX detector concept consists of a small precision tracker, and electromagnetic and hadronic calorimeters, all with near $2\pi$ azimuthal acceptance from the forward beam axis out to $\sim40^\circ$ angle. This detector would be capable of measuring correlations among electrons, pions, protons, and neutrons in electron-nucleus scattering at exactly the energies relevant for DUNE physics} 

\newduneabbrev{mec}{MEC}{meson-exchange currents} {An nuclear effect wherein pairs or larger groups of nucleons within a nucleus are bound together through the exchange of pions or other mesons. Neutrinos and other particles can scatter from these correlated pairs}

\newduneword{imt}{IMT}{Intranuclear momentum transfer}

\newduneword{miniboone}{MiniBooNE}{The Mini Booster Neutrino Experiment, at Fermilab, was designed to fully explore the LSND result}

\newduneabbrev{tms}{TMS}{Temporary Muon Spectrometer}{A muon spectrometer for the Near Detector that will be installed for the initial running period of DUNE, before the \gls{mpd} detector component is ready} 

\newduneabbrev{cvmfs}{CVMFS}{CERN VM File System}{A distributed file system designed for scalable, high-performance distribution of software to interactive and batch computers} 

\newduneword{kerberos}{Kerberos}{A strong authentication system used by the computing resources at Fermilab and CERN} 

\newduneabbrev{mrb}{MRB}{Multi Repository Build System}{A Fermilab-developed build system based on \dword{cmake} that allows development and builds of code from multiple repositories} 

\newduneword{cmake}{Cmake}{CMake is an open-source, cross-platform family of tools designed to build, test and package software}

\newduneabbrev{ups}{UPS}{UNIX product support}{A software tool that sets up a consistent environment of versions of pre-installed products and their dependencies on UNIX-like platforms} 

\newduneabbrev{upd}{UPD}{UNIX product distribution}{A tool for uploading and downloading pre-built software products between local systems and centralized software distribution servers.  UPD is not frequently used on DUNE because newer tools are more convenient} 

\newduneabbrev{sso}{SSO}{single sign-on}{Often used to indicate that a group of services, such as DocDB or the DUNE Wiki share common sign-in credentials and active sessions.  Fermilab services that say "Sign in with SSO username and password" mean to use your Services username and password} 

\newduneabbrev{vo}{VO}{virtual organization}{A database containing a list of member names, certificate distinguishing information, and a list of permissions members have to access computing grid and data resources} 

\newduneabbrev{nas}{NAS}{network attached storage}{Disk storage that is available on computers but shared between them.  Relies on NFS mounts rather than authenticated file transfer protocols.  Usually found on interactive servers to provide space for home directories, app and data storage} 

\newduneabbrev{nfs}{NFS}{network file system}{Industry-standard mechanism for mounting disks over a network.  Provides regular UNIX file and directory access} 

\newduneword{recombination}{recombination}{Electrons freed from Argon atoms will sometimes recombine with the positive argon ions, either the same ones from which they came or nearby ones.  Sometimes called ``quenching"} 

\newduneword{elife}{electron lifetime}{The attachment of electrons drifting through liquid argon to impurity molecules such as oxygen or water is parameterized by an exponential as a function of time with a time constant called the electron lifetime} 

\newduneword{gplane}{grid plane}{The uninstrumented plane of wires in a SP \gls{apa} that borders the drift volume.   It shapes the signals and provides \gls{esd} protection} 

\newduneword{gmesh}{grounding mesh}{A metal mesh attached to the SP \gls{apa} frame between the collection-plane wires and the space inside the frame where the \gls{pd} modules are installed.  It provides electric field uniformity so the collection-plane wires all have similar fields around them} 

\newduneabbrev{bcr}{BCR}{baseline change request}{A DOE project change, part of the change management system process}

\newduneword{ncav}{North Cavern}{the location of two of the planned four DUNE far detector modules at \gls{surf}}

\newduneword{scav}{South Cavern}{the location of two of the planned four DUNE far detector modules at \gls{surf}}

\newduneword{semp}{SEMP}{systems engineering management plan}  

\newduneabbrev{tpcost}{TPC}{total project cost}{The DOE terminology for the total budget and contingency for the entire LBNF/DUNE-US project from CD-0 to CD-4}  
\newduneabbrev{nsint}{NSI}{near site integration}{The scope of work at the near site for the \gls{integoff}} 

\newduneabbrev{opcost}{OPC}{other project costs}{The DOE project costs that support conceptual design, pre-operations commissioning, technical coordination, and power}

\newduneabbrev{moa}{MOA}{memorandum of agreement}{A project management methodology that documents an agreement between \gls{fnal} and the LBNF/DUNE-US Project for how Fermilab will support the project} 

\newduneabbrev{croc}{CROC}{central readout chamber}{central (radial) readout chamber for the ND \gls{gartpc}} 

\newduneabbrev{tki}{TKI}{transverse kinematic imbalance}{The imbalance among final-state particle momenta in the transverse plane to the neutrino direction; different aspects of the imbalance are sensitive to the detail of the nuclear effects in neutrino-nucleus interactions} 

\newduneabbrev{iandi}{I\&I}{Integration and Installation}{One of the three project areas in the LBNF/DUNE-US \gls{doe} project, along with LBNF and DUNE-US } 

\newduneword{hmi}{HMI}{human-machine interface}

\newduneword{lhe}{LHe}{liquid helium}

\newduneword{echain}{energy chain}{mechanical machine elements used to carry and guide power to moving parts of machines or structures, as required for \gls{duneprism} to carry power, data, and utilities to and from each movable near detector component at any arbitrary position along its travel path}

\newduneabbrev{sphd}{SPHD}{single-phase horizontal drift}{LArTPC design in which electrons drift horizontally to wire plane anodes (\glspl{apa}) that along with the front-end electronics are immersed in LAr}

\newduneabbrev{spvd}{SPVD}{single-phase vertical drift}{LArTPC design in which electrons drift vertically to PCB-based anodes at the top and bottom of the LAr volume, with a cathode in the middle}

\newduneabbrev{cru}{CRU}{charge-readout unit}{In the \gls{spvd} design an assembly of the \glspl{pcbp} plus adapter boards; four to a \gls{crp}} 

\newduneword{mpv}{MPV}{most probable value}

\newduneword{pcbp}{PCB panel}{In the \gls{spvd} design, one of four \SI{1.5x1.7}{m} assembled into a \gls{cru}}

\newduneword{anodepln}{anode plane}{a planar array of charge readout devices covering an entire face of a detector module}

\newduneword{msps}{MSPS}{megasamples per second}

\newduneword{gpsdo}{GPSDO}{\gls{gps} disciplined oscillator}

\newduneword{tdaq}{TDAQ}{trigger and DAQ system} 

\newduneword{nios}{NIOS}{Network Identity Operating System} 

\newduneabbrev{corsika}{CORSIKA}{COsmic Ray SImulations for KAscade}{a program for detailed simulation of extensive air showers initiated by high-energy cosmic ray particles}

\newduneabbrev{scd}{SCD}{Scientific Computing Division}{Fermilab's Scientific Computing Division}

\newduneabbrev{garsoft}{GArSoft}{Gaseous Argon Software}{A software toolkit similar to \gls{larsoft}, but targeted at the gaseous argon time projection chamber and calorimeter of \gls{ndgar}}

\newduneword{ndgarlite}{ND-GAr-Lite}{a temporary muon spectrometer consisting of the magnet and steel flux return of \gls{ndgar}, but with a simplified tracking chamber made with scintillating bars}

\newduneword{github}{GitHub}{a commerical web service providing code version management, storage, and browsing via \gls{git}}

\newduneword{git}{git}{a distributed version-control system, commonly used to manage software}

\newduneword{xrootd}{xrootd}{a high-performance data system widely used in HEP to store and to distribute data to jobs.  It allows streaming of data}

\newduneabbrev{gpuaas}{GPUAAS}{GPU As A Service}{a technique that allows many non-GPU-enabled compute nodes to share a GPU resource by sending it work over the network and waiting for results to be returned}

\hypersetup{
    pdftitle={\expshort TDR \thedocsubtitle},
    pdfauthor={\expshort Collaboration},
    final=true,
    colorlinks=false,
    linktocpage=true,
    linkbordercolor=blue,
    citebordercolor=green,
    urlbordercolor=magenta,
    filecolor=black,
    pdfpagemode=UseOutlines,
    pdfborderstyle={/S/U},  
}

\renewcommand\thedoctitle{DUNE Near Detector}

\begin{document}


\pagestyle{titlepage}

\begin{center}
   {\Huge  \thedocsubtitle}  

  \vspace{5mm}

  {\Huge  \tdrtitle}  

  \vspace{10mm}


    \vspace{5mm}



\titleextra

  \vspace{10mm}
  \today
    \vspace{15mm}
    
    {\large{The DUNE Collaboration}
    }
\end{center}

\cleardoublepage

\includepdf[pages={-}]{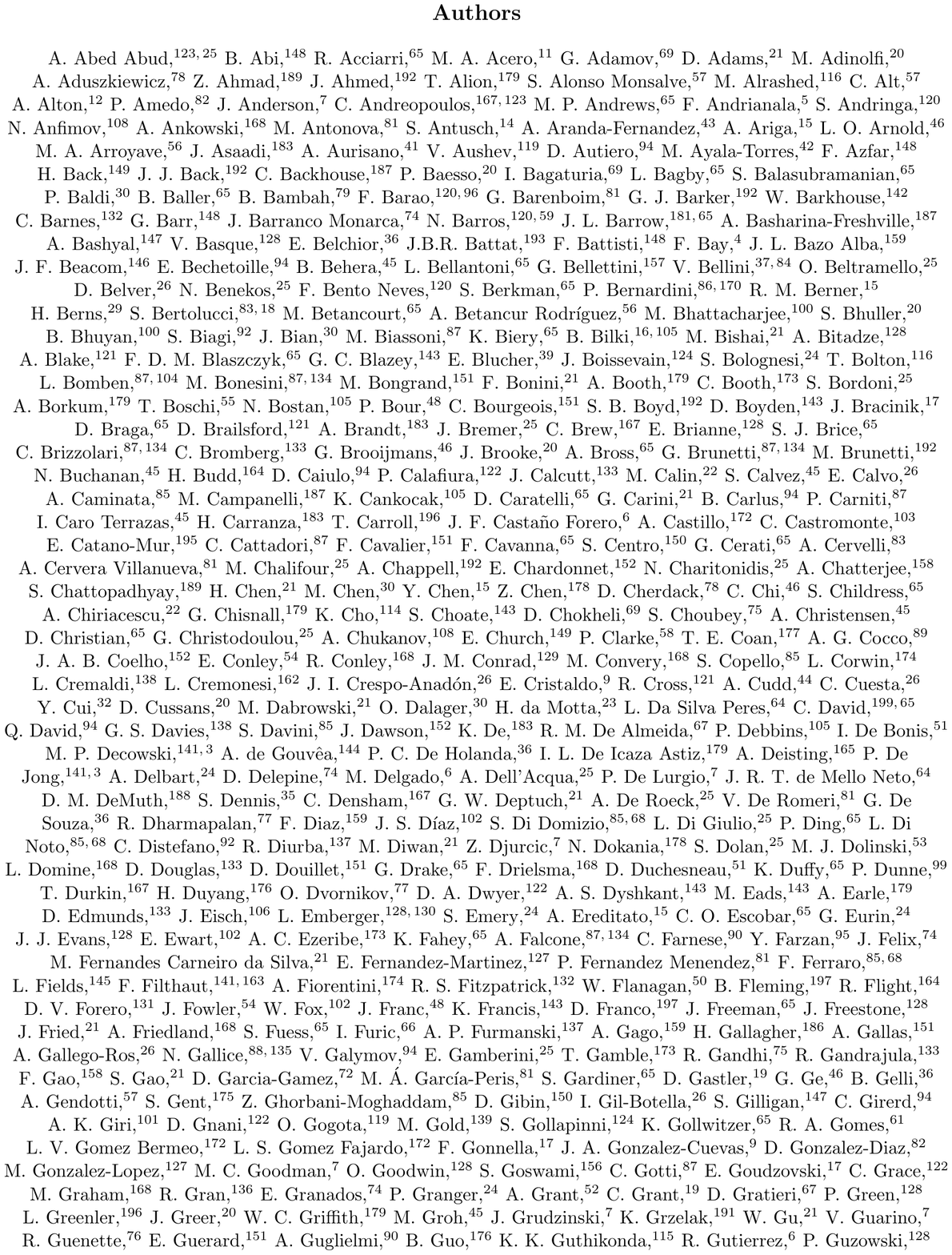}

\cleardoublepage

\renewcommand{\familydefault}{\sfdefault}
\renewcommand{\thepage}{\roman{page}}
\setcounter{page}{0}

\pagestyle{plain}


FERMILAB-PUB-21-067-E-LBNF-PPD-SCD-T

%
%
%
%
This document was prepared by the DUNE collaboration using the
resources of the Fermi National Accelerator Laboratory 
(Fermilab), a U.S. Department of Energy, Office of Science, 
HEP User Facility. Fermilab is managed by Fermi Research Alliance, 
LLC (FRA), acting under Contract No. DE-AC02-07CH11359.
%
%
This work was supported by
CNPq, FAPERJ, FAPEG and FAPESP,              Brazil;
CFI, IPP and NSERC,                          Canada;
CERN;
M\v{S}MT,	                                 Czech Republic;
ERDF, H2020-EU and MSCA,                     European Union;
CNRS/IN2P3 and CEA,                          France;
INFN,                                        Italy;
FCT,                                         Portugal;
NRF,                                         South Korea;
CAM, Fundaci\'{o}n ``La Caixa'', MICINN, GVA, Xunta de Galicia and AEI,  Spain;
SERI and SNSF,                               Switzerland;
T\"UB\.ITAK,                                 Turkey;
The Royal Society and UKRI/STFC,             United Kingdom;
DOE and NSF,                                 United States of America.
%
%
This research used resources of the 
National Energy Research Scientific Computing Center (NERSC), 
a U.S. Department of Energy Office of Science User Facility 
operated under Contract No. DE-AC02-05CH11231.

\textsf{\tableofcontents}

\textsf{\listoffigures}

\textsf{\listoftables}


\iffinal\else
\textsf{\listoftodos}
\clearpage
\fi

\renewcommand{\thepage}{\arabic{page}}
\setcounter{page}{1}

\pagestyle{fancy}

\renewcommand{\chaptermark}[1]{%
\markboth{Chapter \thechapter:\ #1}{}}
\fancyhead{}
\fancyhead[RO,LE]{\textsf{\footnotesize \thechapter--\thepage}}
\fancyhead[LO,RE]{\textsf{\footnotesize \leftmark}}

\fancyfoot{}
\fancyfoot[RO]{\textsf{\footnotesize Conceptual Design Report}}
\fancyfoot[LO]{\textsf{\footnotesize \thedoctitle}}
\fancypagestyle{plain}{}

\renewcommand{\headrule}{\vspace{-4mm}\color[gray]{0.5}{\rule{\headwidth}{0.5pt}}}




\chapter{Introduction and Executive Summary}
\label{ch:intro}


This document summarizes the motivation for and the current status of the design of the \dword{dune} \dword{nd} and accompanying infrastructure.
It is a precursor to the more detailed volume(s) that will make up the \dword{dune} \dword{nd} technical design report (TDR).
This work is done in the context of the \dword{dune} \dword{fd}, physics, and \dword{lbnf} volumes that document the basic \dword{dune} and \dword{lbnf} facilities and experimental configurations, as well as the overall scientific program of the experiment~\cite{Abi:2020wmh,Abi:2020evt,Abi:2020oxb,Abi:2020loh,Acciarri:2016crz,Acciarri:2015uup,Strait:2016mof,Acciarri:2016ooe}.


\dword{dune} will be a world-class, international particle physics experiment that aims to answer fundamental questions about the universe.  It is hosted by the U.S. Department of Energy's Fermi National Acceleratory Laboratory (Fermilab).  It consists of a  \dword{fd} to be located approximately 1.5~km underground at the \dword{surf} in South Dakota, at a distance of 1300~km from Fermilab, and a \dword{nd} that will be located on the Fermilab site in Illinois.  The \dword{fd} will consist of a modular, large, \dword{lartpc} with a total mass of 70~kt and a fiducial mass of roughly 40~kt.  The \dword{nd} is to be located approximately 574~m from the neutrino source for the \dword{lbnf} beam, which will be the world's most intense neutrino beam.  The \dword{nd} will consist of several different components described in detail in this volume: a highly modular \dword{lartpc}, a magnetized gaseous argon \dword{tpc}, and a large, magnetized beam monitor.

The scientific goals of \dword{dune} are described in detail in reference~\cite{Abi:2020evt}.  The driving goals are to:
\begin{itemize}
    \item Conduct a comprehensive program of neutrino oscillation measurements using the intense \dword{lbnf} (anti)neutrino beam;
    \item Search for proton decay in several decay modes;
    \item Detect and measure the \nue flux from a core-collapse supernova within our galaxy, should one happen during the lifetime of the experiment.
\end{itemize}
A rich program of ancillary science goals is enabled by the powerful \dword{lbnf} beam and the detectors that will comprise \dword{dune}. These include:
\begin{itemize}
    \item Other accelerator-based neutrino flavor transition measurements with sensitivity to \dword{bsm} phenomena;
    \item Measurements of neutrino oscillations using atmospheric neutrinos.
    \item Searches for dark matter;
    \item A rich program of neutrino interaction physics, including a wide range of measurements of neutrino cross sections and studies of nuclear effects.
\end{itemize}

Neutrino oscillation physics and several of the ancillary physics topics make use of the \dword{lbnf} beam.  This will be a 1.2~MW wideband neutrino beam with a corresponding protons-on-target of 1.1$\times$10$^{21}$, upgradable to multi-megawatt power.  The expected peak flux for \numu 's is roughly at 2.5 GeV.    

The neutrino oscillation program is the driving force behind the need for, and the design of, the \dword{dune} \dword{nd}.  This program includes measurements of the \dword{cp} violating phase, determination of the mass ordering of the neutrino mass eigenstates, measurement of the mixing angle $\theta_{23}$ and the octant in which it lies, and sensitive tests of the three-neutrino paradigm. The other physics goals listed above are exciting and will be pursued vigorously, but they are considered of secondary importance in terms of the \dword{nd} design.

The \dword{nd} plays many different roles in the oscillation program.
\begin{itemize}
\item The \dword{nd} makes a high-statistics characterization of the beam close to the source.  In the three-neutrino oscillation paradigm, this provides the initial state of the beam which is compared to the observations in the far detector to extract oscillation parameters.  
The use of a \dword{lartpc} in the \dword{nd} that is functionally similar to the \dword{fd} helps to reduce systematic uncertainties associated with detector and nuclear effects.  
\item The \dword{nd} includes a powerful spectral beam monitor that can be used to detect changes in the beam in a timely fashion.  The data are also useful for tuning the beam model and pinpointing the cause for changes in the beam.  Since the beam model is used to extrapolate observations in the \dword{nd} to the expected signal in the \dword{fd}, it is a source of uncertainty that needs to be constrained.
\item The high statistics collected in the \dword{nd}, as well as the similar-to-superior particle ID and kinematic phase space coverage relative to the \dword{fd}, make the \dword{nd} data 
extremely useful for tuning the neutrino interaction model used to move between the beam model and the observed data.  This tuning is an established, powerful technique for reducing the systematic errors in the extracted oscillation parameters.  These data also will provide critically important input for improving the neutrino interaction model which, in turn, can lead to reduced and/or better understood systematic uncertainties.  
\item The \dword{nd} will have the capability of taking data at different off-axis beam positions, which will provide data sets with different beam spectra.  This will allow \dword{dune} to deconvolve the beam and cross section models and constrain each separately.  This capability also provides a powerful handle for understanding the \dword{nd}  response matrix and allows the creation of  \dword{nd} data sets with  flux spectra very similar to the oscillated \dword{fd} fluxes, minimizing errors arising from the near-to-far flux difference, particularly those related to the neutrino interaction model.  
\end{itemize}

The characteristics and capabilities of the \dword{nd} are described in detail in this report.  Some of these characteristics and capabilities have a demonstrably straightforward and quantifiable effect on the \dword{cp} sensitivity of the experiment. Where possible, this is illustrated in this document.  In other cases, the connection is difficult to quantify.  Reasons for this can include: dependence on the understanding of the beam/detectors/models and details of the data taken at the time; the lack of finalized reconstruction algorithms; and imperfect modeling of some of the constraints used in the sensitivity fits. Even though the effects on dword{cp} sensitivity are difficult to quantify, these characteristics and capabilities are included in the design of the \dword{nd} because they are thought to be useful or essential based on the collaboration's collective experience on past and current experiments. 

There are several constituencies for this volume.  It aims to provide non-experts with the conceptual framework to understand why the near detector is necessary and the roles it plays in the context of the experiment as a whole.  It is meant to convince experts that the design is well motivated and likely to help \dword{dune} achieve its scientific goals.
Finally, this volume documents for collaborators and others the current status of the thinking behind, and design of, the \dword{dune} \dword{nd}.  This documentation will reflect the natural, and somewhat uneven, progression of the design across different elements of the \dword{nd}.
The breadth of this mission characterizes the volume and the reader is asked to forgive some variation in the level and tone of the text as it attempts to reach the different audiences.

Going forward, this chapter gives a rather detailed discussion of the motivation for (Sections~\ref{sec:intro-NeedForND} and \ref{sec:intro-OscillationRole}), and basic design of (Section~\ref{sec:intro-BriefOverview}), the \dword{dune} \dword{nd}.  Section~\ref{sec:intro-OscillationRole} relates how lessons learned from current experiments and past experience inform the design features and capabilities of the \dword{dune} \dword{nd}.  
The \dword{nd} requirements are discussed in Section~\ref{sec:intro-requirements}.  This part is intended largely to be used in support of  other sections of the report.  Finally, Section~\ref{sec:intro-organization} provides an overview of the organization of the \dword{dune} \dword{nd} management structure and decision making process.

\section{Need for the Near Detector}
\label{sec:intro-NeedForND}

 A key aim of the \dword{dune} experiment is to measure neutrino interaction rates from which can be extracted the oscillation probabilities for muon  (anti)neutrinos to either remain the same flavor or oscillate to electron (anti)neutrinos. 
Determining these probabilities as a function of the neutrino energy will allow for precision measurements of the free parameters of the \dword{pmns} matrix, as it is known in standard three-neutrino formalism.  
Of particular interest are the unmeasured sign of the atmospheric mass splitting (the so-called mass ordering) and the \dword{cp} violating phase, $\delta_{CP}$.
Measurements of the latter inconsistent with ${\rm sin}$($\delta_{CP}$) = 0 would indicate leptonic CP violation. Oscillation probability measurements inconsistent with the range of predictions allowed by \dword{pmns} formalism would be an indication of physics beyond the Standard Model.

The DUNE experiment will detect neutrinos generated in the \dword{lbnf} beamline at Fermilab~\cite{Acciarri:2016crz}. The \dword{nd} located near the neutrino source at Fermilab will measure the unoscillated neutrino interaction rate. The \dword{fd}, located 1300~km away, 
will measure the neutrino interaction rate after oscillations. A comparison of the measurements at the far and near detectors allows for the extraction of oscillation probabilities.

The role of the \dword{nd} is to serve as the experiment's control. The \dword{nd} establishes the null hypothesis (i.e., no oscillations) under the assumption of the three neutrino paradigm, measures and monitors the beam, constrains systematic uncertainties, and provides essential input for the neutrino interaction model.
The \dword{nd} measures the initial unoscillated \numu and \nue energy spectra, as well as those of the corresponding antineutrinos, $\bar{\nu}_\mu$ and $\bar{\nu}_e$. 
Measuring these spectra as a function of the neutrino energy is necessary as the oscillation probability depends on it.
 \footnote{Aspects of the language commonly used in neutrino physics, and in this document, can be confusing or ambiguous. 
\dword{dune} and \dword{lbnf} aim to achieve the goal set out for long-baseline neutrino oscillations in the U.S. Particle Physics Project Prioritization Panel (P5) report released in 2014~\cite{ParticlePhysicsProjectPrioritizationPanel(P5):2014pwa}: determine leptonic \dword{cp} violation with a precision of three standard deviations or better (i.e., a precision of 3\%), over more than 75\% of the range of possible values of the unknown CP violating phase $\delta_{CP}$ \footnote{In the standard parametrization.}. To achieve this goal, \dword{dune} will need to  pursue aggressively most available avenues to control and reduce the size of the systematic uncertainties encountered in the measurements of neutrino oscillation parameters.  This mitigation of  systematic uncertainty is a core factor considered in the design of the \dword{nd}. 
 \begin{itemize}
     \item The word ``neutrino'' is often used generically to be inclusive of both neutrinos and antineutrinos.  When the particle-antiparticle specificity is important in this document, it will be stated explicitly or obvious from context.
     \item Neutrino energy is not measured directly.  
     The neutrino energy is reconstructed from observed quantities.
     \item In experimental neutrino physics, it is common practice to refer to the neutrino energy (and spectra) when, in fact, what is meant is the reconstructed neutrino energy (spectra), along with all of the flux, cross section, and detector response complexities that implies. In this document, true neutrino energy will be referred to as true neutrino energy and neutrino energy, when standing alone, refers to reconstructed neutrino energy.
     \item Measurements of neutrino spectra include both the energy dependence (a.k.a, ``shape'') and the number of events (normalization) and are often generalized to include other kinematic quantities.
 \end{itemize}      
 }

To first order, a ``far/near'' ratio derived from the simulation can predict the unoscillated neutrino energy spectra at the \dword{fd} based on the \dword{nd} measurements.  The energy spectra at the \dword{fd} are sensitive to the oscillation parameters, which can be extracted via a fit.  The \dword{nd} plays a critical role in establishing what the oscillation signal spectrum should look like in the \dword{fd} because the expectations for the spectra (for a given set of oscillation parameters) are based on precisely measured spectra for $\nu_{\mu}$, $\overline{\nu}_{\mu}$, $\nue$, and $\anue$ interactions in the \dword{nd}.

To achieve the precision needed for \dword{dune}, the experiment must understand and minimize systematic uncertainties. With finite energy resolution and non-zero biases, the reconstructed energy spectrum is an unresolved convolution of cross section, flux, and energy response. The \dword{nd} must independently constrain each of those components and provide information that can be used to model  each component well. The acceptances of the \dword{nd} and the \dword{fd} differ.  The fluxes at the \dword{nd} and \dword{fd} differ due to both geometry and oscillations.  Models of the detectors, beam, and interactions must account for these things and fill in holes and biases left by imperfect understanding.  They are used to estimate the size of many systematic effects.  When imperfect models are not able to match observations, the \dword{nd} must provide the information needed to deal with that and estimate its impact.

In general, this requires that the \dword{nd} significantly outperform the \dword{fd}.
The \dword{nd} must have multiple methods for measuring neutrino fluxes with as much independence  from (or differing dependence on) the cross-section uncertainties as possible. With the necessity of relying on models, the \dword{nd} needs to measure neutrino interactions with much better detail than the \dword{fd}. This includes having a better detection efficiency across the kinematically-allowed phase space of all relevant reaction channels, superior identification of charged and neutral particles, better energy reconstruction, and better controls on experimental biases. The \dword{nd} must also have the ability to measure events in a similar way to the \dword{fd}, so that it can determine the ramifications of the more limited \dword{fd} performance, provide corrections, and take advantage of effects canceling to some extent in the extrapolation from the \dword{nd} to the \dword{fd}. At the same time, the \dword{nd} will operate in an environment with much higher event rates than the \dword{fd} and cannot take the form of a scaled copy of the \dword{fd}. Instead, the \dword{nd} must make measurements of interactions on liquid argon that mitigate the environmental difference so that they can be used confidently to predict the event rates in the \dword{fd}.

The conceptual design of the \dword{nd} is based on the collective experience of the many \dword{dune} collaborators who have significant roles in the current generation of neutrino experiments (\dword{minos}, MiniBooNE, \dword{t2k}, \dword{nova}, \dword{minerva}, and the Short-Baseline Neutrino [SBN] program).  These experiments have provided (and will provide) a wealth of useful data and experience that has led (will lead) to improved neutrino interaction models, as well as
powerful new analyses and reconstruction techniques,  They have also led to a deep appreciation of analysis pitfalls and a better understanding of the error budget. 
These experiments were all done with a lower precision, in a different energy range, or with very different detector technologies relative to \dword{dune}. While the existing and projected experience and data from those experiments will provide a strong base for \dword{dune}, it is not sufficient to enable \dword{dune} to accomplish its physics goals without a highly performing \dword{nd}.  

In addition to the mission described above, the \dword{dune} \dword{nd} will also have a physics program of its own, independent of the \dword{fd}, measuring Standard Model cross sections, as discussed in Chapter~\ref{ch_xsec:sec_xsec}. This cross-section program is coupled intimately to the oscillation measurement insofar as the cross sections will be useful as input to theory and model development and tuning.\footnote{Note that many of the \dword{nd} data samples, particularly those on argon targets, are incorporated into the oscillation analyses directly.}  Other Standard Model measurements, such as measuring the weak mixing angle and parton distribution functions, will also be pursued.  The \dword{dune} \dword{nd} will also be used to look for non-standard interactions, sterile neutrinos, dark photons, and other beyond the Standard Model particles and phenomena. The \dword{dune} \dword{nd} program of beyond the Standard Model physics is discussed more in Chapter~\ref{ch:bsm}. These are important aims that expand the physics impact of the \dword{nd} and the overall \dword{dune} program.

\section{Overview of the Near Detector}
\label{sec:intro-BriefOverview}

The \dword{dune} \dword{nd} has three primary detector components and the capability for two of those components to move off the beam axis. The three detector components serve important individual and overlapping functions with regard to the mission of the \dword{nd}.  Because these components have standalone features, the \dword{dune} \dword{nd} is often discussed as a suite or complex of detectors and capabilities.  The movement off axis provides a valuable extra degree of freedom in the data which is discussed in this report.  The power in the \dword{dune} \dword{nd} concept lies in the collective set of capabilities and the complementary information provided by the components.  
A drawing of the \dword{dune} \dword{nd} in the \dword{nd} hall is shown in Figure~\ref{fig:NDHallconfigs}.  

\begin{dunefigure}[DUNE ND hall; component detectors on- and off-axis]{fig:NDHallconfigs}
{Schematic of the \dword{dune} \dword{nd} hall shown with component detectors all in the on-axis configuration (left) and with the ND-LAr and ND-GAr in an off-axis configuration (right). The SAND detector is shown in position on the beam axis. The beam axis and direction is indicated. }
\includegraphics[width=0.49\textwidth]{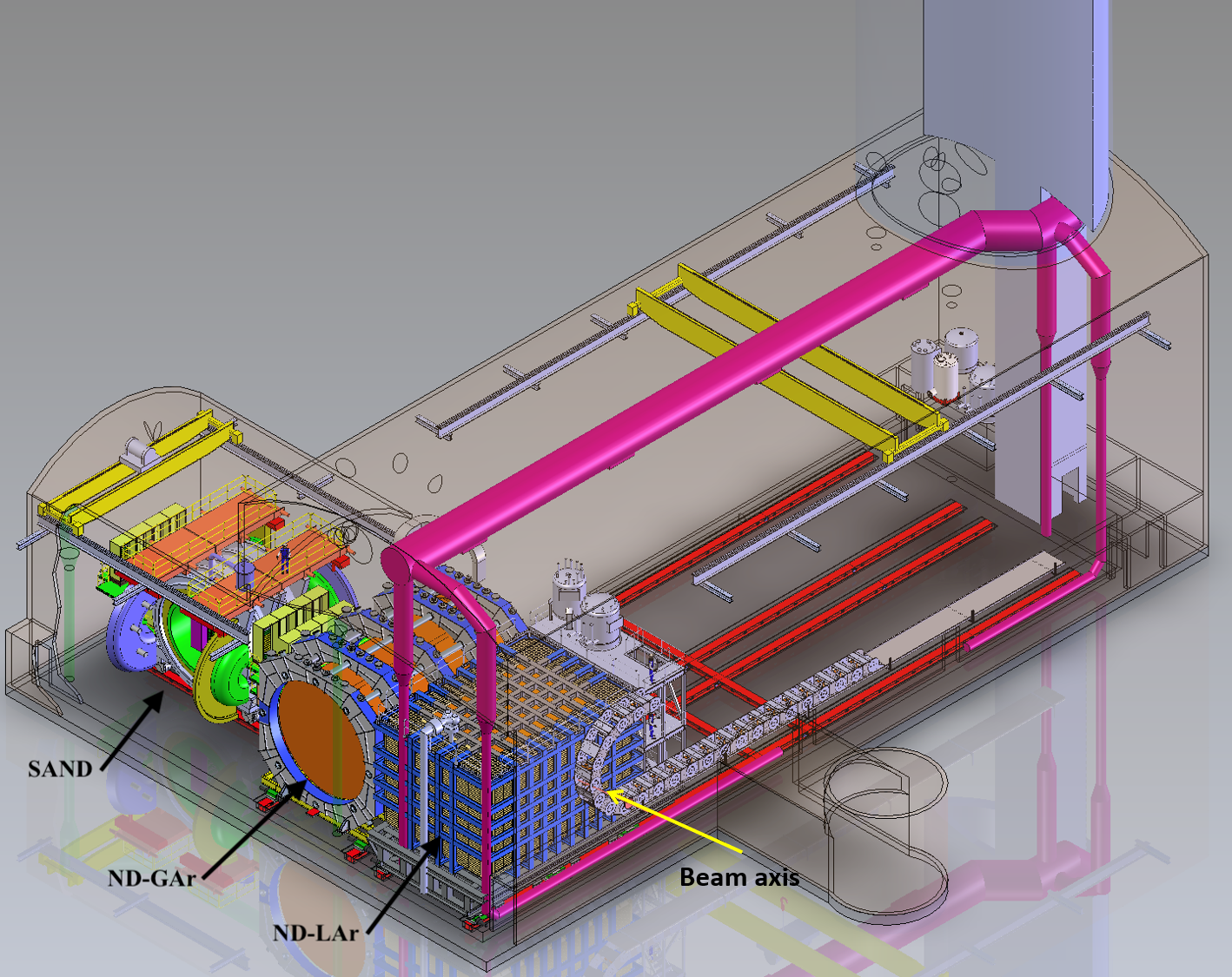}
\includegraphics[width=0.49\textwidth]{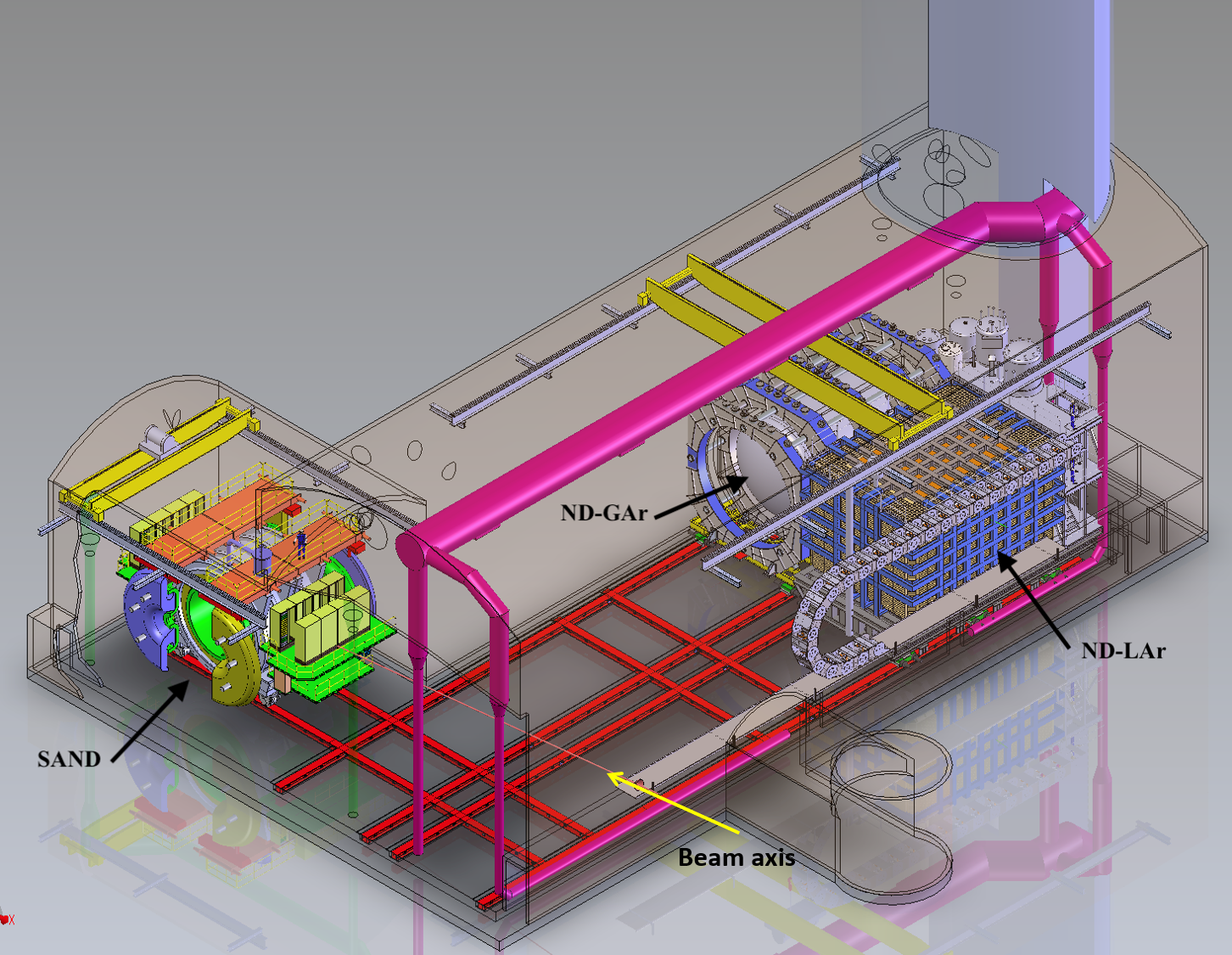}
\end{dunefigure}

A critical component of the \dword{dune} \dword{nd} is a \dword{lartpc} constructed using \dword{arcube} technology.  This component detector of the \dword{dune} \dword{nd} is called \dword{ndlar}.  The particular implementation of the \dword{lartpc} technology in this detector is described in Chapter~\ref{ch:lartpc}.  
This detector has the same target nucleus and uses the same fundamental detection principles as the \dword{fd}.
 The differences between the two are needed because of the expected intensity of the beam at the \dword{nd}.  The use of the same target nucleus and a similar technology reduces sensitivity to nuclear effects and detector-driven systematic uncertainties in the extraction of the oscillation signal at the  \dword{fd}.  \dword{ndlar} is large enough to provide high statistics ($\num{1e8}{\ \numu \text{-CC events/year}}$ on axis) and a sufficient volume to provide containment of the hadronic system.  The tracking and energy resolution, combined with the fiducial mass of \dword{ndlar}, will allow for the measurement of the flux in the beam using several techniques, including the rare process of $\nu$-e$^{-}$ scattering\footnote{This is an important technique since it  is largely independent of nuclear and cross-section uncertainties.
 }. 

 \dword{ndlar} alone begins to lose acceptance for muons above $\sim$\SI{0.7}{GeV/c} due to lack of containment. Because the muon momentum and charge are critical components of the neutrino energy determination, a magnetic spectrometer is needed downstream of \dword{ndlar} to measure both quantities. In the \dword{dune} \dword{nd} concept, this function is accomplished by the \dword{ndgar} detector.
 
 The \dword{ndgar} detector (also sometimes called the multipurpose detector, or MPD) consists of a high pressure gaseous argon TPC surrounded by an \dword{ecal} in a \SI{0.5}{T} magnetic field with a muon system outside of that. The high pressure gaseous argon TPC runs at 10 atmospheres and provides a lower-density medium with excellent tracking resolution to momentum analyze the muons from \dword{ndlar}. In addition, with this choice of technology for the tracker, neutrinos interacting with the argon in the high pressure gaseous TPC constitute a large, independent sample of $\nu$-Ar interactions that can be studied with a very low momentum threshold for charged particle tracking,  excellent tracking resolution, nearly uniform angular coverage, and with systematic uncertainties that differ from the liquid detector. \dword{ndgar} will collect approximately \num{1.6e6} $\numu$-CC events per year of on-axis running with a \SI{1.0}{ton} fiducial volume.

 Since \dword{ndgar} can access lower-momentum protons than \dword{ndlar} and has superior identification of charged pions, events occurring in \dword{ndgar} will be valuable for studying the charged particle activity near the interaction vertex. The misidentification of pions as knocked-out protons (or vice versa) can cause a significant misreconstruction of the neutrino energy 
 and/or a mistake in the event type classification.
 This effect can become quite significant at the lower-energy second oscillation maximum. The gas detector will play an important role in understanding how often the \dword{fd} and \dword{ndlar} make this mistake, since pions are rarely misidentified as protons in the high pressure gaseous argon TPC.  

In addition, the relatively low level of secondary interactions in the gas samples will be helpful for identifying the particles produced in the primary interaction and modeling secondary interactions in denser detectors.  The confusion of primaries and secondaries is known to be an important effect in reconstruction of events in liquid argon TPCs\cite{Friedland:2018vry}. Relative to lower pressure TPCs, the high pressure in the gas TPC of \dword{ndgar} increases the statistics for these studies and improves the particle identification capabilities, while somewhat degrading the momentum resolution. 
\dword{ndgar} is discussed further in Chapter~\ref{ch:mpd}.

\dword{ndlar} and \dword{ndgar} can move to take data in positions off the beam axis.  This capability is referred to as the \dword{duneprism}. As the detectors move off-axis, generally together, the incident neutrino flux spectrum changes, with the mean energy dropping and the spectrum becoming narrower.  Though the neutrino interaction rate drops off-axis, the intensity of the beam and the size of the \dword{lartpc} combine to yield ample statistics at all off-axis positions.   In addition, the statistics in the \dshort{ndgar} are large enough to provide useful data for the PRISM analysis over about half the off-axis range.

The data taken at different off-axis angles will allow the deconvolution of the neutrino flux and interaction cross section, as well as the mapping of the reconstructed versus true energy response of the detector.  This latter mapping is applicable at the \dword{fd} up to the degree to which the near and far \dword{lar} detectors are similar.  Also, it is possible to use information from a linear combination of the different fluxes to create a data sample at the \dword{nd} with an effective neutrino energy distribution that is close to that of the oscillated spectrum at the \dword{fd}.  This data-driven technique will reduce systematic effects coming from differences in the energy spectra of the oscillated signal events in the \dword{fd} and the samples in the \dword{nd} used to constrain the interaction model. Finally, the off-axis degree of freedom provides a sensitivity to some forms of mismodeling in the beam and/or interaction models. The \dword{duneprism} program is discussed further in Chapter~\ref{ch:prism}.

The final component of the \dword{dune} \dword{nd} suite is a magnetized beam monitor called the \dword{sand}. This device monitors the flux of neutrinos going to the \dword{fd} from an on-axis position where it is much more sensitive to variations in the neutrino beam. \dword{sand} consists of an inner tracker surrounded by an
\dword{ecal} inside a large solenoidal magnet. Currently two options are being explored for the inner tracker, one based on a combination of plastic scintillator cubes with \dword{tpc}s and one based on straw-tubes. 
The magnet and \dword{ecal} are repurposed  from the \dword{kloe} detector, which is a cylindrical collider detector previously used to study $\phi$ meson production at the INFN LNF laboratory in Frascati, Italy.  It has a superconducting coil that provides a $\sim$0.6~T magnetic field and an excellent lead-scintillator \dword{ecal} \cite{Franzini:2006aa}.  

\dword{sand} importantly serves as a dedicated  neutrino spectrum monitor that stays on axis when \dword{ndlar} and \dword{ndgar} have moved to an off-axis position. This data will be useful for noting changes in the beam and diagnosing what those changes are so that the beam model can be adjusted as needed (important for the main oscillation measurement).  SAND also provides an excellent on-axis neutrino flux determination using many of the methods discussed in Chapter~\ref{ch:flux}.  The neutrino flux determined using this detector, with  differing detector, target, and interaction systematic uncertainties as compared to the \dword{lartpc}, can be used as an important point of comparison and systematic crosscheck for the flux as determined by \dword{ndlar}.
In addition, the inner tracker is expected to be able to incorporate neutrons in the event reconstruction, in general or for selected event morphologies, depending on the inner tracker choice.
This is expected to be useful for the determination of the flux and for potential improvements in the nuclear model. \footnote{The addition of the neutron reconstruction capability extends the DUNE ND theme of including regions of phase space in neutrino interactions not seen in previous experiments. DUNE is studying neutron reconstruction in each of the component detectors of the ND.}

The different mass numbers, $A$, of the hydrocarbon target relative to argon, in \dword{sand} may prove useful for developing models of nuclear effects and building confidence in the interaction model and the size of numerous systematic uncertainties.  The inclusion of neutrons in the reconstruction  
may provide insights that foster improvements in the neutrino interaction model on carbon.  Though extrapolating such improvements to argon is not straightforward, the development of Monte Carlo neutrino event generators has benefited from data taken with different nuclear targets, including carbon.  It is also thought that the data with the hydrocarbon target will offer a point of comparison to other experiments, such as Hyper-K and MINERvA, that may proved useful for understanding systematic effects and biases. 
\dword{sand} is discussed further in Section~\ref{ch:sand}.



\section{More on the Role of the ND and Lessons Learned}
\label{sec:intro-OscillationRole}

Oscillation experiments need to accomplish three main tasks. First, they must identify the flavor of interacting neutrinos in \dword{cc} events, or identify the events as \dword{nc} interactions. Second, they need to measure the energy of the neutrinos since oscillations occur as a function of baseline length over neutrino energy, \dword{l/e}. Third, they need to compare the observed event spectrum in the \dword{fd} to  predictions based on differing sets of oscillation parameters, subject to constraints from the data observed in the \dword{nd}.  That comparison allows for the extraction of the  measured oscillation parameters and uncertainties.

One effect complicating the connection between the observations in the \dword{nd} and the \dword{fd} is that, as a practical matter, the \dword{fd} uses a heavy nuclear target (argon) rather than hydrogen. Neutrino interactions can be idealized\footnote{This idealization is a common model used by neutrino event generators.} as a three stage process: 
(1) a neutrino strikes a nucleus with a complex internal state, including nucleon-nucleon interactions, (2) scattering occurs with one or more of the nucleons, during which one or more hadrons may be created/ejected, and (3) the resulting hadrons may reinteract with the remnant nucleus as they exit, which is generically referred to as \dword{fsi}.
The presence of the nucleus impacts all three stages in ways that are not fully understood.

The connection between the observations in the \dword{nd} and the \dword{fd} is made using a simulation that convolves models of the neutrino flux, neutrino interactions, nuclear effects, and detector response.
This gives rise to a host of complicating effects that
muddy the simple picture. One issue is that there are backgrounds.
The intrinsic \nue content of the beam is a background to a \nue appearance oscillation signal at the \dword{fd}.  Similarly, \dword{nc} events with a $\pi^0$ that leads to electromagnetic showers from converted photons  can  mimic \nue \dword{cc} interactions, forming a significant background to a \nue appearance oscillation signal at the \dword{fd}.  Both the level of these backgrounds and the detection/identification efficiency are known imperfectly and vary with energy.

Understanding these complicating effects and mitigating the uncertainty they generate are key drivers of the design of the  \dword{nd} complex. For example, the primary target nucleus in the \dword{nd} should be the same as that in the \dword{fd}.  This ensures that the nuclear effects are the same in the two detectors for a given type of neutrino interaction at a given energy.  This reduces systematic uncertainties that arise from nuclear effects in the near-to-far event rate comparison.  Also, it is helpful to have the \dword{nd} technology and functional design be as similar as feasible to those used for the  \dword{fd}. To the extent they are identical, any bias in the efficiency as a function of energy will cancel between the two detectors in a near-to-far comparison.
Since the background misidentification probability tends to be similar between two such similar detectors, it is helpful if the \dword{nd} is more capable than the \dword{fd} at characterizing backgrounds, either due to its technology, or by leveraging the much larger statistics and freedom to take data in alternative beam configuration modes (e.g., movement off the beam axis). Finally, it is useful for the \dword{nd} to be able to measure neutrino interactions well, in terms of particle type and momentum acceptance phase space and interaction morphology identification. This is so that the data are helpful for optimising the neutrino interaction model used to correct for residual differences between the near and far detectors as well as differences in the neutrino energy spectra.



\subsection{An introduction to some of the key complications}
\label{sec:pedagogy}

Since the \dword{fd} uses argon as a target, it is important to use argon as the primary target nucleus in the \dword{nd}.  That said, it is instructive to consider what would happen if the detectors were made of hydrogen.
In a detector made of hydrogen, the initial state is a proton at rest and there are no \dword{fsi}. A variety of processes can happen, depending on the energy. The simplest is \dword{cc} \dword{qe} scattering, that occurs for antineutrinos: $\bar{\nu}_\ell\, p \to \ell^+\, n$. The detector sees a lepton (which establishes the flavor of the neutrino), no mesons, and perhaps a neutron interaction away from the lepton's vertex. Because there are no mesons, the kinematics is that of two body scattering and the neutrino energy can be reconstructed from the lepton's angle (with respect to the $\nu$ beam) and energy. This is true whether or not the neutron is observed.
For $\nu_\ell$ interactions on hydrogen there is no \dword{cc} \dword{qe} process and the simplest scattering channel is single pion production $\nu_\ell\, p \to \ell^-\, \pi^{+}\, p$. In that case the neutrino energy may be reconstructed from the energy of the muon and pion, and their angles with respect to the beam\footnote{With the conservation of four energy-momentum, $E_\nu$ and $\vec{p}_N$ can be computed.The nucleon does not need to be observed. }. In both cases, the neutrino energy can be measured without bias so long as the detector itself measures lepton and meson momenta and angles without bias.  The neutrino energy in complicated scattering channels, such as ones with multiple pions or heavy baryons can be measured in a similar way (at least in principle).

The key feature of a hypothetical hydrogen detector is that there are enough constraints that the neutrino energy can be determined without needing to measure the recoil nucleon.
Additionally, the cross sections for different scattering channels, particularly the simpler ones, can be expressed in terms of leptonic and hadronic currents. The leptonic current is well understood. The structural elements of the hadronic current are known on general theoretical grounds. The current is often represented by form factors that are constrained by electron scattering experiments, beta decay, and neutrino scattering measurements.



The situation is significantly more complicated in a detector with heavy nuclei, as will be the case with \dword{dune}. The nucleons in the initial state of the nucleus are mutually interacting and exhibit Fermi motion. This motion removes the key momentum conservation constraint available in hydrogen due to the target being at rest. Moreover, scattering at lower momentum transfer is suppressed if the nucleon in the final state would have a momentum that is excluded by the Pauli principle. 

The nucleon momentum distribution in heavy nuclei is commonly modeled as a Fermi gas with a cutoff momentum $k_F \approx \SI{250}{MeV/c}$ \cite{Smith:1972xh}.
This picture is overly simplistic.  For example,  there are nucleons with momenta larger than $k_F$ due to short-range correlated nucleon-nucleon interactions (\dword{src}) \cite{Bodek:2014jxa}. Scattering on a nucleon with $p>k_F$ implies that there is a spectator nucleon recoiling against the target with a significant momentum. \dword{src} have been the subject of much investigation but are not fully understood nor fully implemented in neutrino event generators.  It should be noted that there are in use more sophisticated treatments describing the initial state momentum distributions and removal energy of nucleons in nuclei than the Fermi gas model. These ``spectral functions'' will be discussed more in Section~\ref{ch:nu-xsec:nucl-models}.

For the few-GeV neutrinos of interest to \dword{dune}, the typical momentum transfer corresponds to a probe that has a wavelength on a par with the size of a nucleon. In this case, the scattering can occur on two targets in the nucleus which may be closely correlated, known as two-particle-two-hole, or \dword{2p2h} scattering.
This process results in the knock-out of two nucleons. As one or both nucleons may escape detection, particularly if they have low energies, \dword{2p2h} scattering can mimic \dword{qe} scattering from a single nucleon. The presence of a second nucleon in the final state invalidates the assumptions used to calculate neutrino energy based on final-state lepton energy, as is possible for a \dword{qe} interaction.
It is known that \dword{2p2h} scattering contributes significantly to the total scattering cross section at \dword{dune} energies \cite{Ruterbories:2018gub}. The \dword{2p2h} cross section is difficult to compute because it cannot be expressed as the sum over cross sections on individual nucleons. The dependence on atomic number, as well as the observables of the interaction, like the final energies of the two particles, are currently unknown. Finally, it is widely expected that there are modes of scattering from correlated nucleon pairs that result in meson production.
In addition, there are likely higher order processes such as 3p3h, etc. Event generators do not currently include such processes.


Neutrino scattering on nuclei is also subject to \dword{fsi}. \dword{fsi} collectively refers to the processes by which nucleons and mesons produced by the neutrino interaction traverse the remnant nucleus. The hadrons reinteract with a variety of consequences: additional nucleons can be liberated; ``thermal'' energy can be imparted to the nucleus; pions can be created, scattered, and absorbed; and pions and nucleons can undergo charge exchange scattering (e.g., $\pi^- p \to \pi^0 n$).  Event generators include phenomenological models for \dword{fsi}, anchoring to hadron-nucleus scattering data.



The heavy nuclei in a detector also act as targets for the particles that have escaped the struck nucleus. Generally speaking, the denser the detector and the more crudely it samples deposited energy, the more difficult it is to observe low-energy particles. Negatively and positively charged pions leave different signatures in a detector, since the former are readily absorbed while the latter are likely to decay.  Neutrons can be produced from the struck nucleus, but also from follow-on interactions of the neutrino's reaction products with other nuclei. The energy carried away by neutrons is challenging to detect and can bias the reconstructed neutrino energy. 

Finally, it is important to note that due to the relatively broad and high energy neutrino spectrum at DUNE, $\sim$40-50\% of the neutrino interactions will come from deep inelastic scattering rather than the simpler \dword{qe} and single pion production reactions ($\sim$40\% combined) discussed above.  This leads typically to a more complex morphology for events, over and above the heavy nucleus complications,  and greater challenges for the detector, reconstruction, and the modeling.

\subsection{Lessons from Current Experiments}
\label{sec:intro-LessonsLearned}

The past experience of the neutrino community is a driving force in the design of the \dword{dune}  \dword{nd} complex. 
The performance of  current, state-of-the-art long-baseline oscillation experiments  provides a practical guide to many of the uncertainties and potential limitations \dword{dune} can expect to encounter, as well as case studies of issues that arose which were unanticipated at the design stage. 

Neutrino beams are notoriously difficult to model at the precision and accuracy required for modern accelerator-based neutrino oscillation experiments.  Recent long-baseline experiments make use of a  \dword{nd} placed close to the beam source, where oscillations are not yet a significant effect.  The beam model, the neutrino interaction model, and perhaps the detector response model are tuned, or calibrated, by the data recorded in the  \dword{nd}. The tuned model is used in the extraction of the oscillation signal at the  \dword{fd}. Known effects that are not understood or modeled well must be propagated into the final results as part of the systematic error budget.  Unknown effects that manifest as disagreements between the model and observations in the  \dword{nd} also must be propagated into the final results as part of the systematic error budget.  These kinds of disagreements have happened historically to every precision accelerator oscillation experiment.  When such disagreements arise, some assumption or range of assumptions must be made about the source of the disagreement.  Without narrowing down the range of possibilities, this can dominate the systematic uncertainty.

Since the final results depend on the comparison of what is seen in the  \dword{fd} to that in the  \dword{nd}, having functionally identical detectors (i.e., the same target nucleus and similar detector response) is helpful.  In a similar vein, differences between the neutrino spectrum at the  \dword{nd} and the oscillated spectrum seen at the  \dword{fd} lead to increased sensitivity to systematic effects propagated from the  \dword{nd} to the  \dword{fd}.

The \dword{t2k} experiment uses an off-axis neutrino beam that has a narrow energy distribution peaked below \SI{1}{GeV}. This means, relative to \dword{dune}, interactions in \dword{t2k} are predominantly \dword{ccqe} and have relatively simple morphologies.  The data sample has little feed-down from higher-energy interactions.  The \dword{t2k}  \dword{nd} (plastic scintillator and
 \dword{tpc}) technology is very different from its   \dword{fd} (water Cherenkov), though the \dword{nd} contains embedded water targets that provide samples of interactions on the same target used in the \dword{fd}.
The experiment relies on the flux and neutrino interaction models, as well as the \dword{nd} and  \dword{fd} response models to extrapolate the constraint from the  \dword{nd} to the  \dword{fd}.   In recent oscillation results released by \dword{t2k}, the  \dword{nd} data constraint reduces the flux and interaction model uncertainties at the  \dword{fd} from 
11--14\% down to 2.5--4\%
\cite{Abe:2018wpn}. Inclusion of the water target data was responsible for a factor of two reduction in the interaction model systematic uncertainties, highlighting the importance of measuring interactions on the same target nucleus as the \dword{fd}.\footnote{These numbers are not used directly in the analysis but were extracted to provide an indication of the power of the \dword{nd} constraint.}

The  \dword{nova}  experiment uses an off-axis neutrino beam from \dword{numi} that has a narrow energy distribution peaked around \SI{2}{GeV}.  The  \dword{nova}   \dword{nd} is functionally identical to its  \dword{fd}.  However, it is significantly smaller than the  \dword{fd} and it sees a different neutrino spectrum due to geometry and oscillations.  Even with the functionally identical near and  far detectors,  \dword{nova}  uses a model to subtract \dword{nc} background and relies on a model-dependent response matrix to translate what is seen in the  \dword{nd} to the ``true'' spectrum, which is then extrapolated to the  \dword{fd} where it is put through a model again to predict what is seen in the  \dword{fd}~\cite{NOvA:2018gge, WolcottNUINT2018}.  Within the extrapolation, the functional similarity of the near and  far detectors reduces, but does not eliminate, many systematic effects.  Uncertainties arising from the neutrino cross-section model dominate the  \dword{nova}  $\nu_{e}$ appearance systematic error budget and are among the larger uncertainties in the $\nu_{\mu}$ disappearance results.  The  \dword{nd} constraint is significant.  For the $\nu_{e}$ appearance signal sample in recent  \dword{nova}  results, for example, a measure of the systematic error arising from cross-section uncertainties without using the  \dword{nd} constraint is 12\,\% and this drops to 5\,\% if the  \dword{nd} constraint is used \cite{WolcottNUINT2018}.

The process of implementing the  \dword{nd} constraint in both \dword{t2k} and  \dword{nova}  is less straightforward than the typical description implies.  It will not be any more straightforward for \dword{dune}.  One issue is that there are unavoidable near and far differences. Even in the case of functionally identical detectors, the beam spectrum and intensity are very different near-to-far.  For \dword{dune}, in particular, 
\dword{ndlar} is smaller than the \dword{fd} and is divided into modular, optically isolated regions that have a pixelated readout rather than the wire readout of the \dword{fd}.  Space charge effects will differ near-to-far.  All of this imposes model dependence on the extrapolation from near-to-far.  This is mitigated by collecting data at differing off-axis angles with \dword{duneprism}, where an analysis can be done with an \dword{nd} flux that is similar to the oscillated \dword{fd} flux (see Chapter~\ref{ch:prism}). (Data from \dword{protodune} will also be useful to understand the energy-dependent detector response for the \dword{fd}.)  Regardless, near-to-far differences will persist and must be accounted for through the beam, detector, and neutrino interaction models.  

Although long-baseline oscillation experiments use the correlation of fluxes at the \dword{nd} and the \dword{fd} to reduce sensitivity to flux modeling, the beam model is a critical component in understanding this correlation.  Recently, the \dword{minerva} experiment used spectral information in the data to diagnose a discrepancy between the expected and observed neutrino event energy distribution in the \dword{numi} medium-energy beam \cite{JenaNUINT2018}. In investigating this issue, \dword{minerva} compared the observed and simulated neutrino event energy distributions for low-$\nu$ events\footnote{$\nu$ is the energy transfer between the incoming neutrino and the target. A low-$\nu$ cut selects events with small recoil energy.}, as shown in Figure~\ref{fig:minervameflux}.  Since the cross section is known to be relatively flat as a function of neutrino energy for this sample, the observed disagreement as a function of energy indicated a clear problem in the flux model or reconstruction.      
\dword{minerva} believes the observed discrepancy between the data and simulation is best accounted for by what is a mismodeling in the magnetic horn focusing in the beam model combined with an error in the muon energy reconstruction (using range traversed in the downstream spectrometer).  This is notable, in part, because the two identified culprits in this saga would manifest differently in the extrapolation to the far detector in an oscillation experiment. The spectral analysis provided critical information in arriving at the final conclusion.  This experience illustrates the importance of accurate monitoring/measurements of the neutrino beam spectrum.  

\begin{dunefigure}[MINERvA medium-energy \dshort{numi} flux for low-$\nu$ events]{fig:minervameflux}
{Ratio of the observation to simulation for 
the reconstructed \dword{minerva} medium-energy \dword{numi} neutrino event spectrum for low energy-transfer events. 
(This is the updated version of what is shown in \cite{JenaNUINT2018}.)}
\includegraphics[width=0.46\textwidth]{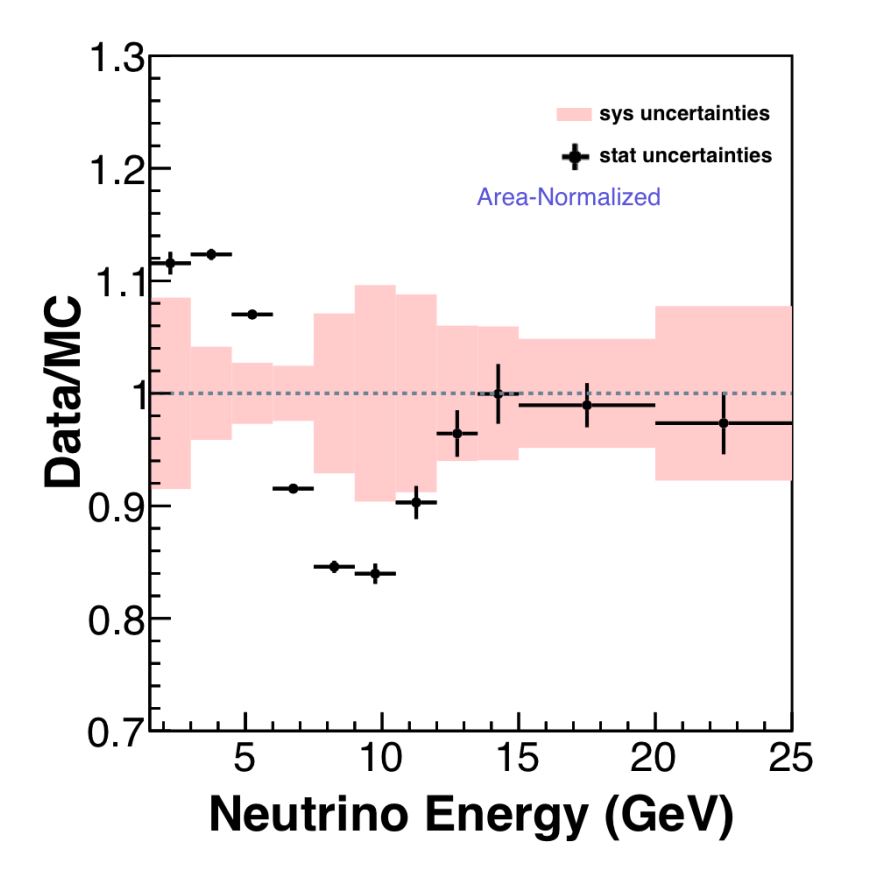}
\end{dunefigure}

Another important issue is that the neutrino interaction model is not perfect, regardless of the experiment and implementation.  With an underlying model that does not describe reality perfectly, even a model tuned to  \dword{nd} data will have residual disagreements with that data.  These disagreements must be accounted for in the systematic uncertainty budget of the ultimate oscillation measurements.  Although the model(s) may improve before \dword{dune} operation, the degree of that improvement cannot be predicted and the \dword{dune}  \dword{nd} complex should have the capability to gather as much information as possible to help improve and tune the model(s) during the lifetime of the experiment.  In other words, the  \dword{nd} needs to be capable of narrowing the range of plausible possibilities giving rise to data-model differences at the  \dword{nd} in order to limit the systematic uncertainty incurred in the results extracted from the  \dword{fd}.   

Recent history provides illustrations of progress and continuing struggles to improve neutrino interaction models.  The MiniBooNE collaboration published results in 2010 showing a disagreement between the data and the expected distribution of \dword{ccqe} events as a function of squared four-momentum transfer, Q$^{2}$ \cite{AguilarArevalo:2010cx,Gran:2006jn}.   They brought the model into agreement with the data by increasing the axial mass form factor used in the model.  K2K \cite{Gran:2006jn} and \dword{minos} \cite{Adamson:2014pgc} made similar measurements.  It has since been shown that the observed disagreement is due to the need to include multinucleon processes and that the use of the large effective axial mass form factor used by these experiments to fit the data leads to a misreconstruction of the neutrino energy. Multinucleon scattering processes had previously been observed in electron scattering experiments but went unappreciated by the neutrino physics community. 

The importance of modeling multinucleon (\dword{2p2h}) processes for oscillation experiments is underscored by the fact that such interactions, when mis-reconstructed as a \dword{ccqe} (1p1h) process, lead to a significant low-side tail in the reconstructed neutrino energy \cite{Martini:2012uc}.  Multinucleon processes also change the hadronic calorimetric response.  The first  \dword{nova}  $\nu_{\mu}$ disappearance oscillation results had a dominant systematic uncertainty driven by the disagreement of their model to the data in their hadronic energy distribution \cite{Adamson:2016xxw}.  In more recent work, the inclusion of multinucleon processes in the interaction model contributed to a substantial reduction of this disagreement \cite{NOvA:2018gge}.

The \dword{minerva} experiment has compiled a significant catalog of neutrino and antineutrino results and recently developed a model tuned to their \dword{qe}-like  (\dword{numi} low energy) data \cite{Ruterbories:2018gub}.  The tune is based on a modern neutrino interaction generator (\dword{genie} 2.8.4 \cite{Andreopoulos:2009rq}, using a global Fermi gas model \cite{Smith:1972xh}  with a Bodek-Ritchie tail \cite{Bodek:1981wr} and the INTRANUKE-hA \dword{fsi} model \cite{Dytman:2007zz}).  
In addition, \dword{minerva} includes a random phase approximation model (RPA) \cite{Nieves:2004wx,Gran:2017psn} for long-range nucleon correlations, as well as scaling down
 non-resonance pion production. 
 The addition of a multinucleon model \cite{Nieves:2011pp, Gran:2013kda, Schwehr:2016pvn}, when scaled by an empirical factor, provides good agreement with \dword{minerva} neutrino data \cite{Ruterbories:2018gub}.
 The same tune as developed on the neutrino data also fits well the \dword{minerva} antineutrino \dword{qe}-like data (with no additional tuning or ingredient).  The required empirical enhancement of the multinucleon contribution to the model implies shortcomings in the interaction model, but the decent fit to data for both neutrinos and antineutrinos implies that the tune is effectively making up for some imperfections in the model. 

More recent versions of \dword{genie} include some of the modifications incorporated by \dword{minerva} in the tune discussed above \cite{Alam:2015nkk}.  This illustrates the dynamic nature of neutrino interaction modeling and the interplay between the experiments and generator developers.  The evolution of the field continues as illustrated with a snapshot of some of the current questions and areas of focus:
\begin{itemize}
    \item There is a pronounced deficit of pions produced at low Q$^{2}$ in \dword{cc}1$\pi^{0}$ events as compared to expectations \cite{BercellieNUINT2018,Altinok:2017xua,Aliaga:2015wva,McGivern:2016bwh,novaminosPC}.  Current models take this into account by tuning to data without any underlying physical explanation for how or why this happens.
    \item The \dword{minerva} tune that fits both neutrino and antineutrino \dword{ccqe} data involves a significant enhancement and distortion of the \dword{2p2h} contribution to the cross section.  The real physical origin of this cross-section strength is unknown.  Models of multinucleon processes disagree significantly in predicted rates.
    \item Multinucleon processes likely contribute to resonance production.  This is neither modeled nor well constrained.
    \item Cross-section measurements used for comparison to models are a convolution of what the models view as initial state, hard scattering, and final state physics.   Measurements able to deconvolve these contributions are expected to be very useful for model refinements.\footnote{This simplification must be approached with caution, as the three stages are physically interdependent.} 
    \item Most neutrino generators make assumptions about the structure of form factors and factorize nuclear effects in neutrino interactions into initial and final state effects via the impulse approximation.  These are likely oversimplifications.  The models will evolve and the systematic uncertainties will need to be evaluated in light of that evolution. 
    \item  Most neutrino detectors are largely blind to neutrons and low-momentum protons and pions (though some $\pi^{+}$ are visible via Michel decay).  This leads to smearing in the reconstructed energy and tranverse momentum, as well as a reduced ability to accurately identify specific interaction morphologies.  The closure of these holes in the reconstructed particle phase space is expected to provide improved handles for model refinement.
    \item There may be small but significant differences between the $\nu_{\mu}$ and $\nu_{e}$ \dword{ccqe} cross sections which are poorly constrained \cite{Day-McFarland:2012}.
    \item It is not possible, with current computing resources, to make ab initio calculations for heavy nuclei.  Assumptions must be made in any nuclear model.
\end{itemize}
Given the critical importance of neutrino interaction models and the likelihood that the process of refining these models will continue through the lifetime of \dword{dune}, it is important the \dword{dune}  \dword{nd} suite be highly capable.   

\subsection{Incorporating Lessons from Current Experiments}
\label{sec:intro-ExperienceMotivatesDesign}

The approach followed in the design of the \dword{dune}  \dword{nd} concept is to provide sufficient redundancy to address areas of known weaknesses in previous/current experiments and known issues in the interaction modeling insofar as possible, while providing a powerful suite of measurements that is likely to be sensitive to unanticipated issues and useful for continued model improvements.  Anything less reduces \dword{dune}'s potential to achieve significantly improved systematic uncertainties over previous experiments in the long-baseline analyses. 

The \dword{dune}  \dword{nd} incorporates many elements in response to lessons learned from previous/current experiments. 
\dword{ndlar} has the same target nucleus and a similar technology to the  \dword{fd}. These characteristics reduce the detector and target systematic sensitivity in the  extrapolation of flux constraints from this detector to the  \dword{fd}.  This detector is capable of providing the primary  sample of \dword{cc} $\nu_{\mu}$ interactions to constrain the flux at the  \dword{fd}, along with other important measurements of the flux from processes like $\nu$-e$^{-}$ scattering and low-$\nu$.  Samples taken with this detector at off-axis angles (\dword{duneprism}) will allow the deconvolution of the flux and cross-section uncertainties and provide potential sensitivity to mismodeling.  The off-axis data can, in addition, be used to map out the detector response function and construct effective  \dword{nd} samples that mimic the energy distribution of the oscillated sample at the  \dword{fd}. By doing this, \dword{duneprism} analyses will minimize the effect of spectral differences near-to-far.  The large \dword{nd} interaction samples can be used to tune and improve the models to mitigate the uncertainties incurred by correcting for  residual differences.  

The \dword{dune}  \dword{nd} provides access to particles produced in neutrino interactions that have been largely invisible in previous experiments, such as low-momentum protons and charged pions and neutrons 
measured in  \dword{ndgar}, and neutrons in the \dword{sand} tracker as well as the \dword{ecal}s of both \dword{sand} and \dword{ndgar}. 
The high pressure gaseous argon TPC provides data on interactions with a minimal amount of distortion due to secondary interactions on the produced particles.  These capabilities improve the experiment's ability to identify specific interaction morphologies, study samples with improved energy resolution, and extract samples potentially useful for improved tuning of model(s) of multinucleon processes. The neutron content in neutrino and antineutrino interactions is different and this will lead to differences in the detector response. For an experiment that is measuring \dword{cpv}, data on neutron production in neutrino interactions is likely to be an important handle in the tuning of the interaction model and the flavor-dependent detector response function model.

\dword{sand} provides dedicated beam spectrum monitoring on axis.  It also provides an independent determination of the on-axis flux with different detector and target systematic uncertainties.
The beam spectrum monitoring is useful for identifying and diagnosing unexpected changes in the beam.  This proved useful for \dword{numi} and is likely to be more important for DUNE given the need to associate data taken at different times and off-axis angles while making measurements depending on spectrum distortions. 


The large data sets that will be accumulated by the three main components of the  \dword{nd}  will allow for differential studies and the use of \dword{tki} variables to precisely identify intranuclear dynamics~\cite{Lu:2015tcr, Furmanski:2016wqo, Abe:2018pwo, Dolan:2018sbb, Lu:2018stk, Dolan:2018zye, 
Lu:2019nmf, 
Harewood:2019rzy, Cai:2019jzk, Cai:2019hpx, Coplowe:2020yea} and the absence thereof~\cite{Lu:2015hea, Duyang:2018lpe, Duyang:2019prb, Munteanu:2019llq, Hamacher-Baumann:2020ogq}. 
Each detector brings its unique strengths to the study: \dword{ndlar} has good tracking resolution and containment and massive statistics; \dword{ndgar} has excellent tracking resolution, very low charged-particle tracking thresholds, and unambiguous track charge sign determination; and \dword{sand} has good containment and can include neutrons on an event-by-event basis.  The neutrino interaction samples acquired by this array of detectors will constitute a powerful laboratory for deconvolving the initial state, hard scattering, and final state physics, which, in turn, will lead to improved modeling and confidence in the final results extracted from the  \dword{fd}.

\section{Near Detector Requirements}
\label{sec:intro-requirements}
As described in Section \ref{sec:intro-BriefOverview}, the reference design of the \dword{nd} consists of several components which together fulfill the needs of \dword{dune}. To articulate requirements, which come in many forms addressing anything from the overall goals of the system to the detailed technical specifications of a subsystem,  a hierarchical system  has been developed to categorize and organize the requirements as follows:

\begin{itemize}
\item  {\bf Overarching:} General goals of the \dword{nd} system that must be fulfilled in order for \dword{dune} to achieve its scientific goals.
\item {\bf Measurements:} Measurements that must be performed with the \dword{nd} in order to fulfill the overarching requirements.
\item {\bf Capabilities:} Capabilities, in terms of detector performance, statistics, {\em etc.} that the \dword{nd} subsystems must have to perform the required measurements.
\item {\bf Technical:} Technical specifications of detectors, in terms of dimensions, mass, tolerances, {\em etc.}  that must be fulfilled in order for the subsystems to have their required capabilities.
\end{itemize}

The overarching and measurement requirements are detector agnostic and independent of the specific implementation of the \dword{nd}. However, at the level of measurement requirements, we specify which subsystem(s) in the \dword{nd} are primarily responsible for performing the measurements, which then lead to implementation and subsystem-specific capabilities requirements, and then the associated technical requirements. The structure is hierarchical such  that higher-level requirements are fulfilled by satisfying the lower level requirements, while each lower level requirement supports specific higher level requirements. The hierarchy also generally has ``nearest neighbor'' relations so that, at each level, the requirements are driven by those one step higher in the hierarchy and are fulfilled by satisfying those one step lower in the hierarchy. Since the fulfillment of the overarching requirements will be primarily verified by full oscillation sensitivity studies, this isolation allows the evaluation of lower level requirements without full sensitivity studies.

A few notes regarding the current state of the requirements:
\begin{itemize}
\item  The requirements focus on the immediate needs of the long-baseline neutrino oscillation analysis. Neutrino interaction/cross section and beyond the Standard Model physics are an important part of the \dword{nd} and \dword{dune} program overall. However the requirements for these physics programs are not reflected in these tables. 
Likewise there are many measurements, particularly neutrino interaction studies,  that would provide important cross checks which are also not in the scope of these requirements.
\item The requirements remain a work in progress and will be continuously developed as simulation tools and other developments continue. In some cases, the requirements, particularly for the higher level overarching and measurement requirements, do not lend themselves to quantitative specifications. In other cases, such specifications are still being studied, in which case there is a blank entry.
\item The fourth level of ``Technical Requirements'' is still in development for each detector system and are not described in the document.
\end{itemize}

While the requirements are ultimately guided and validated through a sensitivity study, a rough sense of the target uncertainties for a CP violation study can be obtained by the variation in the expected number of $\nu_\mu\to\nu_e(\bar{\nu}_\mu\to\bar{\nu}_e)$ candidate events using a beam with forward horn current (\dword{fhc})(reverse horn current (\dword{rhc})) settings. Table \ref{tab:nuerates} (reproduced from the Far Detector Technical Design Report\cite{Abi:2020evt}) shows the expected number of selected  $\nu_\mu\to\nu_e(\bar{\nu}_\mu\to\bar{\nu}_e)$ events with $\delta_{CP}=0$ after seven years assuming the staging scenario and the exposure split evenly between $\nu$ (\dword{fhc}) and $\bar{\nu}$ mode (\dword{rhc}).

In the case of maximal CP violation ({\em e.g.} $\delta_{CP}=-\pi/2$),  the variation in the number of signal events (which depends on other oscillation parameters as well as the beam mode) can be as large as $\sim$40\% relative to the expectation at $\delta_{CP}=0$. However, this deviation in the total number of selected  $\nu_e/\bar{\nu_e}$ events is diluted to $\sim 15\%$  once backgrounds are accounted for. To obtain $3(5)\;\sigma$ significance to $\delta_{CP}\neq  0$ in this case would require total uncertainties to be constrained to $5\%(3\%)$ or better, implying target systematic uncertainties of better than $3\%(2\%)$ in order to prevent systematic uncertainties from dominating the uncertainty. These constraints must consider not only the background ({\em e.g.} misidentified events and  irreducible intrinsic beam $\nu_e$ and ``wrong sign'' $\nu_\mu\to\nu_e$ oscillation events) but also the signal $\nu_\mu\to\nu_e(\bar{\nu}_\mu\to\bar{\nu}_e)$ events in \dword{fhc}(\dword{rhc}), since an accurate modeling of the expected observed signal is needed to extract the oscillation parameters from the number and spectrum of these events. For reference, $3(5)\;\sigma$ significance to maximum $CP$ violation is expected after 3(7) years in the standard \dword{lbnf}/\dword{dune} beam and detector staging scenario, as described in Table 1.3 in \cite{Abi:2020evt}. The significance in the case of maximal $CP$ violation is only one milestone that articulates the sensitivity of \dword{dune} and does not address its performance in other scenarios, like those  in Table 1.3 in \cite{Abi:2020evt} requiring that the expected significance exceed a threshold across some fraction of $\delta_{CP}$ values. Nonetheless, it represents an important test case in targeting scientific milestones in the first several years of the experiment. 

An analogous guideline regarding $\nu_\mu/\bar{\nu}_\mu$ disappearance is more difficult given that the measurement of the spectral distortion plays a much more important role in its sensitivity to $\theta_{23}$ and the mass splitting in a way that cannot be readily approximated in a counting analysis. While the sensitivity to the mass ordering and the CP violation ultimately rests in the $\nu_\mu\to\nu_e/\bar{\nu}_\mu\to\bar{\nu}_e$ events, $\nu_\mu$ disappearance is essential and an inseparable part of the measurement program; a consistent accounting of all the oscillation parameters at play requires a joint analysis across the $\nu_\mu/\bar{\nu}_\mu$ disappearance and $\nu_e/\bar{\nu}_e$ appearance channels. A biased or otherwise compromised measurement of the former would imply that the many systematic uncertainties which are common between the two oscillation channels are not understood. Since $\theta_{23}$ and the mass splitting also directly impact the $\nu_e/\bar\nu_e$ appearance probabilities, they will impact the extraction of $\theta_{13}$ and $\delta_{CP}$ from this channel. Also, an extraction of these parameters serves as a valuable crosscheck with ongoing experiments such as T2K and NOvA.

\begin{dunetable}[DUNE \dword{nd} Electron neutrino event rates]
{ lrr}{tab:nuerates}{$\nu_e/\bar{\nu_e}$ appearance rates: Integrated rate of selected $\nu_e$ CC-like events between 0.5 and 
8.0 GeV assuming 3.5-year (staged) exposures in the neutrino-beam and antineutrino-beam modes. The signal rates are shown for both normal mass ordering (NO) and inverted mass ordering (IO), and all the background rates assume normal mass ordering. All the rates assume $\delta_{CP} = 0$, and NuFIT 4.0 [28, 29] values for other parameters. (Reproduced from the Far Detector Technical Design Report)}
	  	   & \multicolumn{2}{c}{Expected events (3.5 years staged per mode)} \\
				& $\nu$ mode	& $\bar{\nu}$ mode \\ \hline
$\nu_e$ Signal NO (IO)			  & 1092 (497)	& 76 (36) \\
$\bar{\nu}_e$ NO (IO)		& 18(31)		& 224 (470) \\ \hline
Total Signal NO (IO)	& 1110 (528)	& 300 (506) \\ \hline
Beam $\nu_e+\bar{\nu}_e$ background 	& 190	& 117 \\
NC background					& 81	& 38 \\
$\nu_\tau +\bar{\nu}_\tau$	CC background	& 32	& 20 \\
$\nu_\mu + \bar{\nu}_\mu$ CC background	& 14	& 5 \\ \hline
Total background	& 317	& 180 \\
\end{dunetable}

\begin{dunetable}[\dword{nd} Overarching Requirements]
{ll}
{tab:NDoverreq}{Overarching requirements for \dword{nd} .}
Label 		& Description  	 \\  \toprowrule
ND-O0	& 	Predict the observed neutrino spectrum at \dword{fd}\\ \hline
ND-O1	& 	Transfer measurements to \dword{fd}						\\ 
ND-O2	& 	Constrain the cross section model						\\ 
ND-O3	& 	Measure the neutrino flux						\\ 
ND-O4	& 	Obtain measurements with different fluxes						\\ 
ND-O5	& 	Monitor time variations of the neutrino beam						\\ 
ND-O6	& 	Operate in high rate environment					\\ 
\end{dunetable}

\subsection{Overarching Requirements}
\label{subsec:intro-overarching}
The overarching requirements are summarized in Table \ref{tab:NDoverreq}. Within these requirements, ND-O0 represents the ultimate goal of the \dword{nd} in the context of the long-baseline neutrino oscillation measurement, namely to predict the expected observables at the \dword{fd}, which include the number of selected neutrinos of each flavor, their reconstructed energy and other relevant kinematic variables ({\em e.g.} energy transfer), and  backgrounds, as a function of the oscillation parameters. This prediction is compared to the corresponding observations at the \dword{fd} to extract the oscillation parameters. The process is assisted by {\em a priori} information about the neutrino flux, neutrino interaction/cross-section model, and detector response, which provide  a starting point from which the \dword{nd} measurements must further constrain the systematic uncertainties in this model.

The other overarching requirements outline the basic ingredients needed to fulfill ND-O0. 
\begin{itemize}

\item {\bf ND-O1:}  \dword{nd} measurements must be transferable to the \dword{fd}. Since the \dword{fd} are \dword{lartpc}s,  the \dword{nd} must be able to measure interactions on an argon target, and furthermore must have a component that is a \dword{lartpc}.  The transfer must be performed accounting for uncertainties arising from detector modeling, including thresholds, efficiencies, purities, and resolutions for observables that are used in the \dword{fd}, as well as  uncertainties in the flux and cross-section prediction.
\item {\bf ND-O2:} The \dword{fd} performance couples the modeling of  the outgoing particles in terms of the exclusive and differential cross sections to the efficiency to identify these particles. The \dword{nd} detector must sufficiently measure and constrain the uncertainties in this modeling to minimize their impact on the oscillation measurement.
\item {\bf ND-O3:} The {\em ab initio} prediction of the neutrino flux based on Monte Carlo simulation  has significant uncertainties arising from particle production, beam optics, operational variation, {\em etc.} that must be constrained by the \dword{nd}. The various flavor components of the beam must also be suitably constrained.
\item {\bf ND-O4:} Due to the primary role of the neutrino energy in the oscillation physics and the significant model dependence in reconstructing this quantity,  the \dword{nd} must verify that its model predictions and constraints are robust by taking data with different neutrino spectra. 
\item {\bf ND-O5:} The flux and spectrum of neutrinos delivered by the beam can vary due to operational variations as well as unexpected component variances or failures. The \dword{nd}  must detect such variations promptly to minimize impact on overall data quality.
\item {\bf ND-O6:} The \dword{nd} must separate cosmic rays, rock muons, and other beam-induced activity from the activity associated with neutrino interactions in the \dword{fv} , including from other neutrino interactions that may be happening in the \dword{fv} (pile-up).
\end{itemize}

While the overarching requirements are intended to be agnostic of the particular implementation, there are two significant constraints which drive the design and requirements of the \dword{nd}. First is the choice of the \dword{lartpc} technology for the \dword{fd} which drives ND-O1. Second is the intense \dword{lbnf} beam and the relatively shallow near detector hall of the near site conventional facilities, which drive ND-O6. 

\begin{dunetable}[\dword{nd} Measurement Requirements]
{  p{.09\textwidth} p{.23\textwidth} p{0.08\textwidth}p{.32\textwidth}p{.09\textwidth}p{.08\textwidth} }
{tab:NDmeasreq}{Measurement requirements for \dword{nd} .}
Label 		& Description 							& Spec.  	&	Rationale 		& System & Ref. Req. 	 \\  \toprowrule
ND-M1	& Classify interactions and measure  outgoing particles in a \dword{lartpc} with performance comparable to or exceeding that of the \dword{fd}
																	& N/A					&     The \dword{nd} must have a \dword{lartpc} with reconstruction capabilities comparable/exceeding the \dword{fd} in order to effectively transfer measurements.																																														& \dword{ndlar}, \dword{ndgar}		& ND-O1, ND-O2\\ \hline
ND-M2	& Measure  particles in $\nu$-Ar interactions with uniform acceptance, lower thresholds than \dword{lartpc},  minimal secondary interaction effects
																	& N/A					& The ND must measure outgoing recoil particles ($\pi$, $p$, $\gamma$) in $\nu$-Ar interactions to ensure that sensitive phase space is properly modeled.
																									& \dword{ndgar}	& ND-O1, ND-O2\\ \hline	
ND-M3	& Measure the $\nu$ flux using $\nu-e$ scattering
																	& 2\%					& The ND must measure the flux normalization with this ``standard candle''.
																									& \dword{ndlar} & ND-O3\\ \hline
ND-M4	& Measure the neutrino flux spectrum using the 'low-$\nu$' method
											& 5\% for $E_\nu>1$ GeV			& The ND must identify/measure low recoil events which have flat energy dependence in order to measure the spectrum.
																									& \dword{ndlar}, \dword{ndgar} & ND-O3 \\ \hline
ND-M5	& Measure the wrong sign contamination		& FHC <20\%, RHC <5\%			&  The ND must measure and validate the modeling of wrong-sign interactions that dilute the oscillation asymmetries at the \dword{fd}. 
																								& \dword{ndgar}	& ND-O3			 \\ \hline
														
ND-M6	& Measure the intrinsic beam $\nu_e$ component
											& 2\%			& The ND must measure and validate the modeling of this irreducible background 
																								& \dword{ndlar}, \dword{ndgar} & ND-O3 \\ \hline	
ND-M7	& Take measurements with off-axis fluxes with spectra spanning  region of interest		
										& 0.5-3.0 GeV		& The ND must be able to move off the beam axis to take data with different neutrino spectra.
																								& \dword{ndlar}, \dword{ndgar}, \dword{duneprism} & ND-O4  \\ \hline	
ND-M8	&Monitor the  rate of neutrino interactions on-axis	 & <1\% in a week			& The ND must have a component that remains on-axis where beam monitoring is most sensitive and collects a sufficient number of $\nu_\mu$ CC events.	
																			& SAND		& ND-O5			 \\ \hline
ND-M9	& Monitor the beam spectrum on-axis		& N/A			& The ND must use spectrum and position information to detect representative changes in the beam line. 
																			& SAND		& ND-O5				 \\  \hline
ND-M10	& Assess External Background	& N/A			& The ND must measure external backgrounds, which include cosmic and beam-induced activity.
																			& \dword{ndlar}, \dword{ndgar}, \dword{sand}		& ND-O6				 \\                                             
\end{dunetable}

\subsection{Measurement Requirements}
\label{subsec:intro-measurements}
Measurements are the corresponding ``deliverables'' from the \dword{nd} that are needed to fulfill the overarching requirements as shown in Table \ref{tab:NDmeasreq}. Within each entry, there are overarching requirements which match to the particular measurement requirement, as well as the  subsystem(s) (\dword{ndlar}, \dword{ndgar}, \dword{duneprism}, \dword{sand}) that are primarily tasked to perform the measurement. 
\begin{itemize}
\item {\bf ND-M1:}  Due to the intrinsic coupling between the outgoing particles as modeled by the neutrino cross-section model (ND-O2) and the detector response (ND-O1), the \dword{nd} must have a  \dword{lartpc} component that performs comparably  or better than the \dword{fd} in all performance metrics relevant for identifying and reconstructing neutrino interactions at the \dword{fd} for a representative sample of neutrino interactions in order to directly inform how such interactions would appear in the \dword{fd}.   For the critical task of muon spectrometry, due to the limited space in the near detector conventional facilities which results in the inability to make muon momentum measurements by range for forward, high momentum muons from \dword{ndlar}, \dword{ndgar} provides this capability for neutrino interactions observed in \dword{ndlar}. Specific metrics are described in the related capabilities requirements that follow.
\item {\bf ND-M3-6:} These measurements relate to  measuring the neutrino flux as described in ND-O3. ``Standard candles'' such as $\nu-e$ elastic scattering (ND-M3) and  ``low-$\nu$''' events (ND-M4) with small energy transfer must be performed by the \dword{nd} in order to verify and reduce the uncertainties in the flux model. Due to the small cross section of $\nu-e$ interactions, this requirement also drives the fiducial mass and electron identification and reconstruction capabilities of \dword{ndlar}, which will perform this measurement. This will also allow it to perform a measurement of the intrinsic $\nu_e$ content of the beam (ND-M6) that is an irreducible background to $\nu_\mu\to\nu_e$ events at the \dword{fd}. The sign selection capabilities of \dword{ndgar} allow the measurement of the ``wrong sign'' content of the beam ({\em i.e.} neutrinos in \dword{rhc} and vice-versa) which dilute $\nu/\bar{\nu}$ asymmetry measurements at the \dword{fd} (which does not have this separation),  with $\nu_\mu$ events originating in both \dword{ndlar} and \dword{ndgar} and $\nu_e$ events in \dword{ndgar}. Target uncertainties in these measurements are set so that they saturate the systematic error budget for a $5\sigma$ observation of CP violation in the most favorable scenarios.
\item {\bf ND-M2:}  Systematic errors in the \dword{fd} will depend on the accuracy with which thresholds, acceptances, and other detector effects in {\lartpc} ({\em e.g.} secondary interactions) are modeled, which couple to the intrinsic properties of the neutrino interactions in terms of the multiplicity, topology, type, and kinematics of the particles emerging (mainly pions and nucleons) from the interaction,  and impact the performance of \dword{lartpc}s, including \dword{ndlar}. A magnetized low density argon-based detector surrounded by a calorimeter and a muon system (much like a collider detector)  verifies these intrinsic properties are properly modeled prior to the detector effects associated with the dense tracking medium in the \dword{fd} and \dword{ndlar}. 
\item {\bf ND-M7:} The primary means by which the spectrum at the \dword{nd} will be varied (ND-O4) is by  \dword{duneprism}, which exploits the steady decrease and narrowing of neutrino energies as one samples the beam further from the beam axis. Localized variations of the spectrum across the energy range of interest for neutrino oscillation measurements are needed to validate the model across these energies.
\item {\bf ND-M8, 9:} The on-axis neutrino flux which is incident on the \dword{fd} must be continuously monitored for potential variations in the beam line operations, both controlled and inadvertent. The on-axis position also has the largest spectrum variation and flux in the event of any such variation. ND-O5 is fulfilled by the \dword{sand}, which must remain on-axis and have sufficient rate (ND-M8), muon spectrometry, and position capabilities (ND-M9) to perform this monitoring.
\item {\bf ND-M10}: Due to the shallow site and the intensity of the neutrino beam, the \dword{nd} operates in an environment with cosmic rays and a high level of beam-induced background activity. In order to verify that these backgrounds are correctly accounted for and modeled, the \dword{nd} must be able to measure them. 
\end{itemize}

Since the neutrino interactions occur throughout the detector volume, it is impossible to ensure uniform acceptance and efficiency throughout the detector volume. Likewise, since \dword{ndlar} is significantly smaller than the \dword{fd}, the containment of particles necessary for their full reconstruction in the \dword{lartpc} is an issue. In this metric, it is impossible for \dword{ndlar} to match the \dword{fd}. However, due to the very high statistics at \dword{ndlar}, so long as the full phase space of interactions is represented within the fully-contained sample at \dword{ndlar}, high efficiency across the phase space is not necessary. This is what is meant by a ``representative sample'' within the context of ND-M1, namely that some fraction of the events across all areas of relevant interaction phase space (neutrino energy and energy transfer) can be fully contained and reconstructed within \dword{ndlar}, with \dword{ndgar} providing spectrometry for uncontained muons. Within the context of ND-M2, \dword{ndgar}, which uses magnetic spectrometry and does not rely on containment,  provides an important cross check on these assumptions.

\begin{dunetable}[Capabilities Requirements for \dword{ndlar} for ND-M1]
{  p{.12\textwidth} p{.23\textwidth} p{0.14\textwidth}p{.25\textwidth}p{0.10\textwidth} }
{tab:NDlarcapreqm1}{ND-M1 capability requirements for \dword{ndlar} with \dshort{ndgar} acting as a muon spectrometer .}
Label 	& Description 							& Specification  	&	Rationale 		&	Ref. Req. \\ \toprowrule
ND-C1.1	&  Classify events, measure outgoing particles with performance comparable/exceeding that of the \dword{fd} 					
											& N/A			& To  translate measurements,  \dword{ndlar} must reconstruct neutrino events with comparable/better performance than the \dword{fd}.
																			& ND-M1  \\ \hline	
ND-C1.1.1	&   $\nu_e$ identification					
											& Eff. >90\%, Bkg. < 3\%, $E_\nu$ res. <10-15\% 			& \dword{ndlar} must identify and reconstruct $\nu_e$  events as well as FD.
																			& ND-M1  \\ \hline	
ND-C1.1.2	&  $\nu_\mu$ identification					
											& Eff. >95\%, Bkg. < 3\%, $E_\nu$ res. <17\% 			& \dword{ndlar} must identify and reconstruct $\nu_\mu$ events as well as FD.
																			& ND-M1  \\ \hline	
ND-C1.1.3	&  Contained particle reconstruction					
											& Eff.>90\% for $e/\gamma$>20 MeV/c, $\mu/\pi^\pm$>100 MeV/c, $p$>500 MeV/c			& Contained particles should be detected as well as FD.
																			& ND-M1  \\ \hline	
ND-C1.1.4	&  Phase space coverage					
											& <1\% uncovered phase space		& Event topologies and kinematics where no geometric configuration would contain the hadron shower must be limited.
																			& ND-M1  \\ \hline	
\end{dunetable}

\begin{dunetable}[Capabilities Requirements for \dword{ndlar} for ND-M3]
{  p{.12\textwidth} p{.23\textwidth} p{0.14\textwidth}p{.25\textwidth}p{0.10\textwidth} }
{tab:NDlarcapreqm3}{ND-M3 capability requirements for \dword{ndlar}.}
Label 	& Description 							& Specification  	&	Rationale 		&	Ref. Req. \\ \toprowrule
ND-C1.2	&  Sufficiently large sample of $\nu$-e elastic events  identified with high efficiency and low backgrounds
											& $\sim 2\%$		& This is necessary to perform an adequate $\nu-e$ elastic measurement for the flux measurement.
																			& ND-M3 \\ \hline
ND-C1.2.1& Fiducial mass/statistics					& $>2500$ ev/yr 
															& \dword{ndlar} must collect sufficient statistics to allow $<2\%$ statistical uncertainty in the measurement 
																			& ND-M3 \\ \hline
ND-C1.2.2& $\nu-e$  identification					&
															& \dword{ndlar} must be able to distinguish the outgoing electron from other particles ($\mu$, $\gamma$, $\pi^0$)
																			& ND-M3 \\ \hline
ND-C1.2.3& Electron energy resolution				& $5\%$
															& Energy resolution is needed to identify the forward $\nu-e$ events.
																				& ND-M3  \\ \hline
ND-C1.2.4& Electron angular resolution				& core$<5$ mrad, tail$<12$ mrad for $E_e >$2 GeV
															& A tight cut on forward electrons is needed to identify $\nu-e$ events
																					& ND-M3  \\ \hline
ND-C1.2.5& Vertex activity threshold					& 20 MeV 			& Identifying vertex activity is necessary to reject backgrounds
																					& ND-M3  \\ \hline \hline
ND-C1.3.1 	& Timing resolution for scintillation detection & < 20 ns &  Scintillation timing is required to set the $t_0$ for the charge readout and to separate event pile-up
																				& ND-M6	\\
ND-C1.3.2 	& Intermodule  synchronization of scintillation detection. & < 20 ns &  Timing between modules must seamlessly integrate activity observed in the separate modules.
																				& ND-M6	\\
\end{dunetable}

\subsection{Capability Requirements}
\label{subsec:intro-capabilities}
The capability requirements which flow from the measurements are summarized in Tables \ref{tab:NDlarcapreqm1} and \ref{tab:NDlarcapreqm3}  (\dword{ndlar}), \ref{tab:NDgarcapreq} (\dword{ndgar}), \ref{tab:NDprismcapreq}(\dword{duneprism}), and \ref{tab:NDsandcapreq} (\dword{sand}). We visit each subsystem in turn.

The \dword{ndlar} capability requirements are grouped into three parts based on the measurements they support. The muon reconstruction capabilities required to fulfill some of these requirements are separately described.

\subsubsection{ND-M1/ND-C1.1.(1-4): Match \dword{fd} Performance in \dword{ndlar}}
\dword{lartpc}s provide information in the form of  tracking with detailed calorimetry.  Performance of \dword{ndlar} relative to the \dword{fd} is determined primarily by the ability to identify and reconstruct tracks  ($e$, $\mu$, $\pi^\pm$, $p$) and showers ($e/\gamma$). These are driven by containment (size of active volume), sampling (effective voxel size), dynamic range, dead material due to the modular structure, operational parameters (drift field and argon purity), and light collection,  and motivate corresponding technical requirements for \dword{ndlar}. The relative performance can be verified through simulation, supported in some cases by data from prototypes. The capability requirements on $\nu_\mu/\nu_e$ reconstruction and particle tracking follow from what has been currently demonstrated in the \dword{fd}.

Requirements ND-C1.1.1 and ND-C.1.1.2, regarding neutrino flavor selection ($\nu_e$ CC and $\nu_\mu$ CC), are based on the currently achieved \dword{fd} performance in simulations as summarized in Figure 5.14 ($\nu_e$ CC selection efficiency and purity), Figure 5.15 ($\nu_\mu$ CC selection efficiency and purity), Figure 5.9 ($E_\nu$ resolution for $\nu_\mu$ CC), and Figure 5.11 ($E_\nu$ resolution for $\nu_e$ CC) in Volume 2 of the \dword{fd} TDR\cite{Abi:2020evt}. Given the energy dependence of these performance metrics, it is difficult to capture the performance in a single number, and thus the stated requirements should be taken as indicative benchmarks. For the $\nu_\mu$ CC, the energy resolution requirements implies that \dword{ndgar} must measure muon momentum as well as \dword{fd} can through range in LAr.

ND-C.1.1.3 references the reconstruction efficiency for identifying individual contained particles in simulated neutrino interactions that have been achieved in \dword{fd}. These are summarized in Figure 4.26 in Volume 2 of the \dword{fd} TDR\cite{Abi:2020evt}. For track-like particles, the efficiency is primarily a function of the number of observed hits, which correlates to the momentum given the particle mass. For muons and pions, high efficiency ($\sim 90\%$) is achieved for momentum greater than $\sim 100\; \mbox{MeV}/c$ while efficiency for protons reach this level at around $\sim 500 \mbox{MeV}/c$. 

The ability for LArTPCs to accurately reconstruct particles depends on their containment, and thus matching the \dword{fd} performance relies on matching to some extent on its containment capabilities. However, practical constraints on the dimensions  \dword{ndlar} mean {\em a priori} that the fraction of contained (and thus accepted) events will be low given that the typical particle path lengths and the dimensions of  \dword{ndlar} ($\mathcal{O}(\mbox{1 m})$) are similar. For a given event, containment will depend on its geometric configuration, for example the proximity of the interaction vertex  and the direction of the emerging particles with respect to the periphery of the active volume. The aspect ratio of the transverse dimensions of the detector also gives rise to an azimuthal dependence of the acceptance about the neutrino beam direction. However, the enormous statistics expected at \dword{ndlar} ($\mathcal{O}(10^8)$ interactions/year) allow low acceptance to be tolerated so long it is well understood and does not become negligible for any appreciable part of the phase space.  In order that such neutrino events do not contribute appreciably to the error budget of $\sim 2\%$, we require that this uncovered phase space be less than $1\%$ of the expected \dword{fd} sample. This requirement will be studied further in Section \ref{sec:lartpc-dimensions} and is one of the primary drivers of the required detector mass.

The requirements here implicitly assume that the high momentum muons which are not contained within \dword{ndlar} are suitably reconstructed by \dword{ndgar}, as discussed in ND-C2.(1-3).
 
\subsubsection{ND-M3/ND-C1.2.(1-5): $\nu-e$ elastic scattering} 
These requirements relate to the  capabilities needed to identify a sufficiently large and pure sample of $\nu-e$
scattering events that serve as a ``standard candle'' with precisely known cross section. The fiducial mass of the \dword{ndlar} must be sufficiently large for a measurement with $\sim 2\%$ statistical uncertainty, corresponding to $\sim 2500$ events, to limit the impact of flux uncertainties on the total error budget (ND-C1.2.1). A measurement of this precision should be possible each year with nominal beam intensity given the potential variability of the \dword{lbnf} beam line.

The selection of these rare events depends on their characteristic signature of a single forward electron with no associated hadronic recoil activity. \dword{ndlar} must be capable of rejecting non-electron single particle final states such that the remaining background is from $\nu_e$ CC events (ND-C1.2.2). The identification of the (energy-dependent) forward electron signature relies on sufficient resolution on the electron energy (ND-C1.2.3) and angle (ND-C1.2.4). The remaining background arising from $\nu_e$ CC interactions is suppressed by rejecting events with identifiable recoil hadronic activity (ND-C1.2.5). The quantitative requirements are derived from the study described in \cite{Marshall:2019vdy}.

\subsubsection{ND-M4: Low-$\nu$} 
The ``low-$\nu$'' is a flux measurement method that makes use of the fact that the inclusive neutrino interaction cross section is nearly constant with neutrino energy in the limit of low energy transfer ($\nu$) to the nucleus  ($\nu/E_\nu\ll 1$)\cite{MISHRA-Nu0}. By selecting such events, the spectrum of the neutrino flux can be measured. In practice, the method is limited by thresholds in identifying low-$\nu$ events and modelling uncertainties. \dword{minerva} recently reported a measurement of the \dword{numi} low energy flux for $E_\nu>2\;\mbox{GeV}$ using this method with typical uncertainties of 5-10\% and $\nu$ thresholds of 0.3-0.8 GeV in the neutrino energy range most relevant for \dword{dune}\cite{DeVan:2016rkm}.

The role of the low-$\nu$ measurement in the \dword{dune} oscillation measurements is somewhat indirect. 
The low-$\nu$ cross section is approximately independent of energy and this removes the energy dependence of the cross section.   It serves as a critical crosscheck and powerful diagnostic on the modeling of the beam.  
The capabilities of \dword{ndlar} should allow a lower threshold, which would reduce model uncertainties at higher energy and allow the measurement to be extended to lower neutrino energies. As a reference, we target $<5\%$ uncertainty in order to verify the shape of the \dword{lbnf} beam $\nu_\mu$ spectrum beyond the $5-10\%$ uncertainty expected from {\em ab initio} modelling (see Figures 4.5 in \cite{Abi:2020evt}).

\subsubsection{ND-M5: Wrong sign background} 
 The ``wrong-sign'' background is worst in the case of $\bar{\nu}$-mode, where the larger interaction cross sections for neutrinos relative to antineutrinos result in $\nu_\mu\to\nu_e$ events being up to half of the signal $\bar{\nu}_\mu\to\bar{\nu}_e$ (20\% of the total) event rate depending on the oscillation parameters. A $5\%$ measurement of the $\nu_\mu$ event rate in \dword{rhc} is needed to constrain the uncertainty in $\nu_\mu\to\nu_e$ events to be less than $1\%$  of the total event rate for a given set of oscillation parameters. In $\nu$-mode, the opposite is the case, and  a $20\%$ uncertainty is sufficient to achieve the same uncertainty in the relative contribution of  $\bar{\nu}_e$ events from oscillations to the total event rate.
 
ND-C1.1.1 provides the requisite capabilities to identify and reconstruct the $\nu_\mu$ CC events when supplemented with the sign selection capabilities in ND-C2.(1-4) which provides the muon matching and momentum measurement and ND-C3.1 which provides the muon sign selection to separate the ``right'' and ``wrong'' components when applied to muons entering \dword{ndgar} from \dword{ndlar} as described below.

\subsubsection{ND-M6: Intrinsic $\nu_e$ Background} 
The intrinsic $\nu_e/\bar{\nu}_e$ flux that is present in the initial neutrino flux represents a source of irreducible background to the $\nu_e/\bar{\nu}_e$ appearance signal. While its expected rate does not depend significantly on the oscillation parameters,   its relative contribution to the total event rate can be up to 30\% due to the variation in the signal process. A $2\%$ measurement of the rate ensures that the resulting impact on the predicted rate of signal candidates remains below 1\%.                                                                                             

The large mass and $\nu_e$ CC identification capabilities of \dword{ndlar}  allows a precise measurement of the intrinsic beam $\nu_e$ component. The requirements associated with identifying $\nu_e$ events in ND-C1.1 and the more demanding task of identifying $\nu-e$ elastic scattering events allow this capability to also be fulfilled.

\subsubsection{ND-M10/ND-C1.3.(1,2)} 
Timing is the primary means by which beam activity is separated from non-beam background. While pattern recognition using tracking and shower reconstruction with the charge signals from \dword{ndlar} is expected to be a powerful handle to separate activity from different neutrino interactions, the optical signal plays an important role and provides an independent check on this intricate process.

The timing requirements of \dword{ndlar} (ND-C1.3.1) are set such that the pile-up of O(100) events over 10 $\mu$sec, resulting in an average spacing of $\mathcal{O}(100)\;\mbox{ns}$, can be separated using the faster optical signal in the detector. Furthermore, commensurate timing synchronization is required between the  modules (ND-C1.3.2).

\begin{dunetable}[\dword{ndgar} Capabilities Requirements]
{  p{.10\textwidth} p{.23\textwidth} p{0.11\textwidth}p{.30\textwidth}p{0.10\textwidth} }
{tab:NDgarcapreq}{Capability requirements for \dword{ndgar} .}
Label 		& Description 														& Specification  	&	Rationale 		&	Ref. Req. \\ \toprowrule
ND-C2.1	& Acceptance for muons exiting \dword{ndlar}		& $p_\mu>1$ GeV/c, $\theta_\mu<30^\circ$  & 	\dword{ndgar} must detect and analyze muons exiting the \dword{ndlar} without a gap in phase space.	
																								& ND-M1	\\ \hline
ND-C2.2	& Momentum resolution for muons exiting \dword{ndlar}
																								&core$ <4\%$, RMS < $10\%$			
																																& \dword{ndgar} must measure these $\mu$ at least as accurately as the \dword{fd} would.
																																							& ND-M1 \\	\hline
ND-C2.3	& Time resolution for muons exiting \dword{ndlar}
																								& Work-in-Progress			
																																& \dword{ndgar} must determine of the timing of $\mu$ tracks to separate each track from other activity.
																																							& ND-M1 \\	\hline                                                                             ND-C2.4	& Time synchronization with \dword{ndlar}
																								& <20 ns			
																																& \dword{ndgar} must be synchronized with \dword{ndlar} to match activity in the two detectors
																																							& ND-M1 \\	\hline\hline
ND-C3.1	& Sign-select/momentum analyze $e^{\pm},\mu^{\pm}$ across the range of interest 
																								& $[0.1,10]$ GeV/c) 
																																& Precise lepton reconstruction is needed for detailed kinematic studies, beam $\nu_e$, and wrong-sign measurements
																																						& ND-M2, ND-M4, ND-M5, ND-M6 \\ \hline
ND-C3.2	& Detect, identify,  measure momentum of protons emitted from $\nu$-Ar interactions
																							& <10 MeV (5 MeV)	& Low energy proton reconstruction is needed to verify FSI models and LAr response modeling.
																																& ND-M2, ND-M4 \\ \hline	
ND-C3.3	& Detect, identify, sign-select, measure the momentum of $\pi^{\pm}$ emitted from $\nu$-Ar interactions 
																							&<20 MeV (5 MeV)	& Pion multiplicity and spectrum must be measured to ensure accurate $\nu$-Ar and LAr response modeling.
																																&  ND-M2, ND-M4 \\ \hline
ND-C3.4	& Momentum resolution 								&								& Precise momentum resolution of charged recoil particles is needed to study thresholds and measure spectra. 
																																& ND-M2, ND-M4 \\ \hline
ND-C3.5	& Charged particle identification 				&				& Recoil particles must be identified to categorize interactions, tag flavor,  and verify modeling and \dword{ndlar} thresholds.
																											& ND-M2, ND-M4, ND-M6\\ \hline
ND-C3.6	& Detect and measure the $\pi^0$'s with the photons from their decay 
																						&				& $\pi^0$'s must be identified/reconstructed to have a complete model of the pion emission from $\nu$-Ar interactions						
																											& ND-M2, ND-M4\\\hline
ND-C3.7	& Absolute time measurement for tracks
																						&				&  Precise timing is required to provide an absolute reference for the charge signal in the HPgTPC of \dword{ndgar}						
																											& ND-M2, ND-M4, ND-M10\\

\end{dunetable}

\subsubsection{ND-M1,4,5,10/ND-C2.(1-4): Reconstruction of muons from \dword{ndlar} with \dword{ndgar}} 
Since many of the muons from $\nu_\mu$ CC events in \dword{ndlar} will not be contained therein, it is not possible to reconstruct the muon energy using range as they would (in most cases) at \dword{fd}. The energy of these muons could be estimated using their multiple scattering through the liquid argon volume, however, the resolution is far worse than that achieved by range. \dword{ndgar} fills this gap by performing momentum measurements via curvature in the magnetic field. In order to fulfill ND-M1, \dword{ndgar} must provide sufficient acceptance for the muons exiting \dword{ndgar} (ND-C2.1) and measure their momentum with resolution matching that of \dword{fd} (ND-C2.2), so that the overall reconstruction of these events is of the same quality as that achieved in \dword{fd}. 

Section \ref{sec:lartpc-muonreco} describes the acceptance of the \dword{ndlar} for low energy muons $<1\;\mbox{GeV}/c$ and the necessary angular acceptance for higher momentum muons. An acceptance gap remains for those muons which exit the \dword{ndlar} active volume but do not enter the \dword{ndgar} spectrometer that depends on where the interaction occurs in \dword{ndlar}. This gap is discussed further in Section \ref{sec:lartpc-muonreco}. As for ND-C1.1.4, due to the high statistics, low but well-understood acceptances can be tolerated so long as there is no  portion of the phase space where there is no acceptance. ND-C2.3 and ND-C2.4 ensure that \dword{ndgar} have sufficient timing capabilities to match the observed muon to its parent interaction in \dword{ndlar} in the high rate environment. The required time resolution within \dword{ndgar} (ND-C2.3) is still under study, but its synchronization with \dword{ndlar} should be comparable to what is achieved within the modules of \dword{ndlar}.

\subsubsection{ND-M2,4,5,6/ND-C3.(1-7): Low-threshold, uniform acceptance measurements in \dword{ndgar}} 
The density of liquid argon results in secondary interactions and shorter track lengths for the hadrons emerging from a neutrino interaction in \dword{ndlar}. Section \ref{sec:mpd:pionmult} demonstrates how these limitations, in the case of resolving pions emerging from the interaction, impact the ability to correct the $\nu-\mbox{Ar}$ interaction modelling and result in systematic uncertainties that bias the oscillation measurement. This motivates the ability to  reconstruct these final states in a low density tracking volume with magnetic analysis which allows for lower tracking thresholds and negligible secondary interactions. Since magnetic spectrometry does not require containment, events in \dword{ndgar}  will have more uniform acceptance and sign determination, both of which supplement limitations of \dword{ndlar}.

Requirement ND-C3.1 allows \dword{ndgar} to perform sign selection on the primary lepton for neutrino interactions in \dword{ndgar} across the range of interest so that the ``wrong sign'' measurements can be performed. This ability naturally extends to muons entering \dword{ndgar} from interactions in \dword{ndlar}. The proton tracking thresholds  (ND-C3.2) is motivated by the ability to distinguish different nuclear models (Section \ref{sec:mpd:protons}) while the pion thresholds allow the full spectrum of charged pions to be measured (Section \ref{sec:mpd:pionmult}). The momentum resolution (ND-3.4), particle identification (ND-C3.5), and photon reconstruction (ND-C3.6) requirements are still under study, but should allow \dword{ndgar} to fully identify and kinematically analyze the protons and pions in an event. Finally, as a TPC, a fast signal to determine the reference time ($t_0$) in each event is needed (ND-3.7). The details of this requirement are still under study.

\begin{dunetable}[\dshort{duneprism} Capabilities Requirements]
{  p{.09\textwidth} p{.23\textwidth} p{0.14\textwidth}p{.30\textwidth}p{0.09\textwidth} }
{tab:NDprismcapreq}{Capability requirements for \dword{duneprism} .} 
Label 		& Description 							& Specification  	&	Rationale 		&	Ref. Req \\ \toprowrule
ND-C4.1	& Maximum travel distance			 	& $>30.5$ meters 		& This distance is necessary to cover the relevant energy range
																				& ND-M7	\\ \hline
ND-C4.2	& Maintain uniform detector performance across the full off-axis range
											&		<1\% for relevant quantities such as efficiencies, resolutions, {\em etc.}		&	Uniform performance is needed to make comparative measurements across data taken at different locations 
																				& ND-M1, ND-M7 \\ \hline
ND-C4.3	& Place the detector with sufficient granularity and accuracy 
											& 	<10 cm gran., <1 cm acc.			&  The positioning must be granular and precise enough to control spectrum and detector response variations.
																				& ND-M7 \\ \hline
ND-C4.4	& Minimize downtime for motion 			&	$<$8 hours	&  The ramp down, movement, and ramp up cycle must be as short as possible 
																				&  ND-M7 \\ \hline
ND-C4.5	& Regular suite of measurements 			&  $<1$ year	& Given expected annual variations in the beam, a suite of PRISM measurements should be taken each year.
																				& ND-M7 \\ \hline
\end{dunetable}

\begin{dunetable}[\dshort{sand} Capabilities Requirements]
{  p{.09\textwidth} p{.18\textwidth} p{0.18\textwidth}p{.32\textwidth}p{0.09\textwidth} }
{tab:NDsandcapreq}{Capability requirements for \dword{sand}.} 
Label 		& Description 							& Specification  	&	Rationale 		&	Ref. Req \\ \toprowrule
ND-C5.1	& Statistics of identified $\nu_\mu$ CC events &	 $>20$ tons for $p_\mu$,   $> 5$ tons for  $E_\nu$ & SAND must collect and identify enough $\nu_\mu$ CC interactions to perform beam monitoring on a weekly basis.
																			& ND-M8		\\ \hline 				
ND-C5.2	& $E_\nu, p_\mu$ resolution 						& $\sigma(p_\mu)/p_\mu<10\%$ at 5 GeV/$c$, improving to 5\% at 1 GeV/$c$, or $\sigma(E_\nu)/E_\nu < 15\%$							& SAND must have sufficient muon resolution to detect spectral variations in $\nu_\mu$ CC events from a representative set of variations in a week. 
																		&  ND-M9		\\ \hline
ND-C5.3	& Vertex reconstruction				&	<5 cm						& SAND must have the ability to determine the neutrino vertex to separate interactions occurring over distances where the spectrum may vary.
																			&  ND-M9		\\ \hline
ND-C5.4	& Track timing				&			<5 (1) ns in tracker, <400 ps on hits in ECAL				&SAND must have timing to identify and separate activity occurring within the neutrino beam delivery window. Better (1 ns) resolution would further enable directionality capabilities.
																			&  ND-M9, ND-M10		\\ \hline                                                                            
\end{dunetable}

\subsubsection{ND-M7/ND-C4.1-5: DUNE-PRISM} 

The \dword{duneprism} capability requirements pertain to the off-axis measurement requirements in ND-M7. The neutrino energy spectrum intercepted by the detector narrows and peaks at lower energies as the detector moves from the beam axis (``off-axis''), as shown in Figure \ref{fig:prism-1doffaxis}. These variations provide an independent handle on the energy of neutrinos observed by the detector, and by combining distributions of a target variable ({\em e.g.} the reconstructed neutrino energy) obtained at a range of off-axis angles through a weighted linear combination, the expected event distribution for a narrow pseudo-Gaussian neutrino flux or for the oscillated neutrino flux at \dword{fd} can be obtained via the methods described in Section \ref{sec:prism-lincomb}.  

The required range of off-axis travel (ND-C4.1) relates to the neutrino energy range over which the oscillation measurement is performed at \dword{fd}: a larger range, particularly towards lower energies, allows more of the oscillation probability to be measured. A lower threshold of 0.5 GeV is set by the DUNE Global Science Requirements and this translates to ND-C4.1.1. The 30.5 meter range of measurements provide enough variation in the flux to produce a pseudo-Gaussian flux and to model the oscillated \dword{fd} flux down to this threshold (see Section \ref{sec:prism-lincomb}).

The method relies on stable detector performance across the various off-axis measurements. Otherwise, the data taken at the different locations/fluxes would not be equivalent and result in biases in the linear combinations. This leads to ND-C1.4.2, which requires that the variations in detector performance with respect to the targeted variable ({\em e.g.} selection efficiencies, reconstructed neutrino energy) are less than $1\%$.

The linear combinations require suitable granularity and accuracy in  placing detector at the various locations, leading to ND-C4.3. The $<10\;\mbox{cm}$ granularity results from two length scales. First is the length scales over which the neutrino flux changes appreciably. Near the beam axis, the peak neutrino energy changes by about $1\%$ with each 10 cm displacement from the beam axis. Placement granularity at 10 cm thus allows the peak energy to be tuned to the level of the expected systematic uncertainties in energy reconstruction scale. Second is the detector geometry, where the \dword{ndlar} modules are composed of drift volumes with 50 cm length transverse to the beam direction, within which the detector performance may vary. A 10 cm granularity allows a given off-axis displacement to be placed at various points through this drift volume to cross-check the modelling of any performance variations through the drift volume. The placement accuracy and reproducibility should be significantly better than this granularity, motivating the 1 cm requirement.

Requirement ND-C4.4 relates to minimizing downtime during the motion of the detectors to different off-axis locations, during which they are presumed to be inoperable. To this end, we require that the transition between the end of data-taking in one place and the start at another (accounting for ramp down, movement, ramp up and stabilization) can be performed in an eight hour shift. Due to potential changes in the neutrino flux from the \dword{lbnf} beam line resulting from maintenance, component variations, {\em etc.}, we also require with ND-C4.5 that the required suite of off-axis measurements can be performed each year.


\subsubsection{ND-M8,9/ND-C5.1-4: On-axis Beam Monitoring}
In relation to the on-axis beam monitoring performed by \dword{sand} (ND-M8,9), there are four capability requirements. The requirements are studied based on variations in the on-axis neutrino flux resulting from the simulation of representative changes in the \dword{lbnf} beam line informed by past neutrino beam operation as described in Section \ref{sec:beam-monitoring}. The beam monitoring is considered using two observables to detect these variations, namely the observed muon momentum ($p_\mu$) spectrum or the reconstructed neutrino energy ($E_\nu$) in $\nu_\mu$ CC events. Since the variation in the neutrino flux directly impact the neutrino energy spectrum, the latter is more sensitive, with $p_\mu$ diluting the spectral variations. As a result, $p_\mu$-based beam monitoring requires more statistics.

Based on the studies in Section \ref{sec:beam-monitoring}, we find that 5 (20) tons of fiducial mass provides sufficient statistics to detect these respresentative  beam variations with sufficient significance in a one week period using $E_\nu$ ($p_\mu$)-based monitoring, resulting in ND-C5.1. The typical scale of the spectrum variations in both energy and momentum is $\sim 1 \;\mbox{GeV}$, motivating ND-C5.2. Variations in the beam optics or the horn alignment can shift the beam center and can be better detected by analyzing the positional dependence of the spectra. By separately evaluating the $p_\mu,\;E_\nu$ spectrum in four quadrants about the beam axis, the sensitivity to these beam variations is significantly improved. The 5 cm requirement in ND-C5.3 allows this capability in the $\sim 2\;\mbox{m}$ span of the \dword{sand} tracker target; practically, much better vertex resolution is achieved and this requirement is easily met and exceeded.

Finally, the timing requirements ND-C5.4 for the tracking system mirrors that of \dword{ndlar}, namely to ensure that high rate of neutrino interactions occurring throughout \dword{sand} as well as other beam-induced activity elsewhere (in the magnet, hall, other detector systems, etc.) can be separated in time within the $\sim\!20$ ns bunch structure.
For the ECAL, the 400 ns requirement on hits in the ECAL allow directionality of the tracks to be established and thus the capability to distinguish incoming and outgoing activity.

The beam monitoring requirements are based on simulating potential variations in the beam configuration, which results in a new predicted neutrino flux, and assessing the sensitivity of the reconstructed muon and neutrino energy spectra to this change. The criteria will be improved by taking into account the magnitude of their impact on the neutrino flux. For example, if a variation results in a small change in the flux, the requirement to detect the underlying variation should be  relaxed. Conversely, if a variation with an underlying parameter ({\em e.g.} the movement to of a component from its default position and  orientation) gives rise to large changes in the flux, then the variation resulting from a smaller deviation of this parameter should be detectable. The role of measuring the profile of the neutrino interactions and the position dependence of the spectrum need to be better understood. 

\section{Management and Organization of the Near Detector Effort}
\label{sec:intro-organization}

All aspects of the \dword{dune} \dword{nd} are organized and managed by the \dword{dune} collaboration. Stakeholders include the collaborating institutions and Fermilab, as the host laboratory. All collaborating institutions have a representative on the \dword{dune} institutional board. The Collaboration is responsible for the design, construction, installation, commissioning, and operation of the \dword{nd} and prototypes created en route to the construction of the \dword{nd}.

The \dword{dune} Executive Board (EB), described below, is the main management body of the collaboration and approves all significant strategic and technical decisions. The top-level \dword{dune} management team consists of two elected co-spokespersons, the technical coordinator (TC), and the resource coordinator (RC). The TC and RC are selected jointly by the co-spokespersons and the Fermilab director. The management team is responsible for the day-to-day management of the collaboration and for developing the overall collaboration strategy, which is presented for approval to the EB. The EB consists of the leaders of the main collaboration activities. The composition of the EB, currently including the \dword{dune} management team, institutional board chair, physics coordinators, beam interface coordinator, computing coordinator, and leaders of the \dword{fd} and \dword{nd} consortia, described below, is intended to ensure that all stakeholders in the collaboration have a voice in the decision making process.

A Near Detector Design Group (NDDG) was in place through to the conclusion of the \dword{nd} Conceptual Design Report and had the responsibility of developing the design of the \dword{nd} as described in this document. To carry out design and construction work for the \dword{dune} \dword{nd}, \dword{dune} has formed consortia of institutions that have responsibility for different detector subsystems. The structure is parallel to the consortia structure previously developed for the \dword{fd}. Currently, there are two consortia specific to the \dword{nd}:
\begin{itemize}
\item ND Liquid-argon consortium (ND-LAr)
\item ND Beam Monitor consortium (SAND).
\end{itemize}
The Computing consortium will cover the needs of  both the \dword{fd} and \dword{nd}. We have also formed a group (“proto-consortium”) for the ND gas-argon detector (ND-GAr), which is expected to become a full consortium once design and resources have been established. This group is also responsible for the design of a possible Day 1 detector in this location (Temporary Muon Spectrometer, TMS)\footnote{The Day 1 detector and TMS are not discussed further in this document.  They are described in other documents under preparation and may be covered in detail in the \dword{nd} \dword{tdr}.}. It is as yet undecided whether  the DAQ effort should be structured as a separate \dword{nd} consortium or integrated into a joint FD+ND consortium. A task force has been formed to study this question. The physics analysis activities are integrated into the \dword{dune} physics organization coordinated by the Physics Coordinators.

Each consortium has an overall leader, a technical lead, and a consortium board with representatives from each consortium institution. The consortium leaders, as well as one of the leaders of the \dword{ndgar} group and one of the TDR editors are members of the \dword{dune} EB and the Technical Board (TB). The technical leads are members of the \dword{dune} TB. The \dword{nd} sub-groups of the EB and the TB meet regularly with DUNE management to discuss ND-specific issues and plans.  

The consortia have full responsibility for their subsystems and will be responsible for developing a work breakdown structure (WBS), understanding and documenting all interfaces with other systems, preparing final technical designs, and writing their respective sections of the Technical Design Report (TDR). Following approval of the TDR, they will be responsible for constructing their detector systems. 

Figure~\ref{fig:NDorganization} gives a graphical view of the \dword{dune} management structure including the \dword{nd}.  Consortia management structures are shown in the relevant chapters of this document.

\begin{dunefigure}[DUNE ND management ]{fig:NDorganization}
{Organization chart of the DUNE ND management structure. }
\includegraphics[width=1.0\textwidth]{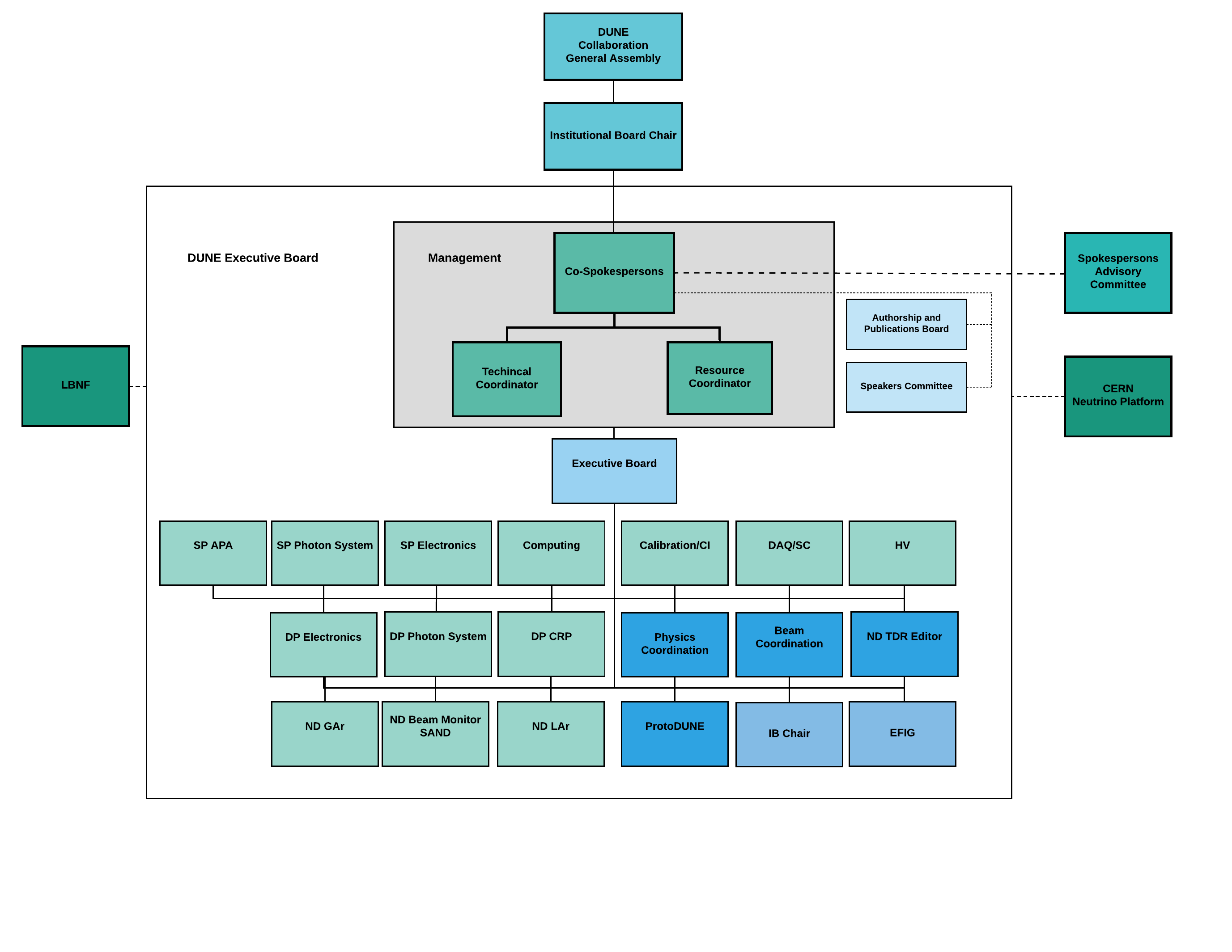}
\end{dunefigure}
\cleardoublepage

\chapter{Liquid Argon TPC - ND-LAr}
\label{ch:lartpc}

\section{Introduction}
\label{sec:concept}

As the target material in the \dword{dune} \dword{fd} modules is \dword{lar}, there needs to be a major \dword{lar} component in the \dword{dune} \dword{nd} complex in order to reduce cross section and detector systematic uncertainties for oscillation analyses~\cite{Acciarri:2016crz, Acciarri:2015uup}. 
With the intense neutrino flux and high event rate at the  \dword{nd}, traditional, monolithic, projective wire readout \dwords{lartpc} would be stretched beyond their performance limits.  To overcome this hurdle, in \dword{ndlar}, it is proposed to fabricate a large TPC out of a matrix of smaller, optically isolated TPCs read out individually via a pixelized readout.  The subdivision of the volume into many smaller TPCs allows for shorter drift distances and times. This and the optical isolation lead to fewer problems with overlapping interactions.  The pixelization of the readout allows for full 3D reconstruction of tracks and enhanced robustness in a high multiplicity environment.  Each of the building-block TPCs is equipped with optical readout that provides the necessary timing to associate tracks and events across the modularized TPC boundaries.

More specifically, a new generation of \dword{lartpc}s, based on \dword{arcube} technology~\cite{argoncube_loi}, is suitable for the high-rate environment expected for the \dword{dune} \dword{nd}. 
The \dword{arcube} technology utilizes detector modularization to improve drift field stability, reducing the \dword{hv} and \dword{lar} purity requirements; pixelized charge readout~\cite{Asaadi:2018oxk, larpix}, which provides unambiguous \threed imaging of particle interactions, drastically simplifying the reconstruction; and new dielectric light detection techniques with \dword{arclt}~\cite{Auger:2017flc} and LCM (Light Collection Module), which can be placed inside the \dword{fc} to increase light yield, and improve the localization of light signals. 
Additionally, these devices use a resistive field shell, instead of traditional field shaping rings, to minimize the dead material introduced through this modularization, maximize the active volume, and to minimize the power release in the event of a breakdown~\cite{bib:docdb10419}.

The \dword{lar} component (\dword{ndlar}) of the \dword{dune} \dword{nd} is made up of a configuration of \dword{arcube} \dwords{lartpc} 
large enough
to provide the required hadronic shower containment and statistics.
\dword{ndlar} is the most upstream of the three subdetectors shown in Figure~\ref{fig:All3Detectors}, where the beam propagates from right to left.
Immediately downstream of \dword{ndlar} is the \dword{gar} component, \dword{ndgar}, which serves \dword{ndlar} as a muon spectrometer. 
Beyond \dword{ndgar}, is the \dword{sand} component of the \dword{nd} that acts as a beam monitor.
The \SI{5}{m} (along beam)$\times$\SI{7}{m} (horizontal, transverse to beam)$\times$\SI{3}{m} (height) dimensions and the
\SI{67}{\tonne} fiducial mass of \dword{ndlar} are optimized primarily for hadronic containment under the assumption that  \dword{ndgar} will measure the sign and momentum of downstream exiting muons.  Figure~\ref{fig:actual-size} shows the arrangement of modules in the crystat for \dword{ndlar}.

\begin{dunefigure}[Illustration of \dshort{ndlar} in the \dword{nd} hall ]
	{fig:All3Detectors}
	{Illustration of the \dword{nd} hall, showing the detector subcomponents. With respect to a beam, which points from right to left in this image, \dword{ndlar} is the most upstream component and immediately downstream is  \dword{ndgar}, which serves \dword{ndlar} as a muon spectrometer. Beyond  \dword{ndgar} is \dword{sand}.}
	\includegraphics[width=\textwidth]{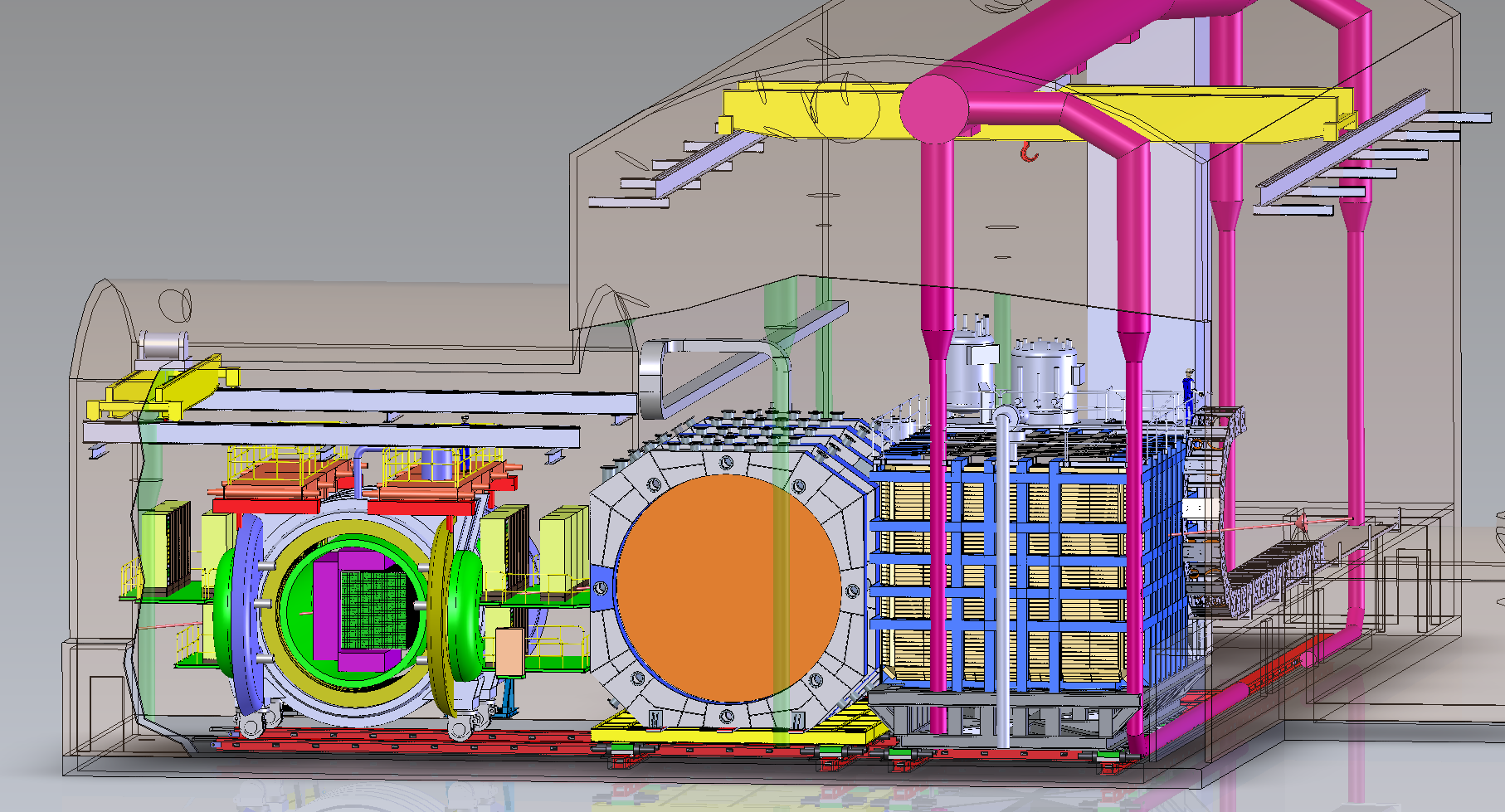}
\end{dunefigure}

\begin{dunefigure}[The current dimensions for the \dshort{arcube} detectors in \dshort{ndlar}]{fig:actual-size}
{The current dimensions for the \dshort{arcube} detectors in \dshort{ndlar}. 
The cryostat is based on the \SI{35}{ton} prototype and \dword{protodune}~\cite{Abi:2017aow}, and is yet to be optimized for the \dword{dune} \dword{nd}.  In this figure, the lower left is the top view, the upper plot is a view of a horizontal cut transverse to the beam, and on the lower right is a horizontal cut along the beam.}
	\includegraphics[width=0.9\textwidth]{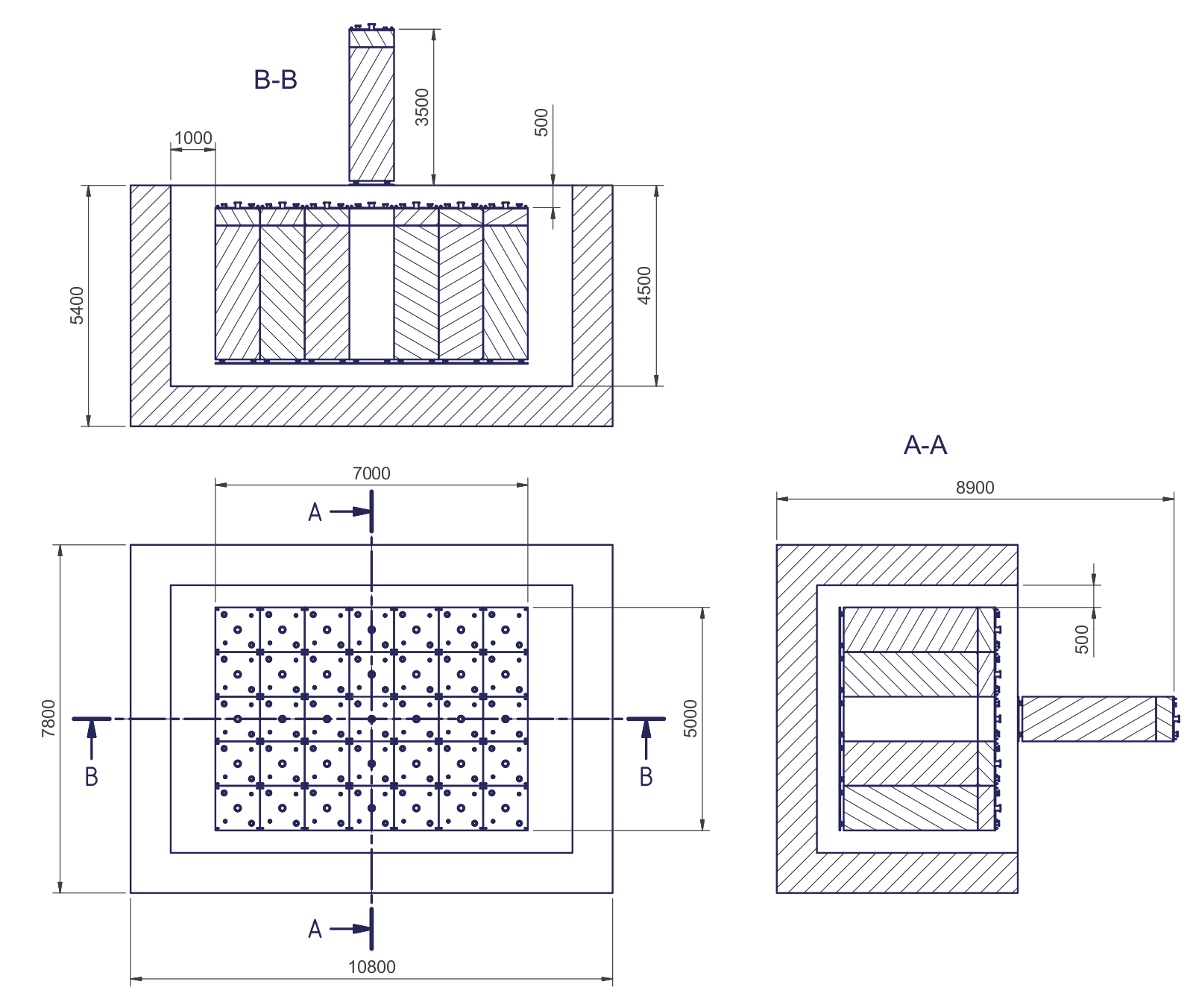}
\end{dunefigure}

Section~\ref{sec:ndlarrequirements} gives a discussion of the physics considerations driving the design of this component of the \dword{nd}.  An overview of the LArTPC structure in \dword{ndlar} is given in Section~\ref{sec:lartpc-ovvw}.
Section~\ref{sec:RandD} describes the \dword{arcube} R\&D program, including a focus on a multi-tonne scale demonstrator that forms the core component of a prototype \dword{dune} \dword{nd}  (\dword{pdnd}).
A dedicated discussion of the  physics goals of \dword{pdnd} is given in Section~\ref{sec:ProtoNDPhysics}.
In Section~\ref{sec:lartpc-dimensions}, a discussion of the optimization of the active volume of \dword{ndlar} 
to achieve good acceptance across the cross-section phase space is given.
Expected  event rates in \dword{ndlar} are presented in Section~\ref{sec:event_rates}.  The level of and mitgation of neutrino interaction pile-up is discussed in Section~\ref{sec:ndlar-pileup}.
Methods to determine the muon and electron momentum resolution scale uncertainties are discussed in Section~\ref{sec:lartpc-p-resolution-scale}.
Finally, techniques to constrain the flux using neutrino-electron elastic scattering and the low-$\nu$ method are described in Section~\ref{sec:lartpc-flux_constraint}.    
 
\section{Requirements}
\label{sec:ndlarrequirements}
{\bf Primary roles of \dword{ndlar}}

\begin{itemize}
\item To fulfill the overarching requirement \underline{ND-O1} and the measurement requirement \underline{ND-M1}, the ND must have a LArTPC. 
The reconstruction capabilities of the \dword{ndlar} have to be comparable to the far detector despite the high intensity of the beam at the near site, in order to effectively transfer measurements. This means that the interactions must be observed in liquid argon with sufficiently high acceptance to cover the phase space of neutrino energy and energy transfer with small uncertainties. The \dword{ndlar} fills this role as described in Section~\ref{sec:lartpc-dimensions}.

\item To fulfill the overarching requirement \underline{ND-O3} and measurement requirement \underline{ND-M3} the \dword{ndlar} must be able to measure the neutrino flux using established techniques with sufficient statistics that it can constrain the flux at the \dword{fd} over periods relevant for oscillation analyses. The \dword{ndlar} fulfills \underline{ND-M3} and the derived requirements by measuring the flux with reliable standard candles, such as the $\nu$-e scattering, providing a normalization measurement. The event rates are shown in Section~\ref{sec:event_rates}.

\item The \dword{nd} must have the ability to reconstruct the neutrino energy (\underline{ND-M1}) as well or better than can be done in the \dword{fd} and measure the wrong-sign contamination of the flux (\underline{ND-M5}). This assumes the presence of a muon range stack or spectrometer downstream of \dword{ndlar}.

\item To fulfill \underline{ND-M4} the \dword{ndlar} must identify and measure low recoil events which have flat energy dependence in order to measure the spectrum, i.e., the low-$\nu$ technique of measuring the spectral shape. The design to fulfill this is described in Section~\ref{sec:lartpc-dimensions} and Section~\ref{sec:lartpc-lownu}.

\item To fulfill measurement requirement \underline{ND-M6}, the \dword{nd} must measure and validate the modeling of the irreducible $\nu_e$ background. The detection thresholds for electromagnetic showers and distinction of electrons from photons in the \dword{ndlar} fulfill this requirement. The performance will be validated as described in Section~\ref{sec:ProtoNDPhysics}.

\item \dword{ndlar} must have the ability to make measurements both on and off the beam axis (overarching requirement \underline{ND-O4} and measurement requirement \underline{ND-M7}).  This allows for the collection of data with different flux spectra enabling the deconvolution of flux and cross section uncertainties and the combination of different fluxes during analysis. The \dword{ndlar} is mobile and can take data up to \SI{30.5}{m} off-axis ($\sim\,$\SI{50}{mrad}) as described in Chapter~\ref{ch:prism}.  These capabilities satisfy requirements ND-C4.1

\end{itemize}

{\bf Derived  \dword{ndlar} capabilities}

\begin{itemize}

\item To fulfill the derived requirements \underline{ND-C1.2 and sub-items} from \underline{ND-M3} the \dword{ndlar} is designed to collect a sufficiently large sample of $\nu$-e elastic events and identify them with high efficiency and low backgrounds to allow <2\% statistical uncertainty in the measurement. As shown in Section~\ref{sec:event_rates} we expect a multiple of the required $\sim$2500 $\nu$-e$^{-}$ scattering events per year in the on-axis location to be accepted.

\item \dword{ndlar} must have sufficient kinematic acceptance and particle identification capabilities to perform differential measurements of many neutrino interaction morphologies as required in the derived requirements \underline{ND-C1.1 and ND-C1.2}. The performance will be validated as described in Section~\ref{sec:ProtoNDPhysics}.

\item The \dword{ndlar} active size must be such that the hadronic recoil from neutrino interactions is contained for a representative sample of such interactions across the relevant phase space of incident neutrino energy and energy transfer.

\item To fulfill the derived requirements \underline{ND-M1, ND-M2, ND-M8, and ND-M9}, all \dword{nd} components must be functional in the presence of beam-related backgrounds and pile-up. The modular design of the \dword{ndlar} addresses this requirement, as demonstrated in Section~\ref{sec:ndlar-pileup}. 

\item The target nucleus of the \dword{ndlar} is argon and the \dword{ndlar} is based on liquid argon \dword{tpc} technology to fulfill \underline{ND-C1.1}.

\item Since auxiliary detectors are not employed, the \dword{ndlar} volume must also contain muons emerging ``sideways'' from \numu CC interactions, i.e., those that do not enter the downstream muon spectrometer, to fulfill \underline{ND-C1.1}.

\end{itemize}


\section{Overview of \dword{ndlar} ArgonCube structure }
\label{sec:lartpc-ovvw}

\dword{ndlar} consists of 35 optically separated LArTPC modules that allow for independent identification of $\nu-{\rm Ar}$ interactions in an intense beam environment using optical timing.
Each TPC consists of a high voltage cathode, a low profile field cage that minimize the amount of inactive material between modules, a light collection system, and a pixel based charge readout. 

The modules are hosted in a common purified bath of liquid argon which is held within a custom designed membrane cryostat. The cryostat and adjacent mezzanine cyogenics are placed on a mobile PRISM platform that allows the entire detector to be shifted off-axis relative to the neutrino beam. The full system is serviced by flexible energy chains that stay connected to \dword{ndlar} in all positions.

Individual TPC modules consist  of a low density profile cathode and field cage, a pixelated charge readout plane and associated low power electronics, a high coverage light readout system, the necessary module support structures including both internal cryogenics and monitoring as well as mechanical interfaces with the cryostat, and dedicated calibration systems. Externally, each TPC module is connected to a high voltage system, as well as the associated warm electronics and power supplies that enable the functioning of readout and calibration systems. 

The LBNF neutrino beam directed at \dword{ndlar} generates intense pulses of few-GeV neutrinos. These neutrinos are mostly muon flavor and the oscillations are negligible at the \dword{nd} distance. The interactions of these neutrinos generate energetic leptons (mostly GeV-scale muons) and a recoiling hadronic component. 
The standalone \dword{ndlar} begins to lose acceptance for muons above $\sim0.7$ GeV/c due to lack of containment. 
 Because the muon momentum and charge are critical components of the neutrino energy determination, a magnetic spectrometer is needed downstream of \dword{ndlar} to measure both quantities. The dimensions of \dword{ndlar} have been chosen to optimize the containment of the complex hadronic showers which can result from neutrino interactions within the active volume of the LArTPC. The corresponding scintillation light, which is detected in tandem with the ionization  produced by charged particles, provides a complementary measurement of the signal position and energy, albeit at much lower granularity, but with substantially better timing resolution ($\sim$20~ns).

One key aspect of \dword{ndlar} operation is the ability to cope with a large number of neutrino interactions  in each spill. As  discussed in Section~\ref{sec:ndlar-pileup}, the LBNF neutrino beam consists of a 10 $\mu$s wide spill, with $\mathcal{O}$(ns) bunch structure, delivered at a $\sim 1$~Hz rate. This means that there will be $\mathcal{O}$(50) $\nu$ interactions per spill in \dword{ndlar}. Given the relatively low expected cosmic ray rate during the beam (estimated to be $\sim$0.3/spill at 60-m depth), this beam related pile-up is the primary challenge confronting the reconstruction of the \dword{ndlar} events. The 3D pixel charge will be read out continuously. The slow drifting electrons (with charge from the cathode taking $\sim 300 \mu$s to arrive across the 50~cm distance) will be read out with an arrival time accuracy of 200~ns and a corresponding charge amplitude within a $\sim 2\mu$s wide bin. This coupled with the beam spill width gives a position accuracy of $\sim$16~mm. While this is already  good spatial positioning, the \dword{ndlar} light system will provide an even more accurate time-tag of the charge as well as the ability to tag sub-clusters and spatially disassociated charge depositions resulting from neutral particles, such as neutrons, coming from the neutrino interaction. Thus the \dword{ndlar} light system has a different role than that in the \dword{fd}, as it must time-tag charge signal sub-clusters to enable accurate association of all charge to the proper neutrino event, and to reject pile-up of charge from other neutrino signals.

\subsection{Field Structures}
\label{sec:lartpc-des-fieldstruc}

The field-shaping structure in a \dword{dune} \dword{ndlar} \dword{tpc} module is used to define a uniform electrostatic field in the liquid argon volume in order to transport ionization electrons -- from the point of creation to the readout pixels on the anode -- without significant distortions. It must achieve a field non-uniformity $<1\,\%$ in the entirety of the active volume and operate reliably under nominal fields of $250\,$V$/$cm and peak fields of up to $500\,$V$/$cm. The footprint of the system has to be minimized in order to optimize the fraction of active volume in the detector as a whole. Additionally, this subsystem should not exceed a local heat density of $100\,$ mW/cm$^2$, which is the typical heat density of electronics used in wire-based LArTPCs and limits liquid argon boiloff.

\begin{figure}[htbp]
\centering
\begin{minipage}[b]{.4\textwidth}
\includegraphics[width=\linewidth]{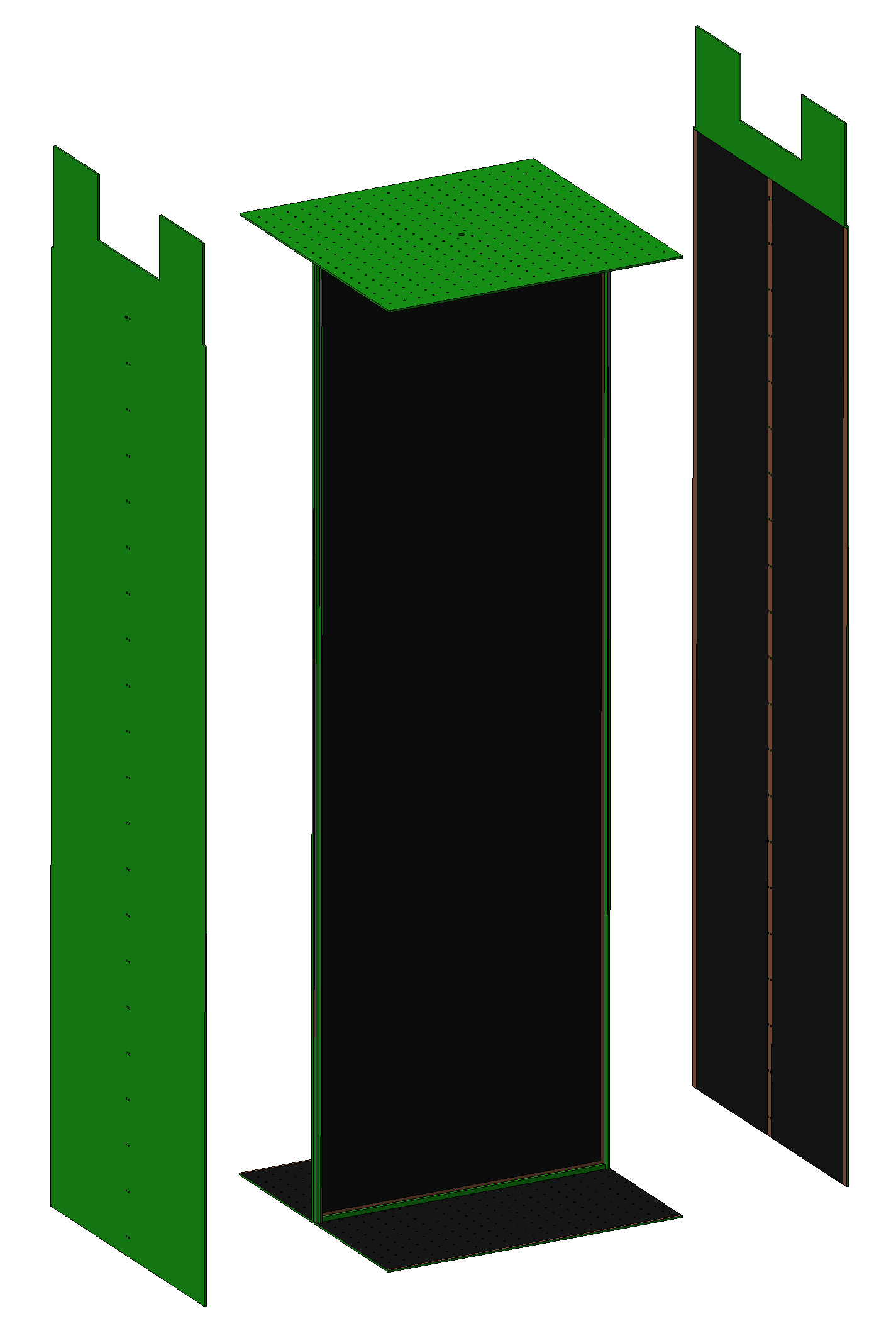}
\end{minipage}
\qquad
\begin{minipage}[b]{.4\textwidth}
\includegraphics[width=\linewidth]{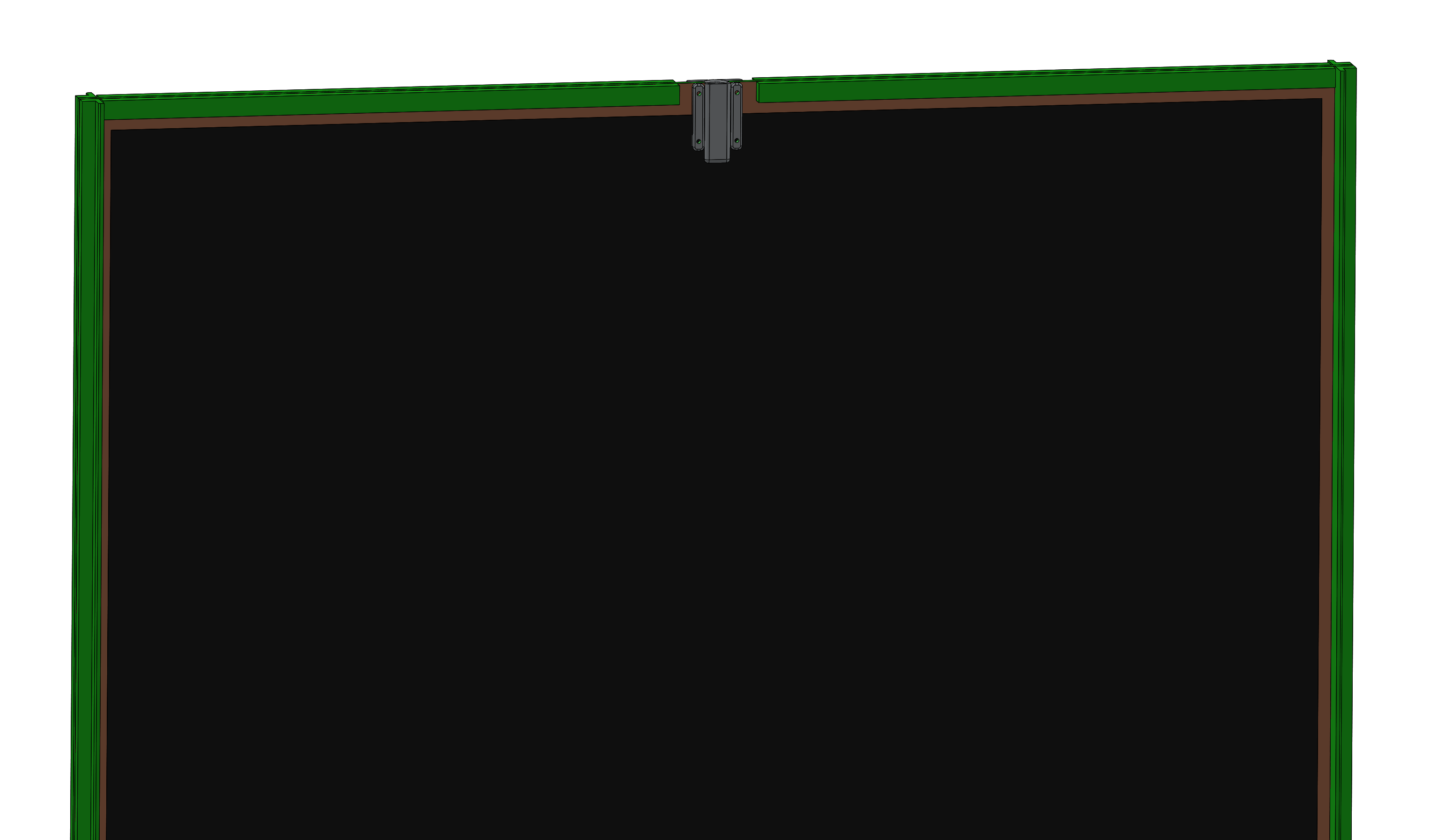}
\includegraphics[width=\linewidth]{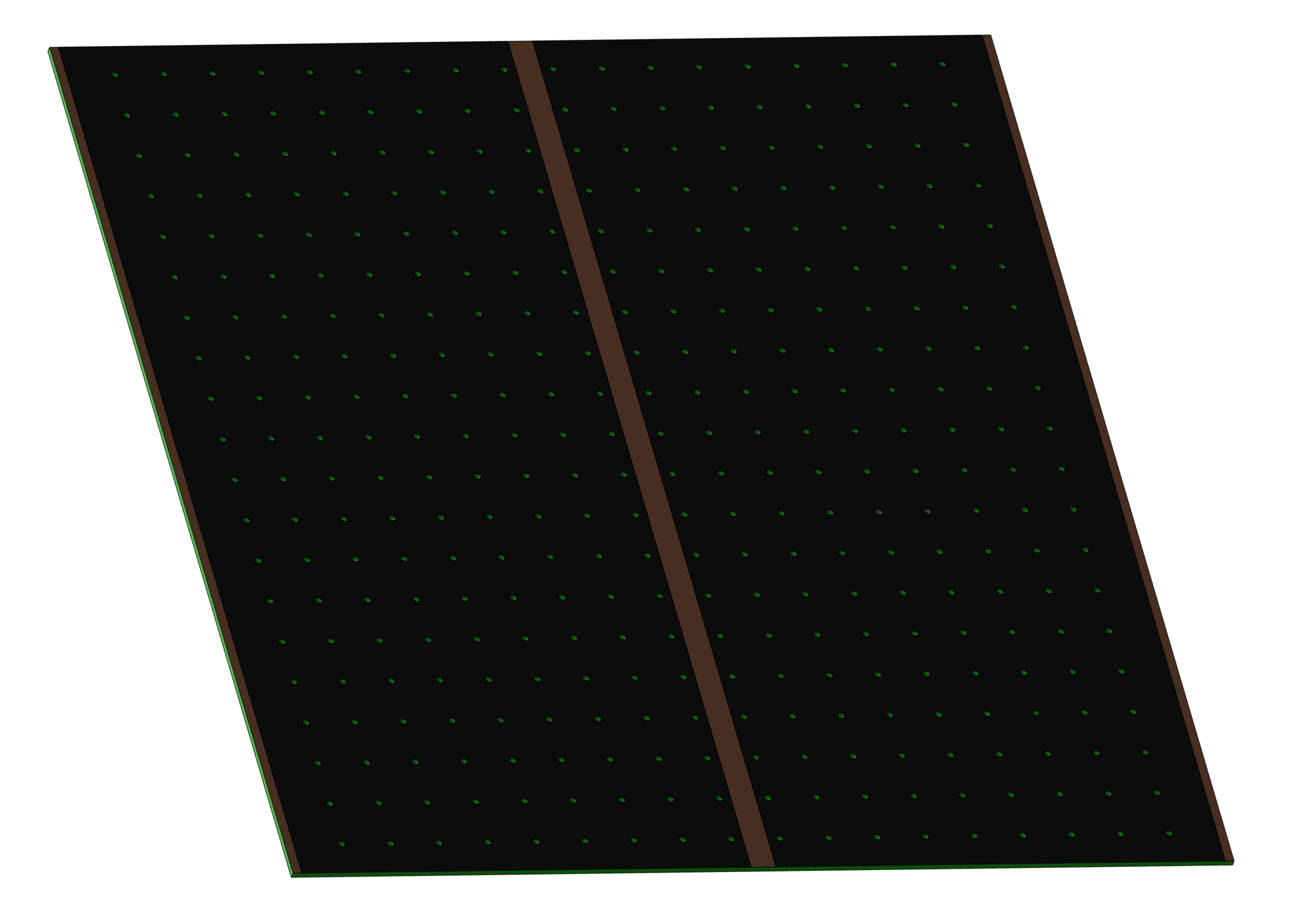}
\end{minipage}
\caption{(Left) Exploded-view drawing of the field-shaping structure of a TPC module. (Top right) High-voltage socket at the top of the cathode. (Bottom right) Bottom field-shell panel.}
\label{fig:field_shell}
\end{figure}

Figure\,\ref{fig:field_shell} shows the schematics of the field-shaping structure as a whole. It is composed of five copper-clad, 6\,mm-thick FR4 panels covered with Dupont Kapton sheets loaded with electroconductive carbon black. The central panel in the figure is the cathode, which splits the \dword{tpc} module into two optically-isolated drift volumes and sets the maximum potential, while the other four panels form the `field shell'.  The field shell is a resistive structure which continuously decreases the voltage from the cathode to the grounded anode. The bottom and top panels of the field shell are perforated with $\sim350$ 4\,mm holes to facilitate liquid argon circulation. This approach to field-shaping has several advantages over a traditional \dword{tpc} field cage:
\begin{itemize}
  \item it extends the achievable active volume by having a smaller the footprint but also by reducing the local field non-uniformity created by field-shaping rings;
  \item the resistive heating is spread over entire panels instead of being localized on the surface of resistors, which reduces significantly liquid argon nucleation;
  \item it does not suffer from single points of failure, as the whole panel drives the resistance;
  \item the field does not spike around rings, considerably reducing the risk of arcing.
\end{itemize}
The field-shaping structure also provides mechanical support for the entire \dword{tpc} module.

The cathode panel is covered on both sides with a layer of 25\,$\mu$m-thick Kapton XC, a material which provides a $\mathcal{O}(1)\,$M$\Omega/\Box$ sheet resistance, identical to the one used in the protoDUNE-SP cathode~\cite{Abi:2017aow}. The use of a resistive material prevents damage to the TPC, including the electronics, in the event of a discharge. The conductivity of the cathode is selected to be sufficient to neutralize the positive argon ions at the same rate as they are collected by the cathode. The high voltage is fed to the cathode plate through a socket placed at the top of the panel. The high voltage is distributed to the field shell through a perimeter of copper cladding connected to the center copper strip of the four remaining panels by metalized G10 corner brackets bolted on with PEEK screws.

The field shell panels are covered with 100\,$\mu$m-thick sheets of Kapton DR8, a variant of Kapton XC which exhibits a higher $\mathcal{O}(1)\,$G$\Omega/\Box$ sheet resistance at room temperature and under low voltage loads. This material is suitable to replace traditional field cages as it provides sufficient bulk resistance to constrain the heat load and limit the necessary power.
Kapton DR8 was extensively studied on $15\times15$\,cm$^2$ panels.
The sheet resistance of this material as a function of temperature and electric field was measured for various samples. 
These results show that at a peak electric field of 500V/cm and at liquid argon temperature, the sheet resistance of DR8 is $\sim3\,$G$\Omega/\Box$. Accounting for a shell aspect ratio of $L/W=1/16$ on either side of the cathode, the bulk resistance of a module is $R\simeq100$\,M$\Omega$. For a cathode voltage of $V=25$\,kV, this corresponds to a total heat load $V^2/R<10$\,W spread over the entire shell, or a local heat density of $<100\mu$W/cm$^2$, which is well below a value that would cause a problem.

\subsection{Charge Readout}
\label{sec:lartpc-des-chargero}

The charge readout system senses and records the signals of liquid argon ionization by charged particles traversing the LArTPC\@.
It must record signals with a spatial granularity at the same level or better than that in the \dword{fd} in order to enable a high-fidelity prediction of the neutrino signal in the \dword{fd}.
The \dword{nd} \dword{lartpc} relies on a novel pixelated anode with 4~mm pixel spacing and 2.5~$\mu$s signal time-binning in order to provide a true 3D record of the ionization signals.
This true 3D imaging is required to overcome signal pile-up in the high-rate environment of the \dword{nd} site, as discussed in Sec.~\ref{sec:ndlar-pileup}\@.

The core element of the charge readout system is the LArPix pixel anode tile, as shown in Fig.~\ref{fig:ndlar-pixeltile}.
These are printed circuit boards adapted to serve as self-triggering charge-sensitive anode surfaces within the \dword{nd} \dword{lartpc}, and instrumented with the custom LArPix low-power cryogenic-compatible application-specific integrated circuit (ASIC)\@. 
A single 34-pin twisted-pair ribbon cable provides power and data connections for each tile.
These cables are connected to a custom PCB-based feed-through mounted on the cryostat lid, directly above each TPC module.
Four PACMAN controllers are mounted in metal enclosures attached to the outside surface of each module feed-through, providing filtered power and noise-isolated data input-output to the tiles.
These controllers in turn receive an external 10~MHz clock and/or sync signal for data synchronization, as well as optional external triggers signals from the light readout system.
A Wiener PL506 24~V power supply delivers power to 20 controllers.
Standard RJ-45 ethernet cables carry data to and from the controllers, and are aggregated in a rack-mounted ethernet switch.
An optical fiber connection transmits data to and from this switch to the \dword{nd} site DAQ system.
Fig.~\ref{fig:larpix-architecture} outlines the charge readout system architecture.

\begin{dunefigure}[LArPix Tile]{fig:ndlar-pixeltile}
{The LArPix pixel anode tile with 10,240 self-triggering charge-sensitive pixels.  Each TPC anode is composed of 20 identical tiles arranged in two columns.}
\includegraphics[width=0.8\textwidth]{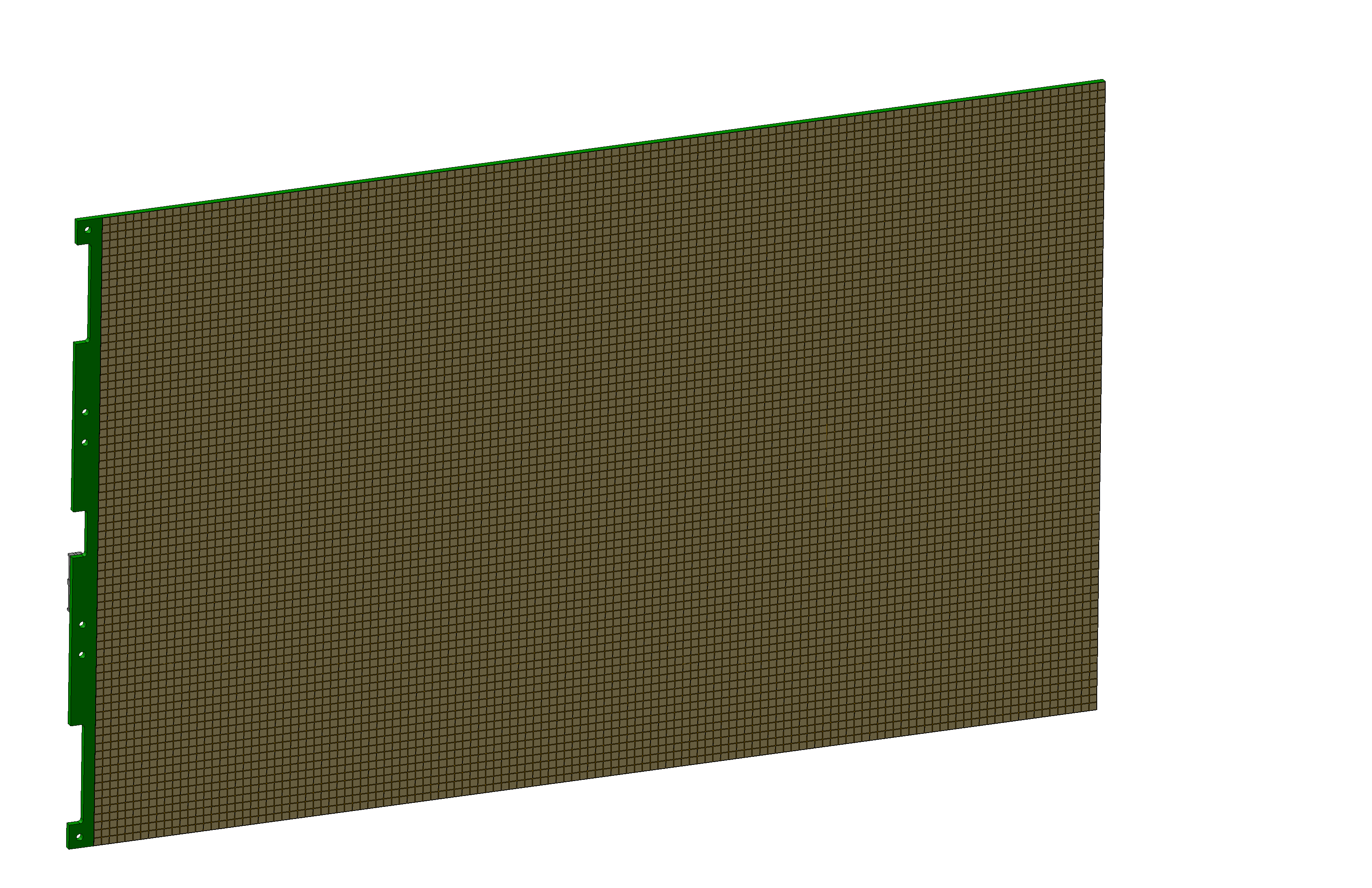}
\end{dunefigure}

\begin{dunefigure}[LArPix Architecture]{fig:larpix-architecture}
{The LArPix system architecture for the \dword{ndlar} detector.}
\includegraphics[width=0.8\textwidth]{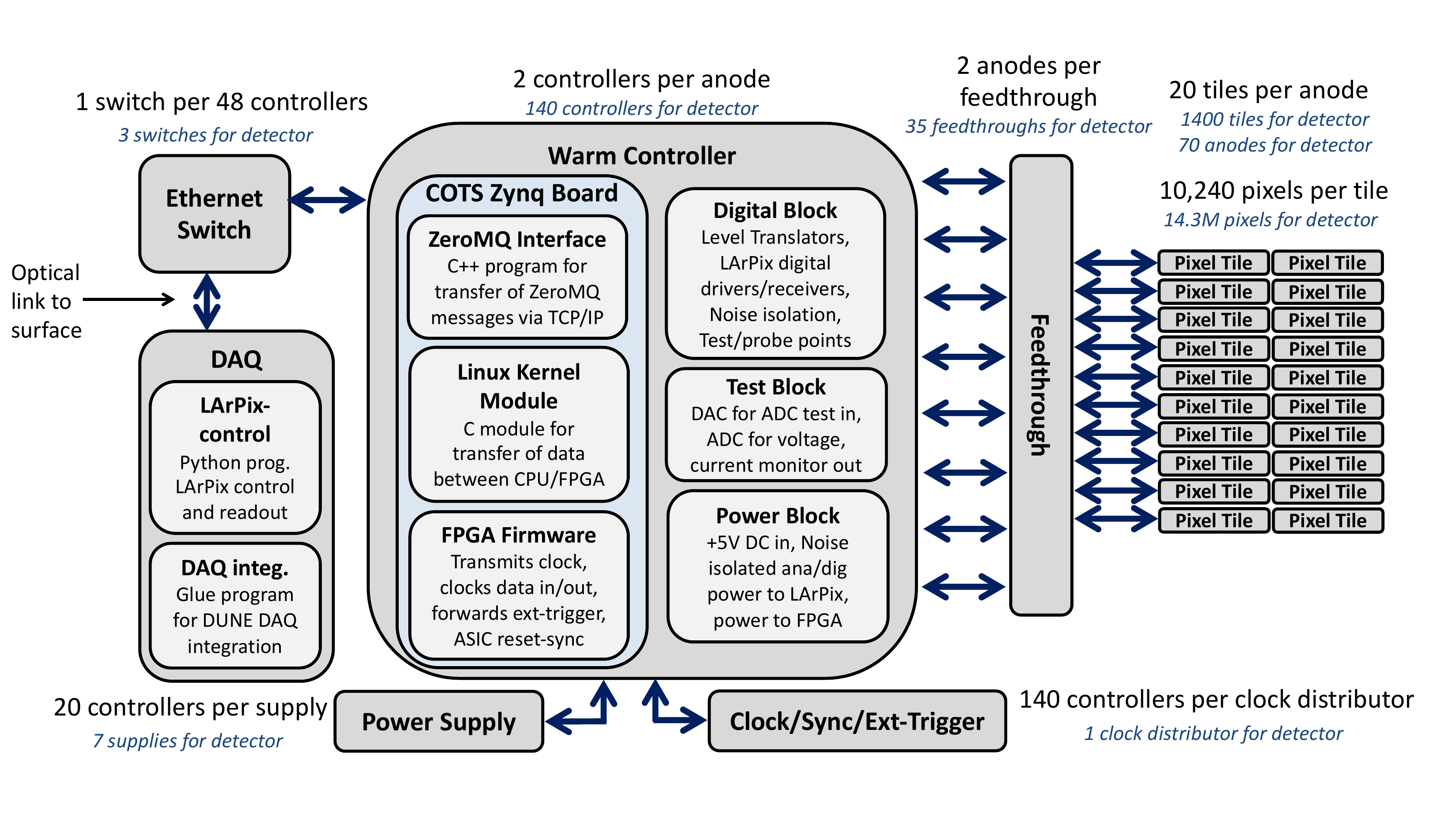}
\end{dunefigure}

The charge readout system design is driven by requirements that fall in three categories: performance, manufacturability, and reliability.
For performance, the charge readout system must deliver 3D spatial granularity at least as good as that in the \dword{fd}.
This drives the pixel spacing of $<$4~mm, and a corresponding density of $>$60,000 channels per square meter.
A noise level of $<$1000~e$^-$~ENC and dynamic range of $>$200,000~e$^-$ matches the \dword{fd} requirements on signal fidelity.
With a measured heat production of roughly 100~$\mu$W per channel, the tile heat density is below the threshold where detrimental boiling of the liquid argon occurs.

The \dword{nd} \dword{lartpc}  requires more than 200~m$^2$ of pixel anode, motivating requirements that facilitate large-scale production and control.
The pixel tile is designed so that it relies on standard multi-layer printed circuit board layout and production techniques, allowing it to be produced and assembled by typical PCB vendors.
Each LArPix ASIC instruments 64 pixels, enabling tiling of ASICs and pixels at the targeted 4~mm pitch without resorting to novel assembly techniques.
The ASICs are packaged to enable assembly via standard pick-and-place and solder re-flow techniques, as well as leveraging vendor-based post-assembly quality control inspection.
Control and readout of approximately 5,000 pixels within a pixel tile has been demonstrated over one I/O channel (four conductors operating at 10~MHz), achieving the high channel density required for the detector with a viable number of cables and feed-throughs.

Given the difficulty to access the detector once the cryostat is filled with liquid argon, the design of the cold-side components must be reliable.
The loss of a few percent of pixels, either individually or for an entire 64-pixel ASIC, does not considerably impact detector performance.
On the other hand, loss of an entire pixel tile would substantially hinder event reconstruction, efficient pile-up rejection, as well as accurate event fiducialization.
(To understand this, consider the ease of interpolating the signal for a missing 4~mm-by-4~mm pixel or 3~cm-by-3~cm ASIC anode region against the relative difficulty of guessing an unknown signal within the 30~cm by 50~cm region of an entire tile.) 
For this reason, the pixel tile and its associated cable and feed-through connections must be very reliable.
Reliability is achieved by minimizing the number of unique parts and the number of active elements.
The tile is also designed to be robust to failure of individual ASICs, and each tile has 4 redundant data I/O connections to the warm-side readout.

\subsection{Light Readout}

\label{sec:lartpc-des-lightro}

The Light Readout System (LRS) provides fast timing information from the prompt scintillation light (at $\sim$\SI{128}{\nano\metre}) emitted by charged particles traversing LAr. 
The optical detection of scintillation photons provides both an absolute reference ($t_0$) and rejection of unassociated charge signals (pile-up) from the specific neutrino signals of interest. Furthermore, the LRS is a dielectric and can be 
placed inside the field-shaping structure to increase light yield and localization of light signals.

The LRS consists of two functionally identical, SiPM-based systems for efficient detection of single UV photons with  large surface coverage: the Light Collection Module (LCM) and the ArCLight module (ArCLight). Readout, front-end electronics, DAQ (ADCs, synchronization and trigger), feedthrough flanges, SiPM power supply, and slow control are part of the system. In addition, the system includes cabling and interconnections between elements. 

The LCM light traps provide high collection efficiency and are to be used for accurate scintillation amplitude and energy reconstruction.  The ArCLight light trap provides good position sensitivity and are used for accurate scintillation position reconstruction.  Both the reconstructed energy and position will be useful for pile-up rejection.

Each of the \num{35} detector modules contains \num{60} LCM and \num{20} ArCLight modules with the alternating arrangement of \num{3} LCM - \num{1} ArCLight. The LRS modules are lined up along the inside of the field cage at 90 degrees to the anode and cathode surfaces.  Surface coverage is shared equally between both types of detectors. Each light module is read out by SiPMs which are located in pairs on a printed circuit board, known as the SiPM-PCB. Each LCM is read out by a single SiPM-PCB and each ArCLight is read out by \num{3} SiPM-PCB boards. 

Three SiPM-PCBs are grouped together by insertion to a single ``E''-shaped PCB, called an E-PCB. The E-PCB is intended to interface SiPM signals to long micro-coaxial cable lines, of length $\sim2$m, which are connected to the feedthrough. The need to transfer the small single photo-electron SiPM calibration signals through the long cable line leads to the requirement that each E-PCB be equipped with \num{6} pre-amplifiers. In total, all light modules in a single TPC-module are driven by  \num{40} E-PCBs. 

A feedthrough PCB with microcoaxial cable connectors provides interconnection between the cold and warm sides of the detector module. A VME 6U electronics crate is located on the warm side of the module, at the top of the cryostat near the feedthrough flange. In total, there are 35 crates, one for each module, positioned on top  of \dword{ndlar}. For each module, forty Microcoaxial cable assemblies routing SiPM signals and delivering power for the SiPMs and pre-amplifiers also connect via feedthroughs to associated VME crates.

Each crate contains front-end electronics boards: SiPM power supply PCB modules based on DACs (SiPM PCB), PCB modules with variable gain amplifiers (VGA PCB), control module PCB, and a patch board that  groups signals and power together into a single cable assembly. Optionally, a trigger module will be placed into the crate to provide the trigger logic that drives the ADCs. All these modules are custom made. Signals from the VGAs connect to the ADCs by means of twisted pair ribbon cables. A network switch provides an optical connection between the ADCs and the DAQ computers. Racks with ADCs, optical switches, and HV power supplies are located at some distance from the cryostat.  

All ADCs will be synchronized by means of White Rabbit (WR) protocol that guarantees subnanosecond precision of clock distribution. The charge clock will be synchronized with the WR 10 MHz clock.  The absolute WR timestamp is 8 ns, which is good enough to improve matching of light-to-charge events.

\begin{figure}[htbp]
\centering 
\includegraphics[width=0.44\linewidth]{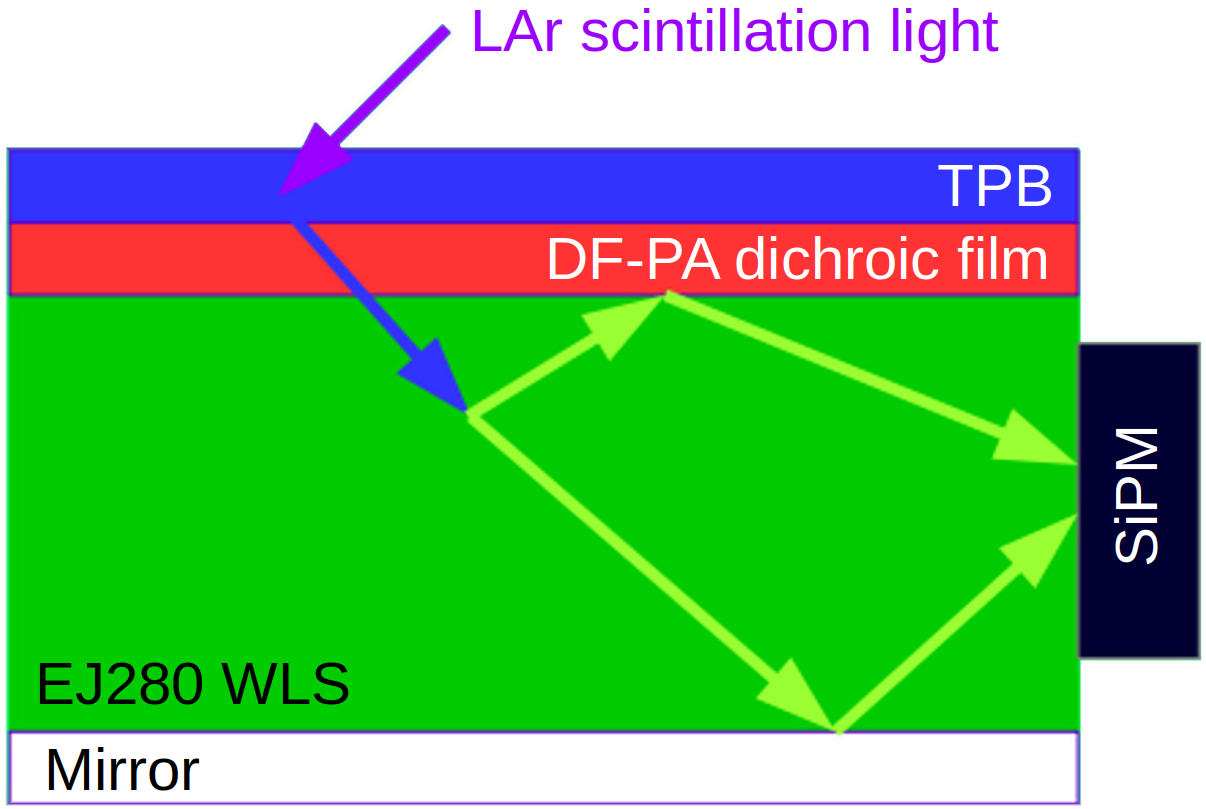}
\qquad
\includegraphics[width=0.5\linewidth]{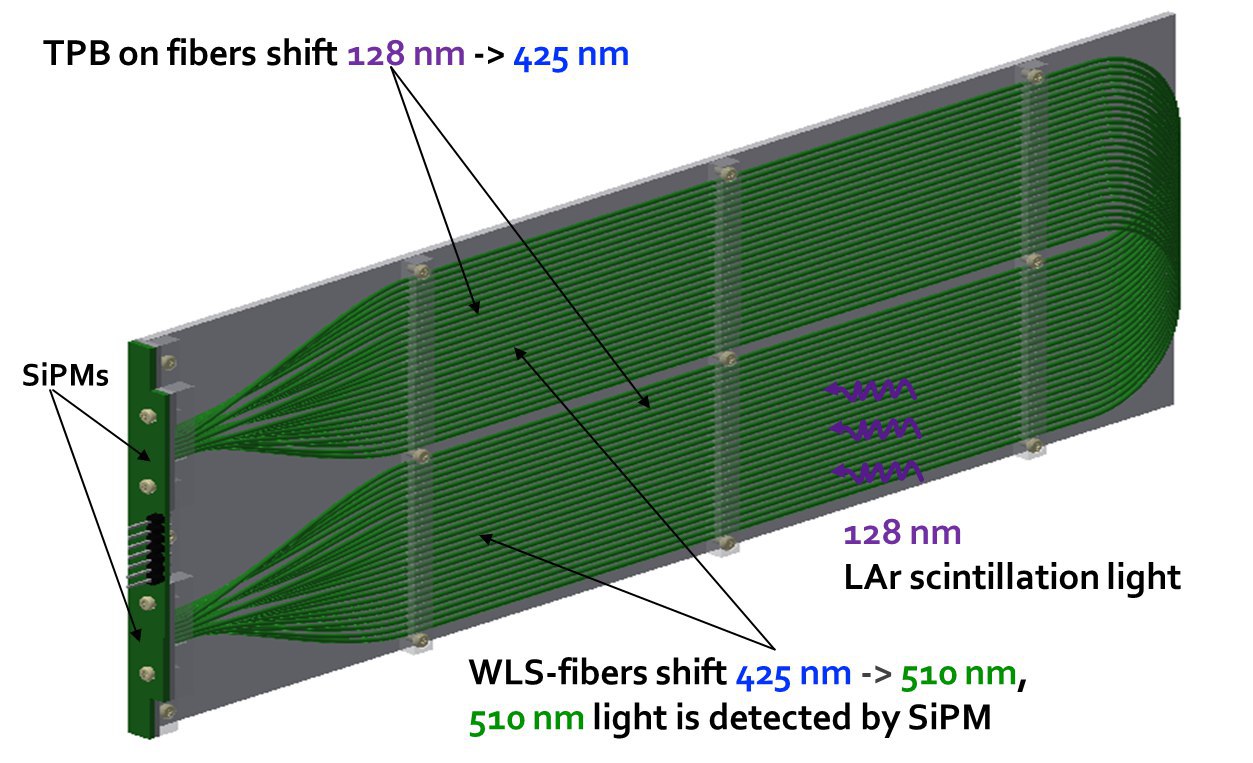}
\caption{\label{fig:fig_modules} Two approaches to light detection: ArCLight (left) and LCM (right).}
\end{figure}

LCMs and ArCLights share the same basic principle. The scintillation vacuum ultraviolet (VUV) light produced by LAr is shifted from \SI{128}{\nano\metre} to visible light by a WaveLength Shifter (WLS).  Tetra-Phenyl-Butadiene (TPB), which is an efficient WLS, coats the surface of the light collection systems. The emission spectrum of TPB is quite broad with a peak intensity of around \SI{425}{\nano\metre} (violet light). The violet light emitted on the surface of the light detection system eventually enters the bulk structure of the detector and is shifted to green light by a dopant (e.g. coumarin) in a bulk material which also acts as a light trap (see Fig.~\ref{fig:fig_modules}).

The ArCLight module (Fig.~\ref{fig:fig_modules}~left), developed by Bern University, uses the ARAPUCA principle of the light trapping. The general idea is to let the violet light go into the shifter bulk to be re-emitted. A reflective coating for the green light is placed on the entire surface except the photosensor window. On the TPB side is a dichroic filter which is transparent for the violet light and reflective for the green light. All other sides are coated with  a mirror film. The green light is trapped and may be detected by the SiPMs. The ArcLight dimensions are \SI[product-units=repeat]{300x\sim500x10}{\milli\metre}.

The LCM 
is a frame cantilevered by a PVC plate that holds WLS fibers which are bent into two bundles.  Each bundle is optically coupled to an SiPM light sensor as shown on the right side of Figure~\ref{fig:fig_modules}. The fibers are grouped and held by spacer bars with holes which are fixed on the PVC plate by means of polycarbonate screws to provide good matching of the thermal contraction. The PVC plate with the WLS fibers is coated with TPB that re-emits absorbed VUV light in the violet. The violet light is shifted inside multi-cladding ($\varnothing$=\SI{1.2}{\milli\metre}) Kuraray Y-11 fibers to green light ($\sim$\SI{510}{\nano\metre}).  That green light is trapped by total internal reflection in the fiber that guides this light towards to the fibers ends that are read out by SiPMs.  The LCM dimensions are \SI[product-units=repeat]{100x\sim500x10}{\milli\metre}.

\subsection{Module Structures}
\label{sec:lartpc-des-modstruc}

The module structure is the connection between the TPCs and the cryostat. It provides the feedthroughs and routing of all infrastructure to and from the detector including: HV, argon (gas and liquid), readout signals, and instrumentation for the temperature and level measurements. The module structure interfaces with the LBNF near site cryogenic infrastructure, as well as all other supporting detector subcomponents.

As it is important to minimize material near the active volume, the module structure is used to provide structural integrity to the TPCs, allowing the TPCs to be constructed with as little material as possible. This maximises the active mass of the detectors by moving structural components away from the active volume.

The module structure must locate each TPC precisely with respect to its neighbouring TPCs. This is important during the installation of module rows into the cryostat. It is also vital for maintaining the required clearances and orientations during the cool down and operation in liquid argon. It minimizes uncertainties when reconstructing events across multiple modules.

The structure is itself modular by design in order to allow individual modules to be tested prior to their integration into a row of 5 modules. This is required for module transportation and handling as the TPC is not a sufficiently rigid body to support itself without the structure above it. This also reduces the requirements on local test facilities, as a single 3 x 1 x 1 m$^{3}$ module is significantly easier to handle and than a full row of five. This allows for commercially available cryostats to be used for testing individual modules prior to integration.

Considering a row of five modules, the cryostat lid above the row is  a section of membrane cryostat  5.7 m long, 1 m wide and 0.8 m deep. I-beams forming the external structure of the cryostat are mounted to the upper edge of this section. Eight titanium support ties pass through the membrane and secure the I-beams to a 25~mm thick steel plate that spans the area below the membrane. This steel plate provides the fixing point for mounting the modules. Above each module is a square steel frame that is attached to the TPC to provide structural rigidity to the module. This frame is precisely located on, and then bolted to, the fixing plate below the membrane. These frames are also used to support individual modules during testing and installation. The eight titanium ties, steel fixing plate, and five steel frames form the structural components. 

There are feedthroughs above each module providing the HV, signal paths and instrumentation lines. Five were chosen to simplify routing in the volume above the modules. Each feedthrough is a single penetration of the membrane 260 mm in diameter, with a cross conflat connection sealing the warm side. The HV feedthrough is located at the centre of the feedthrough, with the warm connection on the top of the cross. It is isolated from other services inside the penetration by a grounded 40~mm steel tube. 
Charge and light readout cabling is routed through separate 60~mm steel tubes inside the membrane penetration. On the warm side, the frontend electronics of both the light and charge readout are mounted directly to either side of a cross conflat connection. Service routing (temperature and level sensors) will use the same principle, with a 40~mm steel tube isolating it through the penetration. On the cold side, the HV cable passes through an opening in the frame and connects directly to the centre of the cathode, all other cabling connects to junction boxes mounted on the frame that connect all sub components. This allows everything below the frame to be tested in isolation from the row.

The module argon supply plays somewhat different roles during cool down and normal operation. During cool down and filling it is used to inject cold argon gas and then liquid to reduce the thermal gradient across the modules as they are filled with liquid from the base.  During operation,  clean and subcooled argon liquid is supplied to the top of the modules where it purifies the TPC volume and provides cooling to the electronics. 

There is a single argon inlet at the end of each row. From the inlet, the argon is routed through vacuum jacketed lines in the ullage volume above the modules to the injection points. The liquid supply is throttled at each injection point to ensure the same volume of argon is supplied to all modules in a given row. The liquid supply to the rows uses symmetric lines to negate the need for a throttling system outside the cryostat. The injection points terminate at diffusers mounted just below the nominal liquid level.
The liquid level is 380 mm below the membrane, in accordance with industrial standards EN-14620.  

During cool down it is vital that all gas volumes are ventable to prevent contamination of the argon during operation.  Therefore, all five feedthroughs are fit with gas bleed ports that returns gas to the condenser and filtration system.

\subsection{High Voltage}
\label{sec:lartpc-des-hv}

The high voltage distribution system sets the required negative potential on the cathode of 
each detector TPC module. The system provides low-pass filters to suppress high-frequency ripple originating from the HV power supply units that operate in switching mode. The system also provides cabling and interconnections between  HV power supply units, potted filters-distributors, and module cathodes. Some elements of slow control, such as voltage/current monitoring as well as temperature, are also included in the system.

\begin{figure}[htbp]
\centering 
\includegraphics[width=0.9\linewidth]{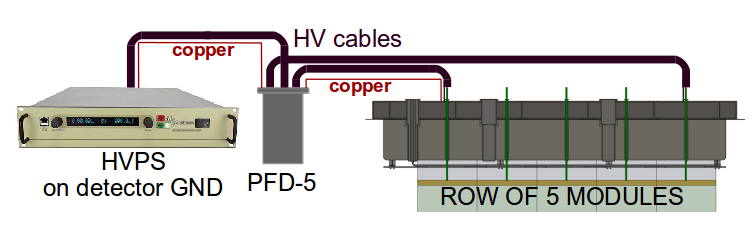}
\caption{\label{fig:gndpath} Electrical schematics of the HV system. Safety ground braid is highlighted in red.}
\end{figure}

Each row of 5 modules is connected to a single HV power supply unit (HVPSU) via a potted filter-distributor unit (PFD-5).
The PFD-5 is connected to the HVPSU via coaxial cable, rated to withstand 100~kV. The PFD-5, in turn, provides 5 sockets for the coaxial cables from the 5 modules. Figure~\ref{fig:gndpath} shows a simplifed illustration, where only two modules are connected. 

The cryostat and the whole HV system reside at detector ground. Current returns from the cryostat to the HVPS is directed via the sheath of the HV cables.  This provides a ground reference connection to the PFD-5 and to the cryostat. To ensure the presence of a safe ground when the HV cables are unplugged from their connectors, additional ground braid is laid out (shown as the red line in Figure \ref{fig:gndpath} and denoted ``copper''). This line must be arranged as close as possible to the bunch of HV cables to minimize the cross section of the ground loop.

\begin{figure}[htbp]
\centering 
\includegraphics[width=0.21\linewidth]{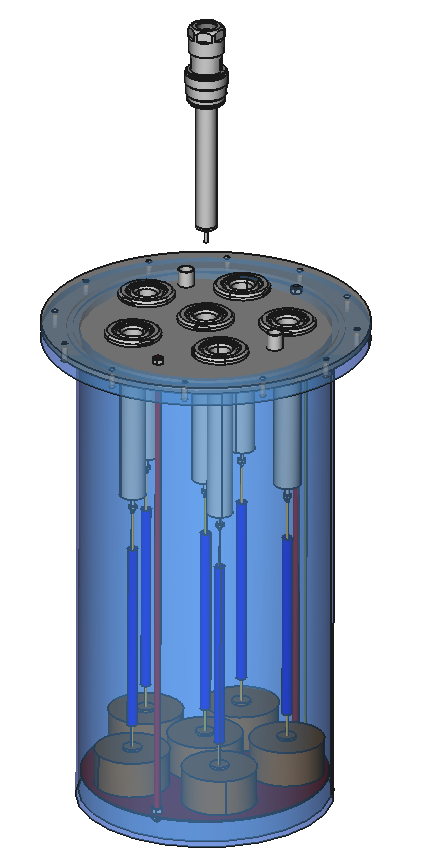}
\includegraphics[width=0.3\linewidth]{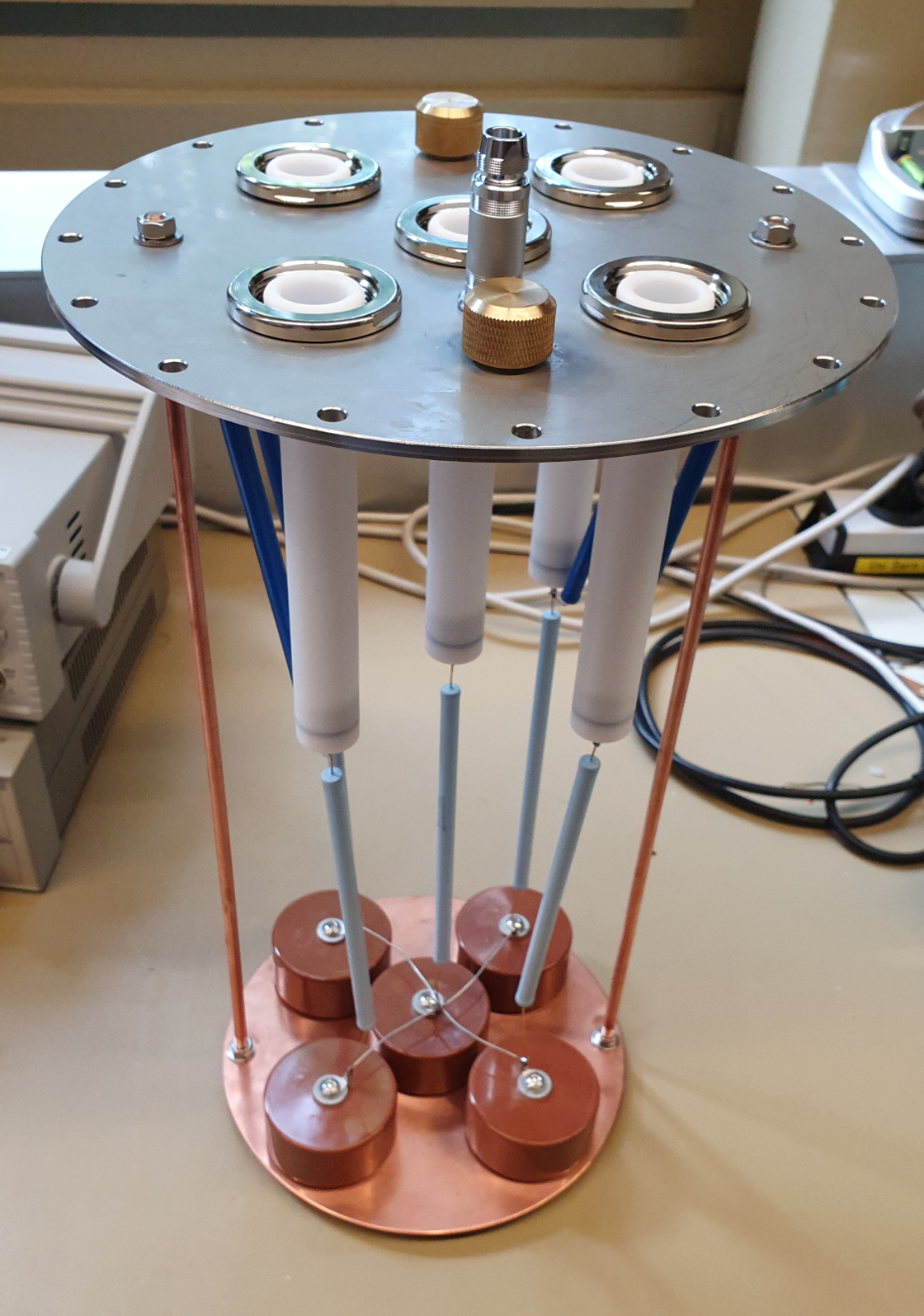}
\includegraphics[width=0.21\linewidth]{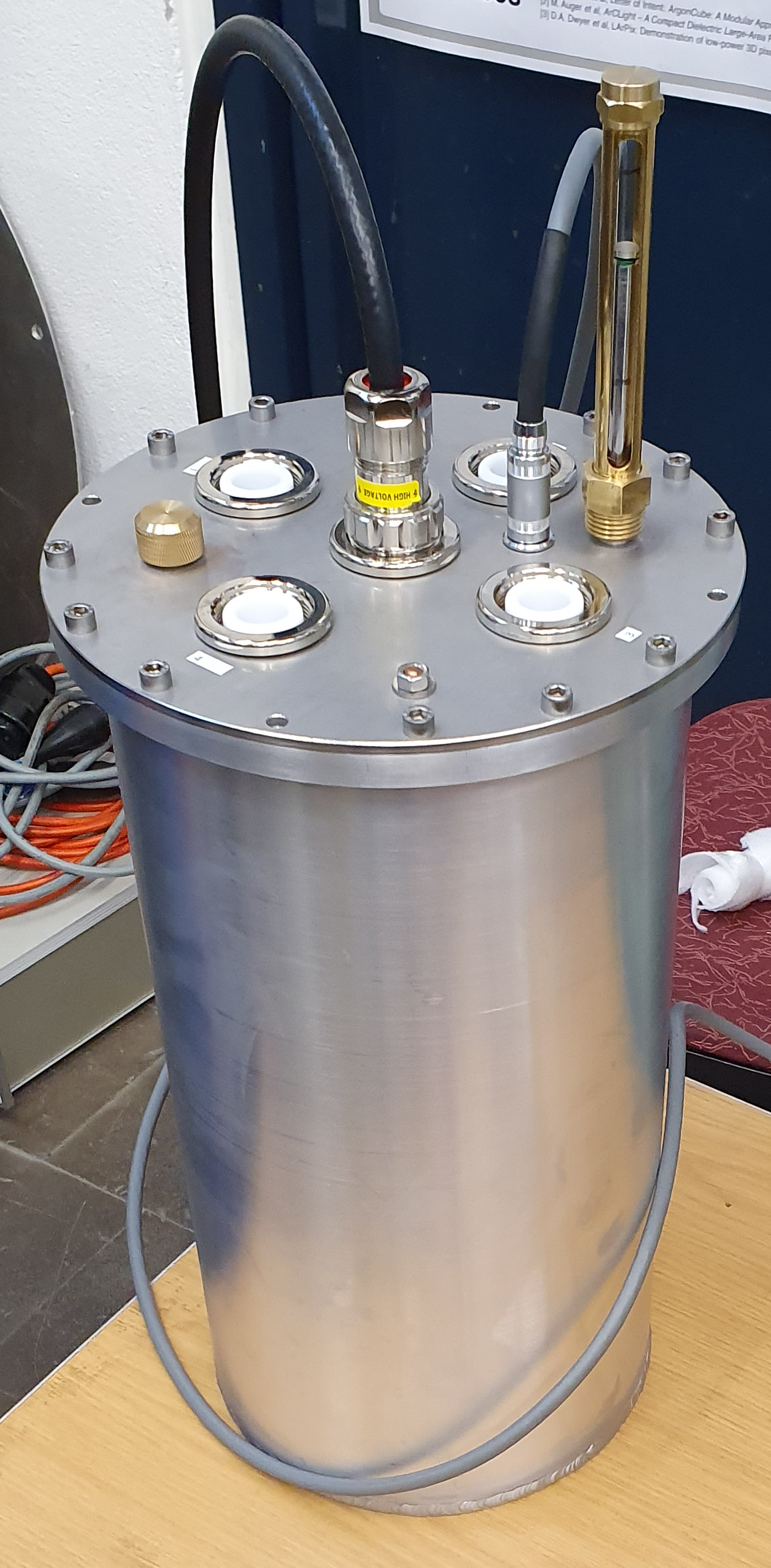}
\caption{\label{fig:pfd5cad} Left: CAD drawing of the PFD-5 Potted Filter-Distributor. The electrical connection between the capacitors at the bottom is omitted for image clarity. Middle: Assembly of the prototype PFD-4 for 2x2 Demonstrator. Right: PFD-4 being tested at 60~kV.}
\end{figure}

The PFD-5 is an oil-filled high voltage low-pass filter with one input and five independent outputs as seen on the left side of  Figure~\ref{fig:pfd5cad}. Each output is equipped with a voltage divider. The output values are digitized with a dedicated controller via Ethernet. In addition, the oil temperature is monitored. 

The principal requirement of the HV system is to provide cathode potentials in the range up to -50~kV, which allows the TPC drift field to  reach values up to 1~kV/cm. At this voltage, the current through the TPC module is expected to be 0.8~mA.
Suppression of the HVPSU output ripple down to 4~mV would result in an equivalent induced pixel charge of 0.016~fC which is below 1\% of the expected charge per pixel from a MIP track ($\sim$4~fC) and corresponds to 100 electrons. Noticeable suppression of the lower frequencies, such as line frequency, is also an asset. 

The change rate of the cathode potential is limited by the maximum allowed induced current on the pixels of the charge readout plane. A ramp rate of 100 V/s or less results in induced current below 1~pA/pixel, which is a safe value.

The long-term stability of the cathode potential is required to be at the level of 0.1\%. This allows the coordinate determination with an accuracy
of better than 0.5~mm at the cathode. This also restricts to below 0.01\% the uncertainty in the ionization charge measurement due to recombination.

The size of the PFD-5 is determined by the maximum operation voltage of 50~kV. The filter components and the final cutoff frequency of 6~Hz is chosen to provide the best performance for a reasonable size.
The HV connectors on the flange of the PFD-5 are required to handle flexible cabling and system testsm as shown on the right of Figure~\ref{fig:pfd5cad}. 

The cathode potential is critical and must be monitored. A fully functional field-shaping shell at constant cathode potential results in a constant consumed current. The cathode currents, therefore, need to be continuously monitored by the slow control system. 

The PFD-5 is filled with high-quality synthetic transformer oil. During operation, up to 30~W of thermal power is dissipated into the oil. The temperature of the oil is  monitored.  It is expected to be below 50 C for natural air cooling for nominal TPC operating parameters at 1~kV/cm. 

In order to monitor the current through the field shell of each module, a pickup circuit is mounted at the anode side of the field shell. A current pickup resistor of Rp=1k provides a voltage signal with 1~V/mA sensitivity, which is routed via the module top flange to a dedicated ADC unit, where it is digitized.

\begin{figure}[htbp]
\centering 
\includegraphics[width=0.4\linewidth]{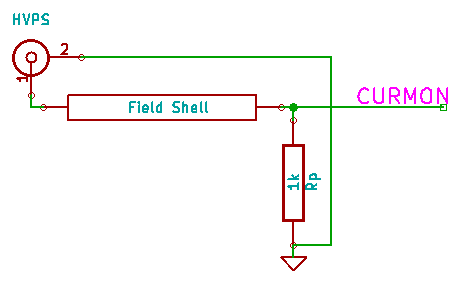}
\caption{\label{fig:hvcurmon} Electrical scheme of field shell current monitoring circuit.}
\end{figure}

The summary of the key design parameters of the HV system is given in Table~\ref{tab:table-hvndlar-params}. 

\begin{dunetable}
[Parameters of the HV distribution and delivery system]
{cc}
{tab:table-hvndlar-params}
{Parameters of the HV distribution and delivery system.}
Parameter & Nominal value  \\ \toprowrule
Output channels & 35 \\ \colhline
Output voltage & <= 50kV \\ \colhline
Current per channel & 0.8 mA  \\ \colhline
Output ripple voltage & < 4 mV  \\ \colhline
Long-term stability & < 0.1\%  \\ \colhline
Voltage ramp/down rate & 100 V/s (< 1 pA/pix)  \\ \colhline
Voltage monitor sensitivity & 0.1 V/kV \\ \colhline
Current pickup sensitivity & 1 V/mA \\ \colhline
PFD-5 Surface temperature & < 50 C  \\ 
\end{dunetable}

\section{The LArTPC Demonstrator Program}
\label{sec:RandD}

To date, the \dword{arcube} R\&D program has been very successful in moving toward a next generation \dword{lartpc}. 
A series of prototypes, with each testing novel aspects of the design, have been operated successfully~\cite{ Ereditato:2013xaa, Zeller:2013sva, art_cold_ero, Asaadi:2018oxk, Cavanna:2014iqa, larpix, bib:docdb10419, Auger:2017flc}. 
With the various technological developments demonstrated in small-scale \dwords{tpc}, the next step in the \dword{arcube} program is to demonstrate the scalability of the pixelated charge readout and light detection systems, and to show that information from separate modules can be combined to produce high-quality event reconstruction for particle interactions. 
To that end, a mid-scale (\SI[product-units=repeat]{1.4x1.4x1.2}{\metre} active volume) modular \dword{tpc}, dubbed the \dword{arcube} 2$\times$2 demonstrator, with four independent \dword{lartpc} modules arranged in a 2$\times$2 grid has been designed, and is under construction.

After a period of testing at the University of Bern, the 
demonstrator will be placed in the \dword{minos} \dword{nd} hall at \dword{fnal} where it will form the core of \dword{pdnd}~\cite{bib:docdb12571}.   
In \dword{pdnd}, the \dword{arcube} demonstrator can be studied in an intense, few-GeV neutrino beam.  
This program aims to demonstrate stable operation and the ability to handle backgrounds, relate energy associated with a single event across \dword{arcube} modules, and connect tracks to detector elements outside of the 
demonstrator.
Further discussion of proposed \dword{pdnd} studies is  in Section~\ref{sec:ProtoNDPhysics}.
The \dword{arcube} 2$\times$2 demonstrator is described below in some detail since the \dword{arcube}  modules to be installed in \dword{ndlar} are anticipated to be very similar.

\subsection{Prototyping Plans}
\label{sec:lartpc-proto}

The prototyping plan for the \dword{nd} LArTPC detector will address a specific set of technical targets between now and the initiation of detector production.  
Prototyping activities fall into two categories: component-level and integration-level prototyping.  
Component prototyping is generally addressed via stand-alone small-scale tests, and the majority of these tests have been completed over the recent years of the ArgonCube R\&D program.
Integration prototyping addresses how these components come together and function coherently within the ND LArTPC design, as well as demonstrating the large-scale production and assembly processes necessary to construct the \dword{nd}.

There are three stages to the integration prototyping plan: the SingleCube Demonstrator, the ArgonCube 2x2 Demonstrator, and the subsequent Full-scale Demonstrator.
The SingleCube Demonstrator is a $\sim$30-liter fully-functional LArTPC designed to validate the integrated performance of the \dword{nd} prototype charge and light readout elements in a field cage of similar mechanical design as that in the \dword{nd}.
The ArgonCube 2x2 Demonstrator is a complete ton-scale LArTPC detector system focused on verifying technical readiness of the ND LArTPC module design before the completion of the \dword{nd} design phase.
The Full-scale Demonstrator (FSD) is a production-scale LArTPC module that will provide an engineering validation of the full-scale component production, assembly, and testing processes before \dword{dune} proceeds to \dword{nd} production.
Fig.~\ref{fig:ndlar-prototypes} shows each of these prototypes.

\begin{dunefigure}[ND-LAr Prototypes]{fig:ndlar-prototypes}
{(Left) The SingleCube LArTPC designed to test a single integrated large-format charge and light readout element. (Center) The mechanical assembly of the first module (Module 0) of the ArgonCube 2x2 Demonstrator, including the cathode, field cage, and anode support panels.  The module is a sub-scale prototype of the \dword{nd} LArTPC module, at 60\% drift length and 40\% module height.  (Right) The engineering model of the full-scale ND LArTPC module (1 m by 1 m footprint and 3.5 m height), shown with the anode panels detached from the field cage.  The pixelated anode tiles (gold rectangles) provide true 3D imaging, while the dielectric light traps (pink and while rectangles) provide high-efficiency scintillation light detection.}
\includegraphics[width=0.3\textwidth]{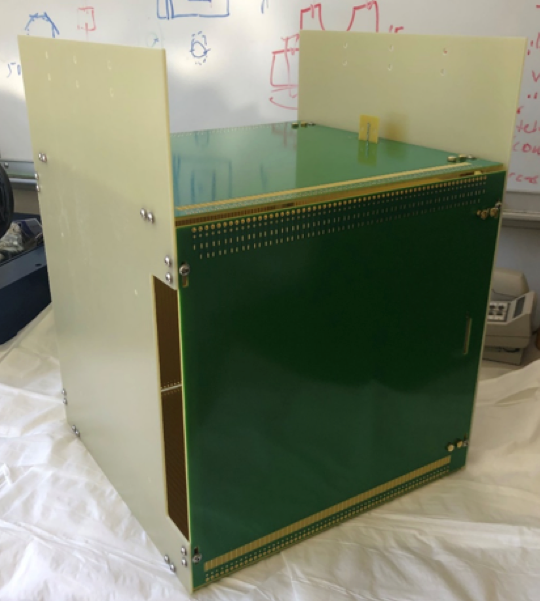}
\includegraphics[width=0.25\textwidth]{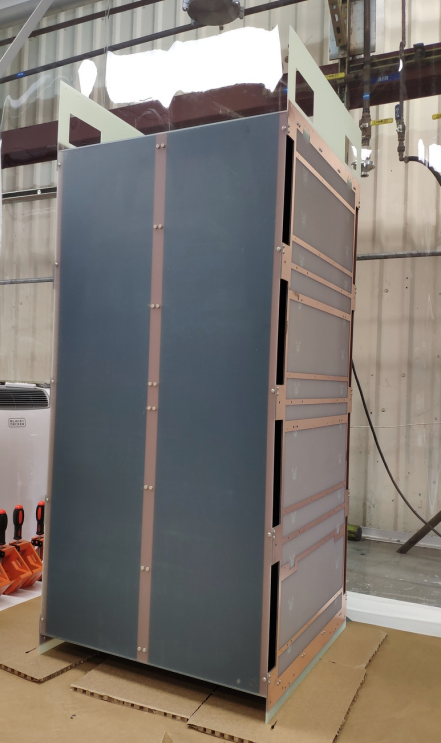}
\includegraphics[width=0.4\textwidth]{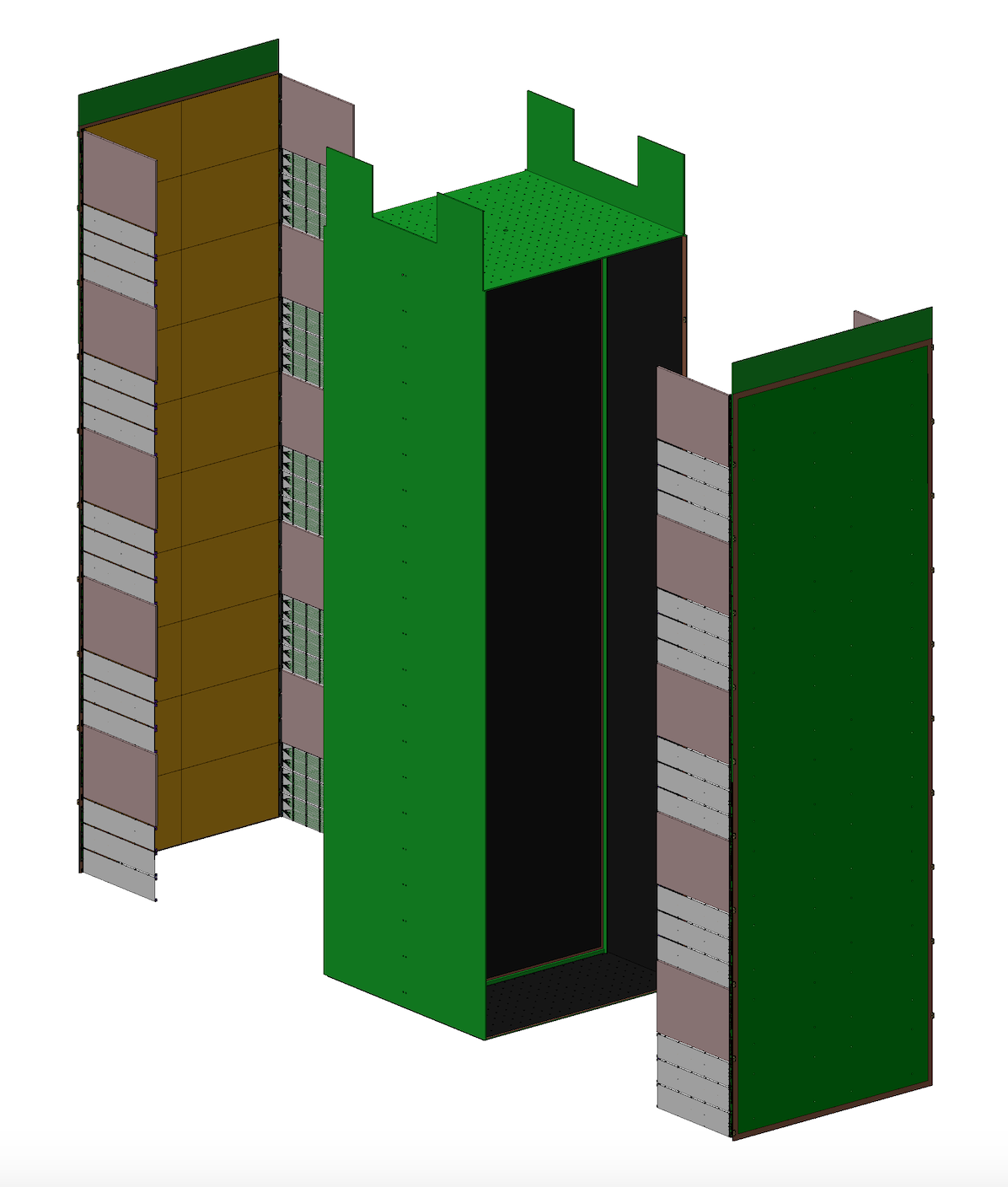}
\end{dunefigure}

\subsection{SingleCube Demonstrators}
\label{sec:singlecube-proto}

The SingleCube Demonstrator is a response to COVID-19 travel restrictions that prevented international partners from traveling to our primary prototyping site at the Univ.~of~Bern\@.
The TPC has a drift length and mechanical interfaces identical to the 2x2 module, but is sized to support only one pixel readout tile and one light readout element (see Fig.~\ref{fig:singlecube-prototype}). 
This facilitates an integrated test of the active detector elements in a smaller liquid argon cryogenic system in advance of their installation in the larger ArgonCube 2x2 Demonstrator module.
Instead of using a field cage based on high-resistivity polyamide film, it relies on a more conventional PCB-based field cage with discrete resistors, easily produced during the pandemic-induced curtailment of activities.
Operation of a SingleCube TPC at Bern in Oct.~2020 provided the first integrated test of the ND LArTPC readout system, successfully imaging cosmic rays and operating stably over the planned week-long run.
This test achieved targets in system noise ($<$1000~e$^-$~ENC), LAr purity ($>$500~$\mu$s), as well as HV field strength (1~kV/cm) and stability (see Fig.~\ref{fig:singlecube-results}).
Five copies of the SingleCube TPC were built at LBNL and distributed to partner institutions for further system testing and refinement.

\begin{dunefigure}[SingleCube Prototype]{fig:singlecube-prototype}
{(Left) Installation of a LArPix tile and ArCLight panel assembly into the SingleCube TPC at the Univ.~of~Bern.  (Right) An overlay of the raw data from 25 typical cosmic ray events collected during the first SingleCube operation run in Oct.~2020\@.}
\includegraphics[width=0.4\textwidth]{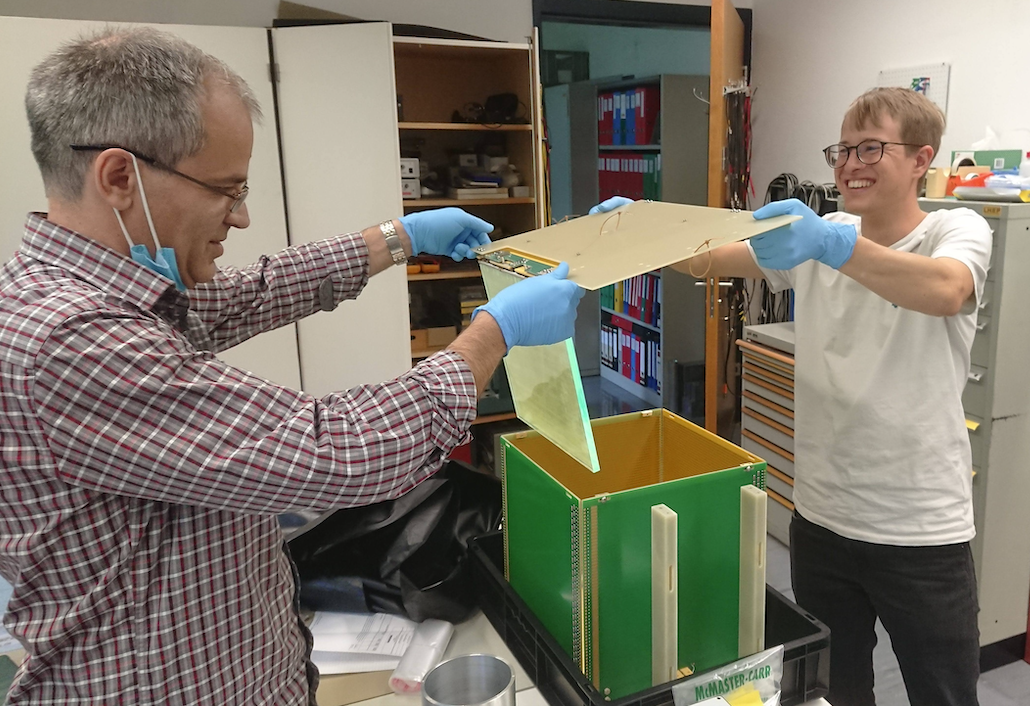}
\includegraphics[width=0.4\textwidth]{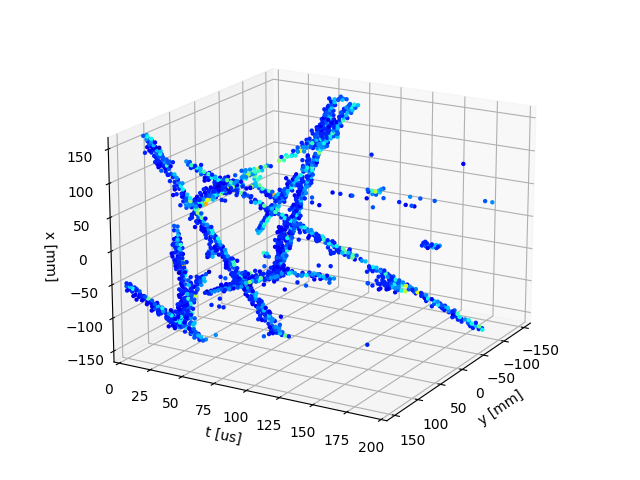}
\end{dunefigure}

\begin{dunefigure}[SingleCube Results]{fig:singlecube-results}
{(Left) The electron lifetime measured using anode-cathode crossing cosmic ray muon tracks during the first operation of the SingleCube TPC. (Right) The distribution of muon energy loss in LAr is consistent with that expected for cosmic ray muon tracks.}
\includegraphics[width=0.4\textwidth]{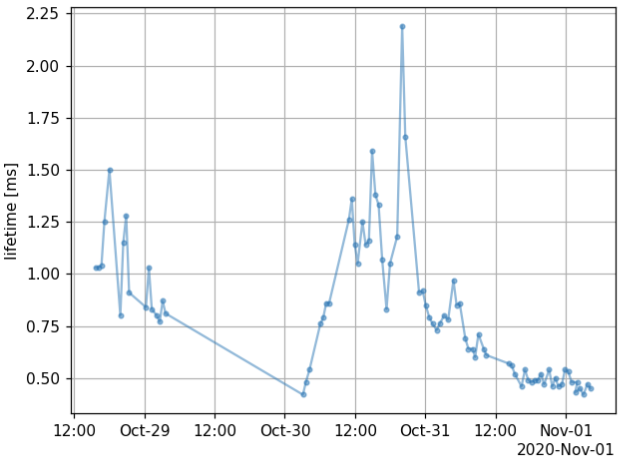}
\includegraphics[width=0.43\textwidth]{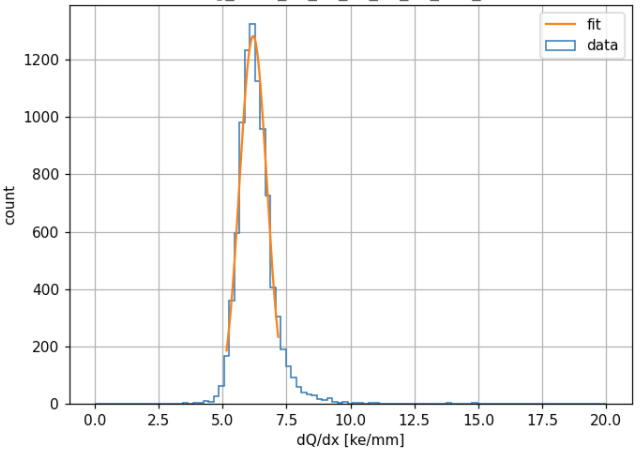}
\end{dunefigure}

\subsection{ArgonCube 2x2 Demonstrator}
\label{sec:2x2Demo}

This demonstrator will consist of four LArTPC modules arranged in a 2x2 grid within a shared high-purity LAr bath.
Each TPC module has a footprint of 0.7 m by 0.7 m, and is roughly 1.4 m tall, as shown in the center panel of Fig.~\ref{fig:ndlar-prototypes}.

The first LArTPC module of this system, called Module 0, will be operated in early 2021 at the Univ.~of~Bern.  
Operation of this first module will achieve the following technical targets of the 2x2 prototyping program necessary for completion of the detector preliminary design by mid-2021:
\begin{enumerate}
    \item Verification of the mechanical robustness (in liquid argon) of the modular LArTPC design, fabricated primarily of fiberglass laminate panels (G10);
    \item Stable delivery of 25~kV baseline (50~kV goal) high voltage to the LArTPC cathode;
    \item Demonstration of an electron lifetime of greater than 500~$\mu$s within the LArTPC;
    \item Demonstration of a pixel charge readout noise of less than 1000~e$^-$~ENC (uncorrelated);
    \item Demonstration of a module scintillation detection efficiency for signals of $>$50~MeV deposited energy.
\end{enumerate}

While the SingleCube TPC test has achieved these performance targets, Module 0 will demonstrate them at a scale comparable to the \dword{nd} TPC module.  
With this large-scale demonstration in hand, the data from Module 0 should also enable the following technical studies:
\begin{enumerate}
    \item 3D imaging and reconstruction of cosmic rays in the modular LArTPC design;
    \item Measurement of the drift field uniformity in the modular LArTPC design.
\end{enumerate}

After evaluation of Module 0,  production will start on the full set of four LArTPC modules to complete the ArgonCube 2x2 Demonstrator.  
Data from operation of these four modules within the 2x2 Cryostat in the surface cosmic ray flux at the Univ.~of~Bern will enable the following technical studies:
\begin{enumerate}
    \item Evaluation of the relative performance of multiple LArTPC modules operating within a common high-purity LAr bath;
    \item Evaluation of the impact of dead volumes using cosmic rays which span multiple LArTPC modules.
\end{enumerate}

\begin{dunefigure}[ArgonCube 2x2 Prototype]{fig:argoncube-prototype}
{(Left) The cryostat for testing Module 0. (Center) The cryostat for the ArgonCube 2x2 Demonstrator. (Right) The 2x2 cryostat and cryogenics system at the Univ.~of~Bern.}
\includegraphics[width=0.2\textwidth]{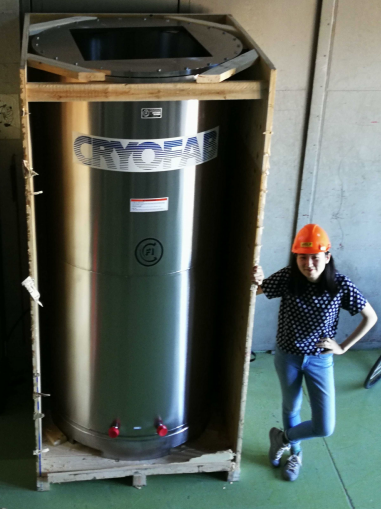}
\includegraphics[width=0.3\textwidth]{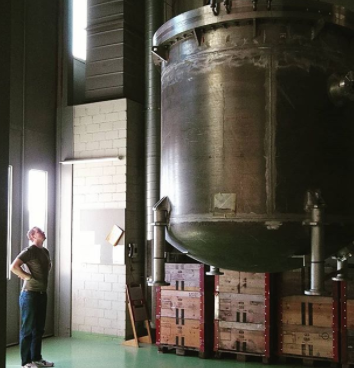}
\includegraphics[width=0.4\textwidth]{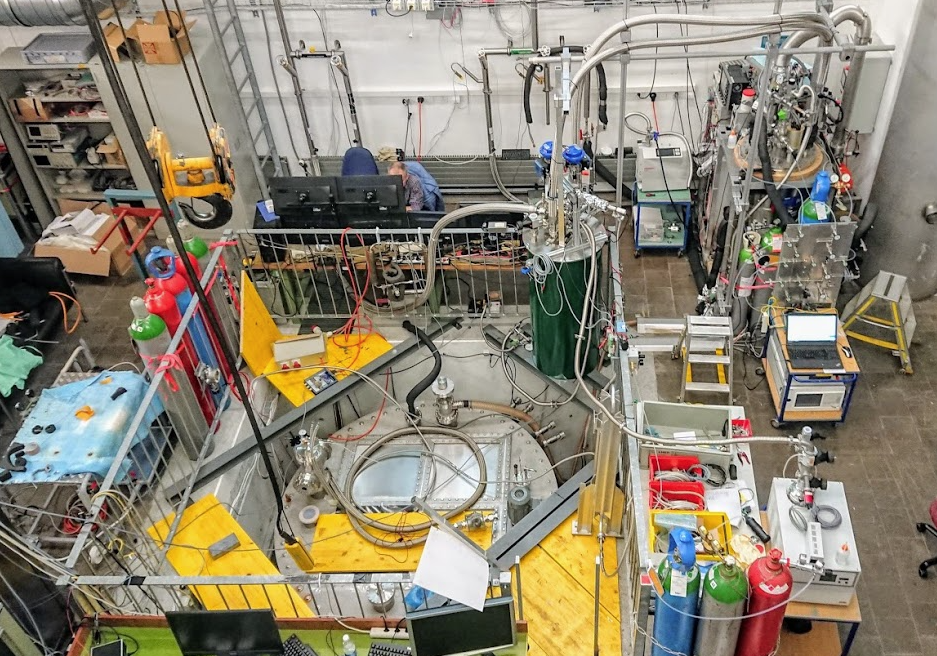}
\end{dunefigure}

After commissioning of the 2x2 at Bern, it will be shipped to Fermilab for installation and operation in the NuMI neutrino beam.  
Data from operation of the 2x2 in this neutrino flux will enable the technical 
study of LArTPC module performance in response to beam neutrino interactions.  Among the goals are to study the following: 

\begin{enumerate}
    \item LArTPC module performance in response to beam neutrino interactions;
    \item Long term operational and stability studies;
    \item Reconstruction of events in multiple modules;
    \item Pile-up studies in the intense beam environment (combination of light and charge signals appropriately in reconstruction);
    \item Connection of tracks from the LArTPC to external detectors (see Section~\ref{subsec:2x2minerva}).
\end{enumerate}

\subsection{Full-scale Demonstrator}

The FSD is an engineering demonstrator for the \dword{nd} LArTPC module design.  
Two phases of FSD operation are foreseen: an initial phase between the completion of the detector preliminary design (mid-2021) and the final design (mid-2022), and a second phase between the completion of the final design (mid-2022) and the start of \dword{nd} production (mid-2023).

The first phase will consist of the construction and operation of one full-scale LArTPC module according to the \dword{nd} design.  
It will be operated in a 1.5-m-diameter and 4-m-tall cylindrical cryostat capable of hosting this one module, and is serviced by a O(10 ton) high-purity LAr cryogenics system.  
The key technical targets of this prototype are:
\begin{enumerate}
    \item Demonstrate that the full-scale LArTPC design continues to meet the key technical specifications described in the preceding section on Module 0 technical targets (e.g. cryo-mechanical stability, HV, LAr purity, charge readout noise, and scintillation efficiency);
    \item Establish and exercise the production and assembly processes for the ND LArTPC modules, including: component production and testing processes, design and production of assembly rigs and lifting fixtures, documented assembly procedures, hazard analyses and safety reviews, etc.;
    \item Identify potential QA/QC issues and use them to refine the QA/QC program in advance of component production;
    \item If appropriate, revise the design to facilitate component production and LArTPC module assembly;
    \item Establish the testing program to be used at the Module Integration Facility (i.e. the ND LArTPC assembly line).  This program will provide validation of the performance of each LArTPC module before these are delivered to the \dword{nd} site for installation and detector commissioning. 
\end{enumerate}

In the second phase, commencing at the conclusion of the final design phase (mid-2022),  another full-scale LArTPC module will be produced according to the final design.  
The assembly and testing program described above will be repeated, and this will serve as a final pre-production validation before we initiate \dword{nd} production in mid-2023.


\section{\dshort{pdnd} physics studies}
\label{sec:ProtoNDPhysics}

Basic detector stability checks will be performed with a period of detector operation at the University of Bern before moving the ArgonCube 2$\times$2 demonstrator module to Fermilab. These tests will include extraction and re-insertion tests of individual modules into the \dword{lar} bath, and checks that the \dword{lar} purity is sufficient. Cosmic muons will be used to validate the technical performance of the modules.  The \dword{lbnf} beamline is an intense source of muon (anti-)neutrinos, with a much higher flux of neutrinos than other accelerator neutrino beams currently in operation~\cite{Strait:2016mof,bib:docdb4559}. A key design requirement for the \dword{dune} \dword{nd}s, and one of the primary concerns motivating \dword{pdnd}, is how well the ND components will perform in a high multiplicity environment. Operating in the NuMI beam will thus allow the verification of these important physics capabilities. Figure~\ref{fig:2x2minerva} shows the deployment of the 2$\times$2 in the MINOS ND hall. The additional components will be described in Section~\ref{subsec:2x2minerva}.

\begin{dunefigure}[The \dshort{arcube} demonstrator as deployed in the MINOS ND hall]{fig:2x2minerva}
	{A drawing of the \dshort{arcube} demonstrator deployed in the MINOS ND hall at Fermilab, forming \dword{pdnd}. One module is shown in the extracted position.  The neutrino beam is incident from the left. Sections of MINERvA are shown upstream and downstream of the demonstrator. A prototype of the gas TPC of \dword{ndgar} is also shown at the downstream end.}
	\includegraphics[width=0.8\textwidth]{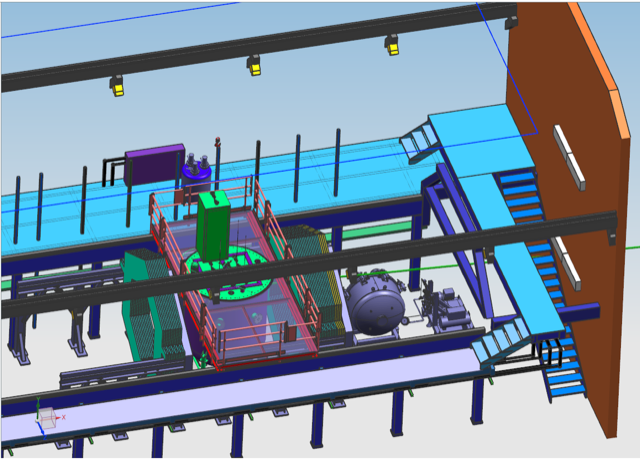}
\end{dunefigure}

In Figure~\ref{subfig:flux}, the \dword{numi} medium energy neutrino and antineutrino fluxes are compared, on an absolutely normalized scale, to the \dword{lbnf} three-horn optimized flux at the 
ND site~\cite{bib:docdb4559}. For the former, the FY2017 delivered protons on target (POT), $5.06\times 10^{20}$, was used to produce a yearly flux and rate~\cite{Convery:2017jlw}, and the nominal POT of $1.1 \times 10^{21}$/year was used for the latter. It is clear that the proposed \dword{lbnf} flux is significantly more intense, but due to the roughly linear relationship between neutrino energy and cross section, the measured rate from the on-axis \dword{numi} beam in the MINOS-ND hall is approximately the same. The rate has been produced with GENIE version v2.12.10\footnote{The version of GENIE used for the studies shown in this section used the ``ValenciaQEBergerSehgalCOHRES'' configuration, which is described in the appendix of reference~\cite{Mahn:2018mai}.}~\cite{genie}. 
Note that the rate is normalized to the active volume of the ArgonCube 2$\times$2 Demonstrator module, showing that significant statistics will be accumulated in a matter of months of \dword{pdnd} operation.

\begin{dunefigure}[Normalized fluxes for $\nu$ beamlines and expected rates for the demonstrator]
{fig:beam_options}
{Comparison of the absolutely normalized fluxes for different neutrino beamlines at Fermilab, and the expected yearly rates in the ArgonCube demonstrator's 
\SI{1.7}{t} active \dword{lar} mass 
as a function of E$_\nu$, produced using GENIE v2.12.10~\cite{genie}.}
\subfloat[Flux\label{subfig:flux}]    {\includegraphics[width=0.35\textwidth]{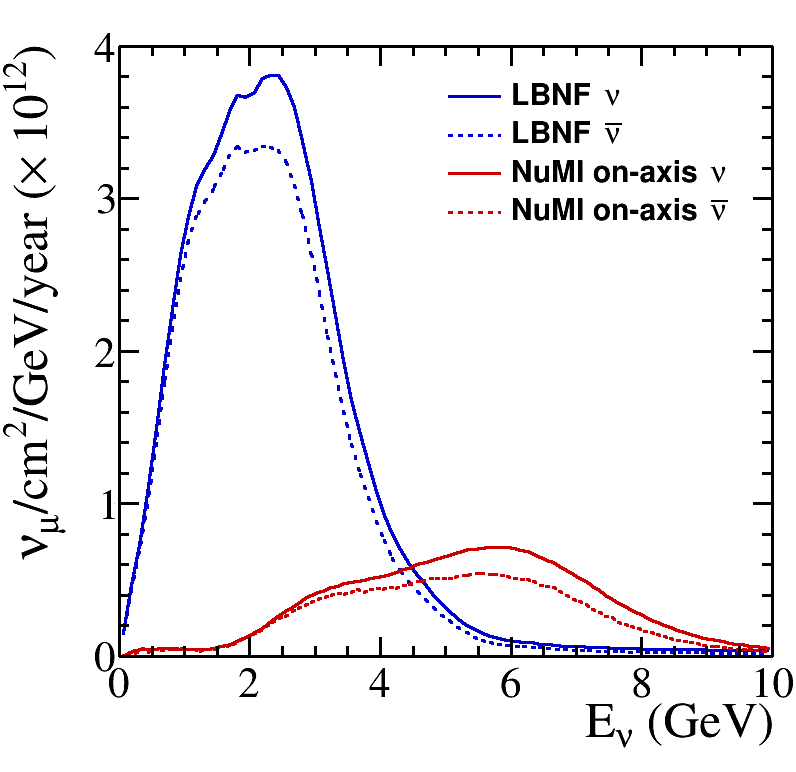}}
	\subfloat[Rate\label{subfig:rate}]    {\includegraphics[width=0.35\textwidth]{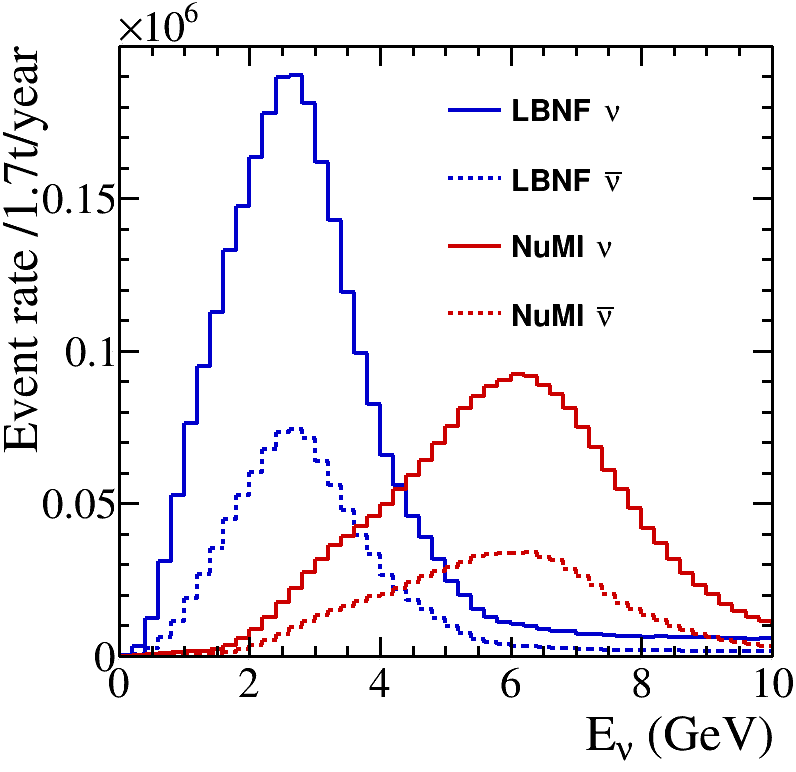}}\end{dunefigure}

In this section, a number of key detector physics questions and tests for the ArgonCube system are identified, which can be answered by \dword{pdnd} and help inform the final design of \dword{ndlar} for the \dword{dune} \dword{nd} deployment. In order to check the feasibility of these studies, two different simulations were used. First, high statistics GENIE samples were produced in order to compare basic properties of neutrino interactions expected in the \dword{lbnf} and \dword{numi} medium-energy (ME) beamlines. Second, GENIE events were used to seed a simple GEANT4 simulation, using the ArgonBox\footnote{\url{https://github.com/dadwyer/argon_box}} software, in order to get a basic understanding of event shape and containment. In the latter simulation, events were simulated in a very large (\SI[product-units=repeat]{200x200x200}{\metre}) box of LAr, and were then distributed randomly inside a volume with the correct spatial dimensions of the ArgonCube 2$\times$2 demonstrator. 
Although the 2$\times$2 geometry was not included in the simulation, this gives an acceptable estimate of the expected event rates and containment for the studies described below, as these do not depend significantly on the detailed geometry of the detector. Note that for all ArgonBox studies shown here, the \dword{numi} on-axis forward horn current (neutrino-enhanced) beam was used. Examples of the ArgonBox simulation with the basic ArgonCube 2$\times$2 Demonstrator geometry superimposed can be seen in Figure~\ref{fig:argonbox_event_display} for a number of different neutrino energies.

\begin{dunefigure}[Sample $\nu_{\mu}$--argon ArgonBox simulated events for different incident $\nu$ energies]
{fig:argonbox_event_display}
{Example $\nu_{\mu}$--argon ArgonBox simulated events for a number of different incident neutrino energies, where the energy deposits in a bulk volume of \dword{lar} are color-coded according to the particle type: $\pi^{\pm}$ --- cyan; $\mu^{\pm}$ --- purple; $e^{+}$ --- green; $e^{-}$ --- yellow; proton --- red; recoiling nuclei --- black. The event vertices are randomly placed within the active volume of the 2$\times$2 Demonstrator module, the geometry for which is superimposed on these images, but which is not simulated by ArgonBox.}
  \subfloat[$E_{\nu}$ = 2.60 GeV] {\includegraphics[width=0.40\textwidth]{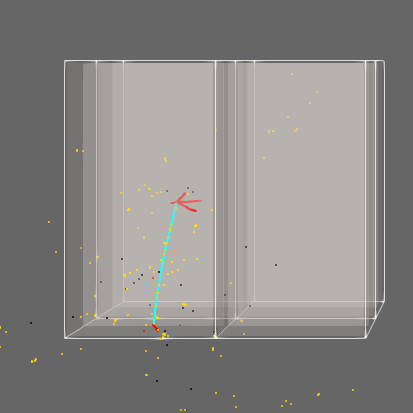}}\hspace{0.04\textwidth}
  \subfloat[$E_{\nu}$ = 3.36 GeV] {\includegraphics[width=0.40\textwidth]{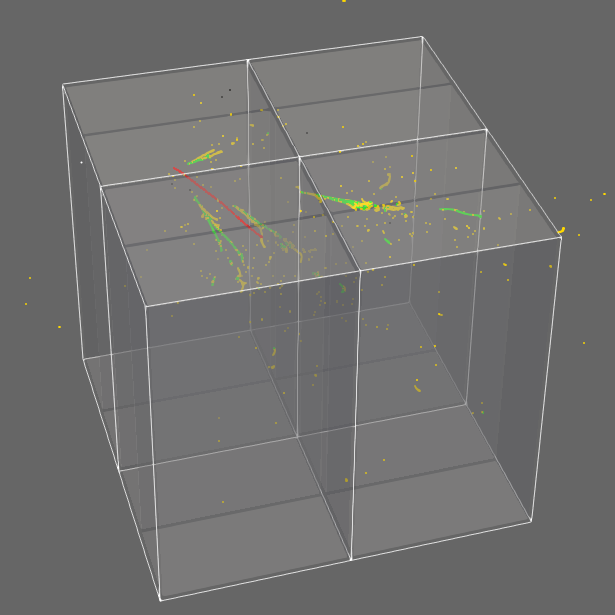}}\hspace{0.04\textwidth}
  \subfloat[$E_{\nu}$ = 9.37 GeV] {\includegraphics[width=0.40\textwidth]{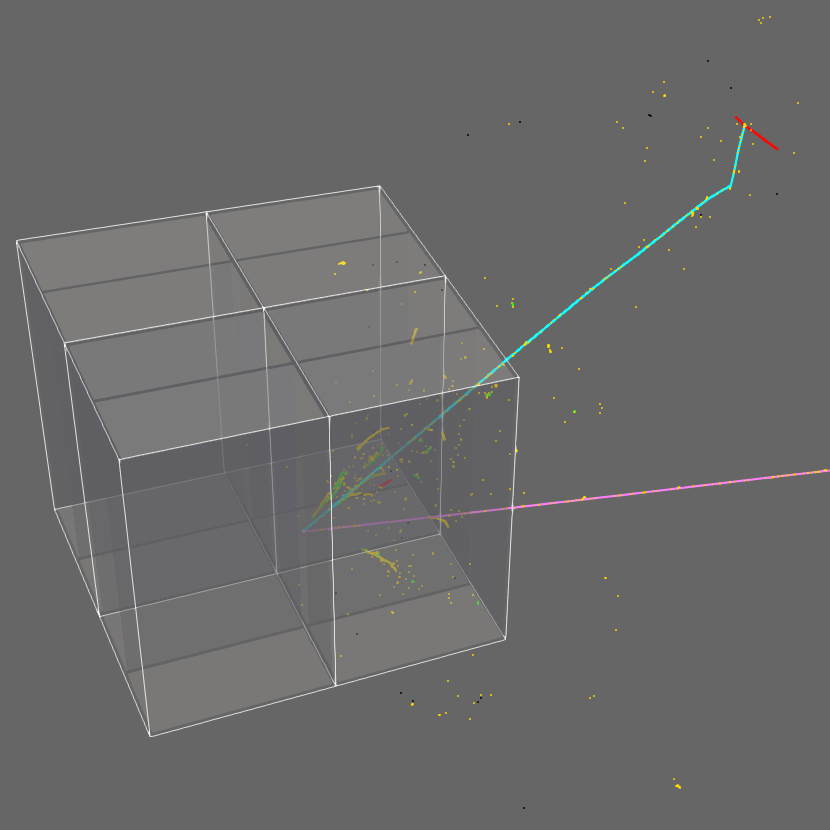}}
\end{dunefigure}

The example event displays shown in Figure~\ref{fig:argonbox_event_display} give a basic idea of how \dword{numi} ME events (in \dword{fhc}) would look in the ArgonCube 2$\times$2 Demonstrator module. Although many of the tracks and showers are not contained, some fraction are, which is discussed in more detail for the detector physics studies described below.

In order to be a relevant test for the full \dword{ndlar} deployment in the \dword{lbnf} beamline, it is useful to verify that the basic properties of the events are similar, despite the \dword{numi} ME beam being somewhat higher energy than the planned \dword{lbnf} beam (as shown in Figure~\ref{fig:beam_options}). Figure~\ref{fig:track_multiplicity} shows the expected multiplicity of ionizing tracks at the vertex for both the \dword{lbnf} and \dword{numi} ME beams, in neutrino and antineutrino mode, produced using GENIE samples. The track multiplicities are similar, which indicates that the scale of the reconstruction problem is similar, and \dword{pdnd} will be a useful benchmark for developing the \dword{ndlar} reconstruction software.

\begin{dunefigure}[Expected yearly rates of min and highly ionizing particles in demonstrator]
{fig:track_multiplicity}
{The yearly rates of minimum and highly ionizing particles expected in the demonstrator's 
\SI{1.7}{\tonne} \dword{lar} mass 
for the \dword{numi} ME and \dword{lbnf} fluxes, produced using GENIE v2.12.10~\cite{genie}.}
  \includegraphics[width=0.5\textwidth]{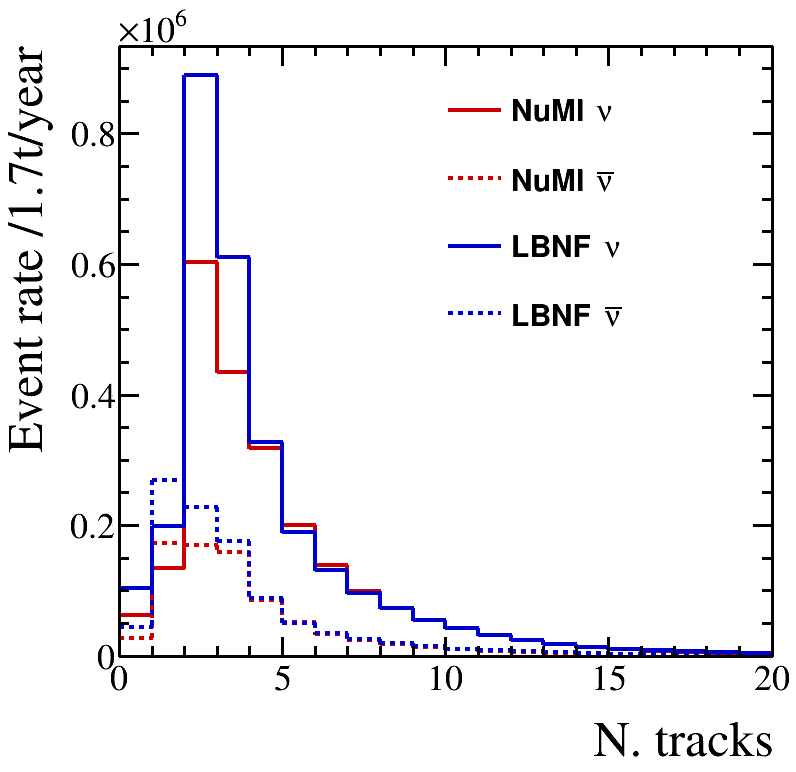}
\end{dunefigure}

In Figure~\ref{fig:momenta}, the momenta of various particles coming from the initial neutrino--argon vertex are compared for the \dword{lbnf} and \dword{numi} ME beams. As expected, the energy distributions of all of the particles are slightly broader for the \dword{numi} ME flux, but there are significant numbers of events in the \dword{numi} sample which have particle kinematics typical of the \dword{lbnf} sample. The \dword{numi} sample is therefore an efficient tool for studying the performance of \dword{ndlar}  in the \dword{lbnf} beam.

\begin{dunefigure}[Expected yearly rates of particles produced at  primary vertex vs $p$ in demonstrator]
{fig:momenta}
{The yearly rates of various particles produced at the primary vertex, as a function of their momentum, as expected in the the demonstrator's 
\SI{1.7}{\tonne} \dword{lar} mass 
for the \dword{numi} ME and \dword{lbnf} fluxes, produced using GENIE v2.12.10~\cite{genie}. Note that every relevant particle from each event is included.}
  \subfloat[$\mu^{\pm}$] {\includegraphics[width=0.30\textwidth]{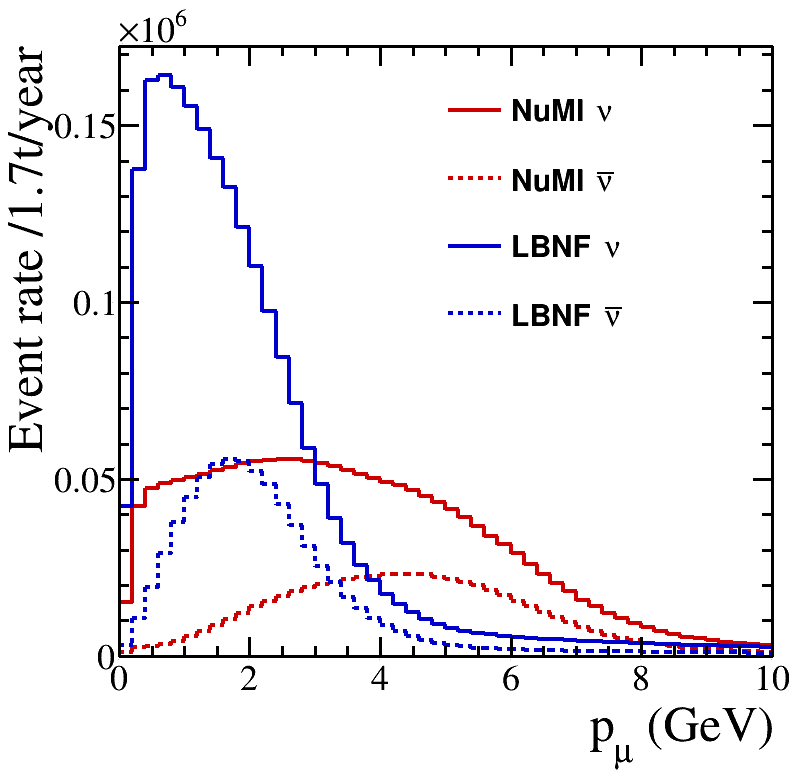}}
  \subfloat[Protons]    {\includegraphics[width=0.30\textwidth]{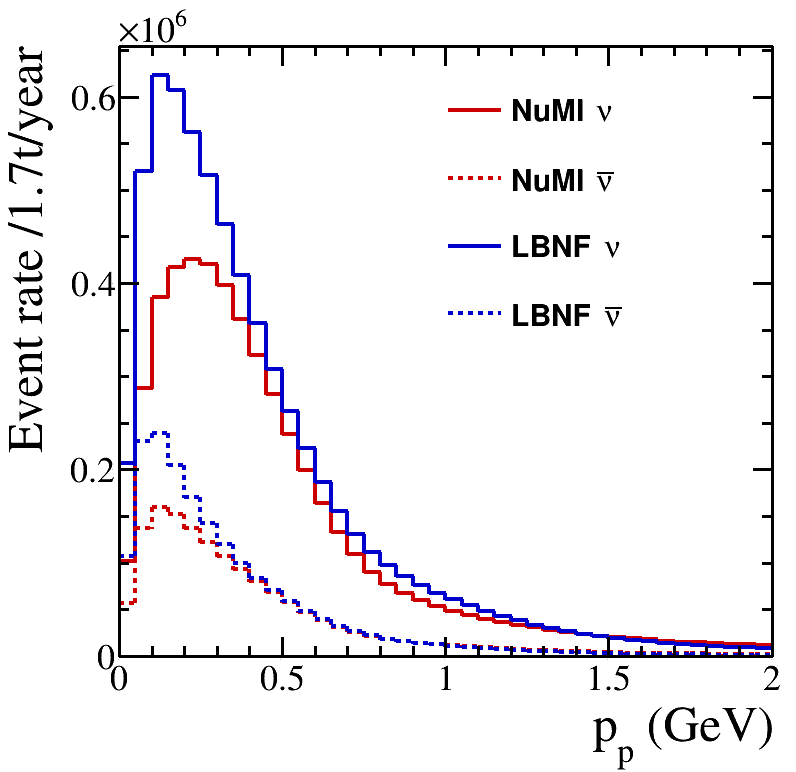}}
  \subfloat[$\pi^{+}$]   {\includegraphics[width=0.30\textwidth]{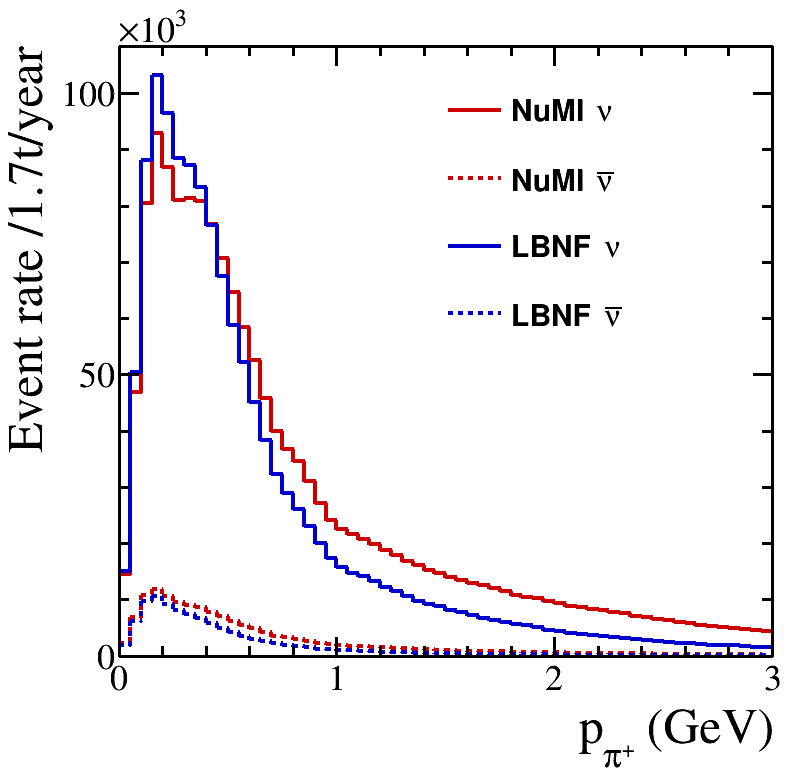}}
\end{dunefigure}

In the full 5 $\times$ 7 module \dword{ndlar} detector and the more intense \dword{lbnf} beamline there will be $\sim$14.7 interactions per \SI{10}{\micro\second} beam spill, making for a very high-multiplicity environment. The entire spill will effectively occur instantaneously in the \SI{250}{\micro\second} drift window. Issues with tracks overlapping from separate neutrino interactions are mitigated by the fully-3D readout, but association of all hits to a specific neutrino interaction can still be challenging. For charged tracks, which are spatially connected to their respective neutrino interaction vertices, this association is straightforward.  However, many events contain photons and neutrons, which produce significant energy deposits that are detached from the rest of the event, and may even occur in a different ArgonCube module. Here, the ArCLight light-readout system, with the ability to measure prompt scintillation light with nanosecond resolution, will play a crucial role to associate particle tracks with the correct interaction vertices. Additionally, the relatively small size of the ArgonCube  2$\times$2 Demonstrator module means that relatively few of the tracks will be contained, making particle identification (PID) studies challenging, except for the cases listed below. Although other detectors are not included in the ArgonBox simulation, the lack of containment and PID capabilities mean that including another subdetector in the \dword{pdnd} setup is essential for any detector response measurements as a function of charge or momentum.

\FloatBarrier
\subsection{Combining light and charge signals}
An important challenge is to develop automated event reconstruction software for the \dword{ndlar} detector. The pixel readout removes the ambiguities present for projective wire readout \dword{lartpc}s, but the reconstruction software for the latter has benefited from several years of development for the MicroBooNE~\cite{microboone} and ICARUS experiments~\cite{icarus}. Recent progress has been made in understanding how to reconstruct pixel readout  via the 
PixLAr experiment (where pixel planes were introduced to the LArIAT experiment~\cite{Cavanna:2014iqa}).  Still, the reconstruction problem for charged particle scattering in a small \dword{lartpc} is much simpler than for the \dword{pdnd} or \dword{dune} \dword{nd} environments. Additionally, the reconstructed track position along the drift direction, and the suppression of cosmic backgrounds within the beam window, will be performed using information from the ArCLight light collection system in the 2$\times$2 Demonstrator and \dword{ndlar}. Verifying that the light and charge signals can be combined in the full-size ArgonCube modules, in a comparably noisy environment to the \dword{dune} \dword{nd}, is an essential test of the ArgonCube design.

\subsection{Neutron tagging}
\label{sec:2x2_neutron}

Neutrons present a particular challenge for neutrino energy reconstruction in DUNE and other long-baseline neutrino oscillation experiments. 
Neutrino oscillations are a function of neutrino energy, but  because neutrons carry away some fraction of the energy, and are not directly observable, the event-by-event energy reconstruction is problematic. 
This is true for neutrons generated at a neutrino vertex and for hadronic showers that fluctuate to neutrons. 
Figure~\ref{fig:neutron_kinematics} shows the expected neutron rate in neutrino interactions as a function of multiplicity and momentum for the \dword{lbnf} and \dword{numi} ME beamlines.

\begin{dunefigure}[Expected yearly rates of neutrons produced at  primary vertex vs $p$ and multiplicity in demonstrator]
{fig:neutron_kinematics}
{The expected yearly rates of neutrons produced at the vertex, as a function of event multiplicity and their momentum, expected in the  the demonstrator's 
\SI{1.7}{\tonne} \dword{lar} mass 
for the \dword{numi} ME and \dword{lbnf} fluxes, produced using GENIE v2.12.10~\cite{genie}. Note that every neutron from each event is included in the momentum distribution.}
	\subfloat[Neutron multiplicity]   {\includegraphics[width=0.45\textwidth]{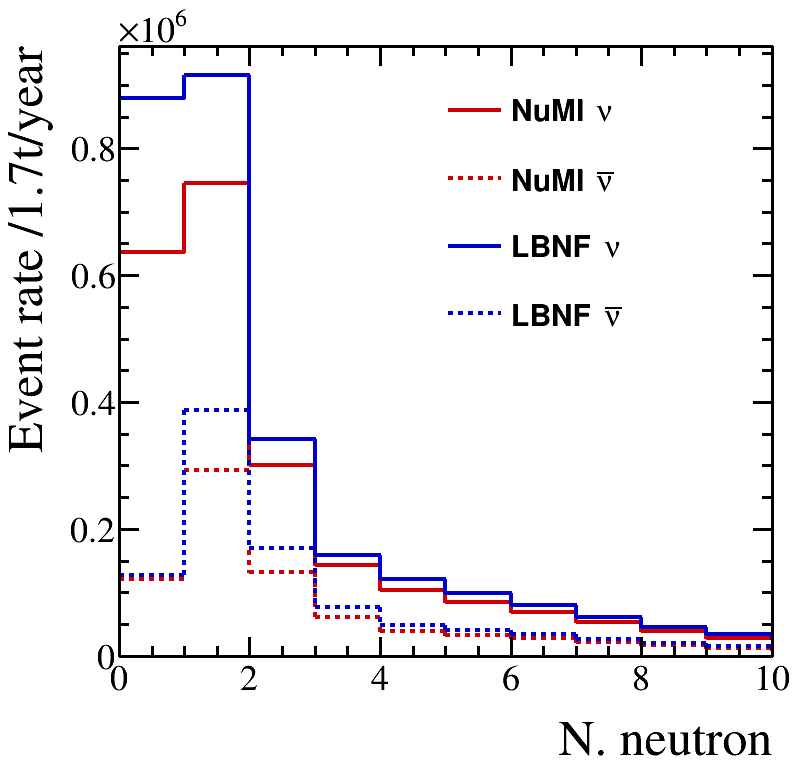}}
	\subfloat[Neutron momentum]       {\includegraphics[width=0.45\textwidth]{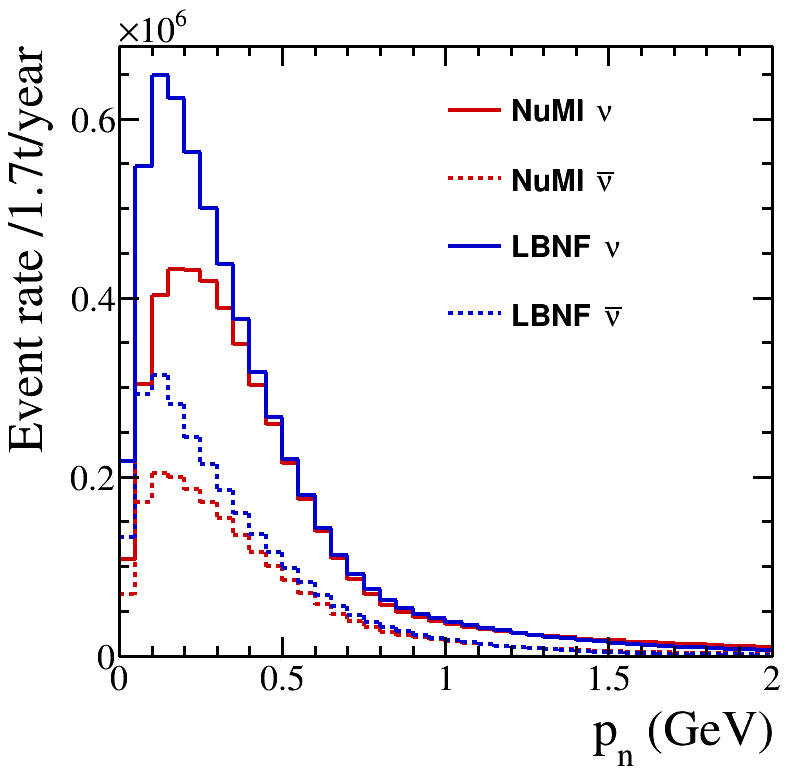}}  
\end{dunefigure}

A common technique for seeing neutrons experimentally is to observe the effects of neutron capture by a nucleus.  In order for this process to happen, the neutrons must thermalize and have a kinetic energy $\mathcal{O}\left(1\right)\,\SI{}{\kilo\electronvolt}$.  For neutrons produced in accelerator neutrino interactions, this can take of $\mathcal{O}\left(1\right)\,\SI{}{\milli\second}$, which is too slow for association with a given neutrino interaction.  Also both the initial direction and kinetic energy information is lost, making the detection of the thermalized neutron less useful for reconstruction of the neutrino event.

The detection of fast neutrons with a kinetic energy $\mathcal{O}\left(1\right)\,\SI{}{\mega\electronvolt}$ to $\mathcal{O}\left(1\right)\,\SI{}{\giga\electronvolt}$ is also possible via the observation of recoiling charged particles after a collision of the neutron with a nucleus. The recoiling particle can be the nucleus as a whole, or, if the neutron exceeds the nuclear binding energy ($\sim$~\SI{5}{\mega\electronvolt} for an argon nucleus), a knock-out proton or heavier nuclear fragments.

For oscillation experiments, fast neutrons  may carry away a significant fraction of the neutrino energy in an event.
It is, therefore, of great interest to investigate the potential of \dword{lar} experiments to tag these missing neutrons with neutron-induced recoils and, where possible, use timing and/or spatial information associated with the recoils to incorporate the neutron into reconstruction.

Neutron tagging will be investigated with \dword{pdnd}. The neutron tagging rate will provide useful information for DUNE sensitivity studies as it provides an opportunity to investigate how well charge and light signals can be combined. In \dword{pdnd}
prompt scintillation light provides an important handle for neutron tagging, allowing for the association of detached energy deposits to the correct neutrino interaction using timing information.  In studies presented here, the values for pixel pitch and ArCLight threshold used  are taken from  
Reference~\cite{argoncube_loi}. Although not identical to those used in the  2$\times$2, they are sufficiently close for the purpose of this work. 

Figure~\ref{fig:NDSpill} shows a simulated beam spill in \dword{ndlar}, highlighting the challenge of associating fast-neutron induced energy deposits to a neutrino vertex using only collected charge.  
By containing scintillation light, prompt light signals can be used to associate fast-neutron induced deposits back to a neutrino vertex anywhere within the detector.
Figure~\ref{fig:Timing} shows the temporal distribution of neutrino vertices within a representative, randomly selected, \dword{lbnf} beam spill in \dword{ndlar}.
The mean separation of neutrino vertices is \SI{279}{\nano\second}, with all fast-neutron induced energy deposits occurring $<$\SI{10}{\nano\second} after each neutrino interaction.      

\begin{dunefigure}[A beam spill in the \dshort{lar} component of the \dshort{dune} ND]{fig:NDSpill}
	{A beam spill in \dword{ndlar}. 
		Fast-neutron induced recoiling proton tracks, with an energy threshold greater than $\sim\,$\SI{10}{\mega\electronvolt}, are shown in white.
		The black tracks are all other energy deposits sufficient to cause charge collected at the pixel planes.}
	\includegraphics[width=.7\textwidth]{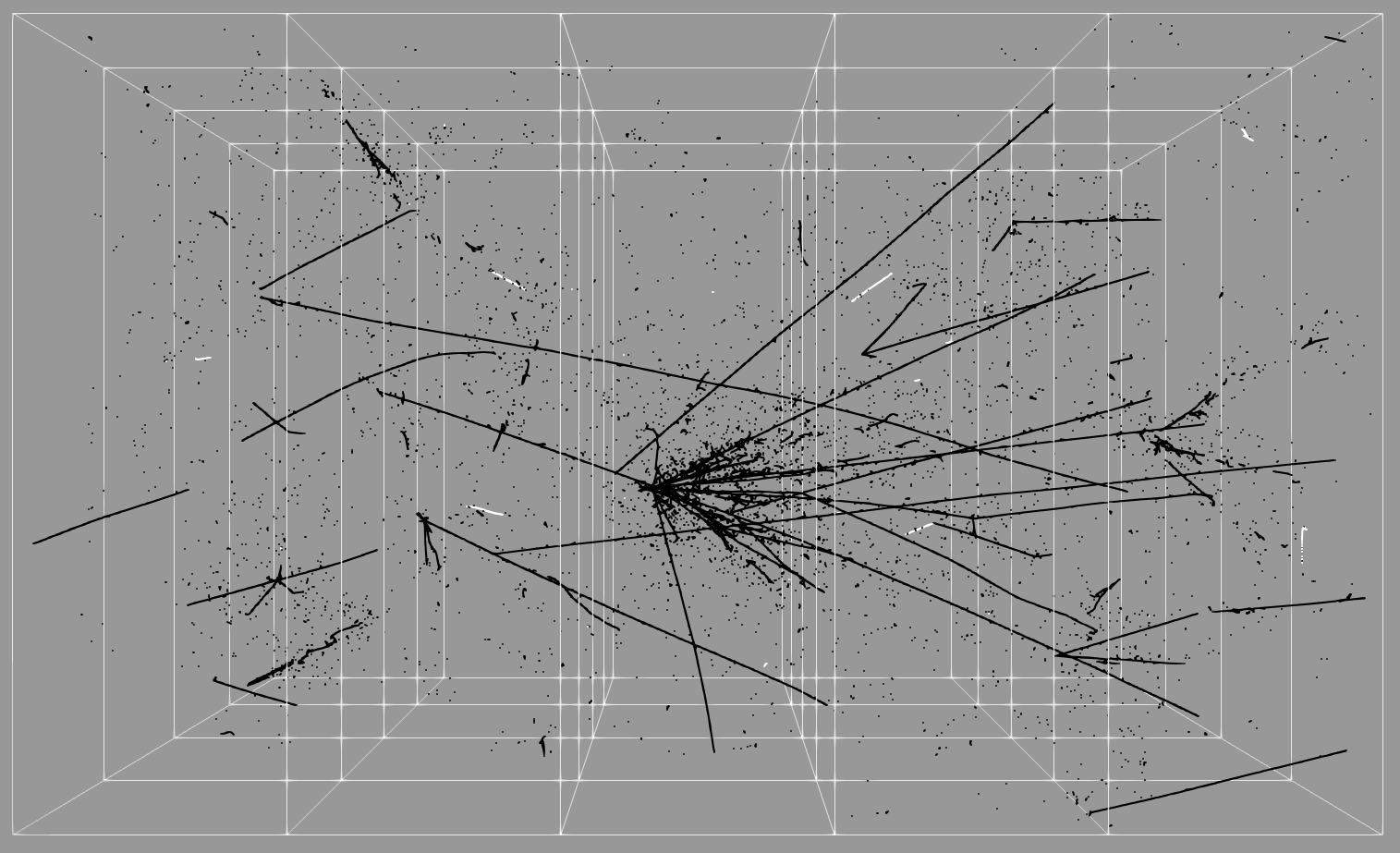}
\end{dunefigure}

\begin{dunefigure}[Temporal distribution of $\nu$ vertices within a beam spill in the ND LAr component] 
	{fig:Timing}
	{The simulated temporal distribution of neutrino vertices (red lines) within a portion of a beam spill in \dword{ndlar}.
		The mean separation of neutrino vertices is \SI{279}{\nano\second}. The filled bins show the number of hits due to recoiling protons, stars indicate a hit due to a recoiling $^{2}$H, $^3$H, $^2$He or $^3$He nucleus.
		All fast-neutron induced energy deposits occur $<$\SI{10}{\nano\second} after each neutrino interaction.}
	\includegraphics[width=0.98\textwidth]{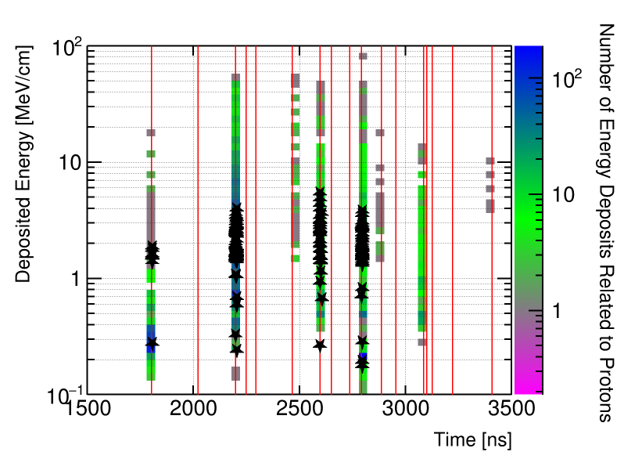}
\end{dunefigure}

Figure~\ref{fig:neutron_recoils} shows the kinetic energy of secondary particles after the interaction of a primary neutron in LAr. While recoiling argon nuclei show typical energies between \SI{100}{\kilo\electronvolt} and \SI{1}{\mega\electronvolt}, recoiling protons show energies $>$\SI{1}{\mega\electronvolt}, up to several \SI{}{\giga\electronvolt}.
Given the LArPix $\sim$~\SI{4}{\milli\metre} pixel-pitch, the minimum reconstructable track length in ArgonCube is also $\gtrsim$~\SI{4}{\milli\metre}. Figure~\ref{fig:proton_length} shows the track length of recoiling protons with respect to the primary neutron kinetic energy. Recoiling protons can, depending on their energy, produce tracks which are up to $\sim$\SI{10}{\centi\metre} long. About 30\% of all recoiling protons are resolvable by the pixelated charge readout, which correspond to protons that are knocked out of a nucleus by primary neutrons with energies $\gtrsim$~\SI{50}{\mega\electronvolt}. The vast majority of neutron recoils contain no direction information, and will be detected only as single pixel hits.

\begin{dunefigure}[KE distribution of secondary particles vs incident neutron KE]
{fig:neutron_recoils}
{Kinetic-energy distribution of secondary particles with respect to incident neutron kinetic energy for neutron interactions in LAr, shown for 100,000 simulated neutrino events (which may have more than one neutron produced at the vertex).}
	\includegraphics[width=0.65\textwidth]{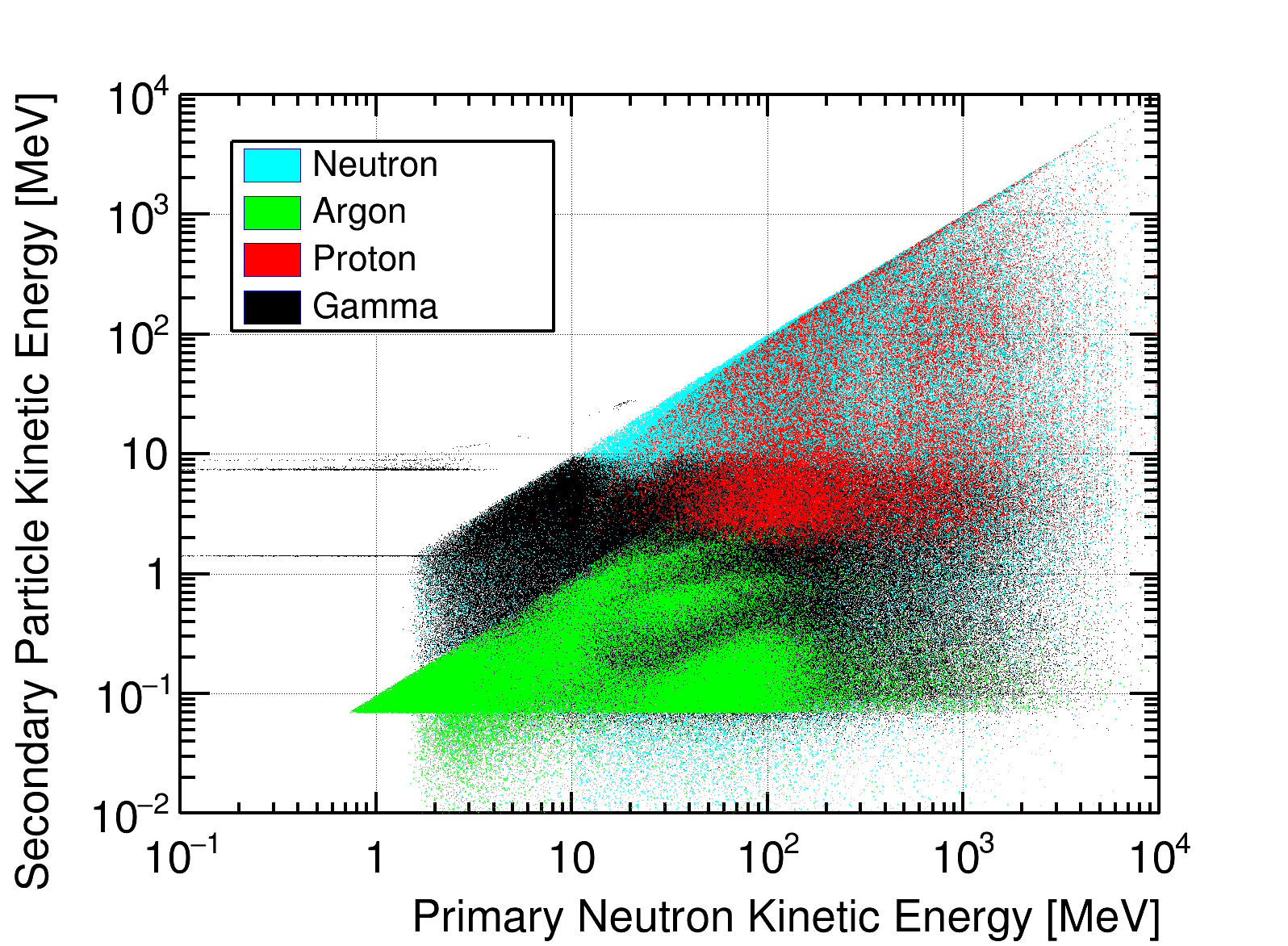}
\end{dunefigure}

\begin{dunefigure}[Track length of recoiling protons for neutrons produced in 100k $\nu$ interactions]
{fig:proton_length}
{Track length of recoiling protons for neutrons produced in 100,000 neutrino interactions. About 30\% of all recoils are resolvable as tracks with the LArPix pixel charge-readout system. The horizontal red line denotes the 3~mm charge-readout pixel pitch, which is considered the minimum length for resolving the corresponding energy deposits as a particle track.}
	\includegraphics[width=0.65\textwidth]{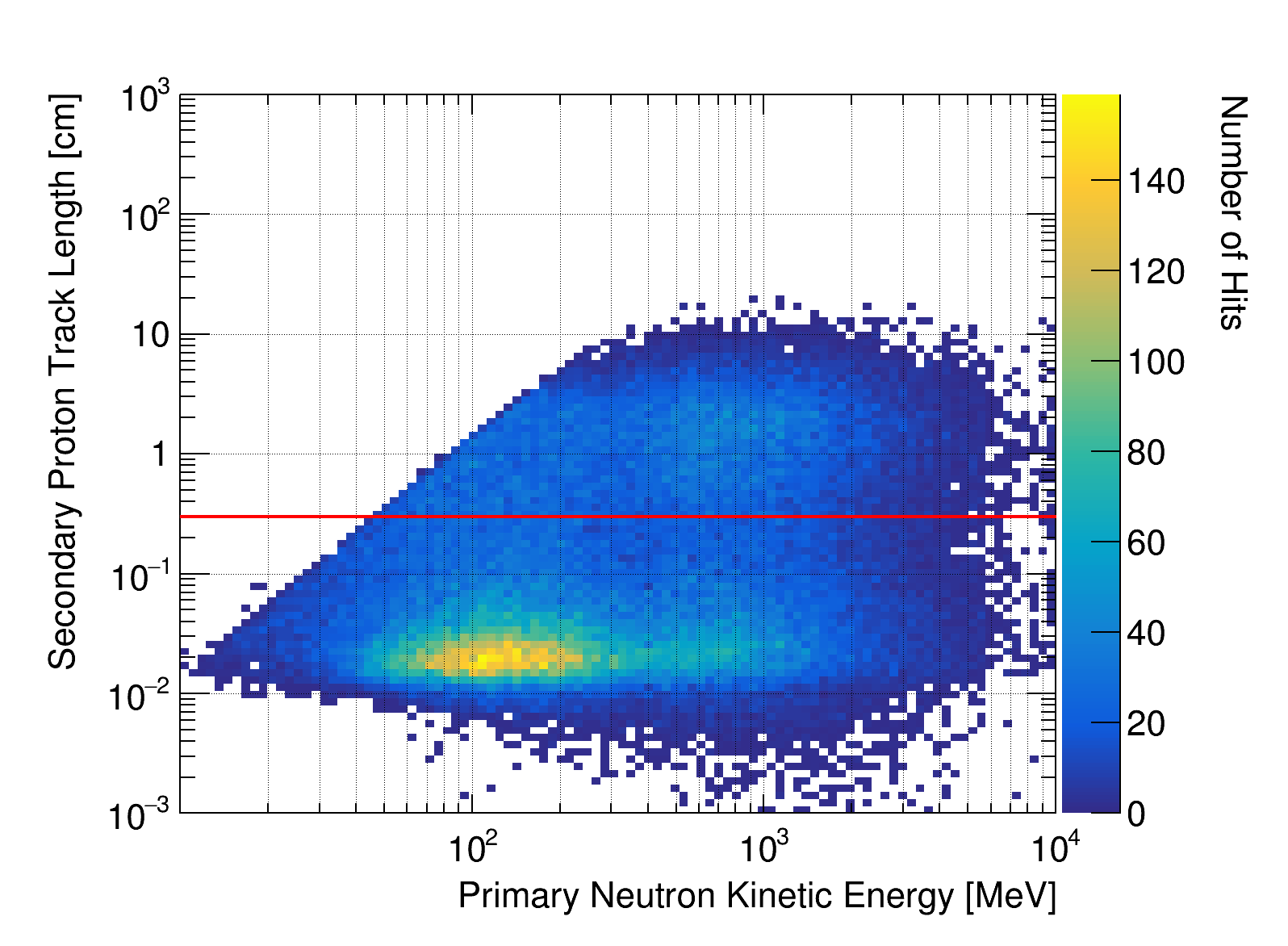}
\end{dunefigure}

Figure~\ref{fig:min_dist_proton} shows the minimum distance between the neutrino vertex and the neutron-induced proton track, as a function of neutron kinetic energy. The majority of proton recoils occur within \SI{1}{m}, so many neutron-induced proton recoils will be contained within the  demonstrator module.

\begin{dunefigure}[Min distance between the $\nu$ vertex and  neutron-induced proton track vs neutron KE]
{fig:min_dist_proton}
{The minimum distance between the neutrino vertex and the neutron-induced proton track, as a function of neutron kinetic energy. Produced with 100,000 initial neutrino events simulated by ArgonBox.}
	\includegraphics[width=0.65\textwidth]{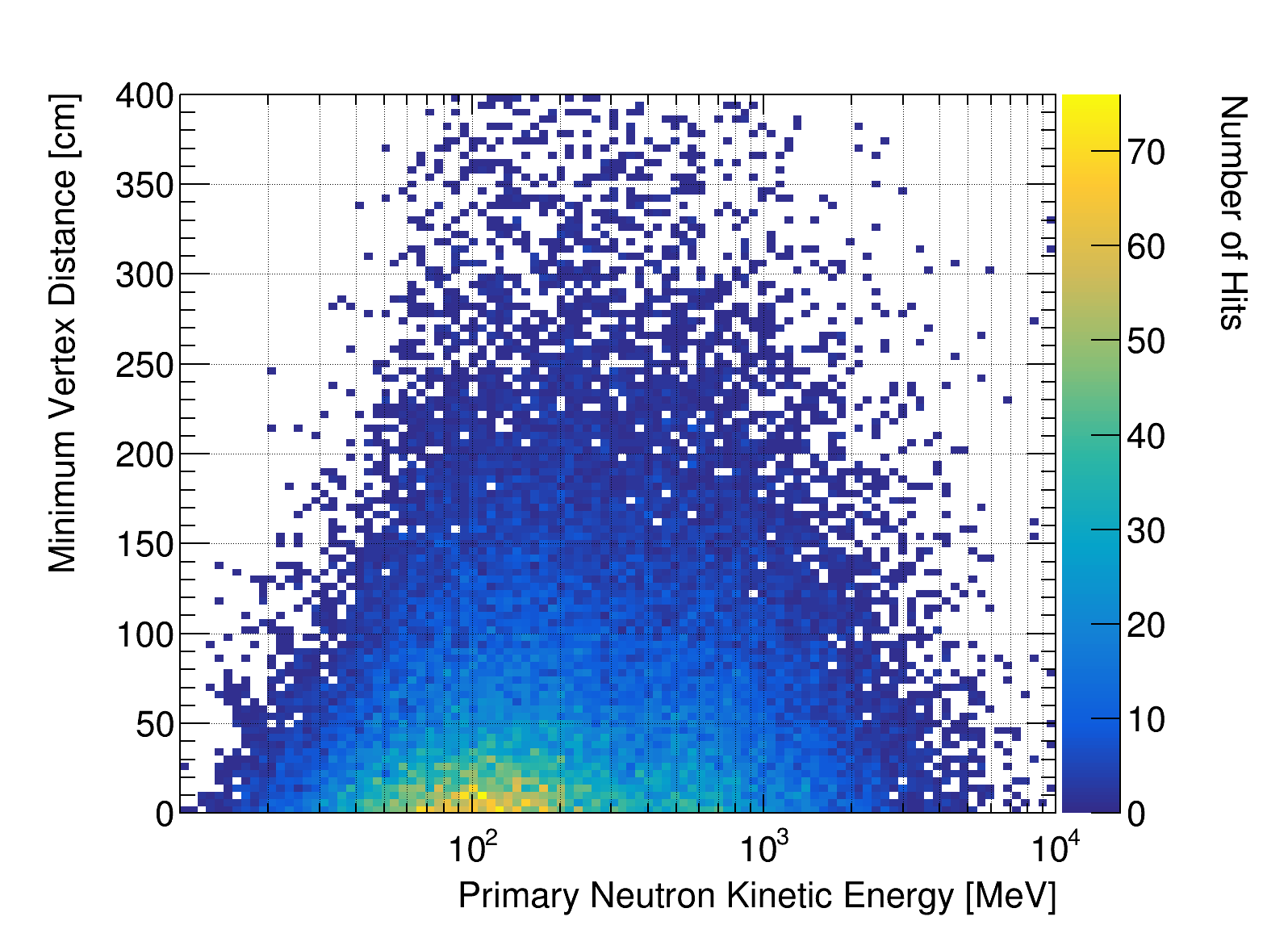}
\end{dunefigure}

\subsection{Reconstruction in a modular environment}
The module walls of the ArgonCube design produce gaps in the active volume for particle tracks traversing multiple modules. This differs from dead wires in classic \dword{lartpc} readouts, as it results in only $\mathcal{O}\left(10\right)\,\mathrm{mm}$ gaps in energy deposits, rather than degrading sensitivity over large areas of charge readout. Algorithms to join such segmented tracks already exist~\cite{Acciarri:2017hat}, but have not been adapted to the ArgonCube design. Simple track matching efficiencies across modules can be calculated using cosmics, which will be an essential first step. However, for events with many tracks produced at the vertex (see Figure~\ref{fig:track_multiplicity}), a detailed study of the reconstruction performance across the module walls will need to be carried out. \dword{pdnd} provides an opportunity to do so, and to develop and understand reconstruction software  before the deployment of \dword{ndlar}.

This problem becomes significantly more complicated for electromagnetic (EM), or hadronic, showers which cross modules. \dword{pdnd} will provide an opportunity to develop reconstruction software and check how well it performs for shower energies in the range of interest for the neutrino interactions expected in \dword{dune}.
At these energies, shower reconstruction in \dword{lar} is a significant challenge due to the disconnected activity arising from the shower development. Additionally, in order to test how well the reconstruction can identify shower depth, a sample of fully contained showers would be extremely useful. Figure~\ref{fig:2x2_shower_containment} shows the efficiency to fully contain EM-showers or proton tracks produced by an interaction within the ArgonCube  2$\times$2 active volume, as a function of initiator particle energy and angle w.r.t the incoming beam direction. Note that if $\geq 90$\% of energy is deposited within the  2$\times$2 active volume, it is classed as contained. 

\begin{dunefigure}[Containment efficiency for EM showers and proton tracks vs initiator particle energy and angle.]
{fig:2x2_shower_containment}
{Containment efficiency for EM showers and proton tracks produced by an interaction within the ArgonCube  2$\times$2 active volume, as a function of initiator particle energy and angle w.r.t the incoming beam direction. Note that if $\geq 90$\% of energy is deposited within the  2$\times$2 active volume, it is classed as contained.}
  \subfloat[EM shower]      {\includegraphics[width=0.45\textwidth]{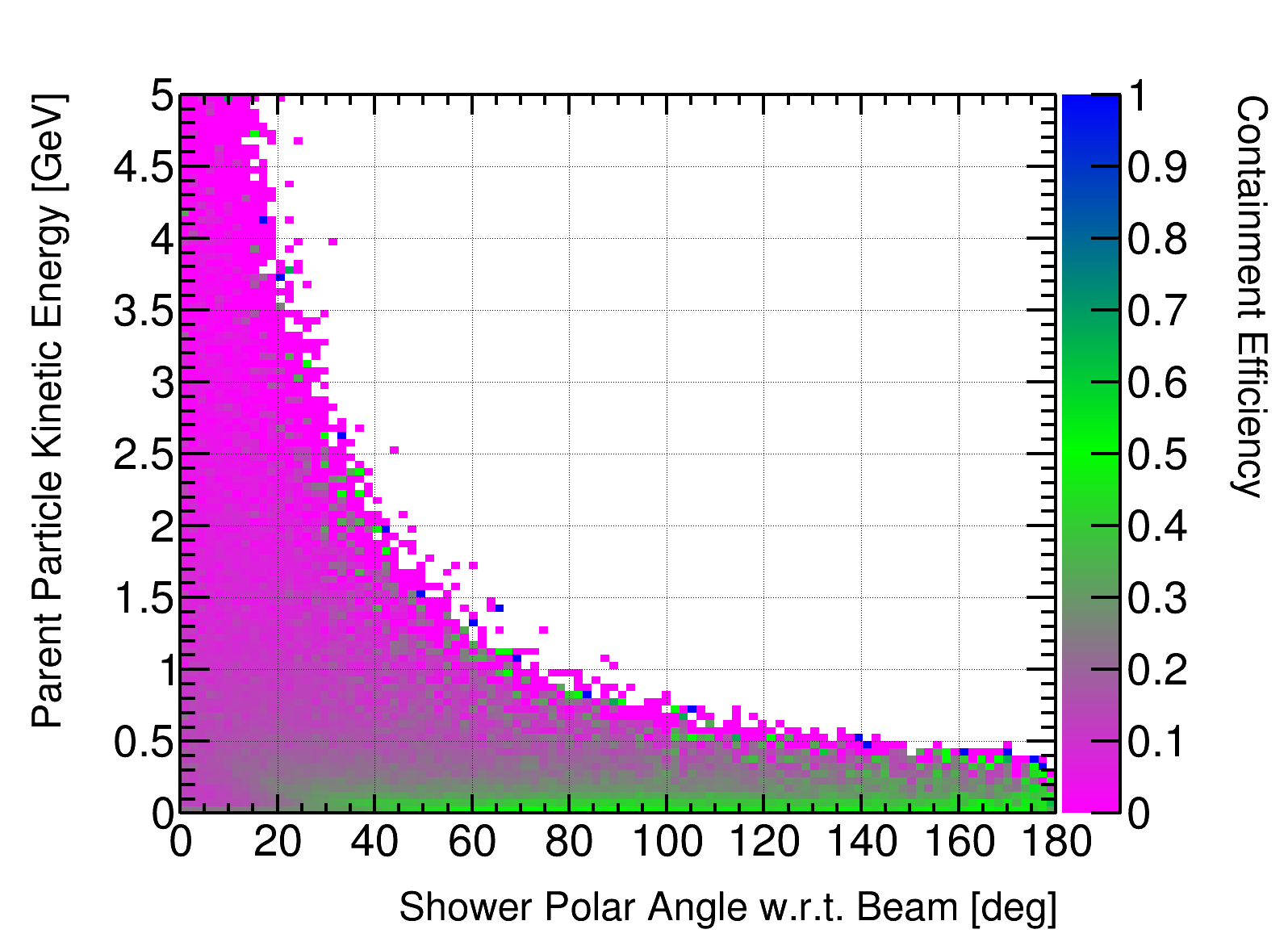}}
  \subfloat[Proton tracks]  {\includegraphics[width=0.45\textwidth]{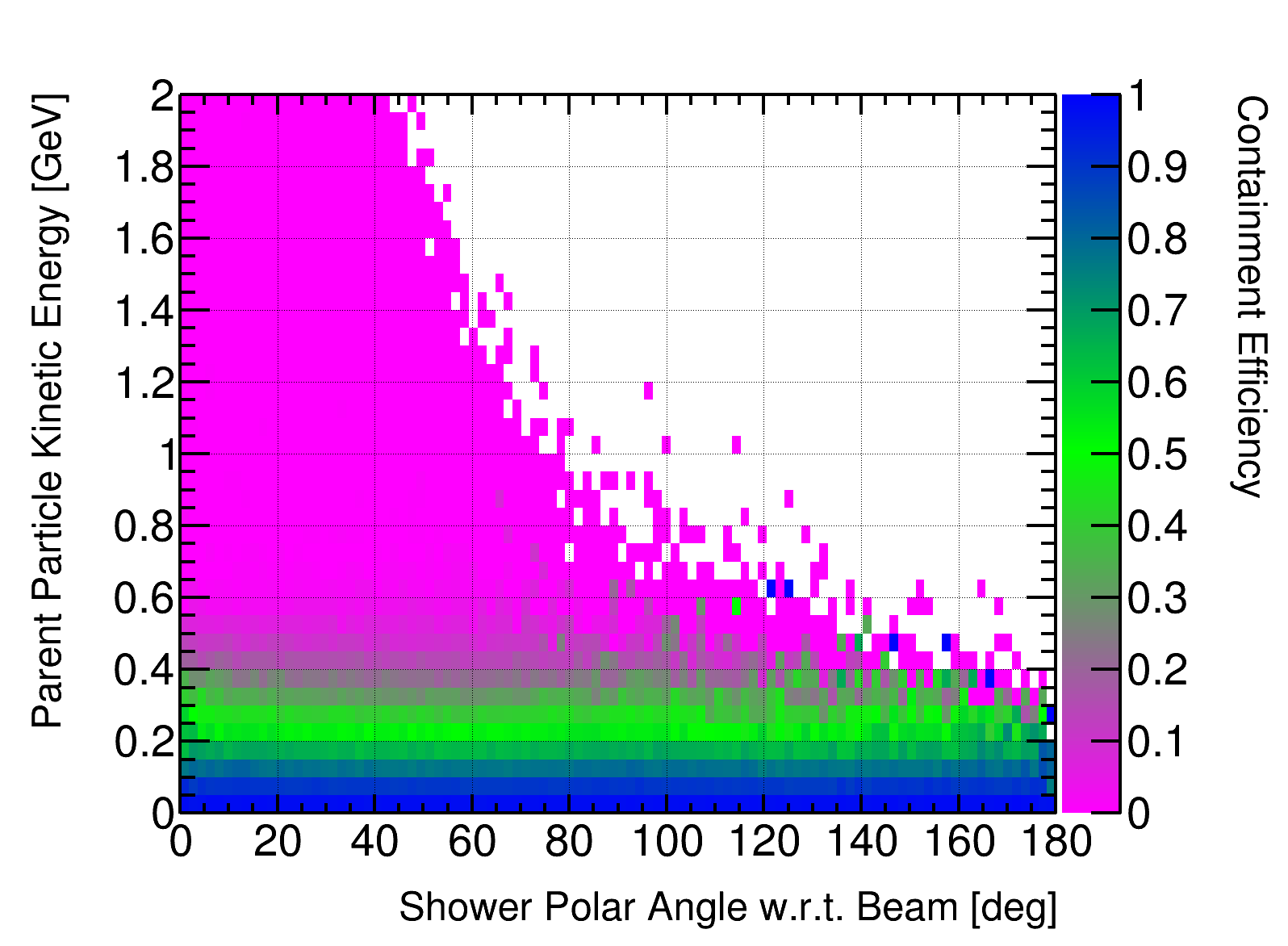}}
\end{dunefigure}

\subsection{Neutral pion reconstruction}
\label{sec:pi0_reco}
A more quantitative measure of how well EM showers can be reconstructed in the modularized ArgonCube detector could be possible using $\pi^{0} \rightarrow \gamma\gamma$ decays (branching fraction $=$ 98.8\%~\cite{Tanabashi:2018oca}), in which both decay photons produce a shower, and are contained in the active volume of the detector. Combining the information on the two showers, and attempting to reconstruct the invariant mass peak of the $\pi^{0}$ provides a measurement of the EM shower resolution. Studies have shown that a 3D-charge readout will improve reconstruction of $\pi^{0}$ showers by removing energy deposits from events crossing the shower~\cite{Goeldi_thesis}. One aim of the \dword{pdnd} program will be to demonstrate this.
Note that Dalitz decay, $\pi^{0} \rightarrow \gamma e^{+} e^{-}$ (branching fraction $=$ 1.2\%~\cite{Tanabashi:2018oca}) may also be an interesting sample in such a high statistics environment, as only a single photon has to convert in the LAr. However, this sample was not considered further in this initial study.

Figure~\ref{fig:pi0_kinematics} shows the expected $\pi^{0}$ production rate in the active volume of the  2$\times$2 in the \dword{lbnf} and \dword{numi} ME beamlines, as a function of $\pi^{0}$ multiplicity in each event and $\pi^{0}$ momentum. 
Figure~\ref{fig:pi0_containment_2x2} shows the efficiency for containing both photon-induced showers from a primary $\pi^{0}$ decay in the  2$\times$2's active volume, shown for all $\pi^{0}$'s produced inside that volume. As expected, the efficiency is low for high energy pions, but it will still be possible to reconstruct a large fraction of the lower momentum $\pi^{0}$'s from Figure~\ref{fig:pi0_kinematics}. Thus, in spite of the photon containment issues and the unavoidable bias toward lower energy EM showers in the 2$\times$2 relative to \dword{ndlar}, it will be a useful exercise to work on $\pi^{0}$ mass peak reconstruction in \dword{pdnd}.

\begin{dunefigure}[Expected yearly rates of $\pi^{0}$'s at the vertex vs $p$ and event multiplicity in demonstrator]
{fig:pi0_kinematics}
{The expected yearly rates of $\pi^{0}$'s produced at the vertex, as a function of event multiplicity and their momentum, expected in the  
demonstrator's \SI{1.7}{\tonne} \dword{lar} 
mass for the \dword{numi} ME and \dword{lbnf} fluxes, produced using GENIE v2.12.10~\cite{genie}. Note that every $\pi^{0}$ from each event is included in the momentum distribution, regardless of containment.}
  \subfloat[$\pi^{0}$ multiplicity]   {\includegraphics[width=0.45\textwidth]{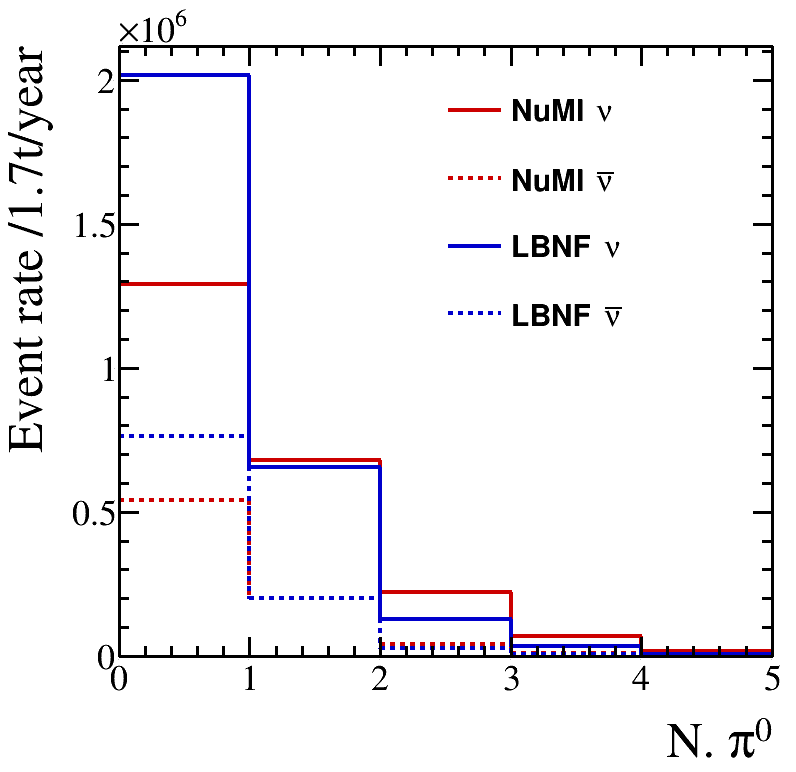}}
  \subfloat[$\pi^{0}$ momentum] {\includegraphics[width=0.45\textwidth]{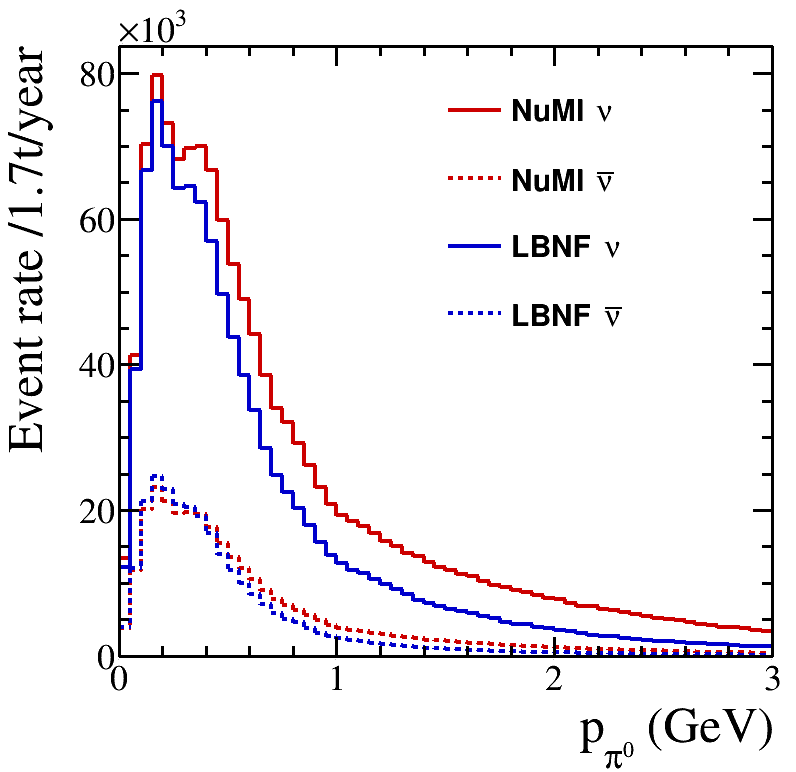}} \end{dunefigure}

\begin{dunefigure}[Efficiency for containing both $\gamma$-induced showers from $\pi^{0}$ decays in the demonstrator, vs $\pi^{0}$ KE and angle]
{fig:pi0_containment_2x2}
{Efficiency for containing both photon-induced showers from $\pi^{0}$ decays in the ArgonCube  2$\times$2 module, as a function of the $\pi^{0}$ kinetic energy and angle w.r.t the incoming neutrino direction. Containment is defined as $\geq$90\% of the energy being deposited in an active volume of a detector, and all primary $\pi^{0}$'s produced inside the  2$\times$2's active volume are included.}
  \includegraphics[width=0.75\textwidth]{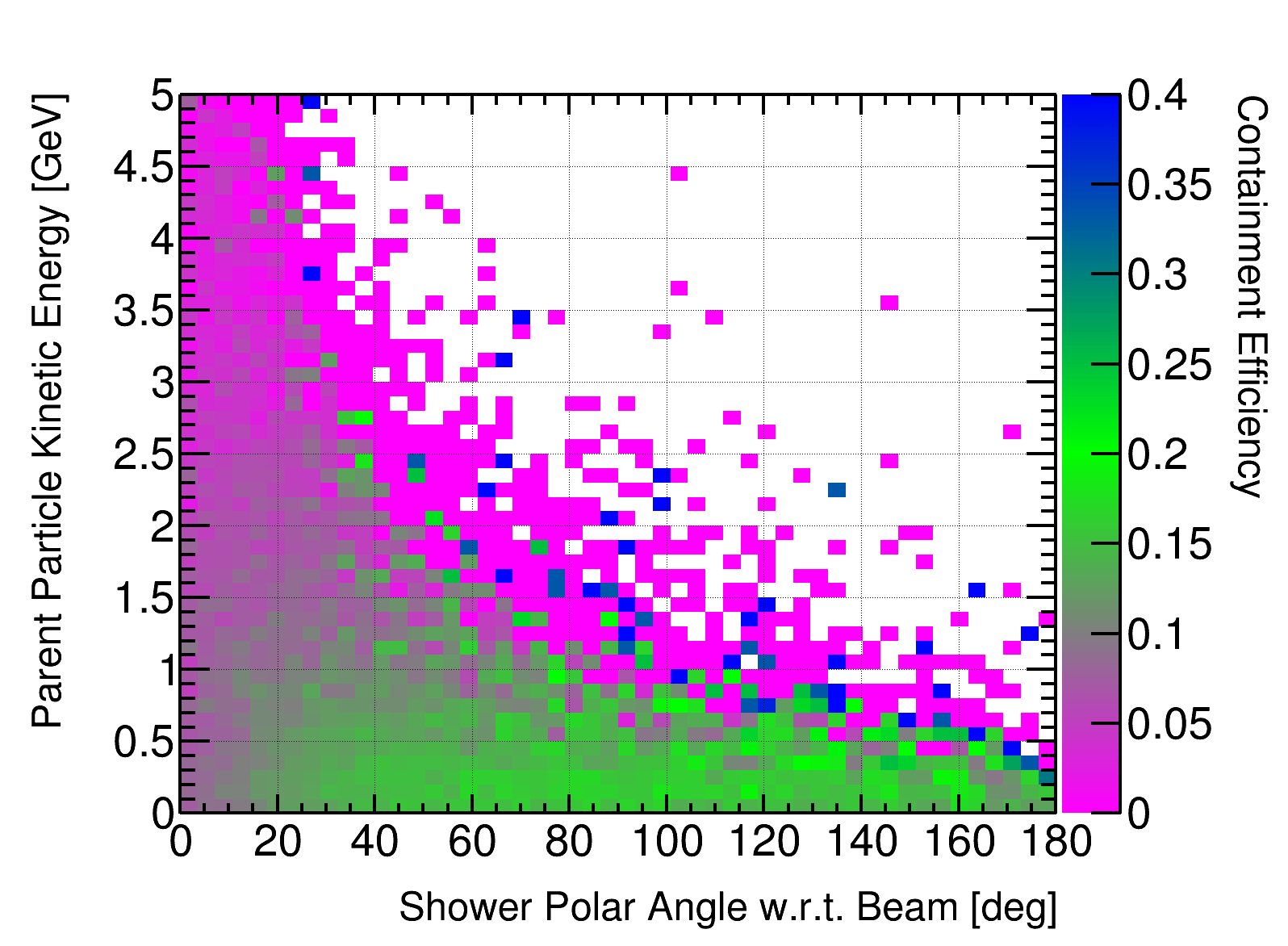}
\end{dunefigure}

Two further issues for this study are apparent. First, events with more than one $\pi^{0}$ introduce a problem in that even if two EM showers are fully contained, they may not come from the same $\pi^{0}$ decay. Second, of those $\pi^{0}$ decays for which both photons are fully contained, the initial $\pi^{0}$ is likely to have a low momentum, which is likely to exclude some fraction of the higher invariant mass events. However, despite these challenges, a measure of EM shower resolution from \dword{pdnd} is expected to be very useful for \dword{dune} \dword{nd} design studies, and warrants further investigation.



\FloatBarrier

\subsection{Additional studies with MINERvA components}
\label{subsec:2x2minerva}

Scintillator planes repurposed from the MINERvA experiment~\cite{Aliaga:2013uqz} will be placed upstream and downstream of the \dword{arcube} 2$\times$2 in \dword{pdnd} to provide upstream and downstream tracking. Additionally, the MINERvA electromagnetic calorimeter and a small number of planes from the MINERvA hadronic calorimeter will be placed downstream to contain electromagnetic showers which exit the 2$\times$2's volume downstream and to identify muons.

As is apparent from Figure~\ref{fig:argonbox_event_display}, many \dword{pdnd} events which have a vertex in the ArgonCube 2$\times$2's fiducial volume will not be contained in it. Although \dword{ndlar} will have a much larger volume, many events will not be fully contained, and in particular, muons will need to be matched with the downstream spectrometer. An example event including an approximate geometry for the downstream tracking component in \dword{pdnd} is shown in Figure~\ref{fig:2x2+MINERvA_event}. The presence of the MINERvA components acting as a downstream spectrometer will provide the opportunity to demonstrate the critical ability to match tracks between the ArgonCube modules, with slow charge and fast light readout, and other, fast detector components (i.e., the fast MINERvA scintillator strips).  In addition, the inclusion of the downstream spectrometer will broaden the phase-space over which events of interest can be reconstructed in \dword{pdnd}.

\begin{dunefigure}[Simulated event for a \SI{7}{\giga\electronvolt} $\nu_{\mu}$--argon CC interaction]
{fig:2x2+MINERvA_event}
{Example simulated event for a \SI{7}{\giga\electronvolt} $\nu_{\mu}$--argon CC interaction, in which particles not contained in the ArgonCube  2$\times$2 exit downstream, and are seen in the repurposed MINERvA detector components. Energy deposits are color-coded according to the particle type: $\pi^{\pm}$ --- blue; $\mu^{\pm}$ --- purple; $e^{+}$ --- green; $e^{-}$ --- yellow; proton --- red; recoiling nuclei --- black. The event vertex was randomly placed inside the active volume of the  2$\times$2 Demonstrator module.}
  \includegraphics[width=0.8\textwidth]{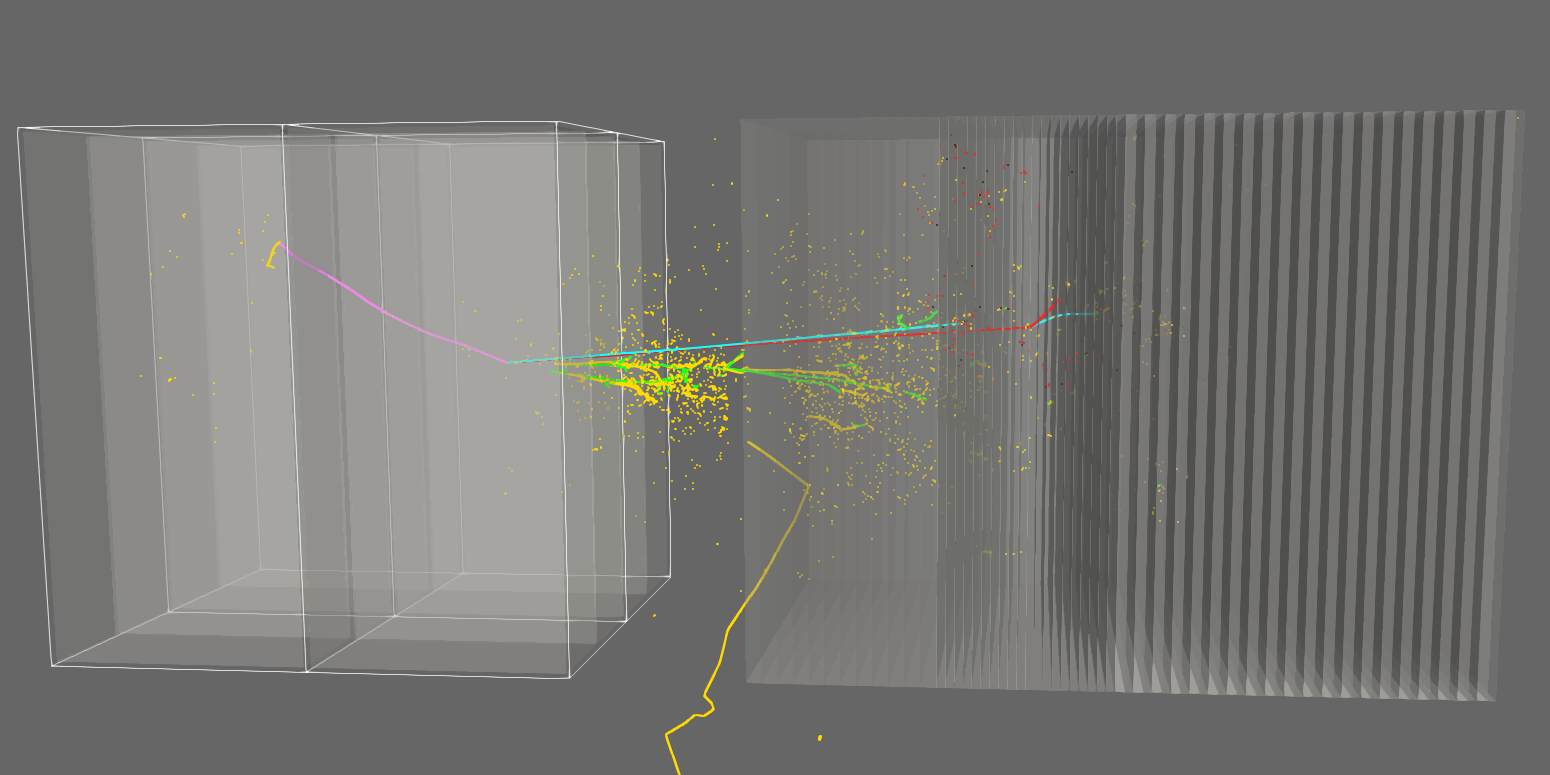}
\end{dunefigure}

The upstream and downstream scintillator tracking planes will also provide important capabilities for testing the stability and performance of the ArgonCube 2$\times$2. 
In particular, a sample of rock muons\footnote{Rock muons are muons formed by beam neutrino interactions in the upstream rock. Often the term means generically beam-related muons formed from interactions outside the detector that pass through the detector. },  tagged independently of the ArgonCube system, constitutes an ideal sample to be used for calibration purposes.  In addition, this sample can be used to test the stability of the calibration, the electric field uniformity, and the reconstruction performance over time.

\section{Acceptance and detector size}
\label{sec:lartpc-dimensions}

The estimated rate of \numu \dword{cc} interactions at the near site is 10$^{6}$ events per year per ton of material. This is sufficiently high that it does not drive the detector size. Instead, the optimization of the \dword{ndlar} total active volume is driven by the requirement that \dword{nd} and \dword{fd} sample the same neutrino interaction phase space, and that the energy resolution of the \dword{nd} be at least as good as that of the \dword{fd}. Given the large size of the \dword{fd}, it has 4$\pi$ acceptance, and both the lepton and hadronic shower are typically fully contained in both $\nu_\mu$ and $\nu_e$ \dword{cc} events. Therefore, the \dword{nd} must be capable of reconstructing events in phase space as close to 4$\pi$ as practical. Equal resolution is achieved by requiring fully-contained hadronic showers, fully-contained electron showers, relying on a downstream spectrometer to analyze exiting muons, and by being able to measure low energy, high angle muons that miss the spectrometer.

Based on the physics requirements, the minimal (and, therefore, optimal considering space and cost) active volume dimensions were found to be \SI{7}{\metre} wide (transverse to the beam), $\times$ \SI{5}{\metre} deep (the direction nearly parallel to the beam) and  $\times$ \SI{3}{\metre} (high), as seen in Figure~\ref{fig:actual-size}. These dimensions were determined in two steps. First, for good hadron containment a detector with a width of \SI{4}{m}, a depth of \SI{5}{m}, and a height of \SI{3}{m} is required. The hadron containment study is described in Section~\ref{sec:lartpc-hadron}. Second, additional detector width, \SI{1.5}{m} on each side, is needed to reconstruct high angle muons that do not enter the downstream spectrometer.  The muon acceptance study is described in Section~\ref{sec:lartpc-muonreco}.

\subsection{Required dimensions for hadronic shower containment}
\label{sec:lartpc-hadron}

Many events will be poorly contained simply because they occur near the edge of the detector, and because final-state particles happen to travel toward the active volume boundary. However, the rate of neutrino interactions in the \dword{nd} is sufficiently high that it is not necessary to analyze every event. The interaction cross section is translationally invariant because the flux is virtually constant across the face of \dword{ndlar}.  It is also invariant under rotations about the neutrino beam axis. These symmetries can be used to sample the events, providing 4$\pi$ coverage of the neutrino-argon interaction cross section phase space with a much smaller detector than that which would be required otherwise. Cross section coverage is defined as the fraction of events for which there is some neutrino interaction point in the detector where the event is well contained, even if the overall acceptance of such an event is small.

To determine the required size of \dword{ndlar} using the metric above, neutrino events were simulated using the DUNE flux with GENIE v2.12.10. Interaction products were propagated through a liquid argon detector volume using a Geant4-based model. For each event, the minimum active volume to contain 95\% of true hadronic energy deposits was determined. Neutrons were excluded from the hadronic energy calculation because only a small fraction of their kinetic energy is visible, even for a detector the size of the \dword{fd}. The minimum active volume was restricted to a rectangular shape. Due to the rotational symmetry about the beam, the two dimensions transverse to the beam axis were considered to be interchangeable, allowing the height to be kept smaller than the width. 


The total cross section coverage as a function of true neutrino energy was determined for detectors of different sizes. Figure~\ref{fig:dune-nd_lartpc-size} shows the coverage as a function of the height and as a function of the length, holding the other two dimensions fixed in both cases. Full coverage of neutrino energy region up to \SI{5}{GeV} is needed to insure coverage of the region most relevant for the oscillation analysis. It is also desirable that the coverage not vary rapidly with the detector dimensions. According to the study, the optimal dimensions for hadron containment were found to be \SI{4}{\metre} wide,  \SI{5}{\metre} deep, and \SI{3}{\metre} high. The longer transverse dimension was chosen to be the width rather than the height because a taller detector might require a costly increase to the hall height.

\begin{dunefigure}[Impact of the \dshort{lartpc} size on hadron containment]{fig:dune-nd_lartpc-size}
   {The cross section coverage, defined in the text, is shown for various \dword{lartpc} heights (left) and widths (right) as a function of true neutrino energy. In each plot, the other two dimensions are held constant at the baseline values while the third is varied. The optimal dimensions for hadron containment are determined to be \SI{4}{\metre} $\times$ \SI{3}{\metre} $\times$ \SI{5}{\metre}.}
	\includegraphics[width=0.48\textwidth]{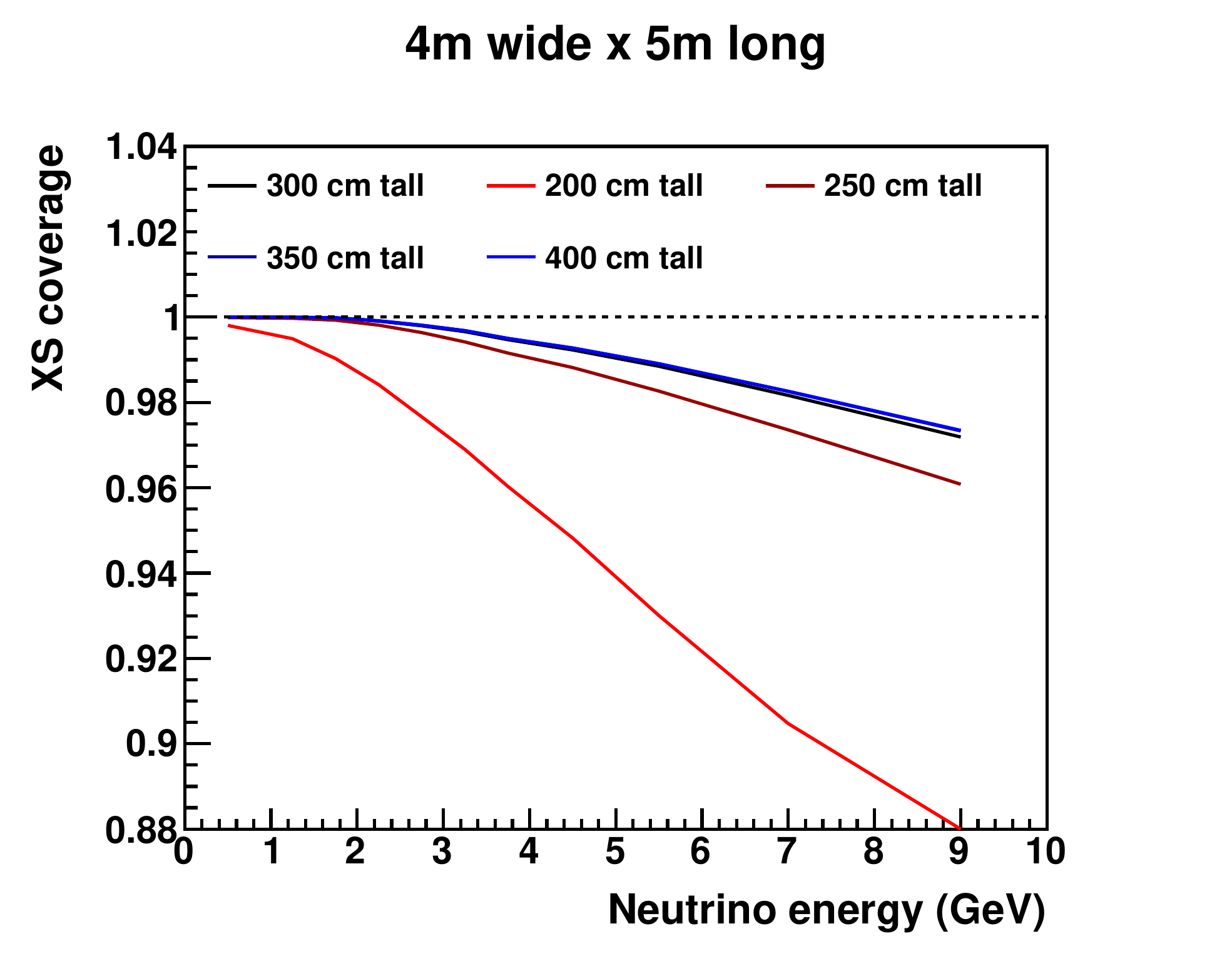}
	\includegraphics[width=0.48\textwidth]{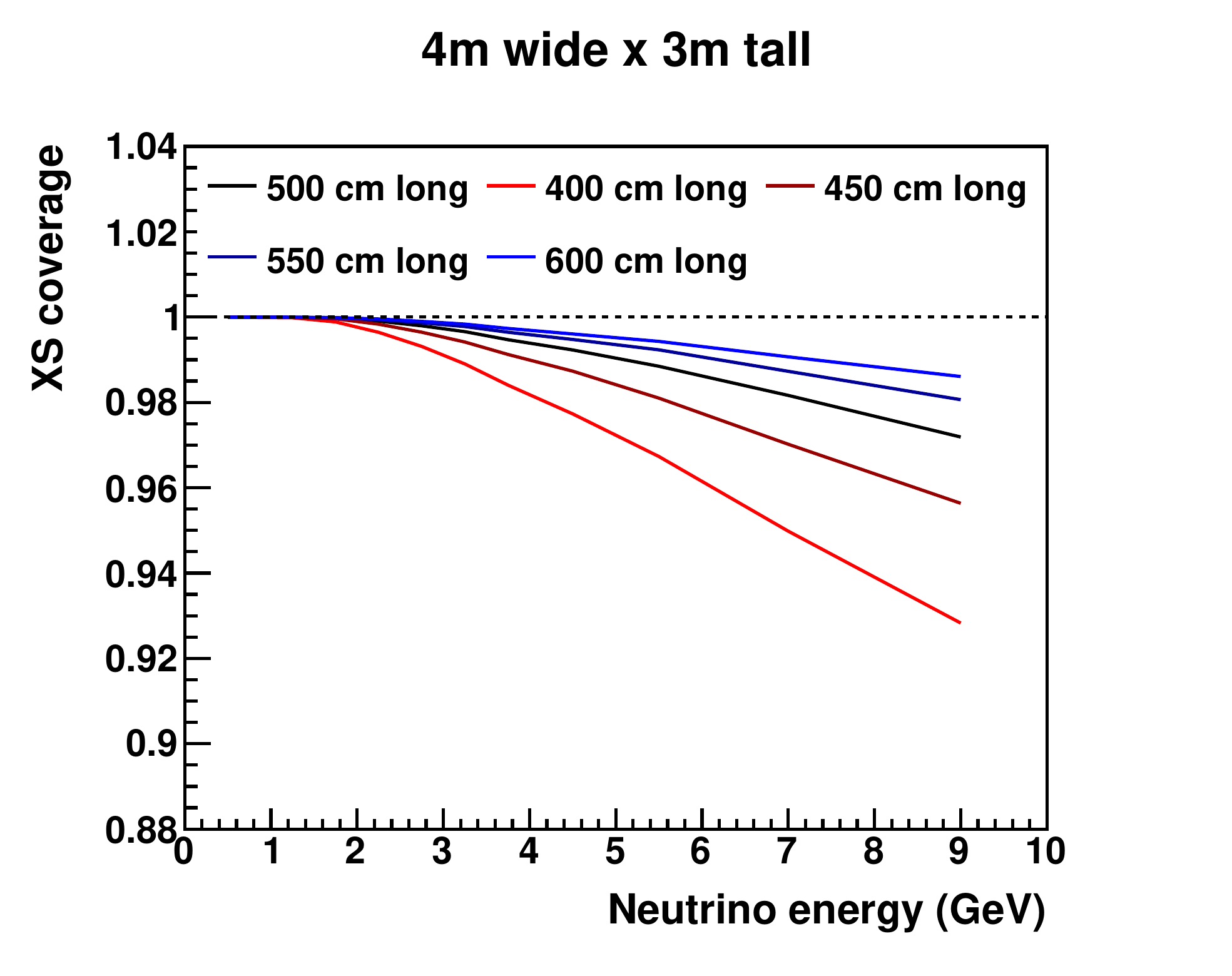}
\end{dunefigure}

\subsection{Muon reconstruction}
\label{sec:lartpc-muonreco}

Muon momentum is reconstructed either by range, when the muon is fully contained in the \dword{lartpc} active volume, or by curvature, when the muon is matched to a track in a downstream spectrometer. \dword{icarus} and \dword{microboone} have demonstrated the use of multiple Coulomb scattering to determine the muon momentum \cite{Abratenko:2017nki}. However, the resolution found using Coulomb scattering is worse than what can be achieved by range at the \dword{fd}, so the \dword{nd} size is determined assuming the use of only the range or curvature methods.

Although a width of \SI{4}{\metre} is sufficient to contain the hadronic component of events of interest, this is not sufficient for muon reconstruction. To contain muons emitted at large angles with respect to the beam, a width of \SI{7}{\metre} is required. By design, the acceptance will be poor for wide-angle muons when the $\nu-\mu$ plane happens to be nearly vertical, but the rotational symmetry allows those same events to be well-reconstructed when the $\nu-\mu$ plane is horizontal.

The muon acceptance for a \SI{7}{\metre} wide, \SI{5}{\metre} deep, and \SI{3}{\metre} high \dword{lartpc} is shown in Figure~\ref{fig:muonacc} as a function of the muon angle and energy for \numu \dword{cc} events in \dword{fhc} mode. The assumed fiducial volume is \SI{6}{\metre} wide,  \SI{3}{\metre} deep, and \SI{2}{\metre} high, which excludes \SI{50}{\centi\metre} from the sides and upstream end, and \SI{150}{\centi\metre} from the downstream end. The downstream spectrometer is assumed to be the \dword{ndgar} described in Chapter~\ref{ch:mpd}.

The acceptance is poor for muons above \SI{1}{\giga\electronvolt} at wide angles, because  many of these muons exit the top or bottom of the \dword{lartpc} and miss  \dword{ndgar} entirely. These events could potentially be recovered by using multiple Coulomb scattering to reconstruct the momentum, which would further increase the efficiency above what is reported here. Also, events in that kinematic region can be reconstructed by range when the muon moves along  the \SI{7}{\metre} dimension of the \dword{lartpc}. The dip around \SI{1}{\giga\electronvolt} in the forward region corresponds to muons that exit the rear of the \dword{lartpc} and stop in either the cryostat or the \dword{ndgar} magnet coil. It is critical to minimize the passive material between the active \dword{lar} and the \dword{ndgar} active region to limit the impact of this dip.

\begin{dunefigure}[Muon acceptance as a function of muon kinematics]{fig:muonacc}
{Muon acceptance shown as a function of true muon kinetic energy and angle with respect to the neutrino beam (left), and projected onto the muon kinetic energy axis for small angles (right). The acceptance includes muons contained in the \dword{lartpc} as well as those that stop in the \dword{ndgar} \dword{ecal} or match to tracks in the \dword{hpgtpc}.}
      \includegraphics[width=0.48\textwidth]{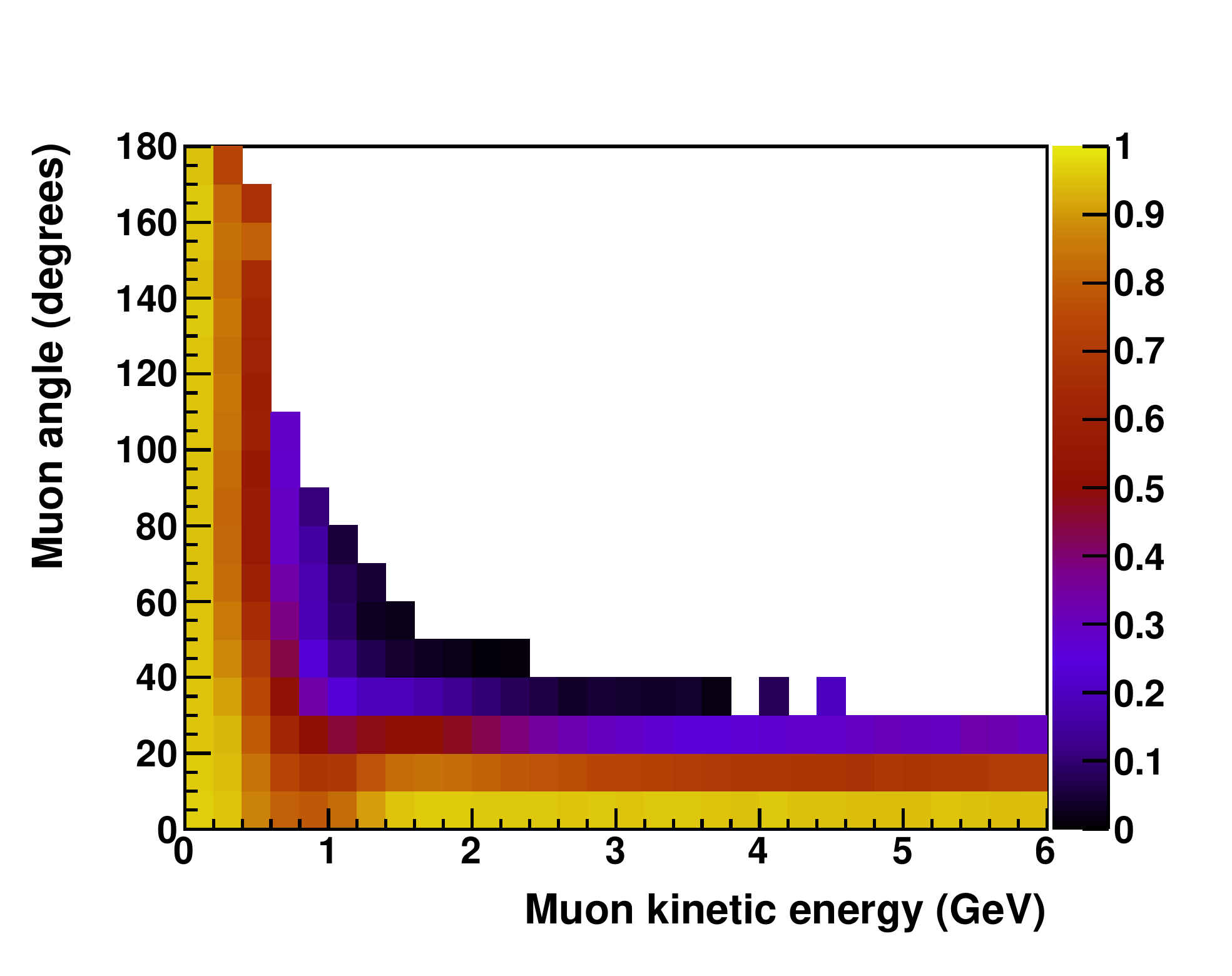}
      \includegraphics[width=0.48\textwidth]{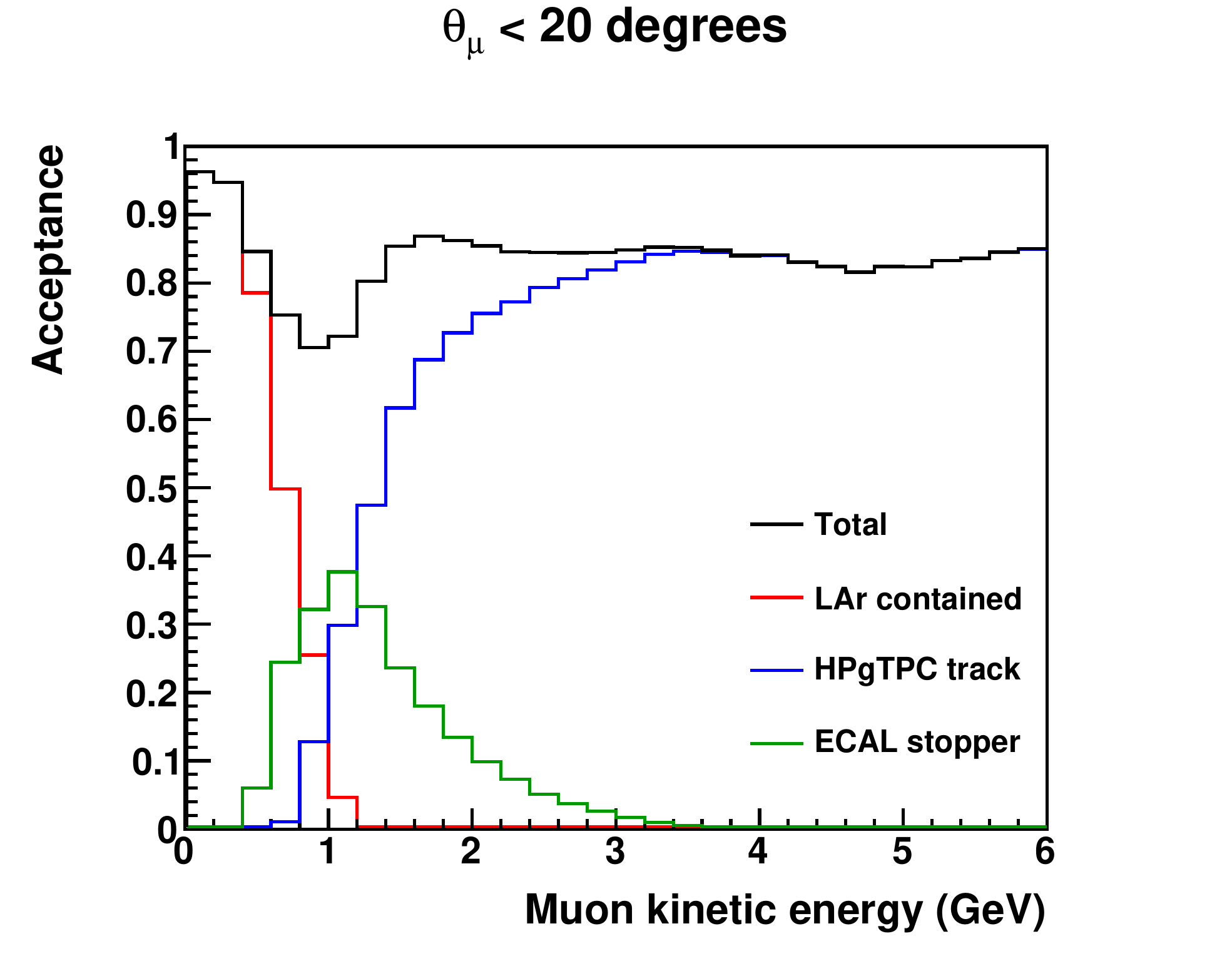}
\end{dunefigure}

\subsection{Acceptance vs. energy and momentum transfer}

It is necessary that the entire cross section phase space have nonzero acceptance with high-quality reconstruction in the full \dword{ndlar} plus \dword{ndgar} configuration, taking into account both the muon and the hadronic system. To explore this, consider the acceptance in slices of neutrino energy as a function of the energy transfer to the nuclear system, $q_{0} = E_{\nu} - E_{\mu}$, and the three-momentum transfer, $q_{3} = \sqrt{Q^{2} + q_{0}^{2}}$, where $Q^{2}$ is the squared four-momentum transfer. This kinematic space has long been used to study nuclear structure in electron-nucleus scattering experiments.

Figure~\ref{fig:q0q3acc} shows the event rate (left figures) and acceptance (right figures) in bins of $(q_3,q_0)$. The rows correspond to two neutrino energy bins. The top row is for $E_\nu$ between \SIrange{1.0}{2.0}{\giga\electronvolt}, which is the region between the first and second oscillation maxima. The second bin is for $E_\nu$ between \SIrange{3.5}{4.0}{\giga\electronvolt}, on the falling edge of the peak of the neutrino energy spectrum. The rate histograms have ``islands''  corresponding to hadronic systems with fixed invariant mass, smeared by Fermi motion. The lower island in $(q_3,q_0)$ corresponds to the quasi-elastic peak while the upper corresponds to the $\Delta$ resonance. 2p2h processes contribute to both peaks and the region between them.  One should note that the axes in the lower row cover a larger range of kinematic space than those in the upper row.

\begin{dunefigure}[Neutrino acceptance as a function of energy and momentum transfer]{fig:q0q3acc}
{Neutrino acceptance shown as a function of energy transfer and momentum transfer ($q_0$ and $q_3$) to the target nucleus. The figures show the event rate (left) and the acceptance (right) for reconstructing the muon and containing the hadronic system. The top row was made for neutrinos with true neutrino energy between \SIrange{1.0}{2.0}{\giga\electronvolt} just below the flux peak, and the bottom was made for neutrinos between \SIrange{3.5}{4.0}{\giga\electronvolt} on the falling edge of the peak.}
      \includegraphics[width=0.45\textwidth]{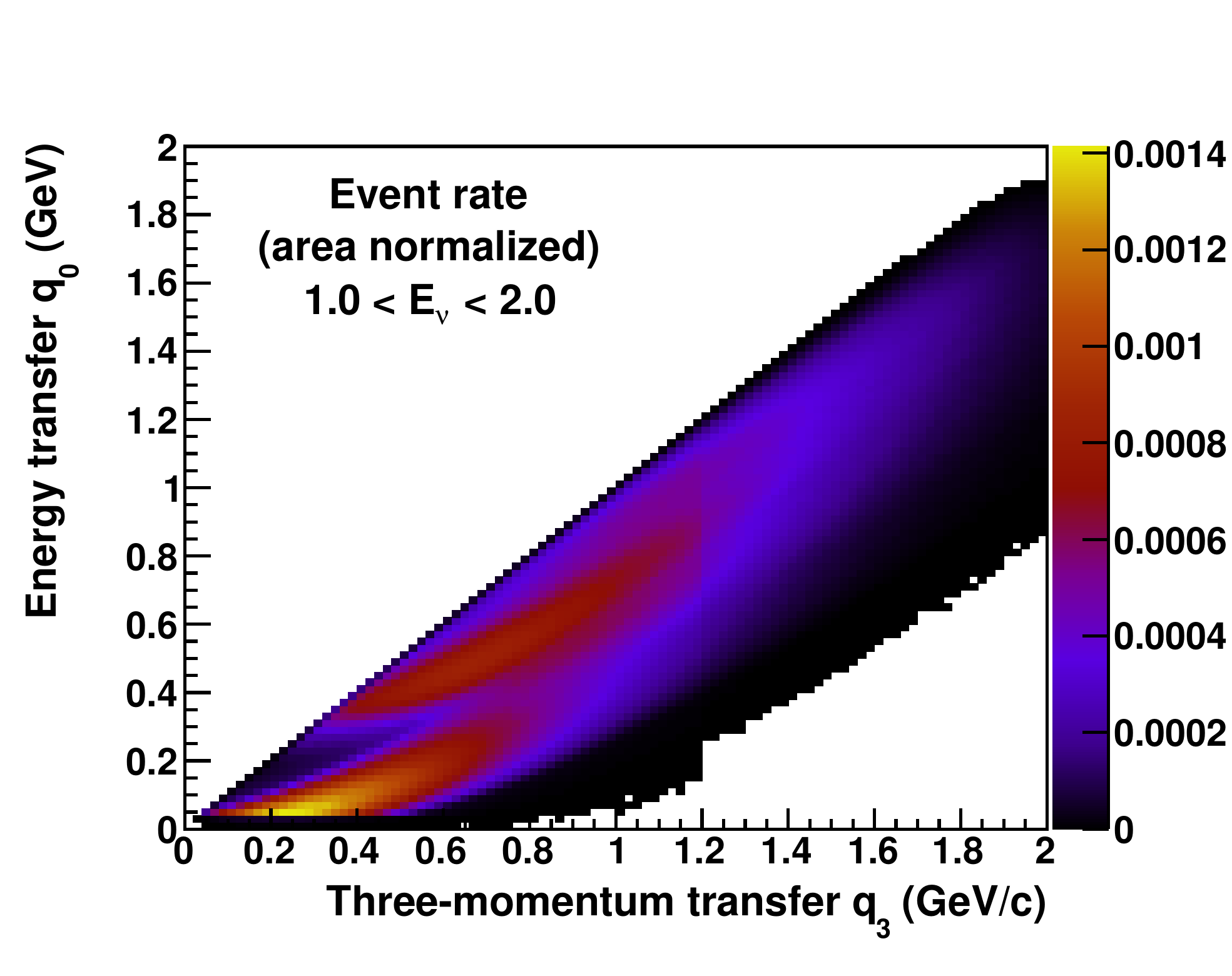}
      \includegraphics[width=0.45\textwidth]{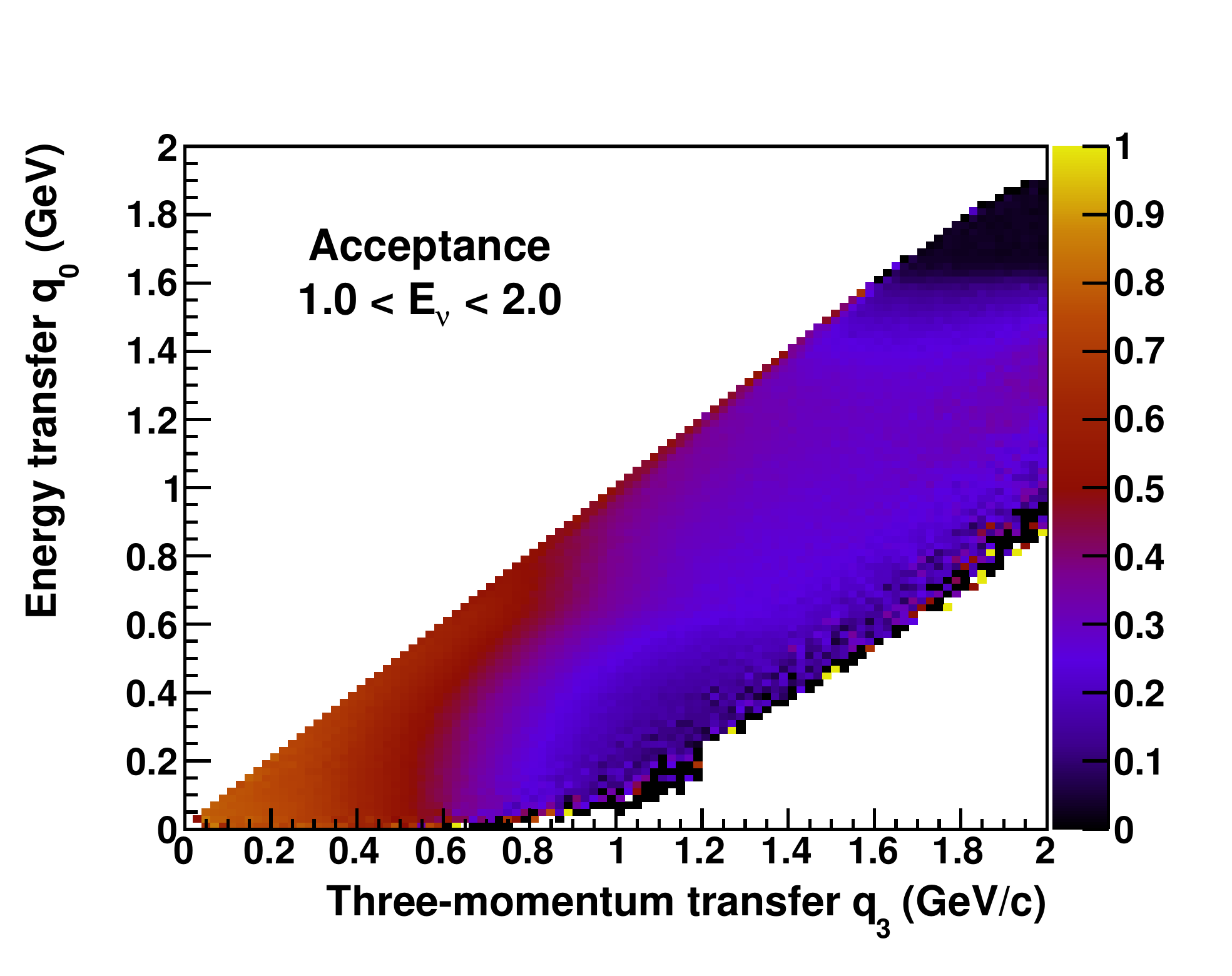}
      \includegraphics[width=0.45\textwidth]{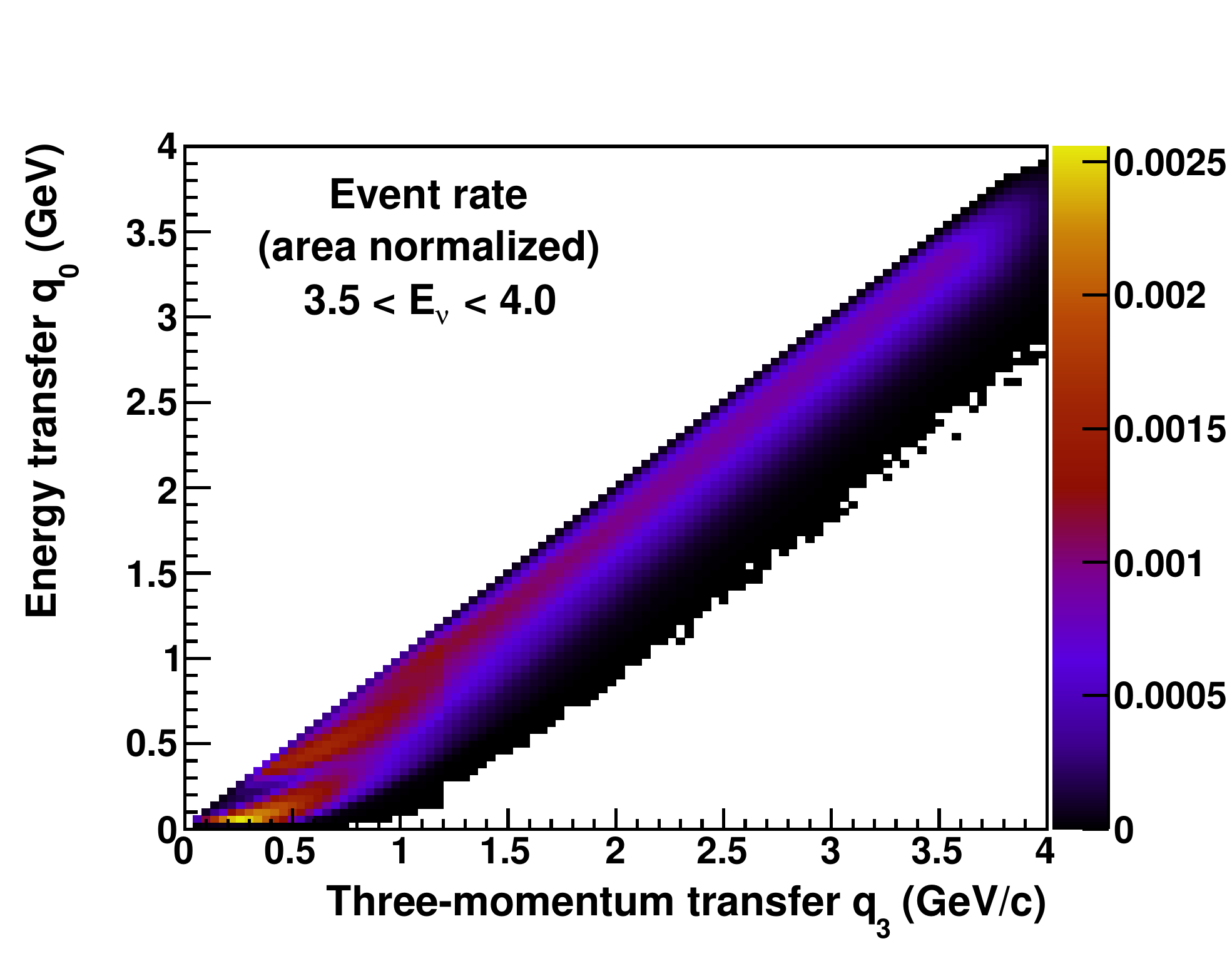}
      \includegraphics[width=0.45\textwidth]{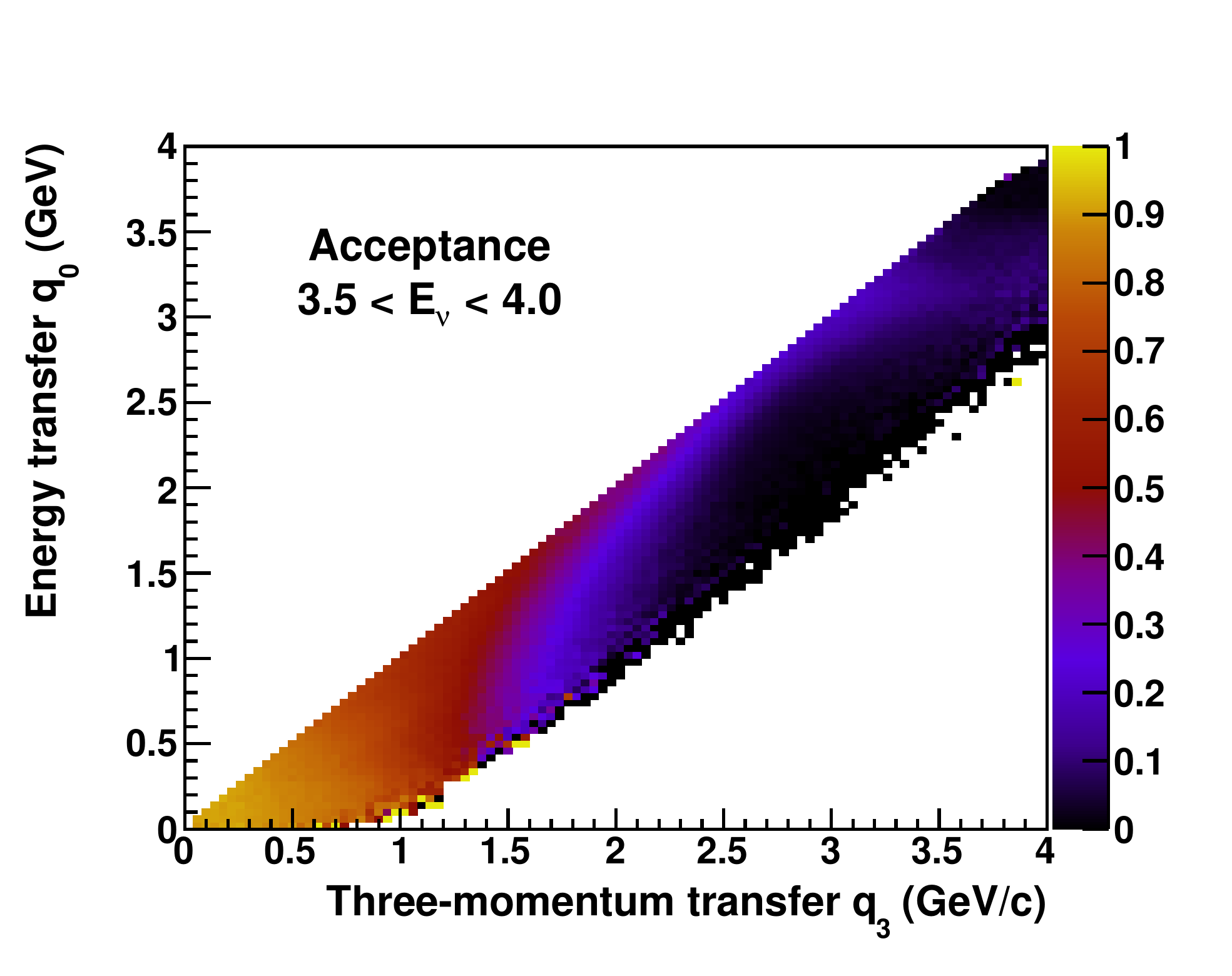}
\end{dunefigure}

Taking the left and right plots together, it can be seen the acceptance is generally very good in the kinematic region where the vast majority of the events occur. Because this acceptance is integrated over the full fiducial volume, it is not expected to be perfect anywhere. The loss of acceptance is due primarily to  geometric effects. Losses typically occur in events with a vertex near one boundary of the detector, where the muon or hadronic system exits out that boundary. However for each lost event there is generally a set of symmetric events that are accepted because the final state is rotated by some angle about the neutrino beam axis ($\phi$ symmetry) or is closer to the centre of the fiducial volume (translational symmetry).

Regions where the acceptance is zero are problematic because they will introduce model dependence into the prediction of the rate at the far detector (which has a nearly $4\pi$ acceptance). Acceptances of even a few \% in some kinematic regions are not necessarily problematic, because the event rate is large enough to accumulate a statistically significant number of events. There is a potential danger if the acceptance varies quickly as a function of the kinematic variables because a small mismodeling of the detector boundaries or neutrino cross-sections could translate into a large mismodeling in the number of accepted events. 

The size of the accepted event set decreases as a function of both $q_0$ and $q_3$ (and therefore $E_\nu$) due to more energetic hadronic systems and larger angle muons. This can be seen clearly in the transition from the colored region to the black region in the $\SI{3.5}{\giga\electronvolt} < E_\nu < \SI{4.0}{\giga\electronvolt}$ acceptance histogram shown in the lower right-hand corner of Figure~\ref{fig:q0q3acc}. The transition is smooth and gradual. 

The acceptance for $\SI{1.0}{\giga\electronvolt} < E_\nu < \SI{2.0}{\giga\electronvolt}$ (shown in the upper right-hand corner of Figure~\ref{fig:q0q3acc}) is larger than 10\% except in a small region at high $q_0$ and $q_3$. Events in that region have a low-energy muon and are misidentified as neutral-current, according to the simple event selection applied in the study. The fraction of events in that region is quite small, as can be seen in the upper left-hand plot of Figure~\ref{fig:q0q3acc}. 

Figure~\ref{fig:q0q3acc_vs_enu} summarizes the neutrino acceptance in the $(q_3,q_0)$ plane as a function of neutrino energy. The vertical axis shows the fraction of events coming from $(q_3,q_0)$ bins with an acceptance greater than $A_{cc}$. The $A_{cc}>0.00$ curve shows the fraction of events for which there is non-zero acceptance. The figure shows that in the oscillation region the fraction of events that occur in a kinematic region with zero acceptance is less than 0.1\%, meaning that there are no acceptance holes. More than two thirds of events in the oscillation region between $\SI{0.5}{\giga\electronvolt} < E_\nu < \SI{5.0}{\giga\electronvolt}$ have at least 10\% acceptance.

\begin{dunefigure}[Neutrino acceptance in the $(q_3,q_0)$ plane as a function of neutrino energy]{fig:q0q3acc_vs_enu}
{This figure summarizes the neutrino acceptance in the $(q_3,q_0)$ plane, as shown in Figure~\ref{fig:q0q3acc}, for all bins of neutrino energy. Here the quantity on the vertical axis is the fraction of events that come from bins in $(q_3,q_0)$ with an acceptance greater than $A_{cc}$. As an example we consider the \SIrange{3.5}{4.0}{\giga\electronvolt} neutrino energy bin. The $A_{cc}>0.1$ curve in that neutrino energy bin indicates that 80\% of events come from $(q_3,q_0)$ bins that have an acceptance greater than 10\%. The $A_{cc}>0.00$ curve shows there are no acceptance holes.}
      \includegraphics[width=0.6\textwidth]{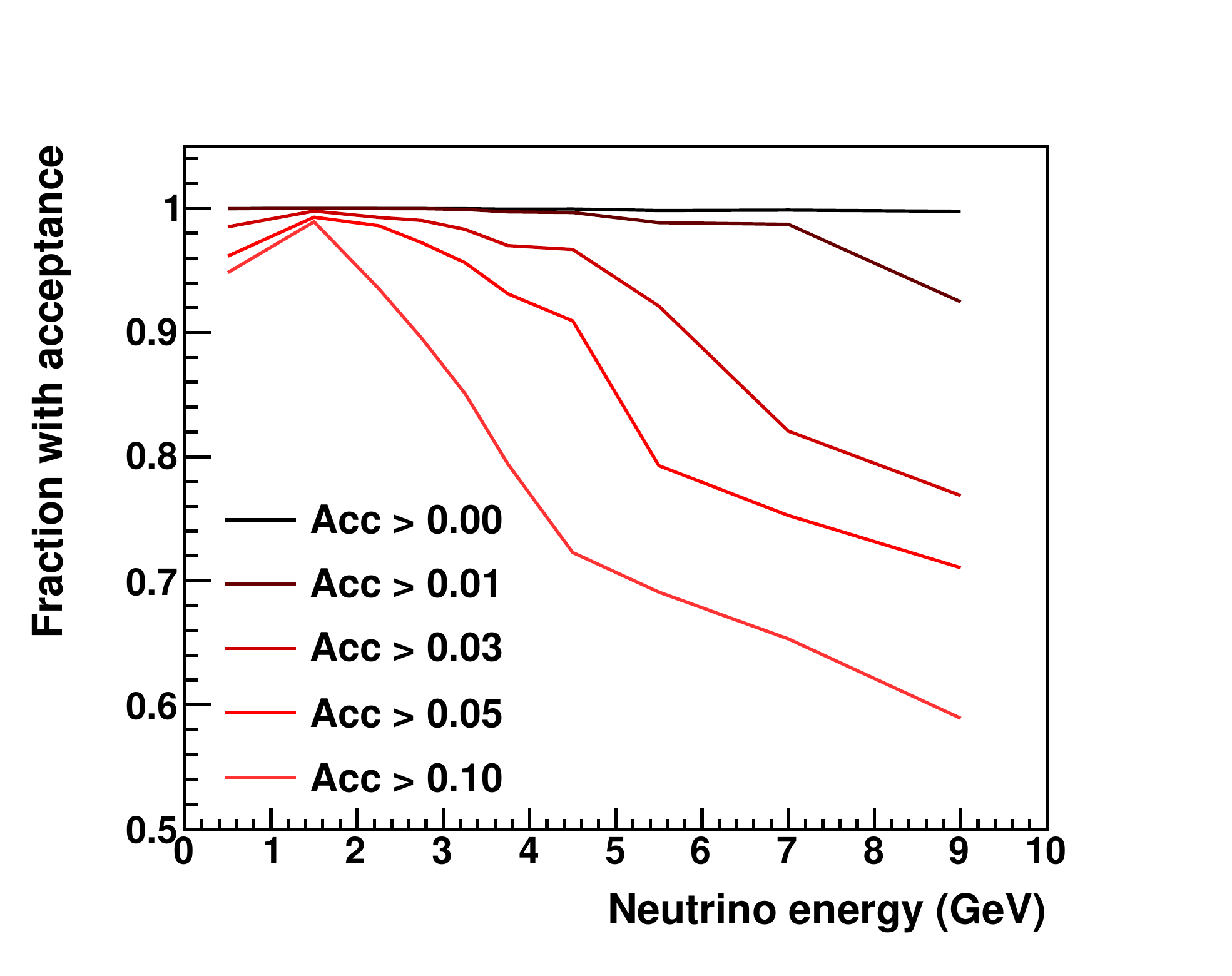}
\end{dunefigure}

Electron reconstruction is not expected to drive the detector dimensions. The radiation length in \dword{lar} is \SI{14}{\centi\metre}, meaning that the miminum \SI{1.5}{\metre} between the fiducial volume and the rear of the active volume corresponds to roughly 11 radiation lengths, which is sufficient to measure the electron energy. The \SI{50}{\centi\metre} buffer on the sides of the active volume is over five times the \SI{9}{\centi\metre} Moliere radius. Thus, the optimal dimensions for \numu \dword{cc} scattering of \SI{5}{\metre} $\times$ \SI{7}{\metre} $\times$ \SI{3}{\metre} is also sufficient for \nue \dword{cc} reconstruction.

\subsection{\dshort{arcube} module dimensions}

The \dword{arcube} module dimensions within \dword{ndlar} are set to maintain a high drift field with a minimal cathode voltage, and to allow for the detection of prompt scintillation light. 
The prompt scintillation light, $\tau<$\SI{6.2}{\nano\second}~\cite{Heindl:2015yaa}, can be efficiently measured with a dielectric light readout with few ns timing resolution, such as the \dword{arclt}~\cite{Auger:2017flc} and LCM that will be used in \dword{ndlar}.
To improve the fidelity of the timing information in the scintillation signal, a short optical path length is desired to reduce light attenuation, and to minimize smearing of the photon arrival time distribution due to Rayleigh scattering, where the scattering length is \SI{0.66}{\metre} at \SI{128}{\nano\metre} in \dword{lar}~\cite{Grace:2015yta}. 
Maintaining a higher electric field serves to suppress the slow ($\mathcal{O}\left(1\right)\,\SI{}{\milli\second}$) scintillation component \efield{}s~\cite{PhysRevB.20.3486} by effectively reducing the ionization density~\cite{PhysRevB.27.5279} required to produce excited states that contribute to the slow component. 

A module with a \SI{1x1}{\metre} footprint split into two \dwords{tpc} with drift lengths of \SI{50}{\centi\metre} requires only a \SI{50}{\kilo\volt} bias to achieve a \SI{1}{kV/cm} electric field.
With \dword{arclt} mounted either side of the \SI{1}{\metre} wide \dword{tpc}, the maximum optical path is only \SI{50}{\centi\metre}.
Reducing the module footprint below this would not yield significant physics improvements, and would only increase the number of readout channels, component count and inactive material.  

As mentioned in Section~\ref{sec:lartpc-des-hv}, the design for a \SI{1}{kV/cm} electric field builds in  robustness, and the electric field can always be reduced to study electron-ion recombination as a function of the electric field strength. 
The field can be set to match that of the \dword{fd}, optimally \SI{0.5}{kV/cm}.
For the given dimensions, at \SI{1}{kV/cm}, the drift window is \SI{250}{\micro\second}, the transverse diffusion is \SI{0.81}{\milli\metre}, and the optimal charge lifetime is \SI{2.4}{\milli\second}; at \SI{0.5}{kV/cm}, the drift window is \SI{333}{\micro\second} and the transverse diffusion is \SI{0.86}{\milli\metre} and the optimal charge lifetime is \SI{3.2}{\milli\second}. 

Figure~\ref{fig:actual-size} shows the overall dimensions of \dword{ndlar} in the \dword{dune}  \dword{nd}. 
With an active volume of \SI[product-units=repeat]{1x1x3}{\metre} per module, the full \dword{ndlar} detector corresponds to seven modules transverse to the beam direction, and five modules along it. 
It should be noted that the cryostat design is currently based on \dword{protodune}~\cite{Abi:2017aow}, and will be optimized for the  \dword{nd} pending full engineering.

\section{Event rates in the \dword{nd} LArTPC}
\label{sec:event_rates}

In the oscillation region of $\SI{0.5} < E_\nu < \SI{4}{\giga\electronvolt}$, the expected event rate in the \SI{50}{\tonne} fiducial volume (\SI{6}{m} wide, \SI{3}{m} deep, and \SI{2}{m} high) of \dword{ndlar} is 59 million \numu \dword{cc} interactions per year in \dword{fhc} mode and 20 million \anumu \dword{cc} interactions per year in \dword{rhc} mode. Of those, over 24 million (10 million) are expected to have a well-reconstructed muon of the appropriate sign, and a well-contained hadronic system in \dword{fhc} (\dword{rhc}). In addition, 450,000 $\nue+\anue$ \dword{cc} interactions are expected per year in \dword{fhc}, and 200,000 in \dword{rhc}. The expected event rate per one year of exposure on axis with the LBNF beam is shown in Figure~\ref{fig:argoncube_eventrate} as a function of true neutrino energy.

\begin{dunefigure}[Neutrino interaction rates in the \SI{50}{\tonne} \dword{arcube} fiducial volume]{fig:argoncube_eventrate}
{The rate of \dword{cc} interactions in the fiducial volume of \dword{ndlar} as a function of true neutrino energy, expressed per year of exposure assuming \SI{1.2}{\mega\watt} beam intensity for the LBNF beam with \dword{fhc} (left) and \dword{rhc} (right) beam polarity.}
	\includegraphics[width=0.48\textwidth]{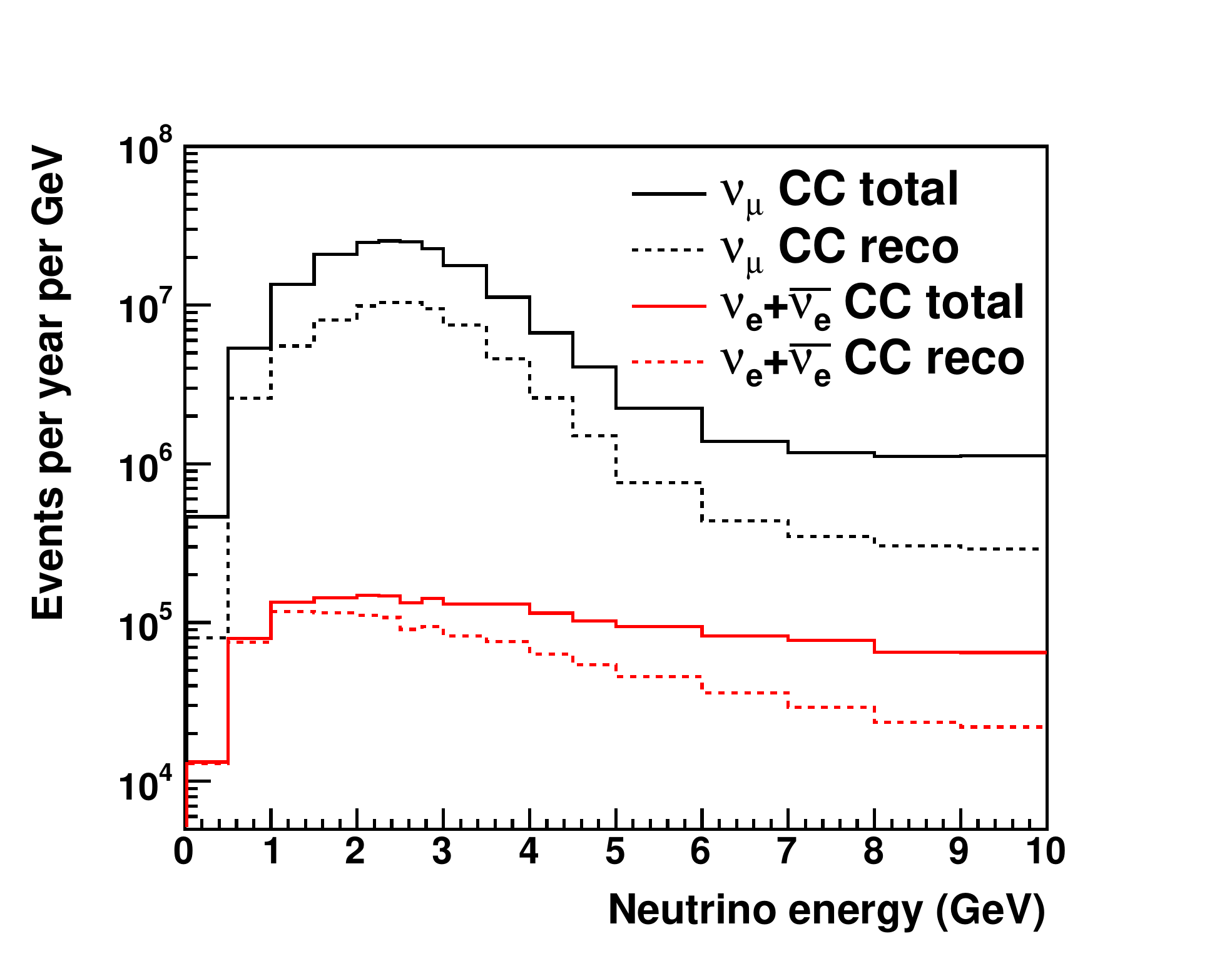}
	\includegraphics[width=0.48\textwidth]{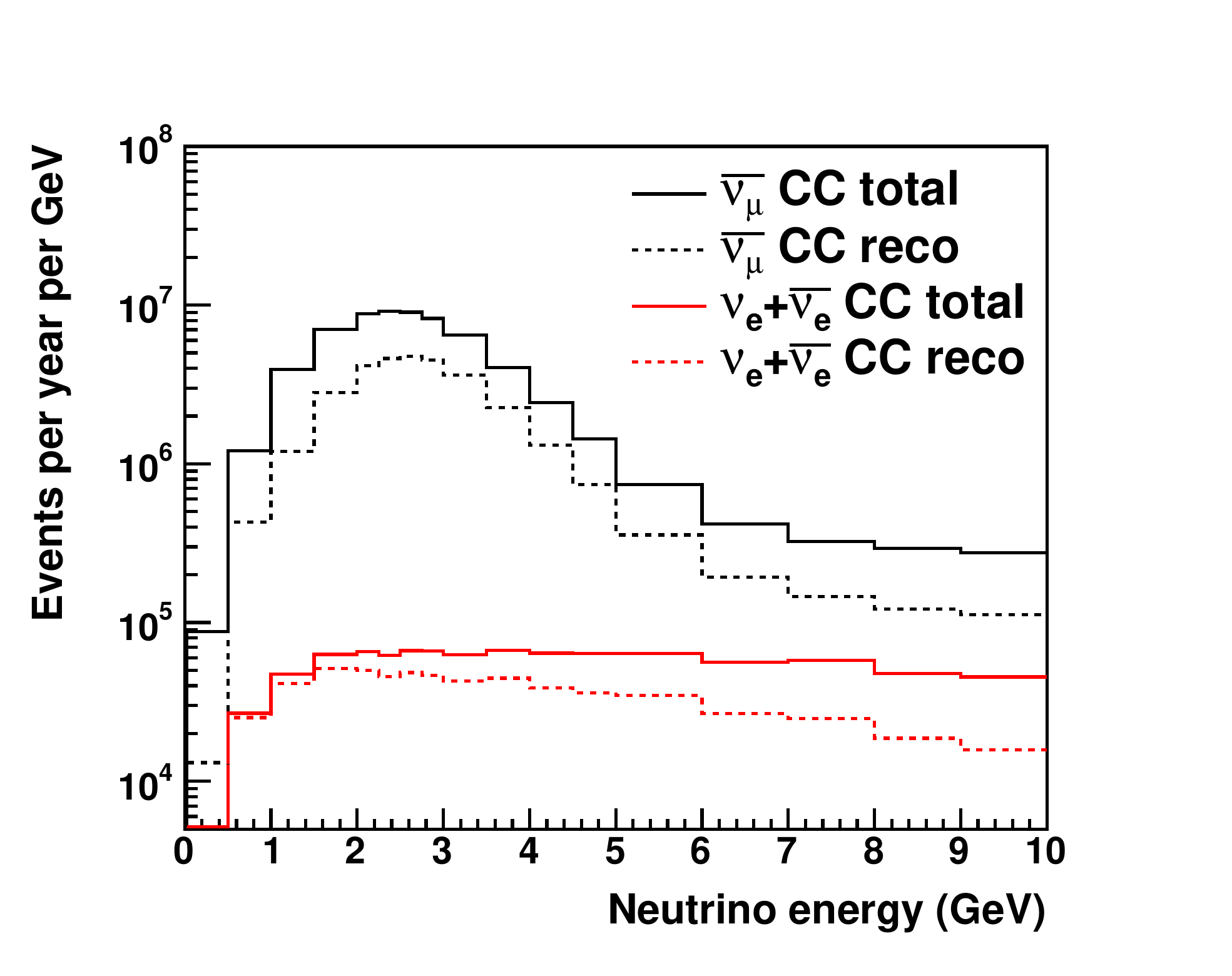}
\end{dunefigure}

Event rates for different final states are given in Tables~\ref{tab:FHC_rates} and~\ref{tab:RHC_rates} for \dword{fhc} and \dword{rhc} beam modes, respectively. The tables are based on a simulation of GENIE version 2.12.10.
The first two columns give the total rate and the estimated number of well-reconstructed events. The second two columns give the same quantities but restricted to the oscillation region.

\begin{dunetable}
	[Event rate table FHC]
	{ccccc}
	{tab:FHC_rates}
	{\dword{fhc} Event rates  in \dword{ndlar} (per year as defined in the text). Accepted is defined as the $\mu$ is either contained or matched to  \dword{ndgar}, and the hadronic shower is contained (\textless\SI{30}{\mega\electronvolt} in the outermost \SI{30}{\centi\metre} of the LAr).}
	\dword{fhc} mode & total & accepted & \SIrange{0.5}{4.0}{\giga\electronvolt} & accepted \\ \toprowrule
	$\nu_{\mu}$ CC & $8.2 \times 10^{7}$ & $3.0 \times 10^{7}$ & $5.9 \times 10^{7}$ & $2.4 \times 10^{7}$ \\ \colhline
$\bar{\nu}_{\mu}$ CC & $3.6 \times 10^{6}$ & $1.4 \times 10^{6}$ & $1.1 \times 10^{6}$ & $4.6 \times 10^{5}$ \\ \colhline
NC total & $2.8 \times 10^{7}$ & $1.6 \times 10^{7}$ & $1.9 \times 10^{7}$ & $1.3 \times 10^{7}$ \\ \colhline
$\nu_{\mu}$ CC$0\pi$ & $2.9 \times 10^{7}$ & $1.6 \times 10^{7}$ & $2.6 \times 10^{7}$ & $1.3 \times 10^{7}$ \\ \colhline
$\nu_{\mu}$ CC$1\pi^{\pm}$ & $2.0 \times 10^{7}$ & $7.5 \times 10^{6}$ & $1.7 \times 10^{7}$ & $6.0 \times 10^{6}$ \\ \colhline
$\nu_{\mu}$ CC$1\pi^{0}$ & $8.0 \times 10^{6}$ & $2.9 \times 10^{6}$ & $6.5 \times 10^{6}$ & $2.2 \times 10^{6}$ \\ \colhline
$\nu_{\mu}$ CC$3\pi$ & $4.6 \times 10^{6}$ & $7.2 \times 10^{5}$ & $1.7 \times 10^{6}$ & $3.8 \times 10^{5}$ \\ \colhline
$\nu_{\mu}$ CC other & $9.2 \times 10^{6}$ & $7.4 \times 10^{5}$ & $1.5 \times 10^{6}$ & $3.1 \times 10^{5}$ \\ \colhline
$\nu_{e} + \bar{\nu}_{e}$ CC & $1.4 \times 10^{6}$ & $6.6 \times 10^{5}$ & $4.5 \times 10^{5}$ & $3.3 \times 10^{5}$ \\ \colhline
$\nu+e$ elastic & $8.4 \times 10^{3}$ & $7.2 \times 10^{3}$ & $5.3 \times 10^{3}$ & $4.2 \times 10^{3}$ \\ 
\end{dunetable}

\begin{dunetable}
	[Event rate table RHC]
	{ccccc}
	{tab:RHC_rates}
	{\dword{rhc} Event rates  in \dword{ndlar} (per year as defined in the text). Accepted is defined as the $\mu$ is either contained or matched to the \dword{mpd}, and the hadronic shower is contained (\textless\SI{30}{\mega\electronvolt} in the outermost \SI{30}{\centi\metre} of the LAr)}
	\dword{rhc} mode & total & accepted & \SIrange{0.5}{4.0}{\giga\electronvolt} & accepted \\ \toprowrule
	$\bar{\nu}_{\mu}$ CC & $2.6 \times 10^{7}$ & $1.2 \times 10^{7}$ & $2.0 \times 10^{7}$ & $9.7 \times 10^{6}$ \\ \colhline
$\nu_{\mu}$ CC & $1.4 \times 10^{7}$ & $3.4 \times 10^{6}$ & $3.1 \times 10^{6}$ & $1.2 \times 10^{6}$ \\ \colhline
NC total & $1.5 \times 10^{7}$ & $9.2 \times 10^{6}$ & $9.3 \times 10^{6}$ & $7.2 \times 10^{6}$ \\ \colhline
$\bar{\nu}_{\mu}$ CC$0\pi$ & $1.2 \times 10^{7}$ & $6.7 \times 10^{6}$ & $1.0 \times 10^{7}$ & $5.6 \times 10^{6}$ \\ \colhline
$\bar{\nu}_{\mu}$ CC$1\pi^{\pm}$ & $7.6 \times 10^{6}$ & $3.5 \times 10^{6}$ & $6.0 \times 10^{6}$ & $2.7 \times 10^{6}$ \\ \colhline
$\bar{\nu}_{\mu}$ CC$1\pi^{0}$ & $2.4 \times 10^{6}$ & $9.6 \times 10^{5}$ & $1.9 \times 10^{6}$ & $7.2 \times 10^{5}$ \\ \colhline
$\bar{\nu}_{\mu}$ CC$2\pi$ & $2.6 \times 10^{6}$ & $8.1 \times 10^{5}$ & $1.6 \times 10^{6}$ & $5.0 \times 10^{5}$ \\ \colhline
$\bar{\nu}_{\mu}$ CC$3\pi$ & $8.3 \times 10^{5}$ & $1.7 \times 10^{5}$ & $3.0 \times 10^{5}$ & $7.5 \times 10^{4}$ \\ \colhline
$\bar{\nu}_{\mu}$ CC other & $1.2 \times 10^{6}$ & $1.4 \times 10^{5}$ & $2.0 \times 10^{5}$ & $4.3 \times 10^{4}$ \\ \colhline
$\nu_{e} + \bar{\nu}_{e}$ CC & $9.3 \times 10^{5}$ & $4.0 \times 10^{5}$ & $1.9 \times 10^{5}$ & $1.5 \times 10^{5}$ \\ \colhline
$\nu+e$ elastic & $6.4 \times 10^{3}$ & $5.7 \times 10^{3}$ & $4.0 \times 10^{3}$ & $3.4 \times 10^{3}$ \\
\end{dunetable}

\section{Neutrino pile-up mitigation}
\label{sec:ndlar-pileup}

The DUNE \dword{nd} complex requires a LArTPC design that is resilient to beam neutrino pile-up, as the incorrect assignment of final state particles can result in mis-classification of neutrino interaction type and/or a bias in reconstructed neutrino energy. For a typical 10~$\mu$s-wide LBNF beam spill at 1.2~MW beam power, a mean of 55 neutrino interactions --- including targets both internal (57\%) and external (43\%) to the LArTPC --- produce ionization and scintillation signals within the 105~m$^3$ active volume. Optically segmenting the detector volume into 70 drift regions results in a mean of 5 scintillation signals per segment per spill. Assuming a scintillation time resolution of 25~ns, the rate of optical signal pile-up is 3\% per module per spill, relative to 30\% for a monolithic detector of equal size. With modest resolutions for both scintillation signal amplitude and position within the module, the corresponding ionization signals in each module can be accurately time-tagged and thereby associated to the correct neutrino interaction. The modular design maintains this capability after the LBNF beam power is upgraded to 2.4~MW.

The ND-LAr detector design entails a $7\times5$ modular array. Each module is 3~m high, 1~m long, 1~m wide and comprised of two TPC drift regions separated by a central cathode plane, each with maximum drift length of 50~cm. Each drift region is optically isolated and independently instrumented for scintillation light detection. This ND-LAr modular design is compared to a far detector-like LArTPC with a central cathode and two drift regions to assess the benefit of modularity relative to a ``monolithic'' LArTPC design. The same instrumented (active) LAr volume and bounding dimensions (5~m $\times$ 7~m $\times$ 3 m) is assumed between modular and monolithic schemes for direct comparisons. The simulation sample used in this study was constructed with LBNF-like neutrino fluxes input into a GENIE neutrino interaction model. The LBNF beam spill microstructure entails six batches each comprised of 84 53.1 MHz bunches. One thousand LBNF-like beam spills were simulated at 1.2 MW beam power and propagated through the ND hall detector geometry including the surrounding rock using a Geant4 simulation. Deposited energy was calculated by the summation of visible energy depositions in the ND-LAr active volume.

\begin{figure}
\centering
\includegraphics[width=0.5\textwidth]{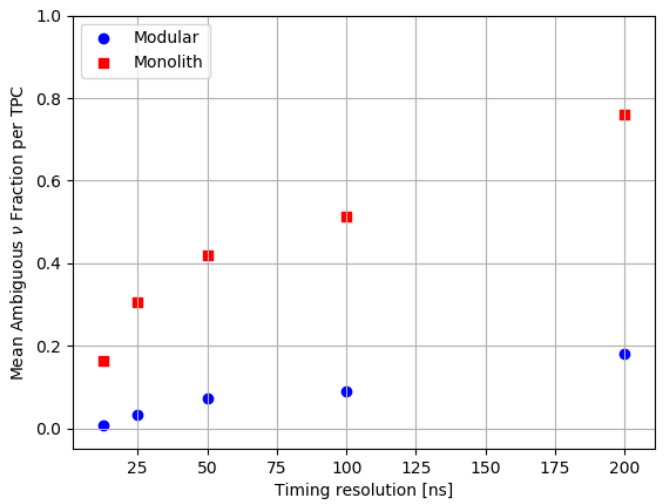}
\caption{Assuming removal of late scintillation light from the prompt component, the ``ambiguous $\nu$ fraction'' reflects the fraction of neutrino interactions per TPC drift region that cannot be resolved for a given timing resolution.}
\label{fig:ndlar-ana-pileup}
\end{figure}

Assuming a 1 MeV threshold for a single visible interaction per TPC, modularity alone reduces the ambiguity in single neutrino interaction selection by a factor of $\approx7$. The 1~MeV visible interaction threshold per TPC has a sub-percent level bias on the visible energy for neutrino interactions with a neutrino vertex residing within the charge instrumented volume and visible energy exceeding 500 MeV.
With roughly 55 independent neutrino interactions producing visible signals per 10~$\mu$s spill, the chance of two interactions being close in time is relatively frequent. The ND-LAr technical requirements call for scintillation light timing resolution of 20 ns. This specification assumes late scintillation light can be effectively subtracted from the prompt component. Scintillation light pile-up is mitigated with modularity. Even at a relatively modest timing resolution of 200~ns, a factor of 4 improvement in individual interaction light identification is gained with the described modularity. At 25~ns timing resolution, 3\% of neutrino interactions within a TPC are within 25~ns of each other with the current modular TPC scheme, whereas 30\% of neutrino interactions are within 25~ns of each other with a monolithic TPC scheme. These results are shown in Figure \ref{fig:ndlar-ana-pileup}.

Note that this comparison does not take into account interactions outside the charge instrumented volume. The monolithic TPC scheme would have sensitivity to light signals from these interactions, as opposed to the modular scheme which would not owing to the light-tight modules. Thus, charge-light signal combinatorics are further complicated in the monolithic TPC design, resulting in an underdetermined linear system.

\section{Muon and Electron Momentum Resolution and Scale Error}
\label{sec:lartpc-p-resolution-scale}

For muons stopping in the \dword{lar} and for those with momenta measured in the downstream tracker (\dword{ndgar}), the energy scale uncertainty from \dword{ndlar} is driven by the material model of the \dword{lar} and passive materials.  
This is expected to be known to better than 1\%.  
Note that the B field in \dword{ndgar} is expected to be known to about 0.5\% from simulation and field maps made with Hall and nuclear magnetic resonance probes, as well as with crosschecks made reconstructing kaon decays as discussed in Section~\ref{sec:mpd:performance}.

For electrons, the energy will be measured calorimetrically, rather than by range.  
The \dword{mip} energy scale (charge/MeV) will be set by rock muons.
The scaling to more dense deposits from EM showers can give rise to uncertainties, i.e., recombination could be different.  
Such uncertainties can be reduced by taking data with \dword{arcube} modules in a test beam.  
Outside of this, a useful calibration sample of electrons up to \SI{50}{\mega\electronvolt} comes from Michel electrons from stopping rock muons. 
The $\pi^0$ invariant mass peak is another good standard candle.  This approach is qualitatively similar to that of MINERvA, which achieved a 2.2\% uncertainty in the electromagnetic energy scale~\cite{Aliaga:2015wva, Park:2015eqa}

\section{Flux constraint with ND-LAr}
\label{sec:lartpc-flux_constraint}

\subsection{Neutrino-Electron Elastic Scattering}
\label{sec:lartpc-nu-e-scattering}

Neutrino scattering on atomic shell electrons, $\nu_{l}(\overline{\nu}_{l}) + e^{-} \rightarrow \nu_{l}(\overline{\nu}_{l}) + e^{-}$, is a purely electroweak process with a known cross section as a function of neutrino energy, $E_{\nu}$, in which all neutrino flavors participate, albeit with different cross sections. 
This process is not affected by nuclear interactions and has a clean signal of a single very forward-going electron. 
\dword{minerva}~\cite{Park:2015eqa,Valencia:2019mkf} has used this technique to characterize the \dword{numi} beam flux normalization (running in both the \dword{numi} low- and medium-energy modes), although the rate and detector resolution were insufficient to constrain the shape of the flux.
This technique has been thoroughly investigated as a cross section model-independent way to constrain the neutrino flux at the \dword{dune} \dword{nd}~\cite{Marshall:2019vdy}, with the highlights given here.

For a neutrino-electron sample, $E_{\nu}$ could, in principle, be reconstructed event-by-event in an ideal detector using the formula
\begin{equation}
  E_{\nu} = \frac{E_{e}}{1 - \frac{E_{e}(1-\cos\theta_{e})}{m_{e}}},
\label{eq:nue}
\end{equation}
\noindent where $m_e$ and $E_e$ are the electron mass and outgoing energy, and $\theta_e$ is the angle between the outgoing electron and the incoming neutrino direction. 
The initial energy of the electrons are low enough to be safely neglected ($\sim$\SI{10}{\kilo\electronvolt}). 
It is clear from Equation~\ref{eq:nue} that the ability to constrain the shape of the flux is critically dependent on the energy and angular resolution of electrons. 
For a realistic detector, the granularity of the $E_{\nu}$ shape constraint (the binning) depends on the detector performance. 
Additionally, the divergence of the beam (few \SI{}{\milli\radian}) at the \dword{dune}  \dword{nd} site sets a limit on how well the incoming neutrino direction can be known.

In work described in Ref.~\cite{Marshall:2019vdy}, the ability for various proposed \dword{dune} \dword{nd} components to constrain the \dword{dune} flux is shown using the latest three-horn optimized flux and including full flavor and correlation information.  
This was used to determine what is achievable relative to the best performance expected from hadron production target models. 
When producing the input flux covariance matrix, it was assumed that an \dword{na61}~\cite{Laszlo:2009vg} style replica-target experiment was already used to provide a strong prior shape constraint. 
Detector reconstruction effects and potential background processes are included, and a constrained flux-covariance is produced following the method used in Ref.~\cite{Park:2015eqa}.

The impact of the neutrino-electron scattering constraint on the flux covariance is shown in Figure~\ref{fig:LAR_nominal_covariances} for a five year exposure of the baseline \SI{1.2}{\mega\watt} \dword{fhc} beam on a \SI{30}{\tonne} \dword{ndlar}  detector (corresponding to $\sim$22k neutrino-electron events).
Note that this represents the baseline detector configuration where \dword{ndlar} will have a \dword{fv} of \SI{60}{\tonne} (for this measurement), but will spend 50\% of the time off-axis for DUNE-PRISM. 
It is clear that the overall uncertainty on the flux has decreased dramatically, although, as expected, an anti-correlated component has been introduced between flavors, as it is not possible to tell what flavor contributed to the signal on an event-by-event basis. 
Similar constraints are obtained for \dword{rhc} running~\cite{Marshall:2019vdy}.

\begin{dunefigure}[\dshort{fhc} flux covariance matrices for nominal \dword{ndlar}]
{fig:LAR_nominal_covariances}
{Pre- and post-fit \dword{fhc} flux covariance matrices for  an effective \SI{30}{\tonne} \dword{lar} detector using a five-year exposure of the baseline \SI{1.2}{\mega\watt} beam.}
  \subfloat[FHC pre-fit]  {\includegraphics[width=0.45\textwidth]{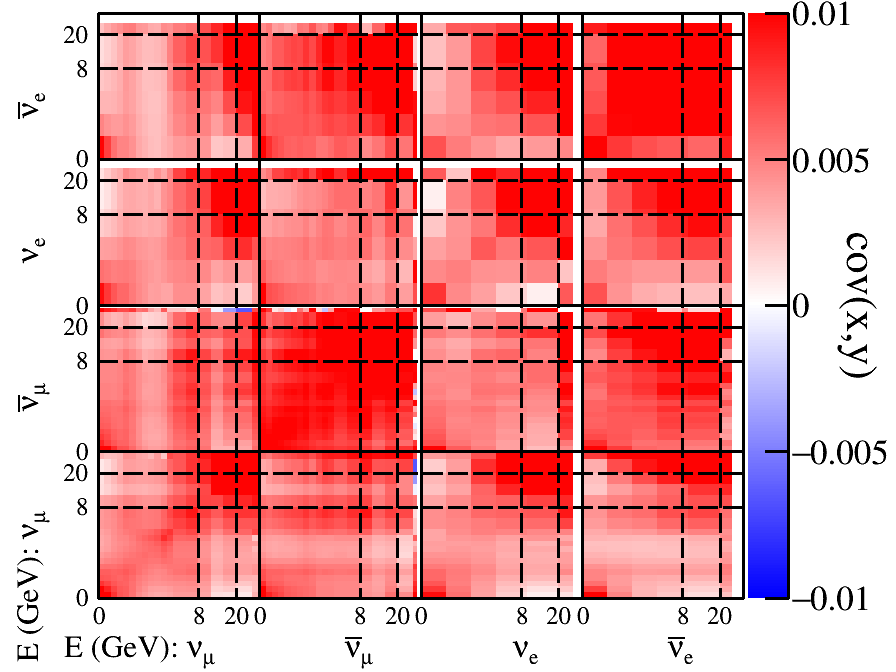}}
  \subfloat[FHC post-fit] {\includegraphics[width=0.45\textwidth]{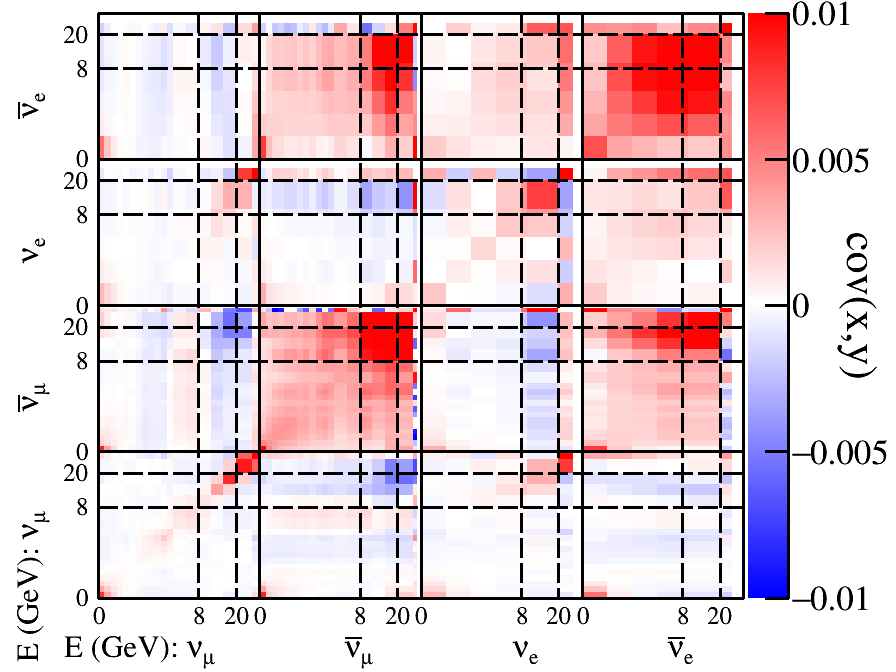}}
\end{dunefigure}

Figure~\ref{fig:nominal_det_constraint} shows the flux uncertainty as a function of $E_{\nu}$ for the $\nu_{\mu}$-\dword{fhc} flux, for a variety of \dword{nd} options. 
In each case, the constraint on the full covariance matrix is calculated (as in Figure~\ref{fig:LAR_nominal_covariances}), but only the diagonal of the $\nu_{\mu}$ portion is shown for ease of interpretation. 
Around the flux peak of $\sim$\SI{2.5}{\giga\electronvolt}, the total flux uncertainty can be constrained to $\sim$2\% for the nominal \dword{lar} scenario, and a lower mass detector (here a 5 t plastic scintillating detector) performs less well, as may be expected.
Clearly the neutrino-electron scattering sample at the \dword{dune} \dword{nd} will be a powerful flux constraint.
However, it is also clear that the ability to constrain the shape of the flux is not a drastic improvement on the existing flux covariance matrix, and none of the possible detectors investigated added a significantly stronger constraint.
The ``perfect'' detector option shown in Figure~\ref{fig:nominal_det_constraint} shows that this technique is also limited by the intrinsic divergence of the beam, and that a detector with better resolution would not perform significantly better than the LAr detector, particularly given the large LAr mass.
The neutrino-electron sample in \dword{ndlar} will make a powerful constraint on the overall neutrino flux normalization, and will be able to produce a constraint on the flux shape at the level of, or slightly better than, the prediction from the beam group.
As such, it will be able to diagnose problems with the flux prediction in a model-independent way, and will be a valuable tool in constraining the systematic uncertainties for the \dword{dune} oscillation program.

\begin{dunefigure}[Rate+shape and shape-only bin-by-bin flux uncertainties]
{fig:nominal_det_constraint}
{Rate+shape and shape-only bin-by-bin flux uncertainties as a function of neutrino energy for a five year exposure of the baseline \SI{1.2}{\mega\watt} beam, with various detector options, compared with the input flux covariance matrix before constraint.}
  \subfloat[Rate+shape]  {\includegraphics[width=0.45\textwidth]{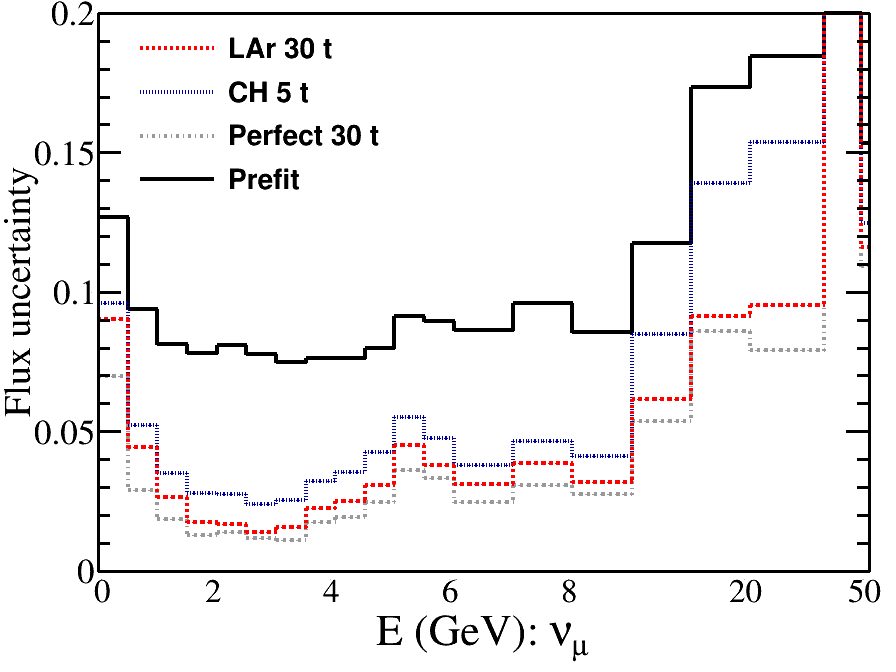}}
  \subfloat[Shape-only]  {\includegraphics[width=0.45\textwidth]{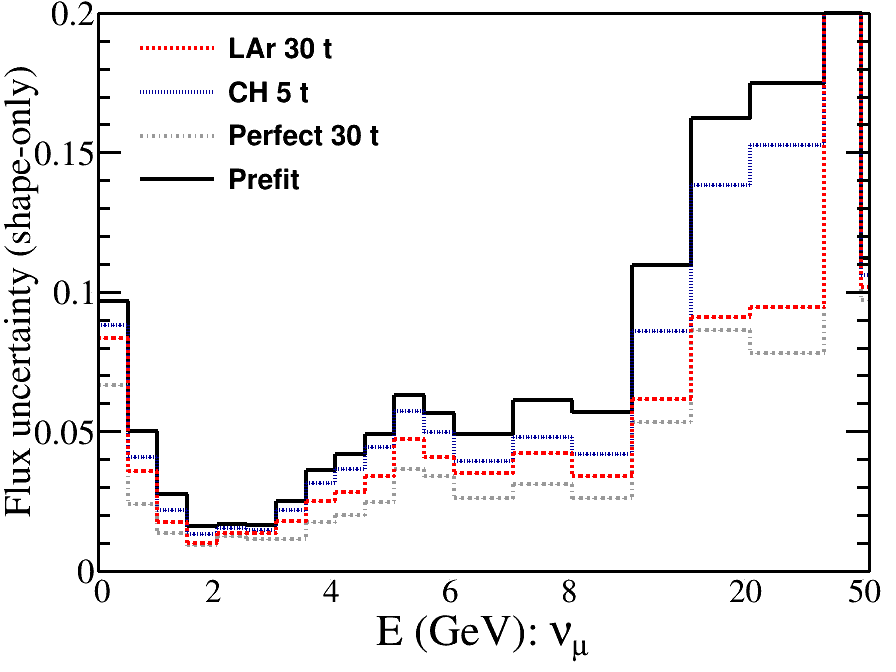}}
\end{dunefigure}

\subsection{Events with low energy transfer to the hadronic system}
\label{sec:lartpc-lownu}

The differential cross section for charged-current neutrino or antineutrino scattering can be written in terms of $\nu$, the total energy transfer to the hadronic system, as

\begin{equation}
\label{eq:lownu}
\frac{d\sigma}{d\nu} = A + B\frac{\nu}{E} - C\frac{\nu^{2}}{2E^{2}}
\end{equation}

\noindent
where $E$ is the neutrino energy, and the coefficients $A$, $B$, and $C$ are integrals of structure functions. The cross section is independent of the neutrino energy in the limit $\nu/E \rightarrow 0$. The energy independence enables a direct measurement of the relative shape of the neutrino flux by selecting a subsample of \dword{cc} events with $\nu$ below some fixed value, $\nu < \nu_{0}$. This technique, called the ``low-$\nu$ method,'' was first proposed by Belusevic and Rein~\cite{Belusevic:1988ab}, later by Mishra~\cite{MISHRA-Nu0} and used by the CCFR~\cite{Seligman:1997fe}, NuTeV~\cite{Tzanov:2005kr}, MINOS~\cite{Adamson:2009ju}, and MINERvA~\cite{DeVan:2016rkm} collaborations.

Extending the low-$\nu$ technique into the neutrino energy range relevant for DUNE oscillation physics is challenging~\cite{Bodek:2012uu}. The cutoff $\nu_{0}$ must be sufficiently small that the $B$ and $C$ terms in Equation~\ref{eq:lownu} are not significant, and sufficiently large to obtain a high-statistics event sample at high neutrino energy. Previous experiments~\cite{Adamson:2009ju, DeVan:2016rkm} have achieved this with a sliding $\nu_{0}$ as a function of neutrino energy, which introduces additional systematic uncertainties. The high rate of \dword{dune} and the large target mass of \dword{ndlar} give a sample of thousands of low-$\nu$ events per year per GeV out to energies of 20 GeV. This sample can be included independently in the long-baseline oscillation analyses as a constraint on the flux shape, and can be combined with the absolute flux measurement from neutrino-electron scattering.

Figure~\ref{fig:lownu} shows a comparison between the input flux and a low-$\nu$ selected sample with a parameterized reconstruction, simulated using GENIE. The flux is integrated over the entire \dword{ndlar} fiducial volume. The hadronic energy is estimated from the visible energy deposits in the active volume of the detector, and required to be less than 200 MeV. The muon energy is estimated assuming a 4\% resolution. The selected events are normalized to the rate per GeV for the full \dword{ndlar} per one year, assuming \SI{1.2}{\mega\watt} beam power. The flux is normalized to the event rate above 5 GeV, corresponding to $\nu/E_{\nu} < 0.04$ so that the cross section is very flat as a function of energy. The low-$\nu$ sample matches the flux shape at high neutrino energy. At low energy, the $B$ and $C$ terms become significant and the cross section is higher, leading to the excess seen in Figure~\ref{fig:lownu}. 

\begin{dunefigure}[\dword{fhc} low-$\nu$ selected sample compared to flux]{fig:lownu}
{The \dword{fhc} flux is compared to a sample of selected low-$\nu$ events from \dword{ndlar}. The shape matches well above 4 GeV, with distortions up to 30\% at 1 GeV. The details of the selection and reconstruction are described in the text.}
	\includegraphics[width=0.6\textwidth]{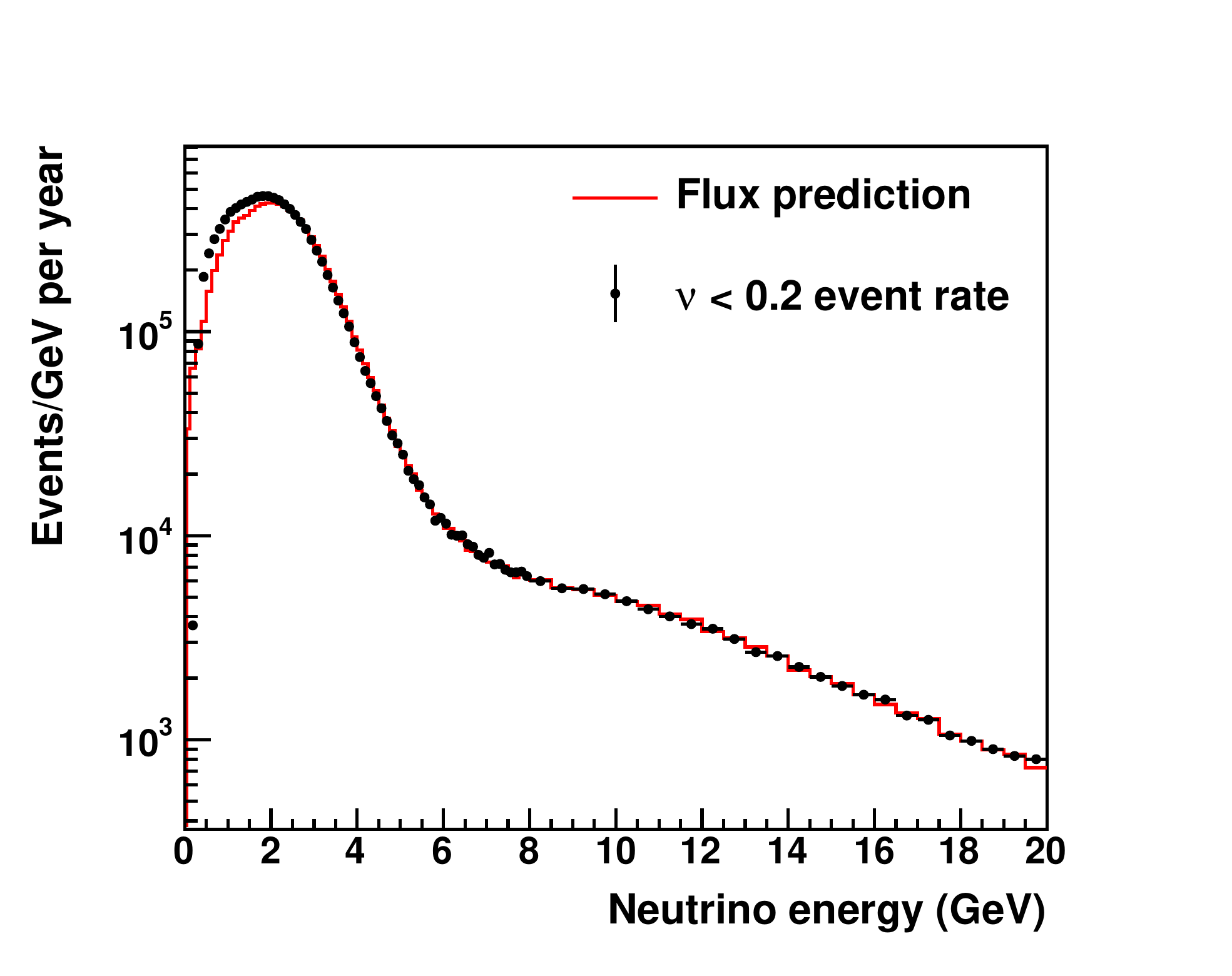}
\end{dunefigure}

Misreconstruction of $\nu$ also contributes to the deviation seen in Figure~\ref{fig:lownu}, as high-$\nu$ events that are selected in the low-$\nu$ sample increase the effect of the energy-dependent terms. In particular, $\nu$ is typically underestimated in events with energetic final-state neutrons. Especially at low neutrino energy, this can lead to events with large true $\nu/E_{\nu}$ appearing as low-$\nu$ events. This effect would be greatly reduced by tagging and rejecting events with fast neutrons, which can be accomplished by incorporating timing information from the photon detector system.

\cleardoublepage

\chapter{Magnetized Argon Target System: \dshort{ndgar}}
\label{ch:mpd}

\section{Introduction}
\label{sec:mpd:intro}
\dfirst{ndgar} is a magnetized detector system consisting of a \dword{hpgtpc} surrounded by an \dword{ecal}, both in a 0.5T magnetic field, and a muon system. A schematic of ND-GAr is shown in Figure~\ref{fig:ND-GAr}.
\begin{dunefigure}[Schematic of ND-GAr.]{fig:ND-GAr}
{Schematic of \dword{ndgar} showing the \dword{hpgtpc}, its pressure vessel, the \dword{ecal}, the magnet, and the return iron.  The detectors for the muon-tagging system are not shown.}
    \includegraphics[width=0.6\textwidth]{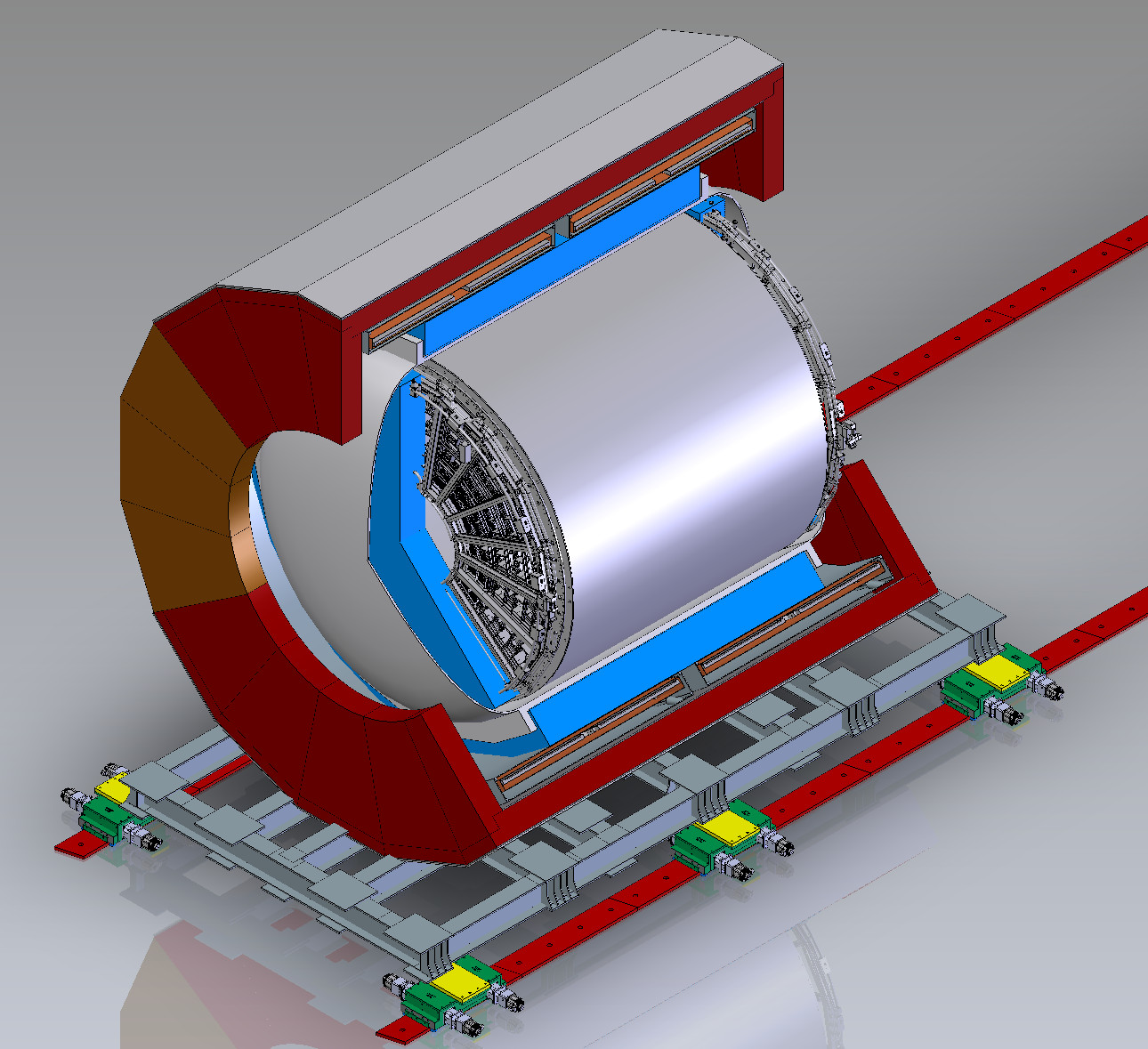}
\end{dunefigure}
\dword{ndgar} extends and enhances the capabilities of the \dword{nd}.  It does this by providing a system that will measure the momentum and sign of charged particles exiting \dword{ndlar}. For neutrino interactions taking place in the \dword{hpgtpc}, it will extend charged particle measurement capabilities to lower energies than achievable in the far or near \dwords{lartpc} and greatly extends the particle ID (PID) performance, particularly for proton-pion separation. These capabilities enable further constraints of systematic uncertainties for the \dword{lbl} oscillation analysis.

This chapter begins with a presentation of the physics requirements in Section~\ref{sec:mpd:reqs}. The \dword{ndgar} reference design is then described in Section~\ref{sec:mpd:reference}. The performance of the reference design is discussed in Section~\ref{sec:mpd:performance} along with some specific performance studies that are closely linked to the physics requirements.

\section{Role in Fulfilling Requirements}
\label{sec:mpd:reqs}

{\bf Primary roles of \dshort{ndgar}}
\begin{itemize}
    \item To fulfill \underline{ND-M1}, \underline{ND-M4}, \underline{ND-M5} and \underline{ND-M7} (and their derived capability requirements \underline{ND-C2.X}, \underline{ND-C3.X}) the \dword{nd} must track, identify the sign, and momentum-analyze muons exiting \dword{ndlar} to measure the energy spectrum of \numu and \anumu charged current interactions that occur in \dword{ndlar}. \dword{ndgar} fills this role and the performance is described in Section~\ref{sec:mpd:larcoverage}.

    \item To fulfill \underline{ND-M2} (and its derived requirements \underline{ND-C3.X}), the \dword{nd} must measure neutrino interactions on argon with a kinematic acceptance and reconstruction precision that equals or exceeds the \dshort{fd} across the energy range relevant to oscillations. This will allow the \dword{nd} to constrain interaction systematic uncertainties and verify the limited acceptance modeling in regions of kinematic phase space not accessible to \dword{ndlar}.  \dword{ndgar} fills this role and the performance is described in Section~\ref{sec:mpd:larcoverage}.
      
    \item To fulfill \underline{ND-M2} (and its derived requirements \underline{ND-C3.X}), the \dword{nd} must also have the ability to clarify the relationship between true and reconstructed energy by studying neutrino interactions on argon with low energy thresholds, good kinematic resolutions, and good particle identification. This will demand that the \dword{nd} be sensitive to particles that are not observed or may be misidentifed in a liquid argon TPC. These include low energy charged tracks, photons, and neutrons. 
\end{itemize}

Fulfilling these requirements leads to a set of derived detector capabilities that are described below.

{\bf Derived \dword{ndgar} detector capabilities}
\begin{itemize}
  
\item The \dword{nd} must be able to make measurements to constrain the muon energy scale with an uncertainty of 1\% or better to achieve the oscillation sensitivity described in volume-II of the \dshort{dune} \dword{fd} TDR\cite{Abi:2020evt}. The associated requirements are \underline{ND-M1} and \underline{ND-M2}. The strongest constraint comes from the calibrated magnetic field of the \dword{hpgtpc} coupled with {\it in-situ} measurements of strange decays. These constraints are described in Sections~\ref{sec:mpd-performance:calib} and \ref{ch:mpd:kshorts}.

  \item The \dword{nd} must be able to measure muons with a momentum resolution good enough to satisfy \underline{ND-C2.2}. The muon resolution of \dword{ndgar} is described in Sec.~\ref{sec:mpd:muresolutions}.
  
    \item To fulfill \underline{ND-C3.2}, the \dword{nd} must have a tracking threshold low enough to measure the energy spectrum of protons emitted due to \dword{fsi} in 
    \dword{cc} interactions. Theoretical studies, such as those reported in \cite{Nieves:2005rq,Lalakulich:2012gm,Mosel-Lalakulich-Gallmeister:2014}, suggest that \dshort{fsi} cause a dramatic increase in final state nucleons with kinetic energies in the range of a few 10s of \si{\MeV}. \dword{ndgar} is suitable for measuring such low energy protons. The kinetic energy threshold in \dword{ndgar} is an interplay between the argon gas density, readout pixel size, and ionization electron dispersion. A threshold of \SI{5}{\MeV} (\SI{97}{MeV/c}) is achievable and satisfies this requirement. The performance study that establishes this threshold is shown in Section~\ref{sec:mpd:protons}.

    \item To fulfill \underline{ND-C3.1} and \underline{ND-C3.3}, the \dword{nd} must be able to characterize the charged pion energy spectrum in \numu \& \anumu \dword{cc} interactions from a few \si{\GeV} down to the low energy region where \dshort{fsi} are expected to have their largest effect.
      \begin{itemize}
      \item Theoretical studies, such as those reported in \cite{Mosel:2014lja}, predict that \dshort{fsi} are expected to cause a large increase in the number of pions with kinetic energies between 20-\SI{150}{\MeV} and a decrease in the range 150-\SI{400}{\MeV}. A kinetic energy of \SI{20}{\MeV} corresponds to a momentum of \SI{77}{\MeV/c}. \dword{ndgar} must be able to measure \SI{70}{\MeV/c} charged pions with an efficiency of at least 50\% so as to keep the overall efficiency for measuring events with three pions at the \SI{70}{\MeV/c} threshold above 10\%. Charged track reconstruction is described in Section~\ref{sec:mpd:reco}.
      \item To fulfill \underline{ND-C3.3} \dword{ndgar} must also have the ability to measure the pion multiplicity and charge in 1, 2, and 3 pion final states so as to inform the pion mass correction in the \dword{nd} and \dword{fd} \dshorts{lartpc}. This capability is most important for pions with an energy above a few \SI{100}{\MeV} since those pions predominantly shower in \dword{lar}. A mock data study showing the impact that multiplicity measurements can have on \deltacp measurements is shown in Section~\ref{sec:mpd:pionmult}.
      \end{itemize}

    \item To fulfill \underline{ND-C3.6}, the \dword{nd} must be able to characterize the neutral pion spectrum in \numu and \anumu \dword{cc} interactions over the same momentum range as for charged pions. Photon and neutral pion reconstruction in \dword{ndgar} is described in Section~\ref{sec:mpd:pizeros}.

    \item To fulfill \underline{ND-C3.5}, the \dword{nd} must be able to identify electrons, muons, pions, kaons and protons. \dword{ndgar} addresses this requirement using a combination of: $dE/dx$ in the \dword{hpgtpc}, $E/p$ using the energy measured in the \dshort{ecal} and the momentum measured by magnetic spectroscopy in the \dword{hpgtpc}, and by penetration through the ECAL and muon system. These capabilities are described in Sections~\ref{sec:mpd:reco}, \ref{sec:mpd:ecalpid}, and \ref{sec:mpd:muon}.

\end{itemize}

\dword{ndgar} is also able to characterize the energy carried by neutrons with kinetic energies in the range \num{50}-\SI{700}{MeV} well enough to be sensitive to 20\% systematic variations. The 20\% specificiation is motivated by plausible model uncertainties. Neutron reconstruction and a preliminary sensitivity study are described in \ref{sec:mpd:neutrons}. Future work on neutron reconstruction will focus on optimization of the calorimeter and studies of the impact on physics sensitivity.




\section{Reference Design}
\label{sec:mpd:reference}
This section describes the components of \dword{ndgar}, detailing the state of their design at this time.  Technical details are presented for the components whose design has progressed beyond the ``concept'' level.
\subsection{High-Pressure Gaseous Argon TPC (\dword{hpgtpc})}
\label{section:mpd_tech_details}

The basic geometry of the \dword{hpgtpc} is a gas-filled  cylinder with a \dword{hv} electrode at its mid-plane, providing the drift field for ionization electrons. The gas is an argon-CH4 mixture, 90\%-10\% (molar fraction), at 10 bar. It is oriented inside the magnet such that the magnetic and electric fields are parallel, reducing transverse diffusion to give better point resolution. Primary ionization electrons drift to the end plates of the cylinder, which are instrumented with multi-wire proportional chambers (MWPCs) to initiate avalanches (i.e., gas gain) at the anode wires.  Signals proportional to the avalanches are induced on cathode pads situated behind the wires; readout of the induced pad signals provides the hit coordinates in two dimensions.  The drift time provides the third coordinate of the hit.

The details of the \dword{hpgtpc} design will be based closely on the design of the ALICE TPC~\cite{Dellacasa:2000bm} shown in Figure~\ref{fig:ALICETPC}. Two readout planes sandwich a central \dword{hv} electrode (\SI{25}{$\mu$m} of aluminized mylar) that generates the drift field, which is parallel to a \SI{0.5}{T} magnetic field. On each side of the electrode, primary ionization electrons drift up to \SI{2.5}{m} to reach the endplates, which are segmented azimuthally into 18 trapezoidal regions instrumented with \dwords{roc} that consist of \dword{mwpc} amplification regions and pad planes to read out the signals. A cross sectional view of an ALICE MWPC-based \dword{roc} is shown in Figure~\ref{fig:ALICE_ROC_MWPC}, with a gating wire grid to eliminate back-drift into the active volume, and an anode wire plane for avalanche amplification of the ionization signals which are subsequently read out by a plane of conductive pads. The \dwords{roc} are built in two sizes: a smaller \dword{iroc} and a larger \dword{oroc}. The trapezoidal segments of the endplates are divided radially into inner and outer sections, and the \dwords{iroc} and \dwords{oroc} are installed in those sections. The existing \dwords{iroc} and \dwords{oroc} in ALICE are scheduled to be replaced by new GEM-based \dwords{roc} for upgraded pile-up capability in the high rate environment of the \dword{lhc} and will be available to \dshort{dune}. The existing \dwords{roc} are more than capable of providing the performance needed by \dshort{ndgar}.  

\begin{dunefigure}[Diagram of the ALICE TPC.]{fig:ALICETPC}
{Diagram of the ALICE \dword{tpc}, from Ref.~\cite{Alme:2010ke}. The drift \dword{hv} cathode is located at the center of the \dword{tpc}, defining two drift volumes, each with \SI{2.5}{m} of drift along the axis of the cylinder toward the endplate. The endplates are divided into 18 sectors, and each endplate holds 36 readout chambers.}
    \includegraphics[width=0.7\textwidth]{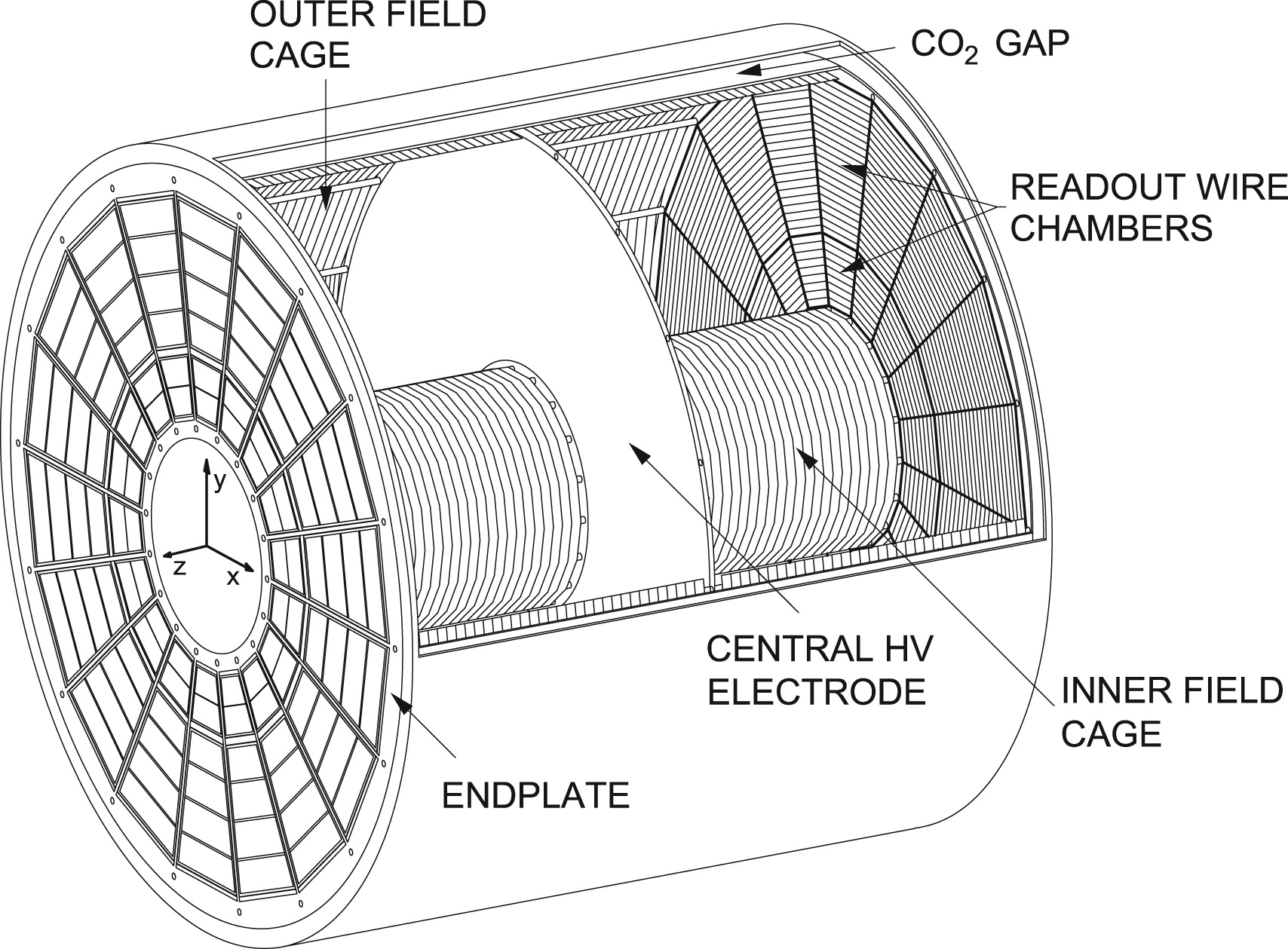}
\end{dunefigure}

\begin{dunefigure}[Schematic diagram of the ALICE MWPC-based ROC with pad plane readout]{fig:ALICE_ROC_MWPC}
{Schematic diagram of the ALICE MWPC-based \dword{roc} with pad plane readout, from Ref.~\cite{Alme:2010ke}.}
    \includegraphics{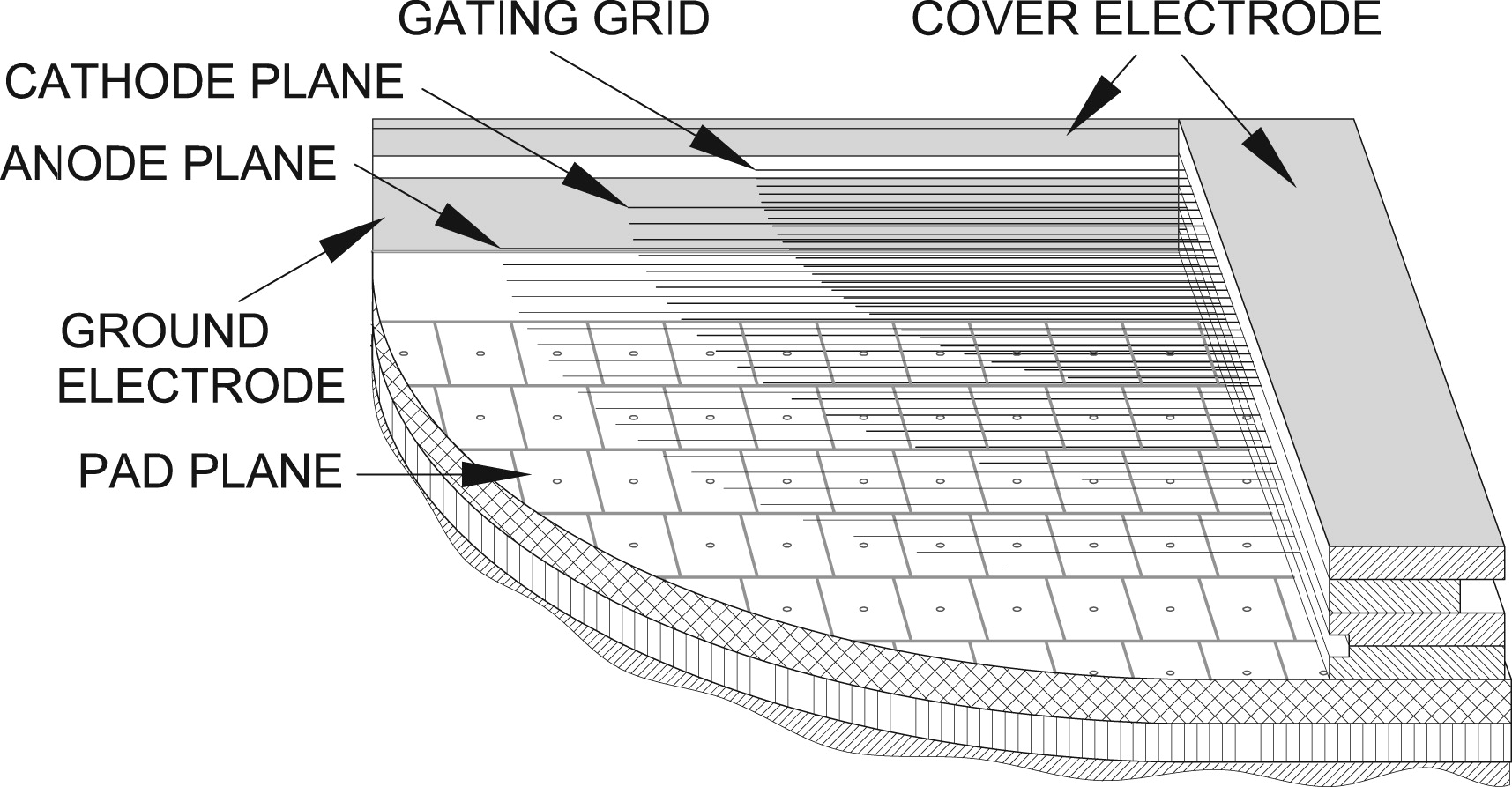}
\end{dunefigure}

In ALICE, which was built to run at a collider accelerator, the innermost barrel region was isolated from the \dword{tpc} and instrumented with a silicon-based inner tracker. For the DUNE \dword{hpgtpc}, the inner field cage labeled in Figure~\ref{fig:ALICETPC} will be removed and new \dwords{croc} will be built to fill in the resulting \SI{1.6}{m} diameter holes in each readout plane. Two possible \dword{croc} layouts are shown in Figure~\ref{fig:CROClayouts}.  With this central region instrumented by newly built \dwords{roc}, the active dimensions of the \dword{hpgtpc} will be \SI{5.2}{m} in diameter and \SI{5}{m} long, which yields an active mass of $\simeq$ \SI{1.8}{t}.

\begin{dunefigure}[Possible \dword{croc} layout options]{fig:CROClayouts}
{Possible design and layout options for new MWPC-based \dwords{croc}. In the layout shown on the left, the irregular hexagons fill the full central hole, with wires intersecting the chamber walls at an 80 degree angle. In the layout shown on the right, some corners of the chambers are constrained to 90 degrees, leading to a 4\% loss in coverage.}
    \includegraphics[height=2.7in]{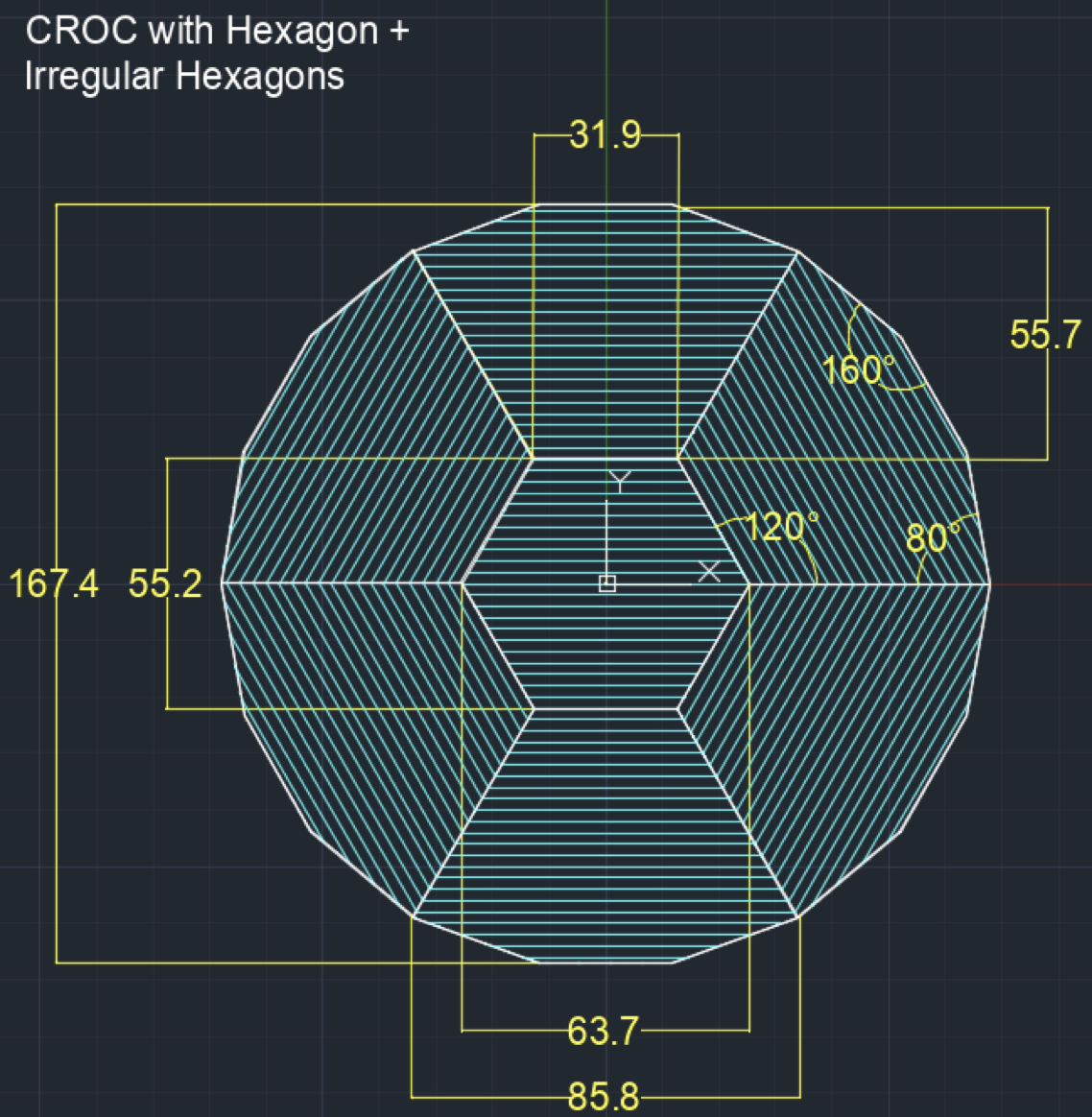}
    \includegraphics[height=2.7in]{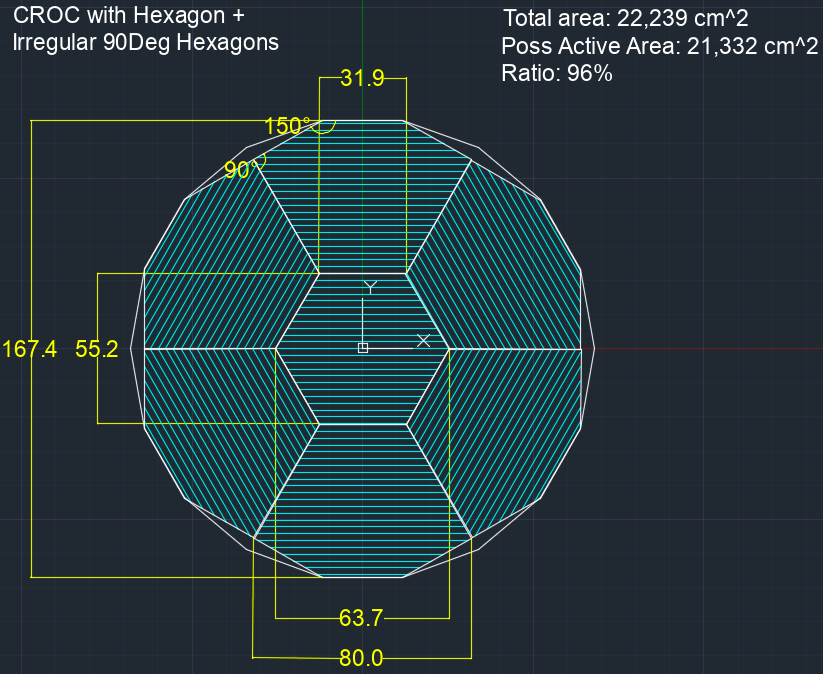}
\end{dunefigure}

While much of the \dword{hpgtpc} concept is based on the ALICE \dword{tpc} design, there are several important differences and requirements.  The major areas of R\&D which are needed for the DUNE \dword{hpgtpc} are concentrated in seven areas: 
\begin{itemize}
    \item {\bf Gas mixture studies:} The ALICE \dword{tpc} operated at atmospheric pressure with a gas mixture of Ne/CO$_2$/N$_2$ or of Ar/CO$_{2}$, which are not the gas mixture and pressure proposed for DUNE. Work is currently in progress to determine the breakdown voltage, gas gain, and diffusion coefficients for the DUNE reference design gas mixture. Work is also in progress to measure the achievable gain with that gas mixture and an ALICE \dword{iroc} at pressures ranging from 1 to 10 atmospheres.  Additional studies will be needed for promising alternative gas mixtures which aim to have unique optical properties for light production and detection, while maintaining wire chamber operational stability.

    \item {\bf Electronics and \dword{daq} development:} While the readout chambers are available from ALICE, the ALICE front end electronics are not.  To achieve a very attractive price point for the front end electronics, and to maximize the synergies with the liquid argon near detector, it is hoped that similar electronics can be used for the \dword{hpgtpc} and the \dword{nd} \dword{lartpc}.  LArPix~\cite{Dwyer:2018phu} development is in progress for the \dword{lartpc}, but some modifications are needed to adapt this for use in the \dword{hpgtpc}, since the \dword{hpgtpc} signal is faster and inverted compared to the liquid argon near detector (as the gaseous argon reads out an induced charge), and the gain in the gas also results in a widened dynamic range. Readout electronics will also need to be developed for the light collection system.
    
    \item {\bf Design of additional \dwords{roc} and mechanical supports:} New central readout chambers will need to be designed to cover the central area of the endcaps, which was not part of the \dword{tpc} in ALICE. This central region would likely be segmented into multiple chambers, rather than a single large chamber, to keep the wire spans in the range of those for the existing \dwords{iroc} and \dwords{oroc}.  A suitable wire spacing and pad layout must be developed for the central region. Prototypes for the new \dwords{croc} will also need to be tested with the appropriate gas mixture. A gas-tight structure must be designed to support the field cage, readout chambers, and supports will need to be developed outside this for the readout electronics, cabling, and services such as water cooling lines.   A concept for supporting the entire detector within the pressure vessel must also be developed.
    
    \item {\bf Field cage and high voltage:} A new field cage and mechanical endcap structures will need to be constructed for the DUNE \dword{hpgtpc} as well.  While ALICE had an inner and outer field cage, the DUNE design will only have an outer field cage.  The ALICE field cage was constructed of parallel mylar strips creating rings surrounding the active volume, as they had very stringent requirements on the material budget. \dword{dune} is investigating a more robust option, in part because the detector is mobile. In ALICE, the field cage elements were housed inside a thin but gas-tight outer field cage vessel to isolate the high voltages of the field cage rings in Ar/CO$_2$ from the grounded containment vessel wall. The gap region between the outer wall of the field cage vessel and the inner wall of the pressure vessel was filled with CO$_2$ gas, which has a higher breakdown voltage than that of Ar/CO$_2$. The DUNE design is complicated by the fact that the \dword{hpgtpc} will be operated at high pressure, which may necessitate a different solution to the field cage isolation, in order not to introduce complications related to strict regulation of differential pressures between two independent gas volumes. It will also be necessary to develop a high voltage feed-through to deliver the $\mathcal{O}(100)$~kV to the drift electrode within the pressure vessel.

    \item{\bf Light collection:} Primary light production in pure argon in the VUV is well understood~\cite{Gonzalez-Diaz:2017gxo}. In pure argon at a pressure of 10~atm, it is estimated that a minimum ionizing particle will produce approximately 400 photons/cm~\cite{ Chandrasekharan:2005edc}, but in typical gas \dword{tpc} operation a quenching gas, or gases, are added that absorb essentially all the VUV photons.  Recent studies have indicated that with the addition of Xe or CF$_4$ gas, among others~\cite{Neumeier:2014gpa}, to an argon mixture, it may be possible to quench the VUV component of the scintillation, allowing for stable wire gain, while producing light in the visible or near-IR.  With suitable instrumentation, this light signal could be used to provide a $t_{0}$ timestamp for events in the gas. Utilizing this light would be a novel development for a gaseous argon \dword{tpc}. R\&D will be needed to understand the potential wavelength-shifting properties and light yield of the argon gas mixtures under study in order to design a photon detection system.  With close coordination among the gas mixture, field cage, and HV groups, a conceptual design will be developed for the collection and readout of light in the gas volume if a suitable gas mixture is identified.
    
    \item {\bf Calibration and slow controls:} To precisely monitor any variations of the drift velocity and inhomogeneities in the drift field, the ALICE \dword{tpc} used a laser calibration system to produce hundreds of beams that could monitor the drift behavior across different slices of the drift region.  The light was transmitted though the field cage support rods.  For the DUNE \dword{hpgtpc}, a conceptual design for a laser calibration system will be developed which might be distributed throughout the drift region as in ALICE, or might only involve light injection from the end caps.  Its design will need to be developed in close collaboration with the HV field cage design.  It should be pointed out that due to the low occupancy in the DUNE \dword{hpgtpc} the impact of space charge on field uniformity is expected to be negligible, in contrast to the operation of ALICE. Many other detector parameters will also need to be continuously monitored, such as temperatures, voltages, currents, as well as gas properties such as drift velocity and diffusion. The \dword{hpgtpc} slow control design will be developed in synergy with the other systems in the DUNE ND hall.
    \item {\bf Gas and cooling systems:} The detector performance depends crucially on the stability and quality of the gas. If the ALICE gas volume designs are adopted, the \dword{hpgtpc} design will likely require two gas systems: one for the Ar/CH$_{4}$ drift gas, and one for the  CO$_{2}$ gas that isolates the field cage vessel HV from the pressure vessel. In this case, the two volumes will need to be kept at similar pressures in order to avoid excessive stresses on the field cage vessel.  For the DUNE \dword{hpgtpc} system, it will be necessary to develop a list of requirements on the control and stability of the CH$_{4}$ level in the drift gas mixture as well as upper limits on O$_{2}$ and H$_2$O contaminant levels in the gas.   The temperature uniformity requirements for the DUNE \dword{hpgtpc} design will also need to be developed.  In addition, the capability to temporarily inject a radioactive gas into the drift region for pad response calibration will need to be developed.
\end{itemize}
%
%

\subsection{\dword{hpgtpc} Pressure Vessel}\label{sec:TPC_PV}

Since the nominal operating pressure for the \dword{hpgtpc} is 10~atm, a pressure vessel will be needed.  The preliminary design of the pressure vessel, presented in Figure~\ref{fig:TPC_PV}, accounts for the additional volume needed to accommodate the TPC field cage, the \dword{roc} support structure and \dword{fe} electronics and the end-cap ECAL (see Section~\ref{sec:ecal-design}).  

\begin{dunefigure}[Pressure vessel preliminary design]{fig:TPC_PV}
{Pressure vessel preliminary design.}
\includegraphics[width=0.6\textwidth]{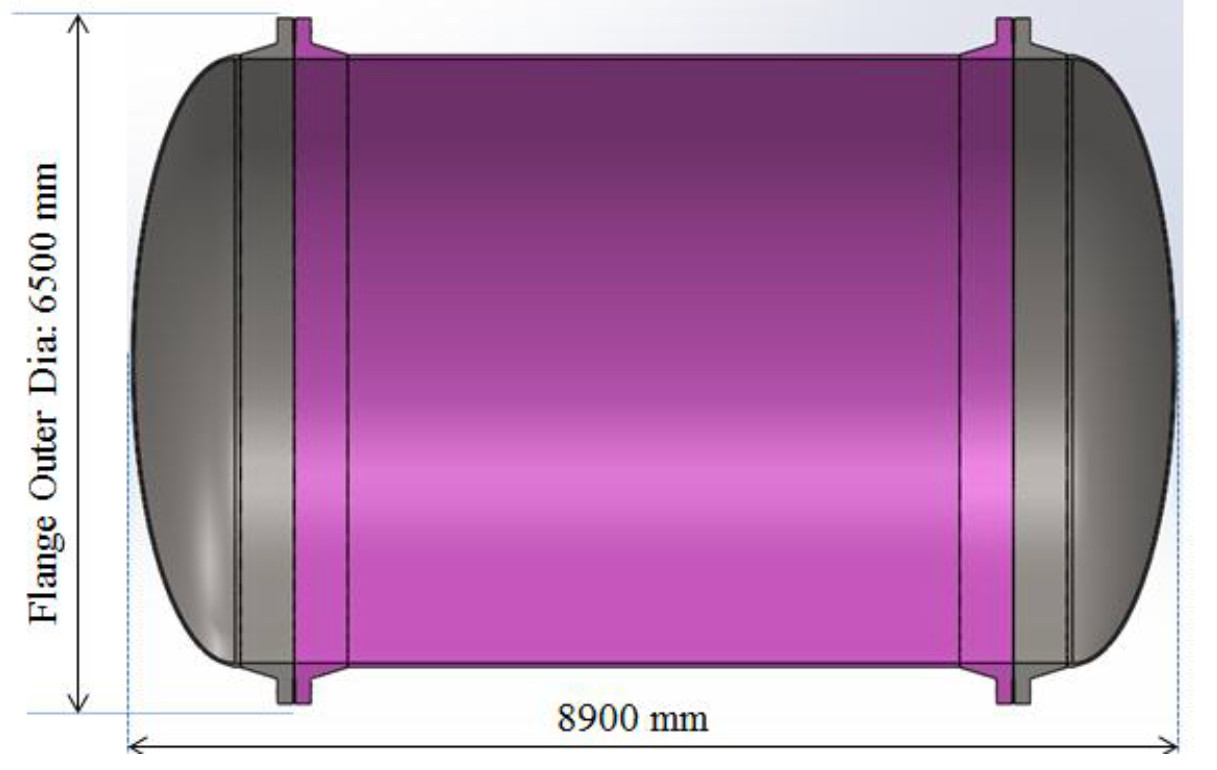}
\end{dunefigure}

The structural design and analysis of the pressure vessel are carried out using the American Society of Mechanical Engineers (ASME) code Section VIII Div-I and II. The materials used in the pressure vessel have been chosen to minimize radiation length, while complying with the code requirements.  Design and analysis includes shell thickness calculations for competitive materials as per UG-27 of ASME, design parameter calculations of the ellipsoidal heads (ASME, Appendix 1). Stresses such as circumferential bending, longitudinal bending, tangential shear, bolt size calculations and flange design have been calculated as per ASME Section VIII Div II, 3D finite element method (FEM) analysis.

The current pressure vessel reference design utilizes 5083 aluminum and has a cylindrical section that is $\simeq$ \SI{6}{m} in diameter and \SI{6}{m} long. It utilizes two semi-elliptical flanged heads. The walls of the cylinder barrel section are $\simeq \SI{4}{cm}$ thick which corresponds to $\simeq$ 0.5X$_0$. It is possible that further reduction of the thickness can be accomplished  with the addition of stiffening rings.  The heads will be constructed out of stainless steel which has minimal impact on the physics because the end-cap \dword{ecal}s are inside the pressure vessel.
\paragraph{Weldments}
An initial analysis of the weldments has been performed.  In this analysis the following points have been considered following ASME Subsection B.
\begin{itemize}
    \item Weld joint categories and joint efficiency consideration
    \item Design of weld joints
    \item Challenges of welding aluminum cum solutions
\end{itemize}
After careful consideration, it has been determined that double-welded butt joints along with a full radiographic examination are the best choice for this application. 
The ASME BPV Code has four categories of welds:
\begin{itemize}
    \item Category A: Longitudinal or spiral welds in main shell.
    \item Category B: Circumferential welds in main shell.
    \item Category C: Welds connecting flanges to main shell.
    \item Category D: Welds connecting nozzles or communicating chambers to main shell.
\end{itemize}
Figure~\ref{fig:welds} gives a schematic of the ASME BPV Code weld categories (A, B, C, and D).
\begin{dunefigure}[ASME BPV Code]{fig:welds}
{ASME BPV Code weld categories.}
    \includegraphics{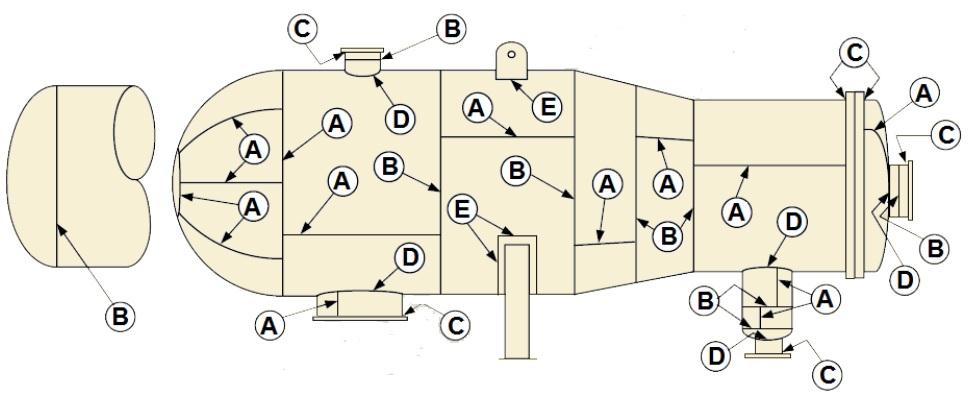}
\end{dunefigure}
Since the central section of the pressure vessel is made of aluminum in order to meet the thickness specification of $\leq$ 0.5X$_\circ$, the challenges presented with aluminum welding are under intense evaluation.  They include:
\begin{itemize}
    \item Thermal conductivity: aluminum is 5 times more thermally conductive than steel. It can cause a lack of penetration in the weld.  
        \begin{itemize}
        \item Solution: Preheating the aluminum work piece.
        \end{itemize}
        
    \item Hydrogen and porosity: H$_2$ is very soluble in liquid aluminum. Once the molten material starts to solidify, it can’t hold the hydrogen in a homogenous mixture anymore. The hydrogen forms bubbles that become trapped in the metal, leading to porosity.  
        \begin{itemize}
            \item Shielding by inert gas.
        \end{itemize} 
        
    \item Melting point: aluminum has lower melting point than steel that can result in burn-throughs.
    However, aluminum oxide has a much higher melting point than aluminum base metal. It acts as an insulator that can cause arc start problems and very high heat is required to weld through the oxide layer. This can cause burn-through on the base material and porosity, since the oxide layer tends to hold moisture.  
        \begin{itemize}
            \item Solution: a welding machine with current control is useful for keeping the aluminum work piece from overheating, causing a burn-through.  Proper cleaning and removing the oxide layers are of utmost importance.
        \end{itemize}
\end{itemize}
\paragraph{3D FE Analysis with distributed mass (300 Ton, \dword{ecal})}

The stress on the cylindrical shell has been analyzed assuming a 300t ECAL load on the shell.  It has been determined via analytical calculation that an aluminum shell thickness of 42mm will meet ASME code.  The stainless steel heads and interface to the cylindrical body have also been studied.  Results from this analysis are shown in Figure~\ref{fig:FEA_head}.
\begin{dunefigure}[PV FEA]{fig:FEA_head}
{FEA for pressure vessel heads.  Left: meshing of stainless steel elliptical heads in comsol multiphysics. Right: stress analysis - maximum von Mises stress 151 MPa is shown (allowable limit is 180 MPa).}
    \includegraphics[width=0.45\textwidth]{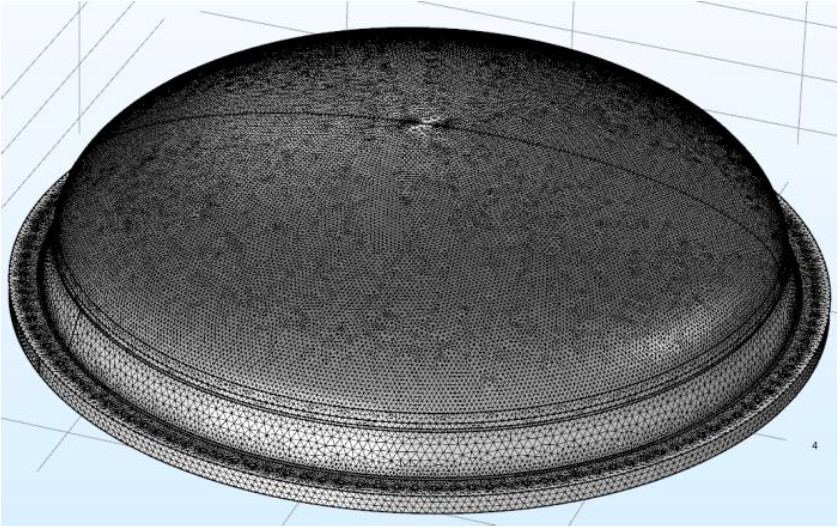}
    \includegraphics[width=0.54\textwidth]{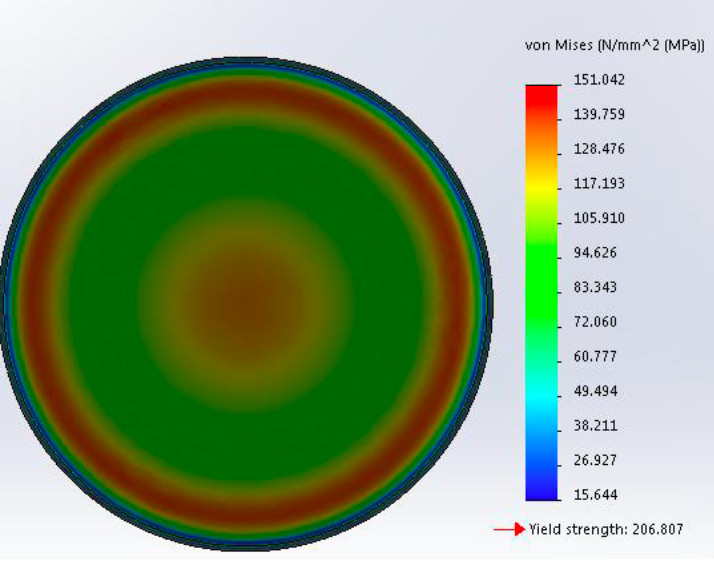}
\end{dunefigure}

\subsection{Electromagnetic Calorimeter (\dshort{ecal})}
\label{sec:ecal-design}

The principal role of the \dword{ecal} is to reconstruct photons produced in neutrino interactions, especially those originating from $\pi^0$ decays. The \dword{ecal} is also capable of measuring electron energies by calorimetry. In addition, it measures the time of entering tracks ($t_0$) which allows the track vertex position along the \dshort{hpgtpc} drift direction to be determined. The detector concept is based on a high-granularity calorimeter that is able to measure both the energy and direction of electromagnetic showers. Those capabilities allow photon induced showers to be associated with interactions observed in the \dword{hpgtpc}, thereby determining the decay vertex of $\pi^0$s. In the case of $\nu_e$ measurements in the \dword{hpgtpc}, the \dword{ecal} will play an important role in rejecting events with $\pi^0$ decays, which represent a background to $\nu_e$ interactions in \dword{ndlar}. The \dword{ecal} can also be used to reject external backgrounds, such as rock neutrons and muons, providing a sub-nanosecond timestamp \cite{Simon:2013zya} for each hit in the detector. The \dword{ecal} is also capable of detecting neutrons that scatter in or near the scintillator layers.  The ECAL performance is discussed in greater detail in Section~\ref{sec:mpd:ecal}.

\paragraph{\dword{ecal} Design}

\begin{dunefigure}[\dshort{ndgar} \dshort{ecal} conceptual design]{fig:ConceptDesign_NDECAL}
{On the left, the conceptual design of \dword{ndgar}. The \dword{ecal} (shown in blue) barrel is located outside the \dword{hpgtpc} pressure vessel and the endcaps are inside. On the right, a conceptual design of the \dword{ecal} system represented by the octagon surrounding the TPC.}
\includegraphics[width=0.55\textwidth]{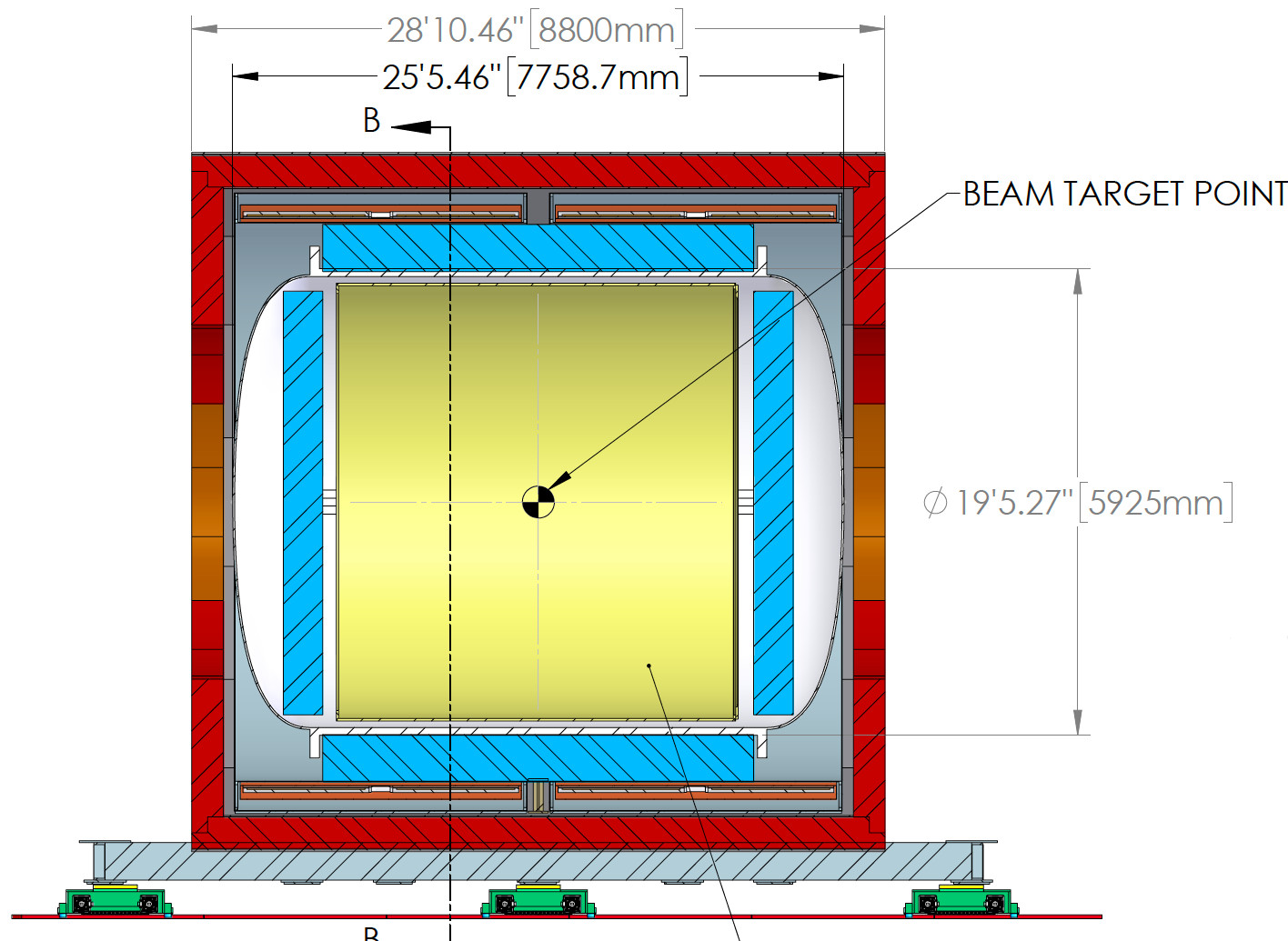}
 \includegraphics[width=0.42\textwidth]{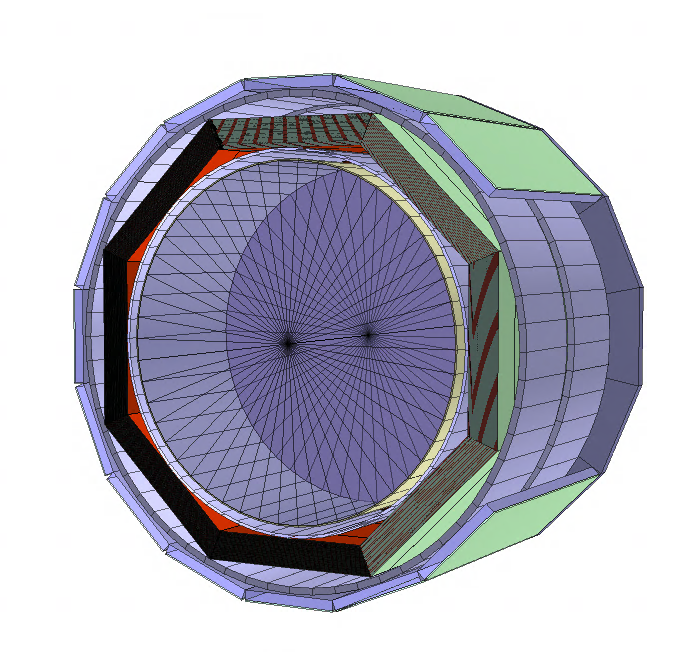}
\end{dunefigure}

The \dword{ecal} reference design, shown in Figure~\ref{fig:ConceptDesign_NDECAL}, is inspired by the CALICE analog hadron calorimeter (AHCAL)~\cite{collaboration:2010hb}.   The barrel has an octagonal shape with each octant composed of several trapezoidal modules.  Each module consists of layers of polystyrene scintillator as active material read out by \dwords{sipm}, sandwiched between absorber sheets. The scintillating layers consist of a mix of tiles with dimensions between $2\times2$ cm$^2$ to $3\times3$ cm$^2$ (see Figure~\ref{fig:ConceptTile_NDECAL}) and cross-strips spanning a full \dword{ecal} module length (between 1.5 to \SI{2.1}{m}, depending on the strip orientation) with a width of \SI{4}{cm} to achieve a comparable effective granularity. The strip design could be very similar to the T2K \dword{ecal} strips~\cite{Allan:2013ofa} using embedded wavelength-shifting fibers, but a solution with no fibers and a more transparent scintillator material is being considered in order to achieve the best possible time resolution. The high-granularity layers are concentrated in the inner layers of the detector, since that has been shown to be the most relevant factor for the angular resolution \cite{Emberger:2018pgr}. With the current design, the number of channels is about 2-3 million. A first design of the \dword{ecal} and the simulated performance has already been studied in \cite{Emberger:2018pgr}. 

This arrangement of the modules with respect to the pressure vessel is under study and optimization. While a location inside of the pressure vessel avoids the negative impact of additional material in front of the calorimeter and reduces the size of the detector, it also increases the radius required for the pressure vessel and, with that, the volume of the magnet.  This is due to the fact that a more complex mounting structure for the \dword{hpgtpc} would be required.  It also results in additional complexity for the \dword{ecal}  services, which would then have to be passed into the high-pressure environment. 

\begin{dunefigure}[Conceptual layout of the \dshort{ndgar} \dshort{ecal}] 
{fig:ConceptTile_NDECAL}
{Conceptual layout of the \dword{ecal} showing the absorber structure, scintillator tiles, \dwords{sipm}, and printed circuit boards (\dwords{pcb}).}
\includegraphics[width=0.55\textwidth]{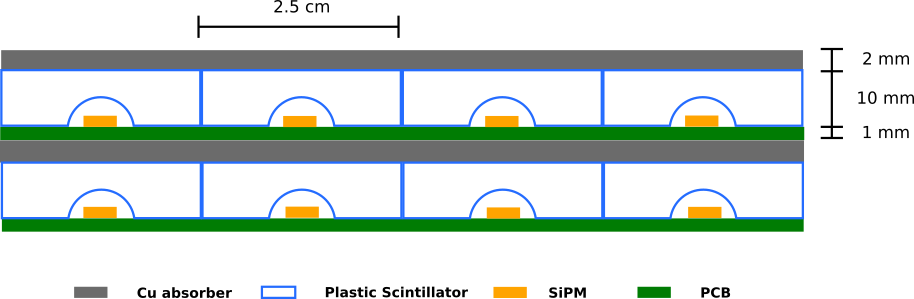}
\end{dunefigure}

The ECAL reference design calls for the barrel to be located entirely outside of the pressure vessel. A study of the influence of the pressure vessel on the \dword{ecal}  energy and angular resolution is shown in Figure~\ref{fig:PressureVessel_ECALPerf}. As the thickness of the pressure vessel wall increases, a large degradation of the energy resolution is observed, in particular at low photon energies. For the angular resolution, a certain amount of additional material is slightly beneficial, while a significant degradation is observed beyond 1 $X_{0}$, especially for lower-energy photons. The current design of the pressure vessel has reduced the required material thickness to $\simeq$0.5 $X_{0}$ in the barrel region. With that thickness the barrel \dword{ecal} can be located outside the pressure vessel without a significant degradation to its performance.

The \dword{ecal} readout design is expected to be very similar to the CALICE AHCAL in the ILD detector \cite{Kvasnicka_2017, Behnke:2013lya}. A dedicated ASIC, the SPIROC \cite{Lorenzo_2013}, could be used for the front-end electronics to read-out SiPMs. The front-end electronics would be embedded in the \dword{ecal} layers. Studies have been performed to evaluate the impact of the additional material by the front-end and have shown that a scenario with the front-end on a \dword{ecal} layer still gives acceptable performance. More details on the expected data rates are available in Chapter~\ref{ch:comp}.

The endcap \dwords{ecal} provide hermeticity and a large solid-angle coverage. They have a design that is similar to the barrel sections. Locating the  endcap \dwords{ecal} outside the pressure vessel may not be practical since that would increase the horizontal extent of the detector. A mixed solution, with the barrel part of the detector outside of the pressure vessel and the endcap \dwords{ecal} located inside, as illustrated in Figure~\ref{fig:ConceptDesign_NDECAL} (left) is foreseen. The detailed layout of the detector, with the goal of minimizing gaps in the acceptance, is subject to further design work. Ongoing studies to optimize the overall detector design, cost and performance are described in section~\ref{sec:ecal-performance}.


\begin{dunefigure}[Influence of pressure vessel on \dword{ecal} performance]
{fig:PressureVessel_ECALPerf}
{Influence of the pressure vessel thickness on the \dword{ecal} energy (left) and angular resolution (right). The points are the ratio of the absolute energy or angular resolution relative to the value without a pressure vessel.}
\includegraphics[width=0.49\textwidth]{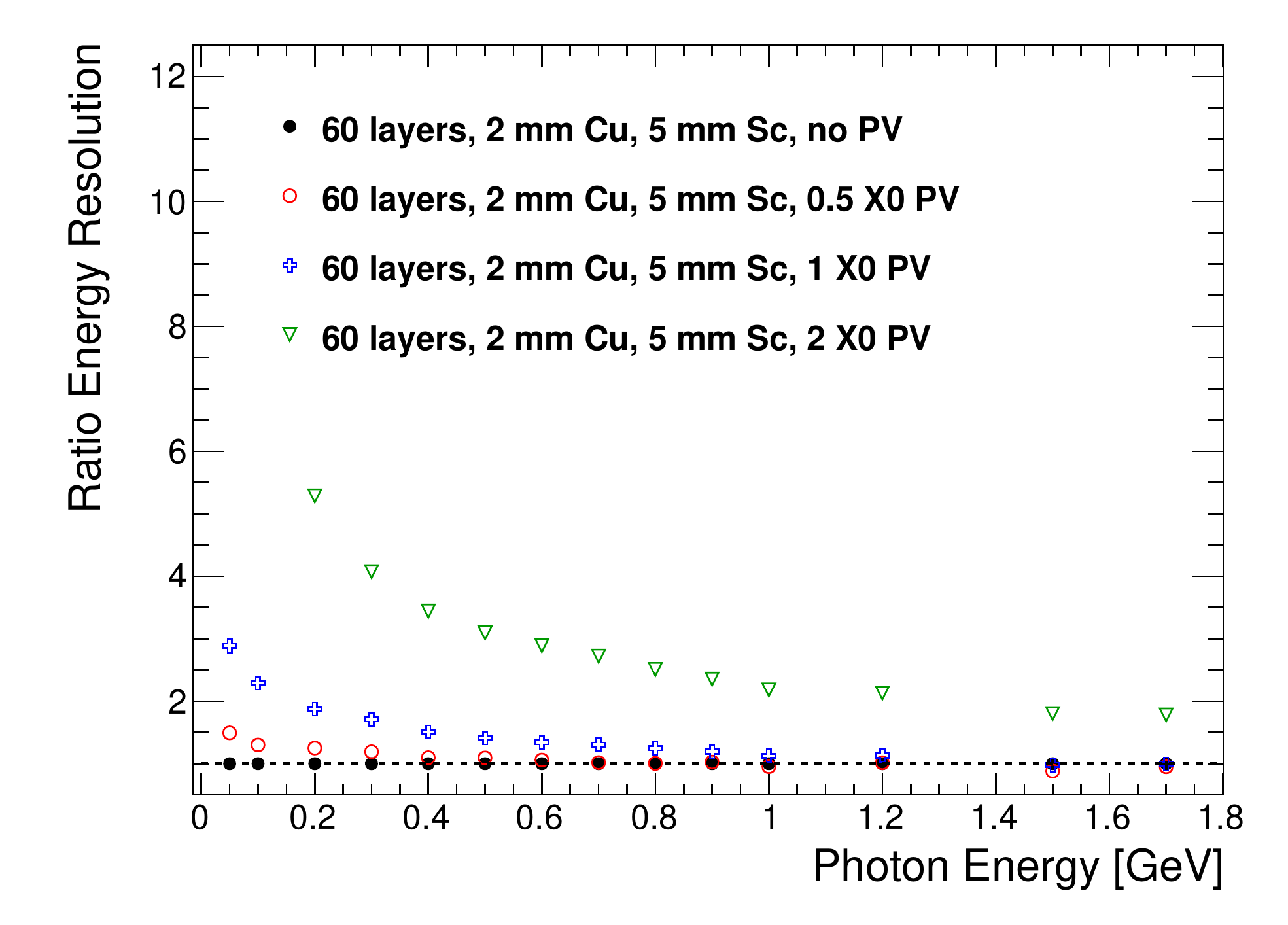}
\includegraphics[width=0.49\textwidth]{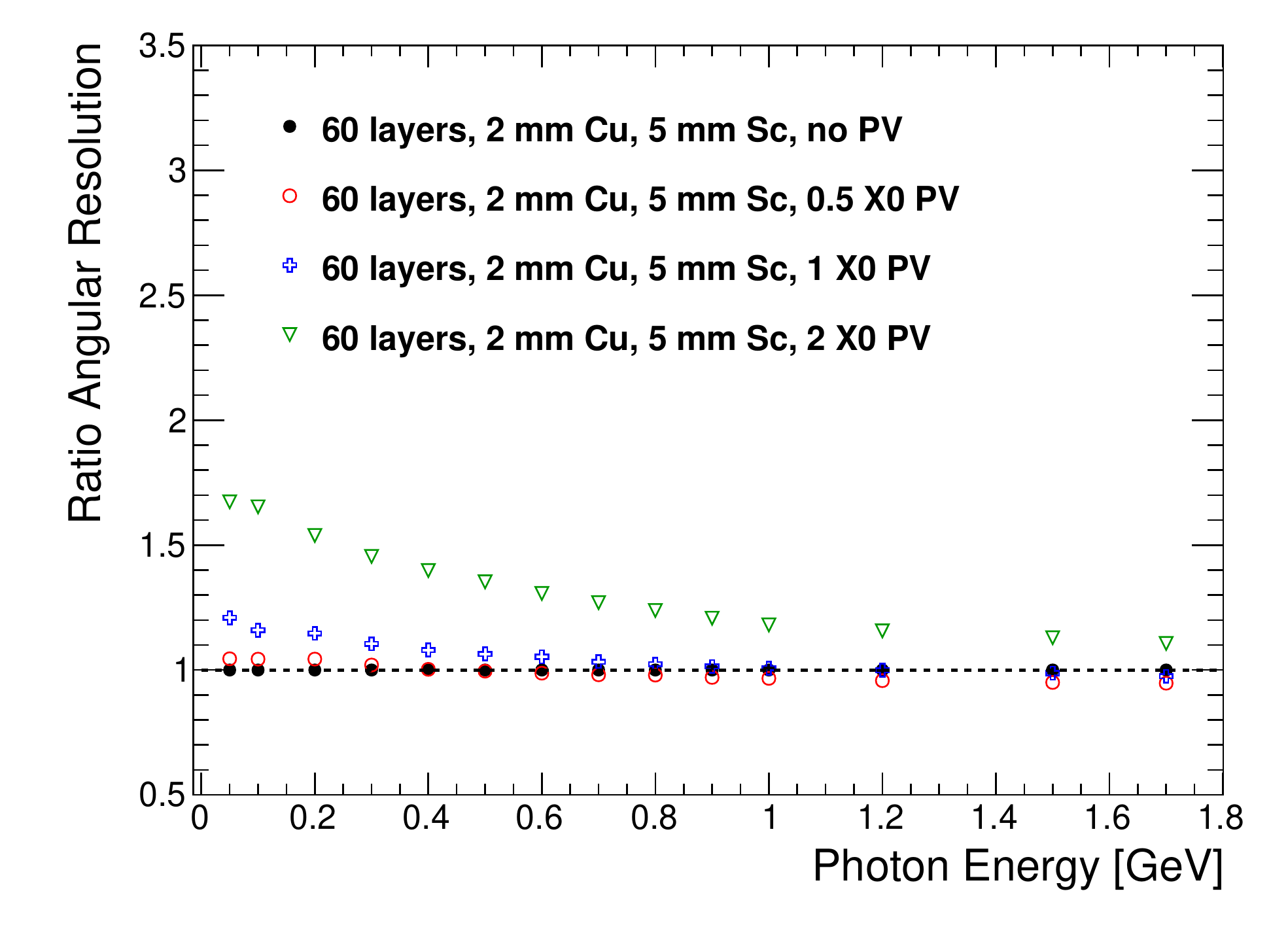}
\end{dunefigure}

\subsection{Magnet}
\label{sssec:nd:appx:mpd-magnet}
The reference design for the \dword{ndgar} consists of two coupled solenoids with flux-return iron which functions as the absorber in the muon tagging system.  The concept is similar to a magnet system built by ASG in Italy for the JINR's Multi-Purpose Detector~\cite{jinrtdr}.

In addition to this reference design, several alternate designs have
been evaluated, comprising some variation on a Helmholtz coils
concept, both with and without a partial return yoke and with and
without trimming coils at the ends. The main advantage of a
Helmholtz-like design is the complete removal of any material in front
of the detectors, except in the exact location of the coils. On the
other hand, the stored energy to reach the design field would be
significantly larger and, in the yoke-less configurations, the stray
field management is complicated. Incorporation of absorber material (either as a return
yoke or as non-magnetic material) is complicated.
The former studies on alternate designs
have not been discarded completely, nevertheless the collaboration
efforts are now focused solely on the solenoid design.
\subsubsection{Reference Design Details}
The reference magnet design is shown in Figure~\ref{fig:dune_mag_ref_design}.
\begin{dunefigure}[Magnet reference design schematic]{fig:dune_mag_ref_design}
{Solenoid arrangement for \dword{ndgar} superconducting magnet.}
\includegraphics[width=0.75\textwidth]{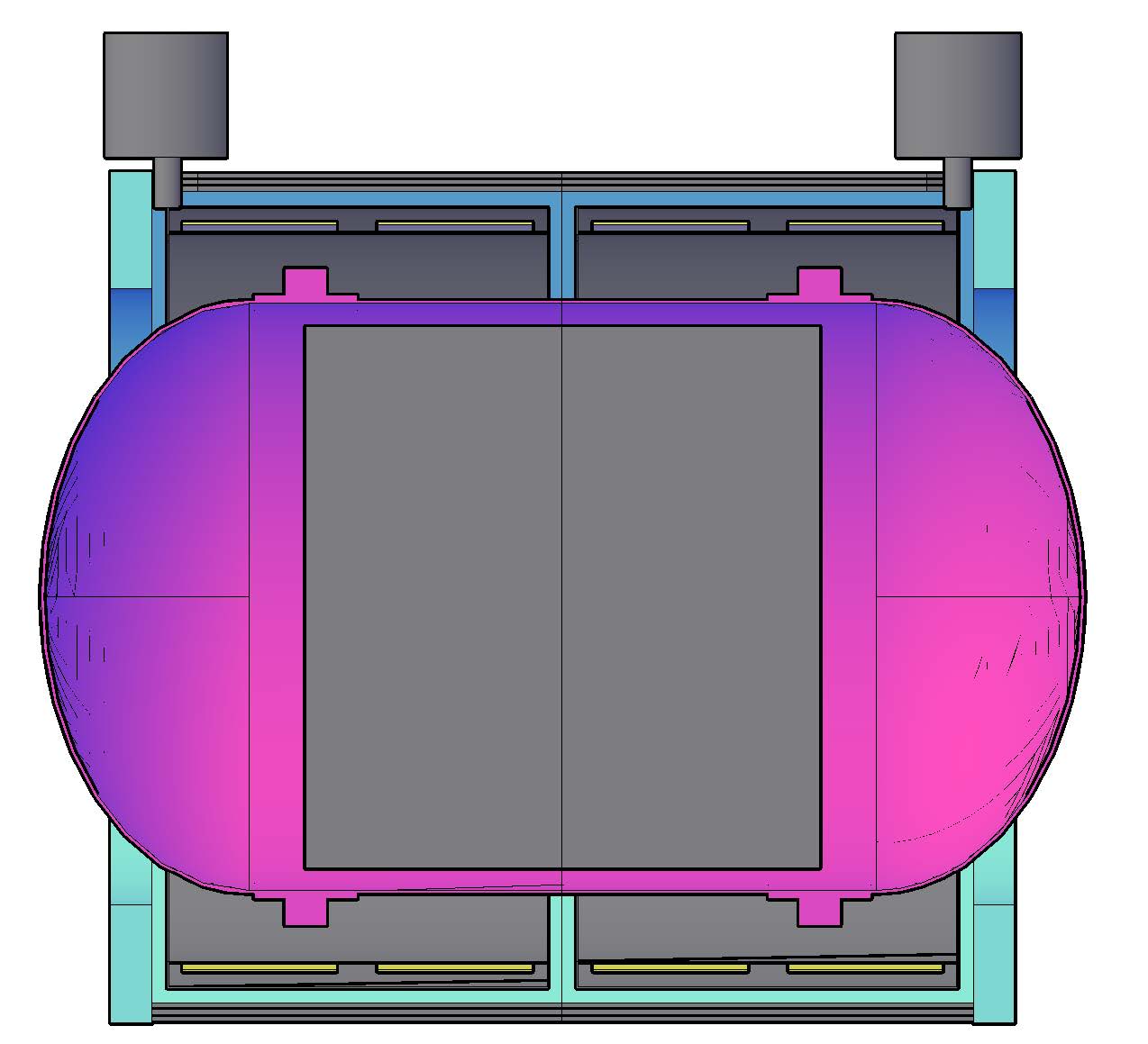}
\end{dunefigure}
The driving concept for this design is to produce a solenoid that is as thin as possible, whose
axis will be perpendicular to the neutrino beam and which has an iron
distribution in the return yoke that minimises the material between
the \dword{ndlar} and the active elements of \dword{ndgar}. This design is
referred to as the Solenoid with Partial Yoke,
SPY.  The need for a return yoke arises when we must
cope with the stray field interacting with the surrounding magnetic
material, in particular with the iron yoke of the SAND magnet. In addition, it
functions as the absorber material in the muon tagging system.
The requirements for the ND-GAr magnet are summarized in
Table~\ref{tab:mpd:magnetreq}: in addition, the inner bore diameter is intended to
host the TPC pressure vessel, whose external diameter is $\sim
5.8\,\,\mathrm{m}$, surrounded by a $4\pi$ calorimeter, $\sim
0.6\,\,\mathrm{m}$ thick. Finally, the allowance for the whole \dword{ndgar}
system, along the neutrino beam direction, is $\sim 8.8\,\,\mathrm{m}$,
between the LArTPC and the wall of the \dword{nd} hall.

\begin{dunetable}[ND-GAr magnet main characteristics and requirements.] {|l|r|c|}
{tab:mpd:magnetreq}{Requirements and characteristics of the ND-GAr reference magnet design.}
\textbf{Parameter} & \textbf{Value} & \textbf{Unit} \\ \toprowrule
Central field & 0.5 & T\\ 
Field uniformity & $\pm20$ & \%\\ 
Inner diameter & $> 7$ & m\\ 
Weight & $\sim 800$ & t\\
Material budget along particle path & $< 0.5$ & $\lambda$ ($\simeq 50$ MeV)\\
\end{dunetable}

The required magnetic field must be perpendicular to the particle path
due to the TPC principle of functioning and directed horizontally due to
the dimensions of the pressure vessel. The pressure vessel end caps,
with their elliptical shape, drive the design of the iron yoke end caps:
to limit the mass of the yoke and the length of the solenoid, SPY is
optimised around the pressure vessel. The value of the longitudinal
component of $B$ along the length of the TPC, for various positions in
the horizontal plane, is shown in Figure~\ref{fig:fie}.
\begin{dunefigure}[SPY field map]{fig:fie}
{Field map for
SPY along the axis of the solenoid ($x$). The field is shown for
different positions in the horizontal plane, ranging from
$z=-2\,\,\mathrm{m}$, in SAND direction, towards $z=2\,\,\mathrm{m}$, in
LArTPC direction. Noticeably there is a small asymmetry due to the
asymmetric distribution of iron in the yoke.}
\includegraphics[width=0.95\textwidth]{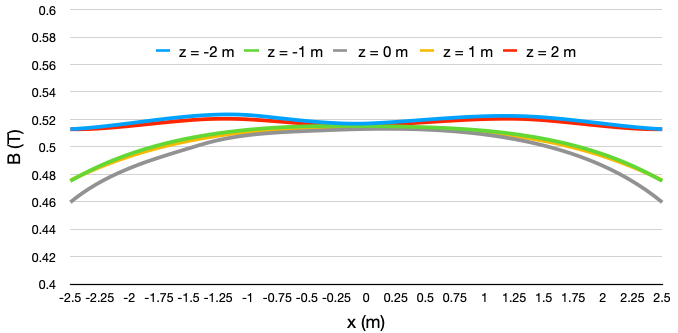}
\end{dunefigure}
The solenoid has been thought as a single layer coil, wound inside a
coil former providing the needed stiffness. The concept is based on
niobium titanium superconducting Rutherford cable stabilised in pure aluminum, as in the state-of-the-art magnets of similar size and
central field. For the design development, a current density on the order of
  $20$ to $50$~A/mm$^{2}$
can be foreseen for this kind of cable.
The overall size of the coil is some \SI{7.3}{m} in diameter and
\SI{7.5}{m} in length. It is unreasonable to foresee a continuous coil
with these dimensions. Therefore the bobbin was split in 4 segments. The
gaps between these sub-coils help to have a more uniform field in the
bore of the solenoid.  In particular, with 4 identical coils of
\SI{1500}{mm}
length each, a good field quality can be achieved.
Presently,  two identical cold masses in two identical cryostats are foreseen, each featuring
two coils.  These will be assembled independently and
powered in series. This simplifies substantially the handling of the
magnet parts, but leaves a large magnetic force
between the two cold masses. Further improvement and
optimisation is needed on this concept.

The present design includes a 16-fold segmented iron yoke, with a large
aperture to allow muons coming from LArTPC to enter the central region with minimal
degradation. The yoke will be thinner in downward direction, along the
neutrino beam axis, to reduce the asymmetric magnetic force acting on
the coils.  It will retain sufficient material to efficiently reduce
the magnetic field reaching SAND and stop pions produced in the interactions of neutrinos inside ND-GAr. The main features of this magnet are summarised in
Table~\ref{tab:mpd:magnetfeatures}.

\begin{dunetable}[Main features of SPY.] {|l|c|c|}
{tab:mpd:magnetfeatures}{Main features of the SPY ND-GAr reference magnet design.}
\textbf{Parameter} & \textbf{Value} & \textbf{Unit} \\ \toprowrule
Central field & 0.5 & T\\ 
Field uniformity & $\pm8$ & \%\\
Stored energy & 48 & MJ\\
Maximum field on cable & 1 & T\\ 
Current density on coil & 30 & A/mm$^2$ \\ 
Magnetic force between cold masses & 0.5 & MN\\ 
Magnetic force on SAND yoke & 20 & kN\\
\end{dunetable}

%

%
%
%
\subsubsection{Backup Design Overview}
In the backup magnet design, illustrated in Figure~\ref{fig:dune_mag_BO}, all five coils have the same inner radius of \SI{3.8}{m}. The center and shielding coils are identical with the same number of ampere-turns. The side coils are placed at $\pm$\SI{3}{m} along axis of the solenoid from the magnet center, while the shielding coils are at $\pm$\SI{5.5}{m}.   The magnet system will have a stored energy of about \SI{110}{MJ}, using a conventional NbTi superconducting cable design either with an Al-stabilized cable (preferred) or a \dword{ssc}-type Rutherford cable soldered in a copper channel.
\begin{dunefigure}[Magnet backup design]{fig:dune_mag_BO}
{5-coil Helmholtz concept for \dword{ndgar} superconducting magnet.}
\includegraphics[width=0.55\textwidth]{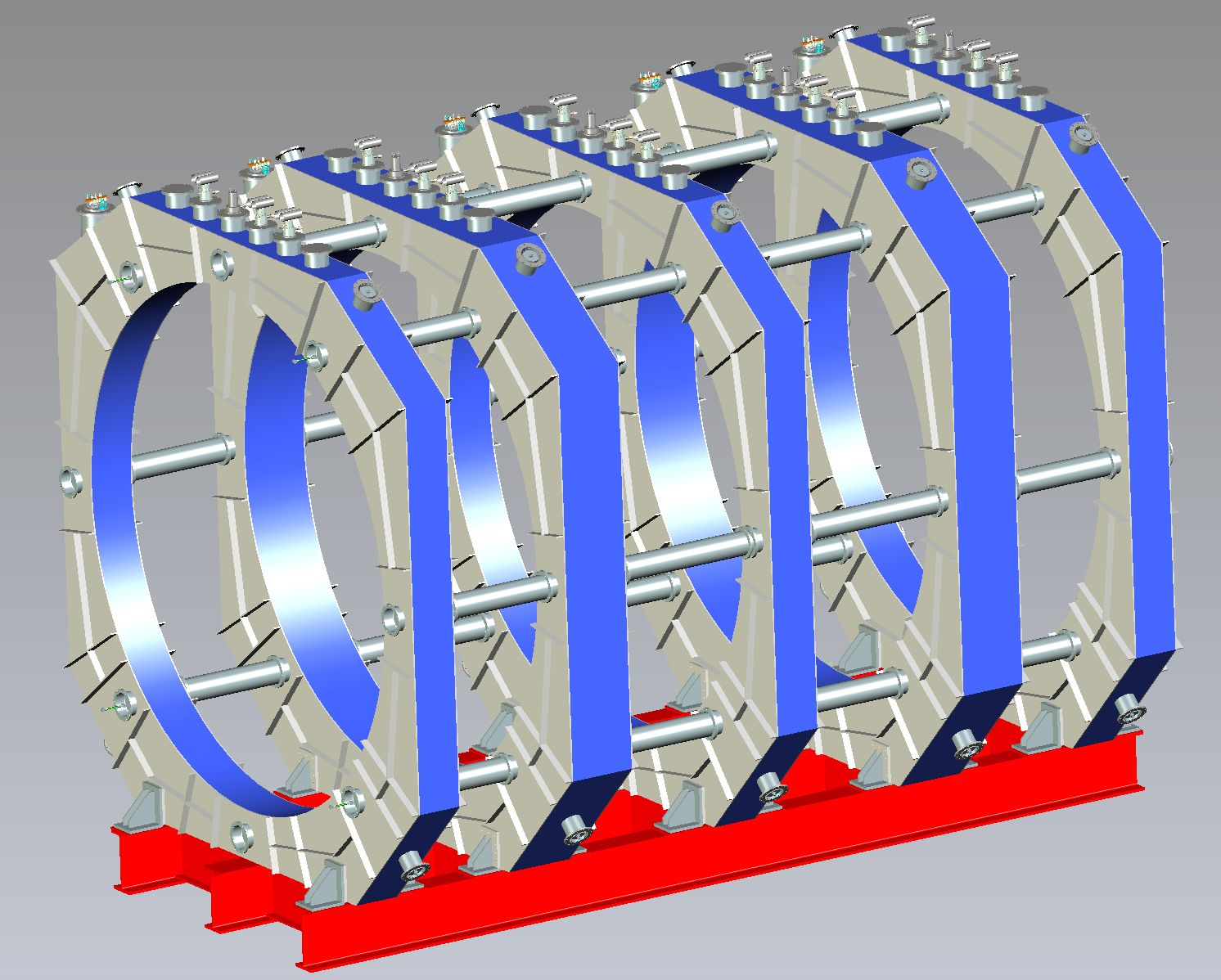}
\end{dunefigure}

Figure~\ref{fig:dune_nd_magnet_sc_fieldmap} shows the magnetic field component along the $z$-axis at different radii in the region where the \dword{hpgtpc} will be located. It is $\simeq$ \SI{0.5}{T} with $\simeq 10\%$ non-uniformities near the ends.
%
\begin{dunefigure}[Field map of the 5-coil Helmholtz magnet along the $z$ axis]{fig:dune_nd_magnet_sc_fieldmap}
{Field map of the 5-coil Helmholtz superconducting magnet along the $z$ axis. The colors represent different radii from the center line. The horizontal scale shows the position along the symmetry axis of the magnet (called z here). The vertical scale shows the component of the magnetic field in the direction of the symmetry axis.}
\includegraphics[width=0.85\textwidth]{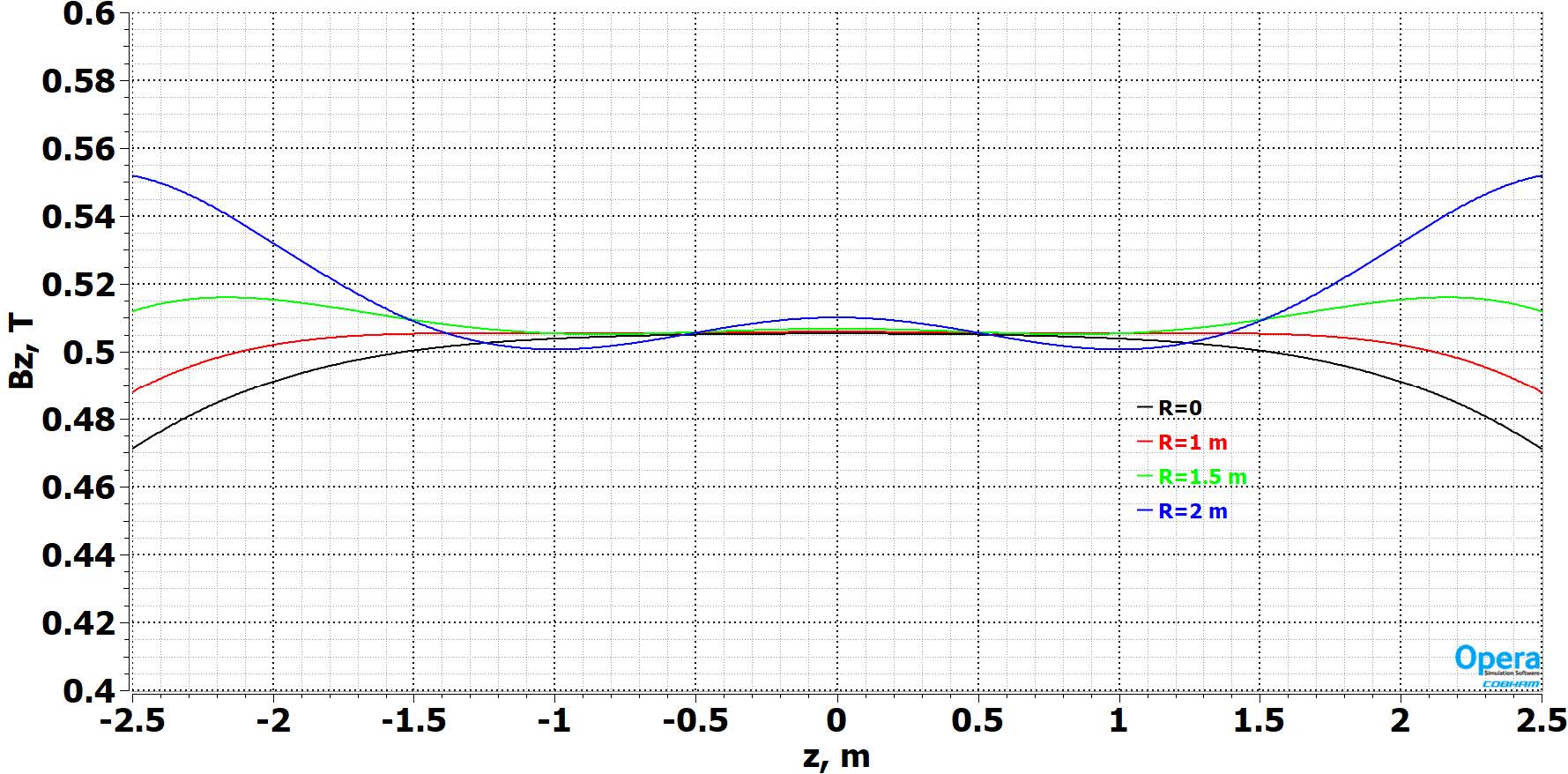} 
\end{dunefigure}
\subsection{Muon System}

The \dword{ndgar} muon system is in a very preliminary stage of design. Its design depends on the particulars of both the \dword{ecal} and magnet systems. The principal role of the muon system is to provide efficient particle identification in order to separate muons and pions punching through the \dword{ecal}. The muon system is absolutely crucial for determining the frequency of wrong-sign interactions, such as $\anumu$ in the \dword{fhc} beam.  This wrong-sign component is small and must be separated from other events like \dword{nc} events with a charged pion.

The preliminary design of the muon system consists of a very coarse longitudinal sampling structure of \SI{10}{cm} iron slabs alternating with few-centimeter-thick layers of scintillator plastic. A minimum of 3 layers is required. This corresponds to a thickness (\dword{ecal} + muon system) of about 3~$\lambda_{\pi}$ which assures that 95\% of the pions will interact. The transverse granularity of the muon detector is still under study. The angular coverage of the muon detector, which is most important in the downstream side of \dword{ndgar}, is also still under study.

%

\section{Expected Performance}
\label{sec:mpd:performance}

\subsection{Event Rates}
\label{sec:mpd:eventrates}

%
%
%


The active volume of the \dword{hpgtpc} is a cylinder with a radius of \SI{260}{\cm} and a length of \SI{500}{\cm}. For the purposes of computing event rates, a fiducial volume is defined by excluding the outer \SI{37}{\cm} of radius and by excluding \SI{30}{\cm} on each end of the cylinder. The resulting fiducial mass is then \SI{1.0}{tons} of argon. The adopted fiducial volume is large enough to find vertices and tracks but for events close to the boundary on the sides, and particularly the downstream edge, the charged particle momentum resolution will be poor, as discussed in Section~\ref{sec:mpd:reco}. Additionally, particle identification will become more difficult due to the shorter track lengths. This will make it necessary for some analyses to make tighter fiducial volume requirements that will result in lower rates. On the other hand, it is unlikely the fiducial volume for many analyses will be cylindrical. For example, the upstream radial requirement can likely be relaxed, increasing the number of events with long, high resolution, tracks. Additionally, energy reconstruction can be done using the calorimeter.

Table~\ref{tab:mpd:eventrates} shows the event rates, assuming a \SI{1.0}{ton} fiducial mass and on-axis running for one nominal year, defined as an exposure of \num{1.1e21} \dword{pot} with a proton beam momentum of \SI{120}{GeV/c}. A total of \num{1.6e6} \numu-CC events per ton per year are expected from the \dword{fhc} beam and a total of \num{5.3e5} \anumu-CC events per ton per year are expected from the \dword{rhc} beam. Table~\ref{tab:mpd:eventrates} also shows the yields of various subprocesses referenced in this chapter. These event rates should not be confused with voxel occupancy of the gas \dword{tpc}, which is quite low. Taking a voxel length to be $\pm3$ times the longitudinal diffusion coefficient of P10 gas for the full \SI{2.5}{m} drift distance, and a voxel area as determined by a pad in the readout chambers, only about 0.03\% of the voxels will have activity during a \SI{10}{microsecond} spill.


It is reasonable to ask if these event yields are large enough to permit detailed physics analyses. An answer can be found by comparing these rates to the ones seen in \dword{minerva}.  \dword{minerva} collected data in the low energy and medium energy \dword{numi} beam configurations. Typical \numu-CC analyses required a fiducial volume in the scintillator tracker as well as the requirement that the muon was measured by the \dword{minos} near detector, located downstream of \dword{minerva}. The near detector acted as a muon spectrometer. Typically events were also required to have $2 <E_\nu < \SI{20}{GeV}$ to select an energy range where the neutrino flux was best known. This resulted in \num{2e5} \numu-CC fiducial events in the low energy \dword{fhc} beam mode. The much larger medium energy exposure resulted in \num{3.5e6} events in the \dword{fhc} mode. Selected event rates for the \dword{rhc} beam are about a factor of two smaller than for the \dword{fhc} beam for equal \dword{pot} exposures. There are also event samples on lead, iron, carbon and water targets that have been analyzed.

The event yields in the \dword{hpgtpc} will be significantly larger than the yields used by the \dword{minerva} low energy beam analyses. They will likely be comparable to the \dword{minerva} yields from the medium energy beam configuration, depending on the amount of on-axis vs. off-axis running and the length of the experiment. To date, \dword{minerva} has 31 cross-section papers using data from the low energy beam and is beginning to publish papers using medium energy beam data. The total number of \dword{minerva} papers is likely to be person-power limited. All the papers feature differential cross-sections, sometimes in multiple kinematic dimensions, and most feature hadrons in the final state. In \dword{minerva}, hadrons often interact in the detector. Those interactions confuse particle identification algorithms and lower the selection efficiency for analyses of exclusive final states. The \dword{hpgtpc} will have a higher efficiency for reconstructing and selecting exclusive event samples with hadrons, due to its much lower density and better tracking resolution and PID performance.  Based on the comparison with \dword{minerva}, the event yield expected in the \dword{hpgtpc} appears to be large enough to enable detailed physics analyses.


\begin{dunetable}[Event yields in \dshort{ndgar}]
{|l|c|l|c|}
{tab:mpd:eventrates}{Expected event yields in the \dword{hpgtpc} of \dword{ndgar}. The rates assume one year of running and a \SI{1}{ton} fiducial mass as described in the text.}
FHC Beam\span\omit & \textbf{RHC Beam}\span\omit \\    
\rowtitlestyle %
Process & \textbf{Events/ton/yr} & \textbf{Process} &  \textbf{Events/ton/yr} \\ \toprowrule
All \numu-CC & \num{1.64e+06}               & All \anumu-CC & \num{5.26e+05} \\ 
\hspace{1em} CC $0\pi$ & \num{5.85e+05}      & \hspace{1em}  CC $0\pi$ & \num{2.36e+05}    \\
\hspace{1em} CC $1\pi^{\pm}$& \num{4.09e+05}  & \hspace{1em}  CC $1\pi^{\pm}$ & \num{1.51e+05} \\
\hspace{1em} CC $1\pi^0$ & \num{1.61e+05}    & \hspace{1em}  CC $1\pi^0$ & \num{4.77e+04}   \\
\hspace{1em} CC $2\pi$ & \num{2.10e+05}      & \hspace{1em}  CC $2\pi$ & \num{5.21e+04}    \\
\hspace{1em} CC $3\pi$ & \num{9.28e+04}      & \hspace{1em}  CC $3\pi$ & \num{1.66e+04}    \\
\hspace{1em} CC $K_s$ & \num{1.20e+04}       & \hspace{1em}  CC $K_s$ & \num{2.72e+03} \\
\hspace{1em} CC $K^\pm$ & \num{4.57e+04}     & \hspace{1em}  CC $K^\pm$ & \num{4.19e+03} \\
\hspace{1em} CC other & \num{1.27e+05}       & \hspace{1em}  CC other & \num{1.62e+04}  \\
\hline
All \anumu-CC & \num{7.16e+04}              & All \numu-CC & \num{2.72e+05} \\
All NC & \num{5.52e+05}                     & All NC & \num{3.05e+05} \\
All \nue-CC & \num{2.85e+04}                & All \nue-CC & \num{1.84e+04}   \\
$\nu e \to \nu e$ & \num{170}               &  $\nu e \to \nu e$ & \num{120}          \\
\end{dunetable}

\subsection{Essential \dshort{ndgar} performance metrics}
The expected performance of \dword{ndgar} is summarized in Table~\ref{tab:TPCperformance}. Details of the \dword{hpgtpc} performance are based upon studies presented in this chapter and experience from operation of the PEP-4~\cite{PEP4_results_Layter,PEP4_Stork,Madaras:1982cj} and ALICE~\cite{Alessandro:2006yt} time projection chambers. Performance of the \dword{ecal} is based on the studies reported in Section~\ref{sec:mpd:ecal} and on experience from the operation of similar \dwords{ecal}. 

\begin{dunetable}[\dshort{ndgar} performance parameters]
  {l|p{0.3\textwidth}|p{0.3\textwidth}}
  {tab:TPCperformance}{Expected \dword{ndgar} performance according to the studies reported in this chapter and also extrapolated from \dshort{alice} and \dshort{pep4} (marked with an *). Here $\perp$ and $\parallel$ refer to the directions perpendicular and parallel to the drift direction. The momentum and angular resolutions were estimated using reconstructed \numu~CC events generated with the LBNF flux. That study is described in Section~\ref{sec:mpd:muresolutions}. The proton energy threshold study is described in Section~\ref{sec:mpd:protons}. The ECAL performance is described in Section~\ref{sec:mpd:ecal}.}
Parameter & Value & Comments \\ \toprowrule
Single hit resolution $\sigma_\perp$ & 250 $\mu$m &  *$\perp$ to TPC drift direction\\ \colhline  
Single hit resolution  $\sigma_\parallel$ & 1500 $\mu$m  &  *$\parallel$ to TPC drift direction \\ \colhline
Two-track separation & \SI{1}{cm} & * \\ \colhline
$\sigma$($dE/dx$)  & 5\% & * \\ \colhline
$\mu$ reconstruction: $\sigma_p/p$ & (2.9\%, 14\%) & (core, tails), \numu~CC events, LBNF flux \\ \colhline
$\mu$ $\sigma_p/p$ vs. track length & (10\%, 4\%, 3\%) & (core),(1,2,\SI{3}{m}), \numu~CC events, LBNF flux \\ \colhline
Angular resolution & \SI{0.8}{\degree} & \numu~CC events, LBNF flux \\ \colhline
Energy scale uncertainty  & $\lesssim$ 1\% & *(by spectrometry) \\ \colhline
Proton detection threshold & \SI{5}{MeV} & kinetic energy\\ \colhline
ECAL energy resolution & 6\%/$\sqrt{E(\si{\GeV})}  \oplus $ & \\ 
 & $1.6\%/{E(\si{GeV})} \oplus 4\%$ & \\ \colhline
ECAL pointing resolution & \SI{10}{\degree} at \SI{500}{\MeV} & \\
\end{dunetable}

\subsection{Kinematic acceptance for muons}
\label{sec:mpd:larcoverage}

\begin{dunefigure}
  [Acceptance in muon angle and energy of \dword{ndlar} and  \dword{ndgar}]
  {fig:LArMPD_muons}
  {Acceptance of the reference \dword{ndgar} design (\SI{500}{cm} diameter, \SI{500}{cm} in length) for muons created in neutrino interactions in the upstream \dword{ndlar}.}
  \includegraphics[trim= 10 0 0 70, clip, width=\textwidth]{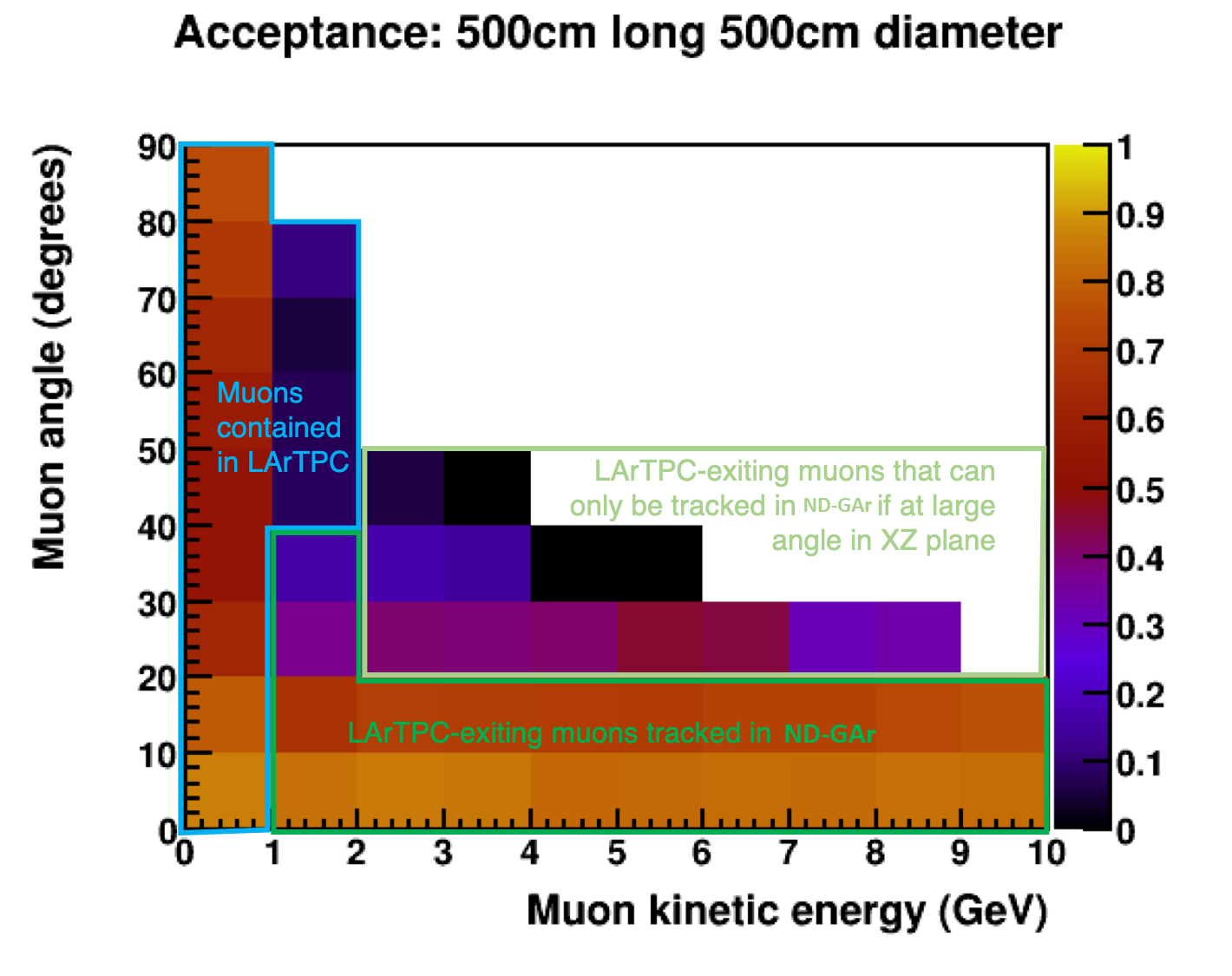}
\end{dunefigure}

\begin{dunefigure}[Acceptance in $E_{\nu}$ vs. $Q^2$ of \dword{ndlar} and \dword{ndgar} compared with the \dword{fd}]{fig:MPDacceptance}
{Comparison of the acceptance of \dword{ndlar} (left),  \dword{ndgar} (middle), and the \dword{fd} (right) for the range of neutrino energy and squared momentum transfer ($Q^2$) of \numu CC interactions expected at DUNE. \dword{ndgar} and the \dword{fd} are well-matched in their acceptance of events across the range of phase space.}
    \includegraphics[width=0.32\textwidth]{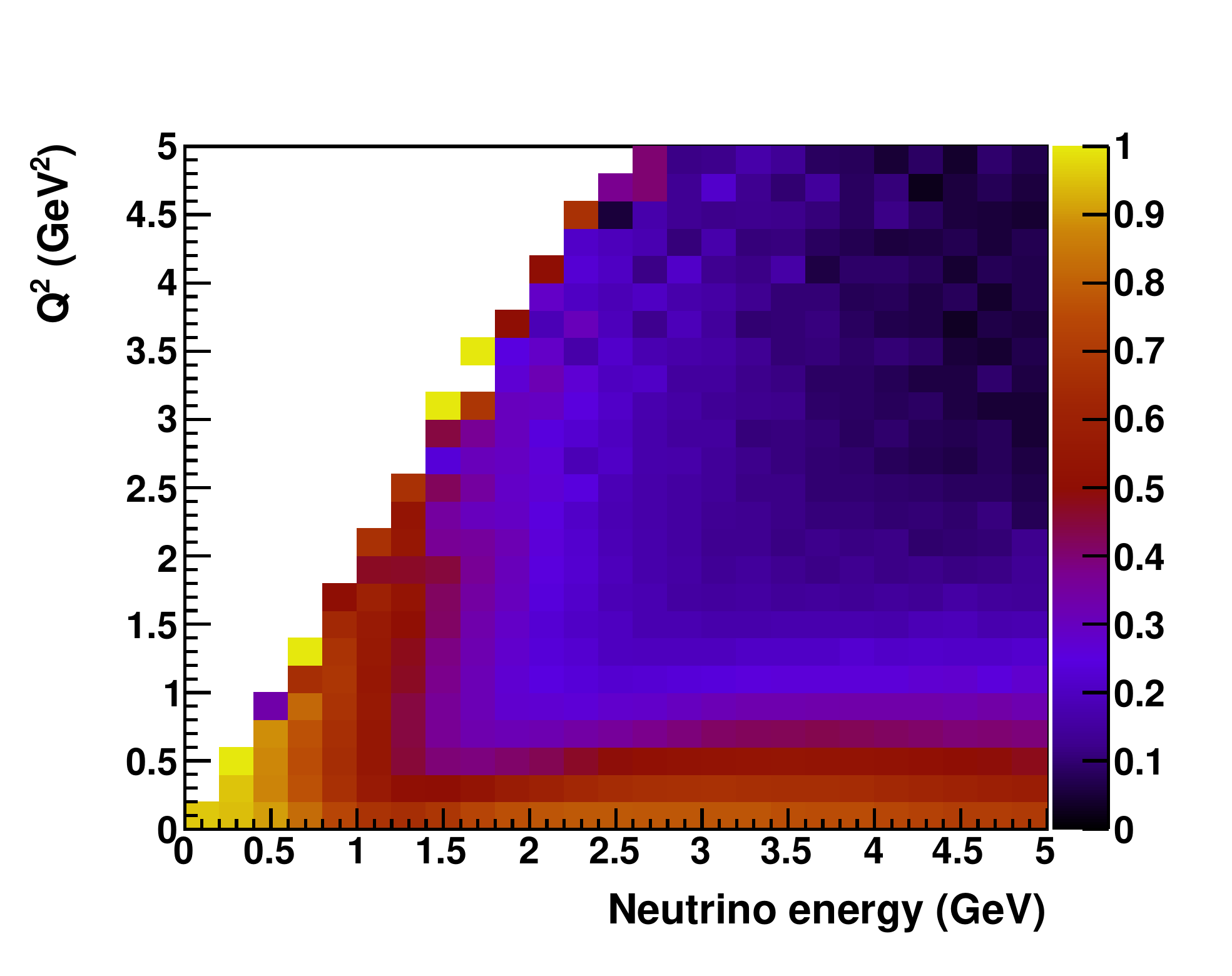}
    \includegraphics[width=0.32\textwidth]{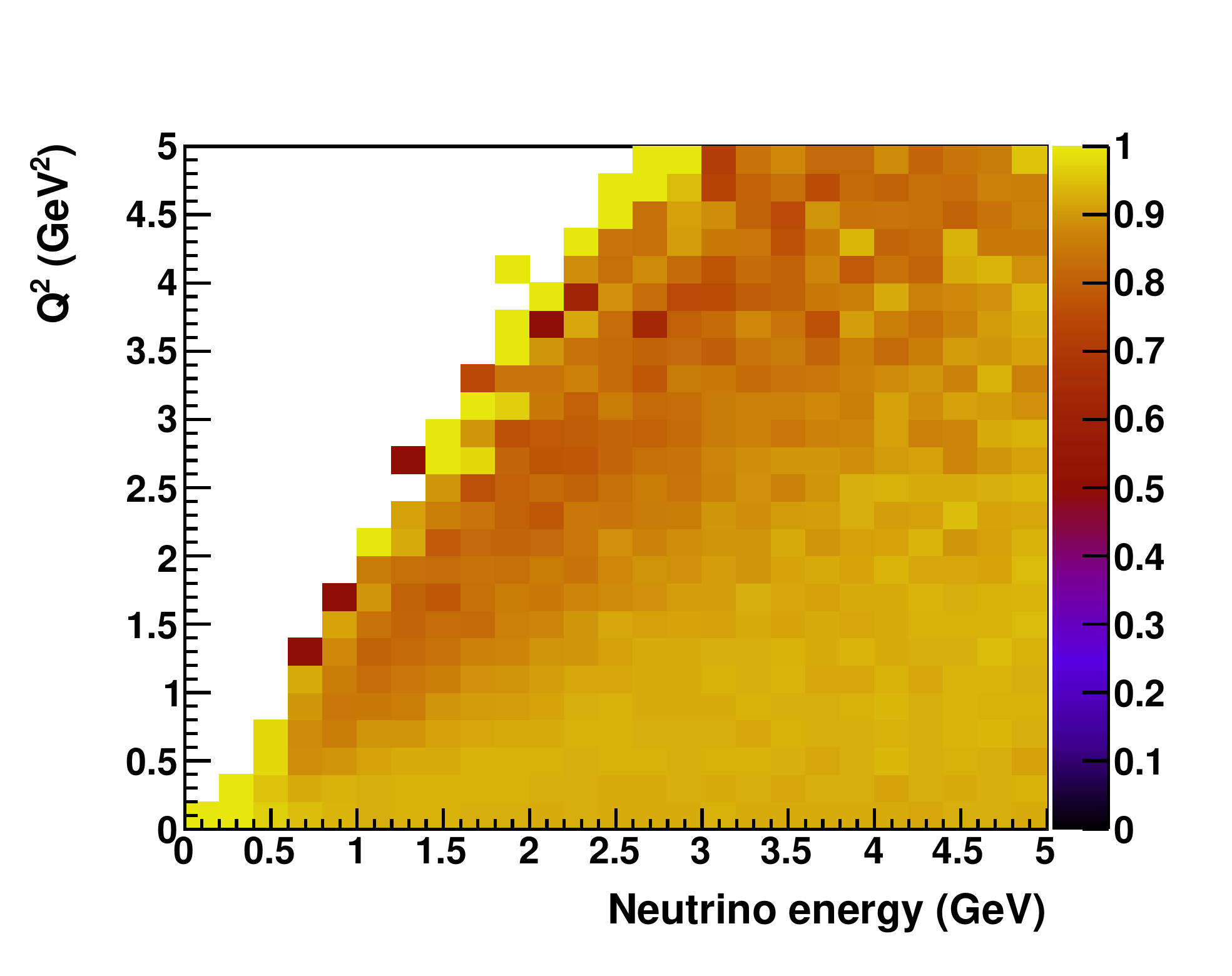}
    \includegraphics[width=0.32\textwidth]{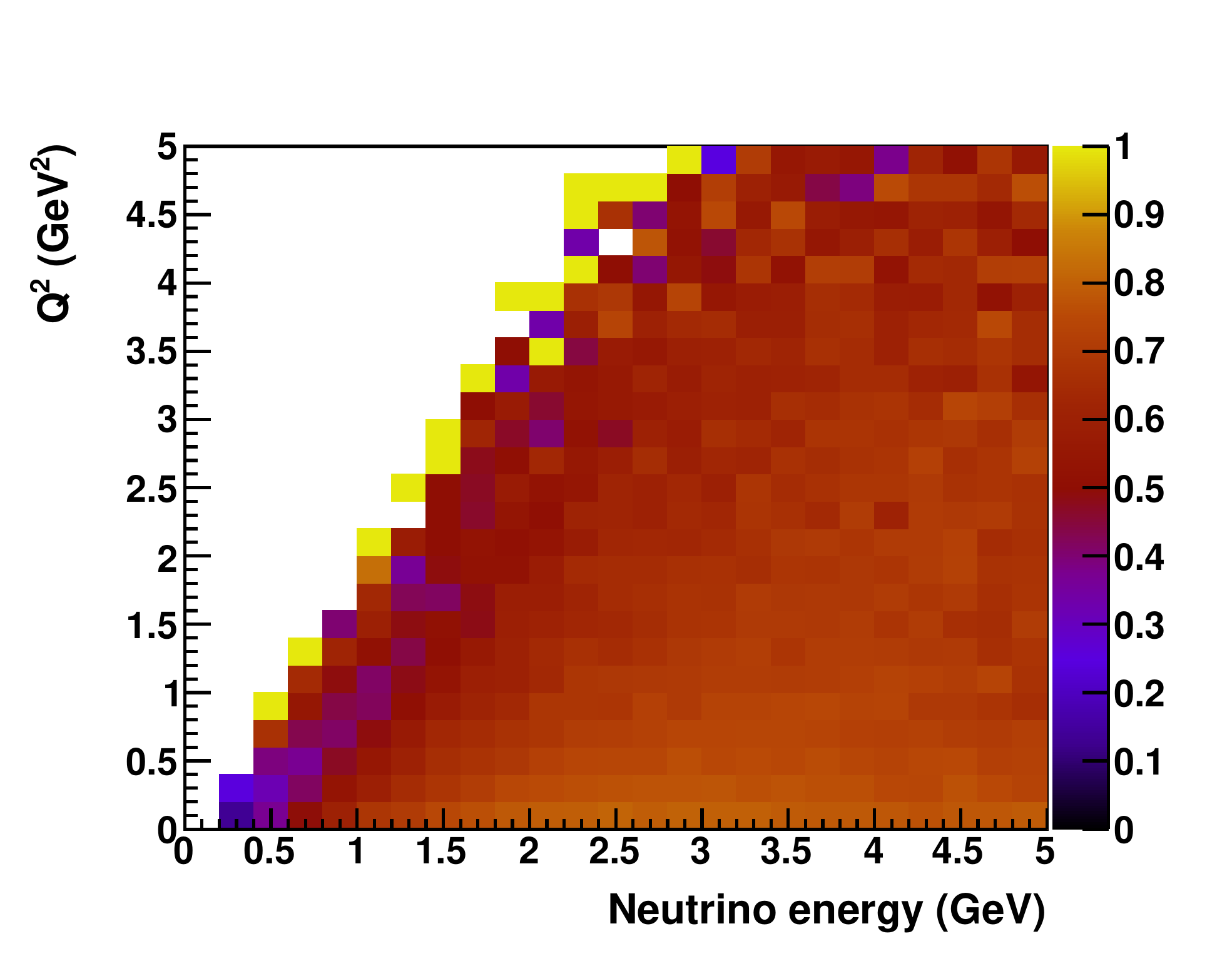}
\end{dunefigure}

\paragraph{Kinematic acceptance of muons produced by \dword{ndlar}:}

\dword{ndlar} will not fully contain high-energy muons or measure lepton charge. The downstream \dword{ndgar} will be able to determine the charge sign and measure the momenta of the muons that enter its acceptance, using the curvature of the associated track in the magnetic field. Figure~\ref{fig:LArMPD_muons} shows the expected distribution of muon angle vs. energy for $\numu$ interactions at the near site. Events with muon kinetic energies below \SI{1}{GeV} are well contained within \dword{ndlar}\footnote{The containment varies by energy within the \SI{1}{GeV} bin. Muon containment in \dword{ndlar} drops quickly above \SI{600}{MeV}, as shown in Fig.~\ref{fig:LArMPD_muons}.}, while events with higher energy muons traveling within \SI{20}{degrees} of the beam direction will exit \dword{ndlar} and enter \dword{ndgar}.

\paragraph{Acceptance comparisons with the \dshort{fd}:}

Figure~\ref{fig:MPDacceptance} compares the muon acceptance for \numu CC interactions  in \dword{ndlar} (aided by \dword{ndgar}, acting as a muon spectrometer), interactions in \dword{ndgar}, and interactions in the \dword{fd}. In each case, the interactions are in a fiducial volume containing liquid or gaseous argon (i.e., not in the \dword{ecal}, support structure, etc). Compared to  \dword{ndlar}, \dword{ndgar} has an acceptance that is much more uniform across the kinematic phase space and much more similar to the \dword{fd}.

\subsection{Magnetic Field Calibration} \label{sec:mpd-performance:calib}

Detector calibrations are a critical aspect of a high performance detector. For \dword{ndgar}, calibration strategies will build upon experience from long-term operations of similar detectors. In particular, careful attention will be paid to calibration of the magnetic and electric fields, the detector environment (temperature, drift velocity, etc.), and the overall energy scale. 

Operation of TPCs in a magnetic field requires either an extremely homogeneous magnetic field precisely aligned with the electric drift field, or precise knowledge of the magnitude and orientation of the magnetic field. The latter requirement will apply to the magnet configuration of \dword{ndgar}.  This was also the case for NA49~\cite{NA49}, where two independent methods for the precise determination of the magnetic field map were adopted: 
\begin{itemize}
  \item {Based on the known configuration and material of the iron yokes and coils, the magnetic field was calculated with TOSCA code.} 
  \item {Detailed field measurements by means of Hall probes on a three-dimensional grid, with $4\times4\times4$ \SI{}{\cubic\cm} spacing were performed.}
\end{itemize}
A comparison of the calculated field map with the measurements allowed a cross check of the TOSCA calculations and of the correction and calibration procedures applied to the measurements. The field maps obtained with the two methods agree within 0.5\%. \dword{ndgar} is expected to achieve a similar level of uncertainty. The field will need to be determined for each position of \dword{ndgar}\footnote{i.e., as the detectors move in the PRISM measurement program.}, as the true field will differ depending on the distance between \dword{ndgar} and \dshort{sand}. A study of the effect of the \dshort{sand} magnet steel on the central field in \dword{ndgar} was undertaken, and found to be \SI{10}{G}, or 0.2\%, with no correction (using {\it in situ} probes and modeling).  The estimate for the overall uncertainty in  \dword{ndgar} central field (due to ferrous material) is $<0.05$\% after corrections, for all positions of \dword{ndgar}.

\subsubsection{Calibration With Neutral Kaons}
\label{ch:mpd:kshorts}

Approximately 2.6\% of events in \dword{ndgar} are expected to have a $K^0_s$ (with a 69.2\% branching fraction to decay to a $\pi^+ \pi^-$ pair).  For a cylindrical fiducial volume \SI{4}{m} in length and \SI{1}{m} in radius, this results in a yield of about 3000 $K^0_s \to \pi^+ \pi^-$ per year of on-axis running. A study to investigate the possibility of using these $K^0_s$ decays as an energy calibration source in \dword{ndgar} has been done. Single $K^0_s$ events were generated in the \dword{hpgtpc} and GArSoft was used to reconstruct the pion track kinematics and the decay vertex (Section~\ref{sec:mpd:reco} discusses the general performance of the reconstruction). The efficiency for reconstructing two tracks of opposite signs with a common vertex was about 33\%, a relatively low number due to the preliminary state of the reconstruction program when the study was done. However, the invariant mass peak was clearly identifiable, as seen in Figure~\ref{fig:mpd:kaon-calib}. As a demonstration of the sensitivity of this calibration technique, the same set of 2226 events are reconstructed assuming two different magnetic fields: the nominal \SI{0.5}{T} field in which the events were generated, and a 1\% biased field of \SI{0.505}{T}. As is clear in the figure, a magnetic field bias as small as 1\% (which is equivalent to a 1\% momentum bias) results in a clear shift of the mean of reconstructed invariant mass. Work is ongoing to repeat the analysis in neutrino events and determine the constraint these events provide on the track energy scale.  There will be a similar number of $\Lambda \to \pi^- p$ decays that may also provide a useful constraint.  





\begin{dunefigure}[Study of $K^0_s$ in  \dword{ndgar}]{fig:mpd:kaon-calib}
  {The invariant mass distribution for reconstructed $K^0_s \to \pi^+ \pi^-$ decays. The $K^0_s$ were generated as single particle events inside the \dword{hpgtpc} and then reconstructed in GArSoft with an efficiency of about 33\%. The kaon mass ($m_{K0}=\SI{497.6}{MeV}$ according to the PDG) is well reproduced by the mean of the distribution, and a 1\% bias in the magnetic field (and therefore also in the momentum) results in a clear shift of the mean.}
\includegraphics[width=0.48\textwidth]{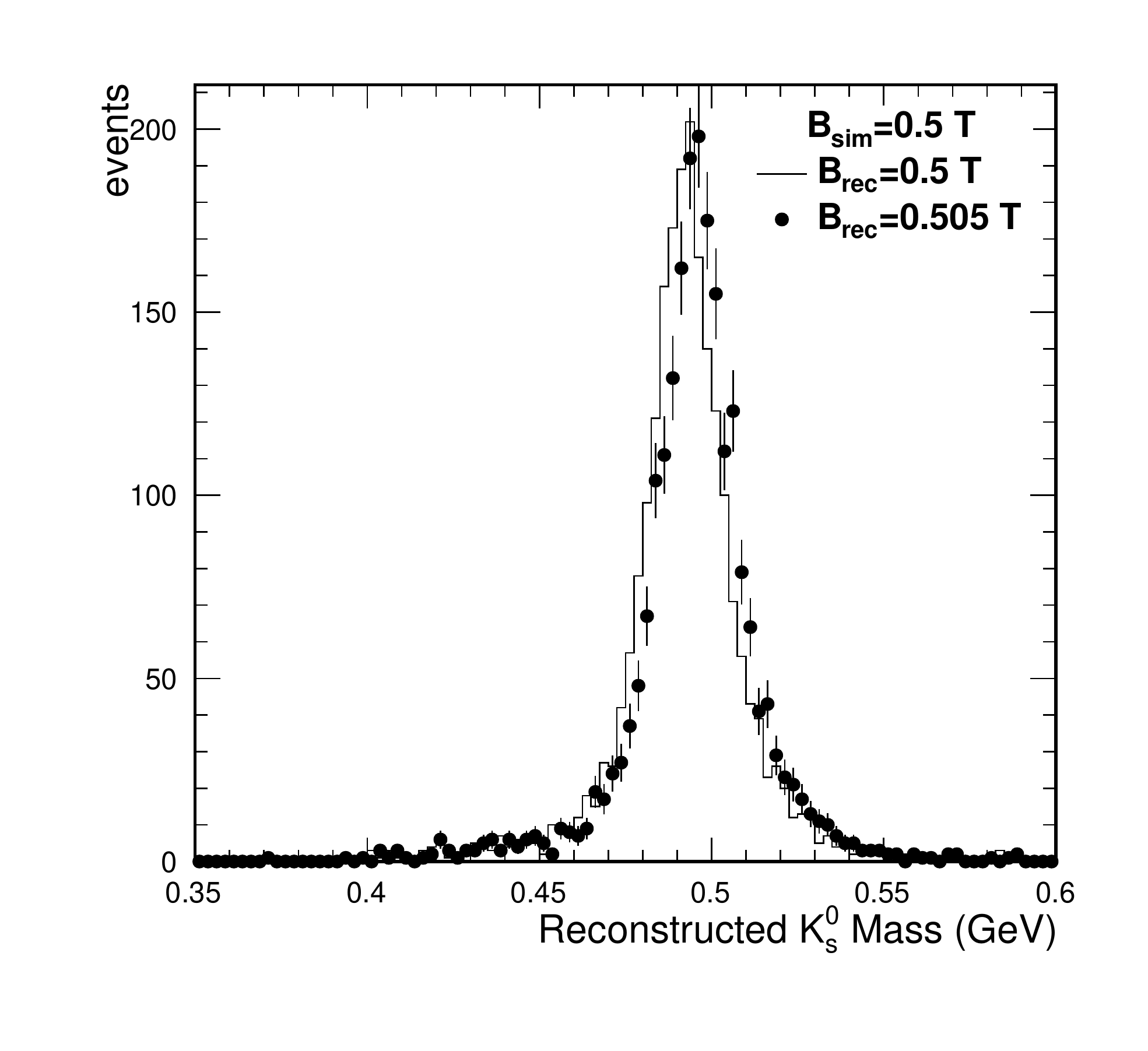}
\end{dunefigure}

\subsection{Track Reconstruction and Particle Identification} \label{sec:mpd:reco}

The combination of very high resolution magnetic analysis and superb particle identification from the \dword{hpgtpc}, coupled with a high-performance \dword{ecal} will lead to excellent event reconstruction capabilities and potent tools to use in neutrino event analysis.  
As an example of this capability, the top panel of Figure~\ref{fig:GAr} shows a $\nu_e\, \mathrm{{}^{40}\!Ar} \xrightarrow{} e^-\, \pi^+\, n\, p\ \mathrm{{}^{38}Cl}$ event in the \dword{hpgtpc} with automatically-reconstructed tracks. In the lower panel, a fully reconstructed $\nu_{\mu}\, \mathrm{{}^{40}\!Ar} \xrightarrow{} \mu^-\, p\, n\, p\, p\, p\, p\, n\, n\ \mathrm{{}^{32}Al}$. 

\begin{dunefigure}[Track-reconstructed $\nu_e$ \dword{cc} event in the \dshort{hpgtpc}]{fig:GAr}
{(Top) Track-reconstructed $\nu_e$ \dword{cc} event in the \dword{hpgtpc}, simulated and reconstructed with GArSoft.  The annotations are from \dshort{mc} truth. (Bottom) Track-reconstructed $\nu_{\mu}$ \dword{cc} event with five protons.
}
    \includegraphics[width=0.99\textwidth]{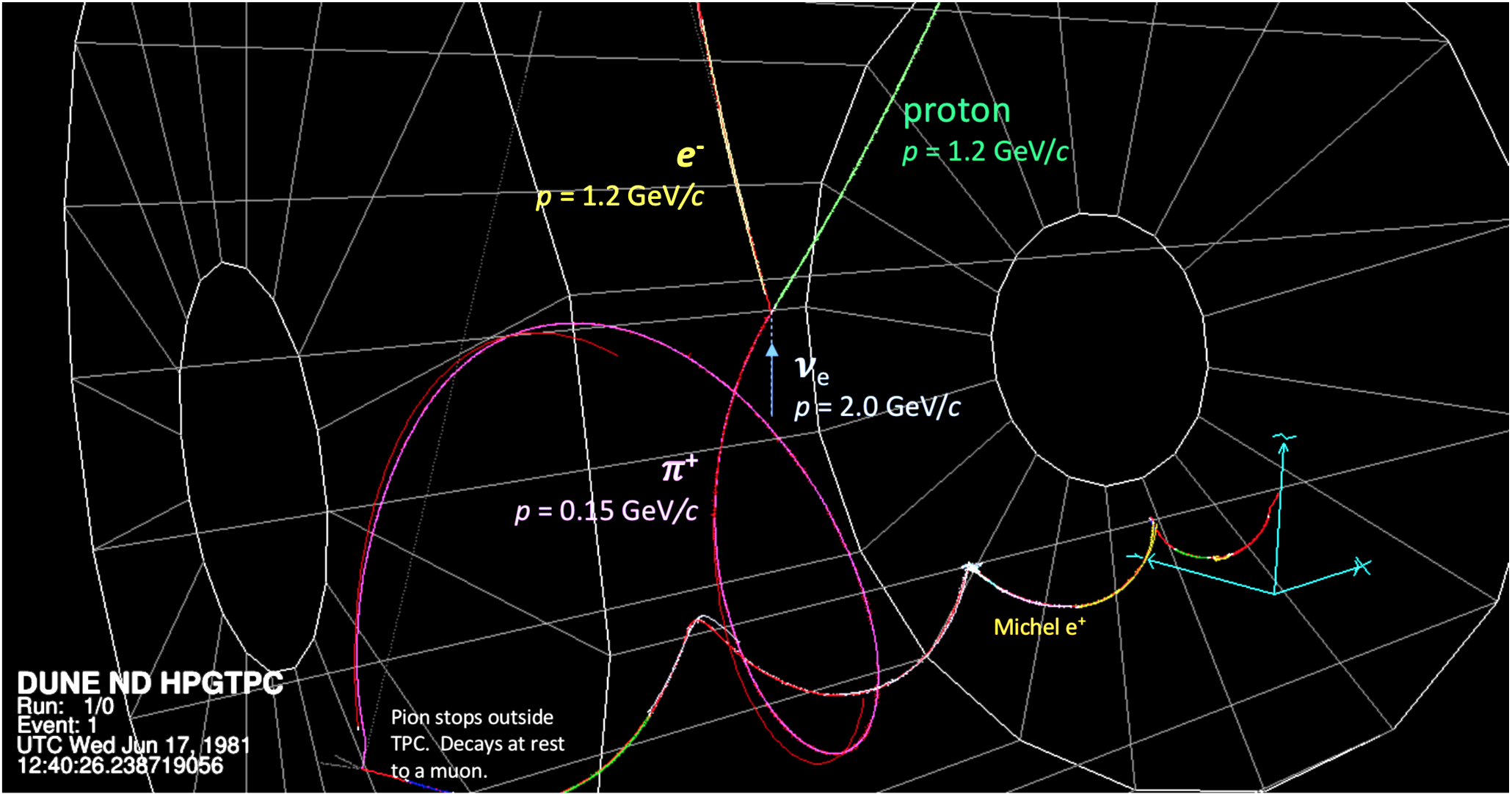}
    \includegraphics[width=0.99\textwidth]{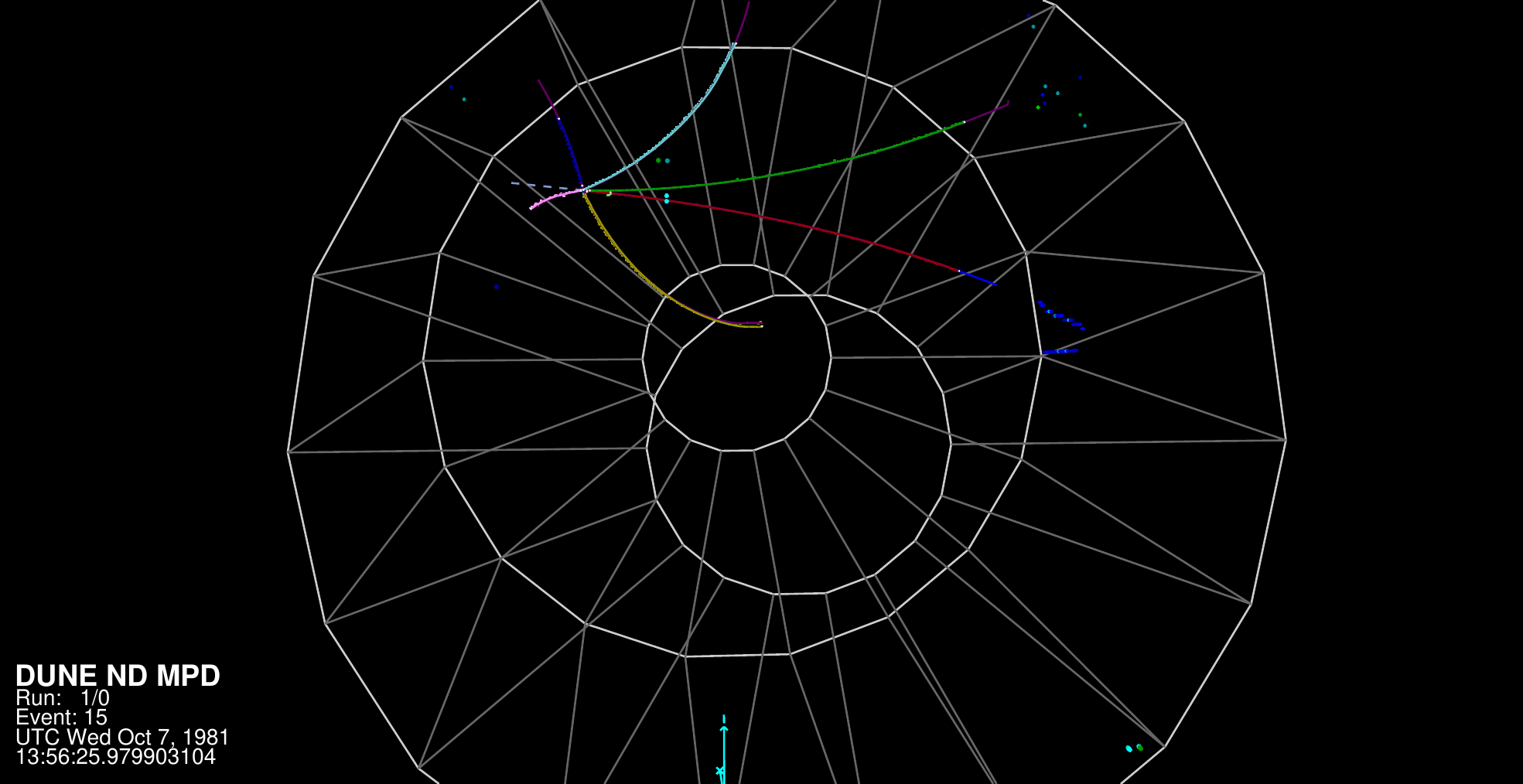}
\end{dunefigure}

Since important components of the hardware and design for the \dword{hpgtpc} are taken from or duplicated from the ALICE detector, the ALICE reconstruction is a useful point of reference in this discussion.
Track reconstruction in ALICE is achieved by combining hits recorded on the ROC pads into tracks following a trajectory that a charged particle traveled through the TPC drift volume.  The \dword{hpgtpc} is oriented so that the neutrino beam is perpendicular to the magnetic field, which is the most favorable orientation for measuring charged particles traveling along the neutrino beam direction.   

The GArSoft simulation and reconstruction package borrows heavily from packages developed for previous \dwords{lartpc}. It is based on the {\it art} event processing framework and {\tt GEANT4}.  It is designed to be able to reconstruct tracks with a full $4\pi$ acceptance.   GArSoft simulates a 10~atmosphere gaseous argon detector with readout chambers filling in the central holes in the ALICE geometry.  GArSoft's tracking efficiency has been evaluated in a large sample of \dshort{genie} $\nu_\mu$ events interacting in the TPC gas at least \SI{50}{cm} from the edges, generated using the optimized \dshort{lbnf} forward horn current beam spectra. The efficiency for reconstructing tracks associated with pions and muons as a function of track momentum $p$ is shown in Figure~\ref{fig:garsoft_efficiency}.  The efficiency is above 90\% for tracks with $p>40$~MeV/$c$, and it steadily rises with increasing momentum. Figure~\ref{fig:garsoft_efficiency} also shows the efficiency for reconstructing all charged particles with $p>\SI{200}{MeV/c}$ as a function of $\lambda$,  the track angle with respect to the center plane.

The tracking efficiency for protons is shown in Figure~\ref{fig:TEpr} as a function of kinetic energy, $T_p$.  Currently, the standard tracking works well down to $T_p \simeq \SI{20}{MeV}$. For $T_p < \SI{20}{MeV}$, a machine-learning algorithm is in development, targeting short tracks near the primary vertex. This algorithm, although currently in a very early stage of development, is already showing good performance, and efficiency improvements are expected with more development. The machine learning algorithm is described in Section~\ref{sec:mpd:protons}.


\begin{dunefigure}[The efficiency to find tracks in the HPgTPC]{fig:garsoft_efficiency}
{(Left) The efficiency to find tracks in the \dword{hpgtpc} as a function of momentum, $p$, for tracks in a sample of \dword{genie} events simulating \SI{2}{GeV} $\nu_\mu$ interactions in the gas, using GArSoft. (Right) The efficiency to find tracks as a function of $\lambda$, the angle with respect to the center plane, for tracks with $p>200\,$MeV/$c$.}
    \includegraphics[width=0.49\textwidth]{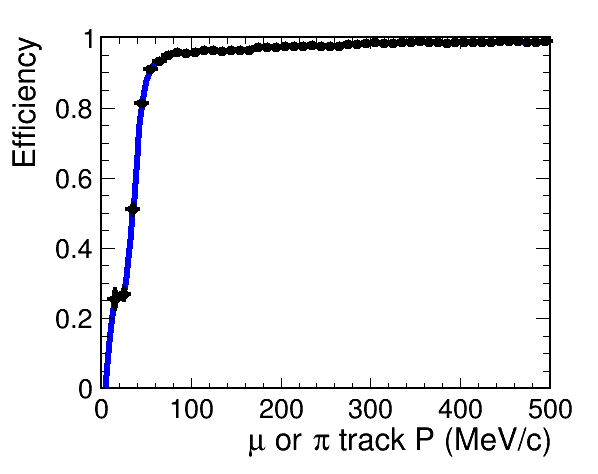}\includegraphics[width=0.49\textwidth]{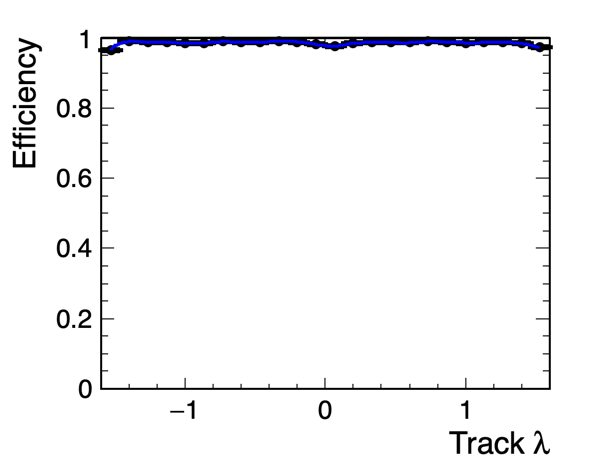}
\end{dunefigure}

\begin{dunefigure}[Tracking efficiency for protons in the HPgTPC]{fig:TEpr}{Tracking efficiency for protons in the \dword{hpgtpc} as a function of kinetic energy.} 
\includegraphics[width=0.65\columnwidth]{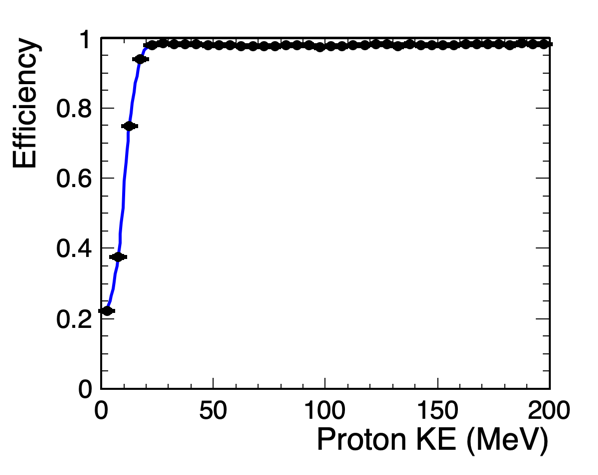} 
\end{dunefigure}

ALICE chose to use neon, rather than argon, for the primary gas in their first run; the decision was driven by a number of factors, but two-track separation capability was one of the primary motivations due to the extremely high track multiplicities in the experiment.  Neon performs better than argon in this regard.  A better comparison for the \dword{hpgtpc}'s operation in DUNE is the two-track separation that was obtained in PEP4~\cite{PEP4_Stork}.  PEP4 ran an 80-20 mixture of Ar-CH$_4$ at 8.5~atmospheres, yielding a two-track separation performance of \SI{1}{cm}.

In ALICE, the ionization produced by charged particle tracks is sampled by the TPC pad rows (there are 159 pad rows in the TPC) and a truncated mean is used for the calculation of the PID signal. Figure~\ref{fig:ALICE_dEdx} (left) shows the ionization signals of charged particle tracks in ALICE for pp collisions at $\sqrt{s} = 7$~TeV. The different characteristic bands for various particles are clearly visible and distinct at momenta below a few GeV.  When repurposing ALICE as the \dword{hpgtpc} component of \dword{ndgar},  better performance is expected for particles leaving the active volume, since the detector will be operating at higher pressure (10~atmospheres vs. the nominal ALICE 1~atmosphere operation), resulting in ten times more ionization per unit track length available for collection. Figure~\ref{fig:ALICE_dEdx} (right) shows the charged particle identification for PEP-4/9~\cite{Grupen:1999by}, a higher pressure gas TPC that operated at 8.5~atmospheres, which is very close to the reference argon gas mixture and pressure of the DUNE \dword{hpgtpc}, and is thus a better indicator of the \dshort{dune} TPC's performance.

\begin{dunefigure}[ALICE and PEP-4 $dE/dx$-based particle identification as a function of momentum]{fig:ALICE_dEdx}
{Left: ALICE TPC $dE/dx$-based particle identification as a function of momentum (from~\cite{ALICE_Lippmann}). Right: PEP-4/9 TPC (80:20 Ar-CH4, operated at 8.5~Atm, from~\cite{Grupen:1999by}) $dE/dx$-based particle identification.}
\includegraphics[width=0.49\textwidth]{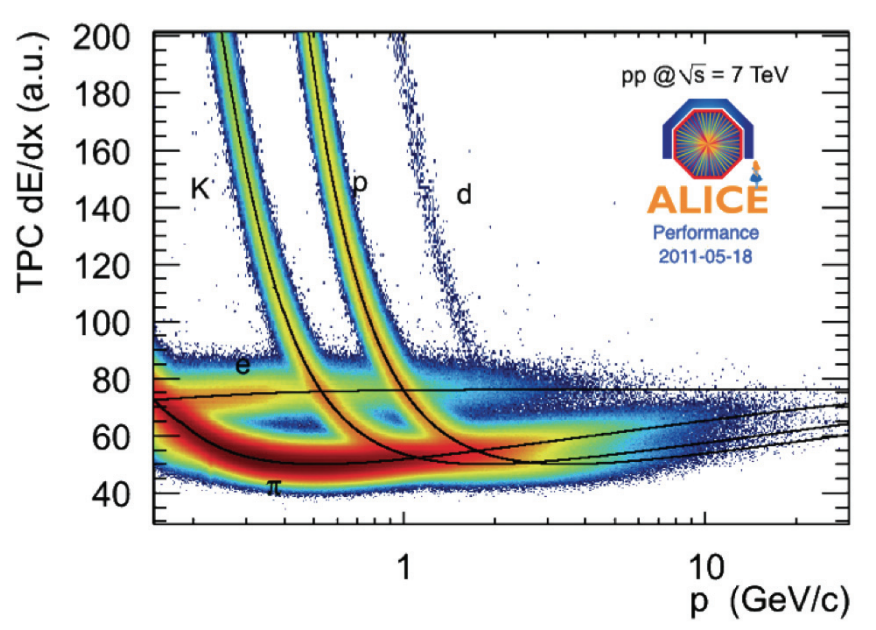}
\includegraphics[width=0.49\textwidth]{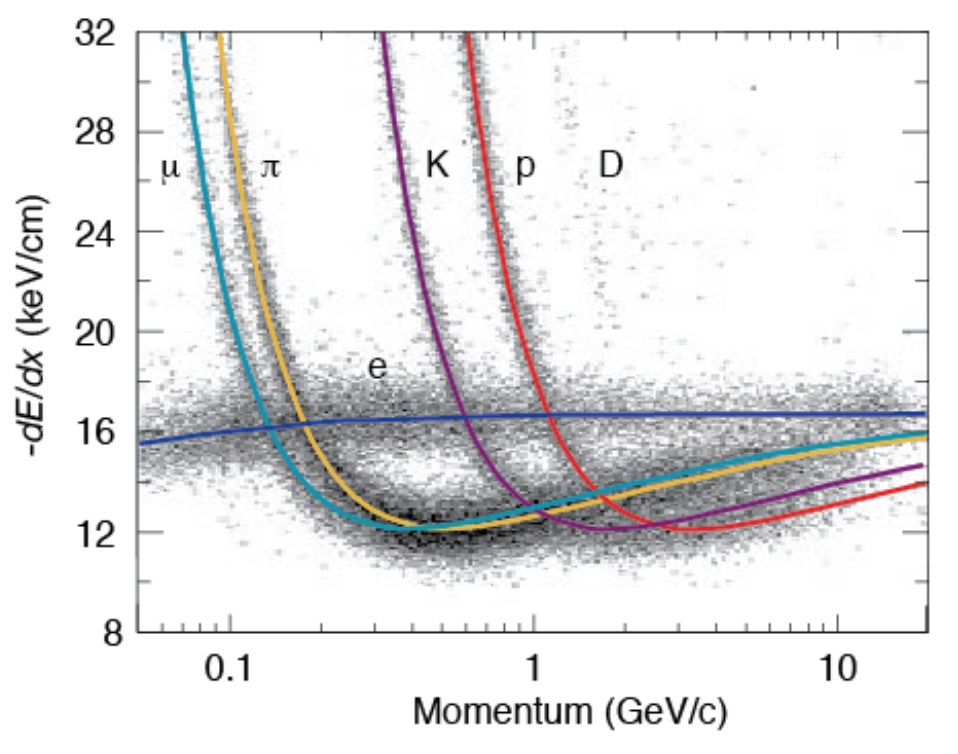} 
\end{dunefigure}

\subsubsection{Momentum and Angular Resolution}
\label{sec:mpd:muresolutions}
The ability to determine the sign of the charge of a particle in the \dword{hpgtpc} tracking volume is limited by the spatial resolution of the measured drift points in the plane perpendicular to the magnetic field, as well as multiple Coulomb scattering (MCS) in the gas. For a fixed detector configuration, the visibility of the curvature depends on the particle's momentum transverse to the magnetic field, its track length in the plane perpendicular to the field, and the number and proximity of nearby tracks.  Because primary vertices are distributed throughout the tracking volume, the distribution of the lengths of charged-particle tracks is expected to start at very short tracks, unless sufficient fiducial volume cuts are made to ensure enough active volume remains to determine particle's charge.  The kinetic energies of particles that leave short tracks and stop in the detector will be better measured from their tracks' lengths than from their curvatures.  Protons generally stop before their tracks curl around, but low-energy electrons loop many times before coming to rest in the gas.

Within the fiducial volume of the \dword{hpgtpc}, charged particles can be tracked over the full 4$\pi$ solid angle.  Even near the central electrode, tracking performance will not be degraded due to the thin (25 $\mu$m of mylar) central electrode.   Indeed, tracks crossing the cathode provide an independent measurement of the event time, since the portions of the track on either side of the cathode will only line up with a correct event time assumed when computing drift distances. The 4$\pi$ coverage is true for all charged particles.  ALICE ran with a central field of 0.5~T and their momentum resolution from $p$--Pb data~\cite{Abelev:2014ffa} is shown in Figure~\ref{fig:ALICE_MOMres}.
\begin{dunefigure}[The TPC stand-alone p$_T$ resolution in ALICE for $p$--Pb collisions]{fig:ALICE_MOMres}
{The black squares show the TPC stand-alone p$_T$ resolution in ALICE for $p$--Pb collisions. From Ref.~\cite{Abelev:2014ffa}.}
\includegraphics[width=0.65\columnwidth]{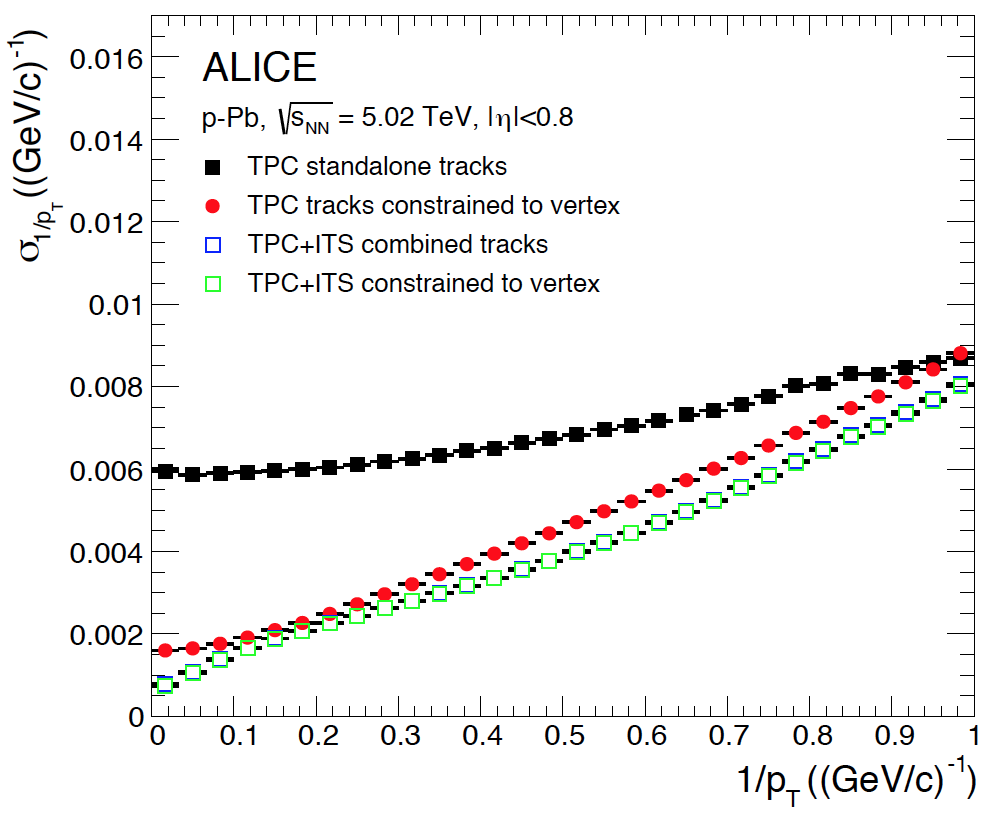}
\end{dunefigure}
%

Figure~\ref{fig:garsoftpamuonres} shows the muon momentum resolution in a sample of \numu~CC events in the \dword{hpgtpc}. The events were generated using the LBNF flux. The fiducial volume was defined by removing events with reconstructed vertices less than \SI{50}{cm} from the radial boundary of the TPC's active area, and less than \SI{30}{cm} from the two end walls. The resolution is $\Delta p/p$ = 2.7\% in the distribution's core and $\Delta p/p = 12\%$ in the tails. As Figure~\ref{fig:garsoftpamuonres} shows, the momentum resolution depends strongly on the track length, and hence the fiducial volume used in the analysis. This resolution differs from ALICE's achieved resolution due to the higher pressure, the heavier argon nucleus compared with neon, the non-centrality of muons produced throughout the detector, and the fact that the GArSoft simulation and reconstruction tools have yet to be fully optimized. The 3D angular resolution for muons in the same study is approximately 0.8~degrees.


\begin{dunefigure}[Momentum resolution for muons measured by the \dword{hpgtpc}]{fig:garsoftpamuonres}
{Top: the momentum resolution for reconstructed muons in GArSoft, in a sample of \numu~CC events. The events were generated using the LBNF flux.  The Gaussian fit to the central core of the $\Delta p/p$ distribution, containing 2/3 of events, has a width of 2.7\%. The tails are well described by a 12\% resolution.  Bottom: The momentum resolution as a function of track length.}
\includegraphics[width=0.6\columnwidth]{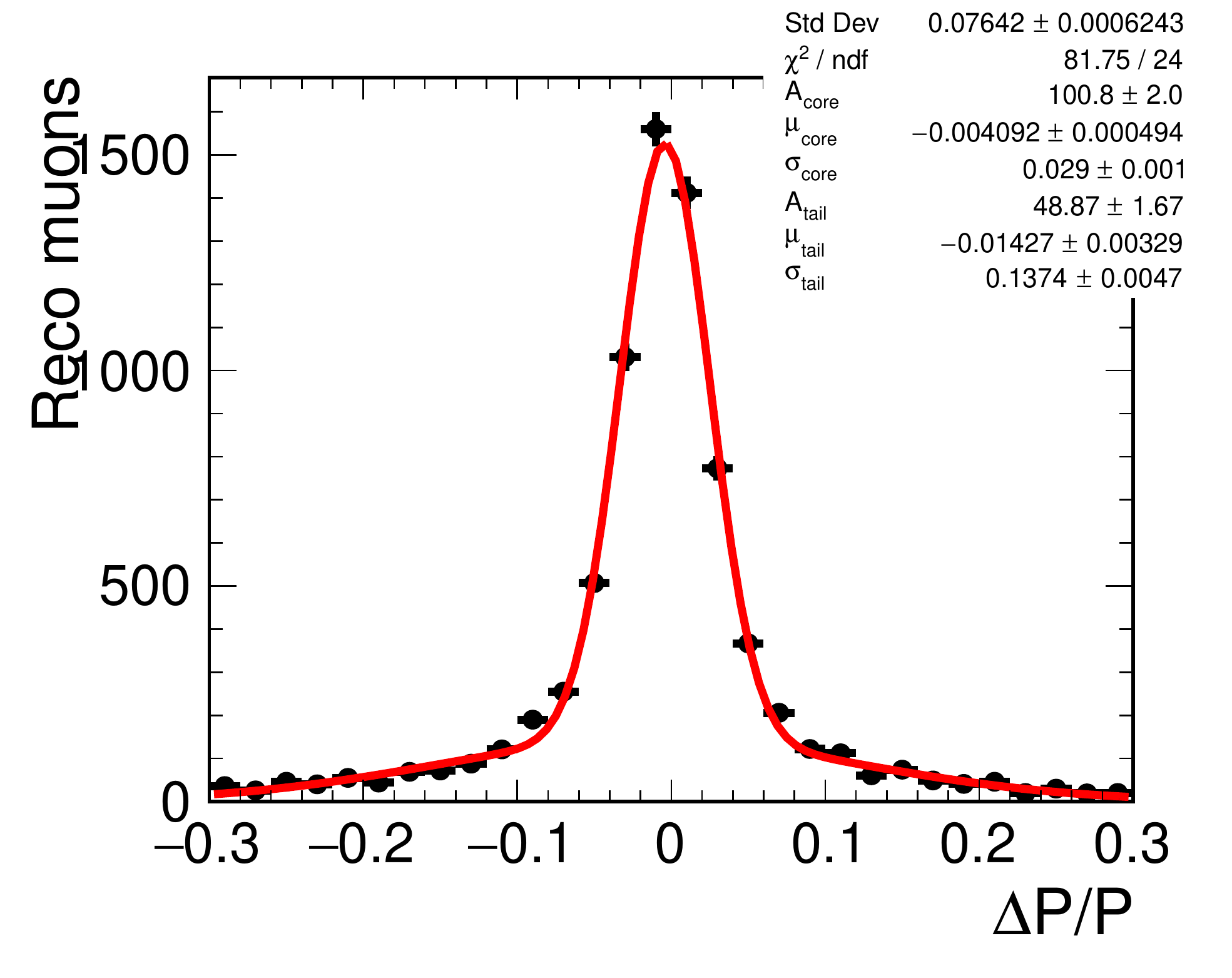}\\
\includegraphics[width=0.6\columnwidth]{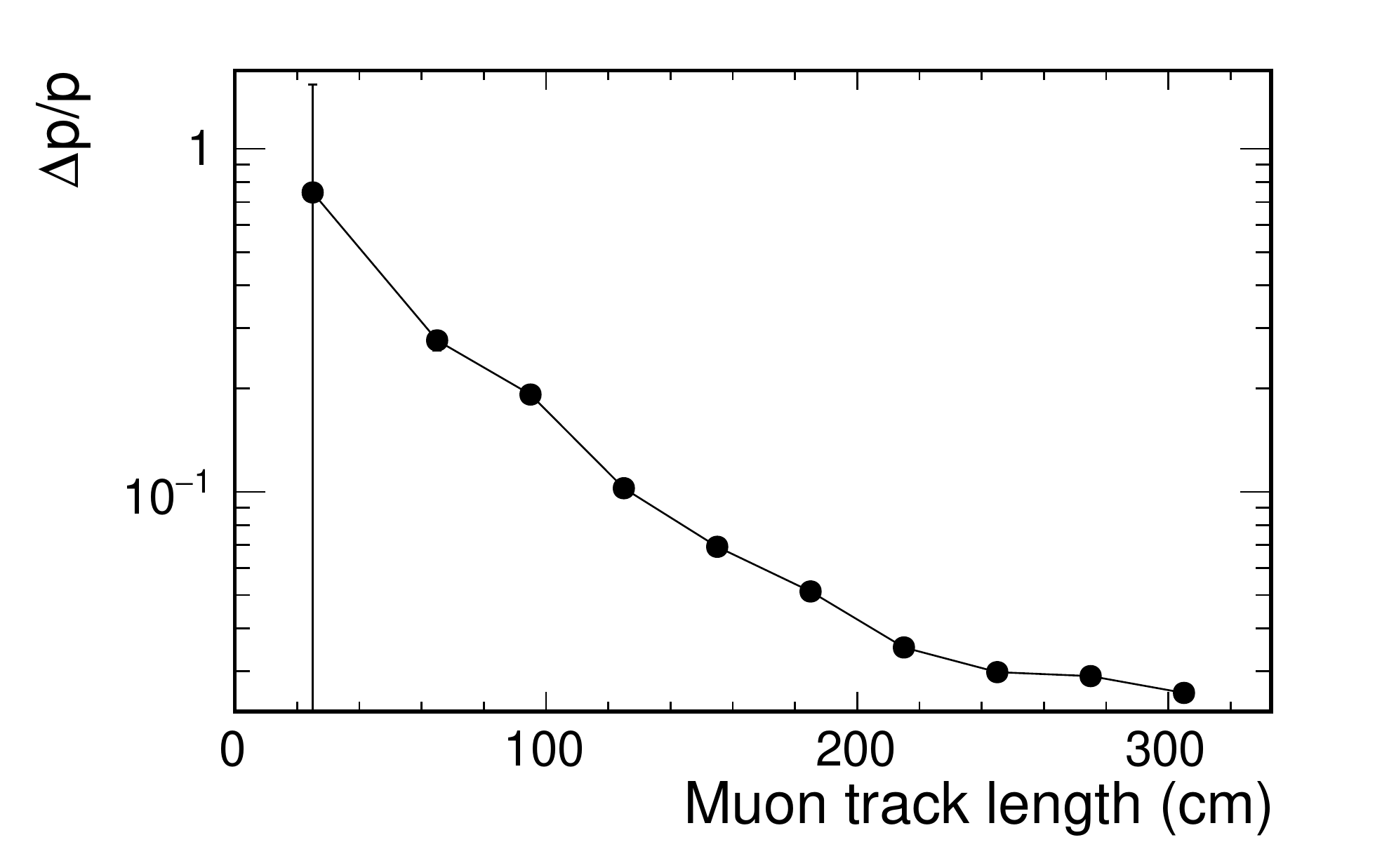} 
\end{dunefigure}

\subsubsection{Low Energy Proton Reconstruction}
\label{sec:mpd:protons}


The target nucleons participating in neutrino interactions reside in a complicated nuclear environment and uncertainties in the initial nuclear state, and in final state interactions, have large effects on final state particle kinematic distributions. In particular, protons with kinetic energies in the range of a few tens of \si{\MeV} are emitted from the struck nucleus due to final state interactions. Those protons create tracks that are too short to fully reconstruct in a \dword{lartpc}. While their energy can be measured calorimetrically, the measurement is likely to be imprecise due to the impact of electron-ion recombination. Furthermore, tracking these protons can inform nuclear modeling. Figure~\ref{fig:protons_FSImodels} shows the low energy proton spectrum predicted by three popular neutrino generators. The three generators disagree significantly below the threshold for proton track reconstruction in a \dword{lartpc} (currently \SI{40}{\MeV}). This energy range is significant, since even a single \SI{20}{\MeV} proton would carry 2\% of the energy of a \SI{1}{\GeV} neutrino. This section discusses how the \dword{hpgtpc} can be used to reconstruct low energy protons and will show that a \SI{5}{\MeV} threshold can be achieved. 

\begin{dunefigure}[Predicted proton energy spectra from three neutrino interaction models]{fig:protons_FSImodels}
{Predicted proton energy spectra from GENIE, NEUT, and NUWRO at low energies (truncated at \SI{100}{MeV}). The dashed vertical line indicates the kinetic energy required to make a 1~cm track in a \dword{lartpc}, and the solid vertical line shows the same for a gaseous TPC at 10~atm. The lower threshold in \dword{ndgar} provides a unique opportunity to distinguish among final state interaction models for the same nuclear target as the ND and FD \dwords{lartpc}.}
    \includegraphics[width=0.5\textwidth]{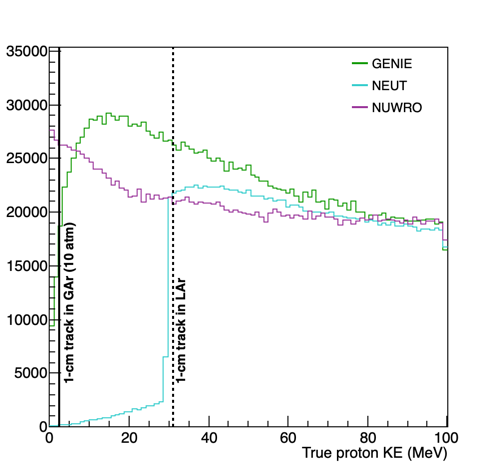}
\end{dunefigure}



Traditional tracking methods struggle to reconstruct very short, interacting tracks. \dshort{dune} has begun using machine learning to augment those methods. Though this effort is still in very early stages, there has been success so far in using a fully connected multi-layer perceptron (MLP) to both regress the kinetic energy of and classify between protons and pions.  Additionally a Random Sample Consensus (RANSAC) based clustering algorithm has been developed to group hits into short tracks for events where there are multiple particles. Together, these two algorithms can be used to measure the kinetic energy of multiple particles in a single event.             

As a demonstration, a test sample of multiple proton events was generated where each event has:
\begin{itemize}
\item 0-4 protons, number determined randomly with equal probabilities;
\item all protons share a common starting point (vertex) whose position in the TPC is randomly determined;
\item the direction of each proton is randomly generated from an isotropic distribution;
\item the momentum of each proton is randomly generated from a uniform distribution in the range 0-\SI{200}{\MeV/c} (0-\SI{21}{MeV} kinetic energy). 
\end{itemize}

The RANSAC-based clustering algorithm assigns individual hits to proton candidates which are passed to a MLP that was trained on a set of individual proton events in the TPC to predict kinetic energy.  Figure~\ref{fig:ML_residuals} shows the kinetic energy residuals, the reconstruction efficiency, and a 2D scatter plot of the measured kinetic energy versus the true kinetic energy for each individual proton with kinetic energy between 3 and 15 MeV in the test sample.  Additionally, the residual for the total kinetic energy in each multi-proton event is given. As can be seen in the figures, even at this early stage, the algorithm reconstructs proton energies to within a few MeV of their true energies. The efficiency is approximately 30\% for \SI{5}{MeV} protons and that is taken as the threshold.  Improvements in both the residuals and the efficiency of finding the protons are expected with further work.
%
%

\begin{dunefigure}[Machine learning residuals for protons in \dshort{ndgar}]{fig:ML_residuals}
{(Top left) Kinetic energy (KE) residual for reconstructed protons with KE in the range \SIrange{3}{15}{MeV}. (Top right) Measured KE vs. true KE for the same energy range. (Bottom right) Reconstruction efficiency as a function of true KE.  (Bottom left) Residual of the total KE of all protons in each event in the test sample. The asymmetric shape is due to protons which were not reconstructed.}
    \includegraphics[width=0.49\textwidth]{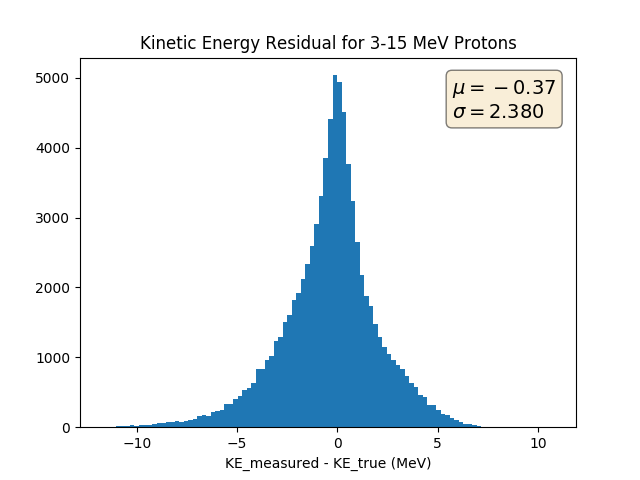}                                                    
    \includegraphics[width=0.49\textwidth]{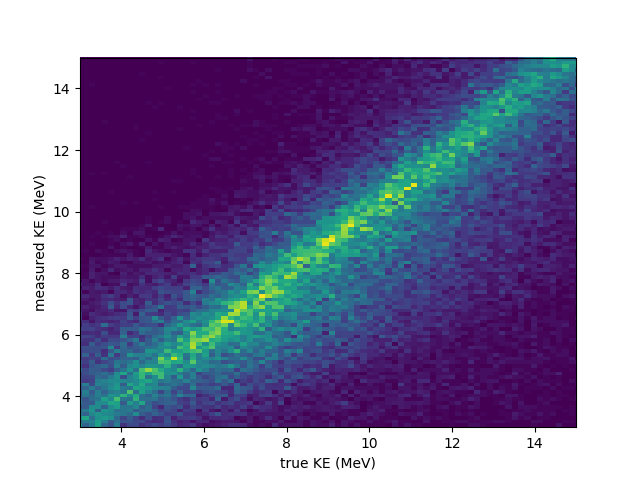}
    \vspace{1mm}
    \includegraphics[width=0.49\textwidth]{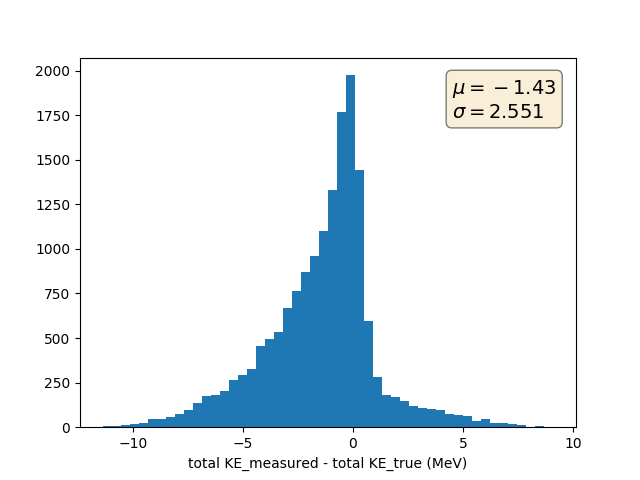}
    \includegraphics[width=0.49\textwidth]{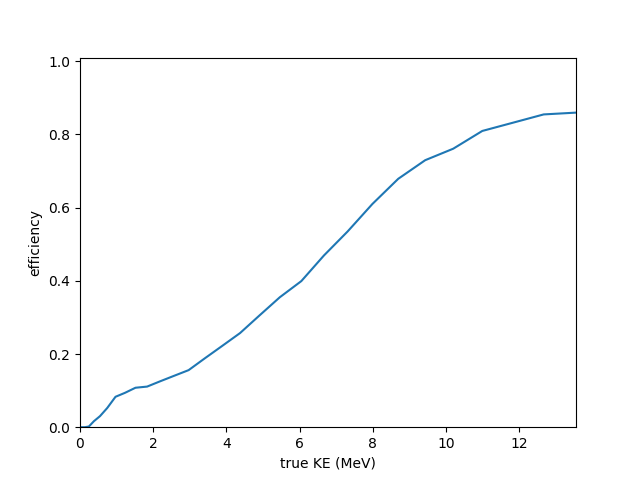}
\end{dunefigure}
%
%

\subsubsection{Pion Multiplicity Measurements}
\label{sec:mpd:pionmult}

The precision tracking capability of the \dshort{hpgtpc} allows \dshort{ndgar} to accurately identify and separate final states in neutrino interactions. Measuring various kinematic distributions for these exclusive final states allows deficiencies with the interaction model to be identified and fixed. The sensitivity of the \dshort{hpgtpc} to model differences is demonstrated via a study using two different neutrino interaction generators: GENIE and NuWro. In this study, the NuWro sample represents data, and the GENIE sample is taken as the default MC generator prediction. Differences between the two are apparent in the reconstructed $Q^{2}$ distributions, especially when these samples are subdivided by final state pion multiplicity.  

A parametrized reconstruction was used to study this. It was applied to \num{400000} simulated neutrino events in the \dshort{hpgtpc} to estimate its reconstruction abilities. This parametrized reconstruction used the \texttt{GEANT4} energy deposits along with additional smearing to estimate reconstructed particle energies. Within this simulation, charged particles were considered to be reconstructed if their track length exceeded \SI{6}{cm}. Protons and pions with momenta less than \SI{1.5}{GeV/c} were considered to be able to be separated perfectly using track $dE/dx$ measurements. For protons and pions with momenta above this threshold, the reconstructed energy of these particles within the ECAL was estimated using an energy resolution of $20 \% \oplus \frac{30\%}{\sqrt{E}}$,
where $E$ is the kinetic energy of the particle in GeV. This estimate of the energy in the ECAL was then compared with the reconstructed energy of the particle from curvature in the magnetic field under the assumption that the particle is a proton or a pion.  Whichever of these values was closer to the reconstructed ECAL energy was used as the reconstructed particle type. Neutral pions were considered to be reconstructed if both of the resulting decay photons had energies greater than \SI{20}{MeV} and if the angle between them was larger than the angular resolution of the ECAL (as described in Section~\ref{sec:mpd:ecal}). Muons were selected, and pions were rejected using the ECAL and the muon system, as described in Section~\ref{sec:mpd:muon}.


The pseudo-reconstruction described above was used to classify the final states. The confusion matrices are shown in Figure~\ref{fig:PiStudyConfusionMatrixBoth}.
	
\begin{dunefigure}[Pion Multiplicity Confusion Matrices]{fig:PiStudyConfusionMatrixBoth}{Confusion matrices for various pion multiplicities within the HPgTPC for both forward and reverse horn currents.}
  \includegraphics[width=0.8\linewidth]{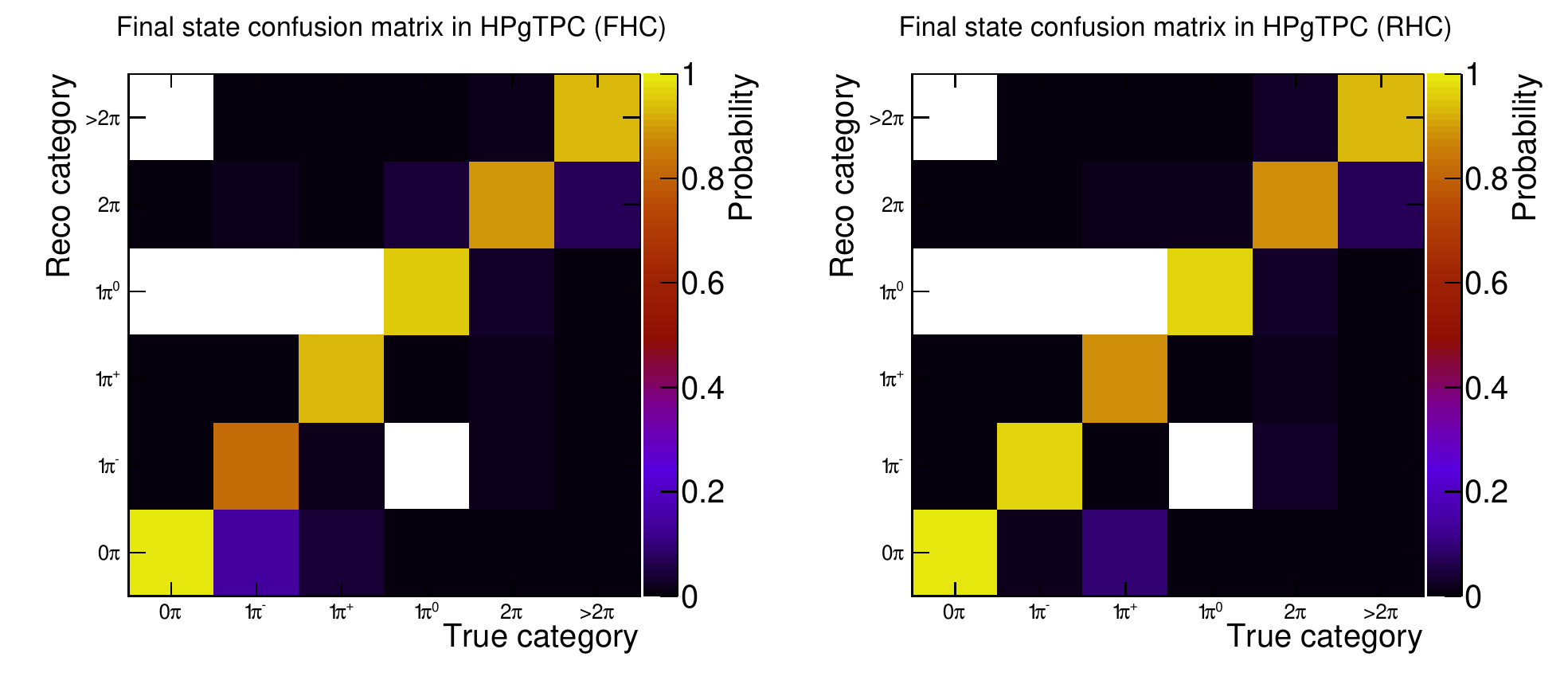}
\end{dunefigure}
	
The reconstructed squared momentum transfer, $Q^{2}_{reco}$ for charged-current muon neutrino events is defined as
\begin{align*}
  Q^{2}_{reco} = 2 E_{\nu, reco} (E_{\mu, reco} - p_{\mu, reco}\text{cos}(\theta_{\mu, reco})) - m_{\mu}^{2} \,,
\end{align*}
where $E_{\nu, reco}$ is the summed energy of the reconstructed lepton and the reconstructed final state hadrons, $E_{\mu, reco}$ and $p_{\mu, reco}$ are the reconstructed energy and momentum of the lepton, respectively, and $\theta_{\mu, reco}$ is the reconstructed angle between the neutrino and the lepton.

A mock data sample was constructed by reweighting GENIE events to resemble NuWro events using a boosted decision tree. The reweighting was done as a function of 18 variables including: neutrino energy, lepton energy, angle between lepton and neutrino, $Q^2$, $W$, $x_{Bj}$ and $y$. The number of and total energy carried by $p,n,\pi^+,\pi^-,\pi^0$ and the number of electromagnetic particles were also used~\cite{Vilela:20191002}. The distribution of $Q^{2}_{reco}$ for both the nominal MC (GENIE) and the mock data (NuWro) was plotted for each final state and the ratio was taken. These ratios are shown in Figure~\ref{fig:PiStudyQ2RecoRatios} for the FHC beam. 


	

        	\begin{dunefigure}[Pion Multiplicity Ratios NuWro/GENIE]{fig:PiStudyQ2RecoRatios}{Left: Reconstructed ratios of the NuWro-reweighted \dshort{hpgtpc} $\nu_{\mu}$ sample to the nominal GENIE sample, separated by pion multiplicity. Right: The true ratios separated by pion multiplicity. Also shown is the same ratio for all reconstructed charged current muon neutrino events. These ratios were made for the FHC beam. The study discussed in the text also uses similar ratios for the RHC beam.}

	  \includegraphics[width=0.45\linewidth]{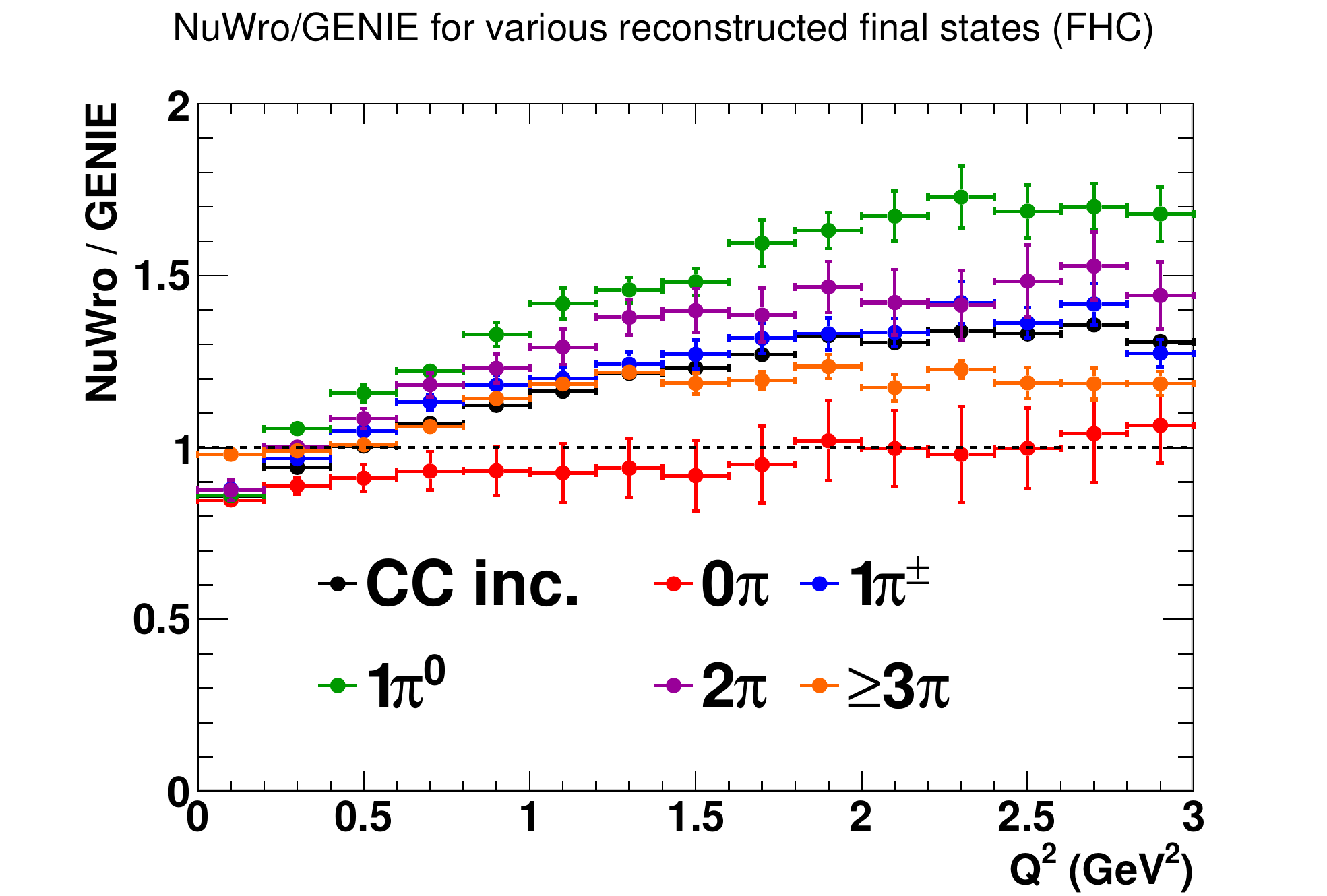}
	  \includegraphics[width=0.45\linewidth]{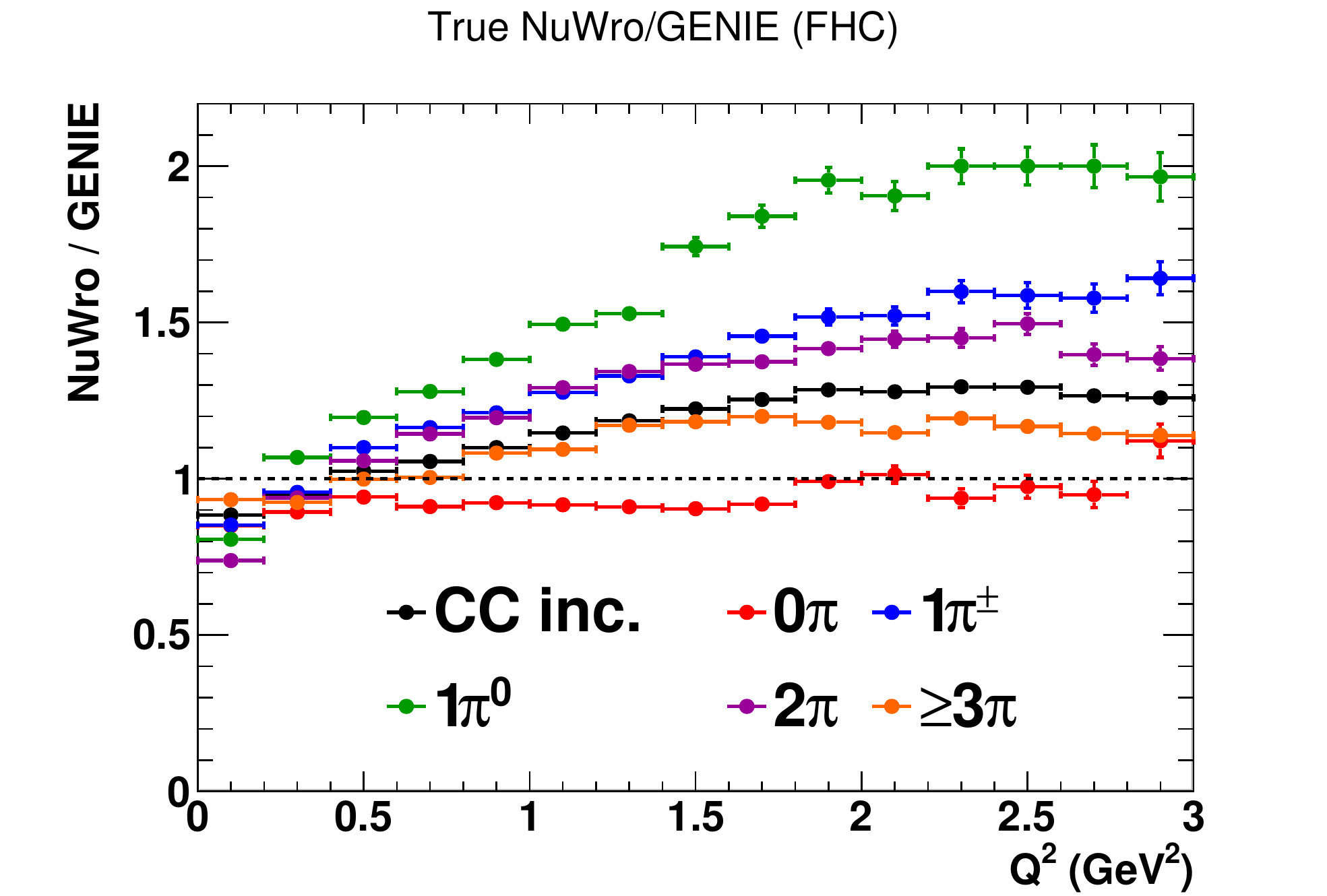}

	\end{dunefigure}

	The errors on the reconstructed ratios in Figure~\ref{fig:PiStudyQ2RecoRatios} were calculated by taking the spread between the ratios in the true categories which is then weighted by the confusion matrix shown in Figure~\ref{fig:PiStudyConfusionMatrixBoth}. This can be thought of as representing the systematic uncertainty on each of these points. Significant differences between GENIE and NuWro can be seen in all but the $0\pi$ curve. The reconstructed ratios on the left reproduce the features of the true ratios shown on the right rather well.
        



\subsubsubsection{Far detector fits with HPgTPC-driven reweighting}

\begin{dunefigure}
  [Bias in the measured $\delta_{CP}$ when fitting with an uncorrected MC]
  {fig:PiStudyBiasNoRwt}
  {The $\delta_{CP}$ bias as a function of true $\delta_{CP}$ when performing a far detector-only fit with NuWro reweighted fake data. The black points show the resulting bias when using the nominal MC to predict what is observed at the far detector in the oscillation parameters fit.}
  \includegraphics[width=0.49\linewidth]{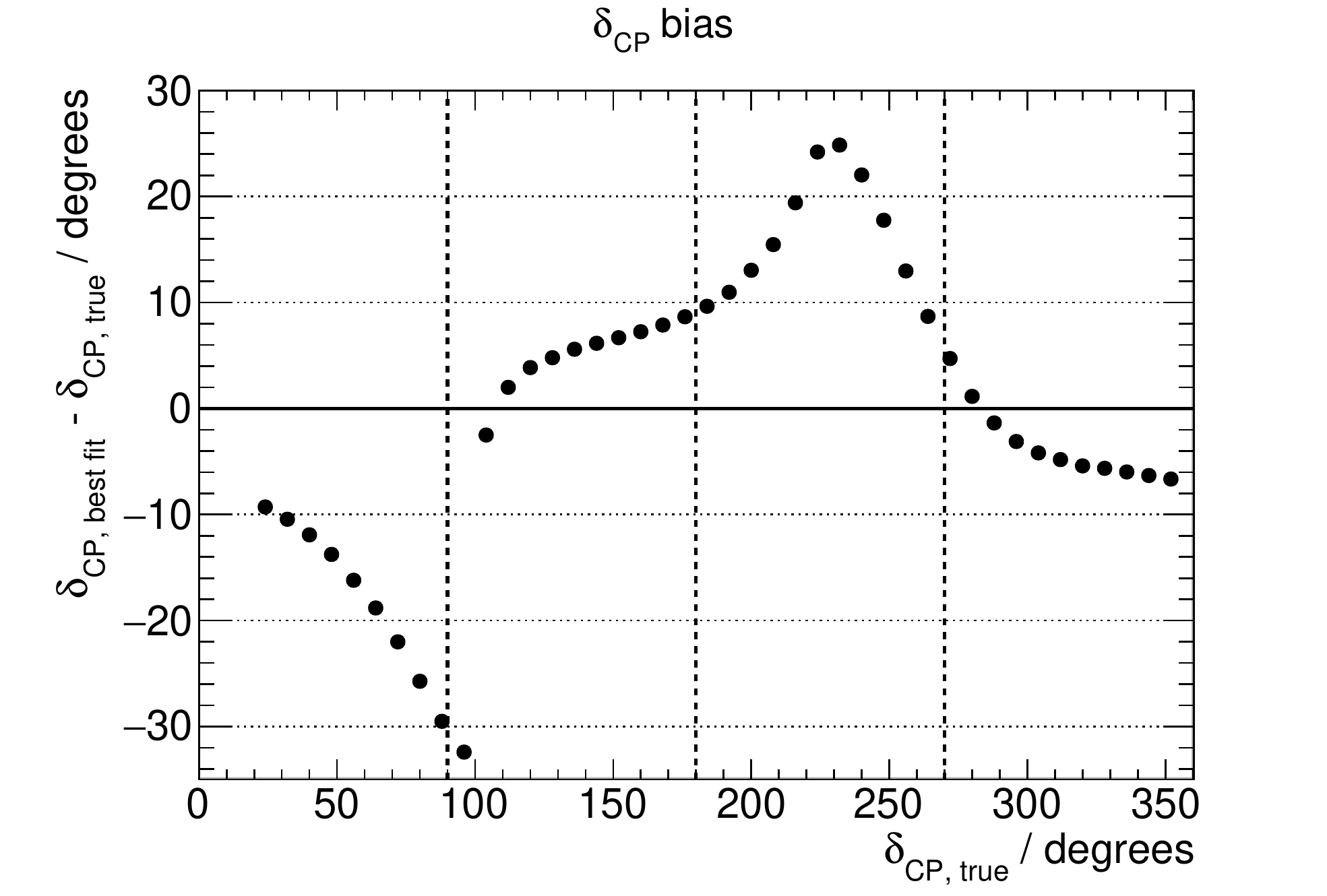}
\end{dunefigure}

Far detector-only fits were done using the nominal GENIE MC as the predicted far detector spectra and the NuWro mock data as the `data'. They resulted in a biased measurement of $\delta_{CP}$, which is shown as the black points in Figure~\ref{fig:PiStudyBiasNoRwt}. The bias in $\delta_{CP}$, defined as the difference between the true value of $\delta_{CP}$ and the best fit value, is as large as $30^{\circ}$ for some true values of $\delta_{CP}$.
	

As an example of how \dshort{ndgar} can rectify this bias, distributions of reconstructed kinematic quantities measured in \dshort{ndgar} were used to reweight far detector Monte Carlo samples. In this study, a two-dimensional distribution of kinematic variables was used for the reweighting. The variables used were the visible energy transfer, $E_{\text{vis}}$, and the visible three-momentum transfer, $p_{\text{vis}}$~\cite{Rodrigues:2015hik}.

$E_{\text{vis}}$ is defined as

\begin{align*}
  E_{\text{vis}} = T_{p} + E_{\pi} \,, 
\end{align*}
where $T_{p}$ is the total kinetic energy of final state protons and $E_{\pi}$ is the total energy of final state pions (including their masses). This definition of  $E_{\text{vis}}$ neglects the energy carried by neutrons. Section~\ref{sec:mpd:neutrons} discusses the capability of \dword{ndgar} to measure the neutron energy spectrum. 

$p_{\text{vis}}$ is defined as

\begin{align*}
  p_{\text{vis}} = \sqrt{2(E_{\text{vis}}+E_{\mu})(E_{\mu} - p_{\mu}\text{cos}\theta_{\mu}) - m_{\mu}^{2} + E_{\text{vis}}^{2}}
\end{align*}

$E_{\text{vis}}$ and $p_{\text{vis}}$ are useful because they are equivalent to the energy transfer, $q_0$, and three-momentum transfer, $|\vec{q}_3|$, from the lepton arm in the limit of no Fermi motion and no final state neutrons. Examples of the NuWro/GENIE ratios of these quantities are shown in figure~\ref{fig:PiStudyQ0Q3RecoRatios}. The examples shown in figure~\ref{fig:PiStudyQ0Q3RecoRatios} are aggregated over all final states.

\begin{dunefigure}
  [NuWro/GENIE event rate ratios in $E_{\text{vis}}, p_{\text{vis}}$ space]
  {fig:PiStudyQ0Q3RecoRatios}
  {Left: Ratio of NuWro-reweighted \dshort{hpgtpc} $\nu_{\mu}$ (FHC) sample to GENIE sample for all reconstructed events. Right: Ratio of NuWro-reweighted \dshort{hpgtpc} $\nu_{\mu}$ (FHC) sample to GENIE sample for all reconstructed events. Bins with $<100$ true MC events are not filled.}
  \includegraphics[width=0.49\linewidth]{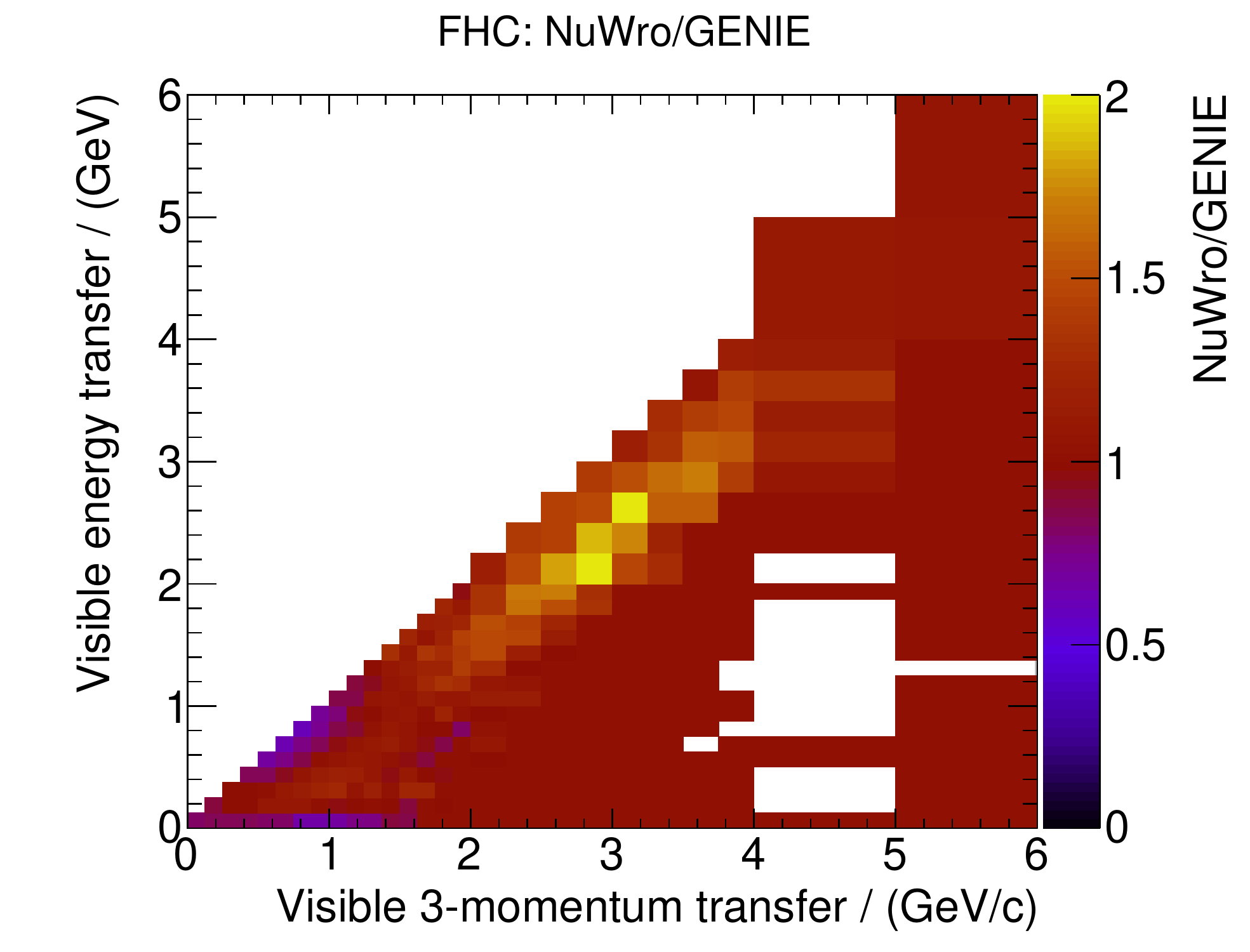}
  \includegraphics[width=0.49\linewidth]{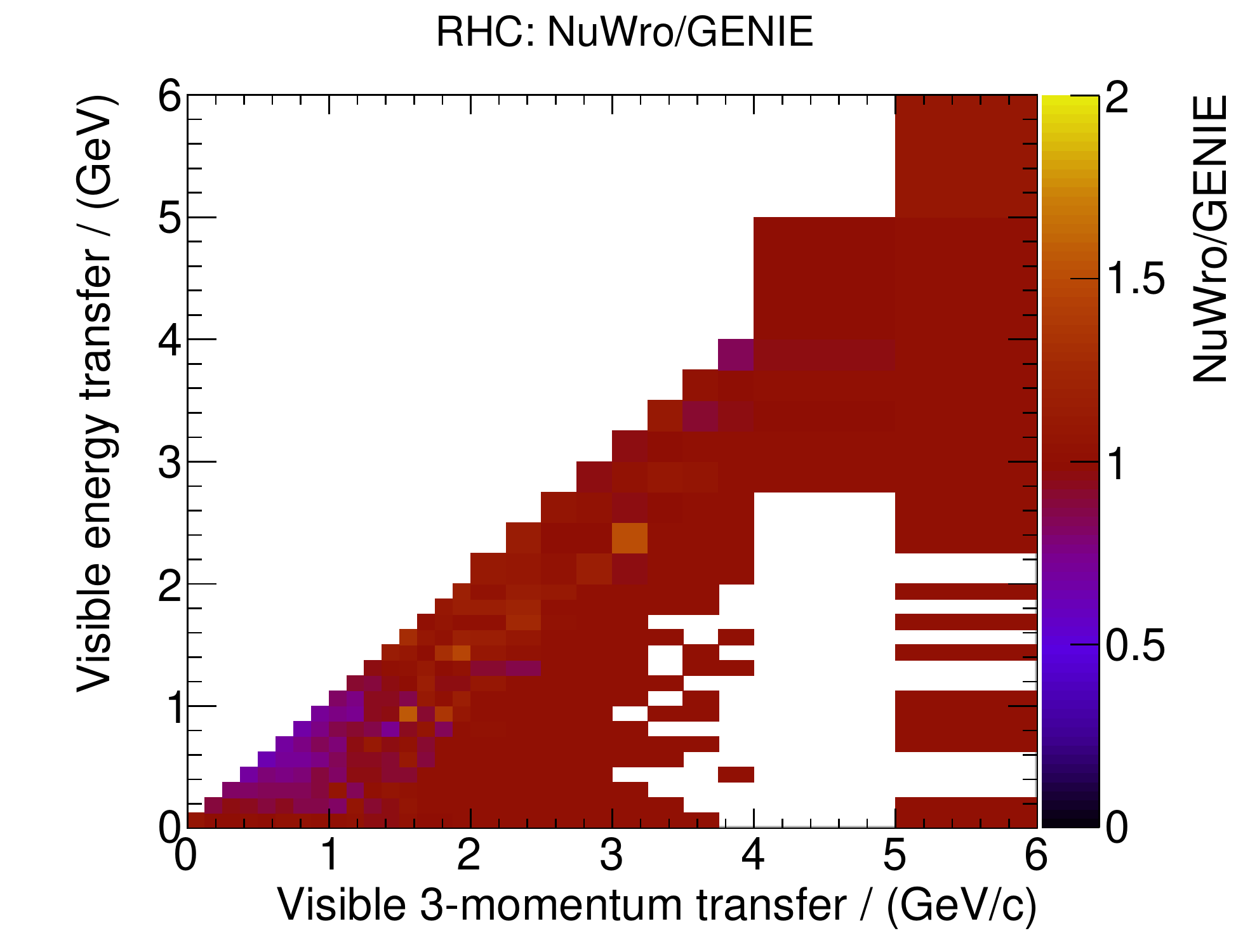}
\end{dunefigure}

These two-dimensional distributions were produced for various reconstructed final states: $0\pi$ with 1 proton, $0\pi$ with $>1$ proton, $1\pi^{+}$, $1\pi^{-}$, $1\pi^{0}$, $2\pi$ and $>2\pi$. Separate histograms were produced and used for the FHC and RHC samples. The ND measurements were used to correct the far detector MC by reweighting events based upon their true values of $E_{\text{vis}}$, $p_{\text{vis}}$ and their true final state. 

The results of this reweighting are shown in Figures~\ref{fig:PiStudyDataMCComp} and~\ref{fig:PiStudyDataMCRatios}. In these examples, the true value of $\delta_{CP}$ is $90^{\circ}$. In Figure~\ref{fig:PiStudyDataMCComp}, the NuWro mock data are shown in black, the nominal GENIE MC is shown in blue and the GENIE MC with the near detector-driven weights is shown in red. The reweighting procedure generally improves the agreement with the \dshort{fd} mock-data, particularly in the FHC beam. It is not surprising that the agreement is still imperfect because there are numerous differences between the mock-data and the MC and this study has only attempted to correct for one class of them. Also, the reconstructed energy does not directly correspond to $E_{\text{vis}}$, $p_{\text{vis}}$, and pion multiplicity, so a successful reweighting in that variable space isn't guaranteed to succeed fully when the reconstructed neutrino energy spectrum is inspected. Finally, the reweighting was done using flux integrated samples but the flux is quite different between the ND and FD due to oscillations.
	
\begin{dunefigure}[FD spectra with and without pion multiplicity weighting] {fig:PiStudyDataMCComp} {Comparison of simulated far detector spectra with and without the ND-derived weighting. The black line shows the spectra for the NuWro reweighted mock data. The blue line shows the nominal MC. The red line shows the nominal MC with the near detector-derived weighting. Errors correspond to 1 years POT.}
  \includegraphics[width=1.0\linewidth]{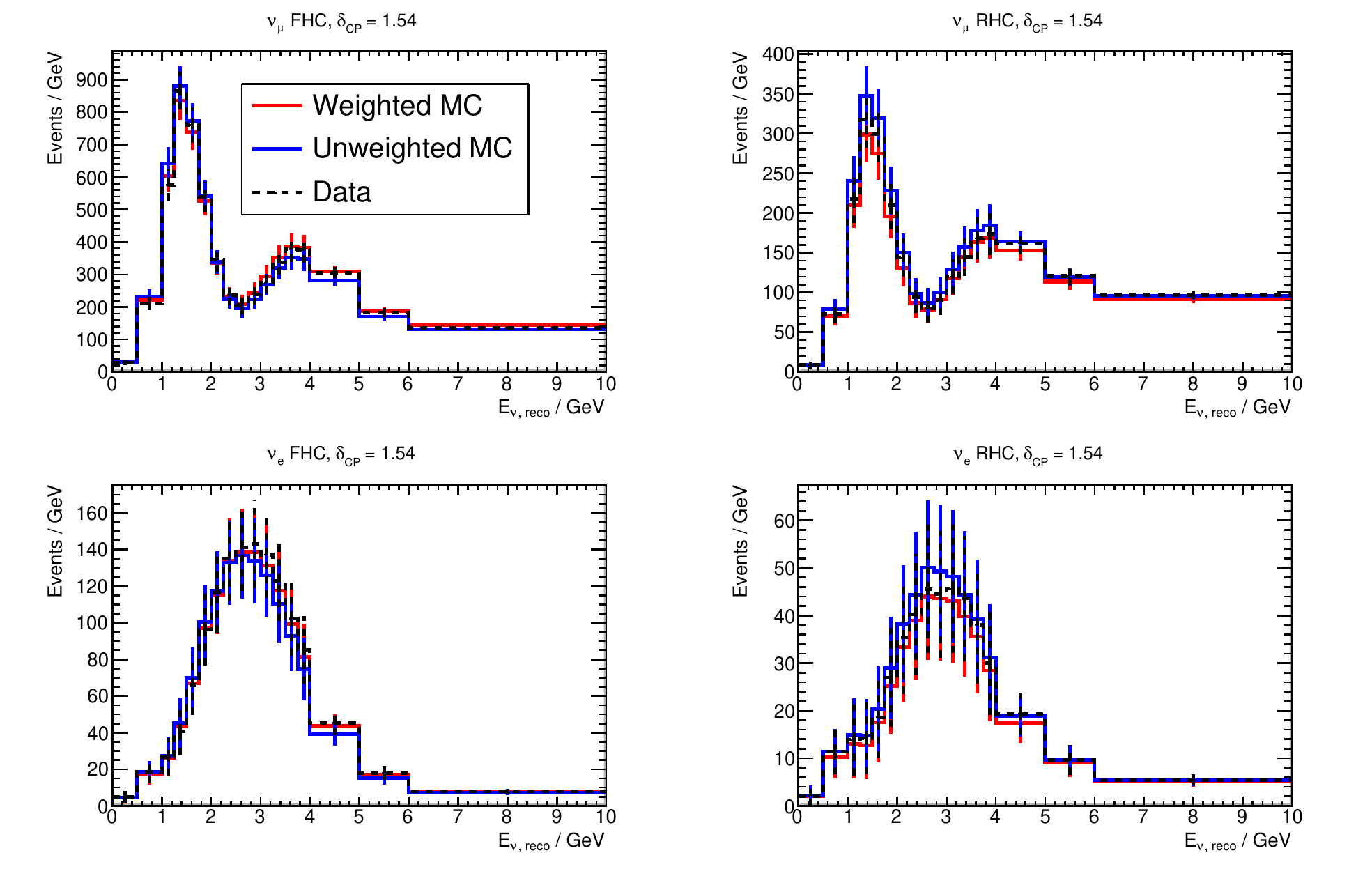}
\end{dunefigure}	

\begin{dunefigure}[Ratios of FD spectra with and without pion multiplicity reweighting]{fig:PiStudyDataMCRatios}
          {Ratio of NuWro mock data to weighted and unweighted MC samples. The blue line is the nominal MC while the red line is the MC with the additional near detector-derived weighting.}
	  \includegraphics[width=0.9\linewidth]{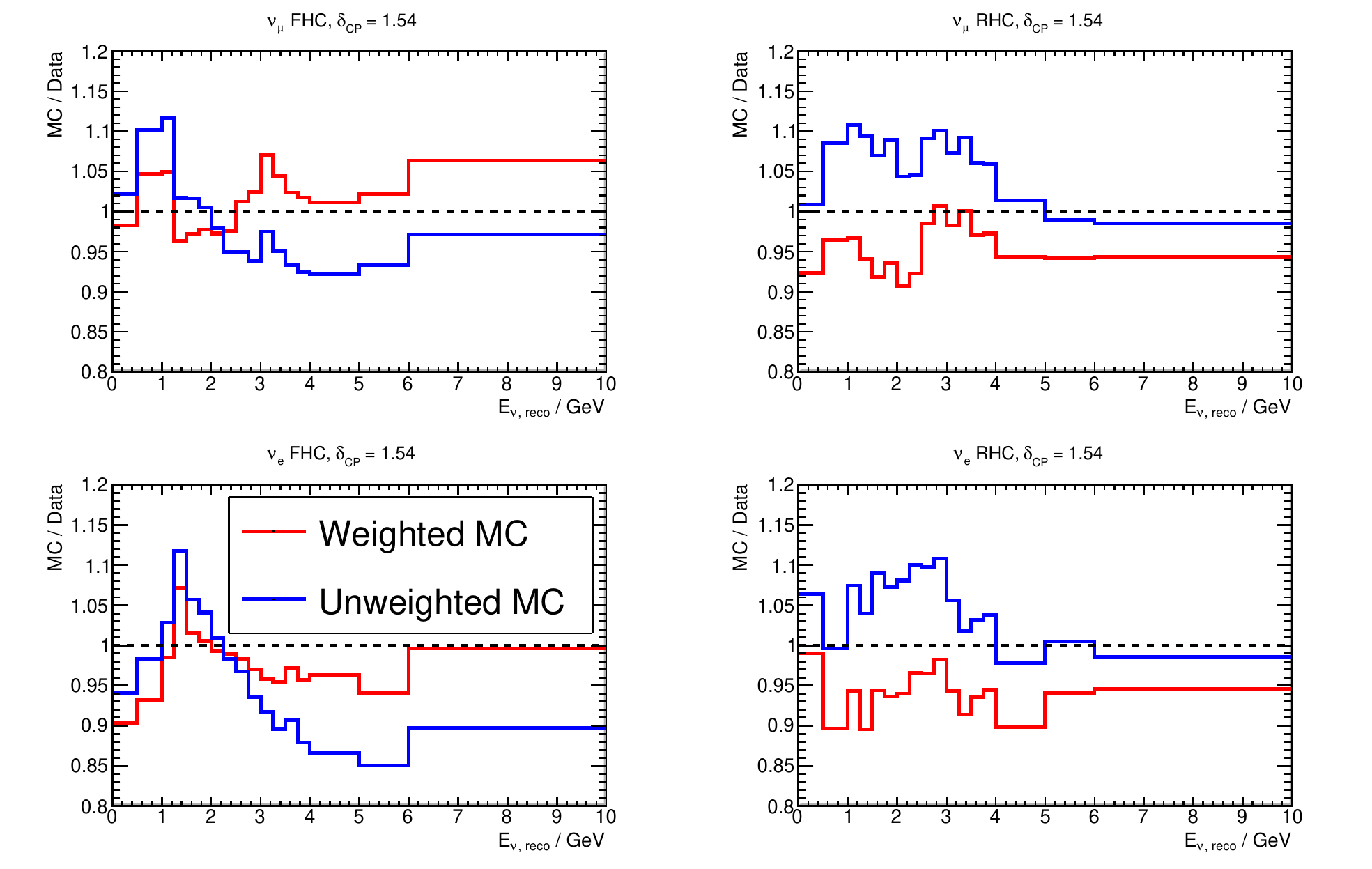}
\end{dunefigure}

Additionally, a sample was reweighted using only a single $E_{\text{vis}}, p_{\text{vis}}$ distribution that is not separated by final state. These distributions are labelled as ``CC inc.'' in Figure~\ref{fig:PiStudyQ2RecoRatios}.  This simulates a situation where the detector does not have the ability to identify final states.

Far detector-only fits were performed at a range of $\delta_{CP}$ values with all cross-section, flux and detector systematics at their nominal values. This was done with three different sets of predicted spectra:
\begin{enumerate}
\item The nominal GENIE MC.
\item The MC reweighted with the $E_{\text{vis}}, p_{\text{vis}}$ weights, separated by pion multiplicity.
\item The MC reweighted with the $E_{\text{vis}}, p_{\text{vis}}$ weights unseparated by pion multiplicity. This is referred to as ``CC inc.''
\end{enumerate}

The resulting bias in $\delta_{CP}$ at different true values of $\delta_{CP}$ for each of these three sets of predicted spectra is shown in Figure~\ref{fig:PiStudyBiasWithGAr}. At most values of $\delta_{CP}$, the bias is reduced when either additional reweighting is used (red and green points). 
As seen in the figure, the pion-separated reweighting provides a larger bias reduction than the unseparated reweighting at most values of $\delta_{CP}$.
	
\begin{dunefigure}
  [Biases in the measured $\delta_{CP}$ with and without pion multiplicity corrections]
  {fig:PiStudyBiasWithGAr}
  {$\delta_{CP}$ bias as a function of true $\delta_{CP}$ when performing a far detector-only fit with NuWro reweighted mock data. The black points show the resulting bias when using the nominal MC. The red points show the resulting bias when the MC is weighted using the ND-derived weights in $E_{\text{vis}}, p_{\text{vis}}$, separated by pion multiplicity. The green points show the bias when the MC is weighted using the ND-derived weights in $E_{\text{vis}}, p_{\text{vis}}$ with no separation by pion multiplicity.}
 
  \includegraphics[width=0.49\linewidth]{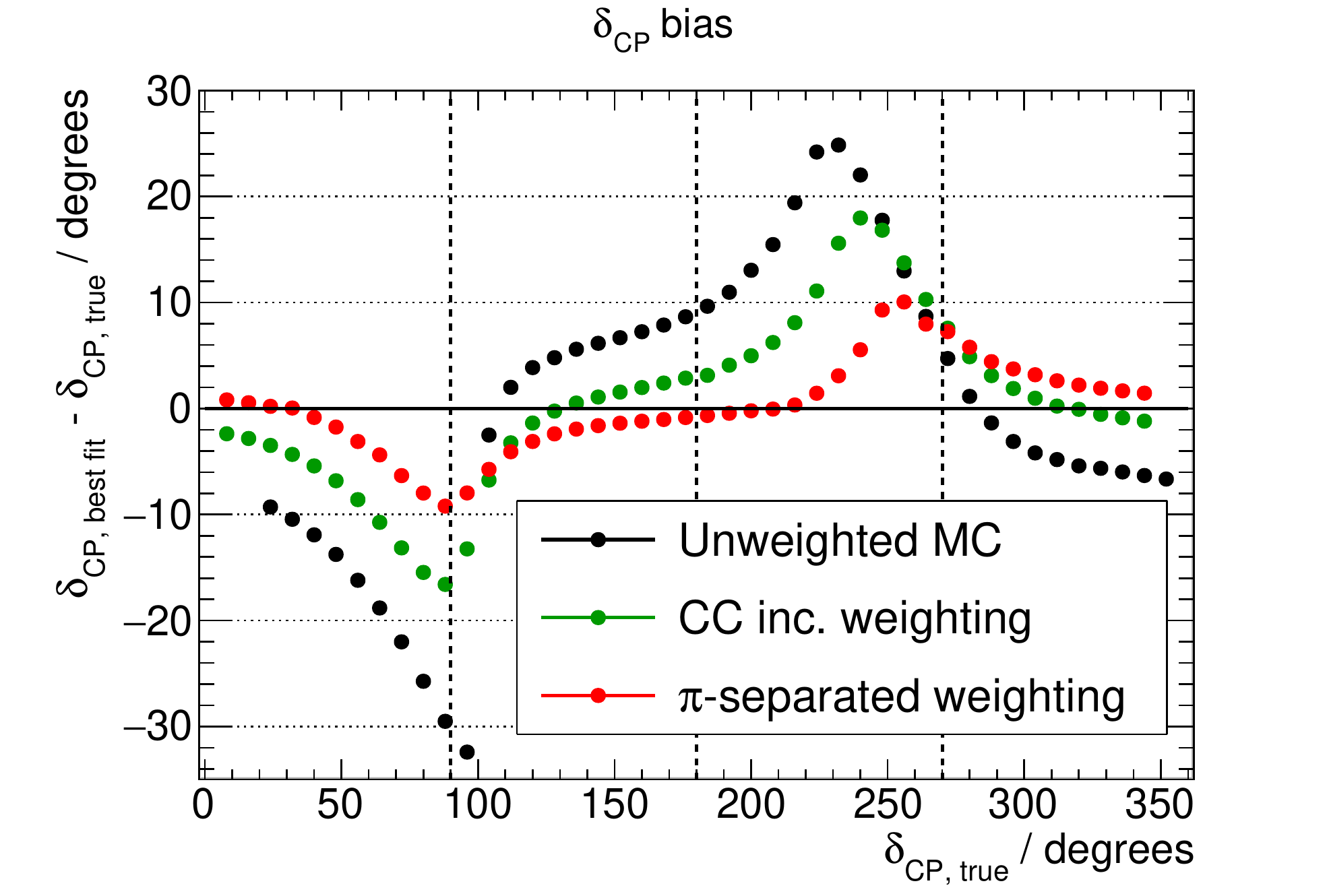}
\end{dunefigure}
			

\subsection{\dshort{ecal} Performance}
\label{sec:ecal-performance}
\label{sec:mpd:ecal}

The expected performance of the calorimeter was studied with Geant4-based~\cite{Agostinelli:2002hh} simulations and GArSoft~\cite{GArSoftwebsite} (commit 91362a8e). In the following, a reference scenario is considered in which the entire \dword{ecal} is located outside the pressure vessel. The simulation shoots single photons from a single point inside the \dshort{hpgtpc} in a cone of 20 degrees towards the downstream \dword{ecal} barrel. The octagonal barrel geometry consists of 60 layers with the following layout:
\begin{itemize}
  \item 8 layers of \SI{2}{\mm} copper + \SI{5}{\mm} of $2.5\times2.5$ cm$^2$ tiles + \SI{1}{\mm} FR4
  \item 52 layers of \SI{2}{\mm} copper + \SI{5}{\mm} of cross-strips \SI{4}{\cm} wide
\end{itemize}
For the present study, copper has been chosen as the absorber material as initial studies have shown that this material provides a good compromise between calorimeter compactness, energy resolution, and angular resolution compared to lead. Digitization effects are accounted for by introducing an energy threshold of 0.25~MIPs ($\sim$\SI{200}{\keV}) for each detector cell/strip, a Gaussian smearing of \SI{0.1}{\MeV} for the electronic noise, SiPM saturation effects, single photon statistics and a Gaussian time smearing of \SI{250}{\pico\second}.

\begin{dunefigure}[Energy and angular resolutions for photons in the \dshort{ndgar} ECAL.]{fig:EResARes_NDECAL}
  {Left: The energy resolution in the barrel as a function of the true photon energy. The energy resolution is determined by a Gaussian fit to the visible energy. Right: The angular resolution in the barrel as a function of the true photon energy. The angular resolution is determined by a Gaussian fit to the 68\% quantile distribution. The fit function is of the form $\frac{\sigma_{E}}{E} = \frac{A}{\sqrt{E}} \oplus \frac{B}{E} \oplus C$.}
\includegraphics[width=0.49\textwidth]{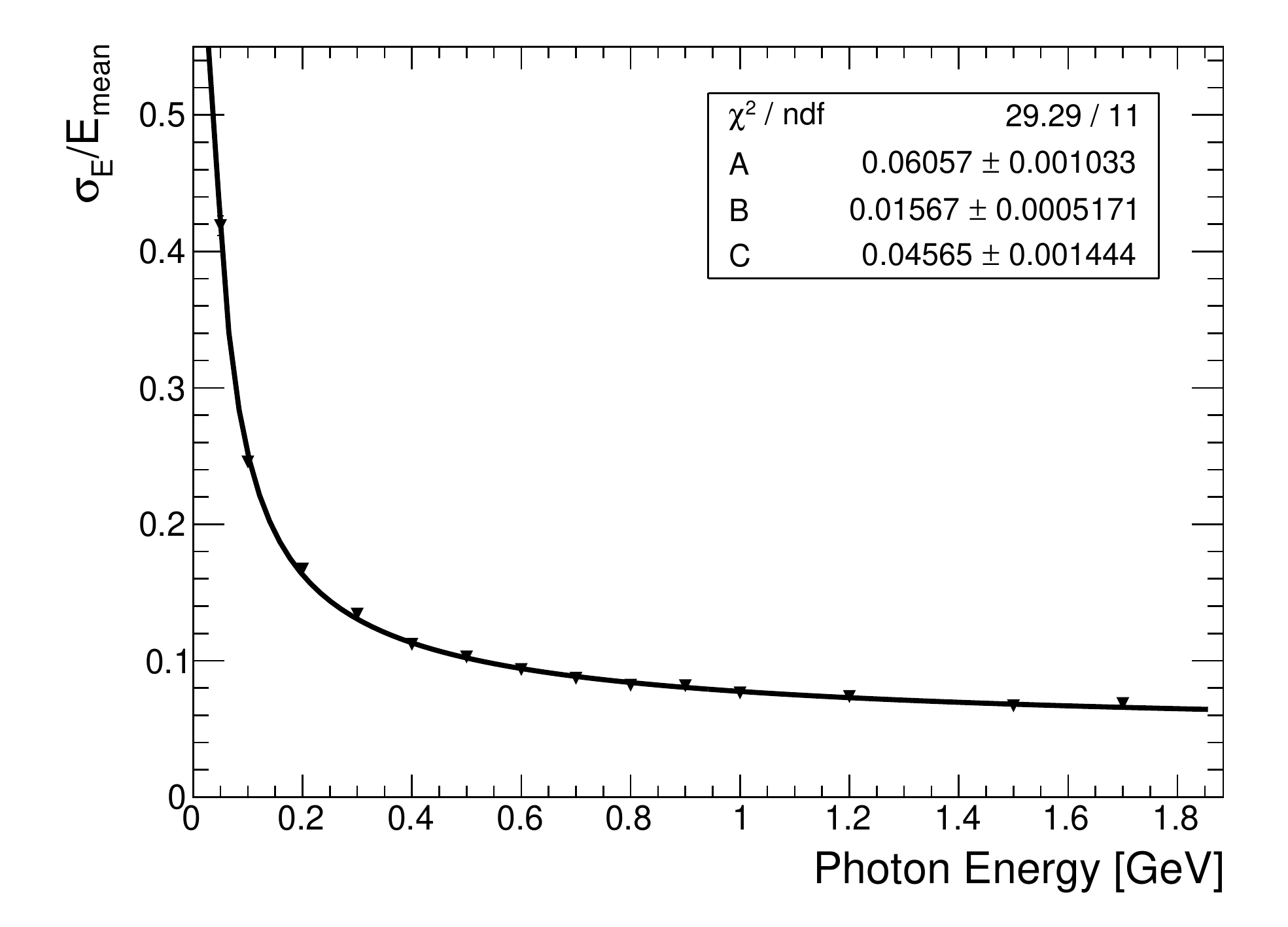}
\includegraphics[width=0.49\textwidth]{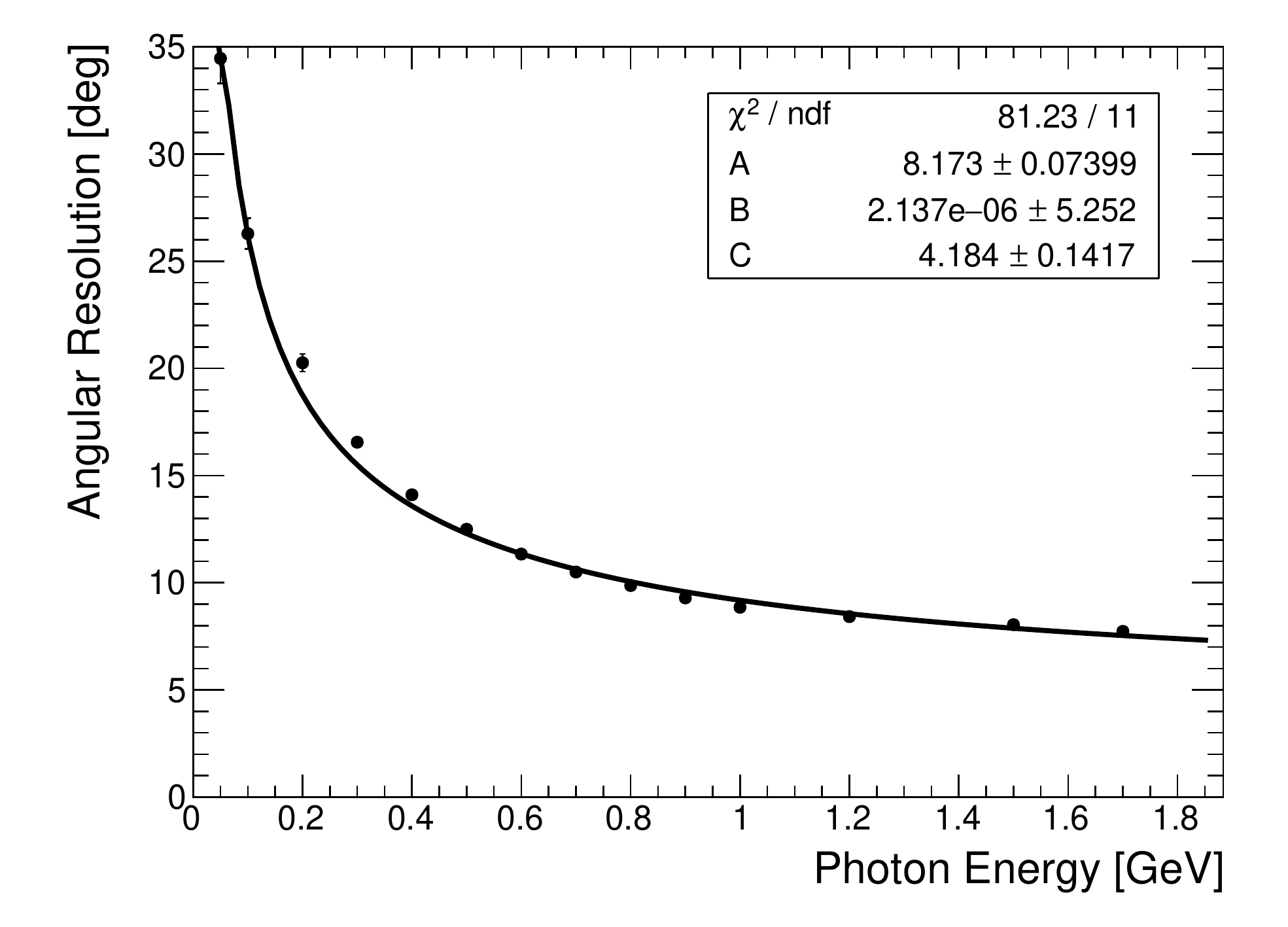}
\end{dunefigure}

\begin{dunefigure}[Momentum spectra of particles created in neutrino interactions in \dshort{ndgar}.]{fig:GENIE_particle_spectra_NDGAR}
  {GENIE prediction of momentum spectra for particles resulting from neutrino interactions in \dshort{ndgar}, shown in log scale on the vertical axis. }
\includegraphics[width=0.7\textwidth]{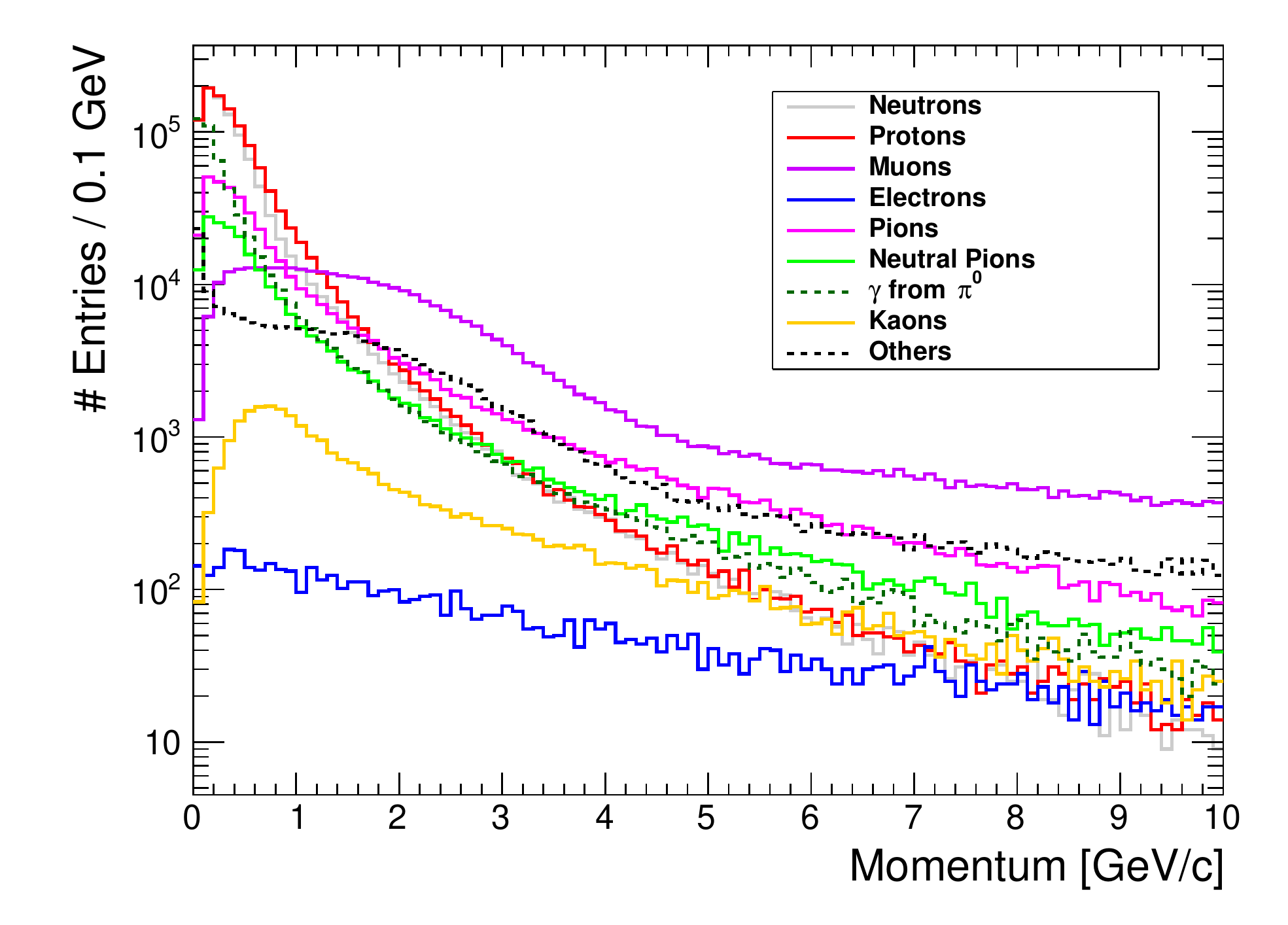}
\end{dunefigure}

\subsubsection{Energy Resolution} The energy resolution is determined by fitting the visible energy with a Gaussian. Photons that converted in the \dword{hpgtpc} are ignored as these require a specific treatment. A fit function of the form $\frac{\sigma_{E}}{E} = \frac{A}{\sqrt{E}} \oplus \frac{B}{E} \oplus C$ is used, where $A$ denotes the stochastic term, $B$ the noise term, $C$ the constant term, and $E$ is in GeV. Figure~\ref{fig:EResARes_NDECAL} shows the energy resolution as a function of the true photon energy. The best fit finds $\frac{\sigma_E}{E}=\frac{6.1\%}{\sqrt{E}} \oplus \frac{1.6\%}{E} \oplus 4.5\%$. For reference, the GENIE prediction of momentum spectra for particles created in neutrino interactions in \dword{ndgar} are shown in Figure~\ref{fig:GENIE_particle_spectra_NDGAR}. It should be noted that due to the lack of non-uniformities, dead cells, and other effects in the simulation, the energy resolution is slightly optimistic. Also, improvements in the reconstruction method could impact the energy resolution.

\subsubsection{Angular Resolution} The angular resolution of the calorimeter has been determined using a principal component analysis (PCA) of all reconstructed calorimeter hits. The direction is taken as the first eigenvector (main axis) of all the reconstructed hits. The angular resolution is determined by first computing the angle between the true and reconstructed photon directions. Then, the central core (68\% quantile) of the distribution is fit with a Gaussian distribution. The mean of the Gaussian is taken as the angular resolution and the error as its variance. Figure~\ref{fig:EResARes_NDECAL} shows the angular resolution as a function of the photon energy. An angular resolution of $\frac{\SI{8.17}{\degree}}{\sqrt{E}} \oplus \SI{4.18}{\degree}$ can be achieved. This could potentially be further improved with a different arrangement of the tile and strip layers, an optimization of the absorber thickness, and an improved reconstruction method. The angular resolution is mainly driven by the energy deposits in the first layers of the \dword{ecal}. Using an absorber with a large $X_{0}$ creates an elongated shower that helps in determining the direction of the shower. In general, high granularity leads to a better angular resolution, however, studies have shown that there is no additional benefit to having cell sizes below $2\times2$ cm$^2$ \cite{Emberger:2018pgr}.

\subsubsection{ECAL Optimization}
The optimization of the \dword{ecal} is underway and is driven by the performance metrics stated in Table~\ref{tab:TPCperformance}, as well as other physics goals, such as the potential for neutron detection and energy measurement.  Optimization studies of the \dword{ecal} design are being done in order to aid in understanding in detail the effects of detector variables, such as geometry, granularity, and passive material on the performance of the detector. An example of the impact of different detector granularity configurations on the angular resolution is shown in Figure~\ref{fig:Optimization_NDECAL_Granularity}. Different ratios of tile layers to strip layers are represented by the various colors. As shown in the figure, a fully tiled \dword{ecal} gives the best angular resolution. The resolution degrades as the ratio of tiles to strips is reduced, up to more than 100\% for energies below 100 MeV in the case of using only strips. Tiles maximize the geometrical information of the shower leading to a better reconstructed axis of the photon shower. However, the method used to determine the axis of the shower (using a principal component analysis) relies mostly on the position information of the hits. In future studies, this method will be further improved by using additional information such as the reconstructed vertex position and timing of \dword{ecal} clusters in order to reduce the reliance on the pure geometrical information given by \dword{ecal} hits, and thus reduce the number of tiled layers in the \dword{ecal}.

\begin{dunefigure}[ECAL Granularity Optimization: Angular resolution curves and ratio]{fig:Optimization_NDECAL_Granularity}{Angular resolution and ratio for different \dword{ecal} detector granularity configurations}
\includegraphics[width=0.49\textwidth]{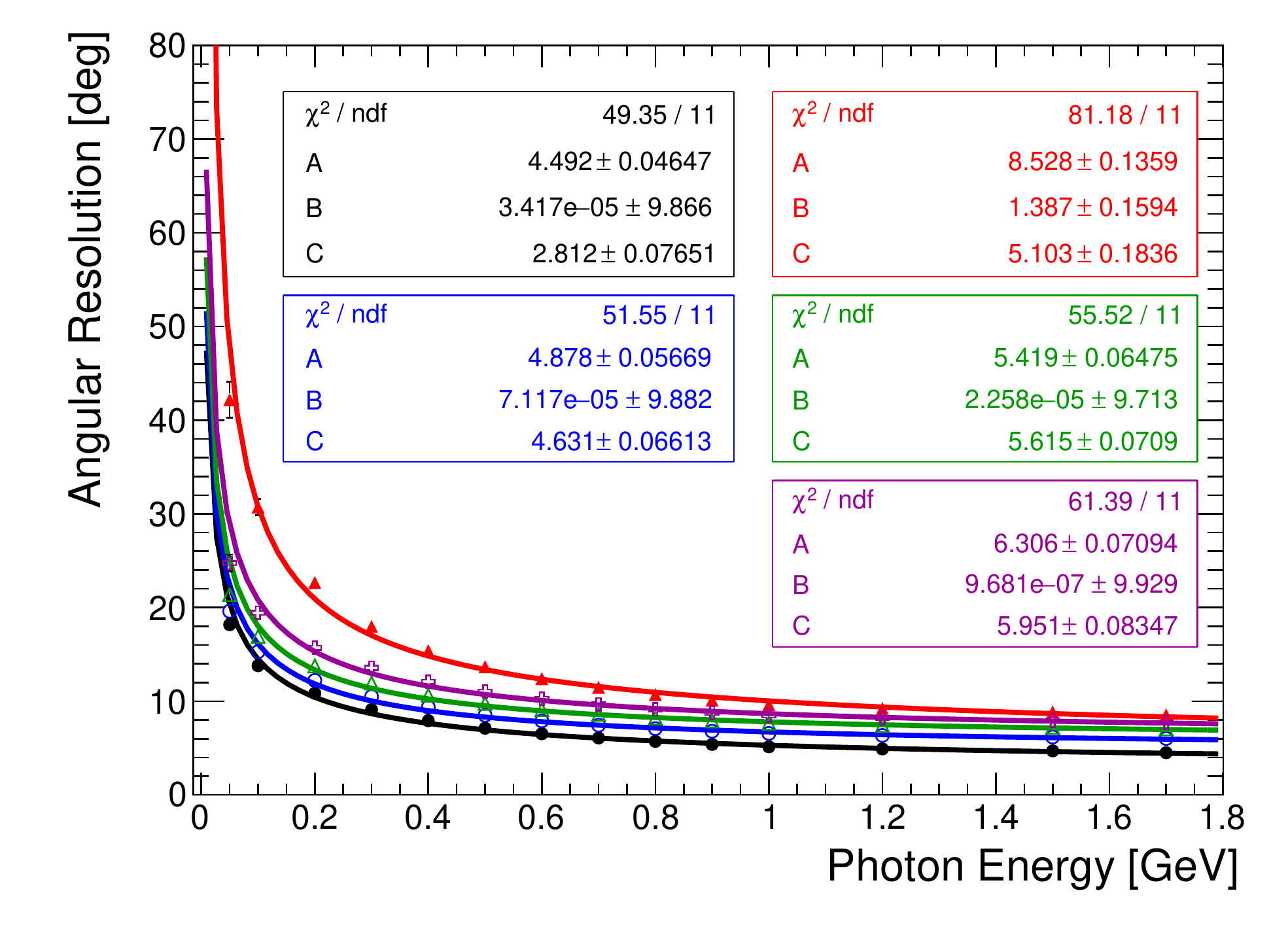}
\includegraphics[width=0.49\textwidth]{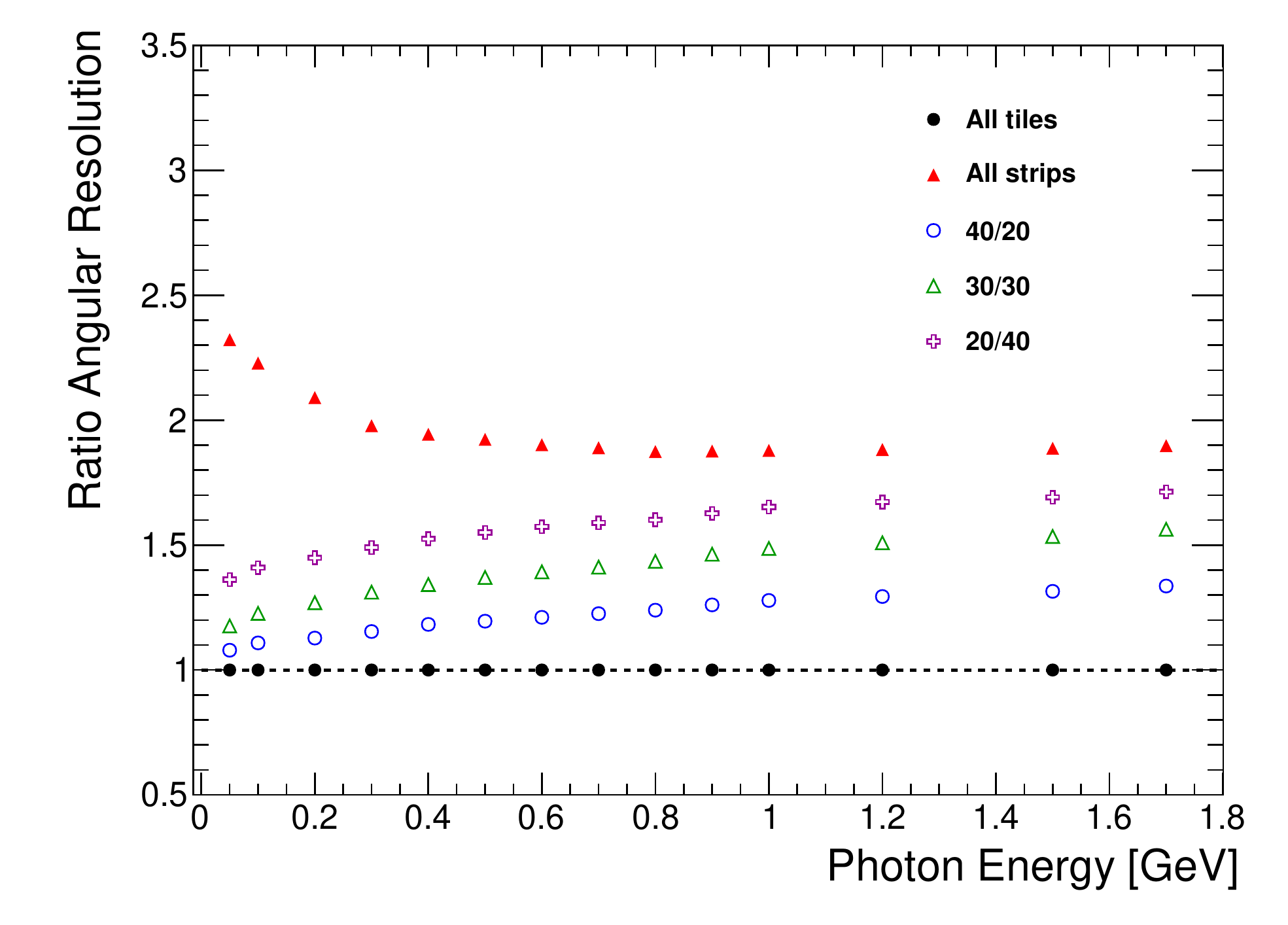}

\end{dunefigure}

Additional optimization studies have also been performed. An example of the impact of different passive material configurations on the energy and angular resolution are shown in Figure~\ref{fig:Optimization_NDECAL_Absorber}. As described in section \ref{sec:mpd:ecal}, the \dword{ecal} absorber is made of \SI{2}{\mm} thick copper (Cu). This study has been done considering lead (Pb) instead in thicknesses from \SI{2}{\mm} to \SI{0.7}{\mm}. \SI{0.7}{\mm} is about equivalent in radiation length to \SI{2}{\mm} copper. Looking at the energy resolution figure, using \SI{2}{\mm} thickness of lead results in a worse energy resolution by about 10-40\% especially at energies below 1 GeV due to sampling fluctuations. However, for thicknesses equivalent to \SI{2}{\mm} copper, the energy resolution is better on average by 20\%. This can be explained by the increase in sampling frequency for energies below 1 GeV and a better containment for energies over 1 GeV. Such an \dword{ecal} design would result in an increase of the number of layers by about 20\%. Then looking at the angular resolution, for the \SI{2}{\mm} case, it results in a worse angular resolution by up to 60-70\%. This can be explained by the Moliere radius and the radiation length of lead that is smaller than that of copper, resulting in showers being more compact longitudinally and more blob-like. Showers in copper are more elongated, which helps determine the direction of the original photon. With smaller lead absorber thicknesses, the angular resolution is degraded by up to 10\% at most. This makes it viable as an absorber material for the \dword{ecal}. However, it would certainly impact the neutron reconstruction efficiency due to using a higher Z material.

\begin{dunefigure}[ECAL Absorber Optimization: Energy and Angular resolution curves and ratio]{fig:Optimization_NDECAL_Absorber}{Energy and angular resolution and ratio for different \dword{ecal} detector absorber configurations}
\includegraphics[width=0.49\textwidth]{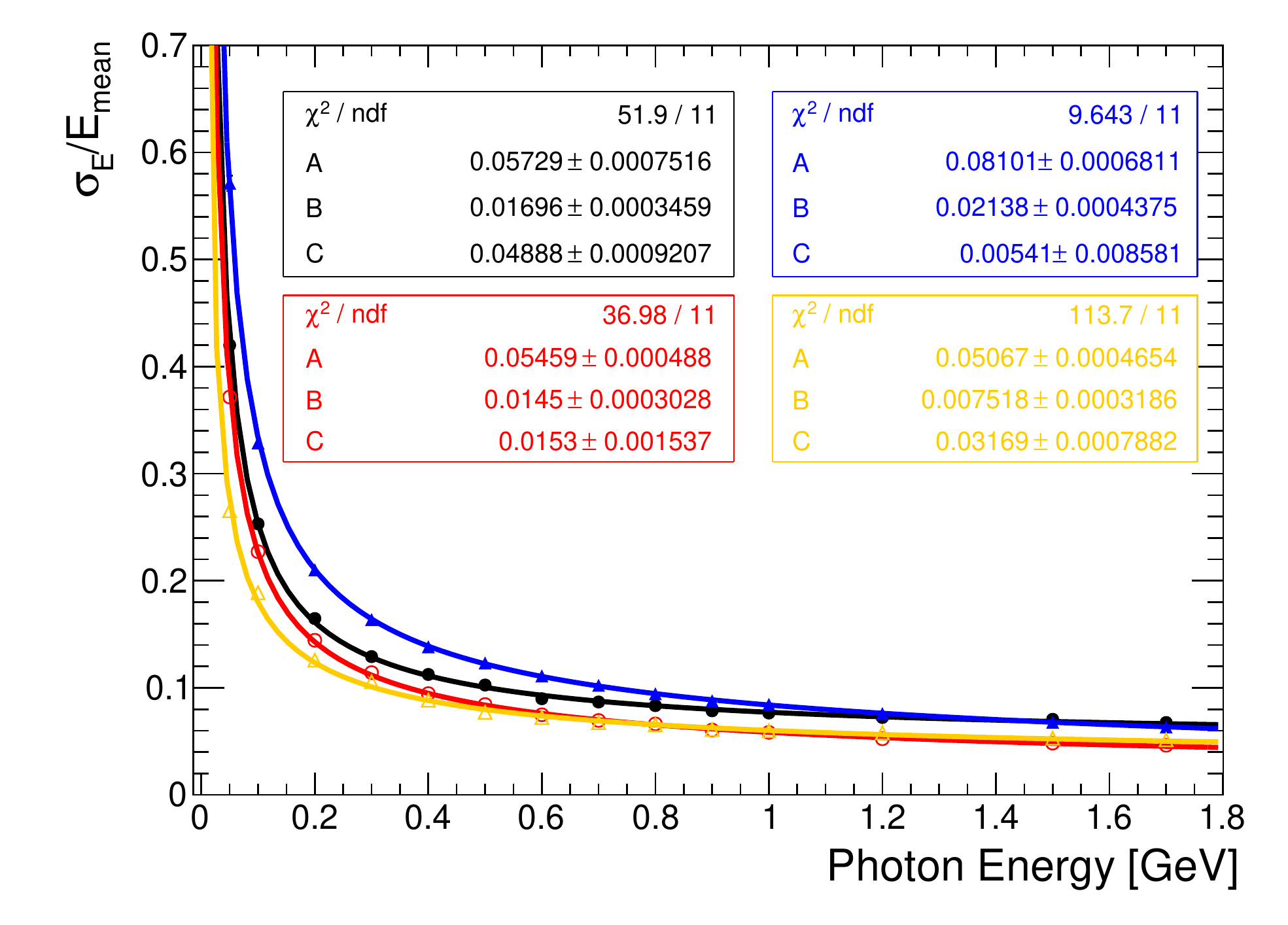}
\includegraphics[width=0.49\textwidth]{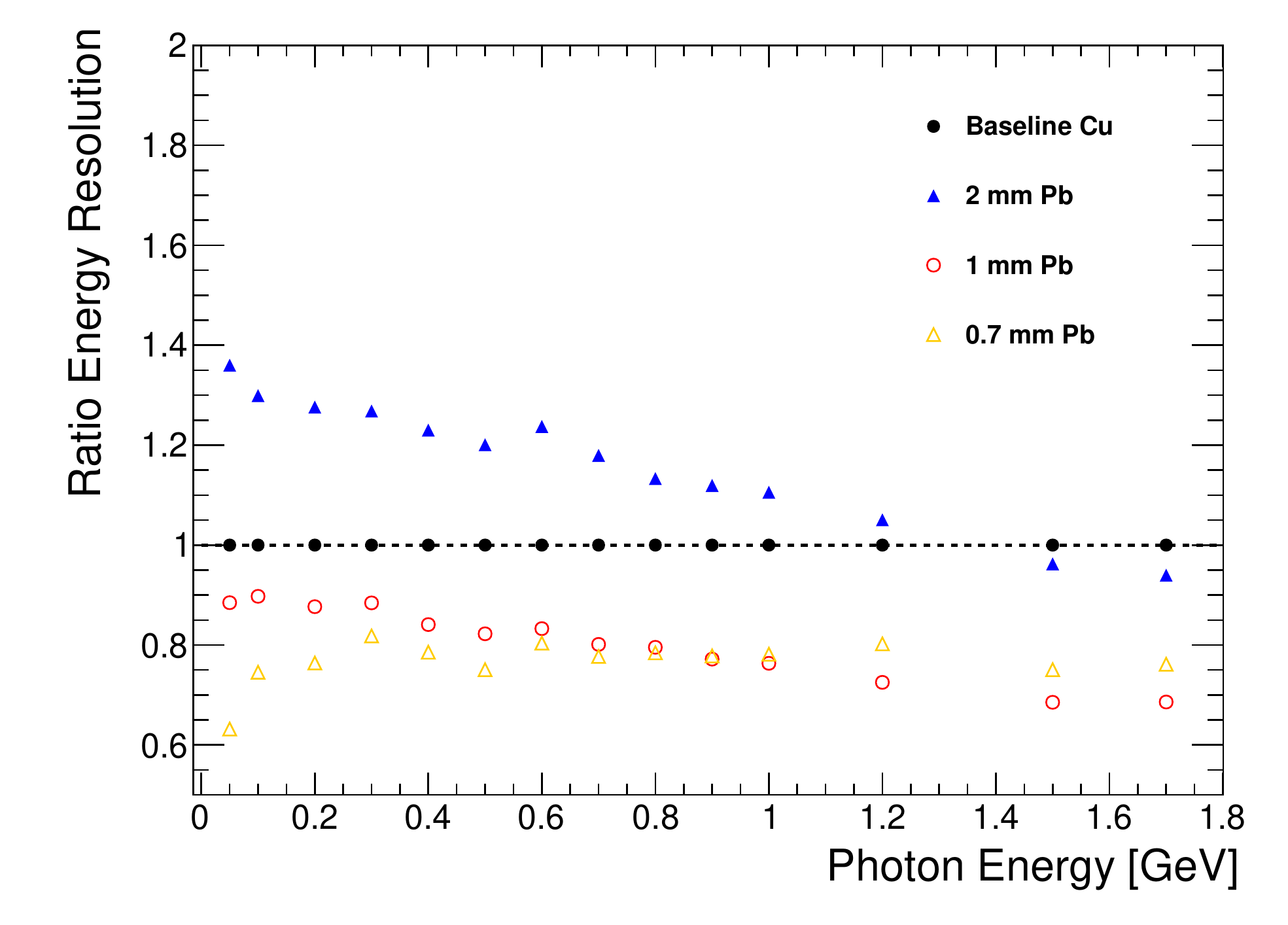}
\includegraphics[width=0.49\textwidth]{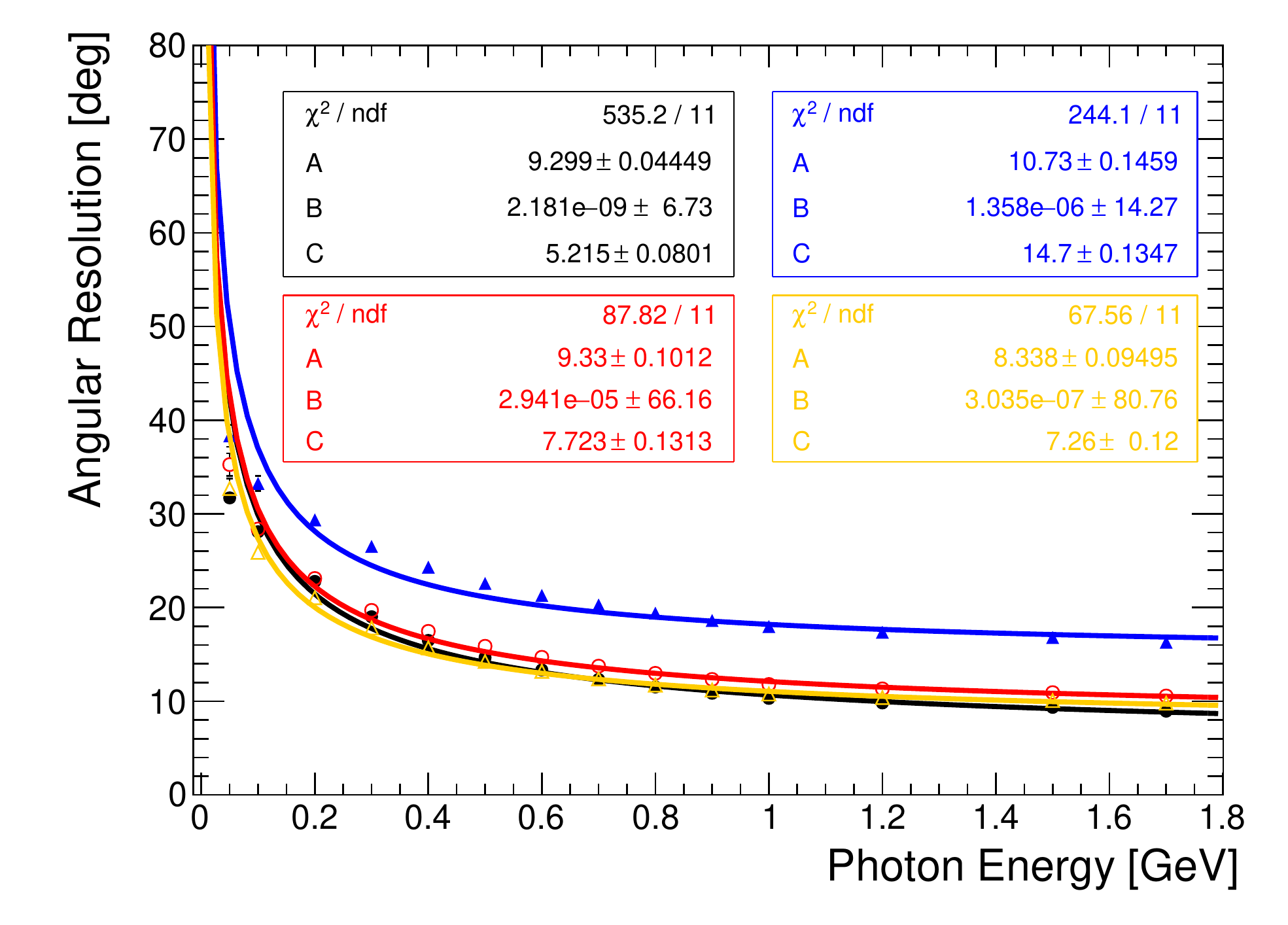}
\includegraphics[width=0.49\textwidth]{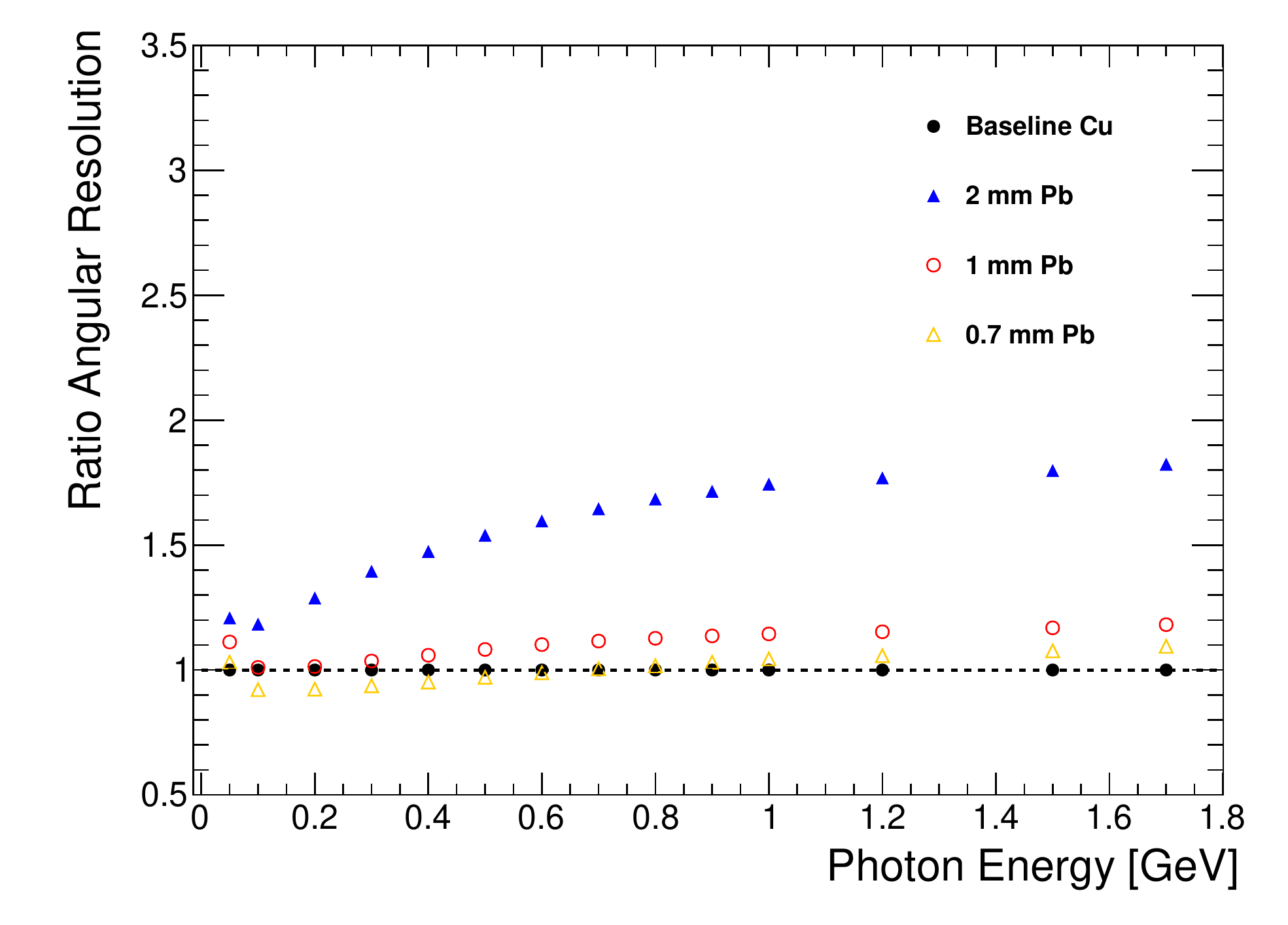}
\end{dunefigure}

\subsubsection{Particle identification}
\label{sec:mpd:ecalpid}

The \dword{ecal} has a complementary role with the \dword{hpgtpc} in terms of particle identification. Apart from identifying neutral pions via invariant mass reconstruction, it will help to separate electrons and photons, positrons and protons and lastly, muons and pions. For the muons and pions, measurement by range for particles stopping in the \dword{ecal} can be done. To separate electrons and photons, a measurement of $\frac{dE}{dX}$ can be done in the \dword{ecal}, as in \dword{minerva} \cite{Valencia:2019mkf}. Figure~\ref{fig:dEdX_ECAL2Planes_BDT}(left) shows the measurement of $\frac{dE}{dX}$ in the first two planes of the \dword{ecal}.

Finally, Figure~\ref{fig:ALICE_dEdx} shows the $\frac{dE}{dX}$ curve for the \dword{hpgtpc}. One can observe the crossing of the positron and proton curves around 1 GeV/c causing a degeneracy in the $\frac{dE}{dX}$ measurement. Here the \dword{ecal} can be complementary. A study has been done using boosted decision trees (BDT). Calorimetric variables such as shower shapes, shower maximum, hit radius, radius containing 90\% of the energy, and the velocity measured by time of flight (track length over the time of the \dword{ecal} cluster) are used as inputs to the BDT. Figure~\ref{fig:dEdX_ECAL2Planes_BDT} (right) shows the classifier output where positrons are signal and protons are background. A cut of 0.1-0.2 separates both with high efficiency and purity. Most of the separation power comes from the ToF measurement.

\begin{dunefigure}[Energy loss in the two first planes of the \dword{ecal} and BDT classification output to separate positrons and protons]{fig:dEdX_ECAL2Planes_BDT}{On the left: $e/\gamma$ discrimination using $\frac{dE}{dX}$ in the two first planes of the \dword{ecal}. On the right, BDT classifier output separating \SI{1}{GeV/c} positrons (signal) and protons (background).}
\includegraphics[width=0.49\columnwidth]{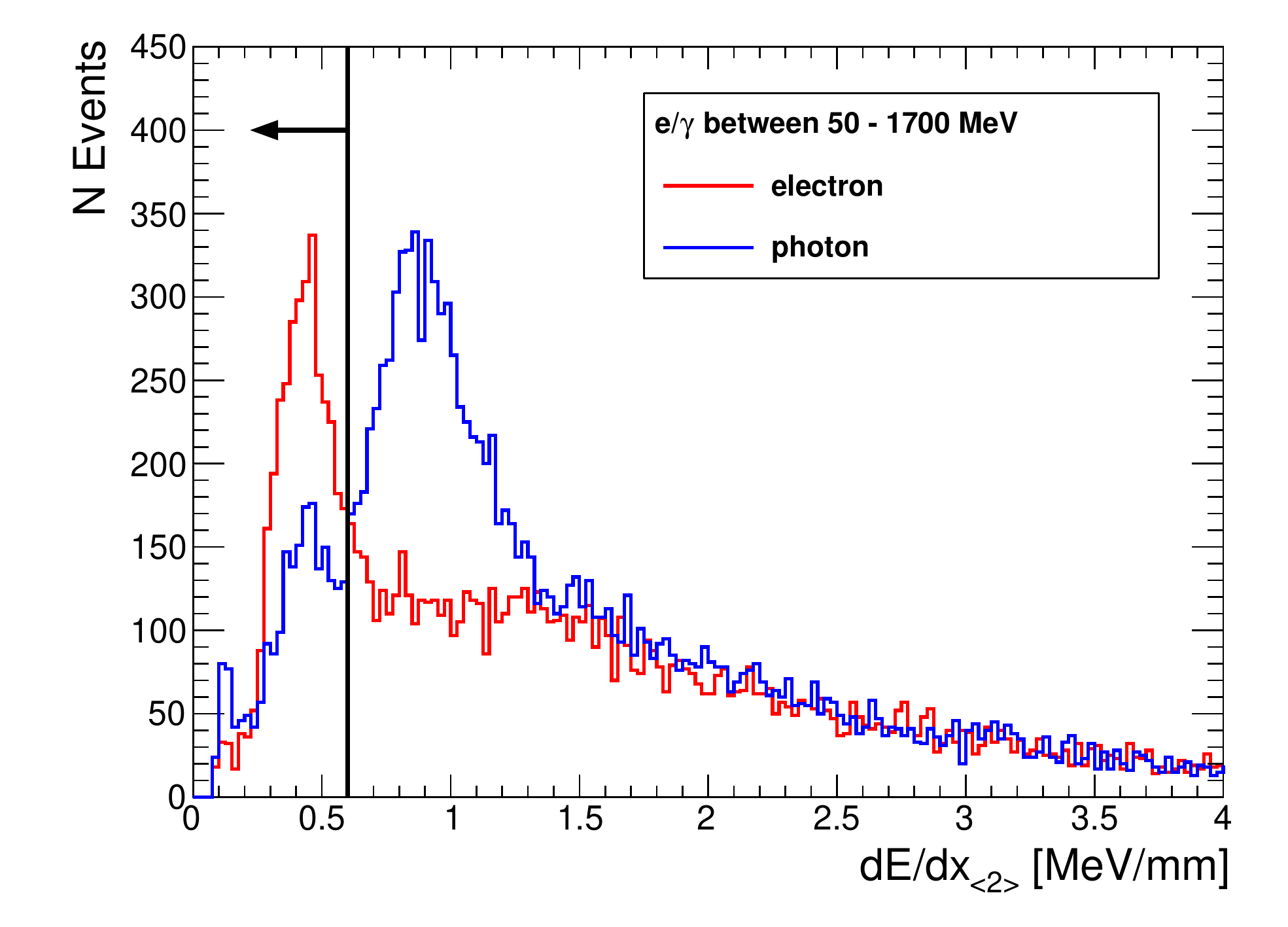}
\includegraphics[width=0.49\columnwidth]{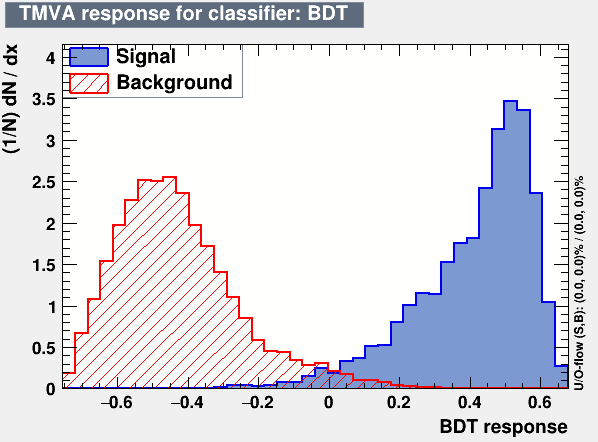}
\end{dunefigure}

\subsubsection{$\pi^0$ reconstruction}
\label{sec:mpd:pizeros}

For identification of neutral pions, both the energy and angular resolution are relevant. First, the reconstruction efficiency of photons in the \dword{ecal} needs to be high for all energies. The reconstruction efficiency of photons in the \dword{ecal} is shown in Figure~\ref{fig:gamma_reco_eff_ecal}. As expected, around 10\% of the photons convert in the \dword{hpgtpc}. Ignoring these, a reconstruction efficiency of 100\% is achieved above \SI{0.4}{GeV} dropping to 98.5\% at \SI{50}{MeV}. In an initial study, the position of the neutral pion is determined by using a $\chi^2$-minimization procedure taking into account the reconstructed energy of the two photons and the reconstructed direction of the photon showers \cite{Emberger:2018pgr}. The location of the decay vertex of the neutral pion can be determined with an accuracy between \SIrange{10}{40}{\cm}, depending on the distance from the downstream calorimeter and the $\pi^0$ kinetic energy. This is sufficient to associate the $\pi^0$ to an interaction in the \dword{hpgtpc}, since the gas will have less than one neutrino interaction per beam spill.
The pointing accuracy to the pion decay vertex may be further improved by a more sophisticated analysis technique and by using precision timing information, and is a subject of current study.
\begin{dunefigure}[Photon reconstruction efficiency]{fig:gamma_reco_eff_ecal}{Reconstruction efficiency of photons in the \dword{ecal} as a function of the photon energy. Photons were generated with a particle gun in the \dword{hpgtpc} volume. The efficiency is the number of events were the photon is reconstructed over the number of simulated photons at each energy. Converted photons in the TPC are ignored and account for around 10\% of events at each energy.}
\includegraphics[width=0.49\columnwidth]{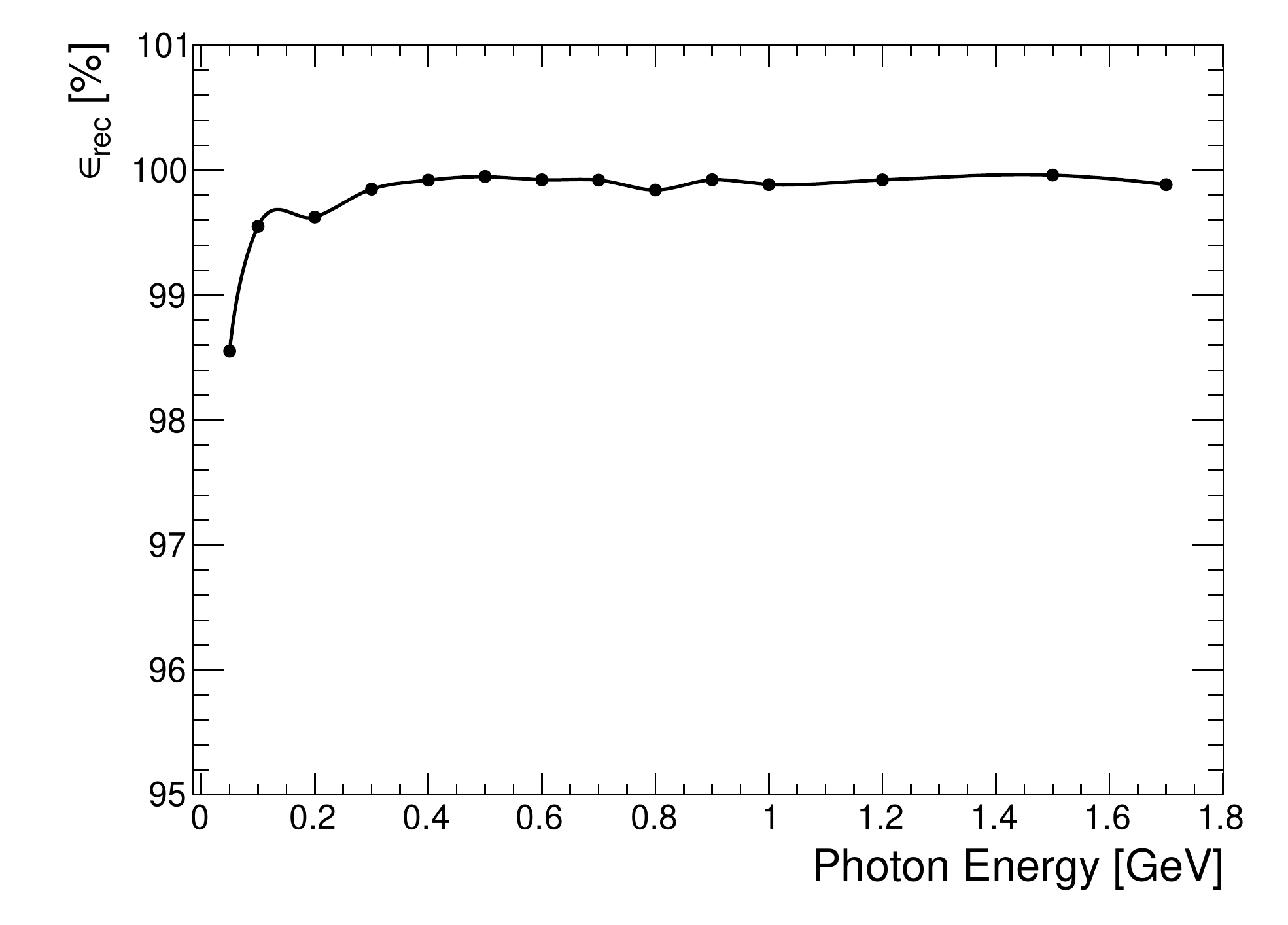}
\end{dunefigure}

\subsubsection{Neutron detection and energy measurement}

\label{sec:mpd:neutrons}


The \dword{ecal} is sensitive to neutrons that interact inside the plastic scintillator or nearby. By precisely measuring the time and position of a neutron-induced hit, it is possible to determine the neutron kinetic energy via \dword{tof}. Typically, knock-out protons have kinetic energies below \SI{10}{\MeV} and therefore travel less than \SI{1}{\mm} in scintillator.  For this measurement to be feasible, it is therefore essential to have good lateral segmentation to be able to distinguish the position of a neutron scatter in a single active layer. As the direction of the neutron cannot be generally determined from the neutrino scattering products, a correlation in time must be used to associate neutrons with their parent neutrino interaction.  In addition, it is difficult to know if a given neutron is primary (from the neutrino interaction) or secondary (backgrounds such as inelastic scattering of charged pions in the \dword{ecal}). 

\begin{dunefigure}[Number of ECAL hits as function of maximum hit energy]{fig:NeutronPhoton2DCut}{Number of hit cells in the ECAL cluster as a function of the maximum hit energy in the cluster for true photons (left) and true neutrons (right). Neutron clusters are selected to the right of the green line $N_{cells}$ = \SI{5}{\per\MeV} (Max cell [\SI{}{\MeV}] - 4).}
\includegraphics[width=0.49\columnwidth]{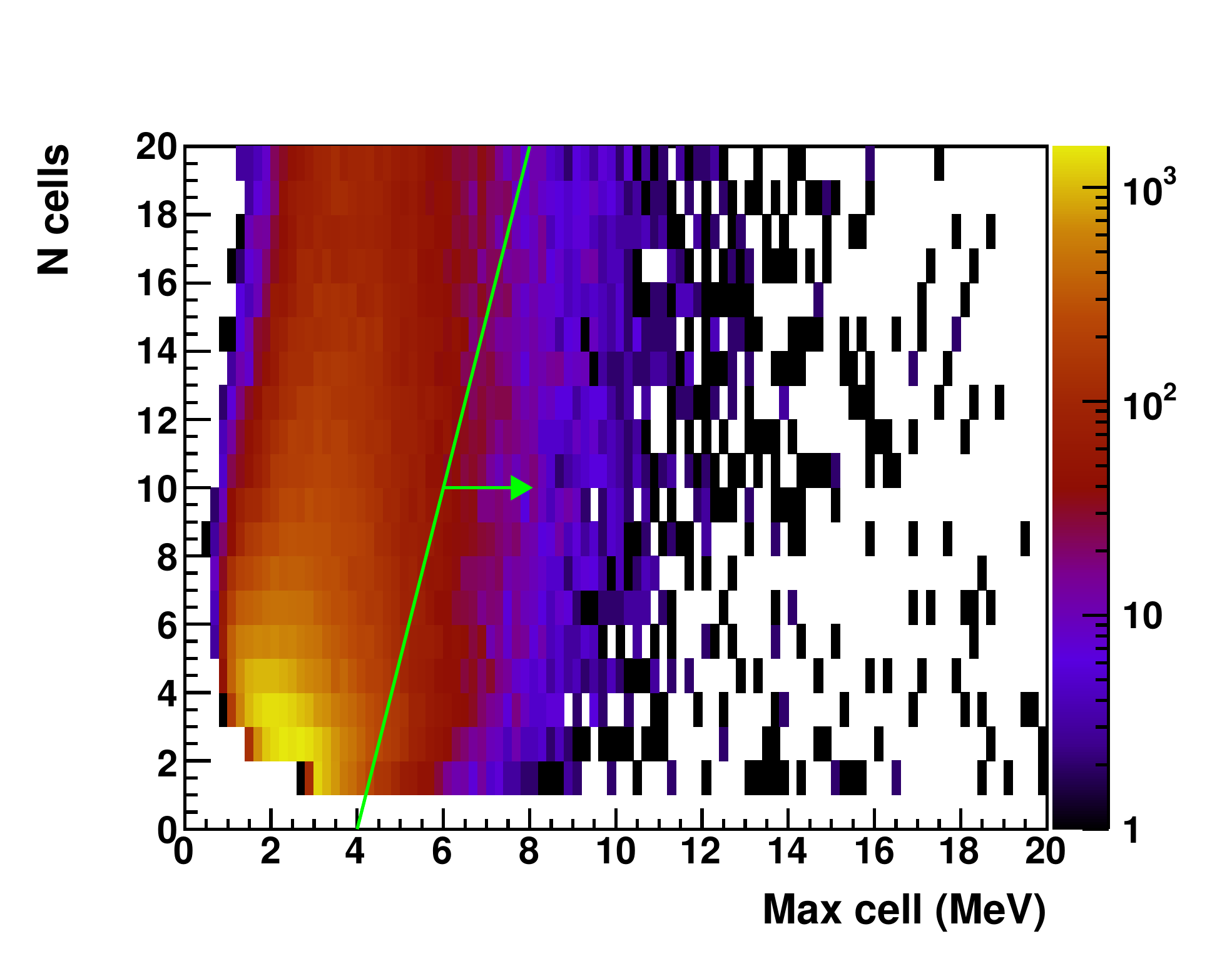}
\includegraphics[width=0.49\columnwidth]{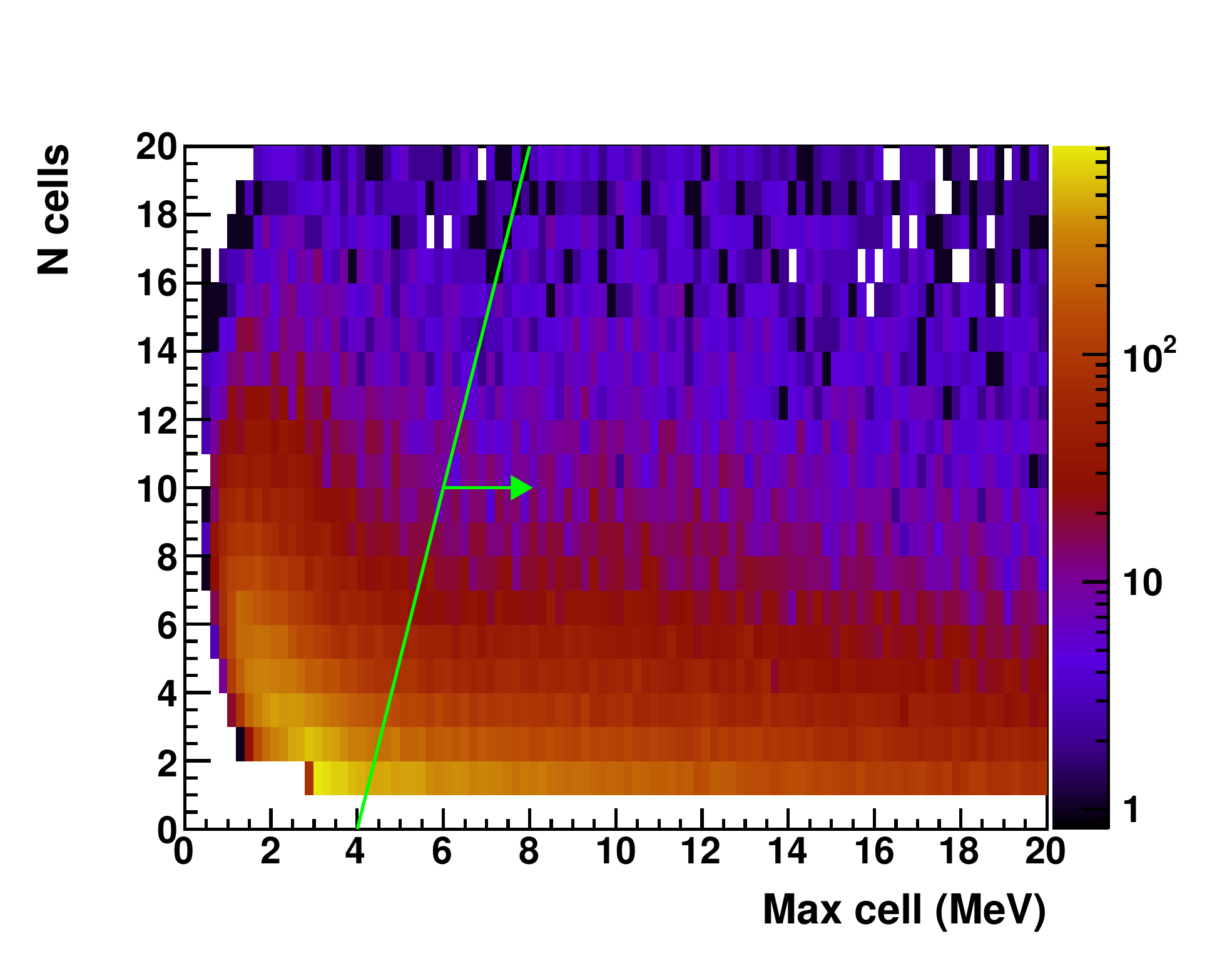}
\end{dunefigure}

For this study (which is described in more detail in Reference~\cite{bib:docdb19033}), the reference 60-layer octagonal \dword{ecal} geometry, described in Section~\ref{sec:ecal-performance}, is surrounded by an approximated magnet made of aluminium \SI{17}{\cm} thick; this magnet configuration has a mass of \SI{100}{tons}. This is sufficient to estimate the neutron background coming from neutrino interactions in the magnet. In addition, neutrino interactions in the rock volume of the \dword{nd} hall are considered\footnote{A volume of \SI{4}{\meter} upstream, \SI{2}{\meter} top and bottom was used. That volume is sufficient to include 95\% of the background rate from the full rock volume.}.

Energy deposits (corrected for scintillator quenching) in the \dword{ecal} are collected by dividing the \dword{ecal} into cells of $2\times2\times0.5$ \SI{}{\cubic\cm}. Hits are formed by aggregating all energy deposits within one cell. Only hits above \SI{0.5}{\MeV} are considered, corresponding to half the expected energy deposit from a minimum ionizing particle (MIP) traversing a scintillator tile. A time resolution of \SI{0.7}{\ns} is assumed for the \dword{ecal} and is applied to both the neutrino interaction vertex and the neutron cluster. A clustering algorithm is used to group hits that occur in-time and within \SI{5}{\cm} of each other, effectively grouping hits from charged single particles. As neutrons can scatter multiple times, for multiple isolated clusters that are within a cylinder of radius \SI{50}{\cm} from the neutrino interaction point, only the nearest candidate is considered, to reduce the possibility of multiple-counting. 

The event selection is performed as follows:
\begin{itemize}
    \item Neutron and photon-induced clusters are separated based on the total number of hits in the cluster, the total energy of the cluster and the maximum hit energy. Neutron clusters are selected by requiring at least \SI{3}{\MeV} of total visible energy and that $N_{hits}$ < \SI{5}{\per\MeV}($E_{max}$-4), where $N_{hits}$ is the number of hits in the cluster and $E_{max}$, the maximum hit energy in the cluster in MeV. This cut requires that clusters have few hits with at least one large energy deposit corresponding to the knock-out proton depositing most of its energy in the scintillator. Figure~\ref{fig:NeutronPhoton2DCut} shows the 2D distributions of the number of hits as a function of the maximum hit energy for both neutron and photon clusters.
    \item Further background can be rejected by requiring that the distance between an isolated cluster and a charged track is over \SI{70}{\cm}. This cut mostly rejects correlated background originating from a charged particle interacting in the pressure vessel or \dword{ecal} and producing a neutron subsequently. The distribution of the cluster distance to a charged track is shown in Figure~\ref{fig:DistChargedTrackCut_ECALVeto}.
    \item To improve the background rejection, especially from uncorrelated background, a veto can be applied using additional activity in the \dword{ecal}. It is expected that an isolated cluster from the uncorrelated background should be relatively close in distance to some other activity in the \dword{ecal}. On the other hand, an isolated cluster from signal is expected to be far from such other activity. Figure~\ref{fig:DistChargedTrackCut_ECALVeto} shows the distance between an isolated cluster and any other activity occurring within \SI{15}{\ns} of it elsewhere in the \dword{ecal}.  Uncorrelated background is rejected by eliminating clusters that are within \SI{2}{\m} of other \dword{ecal} activity. 
    \item Finally, an isolation cut is applied to the candidate neutron cluster. It is required that the candidate is further than \SI{70}{\cm} away from any other isolated cluster. This cut removes further correlated background that passed the first cut. The distribution of the distance to the nearest cluster is shown in Figure~\ref{fig:DistIsolationCut}.
\end{itemize}

\begin{dunefigure}[Cluster distance to charged track distribution and Distance to \dword{ecal} activity distribution (RHC)]{fig:DistChargedTrackCut_ECALVeto}{On the left, the distribution of the cluster distance to a charged track for RHC mode. Events are categorized
based on the origin of the cluster (neutron or photon), whether that particle comes from the signal neutrino interaction or not. Uncorrelated backgrounds are further divided into those coming from the rock and from the rest of the detector hall. A cut is made at \SI{70}{\cm}. On the right, the distribution of the distance of a cluster to any \dword{ecal} activity restricted to be \SI{15}{\ns} apart for RHC mode. The signal peaks at \SI{6}{\m} due to the fact that the signal is preferably forward and more probably observed downstream in the \dword{ecal}. The background is preferably upstream in the \dword{ecal}. A cut is made at \SI{2}{\m}.}
\includegraphics[width=0.49\columnwidth]{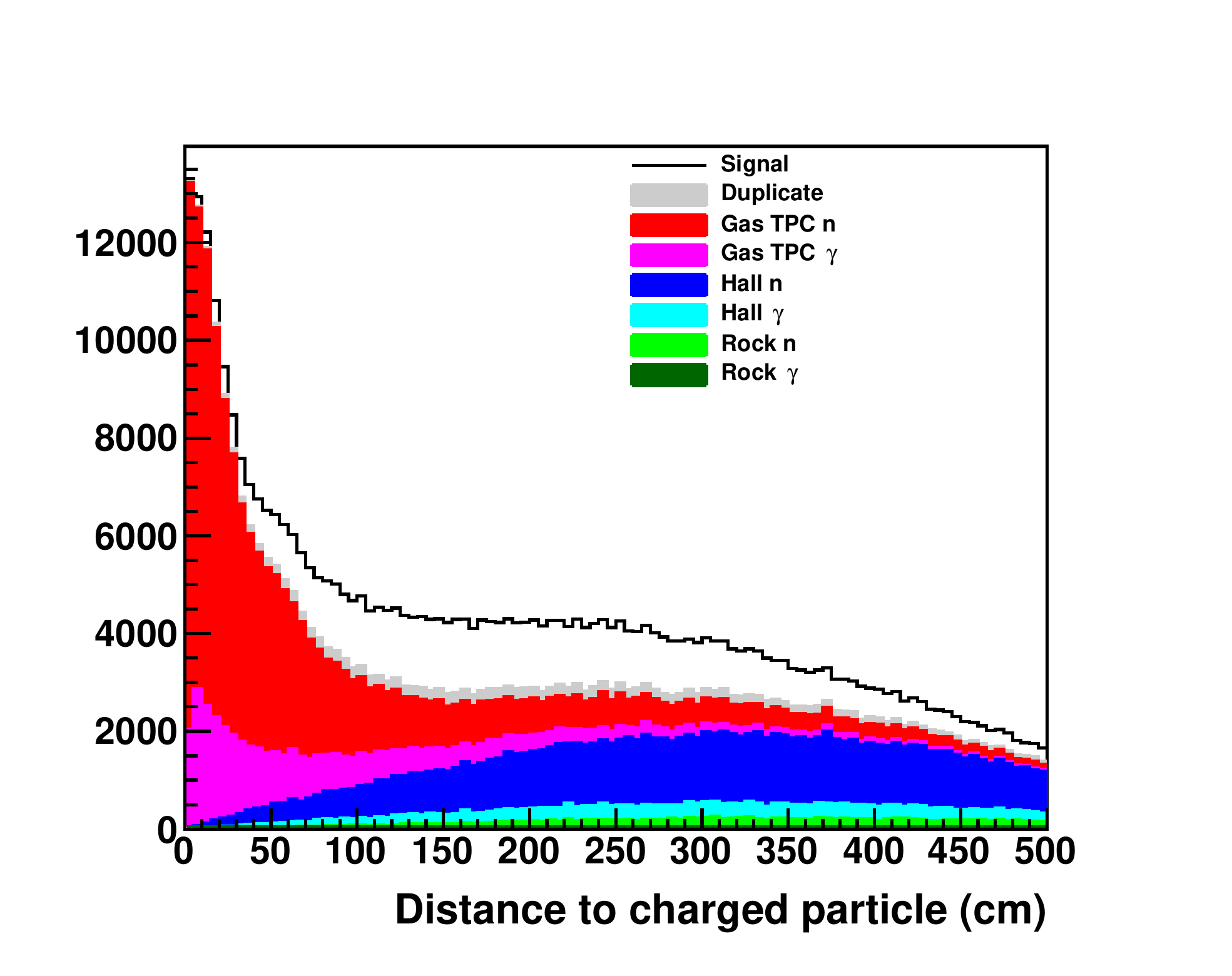}
\includegraphics[width=0.49\columnwidth]{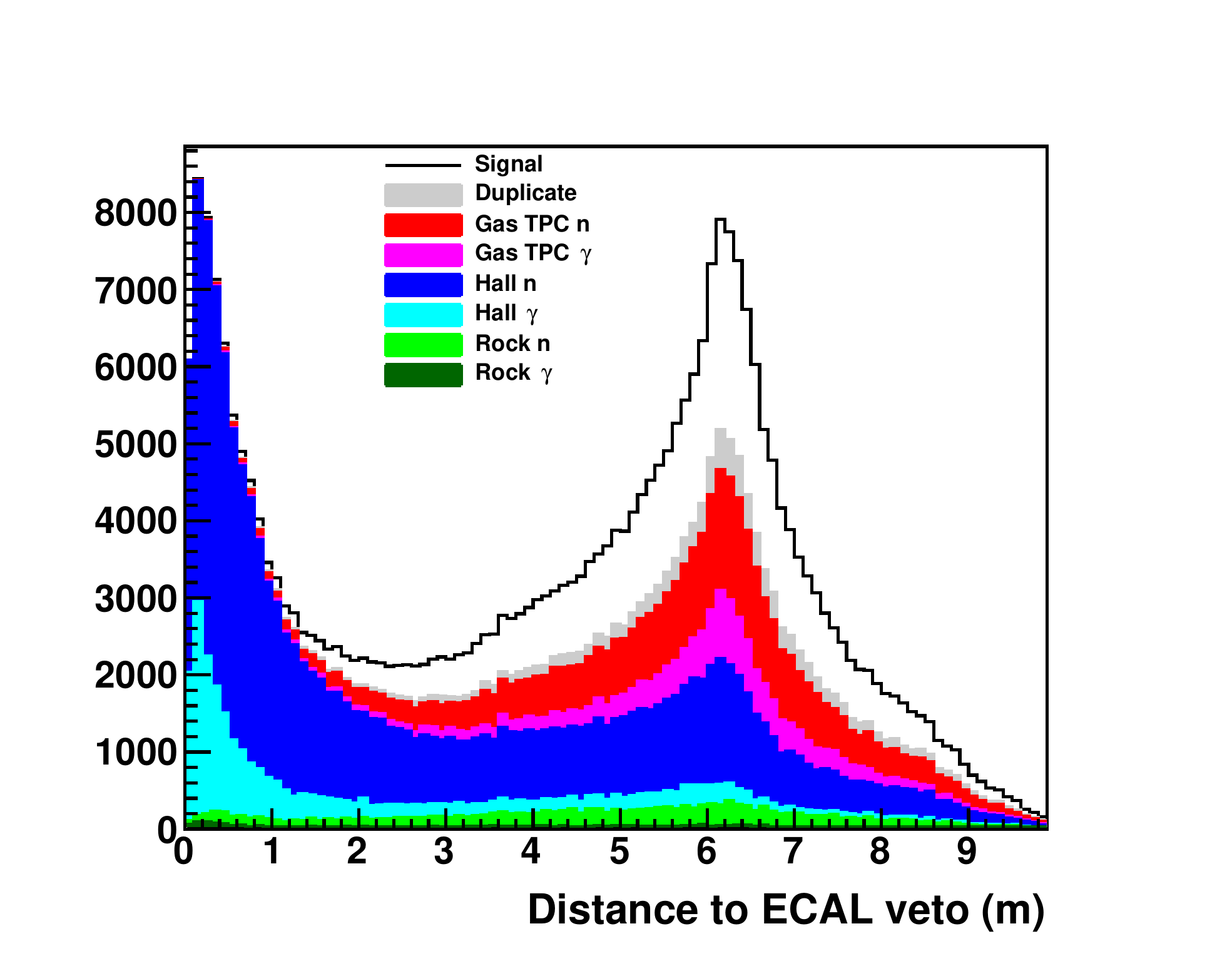}
\end{dunefigure}

\begin{dunefigure}[Distribution of the minimum distance between a cluster candidate and any other isolated cluster (RHC)]{fig:DistIsolationCut}{The distribution of the closest distance of the candidate cluster to any other isolated cluster for RHC mode. A cut is made at \SI{70}{\cm}.}
\includegraphics[width=0.65\columnwidth]{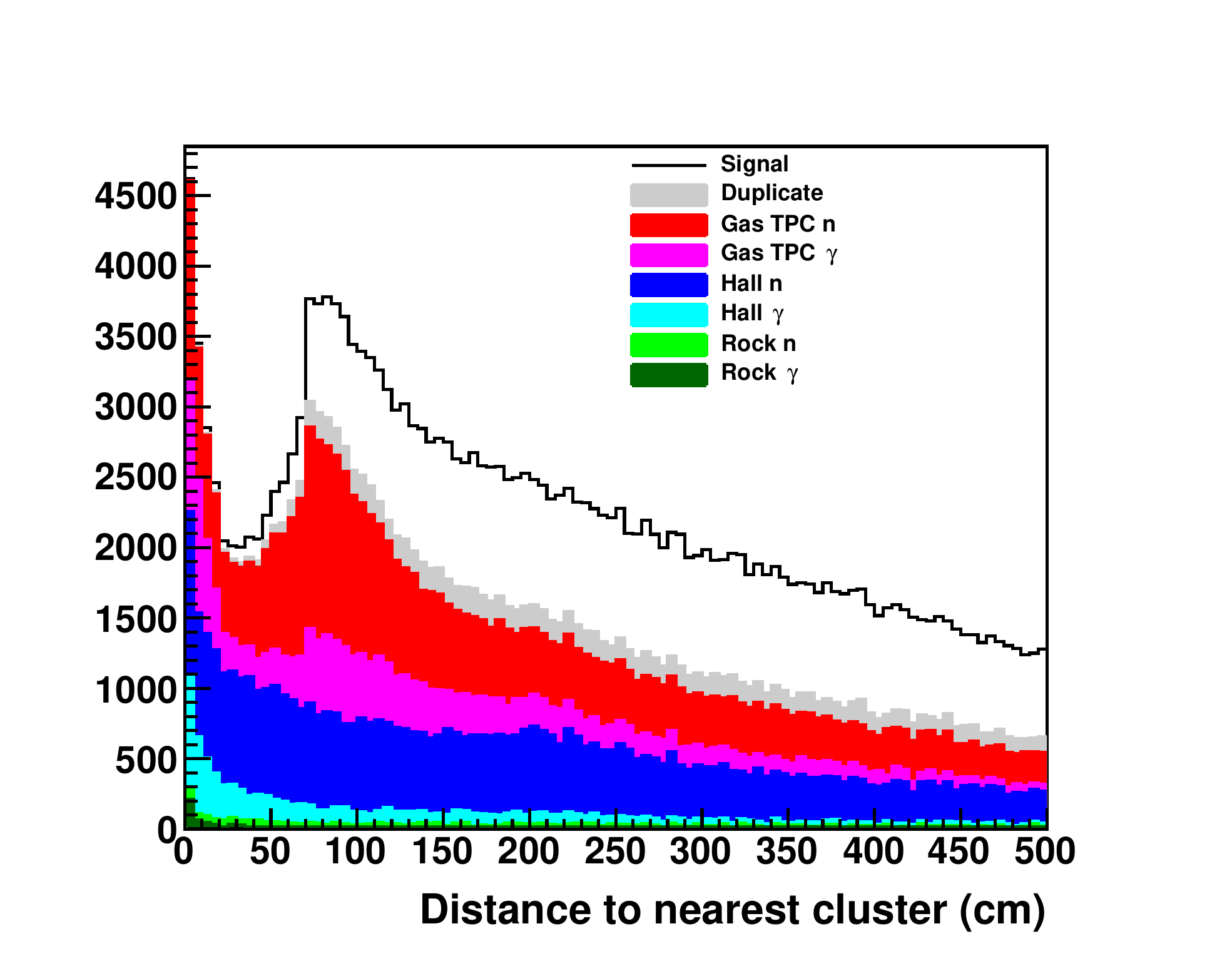}
\end{dunefigure}

The selected samples after the selection cuts are shown in Figure~\ref{fig:SamplesNeutron} for both FHC and RHC modes (both have the same cuts). The purity is lower at low reconstructed kinetic energies due to uncorrelated backgrounds and plateaus above around \SI{100}{\MeV}. The purity is better in RHC mode for several reasons: the background rate is reduced due to the lower cross-section for anti-neutrino scattering (fewer interactions in the surrounding material), in anti-neutrino interactions fewer charged hadrons are produced on average (reducing the amount of correlated background from secondary neutrons) and finally the signal rate is higher due to asymmetry in neutron production caused by the lepton charge. The efficiency and purity as a function of the reconstructed neutron kinetic energy are shown in Figure~\ref{fig:Eff_Purity}.

\begin{dunefigure}[Selected neutron samples]{fig:SamplesNeutron}{On the left, the selected neutron sample in FHC mode. On the right, the selected neutron sample in RHC mode. A significant difference in the purity can be seen at high reconstructed energies due to the facts stated in the paragraph above.}
\includegraphics[width=0.49\columnwidth]{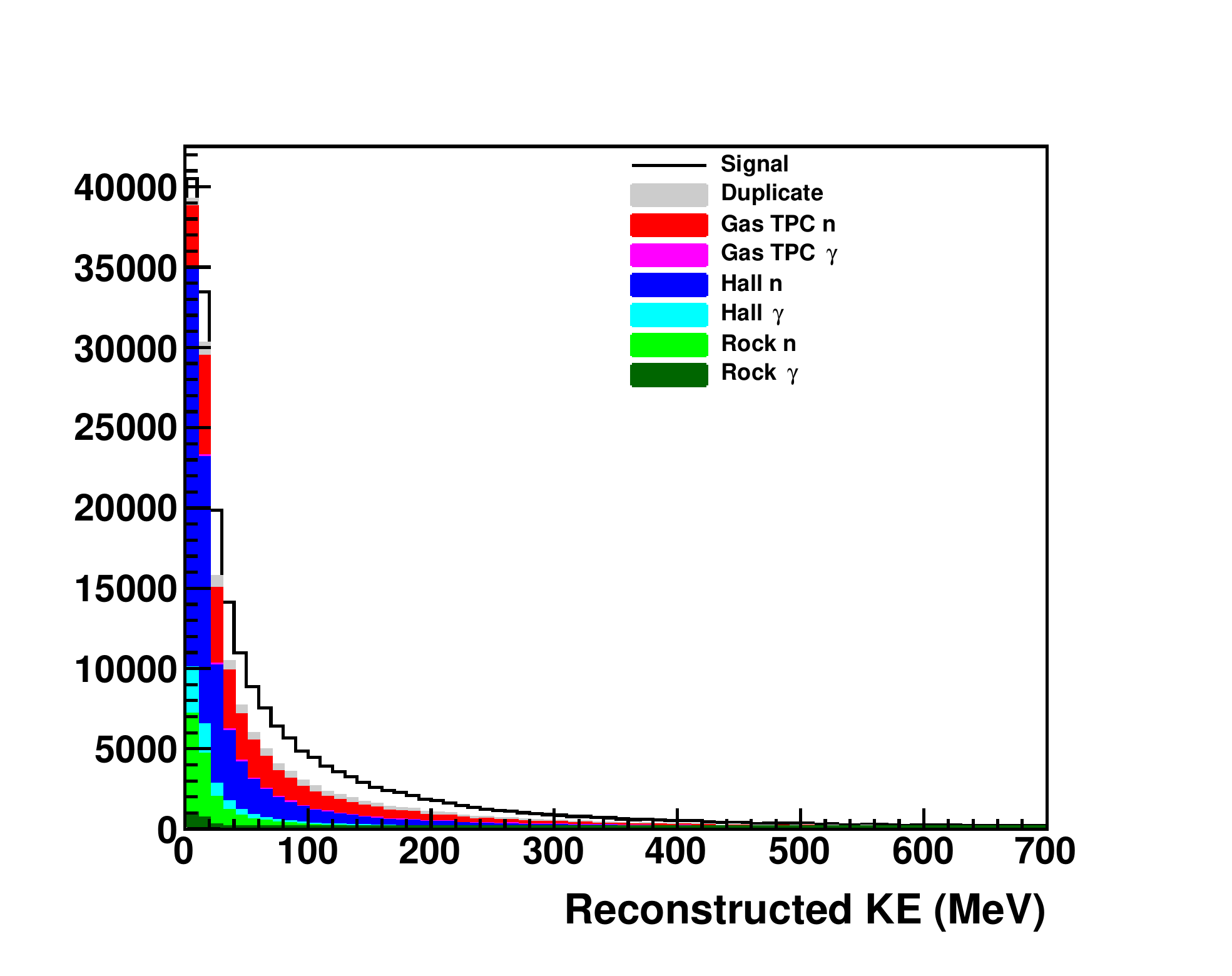}
\includegraphics[width=0.49\columnwidth]{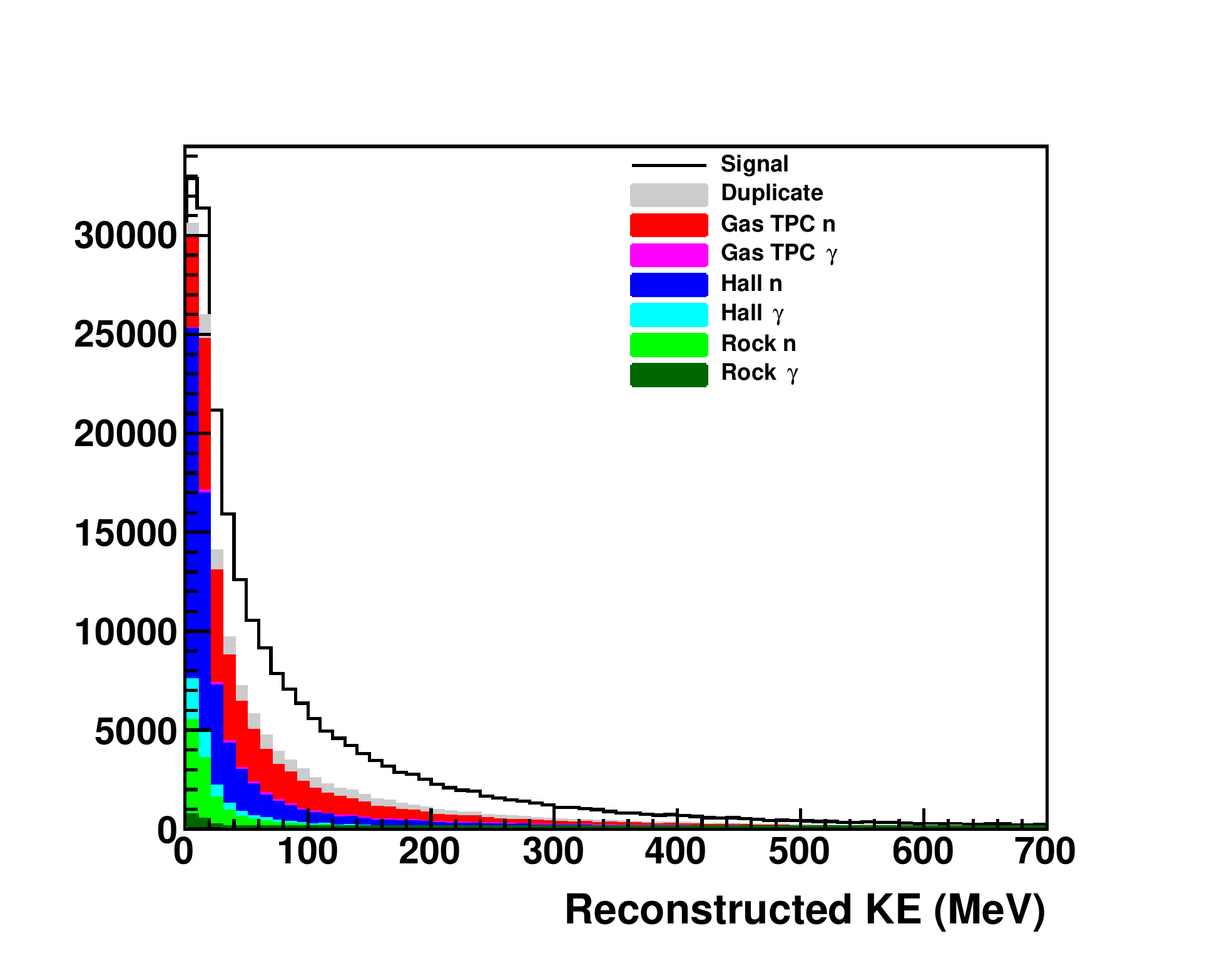}
\end{dunefigure}

\begin{dunefigure}[Efficiency and Purity of neutron samples]{fig:Eff_Purity}{On the left, selection efficiency as a function of the true neutron kinetic energy for the different selection cuts applied in RHC mode. 
On the right, the cumulative purity as a function of the reconstructed kinetic energy in RHC mode. A purity of around 55\% can be achieved for energies above \SI{100}{\MeV}. The fluctuations seen for energies above \SI{300}{\MeV} are due to the low statistics at these higher energies, as seen in Figure~\ref{fig:SamplesNeutron}. Different colors correspond to the same background categories as shown in Figure~\ref{fig:SamplesNeutron}. White is the signal.}
\includegraphics[width=0.49\columnwidth]{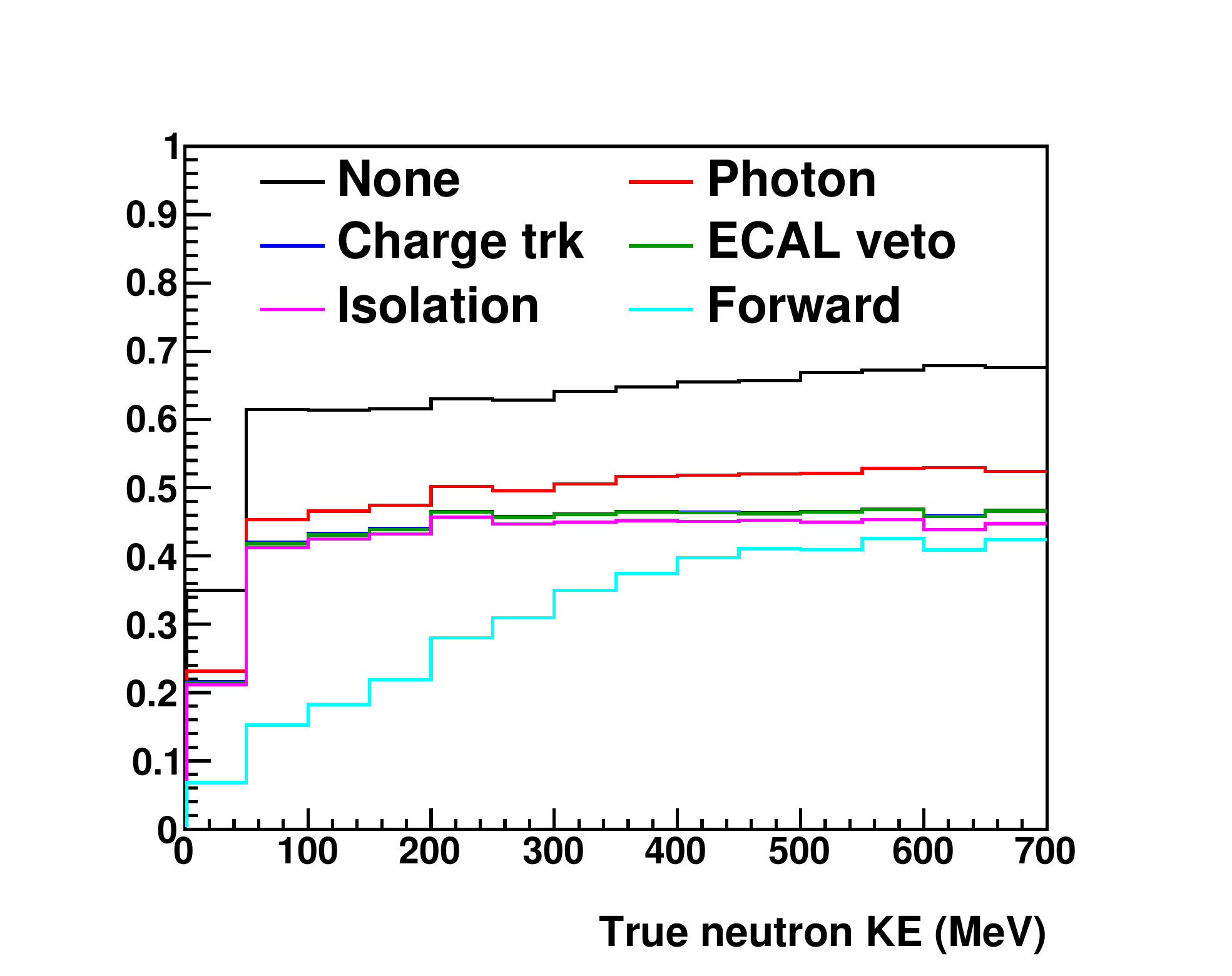}
\includegraphics[width=0.49\columnwidth]{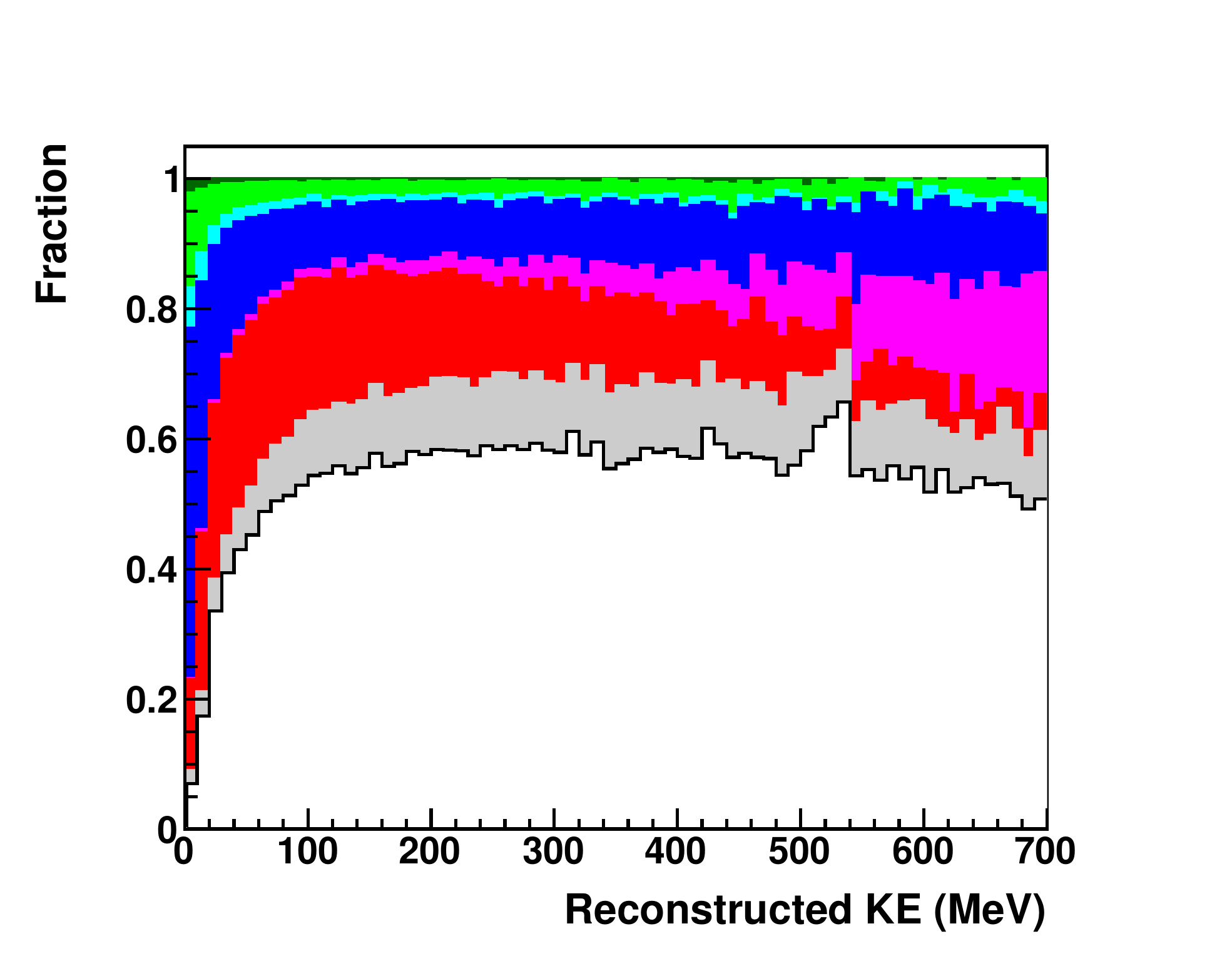}
\end{dunefigure}

In conclusion, this study demonstrates that a direct measurement of neutron production in $\nu$-Ar interactions is possible using a \dword{tof} technique in \dword{ndgar}. A sample purity of 40\% (55\%) for FHC (RHC) events can be obtained with a selection efficiency of around 45\% in both modes. Though the background is significant,  this would provide a unique opportunity to measure neutrons in neutrino interactions on argon. Envisioned improvements to this analysis are directly related to the fraction of passive and active material in the \dword{ecal}, which would significantly improve the energy resolution. Future optimization studies of \dword{ndgar} will take into account both photon ($\pi^0$) and neutron reconstruction performance. It may also be possible to enhance the signal by looking at the momentum balance based on other reconstructed final state particles.

A study of the sensitivity of the neutron energy measurements to changes in the neutron spectrum was done as follows.  A sample of $\anumu$ charged current events were reweighted with a weight based on the fraction of hadronic energy carried by neutrons, $E_n/E_{had}$.  The weighting scheme was defined to preserve the total cross section while increasing $E_n/E_{had}$ by 20\%.  Because $E_n/E_{had}$ and $E_{had}/E_\nu$ are energy dependent, it does not preserve flux. The main effect is to decrease the events that have zero final-state neutrons by $\sim 40\%$ and increase the events with neutrons. A modest change in the spectral shape was also produced.
\begin{dunefigure}[Neutron energy spectrum sensitivity]{fig:neutron:ke-reweight}{Left: Solid curves show true neutron energy distributions, with red being the spectrum reweighted to increase the neutron energy by 20\%. The corresponding points with errors show the measured, background subtracted spectra. Right: The ratio of the true weighted/unweighted neutron distributions (red line) compared with the ratio of the reconstructed distributions (black points).}
\includegraphics[width=0.49\columnwidth]{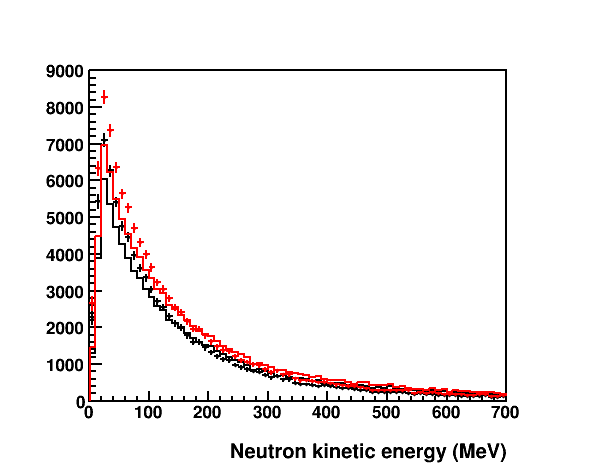}
\includegraphics[width=0.49\columnwidth]{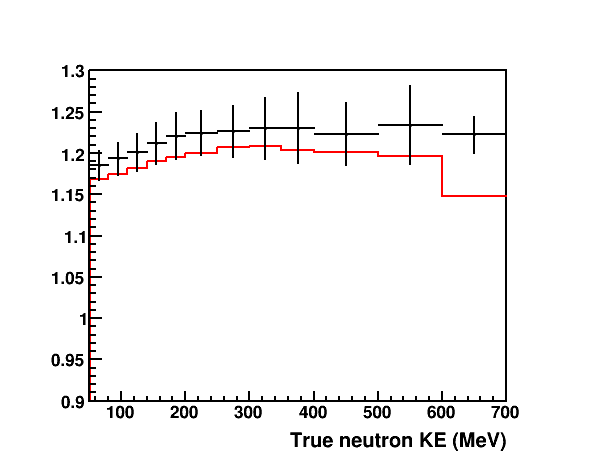}
\end{dunefigure}
The neutron energy spectra were then measured using the procedure outlined above. Backgrounds were subtracted assuming perfect knowledge but statistical errors on the background were propagated. The results of the study are shown in Figure~\ref{fig:neutron:ke-reweight}. The left figure shows the true (lines) and reconstructed spectra (points) for the two cases. The right hand shows the ratio of the true spectra (red line) and the ratio of the reconstructed spectra (points). In an ideal case the two curves would match. Here the reconstructed ratio overshoots the true ratio a bit but largely captures the systematic effect in both magnitude and shape\footnote{The overshoot is because the background due to duplicate neutrons (those that scatter multiple times such that one true neutron is reconstructed as multiple neutrons) is subtracted from MC. So when events with neutrons are weighted up, this background becomes larger and is underestimated, resulting in a slight overshoot of the background-subtracted "data".}. This study indicates that the \dword{ndgar} \dword{ecal} is potentially sensitive to mismodeling of the neutron energy spectrum at a level relevant for \dshort{dune}.


\subsection{Muon system performance}
\label{sec:mpd:muon}
A performance study of the muon system has been conducted. The system consists of three interleaved layers of \SI{10}{\cm} iron and plastic scintillator. The study was performed at the generator level, therefore interactions are based probabilistically on the momentum of the particle. 
In \dword{ndgar}, a muon or pion with a kinetic energy of around \SI{270}{\MeV} will range out in the \dword{ecal}. This enables separation of muons and pions by selecting on momentum and range (number of layers). The momentum cut-off is around \SI{380}{\MeV}/c, where pions start to go through the \dword{ecal}.


A simple sample selection has been done for $\nu_{\mu}$ CC interactions in the FHC and RHC beam modes. The first selection (defined as Raw) simply identifies the highest momentum track in the event as a muon. Then, the selection (defined as \dshort{ecal}) removes tracks that hadronically interact in the \dword{ecal}. If that occurs, the next highest momentum track is taken and categorized as a muon. Finally, the selection (defined as $\mu\text{ID}$) removes tracks that hadronically interact in the muon system. If that occurs, the same procedure is done and the next highest momentum track is categorized as a muon.

\begin{dunefigure}[Purity curves muon samples]{fig:SamplesMuonsPurity}{On the left, the purity curve for the muon sample in FHC mode. On the right, the purity curve for the muon sample in RHC mode.}
\includegraphics[width=0.49\columnwidth]{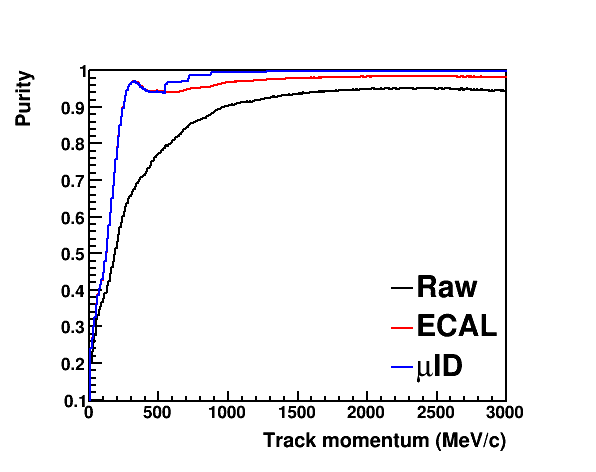}
\includegraphics[width=0.49\columnwidth]{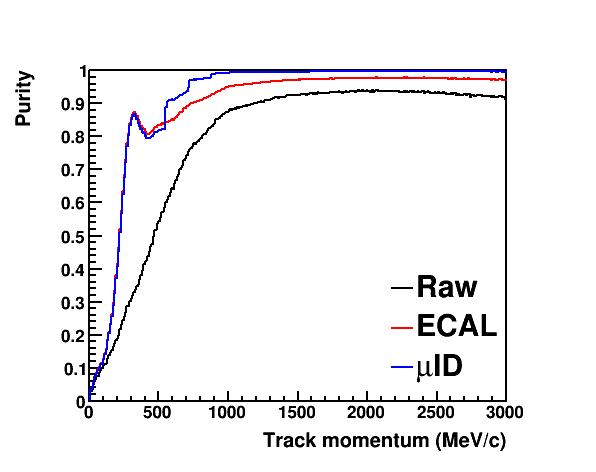}
\end{dunefigure}

As shown in Figure~\ref{fig:SamplesMuonsPurity}, without using the \dword{ecal} or muon system, the purity reaches 90\% at high momentum but falls very quickly at low momentum. The \dword{ecal} improves the purity in the momentum range up to the cut-off around 380 MeV/c, where the purity reaches over 95\%. By adding the muon system, the purity reaches 100\% over 1 GeV/c. The purity can be further improved in the low momentum region by separating muons and pions by range where the purity reaches 100\% up to the cut-off momentum. This is shown in Figure~\ref{fig:SamplesMuonsPuritywRange}. The cut-off momentum can be pushed up to around 480 MeV/c with a slightly thicker \dword{ecal} (80 layers).

\begin{dunefigure}[Purity curves muon samples with range selection]{fig:SamplesMuonsPuritywRange}{On the left, the purity curve for the muon sample in FHC mode. On the right, the purity curve for the muon sample in RHC mode. Both plots include muons and pions that range out in the \dword{ecal}.}
\includegraphics[width=0.49\columnwidth]{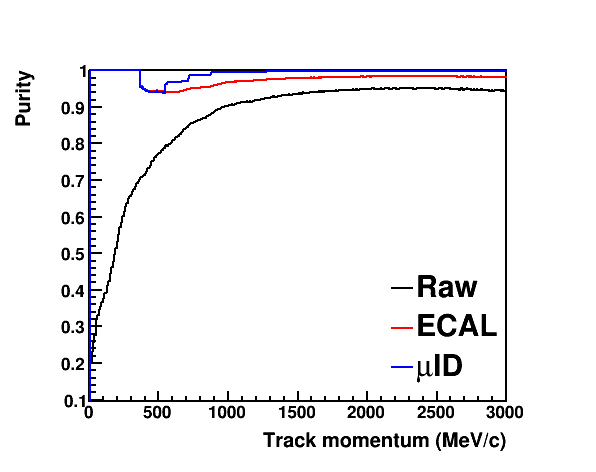}
\includegraphics[width=0.49\columnwidth]{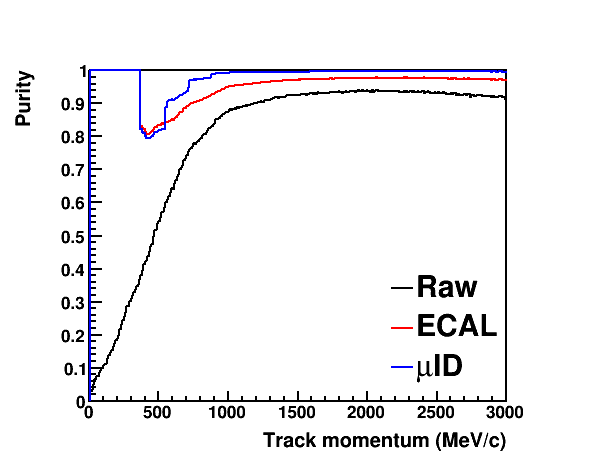}
\end{dunefigure}


Discrimination between backward traveling muons and pions has also been studied. The conclusion is that an upstream muon system is not needed if an upstream \dword{ecal} is present. Most backward traveling pions and muons have a momentum well below \SI{1}{GeV/c}
and will range out in the \dword{ecal}.

In addition, another study has been done in the FHC beam configuration by selecting $\mu^+$ created in $\anumu$ interactions. This selection is challenging because the ratio of muons to pions is much lower than in the case where $\mu^-$ are selected. 
One can observe in Figure~\ref{fig:WrongSignSelection} that without any selection, the purity is very low across the full momentum range. Adding the \dword{ecal} selection, the purity increases by a factor of approximately 2-3 for momenta above \SI{500}{MeV/c}. Below, muons or pions will range out and be easily separated. Finally, adding the muon system, the purity is significantly increased. It increases from 10\% at \SI{500}{MeV/c} to 40\% at \SI{800}{MeV/c} to above 80\% above \SI{1}{GeV/c}. The conclusion is that a muon system is required for even a modest purity \anumu sample in RHC mode.

\begin{dunefigure}[Purity of wrong sign sample selection in FHC]{fig:WrongSignSelection}{
Purity of the wrong sign sample selection of $\mu^+$ created by $\anumu$ interaction in FHC. It includes the ranging out selection of muons and pions in the \dword{ecal}.}
\includegraphics[width=0.49\columnwidth]{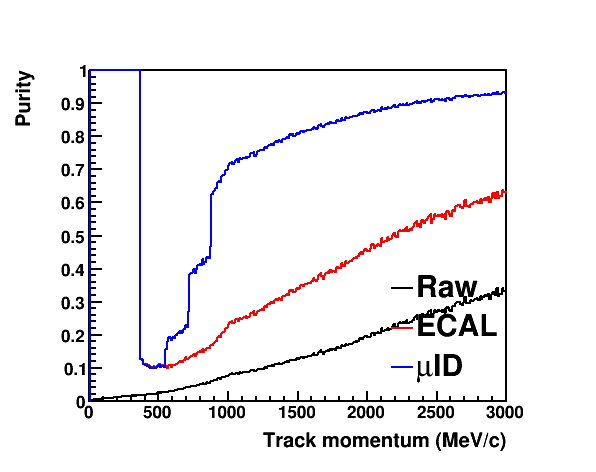}
\end{dunefigure}

The chosen thickness of the muon system will be important to limit the contamination of $\nu_{\mu}$-Ar CC interaction samples with pions. Figure~\ref{fig:PurityVsThickness} shows the purity in FHC and RHC modes for different thicknesses of the muon system. The thickness comes particularly important at high track momentum as most pions or muons below 500 MeV/c will range out in the \dword{ecal} or the first layer of the muon system. The purity above 1 GeV/c goes from 97\% with 10 cm to nearly 100\% with 30 cm independent of the running mode. Additionally, the NC background goes from 0.2\% to 2\% going from 30 cm to 10 cm. A thickness around 15 cm seems to be a good compromise between performance and size.

\begin{dunefigure}[Purity curves for different muon system thicknesses]{fig:PurityVsThickness}{On the left, the zoomed purity curve for the muon sample in FHC mode. On the right, the zoomed purity curve for the muon sample in RHC mode. Both plots shows the purity for different thicknesses of the muon system between 10 and 30 cm.}
\includegraphics[width=0.49\columnwidth]{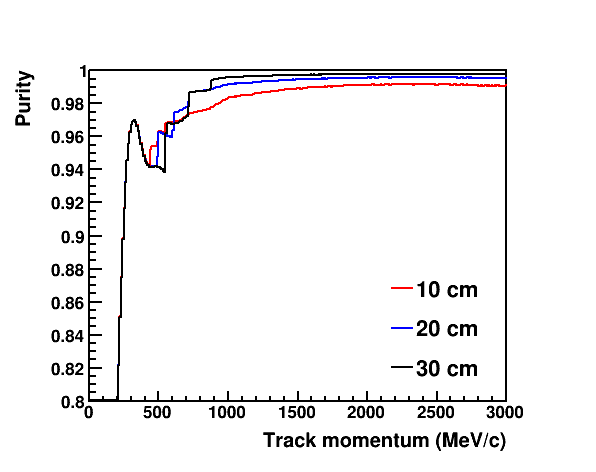}
\includegraphics[width=0.49\columnwidth]{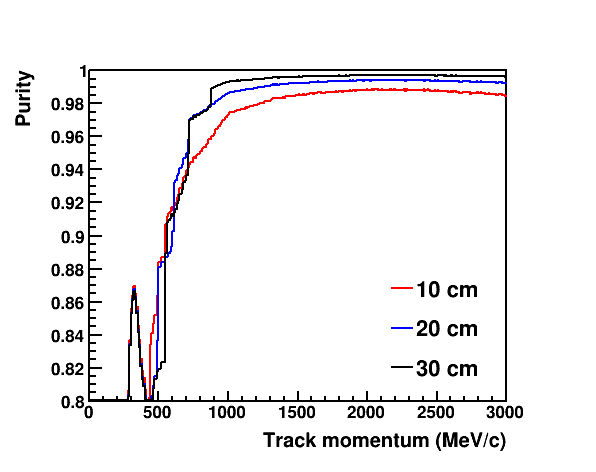}
\end{dunefigure}

\cleardoublepage

\chapter{DUNE-PRISM}
\label{ch:prism}


\section{Introduction to DUNE-PRISM}
\label{sec:prism-intro}

As long-baseline neutrino experiments move into the high precision era, one of the most difficult challenges will be to control systematic uncertainties due to neutrino interaction modeling. The relationship between the observable final state particles from a neutrino interaction and the incident neutrino energy is currently not understood with sufficient precision to achieve DUNE physics goals, due to missing energy from 
undetected and misidentified particles. These effects tend to cause   ``feed-down'' in the reconstructed neutrino energy which produces a low side tail in the reconstructed neutrino energy relative to the true energy. Since neutrino energy spectra at the far and near detectors are very different due to geometry and the presence of oscillations at the far detector, these reconstructed neutrino energy feed-down effects do not cancel in a far/near ratio as a function of true neutrino energy.  This can lead to biases in the measured oscillation parameters.


Bias in the neutrino energy estimate for a given interaction can originate from a number of sources.
The neutrons produced by neutrino interactions can induce a variable number of secondary interactions, and the detector response to these is not well correlated to the kinetic energy carried by the primary neutrons emerging from the argon nucleus. 
The amount of energy carried by neutrons is also expected to be different in neutrino and antineutrino interactions, which could reduce the sensitivity to the measurement of \deltacp{}.
In addition, any undetected or misidentified charged particles will produce biases, since the relationship between observed energy and true particle energy varies by particle. For example, charged pions that are undetected, or are misidentified as protons, will cause the neutrino energy estimate to be incorrect by the difference between the pion mass and the energy of the resulting Michel electron. Such biases in the estimation of incident neutrino energy depend on the kinematics of particles in the final state, and how they couple to the detector and reconstruction.

Constraining neutrino interaction uncertainties is particularly difficult, since no complete model of their interactions is available. If it were possible to construct a model that was known to be correct, even with a large number of undetermined parameters, then the task of a \dword{nd} would be much simpler: to build a detector that can constrain the undetermined parameters of the model. However, in the absence of such a ``correct'' model, this procedure will be subject to unknown biases due to the interaction model itself, which are difficult to quantify or constrain.

In the DUNE neutrino beam, the peak neutrino energy decreases as the observation angle relative to the beam direction increases, as shown in Figure~\ref{fig:prism-1doffaxis}. This property of conventional neutrino beams is used at T2K (44~mrad off-axis) and NOvA (15~mrad off-axis) to study neutrino oscillations in neutrino beams with narrower energy distributions than would be observed on-axis. The DUNE-PRISM (DUNE Precision Reaction-Independent Spectrum Measurement) \dword{nd} capability exploits this effect by making measurements at various off-axis positions with a movable detector, which provides an additional degree of freedom for constraining systematic uncertainties in neutrino interaction modeling. These measurements allow for a data-driven determination of the relationship between true and reconstructed energy that is significantly less sensitive to neutrino interaction models.  It also provides data samples that can be combined to produce a flux at the \dword{nd} that is very similar to the expected oscillated flux at the \dword{fd}.  This can be used to extract the oscillation parameters with minimal interaction model dependence.  

The analysis techniques enabled by DUNE-PRISM, and described in this chapter, will reduce the overall sensitivity of the oscillation parameter extraction to interaction model uncertainties and some detector effects.  It is an important part of the overall \dword{dune} \dword{nd} strategy
to control and reduce systematic uncertainties. It is important to note, however, that even within the context of a DUNE-PRISM analysis the interaction model is still used to make residual corrections and estimate uncertainties and the technique is still sensitive to the beam model.   

\begin{dunefigure}[Neutrino energy vs pion energy and observation angle and neutrino flux vs off-axis angle]
{fig:prism-1doffaxis}
  {Left: the observed neutrino energy in the lab frame from a decay-in-flight pion as a function of pion energy and observation angle away from the pion momentum direction. Right: the predicted DUNE beam muon neutrino flux at the \dword{nd} site as a function of off-axis angle. The arrows indicate the peak neutrino energy for three different off-axis angles.
  }
  \includegraphics[width=0.555\textwidth]{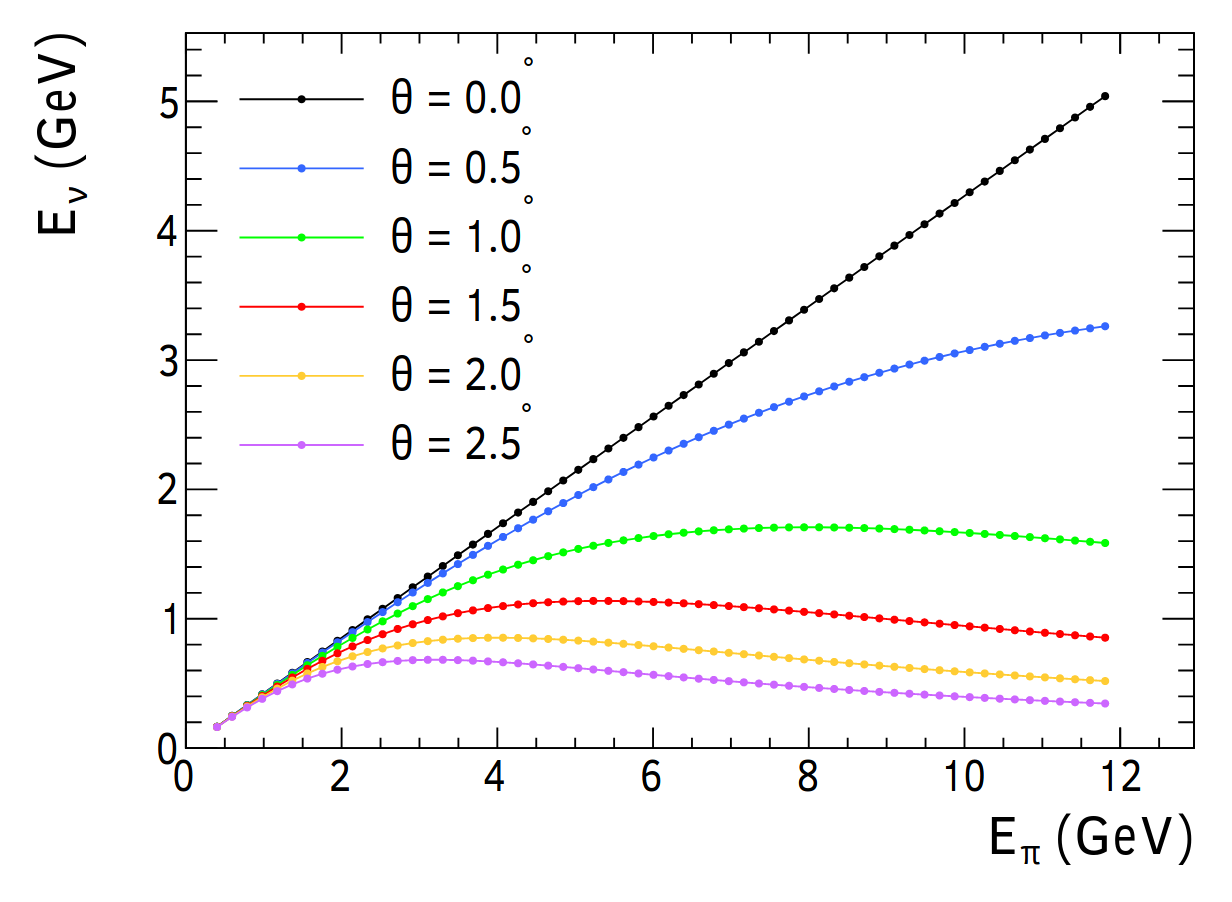}
  \includegraphics[width=0.425\textwidth]{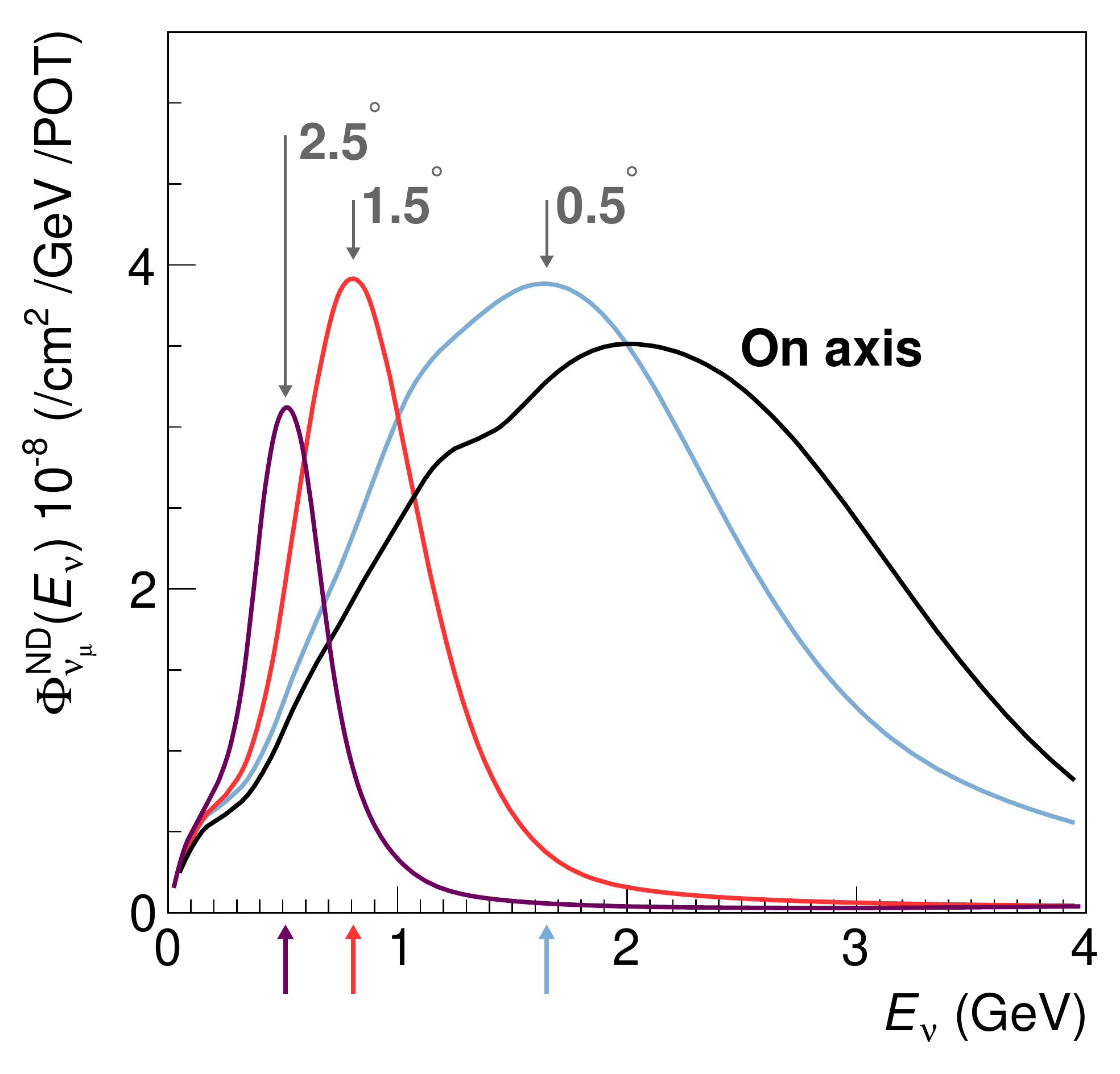}
\end{dunefigure}

This chapter explores the motivation for the DUNE-PRISM program and the plan for implementation of that program. The ways in which the DUNE-PRISM capability fits within the \dword{nd} requirements is discussed in Section~\ref{sec:DP-req}. Following that, Section~\ref{sec:DP-biasstudy} provides a rather detailed discussion of a mock data case study. This study illustrates the power of the off-axis data to uncover interaction modeling problems that might induce an unexpected bias in the extracted oscillation parameters.  Section~\ref{sec:duneprism} describes the nature of the off-axis \dword{lbnf} flux and a potential DUNE-PRISM run plan.  
The use of combined off-axis data samples to explore the detector response matrix, or to construct \dword{nd} samples with fluxes similar to those of the oscillated \dword{fd} samples, is discussed in Section~\ref{sec:prism-lincomb}.  The way the individual sample flux uncertainties propagate to the linearly-combined sample is explored in Section~\ref{sec:prism-flux-systs}.  Finally, Section~\ref{sec:prism-oa} outlines the steps of a complete DUNE-PRISM linear combination analysis.

\section{Requirements}
\label{sec:DP-req}

The requirements for the ND are described in detail in Section~\ref{sec:intro-requirements}. This section briefly recaps those requirements that directly relate to DUNE-PRISM.

The overarching requirement for DUNE-PRISM (ND-O4) is that the experiment take data with different energy spectra to constrain the relationship between the true neutrino energy and the energy reconstructed within the near detector.
A unique feature of DUNE-PRISM is that it provides a direct way of transferring near detector measurements to the far detector (ND-O1), by using data itself to form the predicted far detector oscillated energy spectrum (ND-O0).

The ND-O4 requirement necessitates that the experiment take data with different neutrino spectra in the region of interest, so as to disentangle the effects due to mismodeling of the neutrino flux, neutrino interaction cross-sections, and detector response. 
This is achieved by moving the detector off the beam axis, which lowers the mean of the incoming neutrino spectrum as well as the reducing its spread. Continuously varying energy spectra in the region of interest for neutrino oscillation measurements are needed to validate the model across these energies. 
In a complementary approach, measurements made at various off-axis positions can be combined, with appropriate weights, to predict oscillated spectra at the FD; these can then be used to extract oscillation parameters with minimal interaction model dependence. 
To make these measurements, it is necessary to (a) take data up to 30.5 m off-axis, so as to cover the entire energy range (capability requirement ND-C4.1), (b) maintain uniform detector performance to within 1\% across the full off-axis range (capability requirement ND-C4.2), (c) position the detector with a granularity of better than 10 cm and a desired precision ($<1$~cm) (capability requirement ND-C4.3), so as to control spectrum and detector response variations, (d) limit the downtime while the detector is being moved to a new position (capability requirement ND-C4.4), and (e) make a full suite of measurements within one year to mitigate the effects of expected beam and detector variations (capability requirement ND-C4.5).


\section{Oscillation Parameter Biases From Neutrino Interaction Modeling}
\label{sec:DP-biasstudy}


Long-baseline neutrino experiments use \dword{nd} data to validate and improve the neutrino interaction model used to extrapolate \dword{nd} measurements for a prediction of the oscillated flux at the \dword{fd}. When the \dword{nd} data disagree with the model, additional degrees of freedom are added to the model to force agreement. Ideally, these model modifications will be motivated by known physical effects that are believed to be absent from the model. However, it is usually the case that empirical corrections are needed to match the model to \dword{nd} measurements. If the wrong corrections  are chosen,
it is possible to achieve good agreement with \dword{nd} data while still producing a biased prediction for the \dword{fd}.  Data taken at off-axis positions as part of the DUNE-PRISM program make this unknown mismodeling less likely to happen because the agreement must work across all the different off-axis spectra.

To study such a situation, a ``mock dataset'' was produced by modifying the outgoing particle kinematics in a manner that cannot be reproduced by any choice of parameters in the interaction model used in the DUNE oscillation analysis.
Much of the focus in neutrino-nucleus interaction physics over the past decade has been on understanding the final state lepton kinematics, which are of primary importance for Cherenkov detectors such as Super-Kamiokande, and the detailed composition and kinematics of the hadronic state are less well understood. As an example, Figure~\ref{fig:neuttprotcomparisons} shows the spread of model predictions from three different neutrino interaction simulations, where the predictions of \textsc{GENIE} 2.12.2 and \textsc{GENIE} v3.0.6 (using the \texttt{N18\_10j\_02\_11a} tune from the NO$\nu$A 2020 oscillation results) exhibit large differences in the mean final state proton energy in bins of neutrino energy. For this study, a mock dataset is chosen that assumes that 20\% of the kinetic energy assigned to protons in the nominal model is instead carried away by neutrons (i.e., not detected and included in the reconstruction). Additional modifications are made to the cross section model using a multi-dimensional reweighting method to restore agreement between the model and the \dword{nd} mock data. 
The reweighting can simultaneously modify several model parameters, such as the differential cross sections in proton, muon, and pion kinetic energies, and several angles between the muon and components of the hadronic system. At the end of this procedure, the neutrino interaction model matches the mock data in reconstructed neutrino energy (and an arbitrary number of additional observable distributions), but the relationship between true and reconstructed neutrino energy is different in the mock data and the model.


\begin{dunefigure}[Simulated proton kinetic energy distribution from NEUT, \textsc{GENIE} v2, and \textsc{GENIE} v3]{fig:neuttprotcomparisons}
{The DUNE-flux-averaged mean kinetic energy to protons as predicted by the \textsc{GENIE} v2.12.2, \textsc{NEUT}, and \textsc{GENIE} v3.0.6 simulations. Model differences on the order of the proposed mock data set can be seen over the energy range that drives the experimental sensitivity to neutrino oscillation parameters. The NO$\nu$A CMC refers to the GENIE 'Comprehensive Model Configuration' \texttt{N18\_10j\_02\_11a}, that was developed for the 2020 NO$\nu$A oscillation analysis. }
  \subfloat[][]{
    \includegraphics[width=0.45\textwidth]{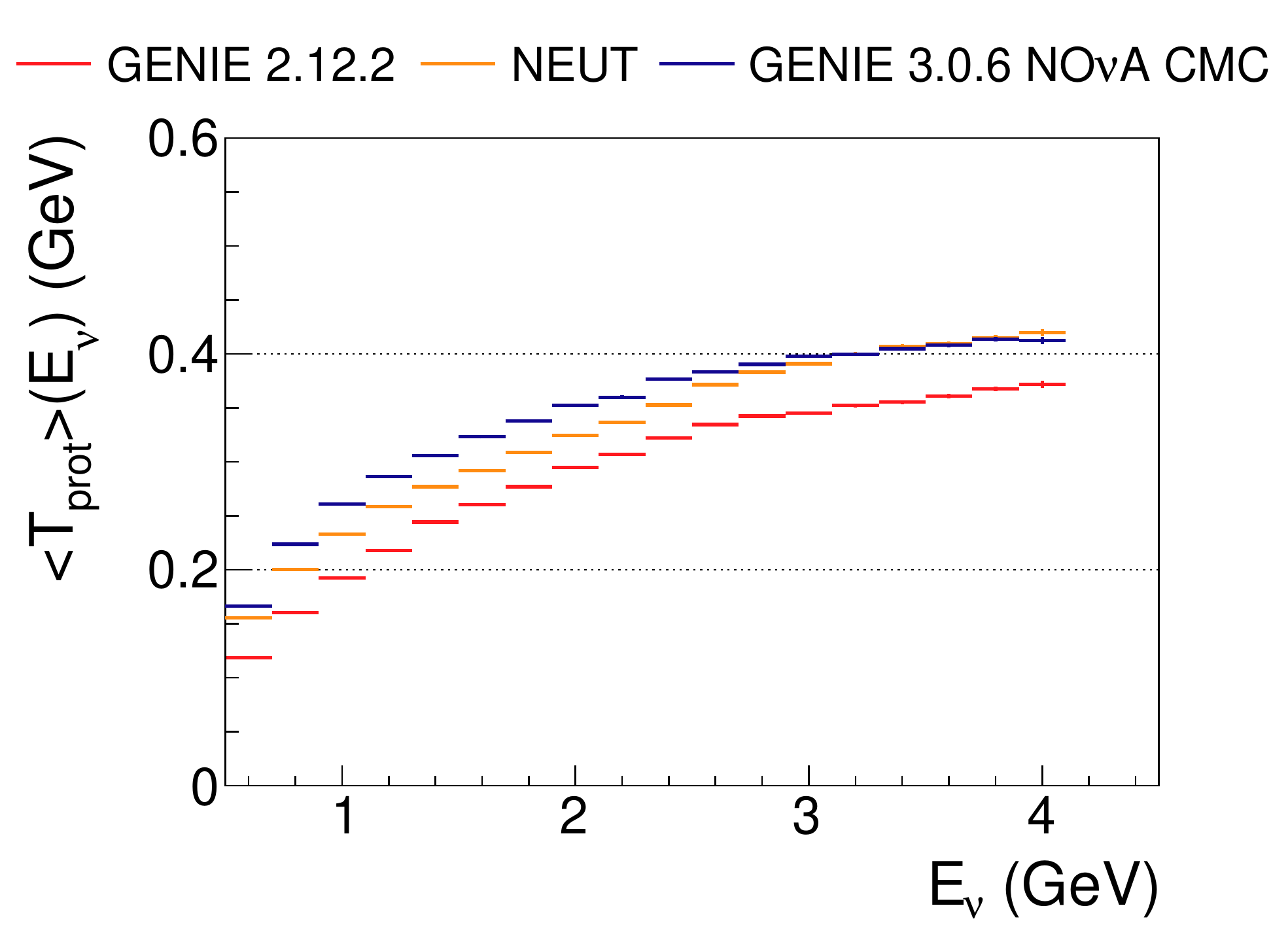}
  }
  \subfloat[][]{
	\includegraphics[width=0.45\textwidth]{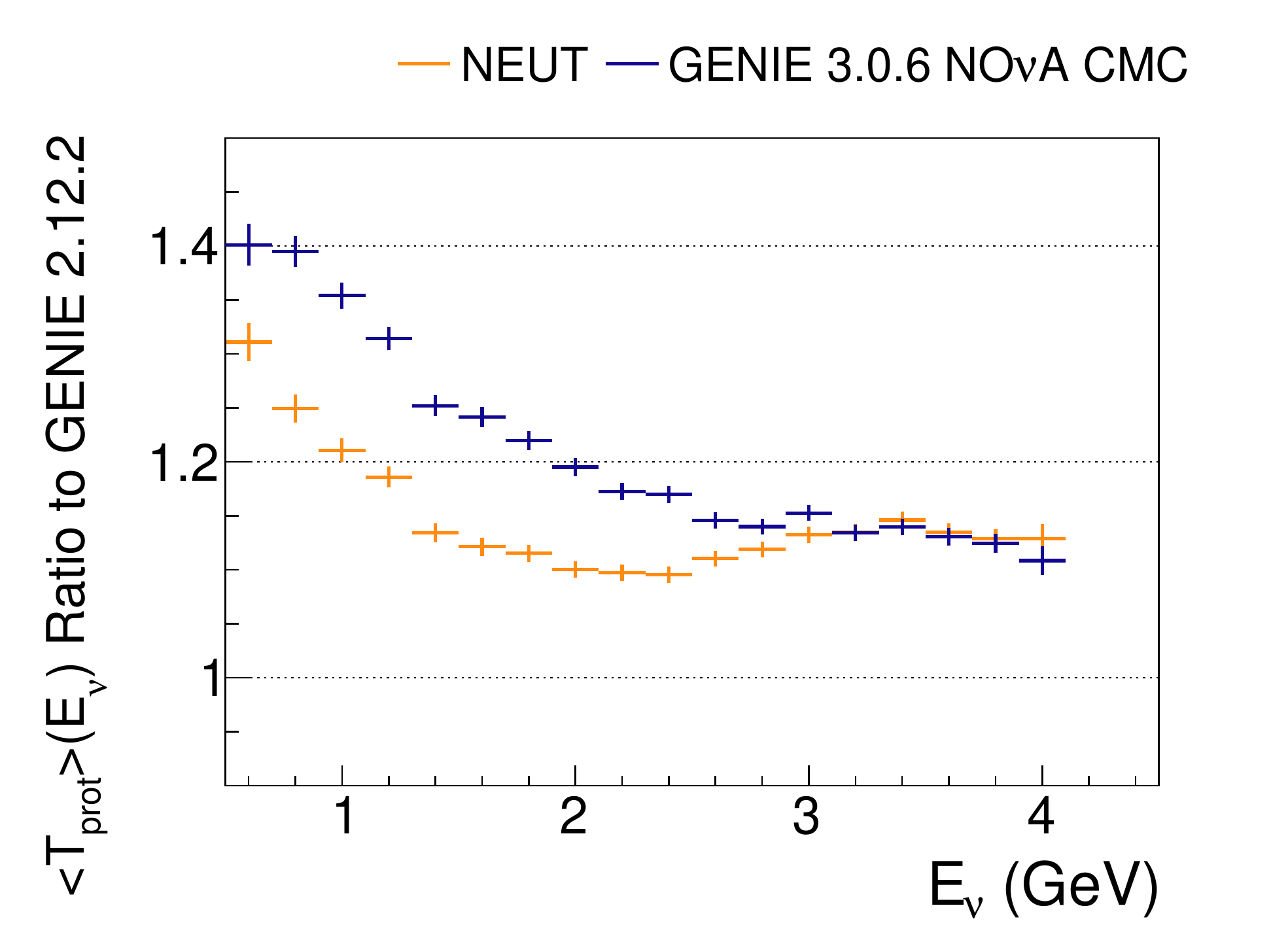}
  }
\end{dunefigure}


\subsection{Reconstructed neutrino energy and event selection}

For the purposes of this study, the neutrino energy estimator is defined as:
\begin{equation}
E_{\rm rec} = E_{\ell}^{\rm true} + E_{\rm proton}^{dep} + E_{\pi^\pm}^{dep} + E_{\pi^0}^{dep} + E_{\rm other}^{dep} + \epsilon \, ,
\end{equation}
\noindent where $E_{\ell}^{\rm true}$ is the true outgoing lepton energy (assuming it is measured perfectly), $E_{i}^{dep}$ is the energy deposit due to particle $i$ and its progeny (assuming no detection threshold and perfect association of energy deposits to particles) and $\epsilon$ takes into account the mass difference between initial state nucleus and final state nucleus and nucleon system, as well as the kinetic energy of the recoiling final state nucleus.

A requirement on the total energy deposit in the veto region, as defined in Section~\ref{sec:prism-oa}, is applied to \dword{nd} events. Events where this energy deposit exceeds 30~MeV are not used. Perfect sign separation is assumed for the outgoing lepton, regardless of its energy, as well as perfect rejection of neutral current events.

\subsection{Mock data set with 20\% missing proton energy}
The mock data set considered here is generated by scaling down the energy deposits due to protons by 20\%, under the assumption that this missing energy is instead carried by some other, unobserved particles, such as neutrons:
\begin{equation}
E_{\rm proton}^{dep} \rightarrow E_{\rm proton}^{\prime dep} = 0.8 \times E_{\rm proton}^{dep} \, .
\end{equation}

This propagates to the reconstructed neutrino energy as:
\begin{equation}
E_{\rm rec} \rightarrow E_{\rm rec}^{\prime} = E_{\rm rec} - 0.2 \times E_{\rm proton}^{dep} \, .
\end{equation}

The effect of this transformation on the on-axis distributions is shown in Figure~\ref{fig:onAxisFakeNoRW}. This transformation can migrate events into the sample if the transformed proton energy deposits drop the energy deposited in the veto region to a value below threshold. It should be noted, however, that this effect is underestimated in the current study, as the range of the protons is unchanged in the transformation. A more detailed approach would involve scaling down true proton energy at generator level and running the transformed vectors through the Geant4 simulation, leading to shorter proton ranges and a larger migration into the accepted sample. 


\begin{dunefigure}[Reconstructed and deposited energy in protons, mock data compared to nominal]{fig:onAxisFakeNoRW}
{Effect of scaling energy deposits due to protons down by 20\% on the $E_{\rm rec}$ (left) and $E_{\rm proton}^{dep}$ (right) distributions.}
  \subfloat[][$E_{\rm rec}$]{
    \includegraphics[width=0.45\textwidth]{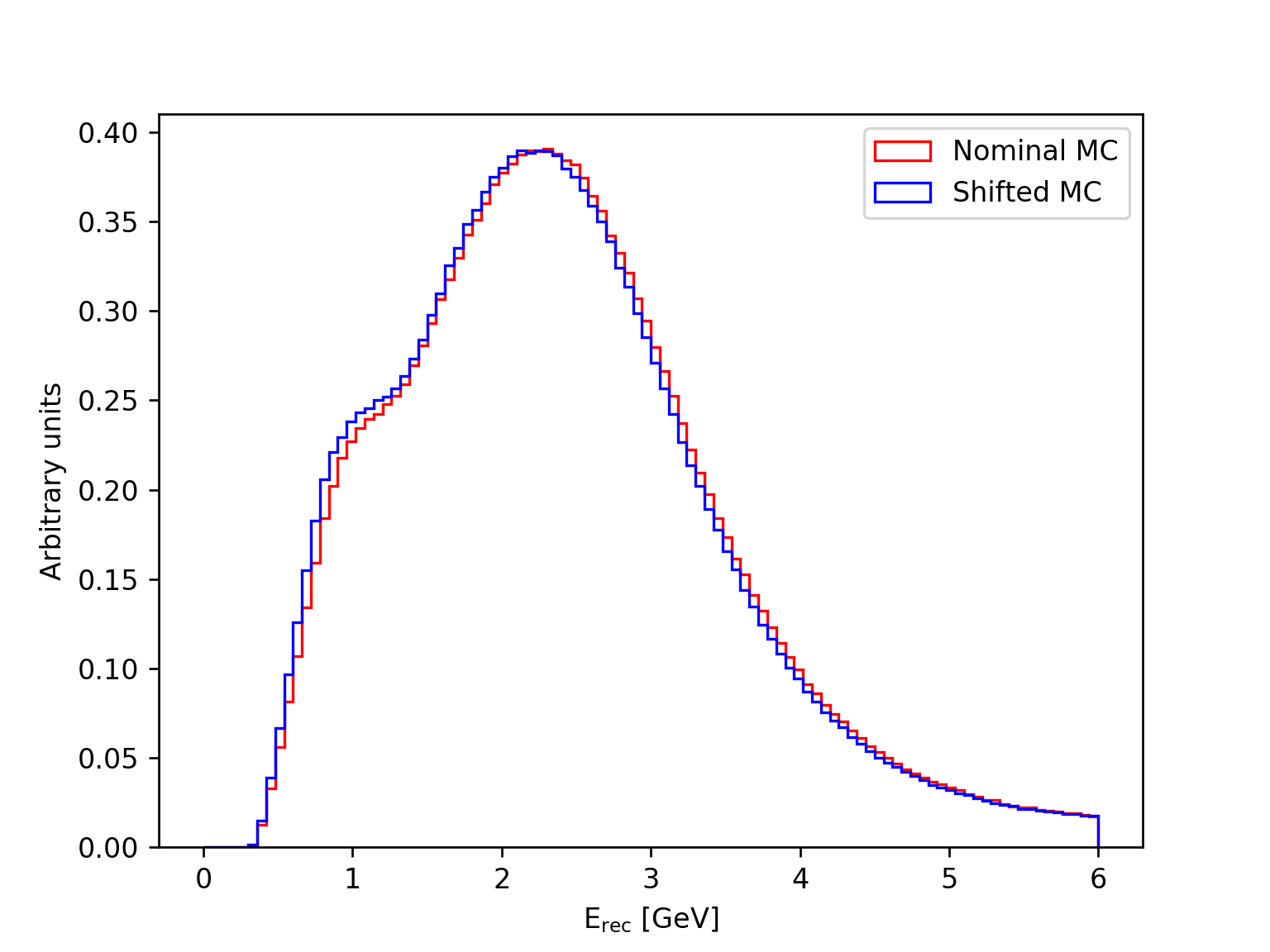}
  }
  \subfloat[][$E_{\rm proton}^{dep}$]{
	\includegraphics[width=0.45\textwidth]{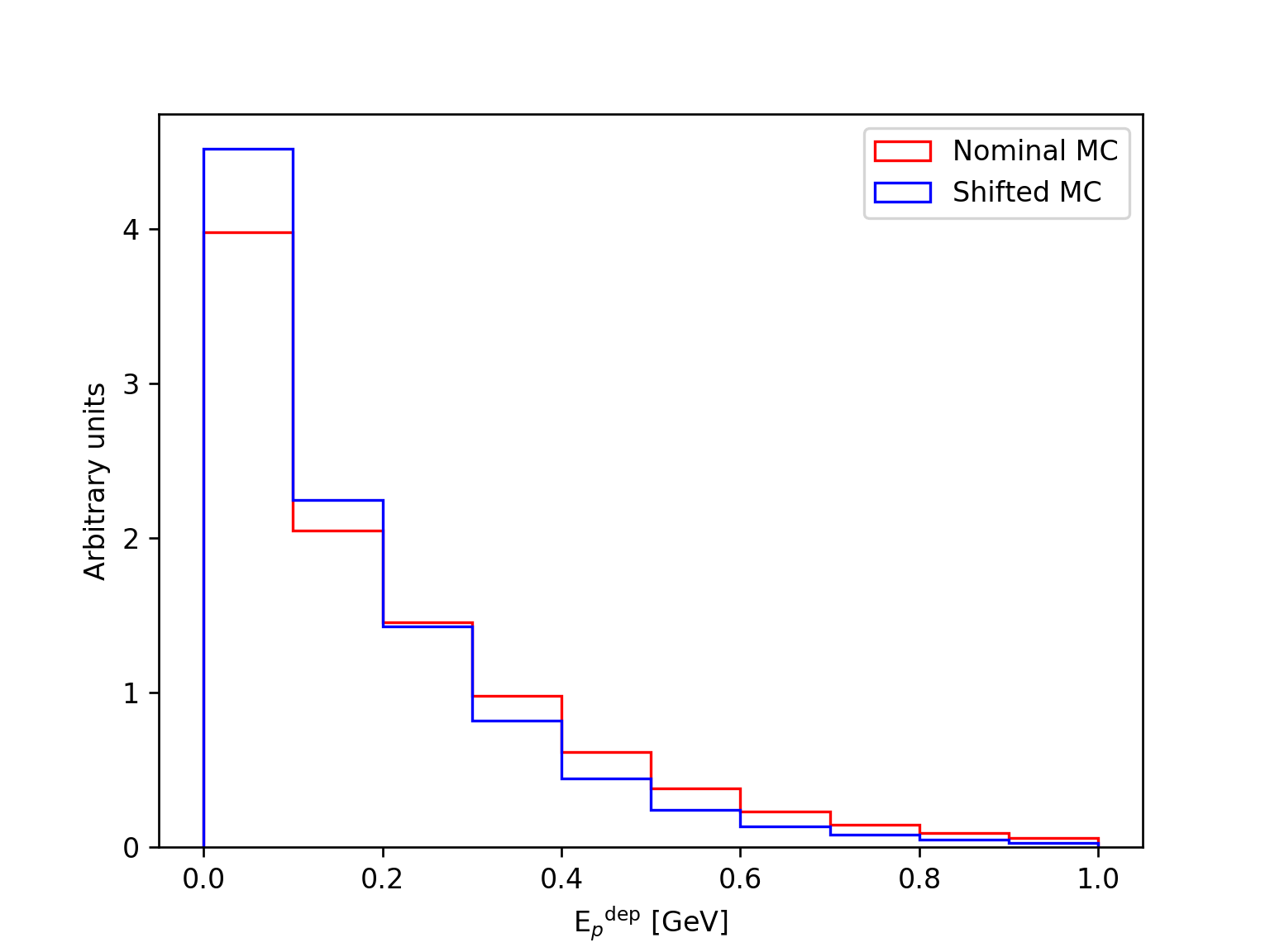}
  }
\end{dunefigure}

\subsection{Multivariate event reweighting}
The nominal neutrino interaction model (red in Figure~\ref{fig:onAxisFakeNoRW}) could be made to agree with the mock data (blue in  Figure~\ref{fig:onAxisFakeNoRW}) by modifying the cross section model in a variety of different ways (i.e. not just the ``correct'' choice of reducing the proton kinetic energy by 20\%). In an actual experiment, this type of a data/MC discrepancy is typically addressed by adding some (possibly incorrect) flexibility to the cross section model, which would correct the data/MC disagreement via a fit to on-axis \dword{nd} data.

To demonstrate this, a multivariate event reweighting method \cite{Rogozhnikov:2016bdp} is used, in which a gradient boosted decision tree algorithm is trained on a subset of the available data to predict weights which, when applied to the mock data set, will make a multidimensional distribution of observables agree with the nominal set.

The variables considered are $E_{\ell}^{\rm true}$, $E_{\rm proton}^{dep}$, $E_{\pi^\pm}^{dep}$ and $E_{\pi^0}^{dep}$, as defined above. The training data set comprises 75\% of the total MC and 200 trees are grown to a depth of 3 using the mean squared error as the splitting criterion while requiring that the minimum number of samples in a leaf is larger than 1000. For the boosting, a learning rate of 0.1 is used and the loss regularization is set to 1, though the latter parameter was found not to have a significant impact on the outcome of the training procedure.

The multidimensional distributions for the nominal sample and the mock data sample before and after reweighting are shown in Figure~\ref{fig:CornerRWFHC} for \dword{fhc} and Figure~\ref{fig:CornerRWRHC} for \dword{rhc}. In both cases it is evident that the reweighting procedure recovers the agreement with the nominal MC in all the projections shown.

\begin{dunefigure}[Multidimensional distributions for the nominal and fake data sets for \dword{fhc}]{fig:CornerRWFHC}
{Multidimensional distribution of observables in \dword{fhc} for the nominal sample (red), the mock data set before (blue) and after (green) reweighting. The histograms on top of the columns show one-dimensional projections of the distribution and the others show contours in pairwise two-dimensional projections.}
    \includegraphics[width=1.\textwidth]{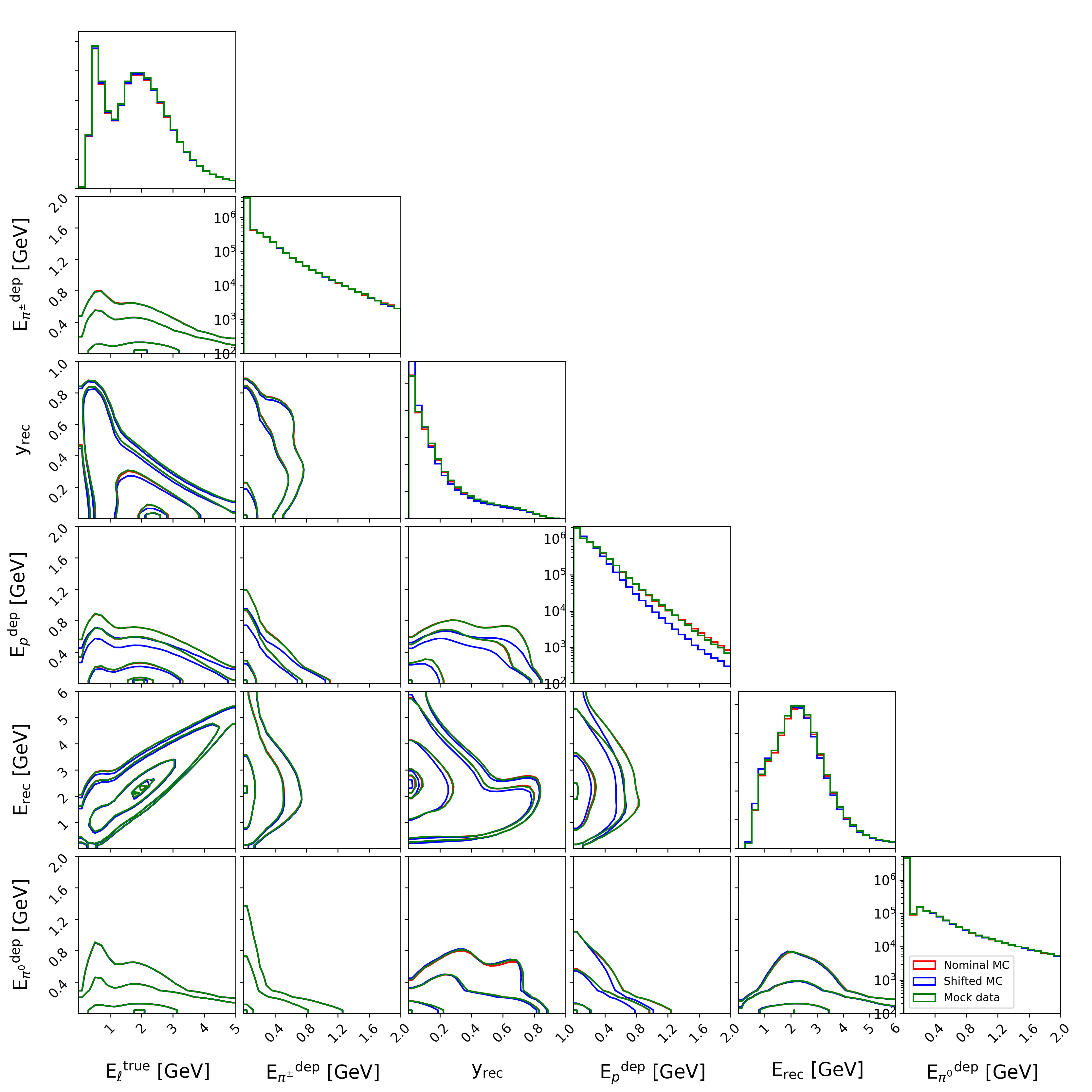}
\end{dunefigure}

\begin{dunefigure}[Multidimensional distributions for the nominal and fake data sets for \dword{rhc}]{fig:CornerRWRHC}
{Multidimensional distribution of observables in \dword{rhc} for the nominal sample (red), the mock data set before (blue) and after (green) reweighting. The histograms on top of the columns show one-dimensional projections of the distribution and the others show contours in pairwise two-dimensional projections.}
    \includegraphics[width=1.\textwidth]{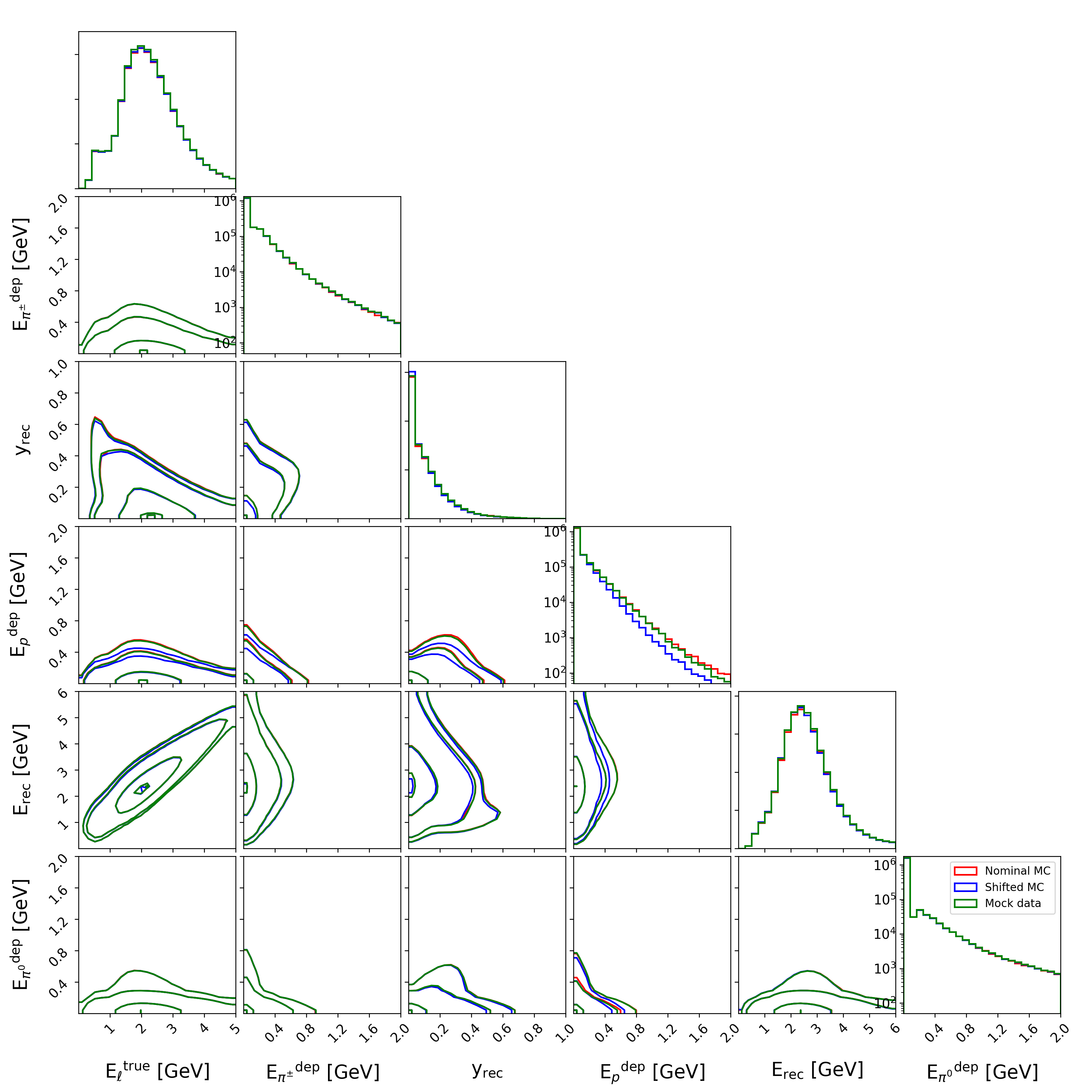}
\end{dunefigure}

The fractional difference between the true and reconstructed neutrino energy in the nominal sample and mock sample before and after reweighting is shown in Figure~\ref{fig:FakeDataErecBias} for both \dword{fhc} and \dword{rhc}. A clear bias is introduced by scaling the deposited proton energy and the bias is largely unchanged by the reweighting procedure. As would be expected, the effect of scaling the proton energy is larger on \dword{fhc} events than on \dword{rhc}. This demonstrates that it is indeed possible to achieve good agreement in the multidimensional distribution of a number of observables at an on-axis \dword{nd} via reweighting, while a significant bias in neutrino energy estimation goes unnoticed.

For this particular set of reweighted distributions, it is possible to identify residual differences between the interaction model and the observed ND data in some other observable distributions. However, this mock data exercise can be repeated for ever larger sets of observable distributions to mitigate detectable differences between the model predictions and ND mock data.

\begin{dunefigure}[Fractional difference, true vs reconstructed neutrino energy, nominal and fake data]{fig:FakeDataErecBias}
{Fractional difference between true and reconstructed neutrino energy for the nominal sample (red) and mock data sample before (blue) and after (green) reweighting.}
  \subfloat[][FHC]{
    \includegraphics[width=0.45\textwidth]{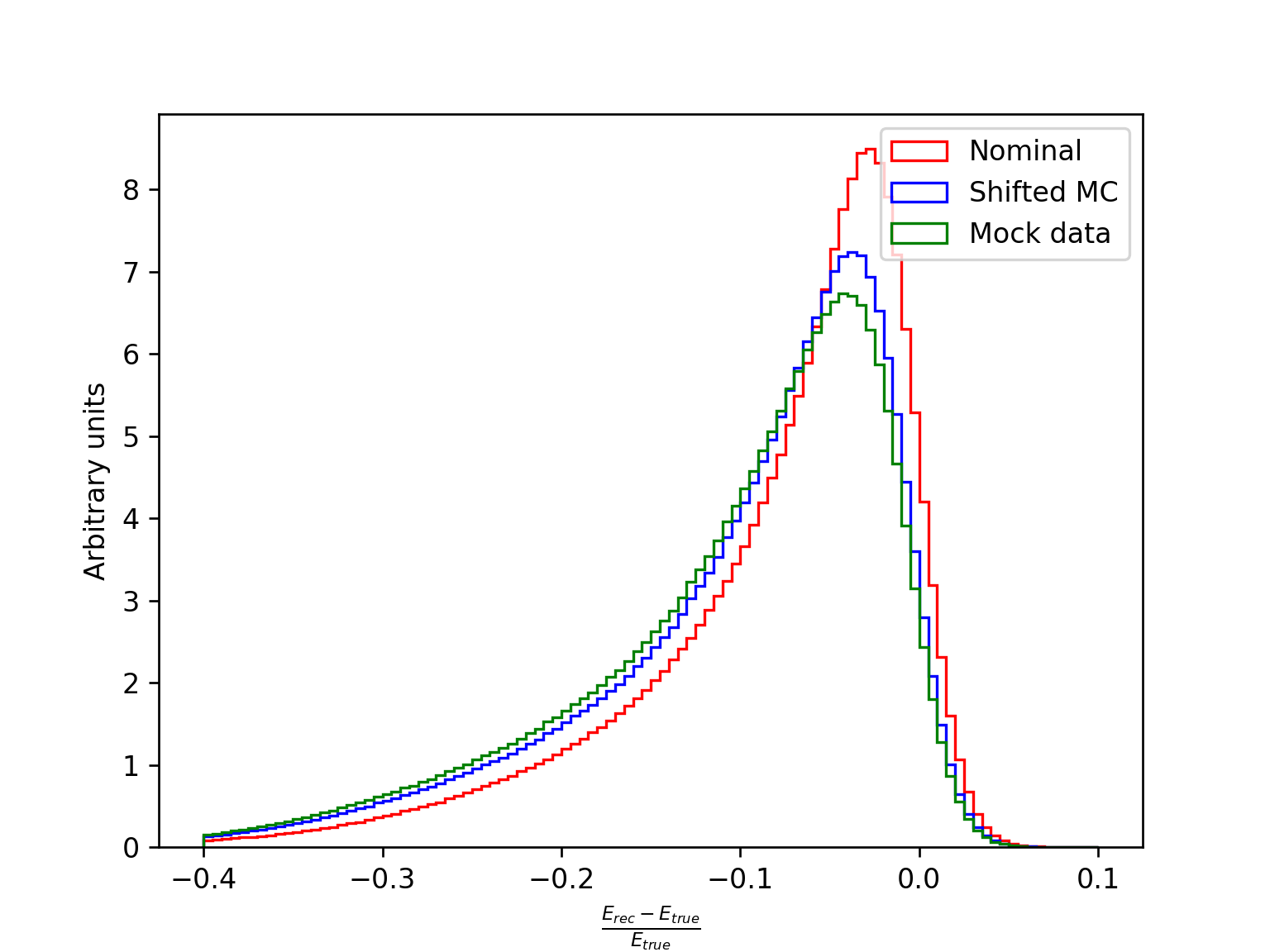}
  }
  \subfloat[][RHC]{
	\includegraphics[width=0.45\textwidth]{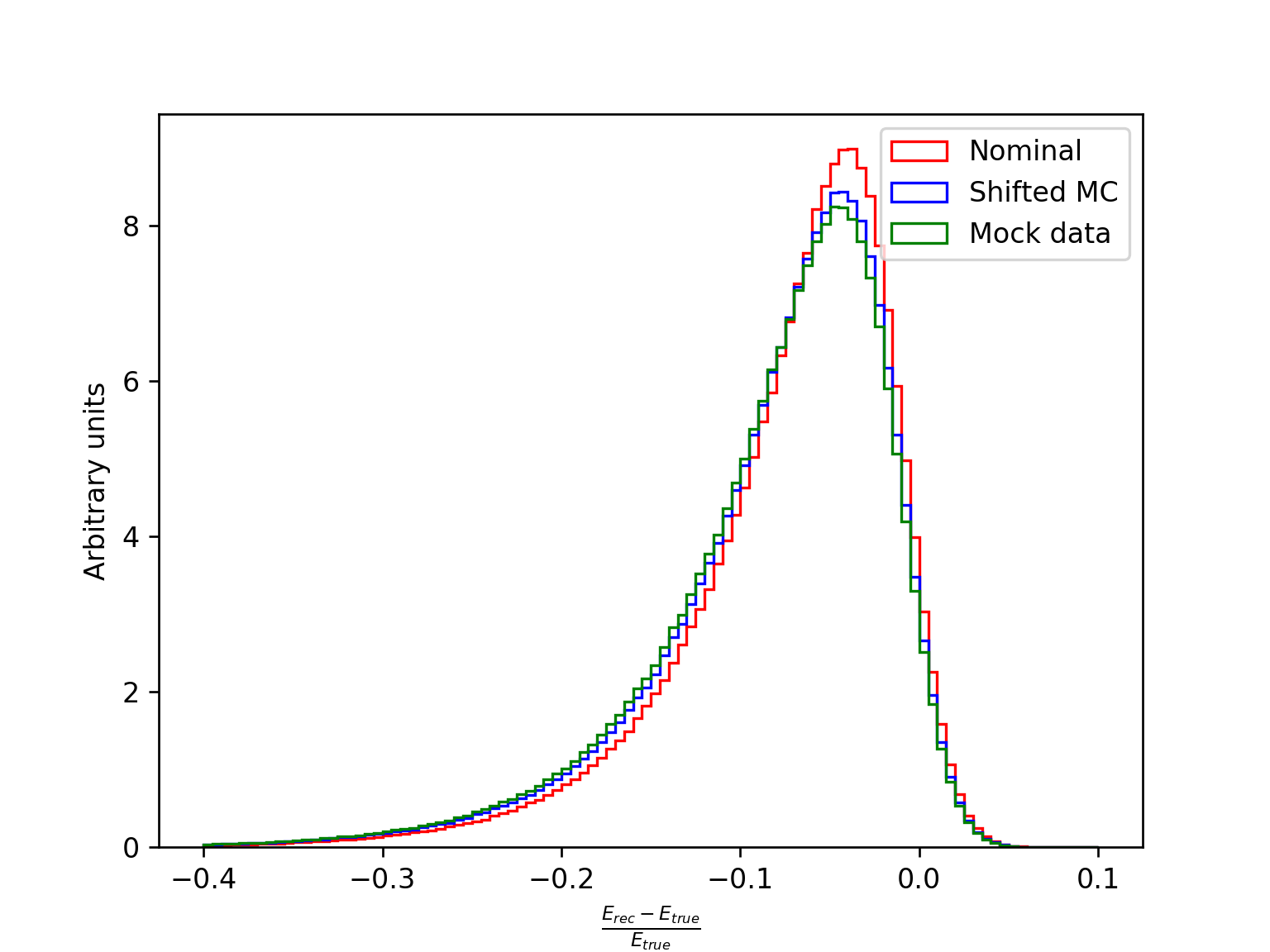}
  }
\end{dunefigure}

\subsection{Propagation of the model in true kinematic variables}
To investigate the impact of this type of mismodeling on oscillation analyses, and also to understand how measurements at off-axis positions might help resolving the degeneracy, it is necessary to propagate the reweighting scheme to the \dword{fd} and different off-axis positions.

An additional BDT is trained to learn the weights predicted by the BDT described above as a function of true kinematic variables: neutrino energy, total proton kinetic energy and inelasticity. The hyper-parameters for this second BDT are similar to the former's, except for the loss function, which for this regression task is chosen to be the mean squared error. The existing MC events are used to relate the output of the inital BDT for each event, which is itself a function of reconstructed quantities, to the true kinematic variables describing the same events. The resulting weights as a function of true kinematics are shown in Figure~\ref{fig:RW_IntVars_FHC} for \dword{fhc} events, and Figure~\ref{fig:RW_IntVars_RHC} for \dword{rhc} events. This process allows for the re-weighting scheme that produces good agreement between mock data and the nominal MC in the on-axis \dword{nd} to be applied as a function of true kinematic variables to the \dword{fd} and off-axis \dword{nd} samples.

When applying this model to \dword{fd} data, where no (or very little) sign selection will be possible, the weighting scheme measured in \dword{fhc} mode is applied to true neutrino events and the scheme measured in \dword{rhc} mode (assuming perfect charge separation at the \dword{nd}) is applied to true anti-neutrino events.

\begin{dunefigure}[Event weights for \dword{fhc} binned vs interaction mode, three pairs of true kinematic variables]{fig:RW_IntVars_FHC}
{\dword{fhc} event weights binned in interaction mode and three pairs of true kinematic variables: true neutrino energy \emph{vs} Q$^2$ (left); true neutrino energy \emph{vs} true proton kinetic energy (middle); and q$_0$ \emph{vs} q$_3$ (right).}
    \includegraphics[width=1.\textwidth]{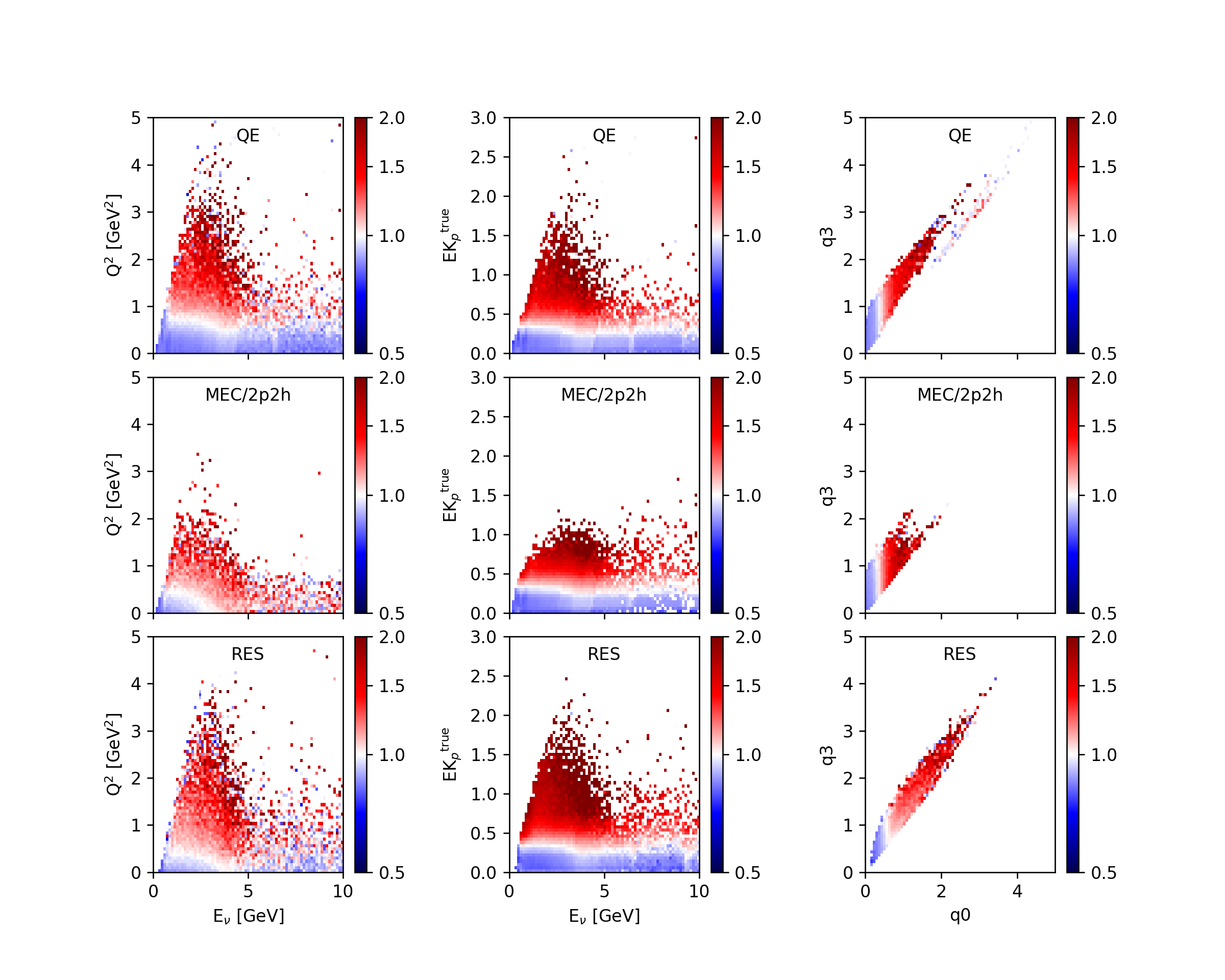}
\end{dunefigure}

\begin{dunefigure}[Event weights for \dword{rhc} binned vs interaction mode, three pairs of true kinematic variables]{fig:RW_IntVars_RHC}
{\dword{rhc} event weights binned in interaction mode and three pairs of true kinematic variables: true neutrino energy \emph{vs} Q$^2$ (left); true neutrino energy \emph{vs} true proton kinetic energy (middle); and q$_0$ \emph{vs} q$_3$ (right).}
    \includegraphics[width=1.\textwidth]{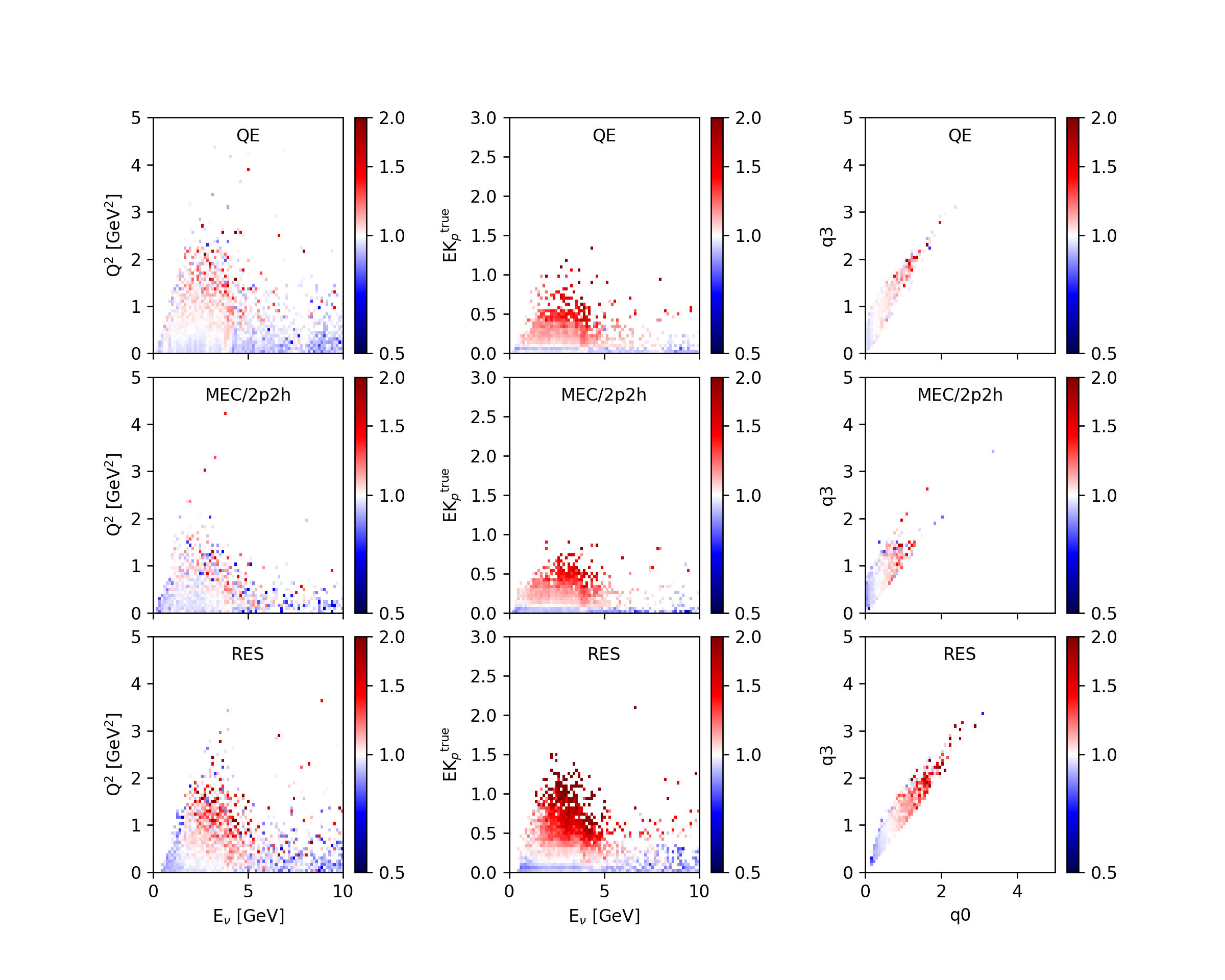}
\end{dunefigure}

\subsection{Effect on Measured Oscillation parameters}\label{sec:dposcbiases}

By construction, the reconstructed energy distribution in the on-axis \dword{nd} for both the mock data and the default model prediction agree.
This procedure produces a situation analogous to that encountered when an experiment empirically (and incorrectly) adjusts its neutrino interaction model to achieve agreement with \dword{nd} data (even if, in this procedure, the mock data are adjusted rather than the default interaction model).

To study the impact on the measured oscillation parameters, the mock data are prepared for both the near and far detectors, and the same near+far fit described in Section~\ref{sec:prism-oa} is performed. Since the high statistics \dword{nd} \numu samples agree by construction, the \dword{nd} MC does not change during the fit, and the flux and cross section parameters are strongly constrained to remain very close to their pre-fit values. The \dword{fd} event distributions, however, are modified by the fit by variations in the oscillation parameters, and, to a lesser extent, detector systematic uncertainties. The \dword{fd} distributions before and after the fit are shown in Figure~\ref{fig:duneprismeventrates}. As shown in the figure, adjustments to the oscillation parameters can produce good agreement in the \dword{fd} $\nu_e$ and $\nu_\mu$ distributions, thereby produce a satisfactory goodness of fit, but with incorrect oscillation parameters. The extracted 90\% confidence limit regions for fits to both the nominal Monte Carlo and the mock data sample are shown in Figure~\ref{fig:duneprismsensitivity}. The measured oscillation parameters extracted from the mock data are inconsistent with the true oscillation parameters by much more than the  uncertainty returned by the fit, despite good data/MC agreement at the \dword{nd}. 

\begin{dunefigure}[Predicted distributions of reconstructed neutrino energy for selected $\nu_\mu$ and $\nu_e$ events]
{fig:duneprismeventrates}
{Predicted distributions of reconstructed neutrino energy for selected $\nu_\mu$ (top) and $\nu_e$ (bottom) events, in \dword{fhc} (left) and \dword{rhc} (right) beam modes in 7 years. The black curve shows the nominal GENIE prediction, the red points are the mock data samples, and the blue curve is the post-fit result, where systematic and oscillation parameters are shifted to match the mock data. The ND spectra match the pre-fit prediction by construction and are not shown.}
    \includegraphics[width=1.\textwidth]{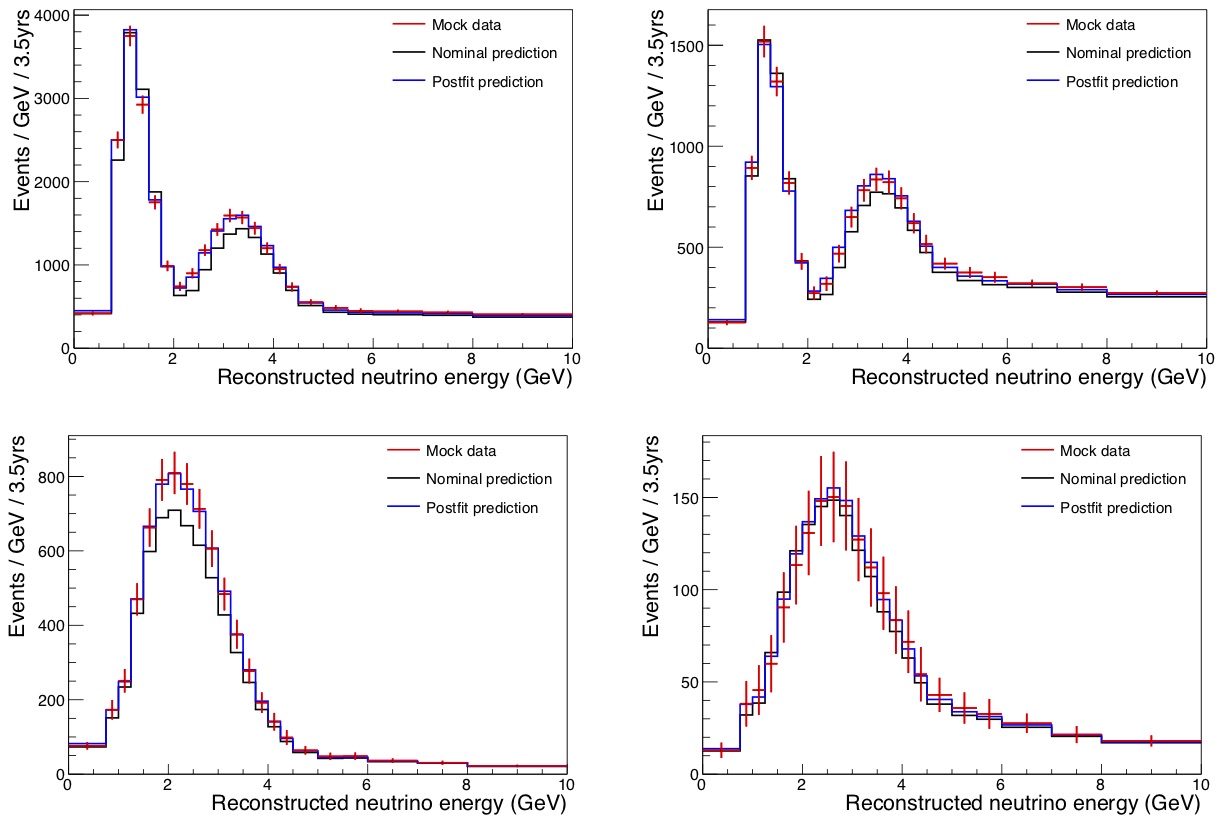}
\end{dunefigure}

\begin{dunefigure}[Results of fit to nominal and mock data]{fig:duneprismsensitivity}
{Results of a fit to both the nominal MC (dashed) and the mock data samples (solid). The true values of $\Delta m^2_{32}$ and $\sin^2\theta_{23}$ are given by the star, and the allowed 90\% C.L. regions are drawn around the best-fit point, for 7, 10, and 15 years of exposure.}
    \includegraphics[width=0.6\textwidth]{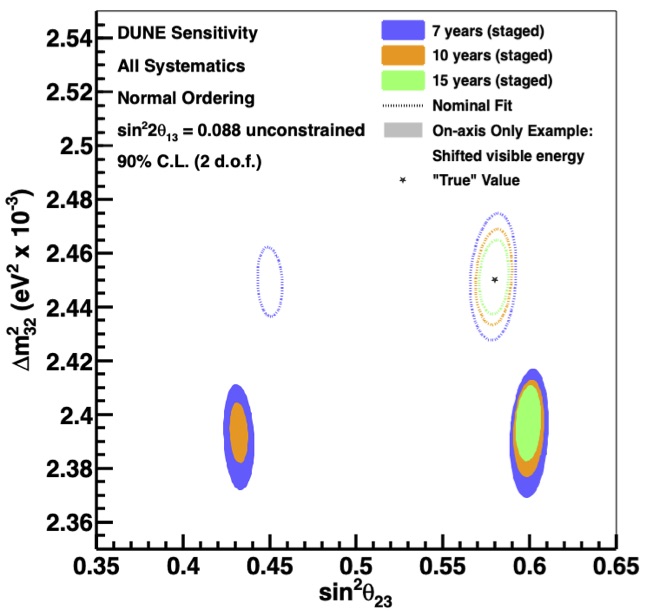}
\end{dunefigure}

\section{The DUNE-PRISM Measurement Program}\label{sec:duneprism}%

The DUNE-PRISM measurement program exploits an intrinsic property of standard neutrino beams in which the peak energy of the neutrino spectrum decreases and the size of the high energy tail is reduced when the detection angle relative to the beam axis is increased, as shown in Figure~\ref{fig:prism-1doffaxis}. This feature of neutrino beams has motivated the T2K and NO$\nu$A experiments to point their beams 2.5$^\circ$ and 0.8$^\circ$ away from their respective far detectors in order to reduce neutral current backgrounds from high energy neutrinos, and to tune the peak of the neutrino energy spectrum to match the peak of the oscillation probability. By constructing a \dword{nd} that spans a wide range of off-axis angles, it is possible to sample a continuously varying set of neutrino energy spectra. The DUNE-PRISM measurement program consists of moving \dword{ndlar} and \dword{ndgar} laterally through the underground \dword{nd} hall to span an off-axis angle range of 0$^\circ$ to 3.2$^\circ$.

The information provided by the off-axis measurements can be implemented in DUNE oscillation analyses using 2 complementary approaches:
\begin{enumerate}
\item \underline{To identify problems in neutrino interaction modeling.} By comparing \dword{nd} data to MC at off-axis locations with different energy spectra, the neutrino interaction model will be over-constrained, and the potential for biases in the measured oscillation parameters can be identified. Figure~\ref{fig:prism-oaevents} demonstrates that the oscillation parameter biases seen in the mock data study described above can be clearly identified off-axis by the disagreement between the data and the predicted rate. 

This use of off-axis data is similar to the approach used by existing long-baseline neutrino experiments, in which data are iteratively compared to improvements in the neutrino interaction model until satisfactory agreement is achieved. However, since DUNE-PRISM provides data across a wide range of neutrino energy spectra that span the important oscillation features in the \dword{fd} energy spectrum, this requirement becomes more stringent. Any neutrino interaction model that can simultaneously reproduce all relevant final state kinematic distributions over all the sampled initial energy spectra is expected to accurately predict the various oscillated \dword{fd} energy spectra.

\item \underline{To overcome problems in neutrino interaction modeling.} It is possible that no first-principles neutrino interaction model with sufficient precision to achieve DUNE physics goals will be available on the timescale of the experiment.
In this scenario, the most important novel feature of DUNE-PRISM is that measurements at different off-axis positions within the detector can be linearly combined to statistically determine any observable for a given choice of incident neutrino flux.
In particular, it is possible to match any given oscillated \dword{fd} neutrino energy spectrum using a linear combination of off-axis \dword{nd} neutrino energy spectra. By applying this linear combination to any measured quantity (e.g. calorimetrically reconstructed neutrino energy, $E_{rec}$), it is possible to directly measure the expected \dword{fd} $E_{rec}$ for any chosen set of oscillation parameters. Using this method, any unknown cross section effects that produce a mismatch between $E_{true}$ and $E_{rec}$ are naturally incorporated into the \dword{fd} prediction.

It is also possible to use this technique to produce Gaussian incident neutrino energy spectra, which allows for a direct measurement of the relationship between true and reconstructed neutrino energy. To construct such a spectrum for a desired neutrino energy, the linear combination primarily utilizes measurements from the off-axis region that peaks at the chosen mean energy, and then subtracts contributions from the more on-axis (off-axis) detector locations to reduce the high (low) energy tails of the energy spectrum. These constructions can produce strong constraints on neutrino interaction models, and provide the first ever mechanism to measure neutral current interactions as a function of neutrino energy, which will provide direct constraints on backgrounds to the oscillation measurement.

\end{enumerate}

The neutrino energies that can be directly sampled by moving the detector off-axis range from 0.5~GeV, as determined by the maximum off-axis position that the detector can access, to just above the on-axis flux peak of 2.5~GeV. However, in order to  constrain the feed-down in reconstructed neutrino energy above the first oscillation maximum, additional information at higher energies is needed. It is possible to achieve this with a short special run in which the current supplied to the magnetic focusing horn is lowered by 13~kA relative to the nominal horn current as described in the sections below.

\subsection{Event Rates and Run Plan}

The DUNE-PRISM measurement program requires data taking at several off-axis positions. There are additional motivations for taking more data in the on-axis position than any of the individual off-axis positions, since this is the position at which the \dword{nd} flux is most similar to the unoscillated flux at the \dword{fd}, and measurements such as $\nu$-e$^{-}$ scattering, which can constrain the flux uncertainties at the \dword{fd}, require high statistics. To demonstrate the feasibility of fulfilling both of these needs, a sample run plan is given in Table~\ref{tab:runplantable}. Despite only spending approximately 2.5 running weeks at each off-axis position, sufficient statistics are accumulated, even at the most off-axis position. Also itemized in the table, a small fraction of the on-axis data, corresponding to 1 week per year, will be collected with a lower horn current setting (280 kA instead of 293 kA).

\begin{dunetable}
[Example 10 month PRISM run plan and event rates]
{|c|c|c|c|c|c|c|}
{tab:runplantable}
{A sample run plan is outlined, for which approximately half of the assumed 29 week beam-year the detector is in the  on-axis position, one week is spent on axis but with a lower horn current (280 kA), and the remaining time is evenly divided between off-axis positions. The fiducial volume assumed is 4\,m wide, with the largest off-axis position sampled at 32.5\,m, The table shows the rate of $\nu_\mu$ CC events before ($N_{\nu_\mu CC}$) and after (N$_{Sel}$) the \dword{nd} event selection, the fraction of wrong-sign background (WSB), and the fraction of neutral current events (NC). The total number of $\nu_\mu$ CC interactions in the gas is also provided.} 
 \span\omit  & \multicolumn{4}{|c|}{ND-LAr} & ND-GAr\\ \hline
 \multicolumn{2}{|c|}{} & All int. &\multicolumn{3}{|c|}{Selected} & All int. \\
\hline
Stop&Run duration&N$_{\nu_{\mu}CC}$ & N$_{Sel}$ & WSB & NC & N$_{\nu_{\mu}CC}$ \\
\hline
On axis (293 kA)\,m & 14 wks. & 21.6M & 10.1M & 0.2\% & 1.3\% & 580,000\\
On axis (280 kA)\,m & 1 wk. & 1.5M & 690,000 & 0.3\% & 1.3\% & 40,000\\
4 m off axis\,m & 12 dys. & 2.3M & 1.2M & 0.3\% & 1.0\% & 61,000\\
8 m off axis\,m & 12 dys. & 1.3M & 670,000 & 0.5\% & 0.9\% & 35,000\\
12 m off axis\,m & 12 dys. & 650,000 & 330,000 & 0.8\% & 0.7\% & 17,000\\
16 m off axis\,m & 12 dys. & 370,000 & 190,000 & 1.1\% & 0.7\% & 10,000\\
20 m off axis\,m & 12 dys. & 230,000 & 120,000 & 1.3\% & 0.7\% & 6,200\\
24 m off axis\,m & 12 dys. & 150,000 & 75,000 & 1.8\% & 0.7\% & 4,100\\
28 m off axis\,m & 12 dys. & 110,000 & 50,000 & 2.1\% & 0.8\% & 2,900\\
30.5 m off axis\,m & 12 dys. & 87,000 & 39,000 & 2.3\% & 0.7\% & 2,300\\
\end{dunetable}

The event rate distributions for this run plan are shown in Figure~\ref{fig:prism-oaevents}. It can be seen that on-axis, the default GENIE simulation matches the mock dataset well, by construction,
even though the mapping between neutrino energy and observables has been distorted by the missing proton energy. At further off-axis positions, the default GENIE simulation is no longer a good predictor for the mock dataset, thus clearly identifying that the model is incorrect and not suitable for producing oscillated \dword{fd} predictions.

\begin{dunefigure}[Predicted \dword{nd} event rate per run-plan year at various off axis positions]{fig:prism-oaevents}
  {The predicted \dword{nd} event rate per run-plan year at various off axis positions for the default GENIE simulation (orange line) and the modified 'missing proton energy' mock dataset (black points). The model deficiencies that gave rise to the biased oscillation parameter measurements, and were not detected on-axis, can clearly be seen in the off-axis positions. The barely visible content in the histograms is the background.}
  \includegraphics[width=0.49\textwidth]{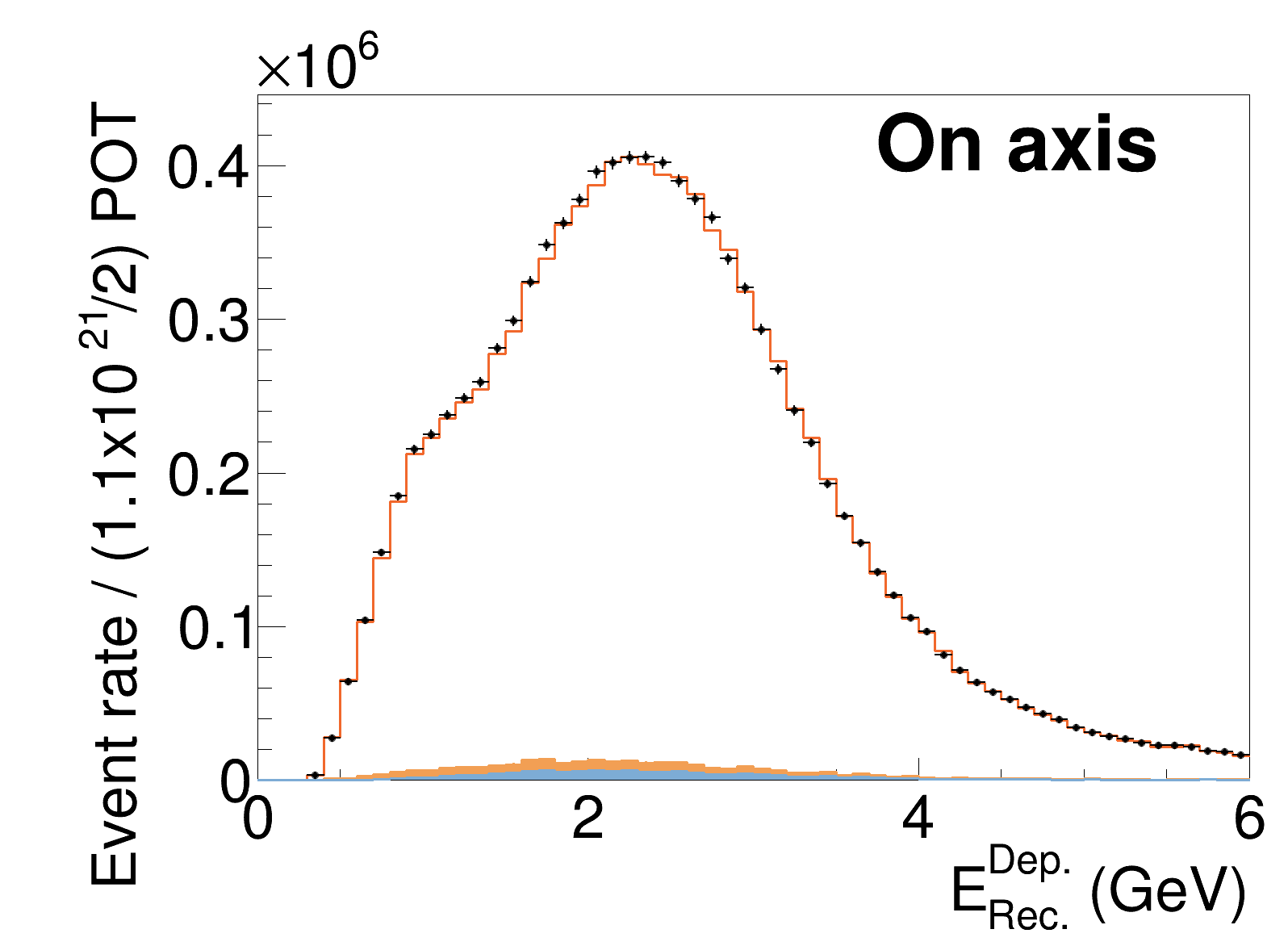}
  \includegraphics[width=0.49\textwidth]{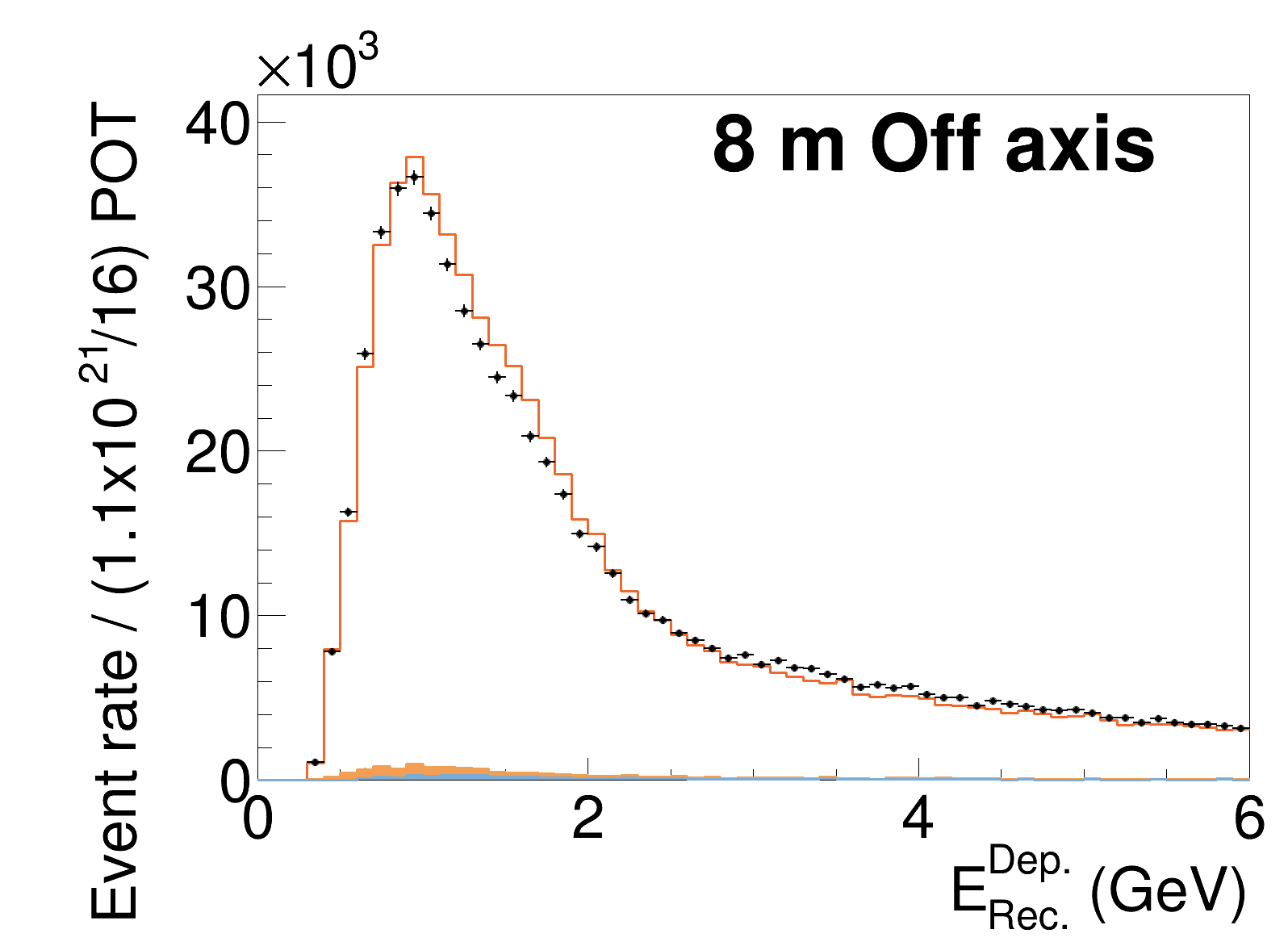}
  \includegraphics[width=0.49\textwidth]{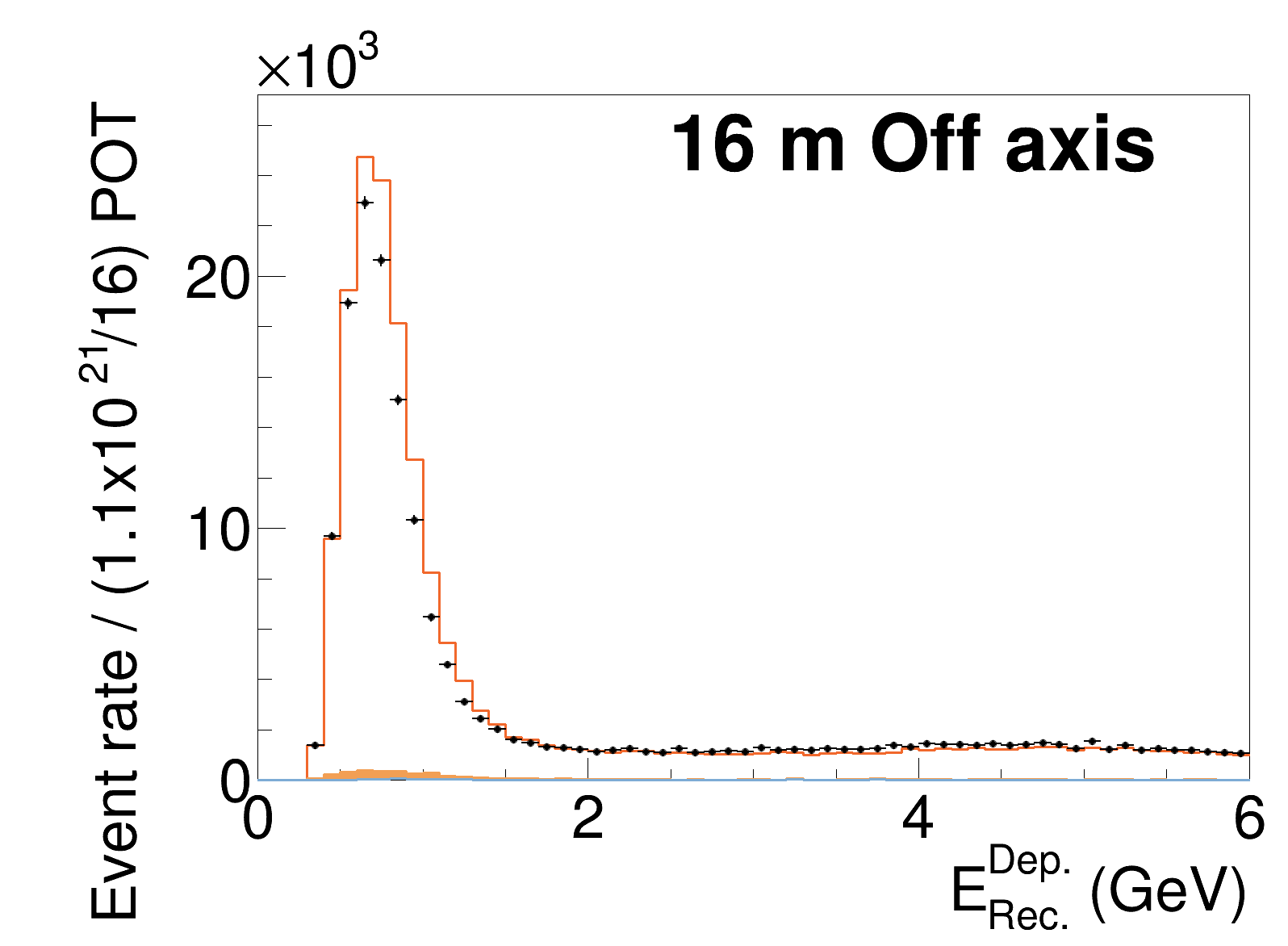}
  \includegraphics[width=0.49\textwidth]{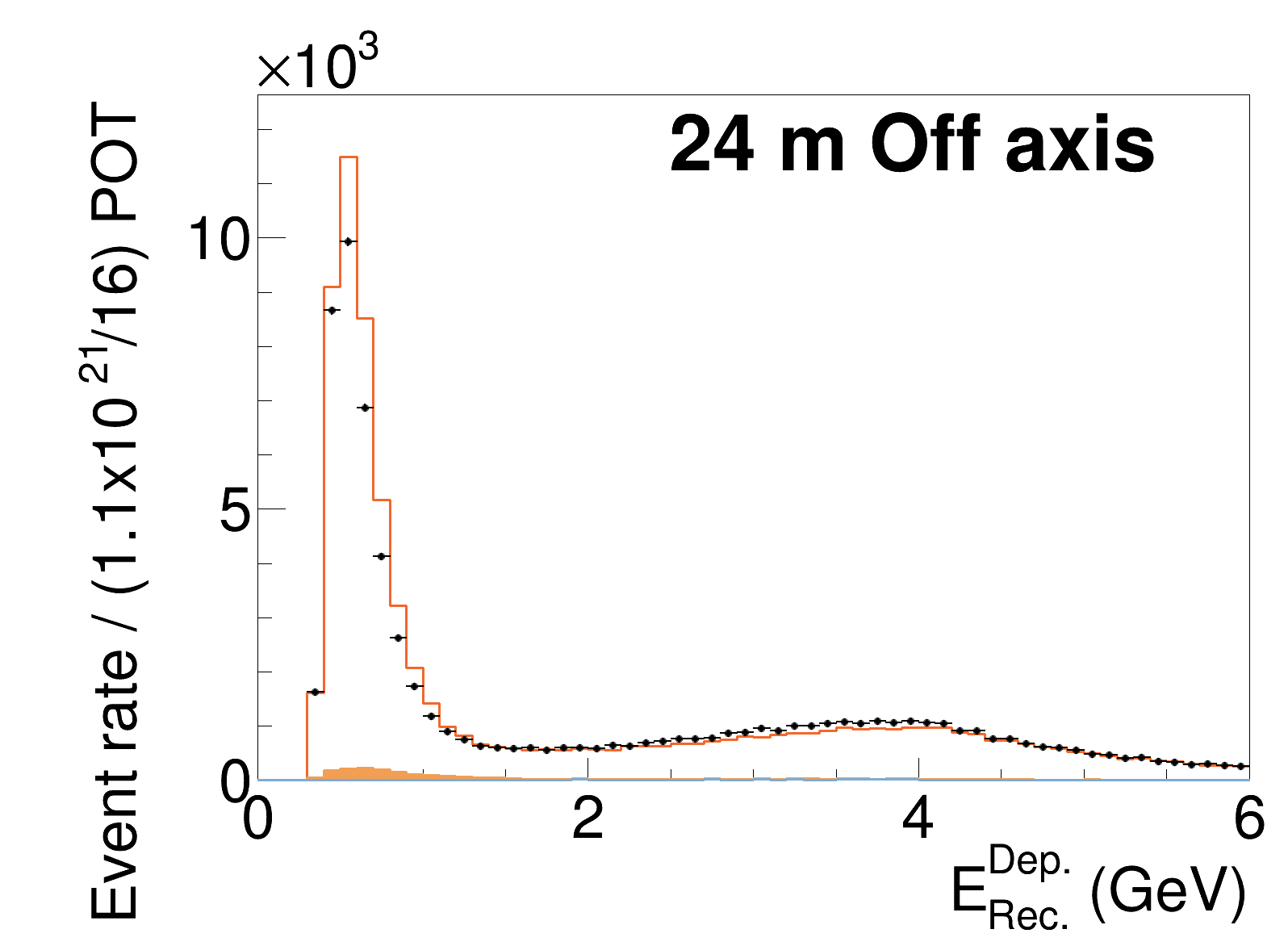}
\end{dunefigure}

\subsection{The LBNF Neutrino Flux at the Near Site}
\label{sec:prism-flux}


The predicted muon neutrino flux for the \dword{fhc} DUNE/LBNF beam as a function of off-axis position and neutrino energy is shown in Figure~\ref{fig:prism-flux2d}. Taking measurements at different off-axis positions gives sensitivity to different parts of the neutrino cross section, which will be important for understanding the different interactions and disentangling nuclear effects from neutrino interaction cross sections. Figure~\ref{fig:prism-flux2d} (right) shows how measurements at the more extreme off axis positions can be used to study quasi-elastic (1p1h) and multi-nucleon (2p2h) interactions with much higher purity than possible at the on-axis position. At medium off-axis displacements ($\sim15$ m), high-purity samples of resonant pion production interactions will be achievable, with a good understanding of contamination from the QE and QE-like reactions possible from measurements taken at far off-axis positions. In this way, the ability to take measurements in the same beam, but with different spectral exposures will be a powerful new technique for constraining neutrino interactions and nuclear effects on argon, and directly inform oscillation analyses in the DUNE beam.


\begin{dunefigure}[\dshort{nd} flux prediction vs energy and off-axis detector position]{fig:prism-flux2d}
  {Left: the \dword{dune} \dword{nd} flux prediction is shown as a function of energy and off-axis detector position, rather than off-axis angle. Right: the total muon neutrino charged-current cross section as a function of true neutrino energy overlaid on a number of representative flux positions at different off axis displacements at the near site. The total is separated into contributions from elastic, and elastic-like (1p1h+2p2h), resonant pion production (Res $1\pi$), and deep inelastic scattering (DIS) channels. Samples taken at different off axis positions will contain varying contributions from different interaction channels.
}
  \includegraphics[width=0.4\textwidth]{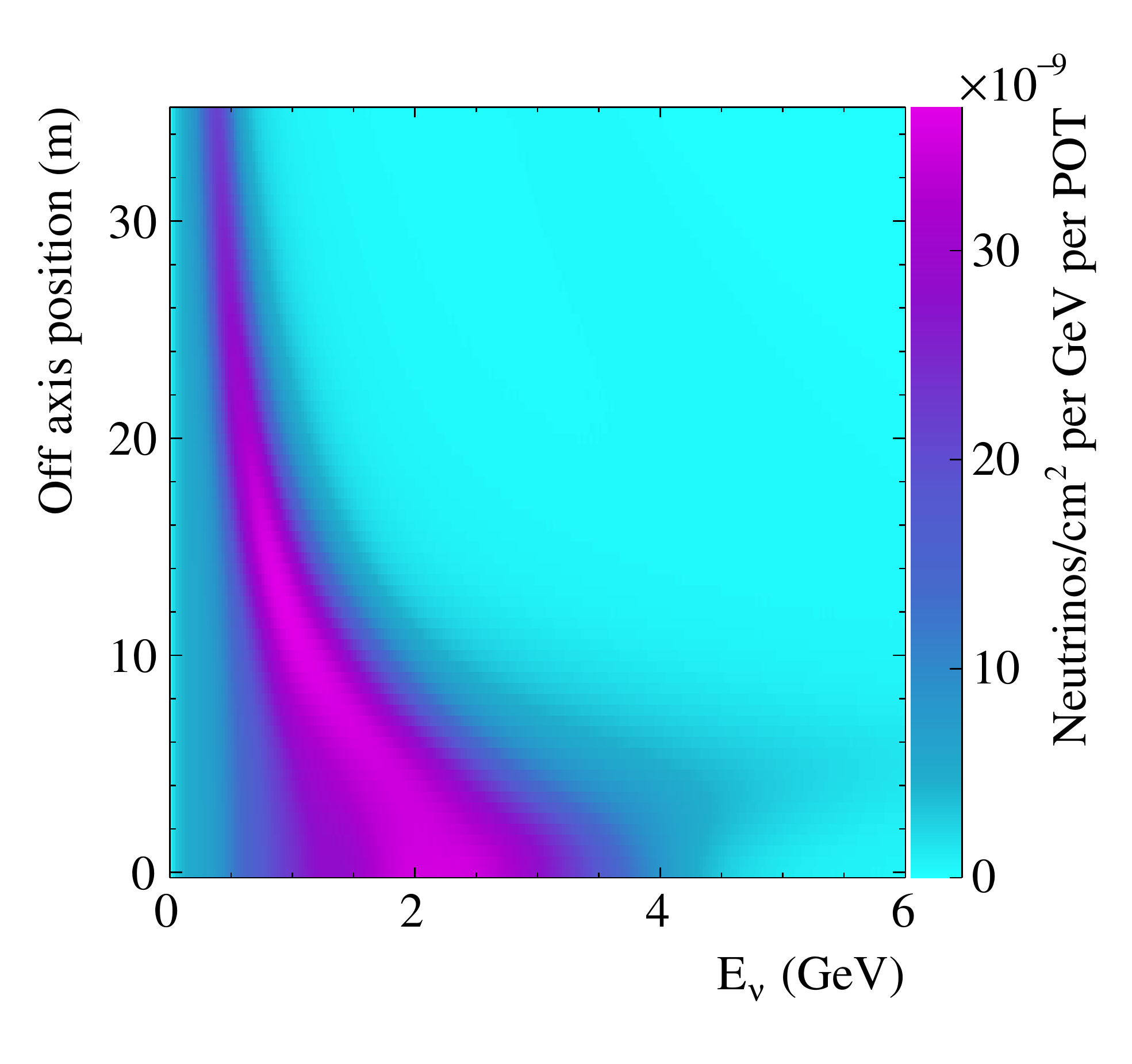}
  \includegraphics[width=0.58\textwidth]{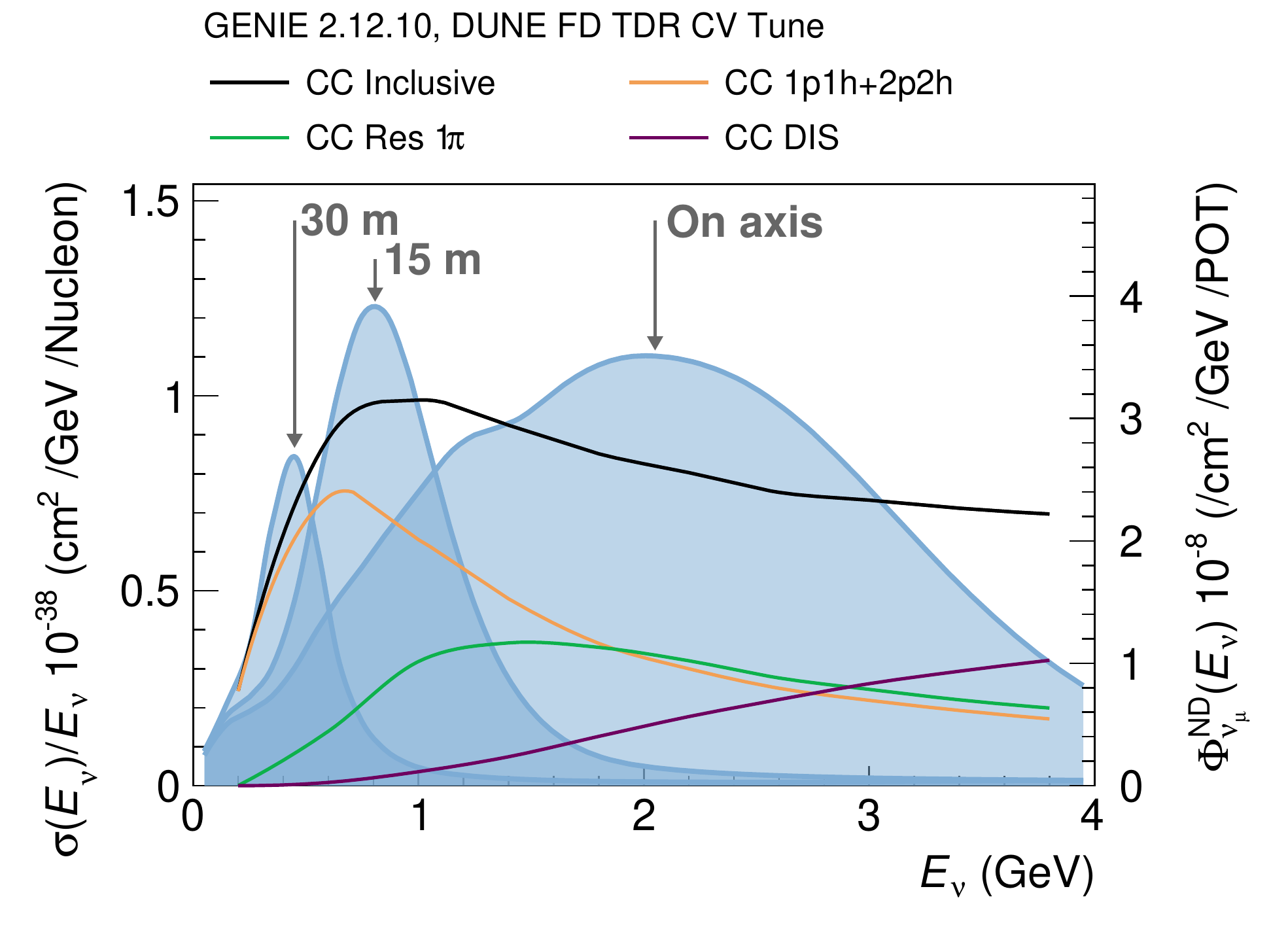}
\end{dunefigure}

\section{Flux Matching}
\label{sec:prism-lincomb}

The $\nu_\mu$ fluxes at each off-axis position constitute a set of states with varying peak energies across the range of DUNE neutrino energies, but with overlapping tails. By taking linear combinations of these energy spectra, it is possible to construct a spectrum that is similar to the 
oscillated energy spectrum at the \dword{fd} for any choice of oscillation parameters. The simplest expression for a set of coefficients which produces matching near and \dword{fd} fluxes is obtained by solving the equation:

\begin{align}
    \left|N \vec{c} - \vec{F}\right| = 0
    \label{simple}
\end{align}

Where $N$ is the \textit{\dword{nd} flux matrix}, which is shown in Figure~\ref{fig:prism-flux2d} (left), and has dimension $N_{\mathrm{Energy \, bins}} \times N_{\mathrm{Off-axis \, bins}}$. $\vec{c}$ is the vector of coefficients (length $N_{\mathrm{Off-axis \, bins}}$), and $\vec{F}$ is the \dword{fd} flux treated as a vector (length $N_{\mathrm{Energy \, bins}}$).  By minimizing the norm of the residual vector, the resulting linear combination matches the \dword{fd} flux very well.  The expression for the coefficients is then just:

\begin{align}
    \vec{c} = N^{-1} \vec{F}
    \label{simpleinv}
\end{align}

Where $N^{-1}$ is the left-inverse of the non-square matrix $N$.  This produces very good matching between fluxes, but the resulting coefficient values have a very large variance, and do not vary smoothly from one off-axis stop to the next, as seen in Figure~\ref{fig:unreg}.

\begin{dunefigure}[\dshort{fd} flux and \dshort{nd} linear combination flux match and corresponding coefficients; Eqn~\eqref{simpleinv}]
{fig:unreg}
{The \dword{fd} flux and the \dword{nd} linear combination flux match (left) and corresponding coefficients (right) are shown using the solution of Equation~\eqref{simpleinv}.}
    \includegraphics[width=0.55\textwidth]{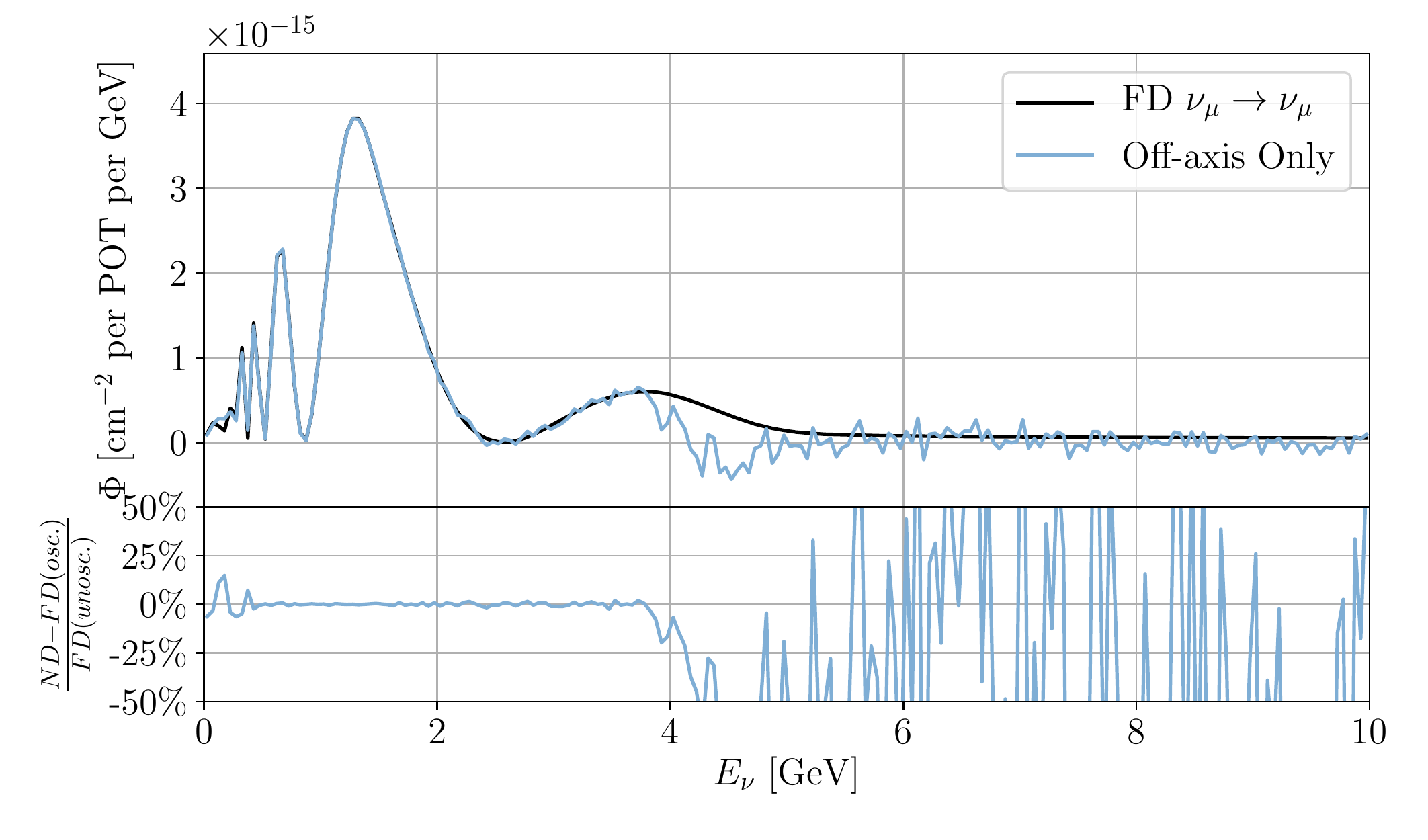}
    \includegraphics[width=0.43\textwidth]{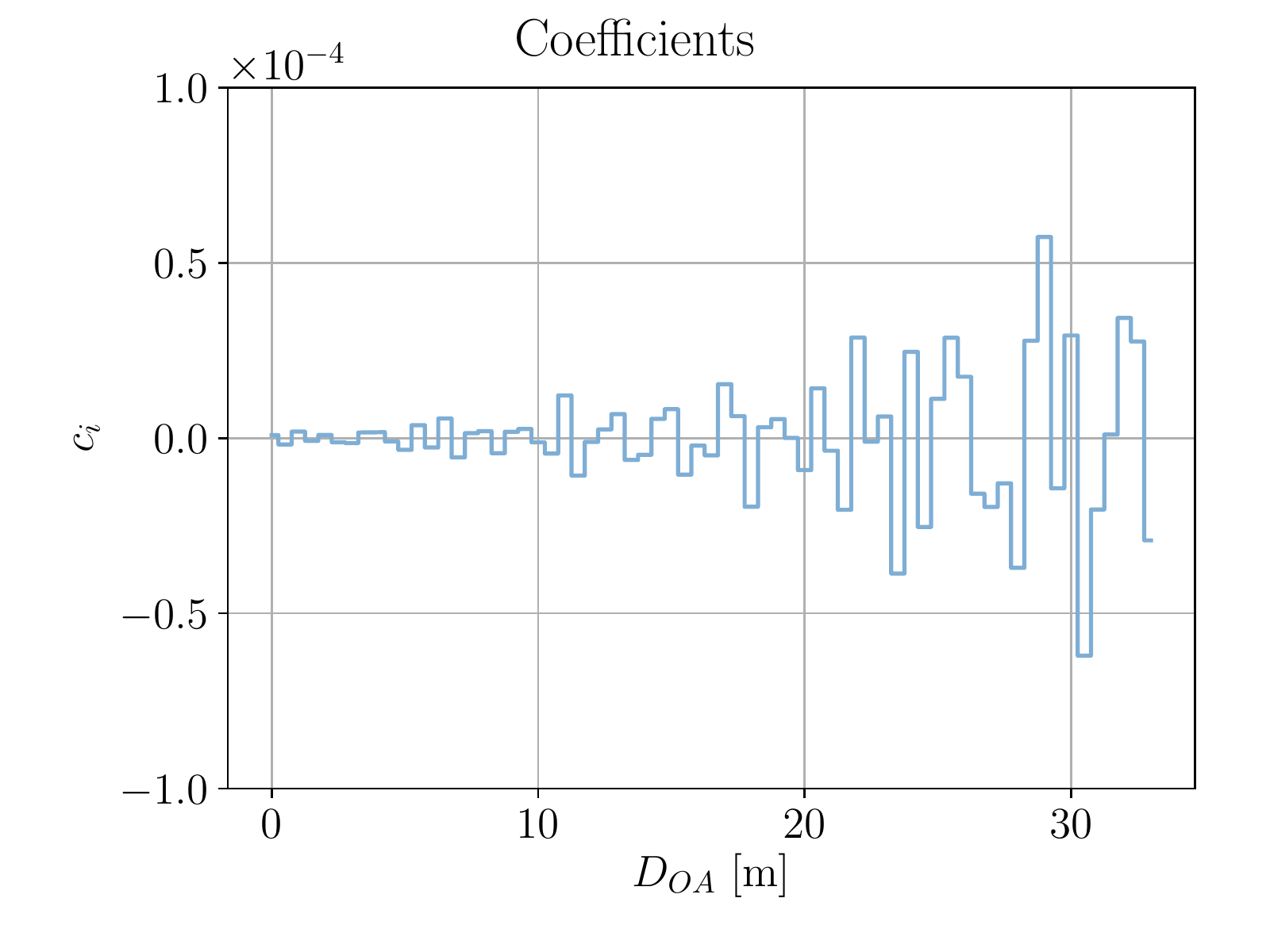}
\end{dunefigure}

However, there are many potential solutions to Equation~\eqref{simple} that can reduce the variance of the resulting coefficients, $\vec{c}$.
One such solution is obtained by imposing a regularization condition on the coefficient vectors. In particular, we wish to minimize the change from one coefficient to the next, so we have the regularization condition:

\begin{align}
    \left| A \vec{c} \right| = 0
    \label{diffreg}
\end{align}

Where $A$ is a difference matrix:

\begin{align}
    A = \begin{pmatrix}
        1 & -1 & 0 & 0 & \ldots & 0 & 0 \\ 
        0 & 1 & -1 & 0 & \ldots & 0 & 0 \\ 
        0 & 0 & 1 & -1 & \ldots & 0 & 0 \\
        0 & 0 & 0 & 1 & \ldots & 0 & 0 \\
        \vdots & \vdots & \vdots & \vdots & \ddots  & \vdots & \vdots \\
        0 & 0 & 0 & 0 & \ldots & 1 & -1 \\ 
        0 & 0 & 0 & 0 & \ldots & 0 & 0   
    \end{pmatrix}
\end{align}

The left-hand side of equation \eqref{diffreg} is referred to as the \textit{solution norm}. Simultaneously minimizing both of these norms is accomplished using the method of Tikhonov regularization~\cite{nla.cat-vn720231}. This is a general method for solving ill-posed problems by incorporating additional regularization conditions. In this case, the expression for the coefficients takes the form:

\begin{align}
    \vec{c} = \left[N^\top P N + \Gamma^\top \Gamma\right]^{-1} N^\top P \vec{F}
    \label{tikhonov}
\end{align}
where the new elements are $\Gamma = \lambda A$, a tunable version of the regularization matrix scaled by $\lambda$, and $P$, which functions as an $N_{\mathrm{Energy \, bins}} \times N_{\mathrm{Energy \, bins}}$ inverse-covariance matrix.  In this application, $P$ will be purely diagonal, and used to apply weighting to regions of the flux so that, for example, very low energy regions do not overly constrain the coefficients. By choosing a regularization parameter, such as $\lambda = 10^{-8}$, the coefficients vary more smoothly without significantly degrading the quality of the flux match, as shown in Figure~\ref{fig:reg}.

\begin{dunefigure}[\dshort{fd} flux and \dshort{nd} linear combination flux match and corresponding coefficients; Eqn~\eqref{tikhonov}]
{fig:reg}
{The \dword{fd} flux and the \dword{nd} linear combination flux match (left) and corresponding coefficients (right) are shown using the solution of Equation~\eqref{tikhonov}.}
    \includegraphics[width=0.55\textwidth]{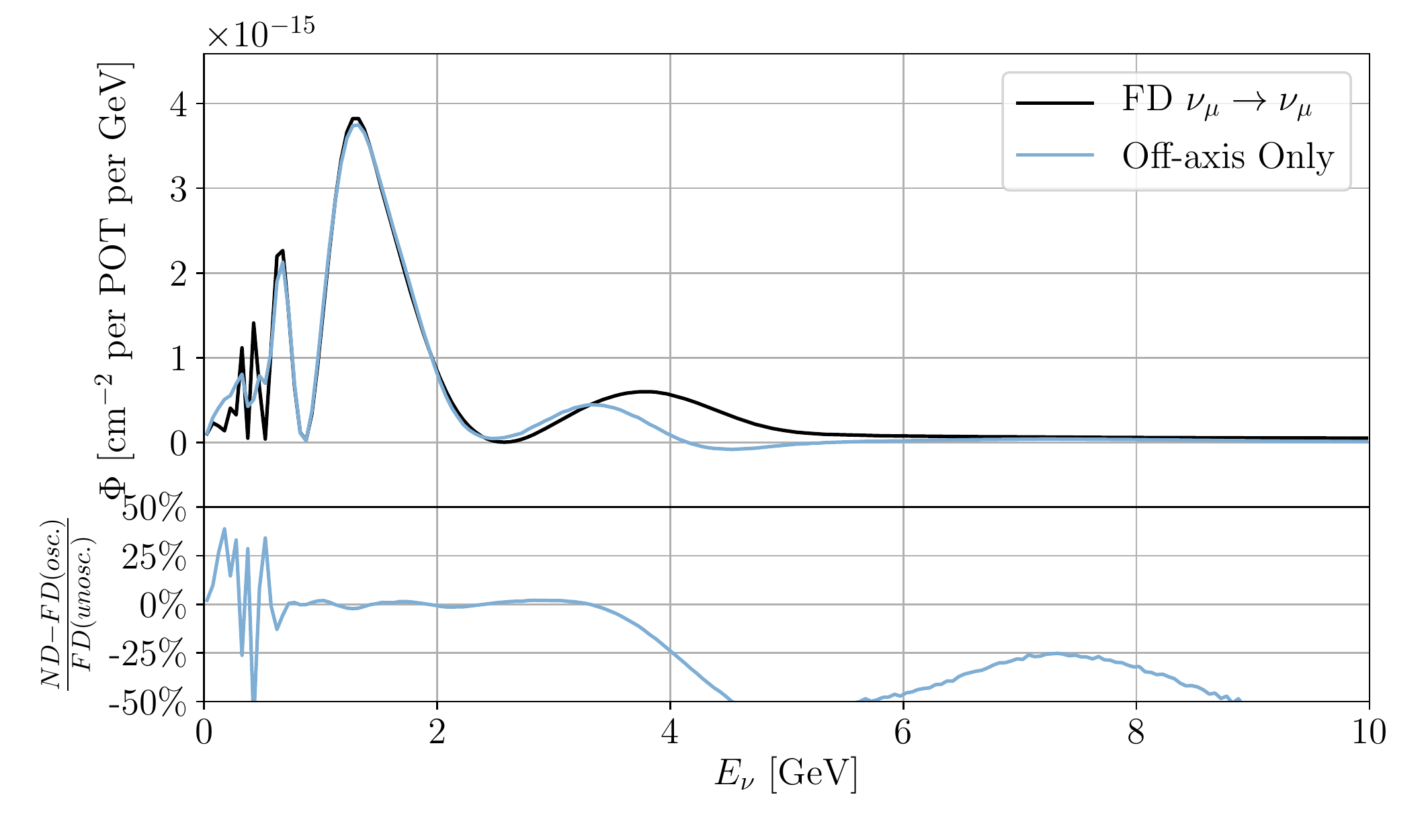}
    \includegraphics[width=0.43\textwidth]{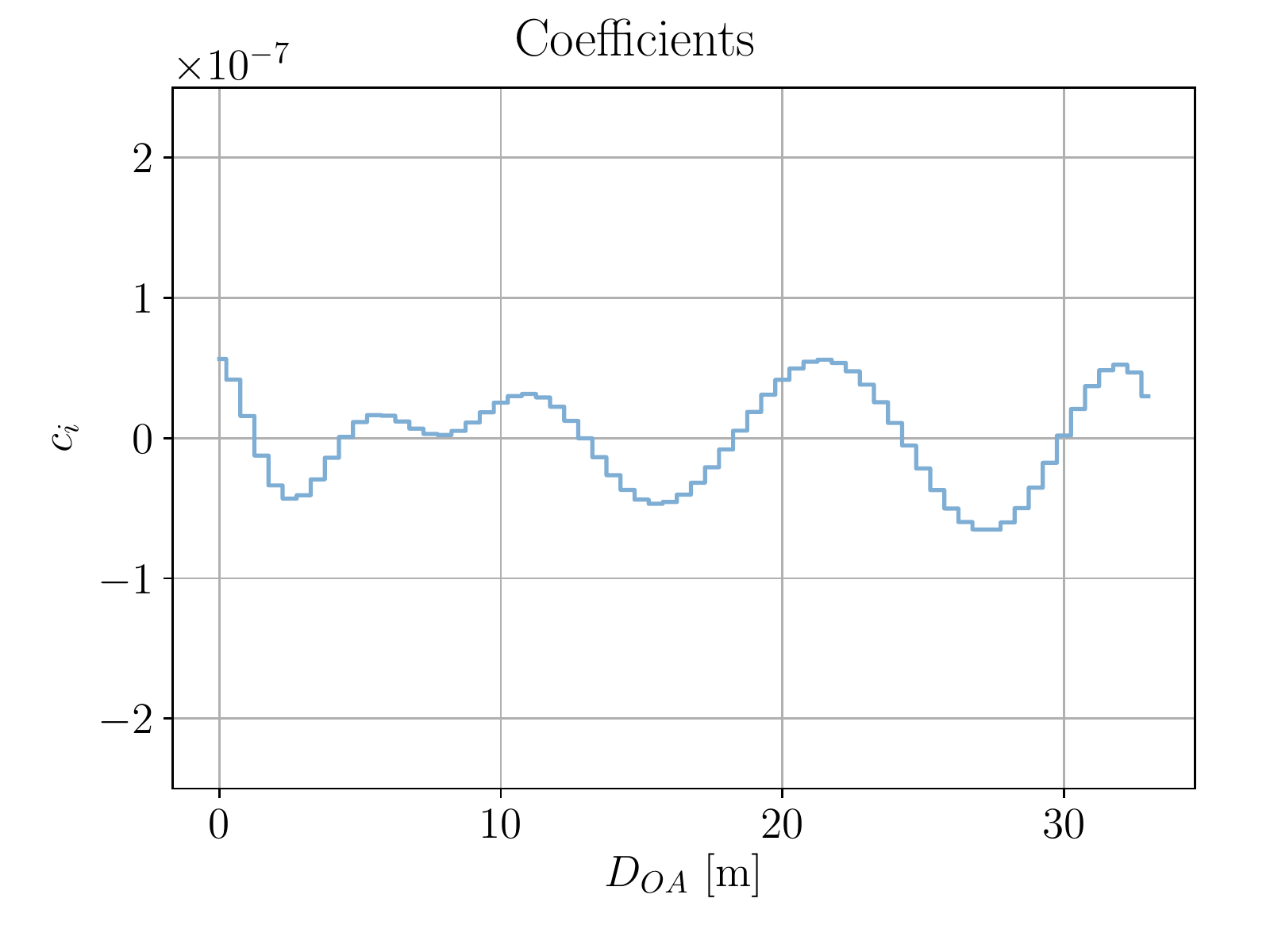}
\end{dunefigure}

The regularization parameter can be further optimized by constructing an "L-curve", the parametric plot produced by comparing the norm of the residuals (left-hand side of equation \ref{simple}) and the norm of the regularization term (left-hand side of equation \ref{diffreg}).  The optimal value for $\lambda$ is found when decreasing $\lambda$ stops reducing the residual norm and instead causes the regularization norm to increase~\cite{hansenlcurve}.

\subsection{Incorporating Horn Current Fluxes}

The linear combination of off-axis fluxes shown in Figure~\ref{fig:reg} is unable to match the target oscillated far detector flux in the 3~GeV to 5~GeV region. This is because the highest energy flux available in the linear combination is the on-axis flux, which peaks around 2.5~GeV. In order to improve the flux matching, an additional flux with unique information around 4~GeV is needed, and this can be obtained through a small variation in the current of the magnetic focusing horns in the DUNE beamline.

Additional fluxes can be added to the treatment described in the previous section as extra rows in the ND flux matrix. The resulting coefficient vector then has a corresponding entry for each additional flux. Regularization for these fluxes can be done independently, by treating $\Gamma$ as a block-diagonal matrix. In this treatment, no regularization is used for these extra fluxes:

\begin{align}
    \Gamma = \begin{pmatrix}
        \lambda_1 A & 0 \\
        0 & 0
    \end{pmatrix}
\end{align}

Preliminary studies have indicated that the presence or strength of regularization for this term has very little effect on the value of the corresponding coefficient.  The rest of the method remains unchanged.

By incorporating an additional flux with a horn current of 280~kA, more information is obtained in the 4~GeV region, where the off-axis fluxes are insufficient to fit the \dword{fd} oscillated flux, as shown in Figure~\ref{fig:prism-oaandhorn}.  The result of the flux matching with this additional horn current flux is shown in Figure~\ref{fig:oaandhorn_fit} alongside the off-axis only flux matching result.  The linear combination of the off-axis fluxes, including the 280~kA horn current flux, now provides a good fit across the entire high energy portion of the \dword{fd} energy spectrum.

\begin{dunefigure}[DUNE ND Neutrino Flux at three off-axis positions compared to on-axis flux with lower horn current]{fig:prism-oaandhorn}
  {The ratio to the nominal on-axis flux is shown for three off-axis positions, and for on-axis running where the nominal \SI{293}{kA} horn current has been lowered to \SI{280}{kA}. The modified horn current provides an additional constraint just above the first oscillation maximum, with no effect at the lower energies sampled by the off-axis fluxes.}
  \includegraphics[width=0.6\textwidth]{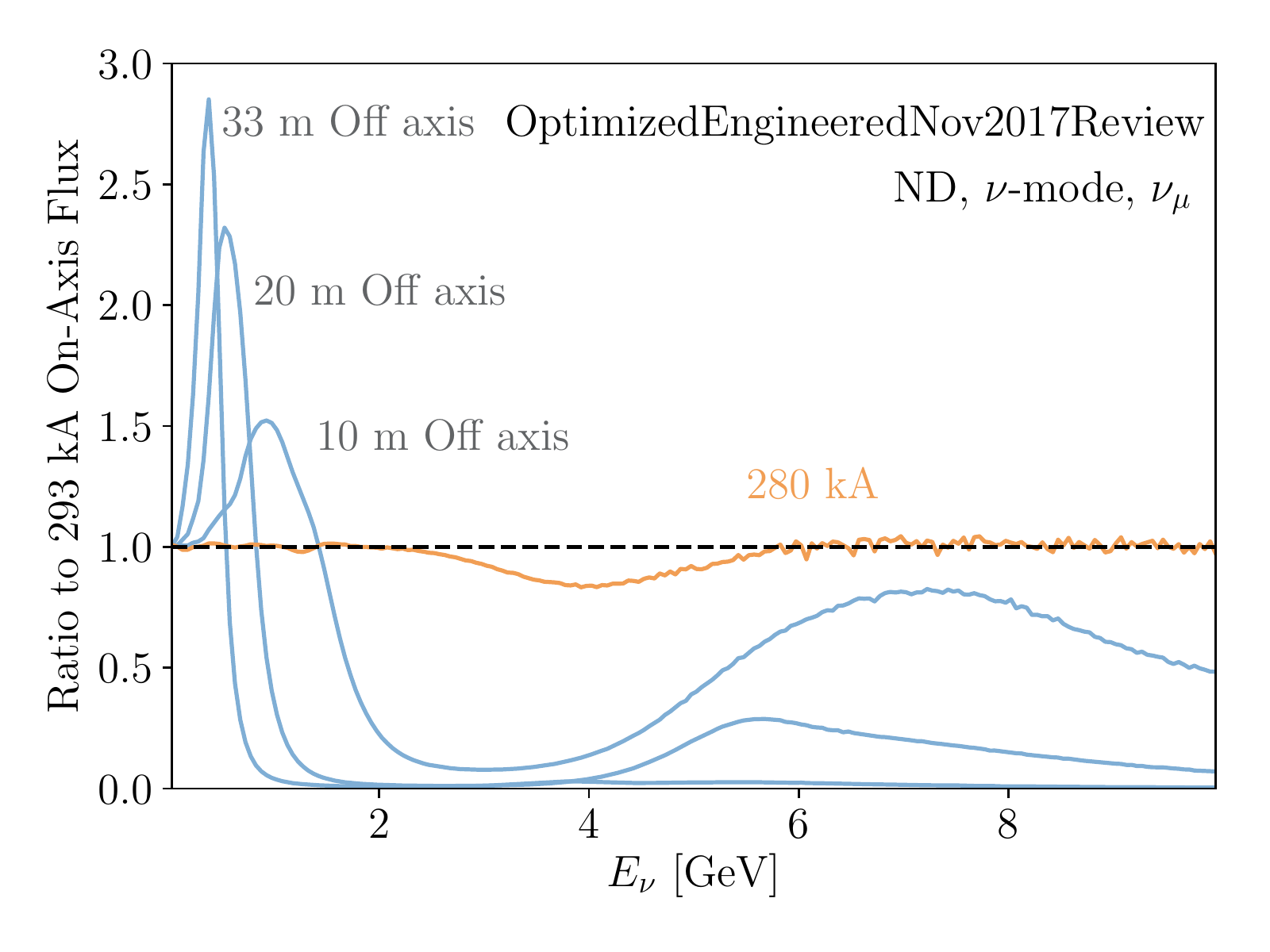}
\end{dunefigure}

\begin{dunefigure}[\dshort{fd} flux and \dshort{nd} linear combination flux match and corresponding coefficients, utilizing additional horn current flux; Eqn~\eqref{tikhonov}]
{fig:oaandhorn_fit}
{The \dword{fd} flux and the \dword{nd} linear combination flux match (left) and corresponding coefficients (right) are shown utilizing an additional horn current flux.}
    \includegraphics[width=0.55\textwidth]{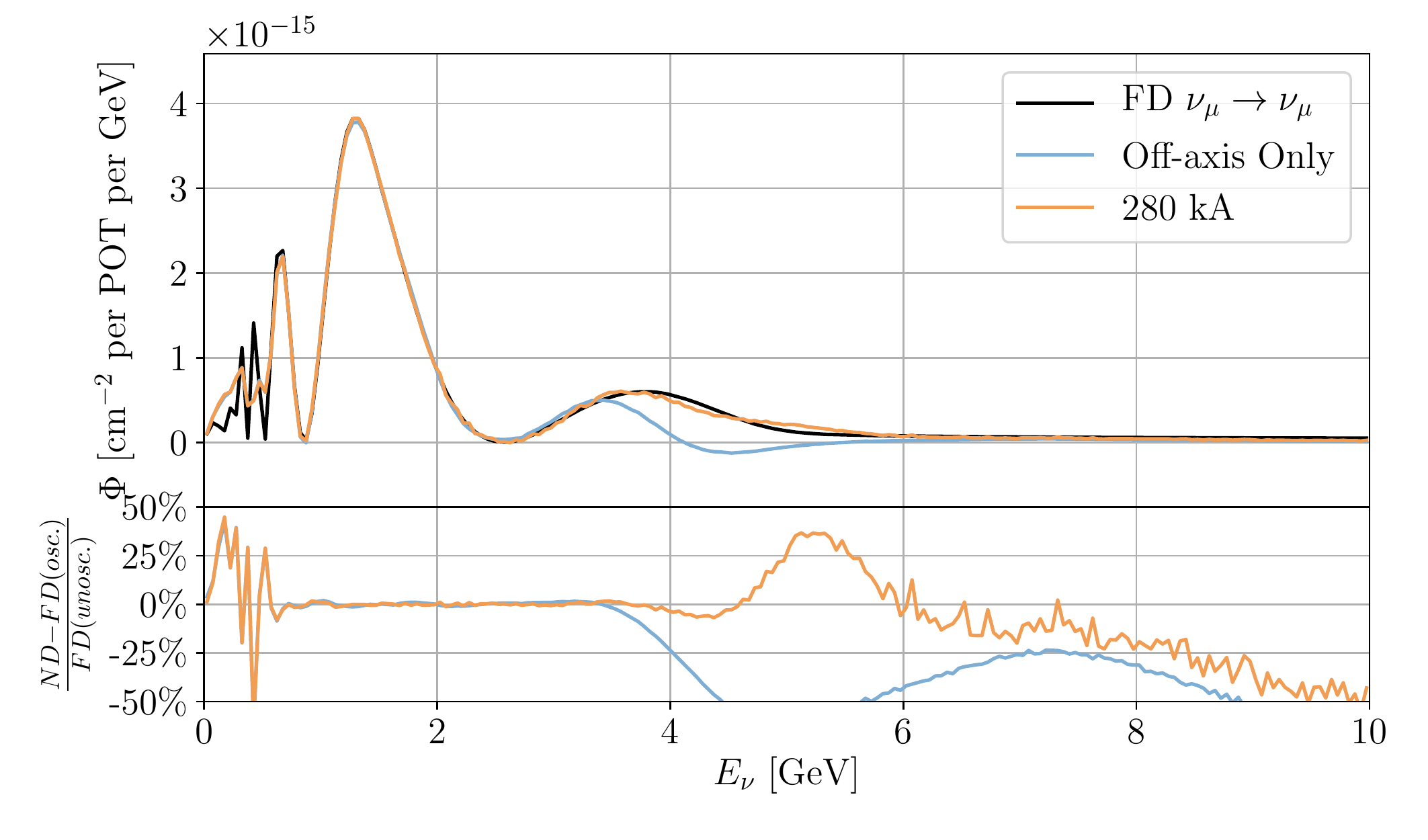}
    \includegraphics[width=0.43\textwidth]{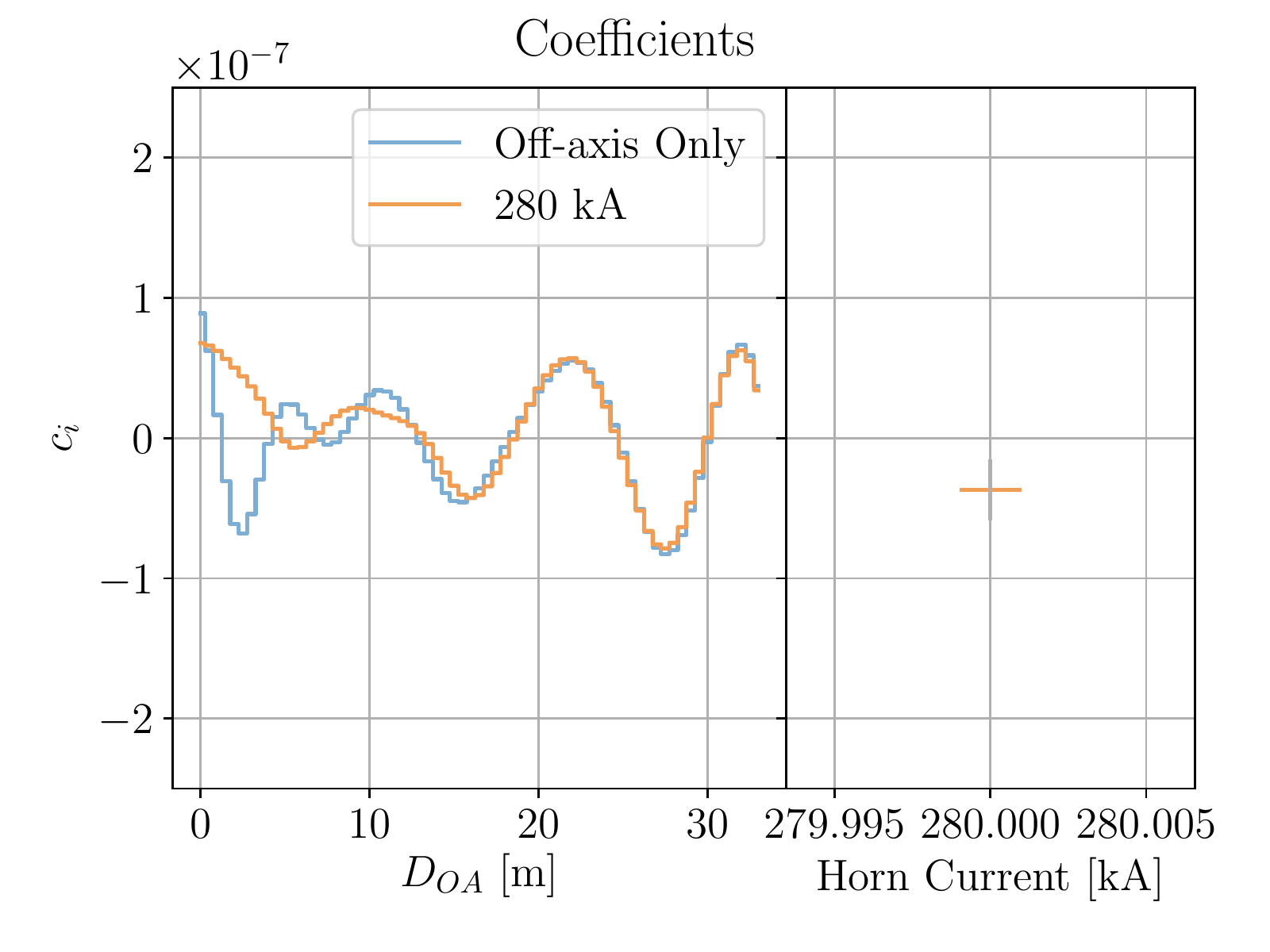}
\end{dunefigure}

The result is a method that provides good flux matching over multiple sets of oscillation parameters, as shown in Figure~\ref{fig:disapp_fits}.


\begin{dunefigure}[\dshort{fd} predicted muon neutrino spectra under a variety of oscillation hypotheses]
{fig:disapp_fits}
{\dword{fd} predicted muon neutrino spectra under a variety of oscillation hypotheses. Left: current best results for muon neutrino disappearance parameters, colors showing chosen oscillation hypotheses. Right: solid lines correspond to the \dword{fd} oscillated flux predictions in color coordination with points on the left plot.  The dashed lines are the best match spectra for oscillated \dword{fd} fluxes constructed from linear combinations of \dword{nd} fluxes (33 m off-axis + 280 kA special horn current run).}
    \includegraphics[width=0.49\textwidth]{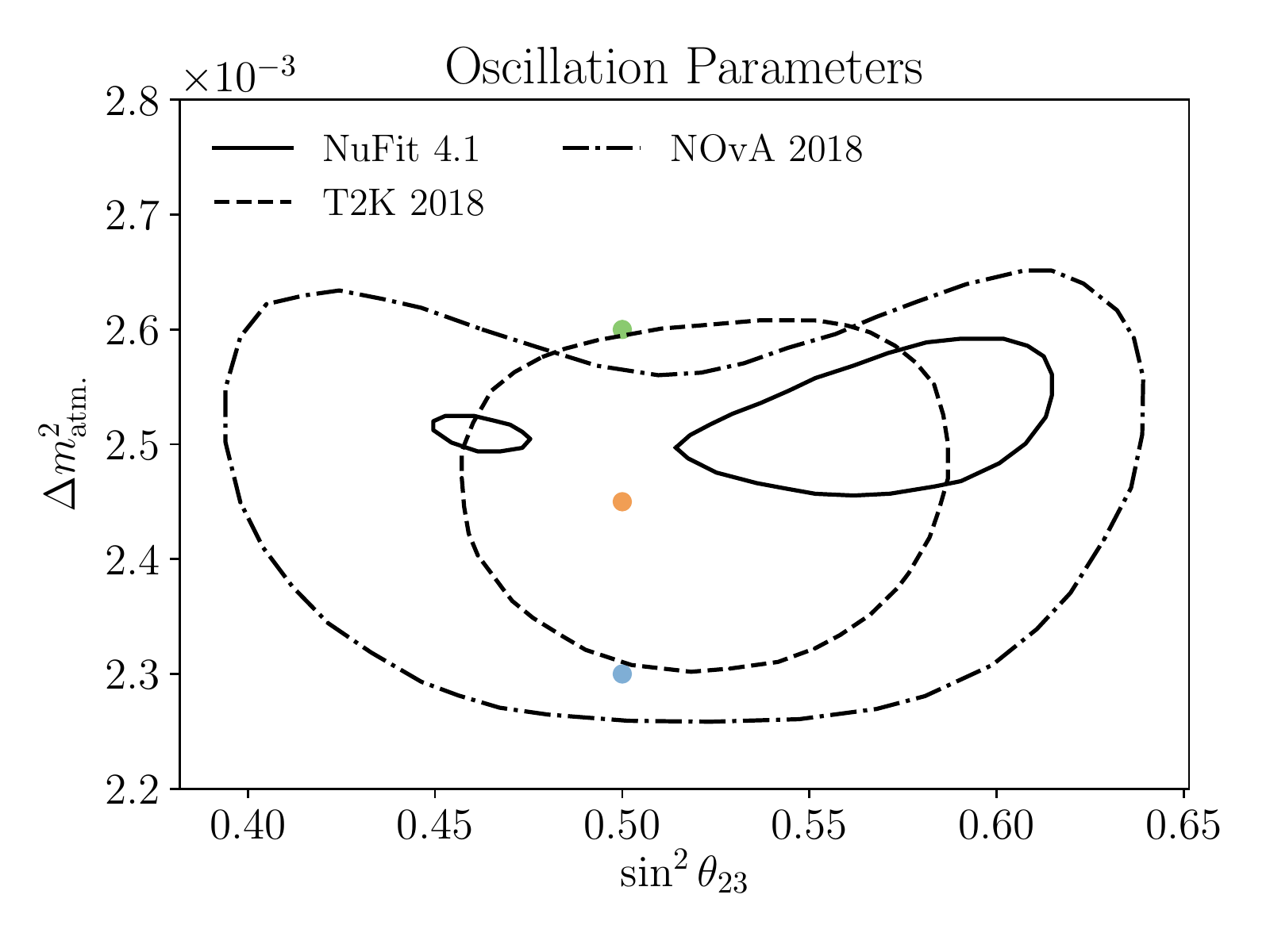}
    \includegraphics[width=0.49\textwidth]{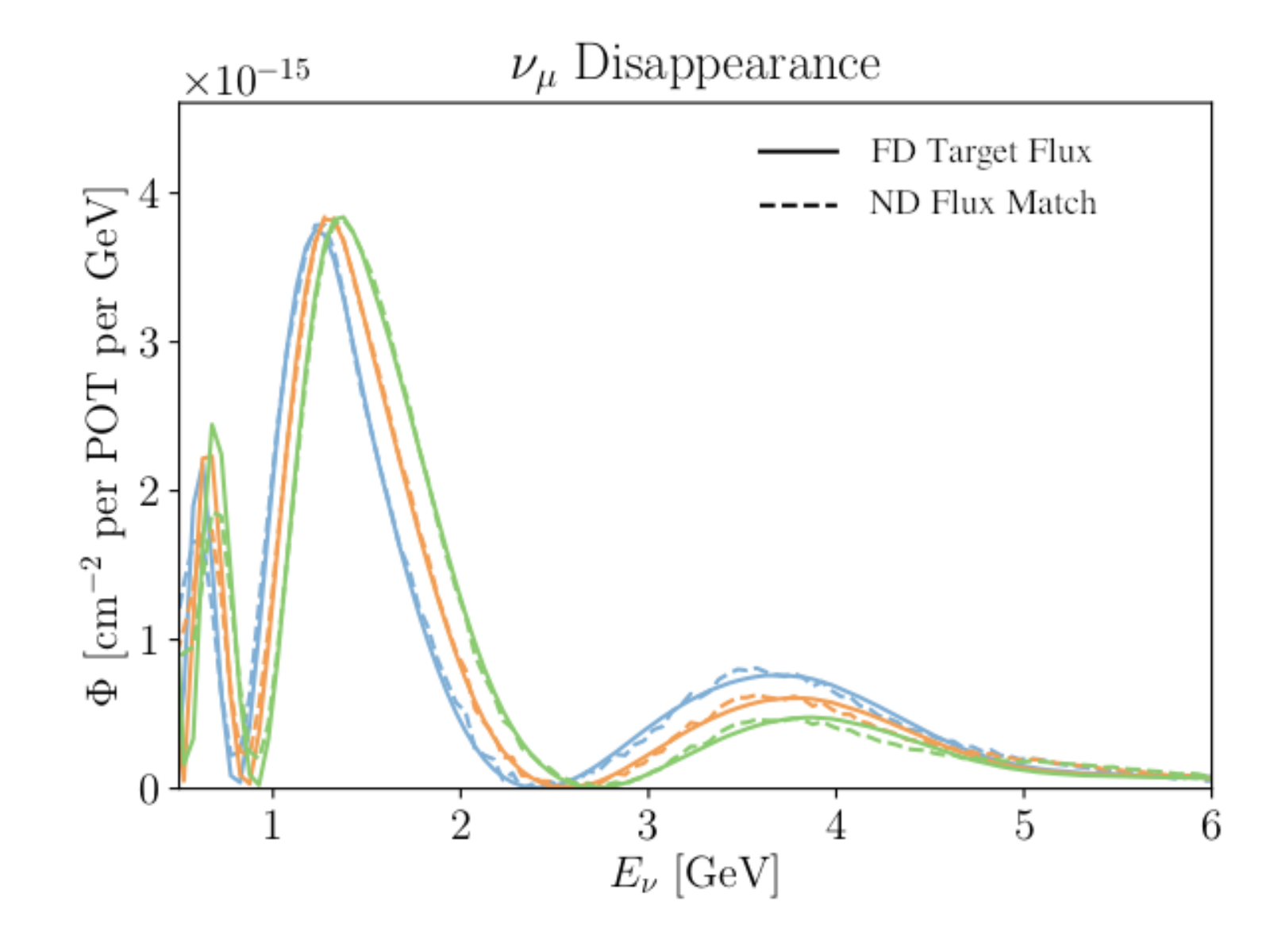}
\end{dunefigure}

\subsection{Electron Neutrino Appearance Flux Matching}

The $\nu_e$ appearance analysis can proceed similarly to the $\nu_\mu$ disappearance analysis, with the additional complication of $\sigma(\nu_e)/\sigma(\nu_\mu)$ cross section corrections. Thus, the analysis requires several steps:

\begin{enumerate}
\item A fit of the \dword{nd} $\nu_\mu$ off-axis fluxes to the \dword{fd} oscillated $\nu_e$ spectrum. This generates a \dword{fd} prediction under the assumption that $\nu_e$ and $\nu_\mu$ interactions have identical cross sections.
\item A fit of the \dword{nd} $\nu_\mu$ off-axis fluxes to the \dword{nd} intrinsic $\nu_e$ spectrum. This allows for a double-differential comparison in lepton kinematics and hadronic energy of the $\nu_\mu$ and $\nu_e$ cross sections and detector response, from which a correction can be extracted for step 1. The intrinsic $\nu_e$ data used in this step are integrated over the off-axis range between 16~m and 32~m to achieve sufficient $\nu_e$ statistics, and to reduce the contamination from $\nu_\mu$ backgrounds.
\item A direct measurement of the neutral current backgrounds using data from the on-axis \dword{nd} position, which has an NC energy spectrum similar to that of the \dword{fd}. The NC backgrounds will also be constrained from Gaussian flux matched data as described in Section~\ref{sec:prismgauss}
\end{enumerate}

Figure~\ref{fig:dposcnue} shows the step 1 fits for $\nu_e$ and $\bar{\nu}_e$, and Figure~\ref{fig:dpintrinsicnue} shows the step 2 fit to the intrinsic $\nu_e$ distribution. In all cases, the $\nu_\mu$ (or $\bar{\nu}_\mu$) flux can largely reproduce the target $\nu_e$ distribution.

\begin{dunefigure}[Matching of \dshort{nd} $\nu_\mu$ flux to the oscillated \dword{fd} $\nu_e$ flux]
{fig:dposcnue}
{The flux matching of the \dword{nd} $\nu_\mu$ flux to the oscillated \dword{fd} $\nu_e$ flux (left) is shown assuming various sets of oscillation parameters (right).  Target fluxes are shown with solid lines and resulting fits are shown in dashed lines.}
    \includegraphics[width=0.49\textwidth]{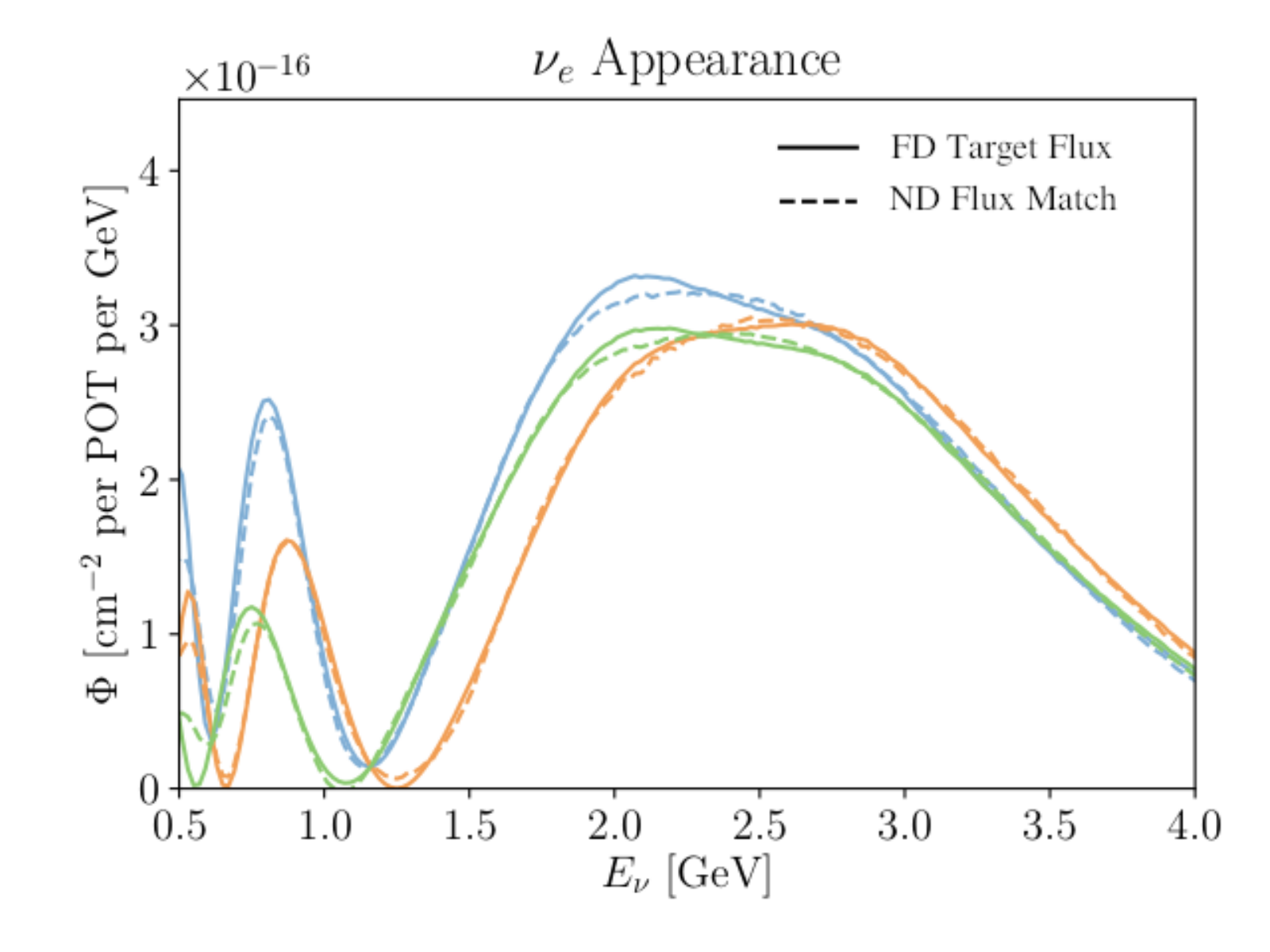}
    \includegraphics[width=0.49\textwidth]{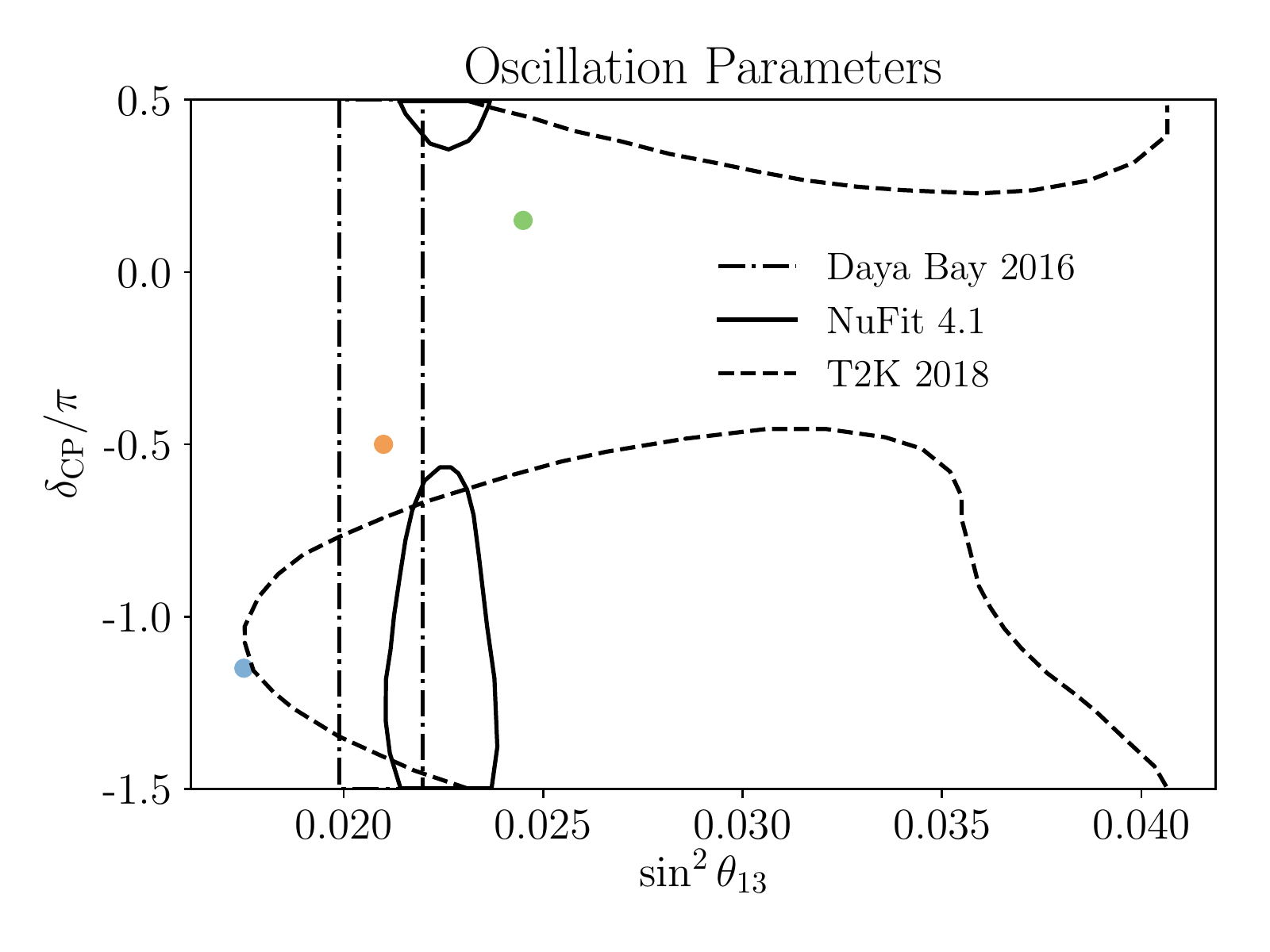}
\end{dunefigure}

\begin{dunefigure}[Matching \dshort{nd} off-axis $\nu_\mu$ fluxes to average \dword{nd} intrinsic $\nu_e$ flux]
{fig:dpintrinsicnue}
{Left: the matching of \dword{nd} off-axis $\nu_\mu$ fluxes to the average \dword{nd} intrinsic $\nu_e$ flux between 16.5 and 33 m is shown. Right: the linear combination coefficients used for the fitted flux on the (\emph{left}).}
    \includegraphics[width=0.49\textwidth]{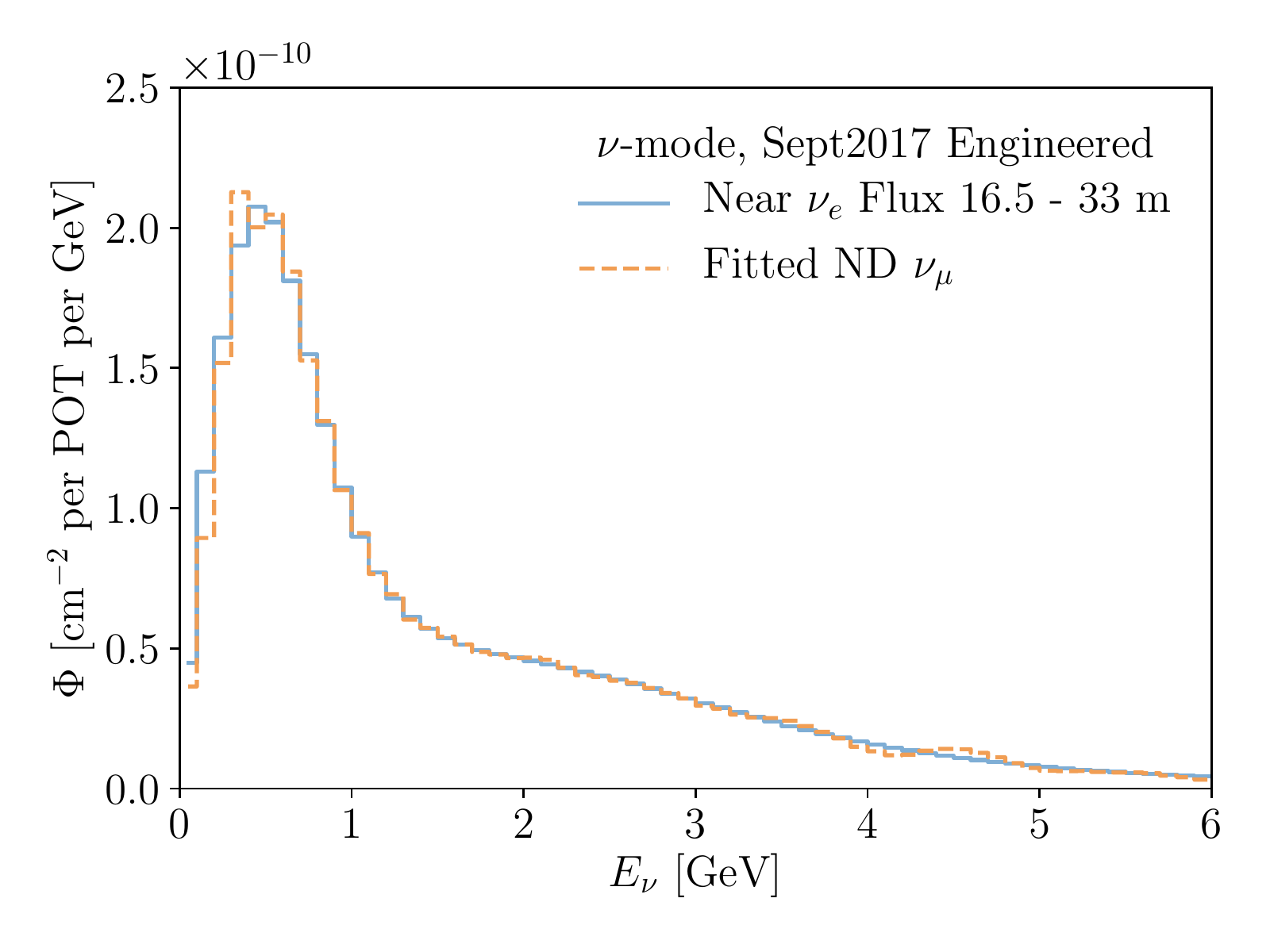}
    \includegraphics[width=0.49\textwidth]{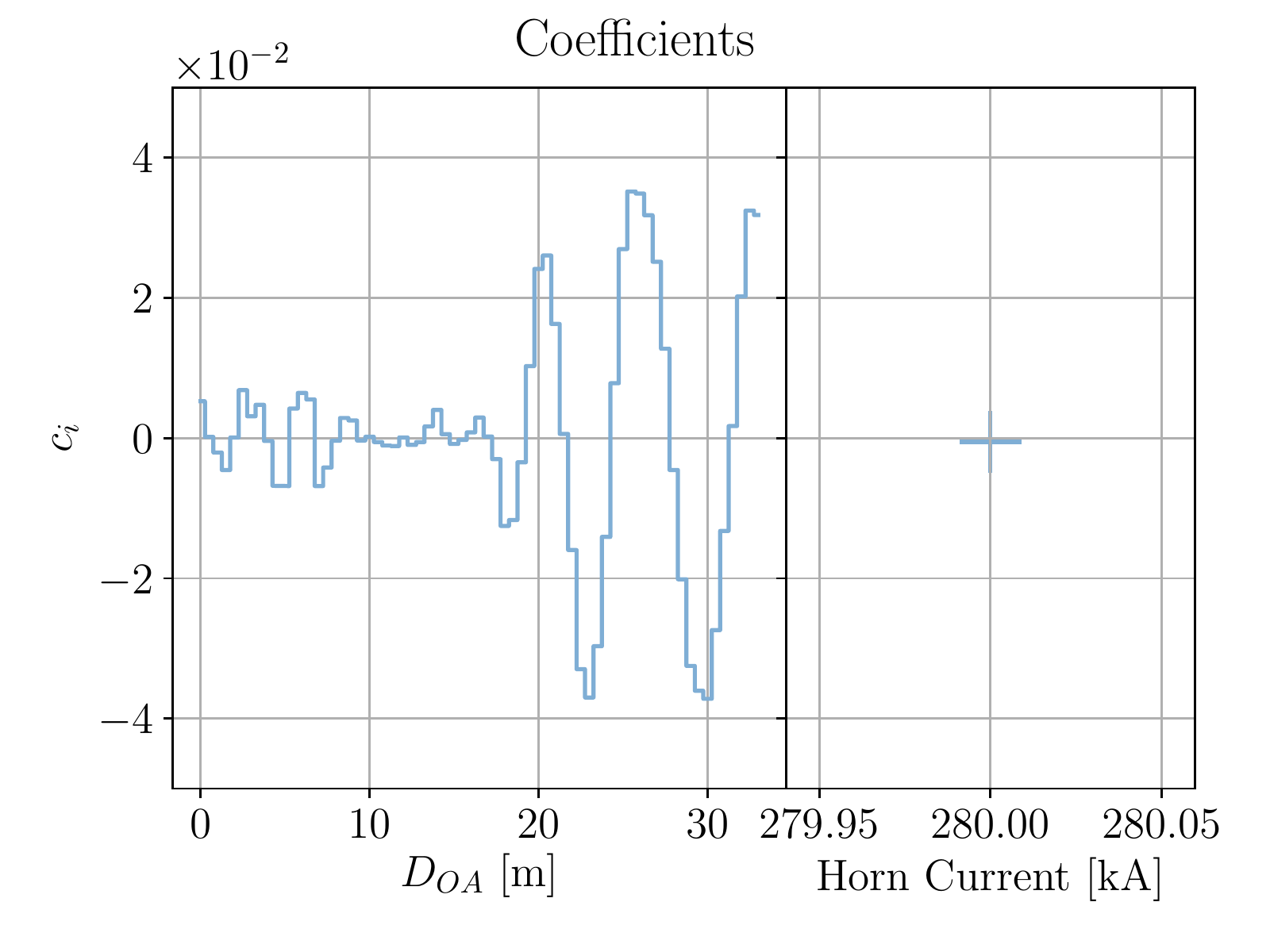}
\end{dunefigure}

\subsection{Gaussian Flux Matching}\label{sec:prismgauss}

Rather than using linear combinations to match an oscillated \dword{fd} flux, it is also possible to match a Gaussian energy distribution to enable measurements of final state particle kinematics for a known incident neutrino energy. This is particularly useful for constraining neutral current backgrounds to the $\nu_e$ appearance measurement as a function of neutrino energy, which is a novel, unique capability of the DUNE-PRISM near detector off-axis measurement program.

Figure~\ref{fig:gaussianfluxes} shows examples of constructed Gaussian energy distributions at neutrino energies of 0.5, 1, 2, and 3~GeV, each with a width that is 10\% of the mean. These combinations can be used to constrain background processes, including backgrounds from neutral current interactions, as a function of incident neutrino energy. This allows for a direct measurement of the reconstructed energy distribution as a function of true neutrino energy.

\begin{dunefigure}[DUNE-PRISM Gaussian fluxes]{fig:gaussianfluxes}
  {Linear combinations of ND $\nu_\mu$ off-axis fluxes used to produce Gaussian incident neutrino energy spectra.  These are shown for four different selected energies, each with a 10\% width.}
  \includegraphics[width=0.58\textwidth]{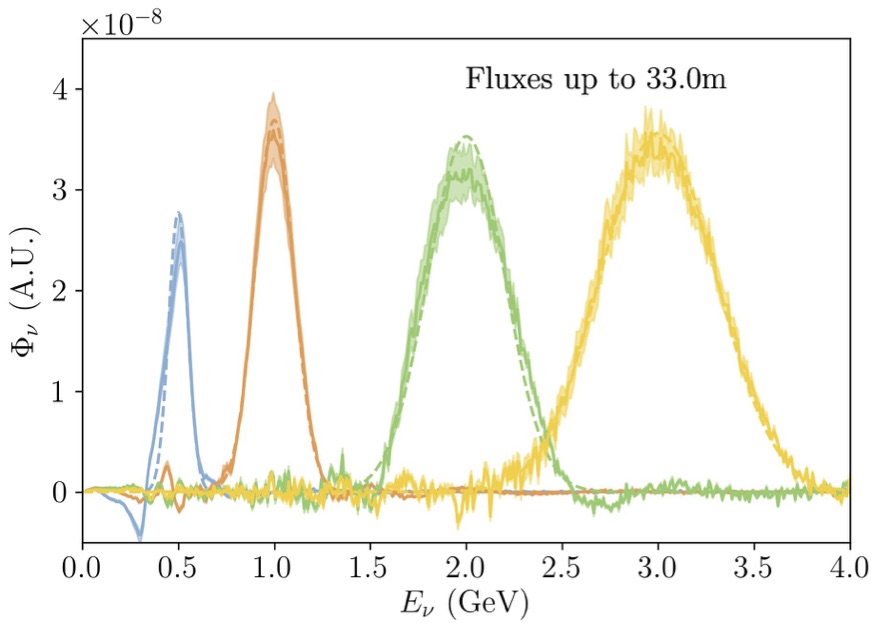}
\end{dunefigure}

\section{Flux Systematic Uncertainties}
\label{sec:prism-flux-systs}

Although the use of \dword{nd} linear combinations to produce oscillated \dword{fd} energy spectra predictions substantially reduces the dependence of the oscillation analysis on neutrino interaction modeling, this analysis strategy is still susceptible to systematic uncertainties in the neutrino flux prediction. However, unlike many important neutrino interaction modeling uncertainties, most of the flux uncertainties largely cancel when comparing the linearly combined \dword{nd} fluxes to a given oscillated \dword{fd} flux.

\subsection{Impact on Linear Combination Analysis}

To illustrate the impact flux uncertainties will have on the linear-combination-based oscillation analysis (as described in Section~\ref{sec:prism-oa}), several throws of the systematic uncertainties in the hadron production in the LBNF beam line are  simultaneously applied to the linearly combined \dword{nd} fluxes and the corresponding oscillated \dword{fd} flux, without changing the linear coefficients. Figure~\ref{fig:dpsystvariation} shows a particularly large flux variation that modifies the \dword{fd} flux by more than 10\% of the unoscillated flux. Despite such a large variation, the change in the \dword{nd} linear combination largely tracks the change in the \dword{fd} oscillated flux, and the resulting systematic uncertainty from this variation, given by the difference between the variations in the near detector linear combination and the far detector fluxes, is at the percent level.


\begin{dunefigure}[Cancellation of hadron production systematic variation]
{fig:dpsystvariation}
{Cancellation of hadron production systematic variation.  The top plot shows the difference between the systematically varied and nominal flux (FD) and the flux formed from the linearly combined \dword{nd} fluxes.  The bottom plot shows the cancellation achieved as the two variations affect the target flux and the combination flux in the same way.  Note that the red dashed lines indicate the region where the target flux is fit.}
    \includegraphics[width=0.5\textwidth]{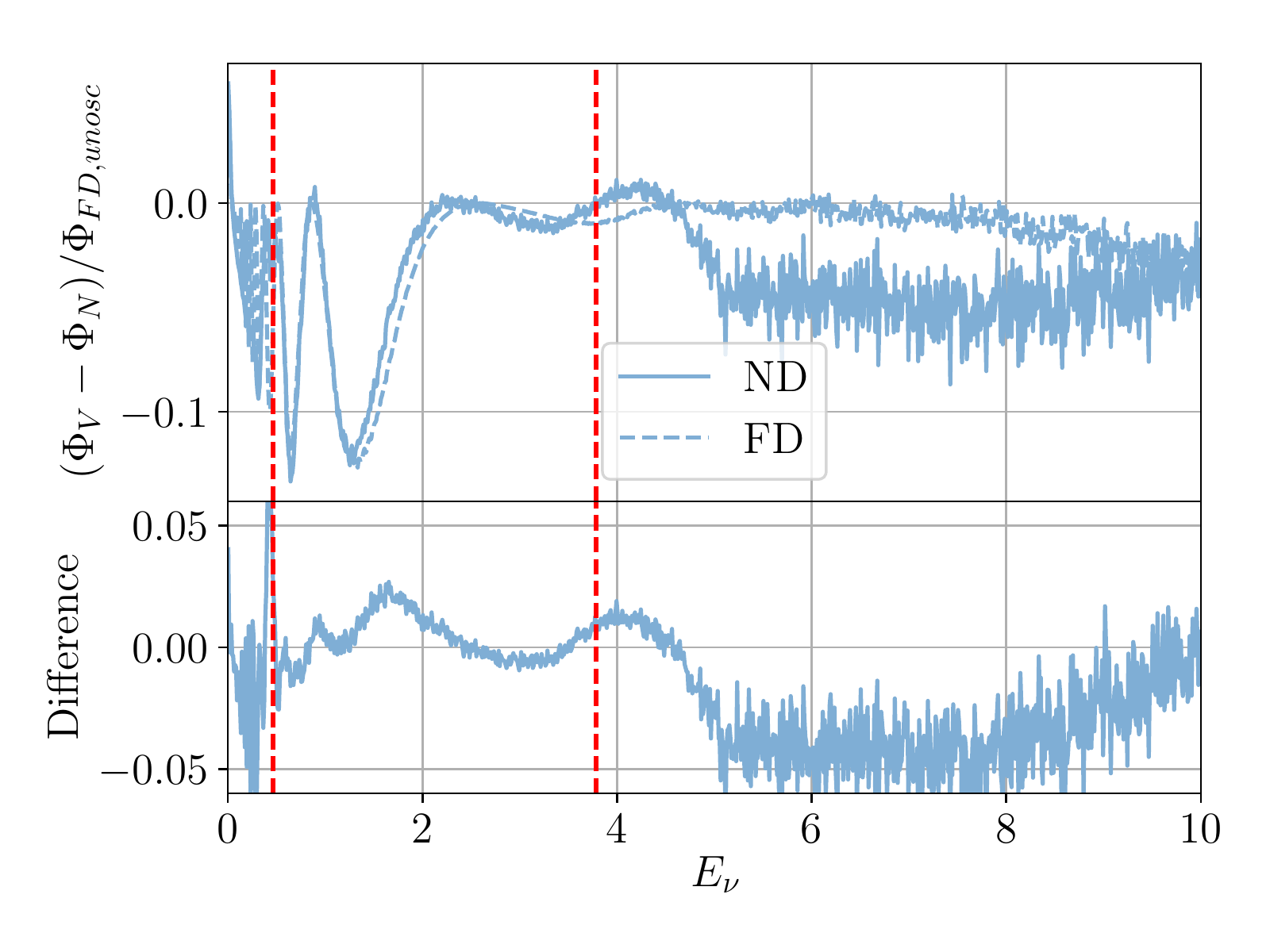}
\end{dunefigure}

This exercise can be repeated for a large number of flux throws to produce a 1$\sigma$ uncertainty band due to hadron production uncertainties in the beam line. Figure~\ref{fig:fluxerrorgrid} shows the resulting error bands for several choices of the the oscillation parameters. The total uncertainty is constrained to the percent level throughout the oscillation region.


\begin{dunefigure}[Cancellation of hadron production systematic variations; various oscillation hypotheses]
{fig:fluxerrorgrid}
{Cancellation of hadron production systematic variations assuming various oscillation hypotheses (top right).  Shown are median values (solid line) and 68\% containment intervals (bands).  The figure colors correspond to the oscillation hypothesis points on the upper right panel.}
    \includegraphics[width=0.9\textwidth]{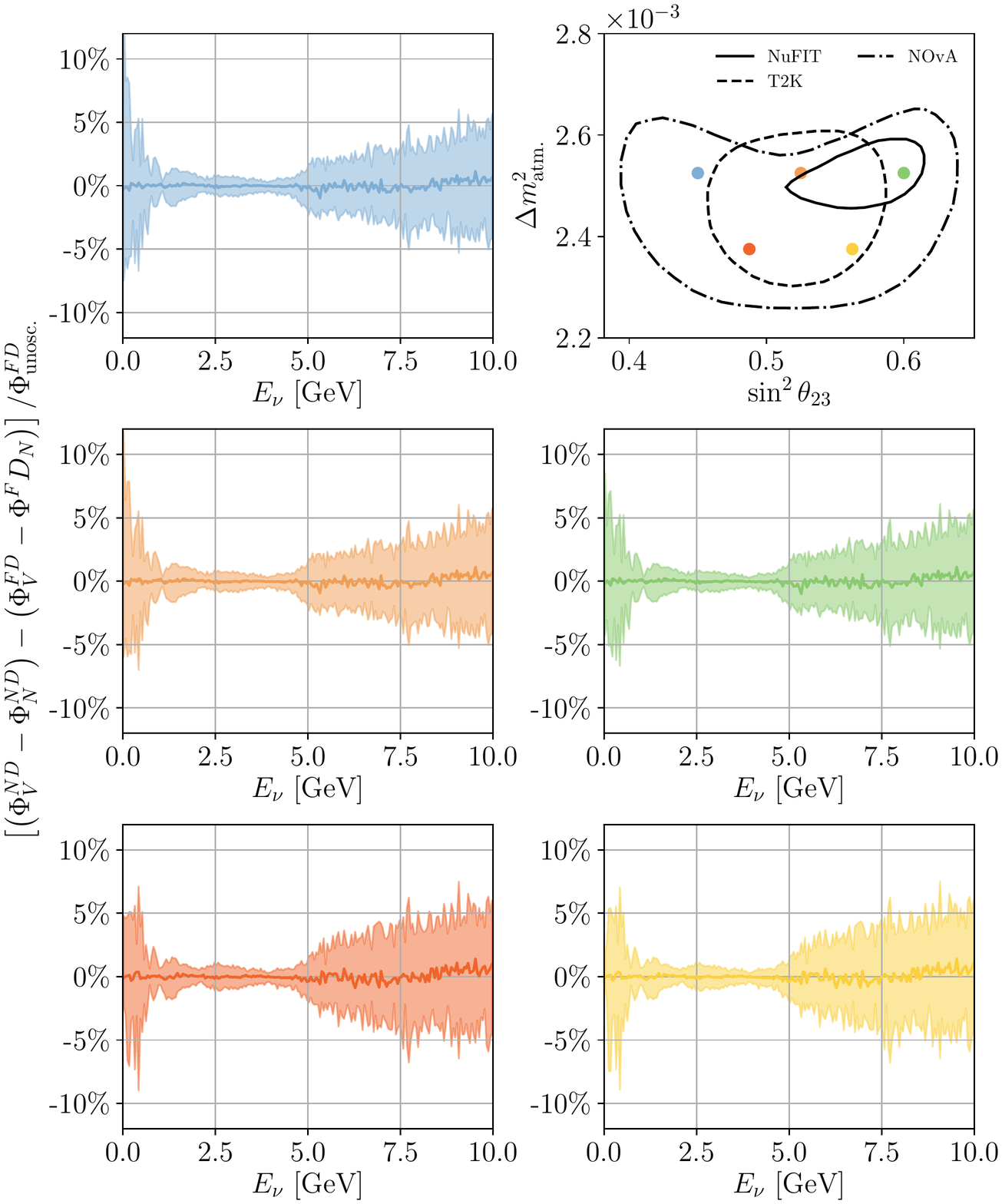}
\end{dunefigure}

\subsection{Systematic Uncertainties on Gaussian Fluxes}

A similar evaluation of the impact of flux systematic uncertainties can be performed for the matched Gaussian fluxes. Systematic variations are applied to the \dword{nd} fluxes, and the resulting variations in the Gaussian mean, width, and normalization are extracted, as shown in Figure~\ref{fig:gaussianfluxerrors}. The chosen true neutrino energy for each Gaussian energy spectrum is precisely controlled to better than 0.5\% accuracy. Given the limitations of the available energy spectra at high and low energies, it is difficult to produce Gaussian energy spectra with widths precisely 10\% of the mean energy, and the resulting widths can range up to 11\% or 12\% of the mean energy. However, the variation of these widths due to hadron production uncertainties is small. Finally, since measurements with Gaussian fluxes do not involve any a priori near/far cancellation, the full flux normalization uncertainty is propagated to the constructed Gaussian fluxes.

\begin{dunefigure}[DUNE-PRISM Gaussian flux uncertainties]{fig:gaussianfluxerrors}
  {The variation in the mean, width, and normalization for the linearly combined Gaussian neutrino energy spectra from 100 throws of the hadron production uncertainties is shown as a function of the chosen Gaussian mean neutrino energy. ``PPFX" is the name of the beam modeling framework that allows for the variation of uncertainties.}
  \includegraphics[width=0.32\textwidth]{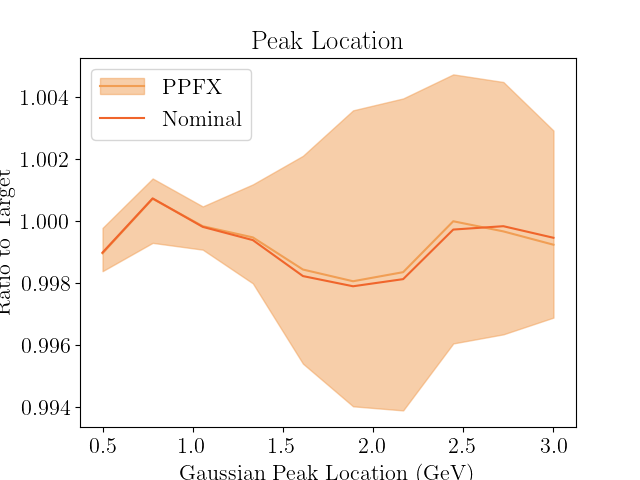}
  \includegraphics[width=0.32\textwidth]{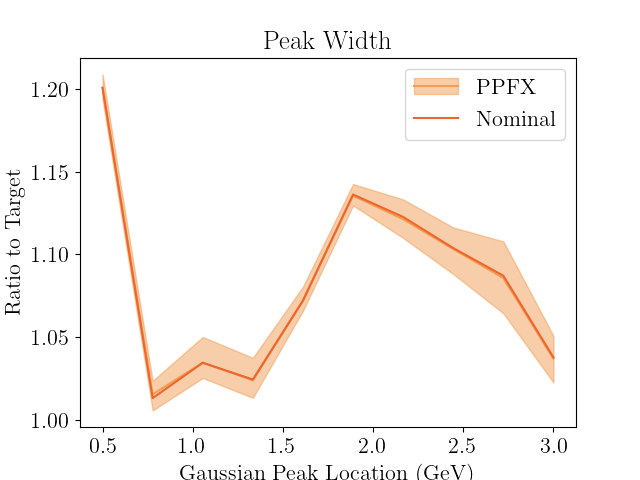}
  \includegraphics[width=0.32\textwidth]{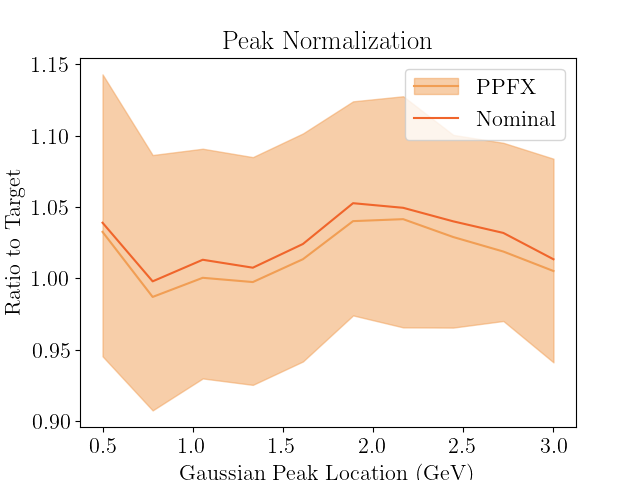}
\end{dunefigure}

\section{Linear Combination Oscillation Analysis}
\label{sec:prism-oa}

A complete DUNE-PRISM linear-combination analysis is not yet possible, since this requires full \dword{nd} simulation and reconstruction tools. However, this section will outline the steps of such an analysis, and demonstrate that such an analysis is expected to avoid the oscillation parameter biases outlined in Section~\ref{sec:dposcbiases}.

\subsection{Observation Weights}

For each off-axis detector position, the active region of the detector can be logically divided into ``flux windows'', which are sequential fiducial volume slices along the off-axis dimension. 
Given a set of flux windows, the procedure for determining the coefficients,
$c_i$, to use in the linear combination of interactions occurring
within each flux window is described in Section~\ref{sec:prism-lincomb}.
The flux windows are defined by an extent in absolute off-axis position,
relative to origin of the \dword{nd} coordinate system. In the flux fits used
here, each window is $50\,\textrm{cm}$ wide and contiguously span the region between on-axis and $33\,\textrm{m}$ off-axis.
These definitions may be further optimized to
improve the fidelity of the oscillated flux fits. Selected events are separated
into samples based on the flux window in which they occurred. Perfect
reconstructed vertex resolution is assumed for this process.
%

For a perfect fit of the \dword{fd} oscillated flux, in the absence of any event
selection differences between the near and far detectors, the predicted far
detector rate, $\mathcal{R}_\textsc{fdp}\left(\vec{y}\right)$, is given by

\begin{equation}
  \mathcal{R}_\textsc{fdp}\left(\mathcal{H}_\textsc{oa}, \vec{y}\right) = M_\textsc{nf}\sum_{i} c_{i}\left(\mathcal{H}_\textsc{oa}\right)\mathcal{R}_{\textsc{nd},i}\left(\vec{y}\right).
  \label{eq:PRISM_perfworld}
\end{equation}

In Equation~\ref{eq:PRISM_perfworld}, $c_{i}\left(\mathcal{H}_\textsc{oa}\right)$
is the set of observation weights determined under oscillation hypothesis,
$\mathcal{H}_\textsc{oa}$, $\mathcal{R}_{\textsc{nd},i}\left(\vec{y}\right)$ is the
distribution of selected \dword{nd} events occurring in flux window, $i$, and
$M_\textsc{nf}$ is fiducial mass ratio of the \dword{fd} to the \dword{nd}.
Here, $\vec{y}$ denotes some projection of the observed event kinematics. Perhaps the most
obvious projection is reconstructed neutrino energy, $E_{\nu,\textrm{rec.}}$, as a
proxy for true neutrino energy, which affects oscillation probabilities.
However, it may be the case that oscillation parameter sensitivity could be
greater in some other, potentially multi-dimensional projection (\textit{e.g.}
the T2K oscillation analysis is performed in $\vec{y} = (p_{\mu},\theta_{\mu})$).
The DUNE-PRISM technique places no assumptions on the projection used, only that it can be made similarly for selected events at both the near and \dword{fd}.
Any relative normalization of the near and far neutrino fluxes has
been absorbed into the observation weights.

\subsubsection{Event Selection}

The event selection used here aims to select muon-neutrino, charged-current
interactions with well-contained hadronic showers. The selection is given
access to the full, generator-level, final-state, primary lepton truth
information and the GEANT4-simulated energy deposits left by the hadronic
shower. Events that leave less than $30\,\textrm{MeV}$ of hadronic energy in
the outer 30~cm of the liquid argon active volume are selected as having well-contained hadronic showers to ensure that no hadronic energy was lost. To minimize the dependence on cross section modeling, events with vertices in the outer 1.5~m in the off-axis dimension are excluded. Finally, for this
study, no selection criteria were placed on the final-state muon kinematics.

The aim of the DUNE-PRISM analysis is to compare a prediction of some far
detector observable (in this case $E_{rec}$) built from combinations of \dword{nd} measurements
to \dword{fd} data to better understand neutrino oscillations. The comparison
between the \dword{fd} prediction and the \dword{fd} data is only reasonable
if any differences between the event samples at the near and far detectors are
accounted for. These differences are broadly separable into selection efficiency
and purity differences.

\subsubsection{Impact of Backgrounds}

In a real data analysis, background rates at the near and far detectors will differ because of the effects of oscillation, the off-axis position of the \dword{nd}, and differences in the detector technology, geometry, and environment (more cosmic rays and entering beam-related backgrounds at the \dword{nd}). Anticipated sources of irreducible background include: neutral current interactions that produce charged pions (that get mistaken for a single muon event); neutral current interactions that produce a neutral pion (where the electromagnetic shower from the neutral pion decay is mistaken for an electron shower); and 'wrong sign' or 'wrong lepton' events where the lepton flavor or charge are not accurately determined. The expected rates of each of these backgrounds will be studied and tuned with dedicated near detector samples, utilizing the Gaussian fluxes described in Section~\ref{sec:prismgauss}. 

Any irreducible sources of background present in the \dword{nd} signal samples must be subtracted prior to performing the linear combination. Similarly, the background in the \dword{fd} must be added to complete the \dword{fd} event rate prediction. This is included in Eq.\,\ref{eq:PRISM_perfworld} by the subtraction of the predicted total background rate for the \dword{nd} ($\mathcal{B}_{\textsc{nd},i}\left(\vec{y}\right)$), and the addition for the \dword{fd} ($\mathcal{B}_{\textsc{fd}}\left(\vec{y}\right)$). The background term can be simply written as,

\begin{equation}
  \mathcal{P}_\textsc{fdp}\left(\vec{y}\right) = \mathcal{B}_{\textsc{fd}}\left(\vec{y}\right) - M_\textsc{nf}\sum_{i} c_{i}\left(\mathcal{H}_\textsc{oa}\right)\mathcal{B}_{\textsc{nd},i}\left(\vec{y}\right).
  \label{eq:PRISM_purity}
\end{equation}

\subsubsection{Efficiency Correction}

Any difference in the event selection performance between the near and far detector must be corrected in any near+far oscillation analysis. To mitigate the impact of cross section modeling on the \dword{nd} efficiency correction, a geometric efficiency is being developed, in which each selected \dword{nd} event is translated throughout its off-axis flux window, and rotated about the beam axis, to determine the fraction of such configurations in which the event would have been detected. In this way, an empirical event-by-event efficiency correction can be determined that does not depend on the neutrino interaction simulation.

In the same way, it also possible to determine the efficiency in the \dword{nd} of \dword{fd} events. \dword{fd} events that fall below a particular efficiency threshold are identified as those that are not directly constrained by \dword{nd} data. Typically such events have large hadronic showers that cannot be contained within the size of \dword{ndlar}, and must be treated separately from the ND-constrained events. Figure~\ref{fig:geoeff} shows the true neutrino energy spectrum of all the events in the \dword{nd} fiducial volume, the selected events, and the selected events with the geometric efficiency correction applied. The remaining gap between the efficiency-corrected distribution and the true event rate are from events that are never detected in the \dword{nd}, which constitute 5\% to 10\% of the event sample throughout the oscillation region, and increase as the neutrino energy increases.

\begin{dunefigure}[Geometric efficiency correction]{fig:geoeff}
  {The true neutrino energy distribution is shown for all interactions within the \dword{nd} fiducial volume, events that pass the selection criteria, and selected events with the event-by-event geometric efficiency correction applied.}
  \includegraphics[width=0.6\textwidth]{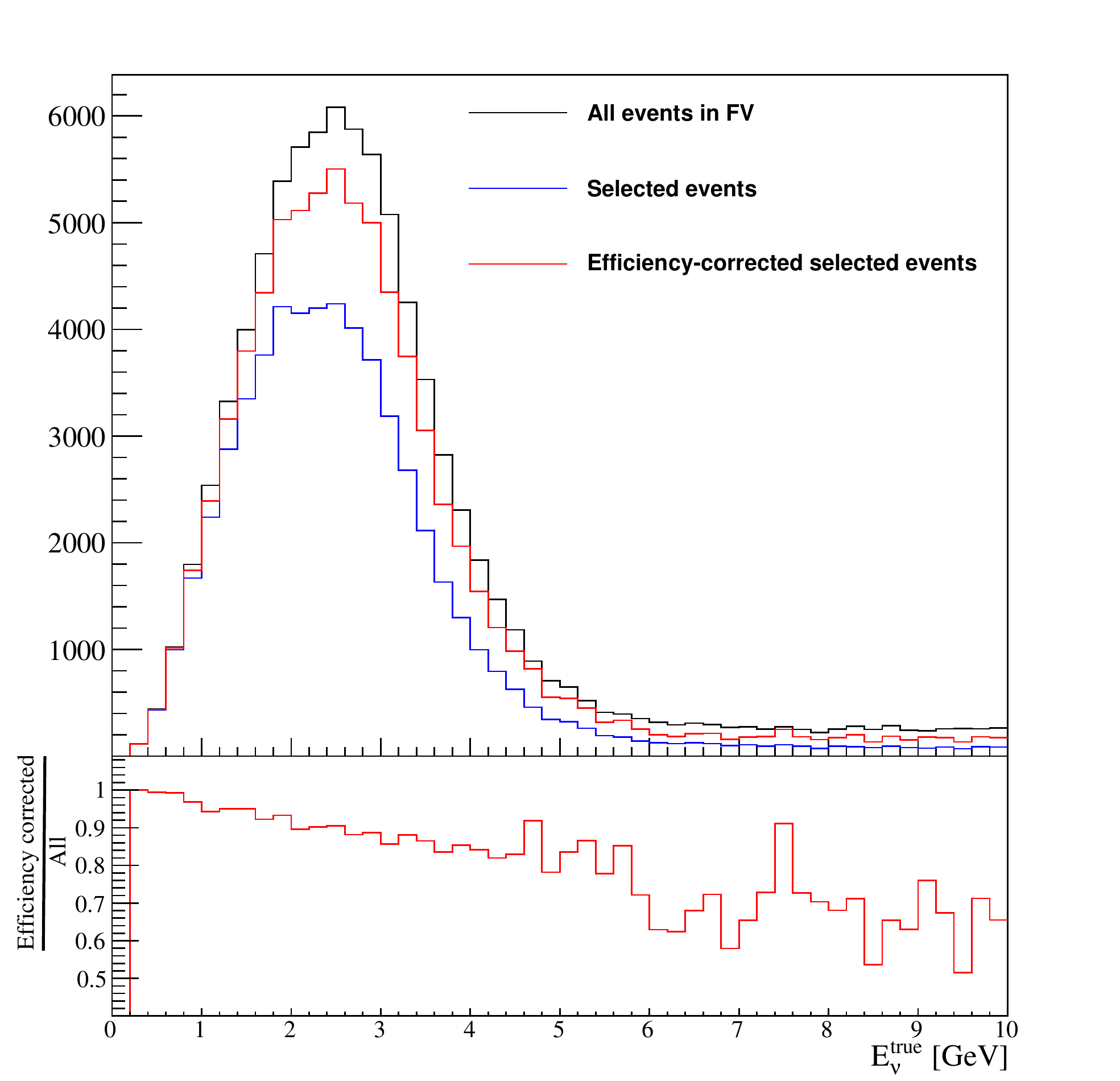}
\end{dunefigure}

%

The geometric efficiency correction has not yet been implemented in the existing analysis, and so a toy efficiency correction was developed that attempts to emulate how such a shower-topology-dependent correction may work.
The correction weights events based on the simulation-derived selection
efficiency as a function of $E_{\textrm{\tiny Hadr. avail.}}$  This is a
simulation truth-level proxy for the energy available to the hadronic shower,
and is defined as
\begin{equation}
  E_{\textrm{\tiny Hadr. avail.}} = \sum_{p,o} T_{p} + E_{o},
\end{equation}
where $p$ iterates over final state protons and $o$ iterates over all other
non-neutron final state hadrons.
The correction is performed in $(x_\textrm{\tiny Det},E_{\textrm{\tiny Hadr. avail.}})$, where
$x_\textrm{\tiny Det}$ is the $x$ position of the interaction within the non-veto
detector volume.

The correction was calculated from the full event sample and thus marginalizes
over the detector stops. As the goal was to characterize the selection efficiency
using only the hadronic shower energy and proximity to the
side veto regions, the absolute off-axis position was not included by design.

Herein $E_{\nu,\textrm{rec.}}$ is defined as
\begin{equation}
  E_{\nu,\textrm{rec.}} = E_{\mu} + E_{\textrm{\tiny Dep},\textrm{\tiny FV}} + E_{\textrm{\tiny Dep},\textrm{\tiny veto}},
\end{equation}
where $E_{\textrm{\tiny Dep}} = E_{\textrm{\tiny Dep},\textrm{\tiny FV}} + E_{\textrm{\tiny Dep},\textrm{\tiny veto}}$ is the total GEANT4-simulated energy deposit.

With the addition of an event-by-event efficiency correction, Eq.~\ref{eq:PRISM_perfworld}
becomes

\begin{equation}
  \mathcal{O}_\textsc{fdp}\left(\mathcal{H}_\textsc{oa}, \vec{y}\right) = M_\textsc{nf}\sum_{j} \mathcal{C}\left(\mathcal{H}_\textsc{oa}, \vec{x}_{j}\right)\mathcal{E}\left(\vec{x}_{j}\right)\mathcal{Y}_{\textsc{nd}}\left(\vec{x}_{j}\right),
  \label{eq:PRISM_effcorr}
\end{equation}
where $j$ iterates the selected \dword{nd} data events,
$\mathcal{C}\left(\mathcal{H}_\textsc{oa}, \vec{x}_{j}\right)$ gives the
observation weight, and $\mathcal{E}\left(\vec{x}_{j}\right)$ the efficiency
correction weight, for event $j$. Each event, with full observed kinematics
$\vec{x}_{j}$ is projected into analysis bins,
$\mathcal{R}\left(\vec{y}\right)$, by $\mathcal{Y}_{\textsc{nd}}\left(\vec{x}_{j}\right)$.

\subsubsection{Flux Matching Correction}

Any discrepancy between \dword{nd} linearly combined fluxes and the
corresponding predicted oscillated far flux must be corrected. While the
flux matching shown in Section~\ref{sec:prism-lincomb} reproduces the predicted \dword{fd} flux well
throughout the oscillation-sensitive region of the neutrino energy spectrum, the correspondence is somewhat degraded in both the high- and low-energy regions. The fractional difference between the \dword{nd} linear combination and the \dword{fd}
oscillated flux as a function of true muon neutrino energy is given by
$\Delta\Phi\left(E_\nu\right) = 1 - \left(\Phi_\textsc{lc}\left(E_\nu\right)/\Phi_\textsc{fd}\left(E_\nu\right)\right)$.
%

This can be added to Eq.\,\ref{eq:PRISM_effcorr} by a flux correction term,
\begin{equation}
  \mathcal{F}_\textsc{}\left(\vec{y}\right) = \sum_{k} \Delta\Phi\left(E_{\nu,k}\right)P_\textrm{osc.}\left(\mathcal{H}_\textsc{oa}, E_{\nu,k}\right)\mathcal{Y}_{\textsc{fd}}\left(\vec{x}_{k}\right),
  \label{eq:PRISM_fluxcorr}
\end{equation}
where $k$ iterates the simulated selected \dword{fd} events, $E_{\nu,j}$ is
the true neutrino energy for event $k$, and $P_\textrm{osc.}$ is the oscillation
weight.

\subsubsection{The Far Detector Prediction}

The full \dword{fd} prediction is then given by
$\mathcal{R}_\textsc{fdp}\left(\vec{y}\right) = \mathcal{O}_\textsc{fdp}\left(\vec{y}\right) + \mathcal{P}_\textsc{fdp}\left(\vec{y}\right) + \mathcal{F}_\textsc{}\left(\vec{y}\right)$
The result is shown in Figure~\ref{fig:prism-finalplots} for five different values of the oscillation parameters.

For the mock data set described above, the DUNE-PRISM prediction matches the \dword{fd} reconstructed neutrino energy distribution. The prediction is dominated by measured \dword{nd} data, which naturally contain any unknown neutrino-argon cross section effects, including the effect contained within the mock data. This is in contrast to the \dword{nd} constrained model extrapolation, which shows the bias that gives rise to the incorrect oscillation parameters shown in Figure~\ref{fig:duneprismsensitivity}.


\begin{dunefigure}[DUNE ND neutrino flux and cross section overlay]{fig:prism-finalplots}
  {The DUNE-PRISM muon neutrino \dword{fd} prediction for a range of disappearance hypotheses. The hypothesis used is noted above each sub-figure and compared to the world data in the \emph{top right} sub-figure. The 'data' used in each of these figures is the missing proton energy mock dataset. For each prediction: the linear combination of \dword{nd} 'data' measurements is shown in green; the \dword{fd} simulation correction that accounts for the imperfect flux match is shown in gray; the predicted \dword{fd} 'data' is shown as black points; and the \dword{fd} simulation is shown in dark red. The statistical uncertainties on the PRISM predicion are shown, these are determined from the MC sample size at the \dword{nd}, and are large due to a lack of \dword{nd} MC. It can be seen that the GENIE-based \dword{fd} prediction (dark red) is a poor predictor for the \dword{fd} data, whereas the linear combination of \dword{nd} data correctly predicts the \dword{fd} spectrum, despite the presence of an unknown cross section modeling problem in both the near and \dword{fd} data. }
  \includegraphics[width=0.45\textwidth]{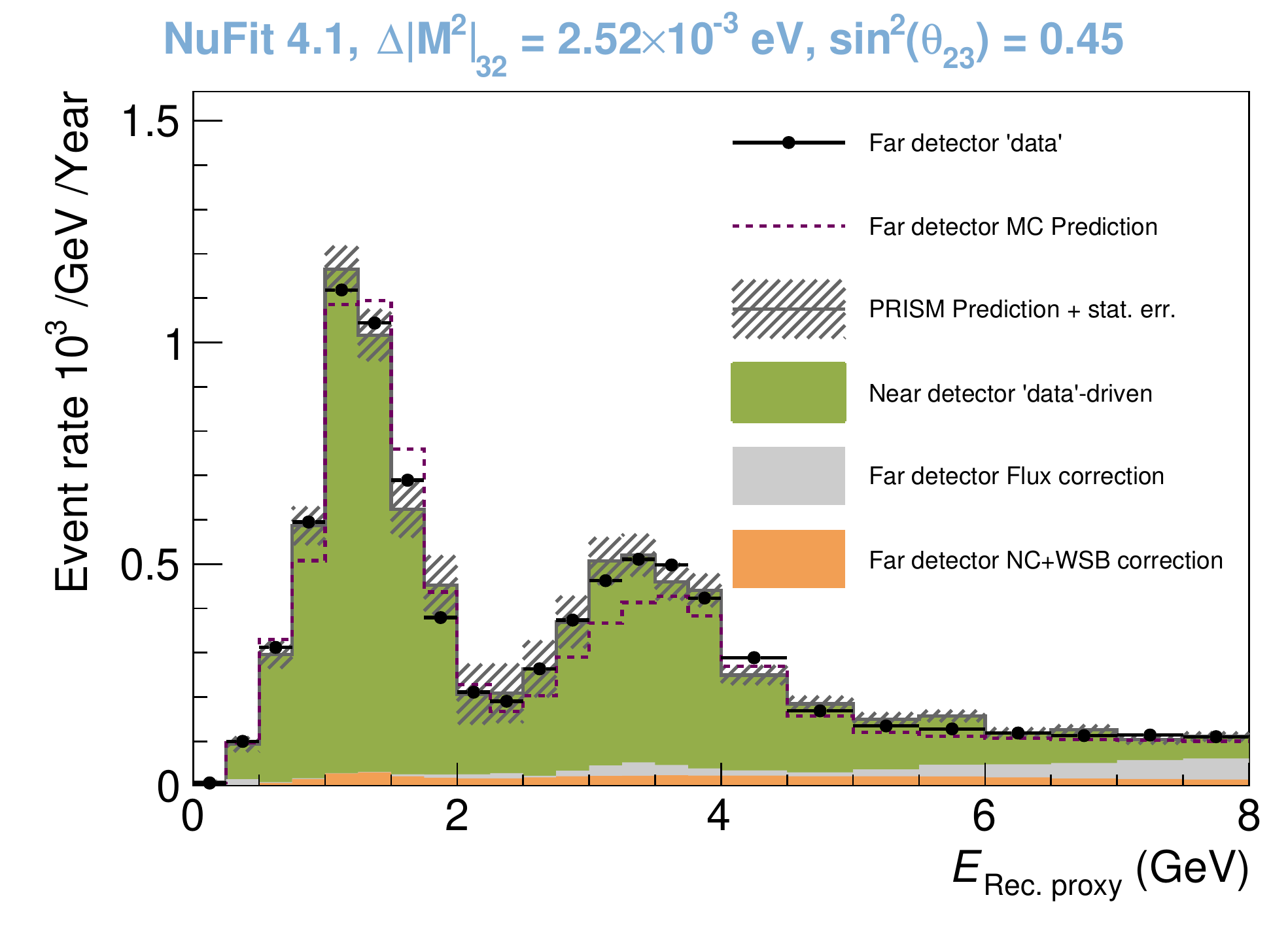}
  \hspace{0.05\textwidth}\includegraphics[width=0.38\textwidth]{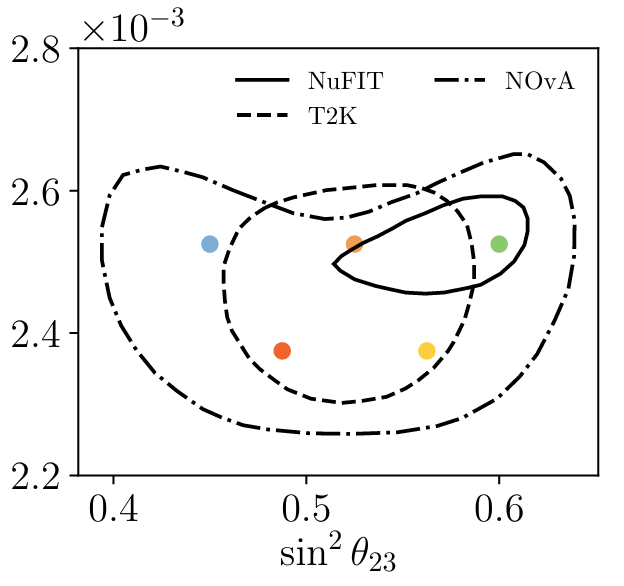}
\includegraphics[width=0.45\textwidth]{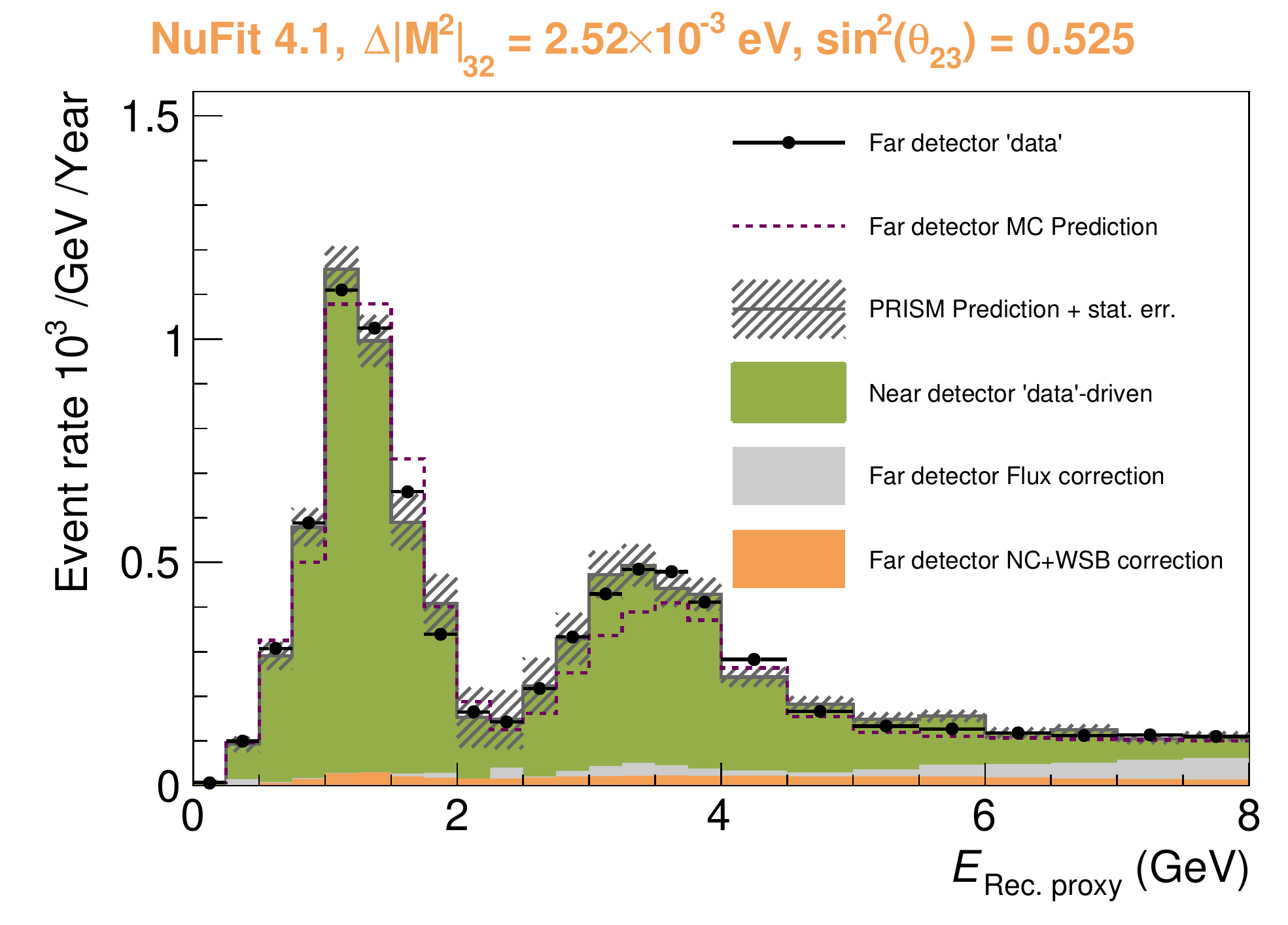}
\includegraphics[width=0.45\textwidth]{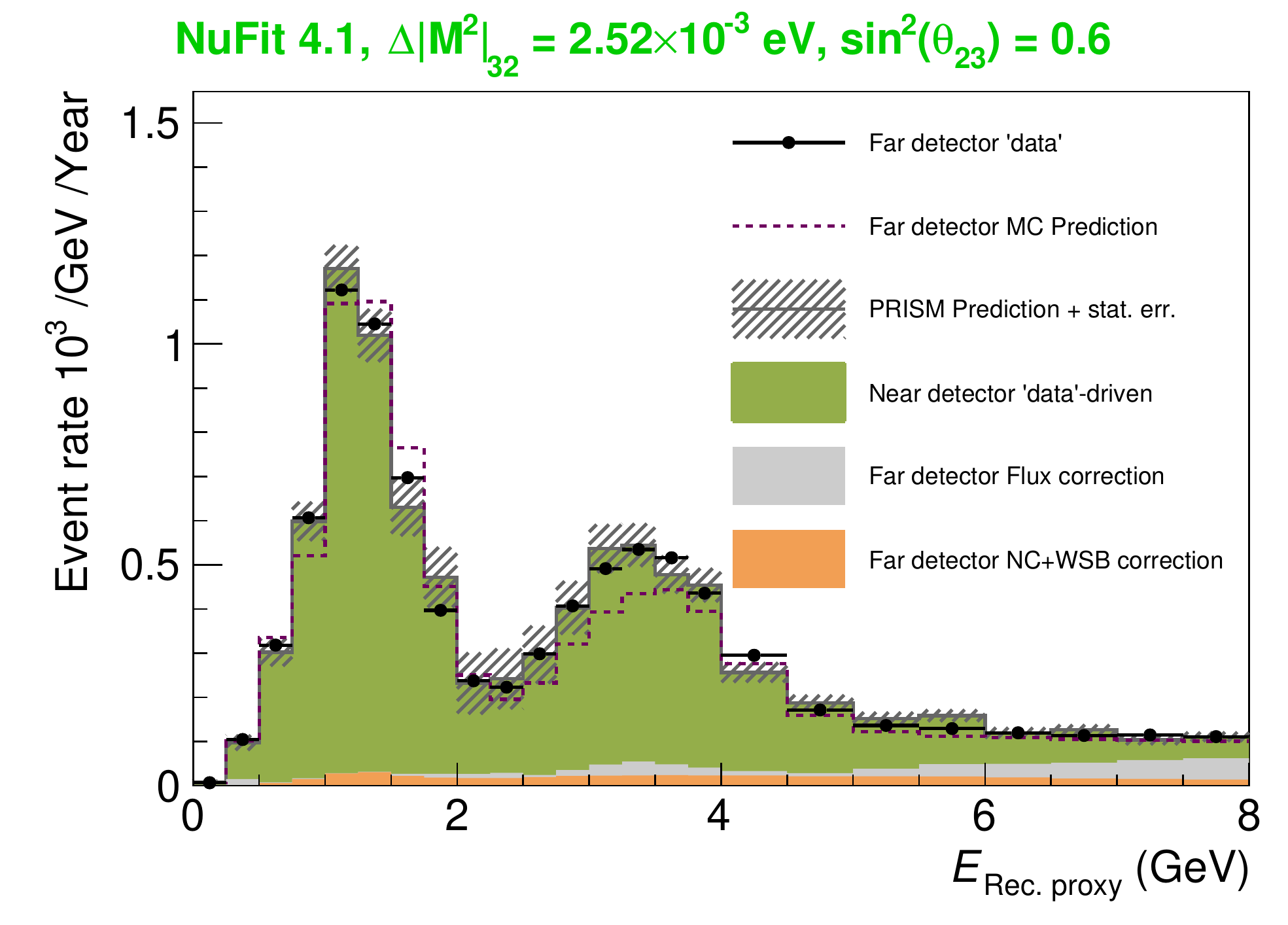}
\includegraphics[width=0.45\textwidth]{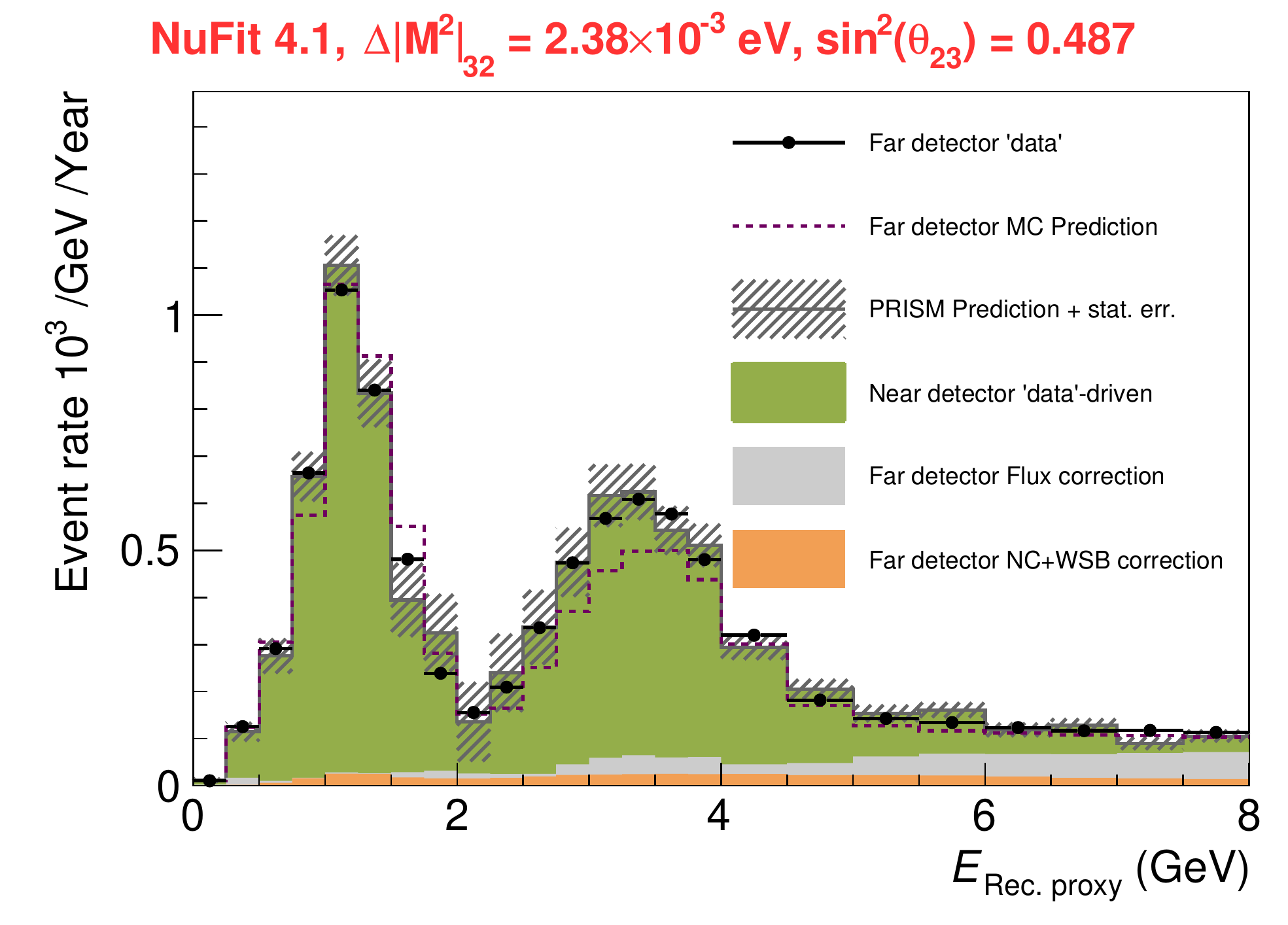}
\includegraphics[width=0.45\textwidth]{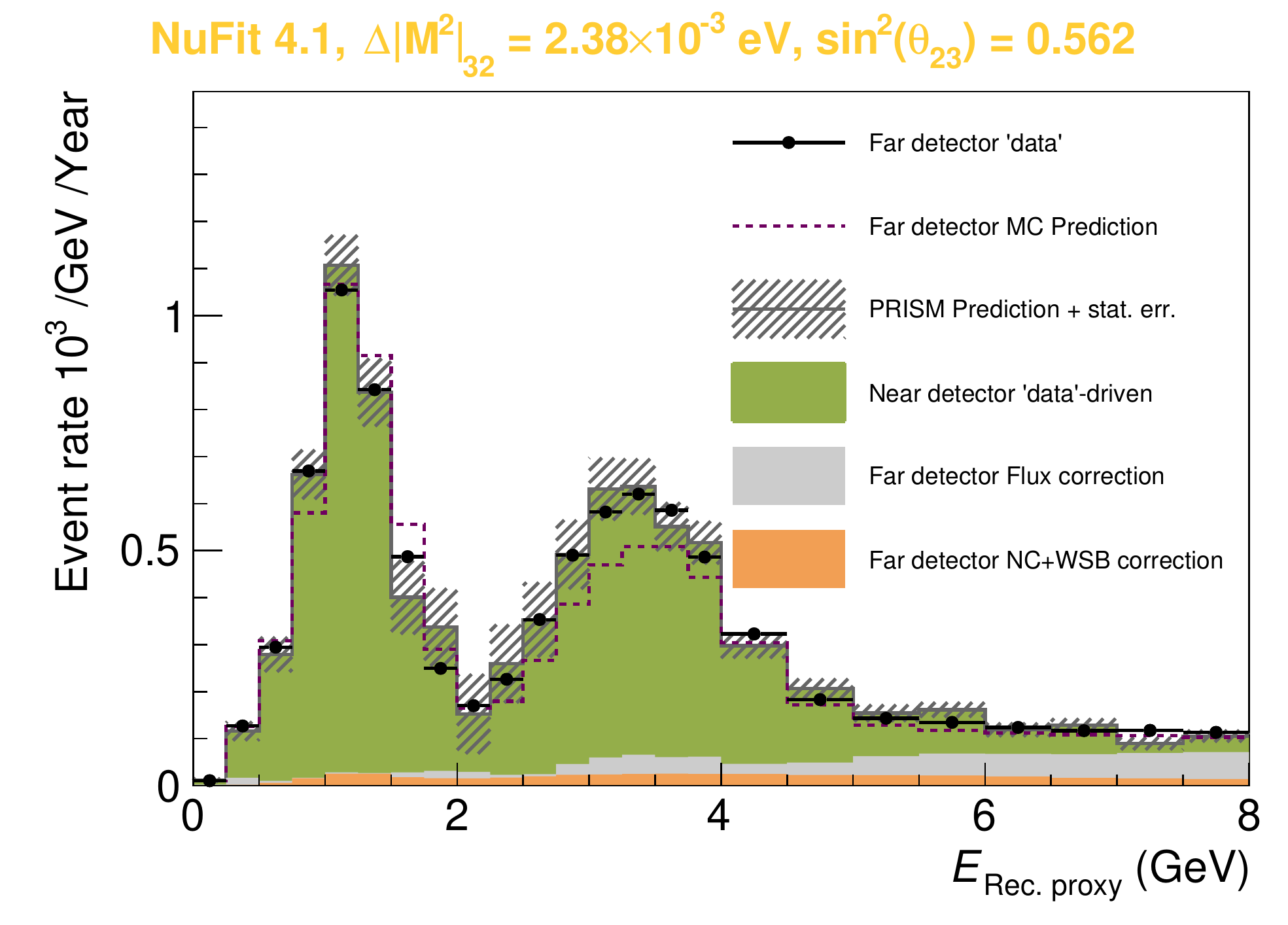}
\end{dunefigure}

\cleardoublepage

\chapter{System for on-Axis Neutrino Detection - SAND}
\label{ch:sand}

\section{Overview }
\label{sec:intro}

All DUNE accelerator-based physics studies use flux uncertainties that assume parameters such as horn positions and currents are known to certain tolerances. Beamline instrumentation is being developed to monitor these parameters but many potential deviations from the tolerances are best identified by monitoring of the neutrino energy spectra in the \dword{nd} for the distortions those deviations cause. Typical sources of beamline distortion are most easily seen and diagnosed in neutrino energy spectra measured on the beam axis and are diluted in off-axis spectra. However, the \dshort{duneprism} measurement program (Ch.~\ref{ch:prism}) calls for the \dshort{ndlar} and \dshort{ndgar} to spend approximately 50\% of the time collecting data at off-axis positions. \dshort{duneprism} relies on the well understood relationship between the off-axis angle and the neutrino energy spectrum. It is essential to \dshort{duneprism} that the beam remains stable while data are taken at different positions or, failing that, that distortions in the beam can be quickly identified and (eventually) modeled well. As a consequence, DUNE needs a continuous on-axis beam monitoring system, a role filled by \dword{sand}.

\section{Role in Fulfilling Requirements}
\label{sec:sandrequirements}

\paragraph{Fulfilling ND-M8}

To fulfill ND-M8 (and the overarching requirement ND-O5) \dshort{sand} monitors the beam on-axis. It must also have a target mass that is large enough for the interaction rate of neutrinos to provide statistically significant feedback on changes in the beam over a short time period. The requirement \underline{ND-C5.1} defines the short time period as one week. 

\paragraph{Fulfilling ND-M9}


To fulfill ND-M9 \dshort{sand} must measure the muon/neutrino energy and vertex distribution to detect representative changes in the beamline. There are two follow-on requirements: \underline{ND-C5.2} requires that \dshort{sand} have sufficient muon or neutrino energy resolution in \numu events to detect spectral variations; \underline{ND-C5.3} requires that the muon vertices in \numu CC events be measured well enough to divide the sample spatially relative to the beam center.

Fulfilling these requirements demands that \dshort{sand} is able to reconstruct the vertices in \numu CC interactions. The muons emanating from those vertices must be reconstructed with
good momentum resolution over a broad momentum range (roughly $0.5 \lesssim p_\mu \lesssim \SI{10}{GeV/c}$). This necessitates a tracking detector with a magnetic field. As such, \dword{sand} reuses the KLOE solenoidal superconducting magnet. The magnet provides a \SI{0.6}{T} magnetic field and a large tracking volume (Sec.~\ref{sec:magnet}). KLOE includes a 4$\pi$ \dword{ecal} that is useful as a target mass for the beam monitoring mission but also provides additional capabilities (Sec.~\ref{sec:kloe-calo}). The inner magnetized volume ($\sim$~43 m$^3$) is instrumented with a target and tracking system (``target/tracker''). There are two designs for the target/tracker, a reference and an alternative. Both feature hydrocarbon target masses and naturally provide for some additional capabilities. 

The performance studies that demonstrate how \dshort{sand} fulfills the beam monitoring requirements are described in Sec.~\ref{sec:beam-monitoring}. Fulfilling the requirements also leads to a set of derived detector capabilities that are described below.

\paragraph{Derived \dword{sand} detector capabilities}

Because \dword{sand} is required to measure the sign and momentum of muons it is also capable of similar measurements of charged hadrons. The target/tracking systems provide particle identification by dE/dx. The ECAL is able to measure photon and electron energies by calorimetry, and adds to the particle identification capability of the apparatus. These capabilities stem from the beam monitoring requirements but allow \dword{sand} to conduct a neutrino interaction measurement program that augments DUNE's oscillation physics mission. In particular \dword{sand} adds the following capabilities:

\begin{itemize}
\item \dword{sand} is able to provide an independent measurement of the interaction rate and energy spectra of the $\nu_\mu$, $\bar \nu_\mu$, and $\nu_e$, $\bar \nu_e$ beam components.  The capability of \dword{sand} to identify and reconstruct different types of interactions will enable complementary measurements of both the normalization and energy dependence of the flux. This redundancy can be used to improve confidence in the extrapolation of the neutrino and anti-neutrino fluxes to the far detector.

\item As discussed at length in Chapter~\ref{ch:intro}, nuclear effects present a significant source of uncertainty for \dword{dune}.  There are large uncertainties in the modeling of (anti)neutrino-nucleus cross sections. In particular, final state interactions are not well modeled but change the composition of hadrons in the final state and the hadrons' energies. The choice of argon as the primary target nucleus in the \dword{nd} is to mitigate the effect of these uncertainties in the \dword{nd} to  \dword{fd} comparison.  That said, things will not cancel perfectly in the near-to-far extrapolation, even with the implementation of \dshort{duneprism}.  \dword{sand} enables a program of measurements on nuclei other than argon (carbon and hydrocarbons) that may help constrain systematic uncertainties arising from nuclear effects.



\item The hydrocarbon in the target/tracker of both the reference and alternative designs results in a large event sample on carbon and also a smaller but still significant event sample on hydrogen. For some interaction channels, hydrogen enriched samples can be selected using transverse kinematic imbalance, or TKI, techniques~\cite{Lu:2015tcr, Furmanski:2016wqo, Abe:2018pwo, Dolan:2018sbb, Lu:2015hea, Lu:2018stk, Dolan:2018zye, Lu:2019nmf, Harewood:2019rzy, Cai:2019jzk, Cai:2019hpx, Coplowe:2020yea,Duyang:2018lpe,Duyang:2019prb}. The isolation of a sample enriched in neutrino-hydrogen interactions is very valuable since uncertainties due to nuclear effects are only present in the background and may potentially be mitigated by kinematic sidebands or the use of carbon targets with acceptance identical to the hydrocarbon ones. These targets are foreseen to allow a model independent background subtraction.

\item \dword{sand} is able to combine information from the ECAL and tracker/target to tag neutrons and measure their energy. The use of this information will improve the neutrino energy resolution and reduce the bias in the neutrino energy measurement, leading to a reduction in the related systematics. Neutron measurements can also improve the reconstruction of event kinematics.




\end{itemize}

\section{The Overall Design of SAND}
\label{sec:detector-design}
\label{sec:reference}

\dword{sand} is largely based on a reuse of the magnet and calorimeter from the KLOE experiment. The KLOE detector was designed primarily for the study of CP violation in neutral kaon decays at the DA$\Phi$NE $\phi$-factory. KLOE took data from April 1999 to March 2018. Throughout that time, the detector performance was stable. In the KLOE experiment, the inner volume of the magnet and ECAL was occupied by a large drift chamber.  In the \dword{dune} \dword{nd}, this volume will be instrumented with a target/tracking system. The detector itself will be installed so that neutrino beam enters through the side of the barrel, perpendicular to the magnetic field. Two potential designs for the target/tracking system are considered: a reference design, and an alternative design.


\paragraph{The reference design} uses a \dword{3dst} system (Sec.~\ref{sec:3dst}) as an active target inside of the magnet's tracking region. It is surrounded on the top, bottom, and downstream sides by low-density tracking chambers that measure the charge and momentum of outgoing particles. The tracking chambers will be \dshorts{tpc} (Sec.~\ref{sec:tpc}), straw tubes trackers (STT) (Sec.~\ref{sec:sand:STT}), or a mix. These two variants on the reference design are called {\bf 3DST+TPCs} and {\bf 3DST+STT}. The reference design is illustrated in Figure~\ref{fig:sand-geometry}.

\paragraph{The alternative design} does not use the \dshort{3dst} and surrounding tracking chambers. It instead fills most of the magnetic volume with orthogonal XY planes of \dword{stt} (the same technology as for the reference design) interleaved with various thin carbon and hydrocarbon  layers to add mass and act as additional targets for neutrino interactions. This variant is called {\bf STT-only}.

A thin LAr target is also foreseen in both designs. That target would be located inside the magnetic volume between the tracking region and the upstream inner edge of the ECAL.

This chapter describes the main features of each of the proposed components of SAND  and summarizes  the existing/ongoing studies to evaluate their performance (Sec.~\ref{sec:sand:physics}).

\begin{dunefigure}[The 3DST+TPCs design option for \dword{sand}.]{fig:sand-geometry}
{Drawing of the \dword{sand} system showing 3DST+TPCs configuration with the \dword{3dst} in the center (light green), low-density tracker (TPC or STT, Magenta),  \dword{ecal} (green), the magnet coil (gold), and the return yoke (gray). }
	\includegraphics[width=3.5in]{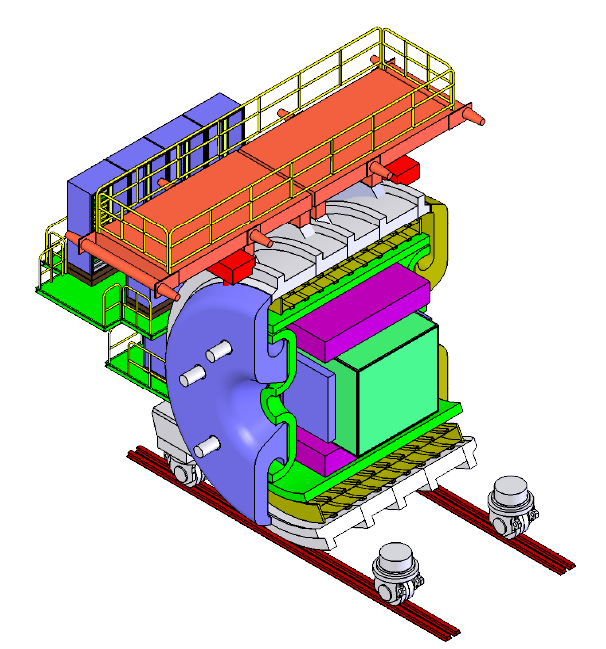}
\end{dunefigure}

\subsection{The Superconducting Magnet}
\label{sec:magnet}

The KLOE superconducting magnet~\cite{Smith:1997qhd}, shown in Figure~\ref{fig:kloe_layout}, was designed in conjunction with its iron yoke to produce 0.6 T over a 4.3 m long, 4.8 m diameter volume. The coil is operated at a nominal current of 2902 A and the stored energy is 14.32~MJ~\cite{Modena:1997tz}. The coil is located inside a cryostat (outer diameter: 5.76 m, inner diameter: 4.86 m, overall length: 4.40 m) positioned inside the return yoke (Figure~\ref{fig:kloe_layout}~Right). The overall cold mass is $\sim$8.5 tons and the mass of the KLOE return yoke is 475 tons.

\begin{dunefigure}[The KLOE magnet and \dword{ecal}.]{fig:kloe_layout}
{The KLOE detector: (left) 3D engineering CAD model of the magnet and (right)  vertical cross section (taken from~\cite{Adinolfi:2002zx}). }
\includegraphics[width=0.48\textwidth]{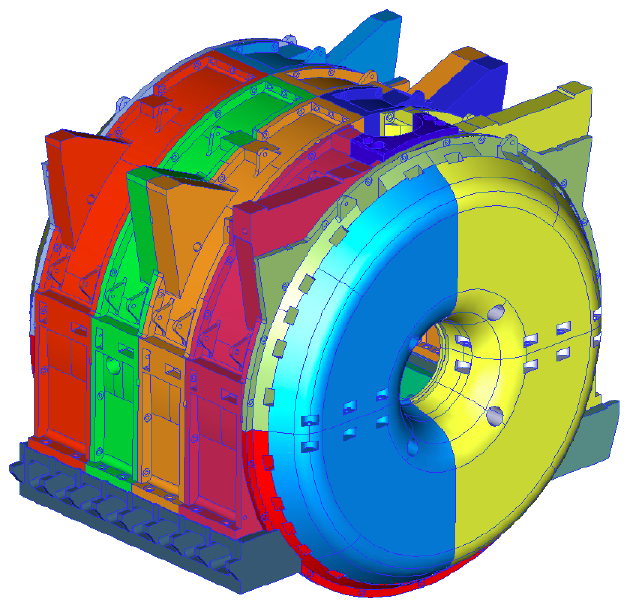}
\includegraphics[width=0.47\textwidth]{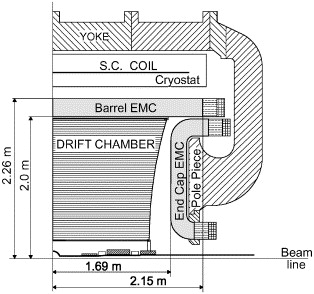}
\end{dunefigure}

The cooling of the coil is performed by thermo-siphoning cycles: gaseous He at 5.2 K is injected at 3 bar (absolute pressure) from the cryogenic plant and liquefied through Joule-Thomson valves into a liquid He reservoir in thermal contact with the coil. The current leads are directly cooled by the liquid He while the radiation shields are cooled by gaseous He at 70~K from the cryogenic plant. 
The heat loads are, respectively:
\begin{itemize}
\item 55 W at 4.4 K for the magnet coil;
\item 0.6 g/s of liquid He for the current leads; 
\item and 530 W at 70 K for the thermal radiation shields.
\end{itemize}

\noindent
The coil, cryostat and cryogenic system were developed by Oxford Instruments A.T.G., UK. In particular, the coil support cylinder is a rolled and welded cylinder of 5083 aluminum with cooling channels welded to the outside. The coil is a two layer conductor wound on flat with a full vacuum impregnated insulation system. The conductor is a composite consisting of a (Nb-Ti) Rutherford cable co-extruded with high purity aluminum. The left part of Figure~\ref{fig:kloe_magnet} shows a picture of the  magnet in the LNF Assembly Hall in Frascati. 
\begin{dunefigure}[The KLOE magnet and longitudinal B field.]{fig:kloe_magnet}
{(left) A part of the KLOE magnet near one of the end caps. (right) The magnitude of the longitudinal (i.e., along the magnet symmetry axis) component of the magnetic field (in Gauss) as a function of the position along the magnet axis (in cm), measured from the center ($z=\SI{0}{cm}$) toward the end cap ($z\approx \SI{200}{cm}$). Data and MC are compared. The field was measured and simulated on the magnet symmetry axis.}
\includegraphics[width=0.38\textwidth]{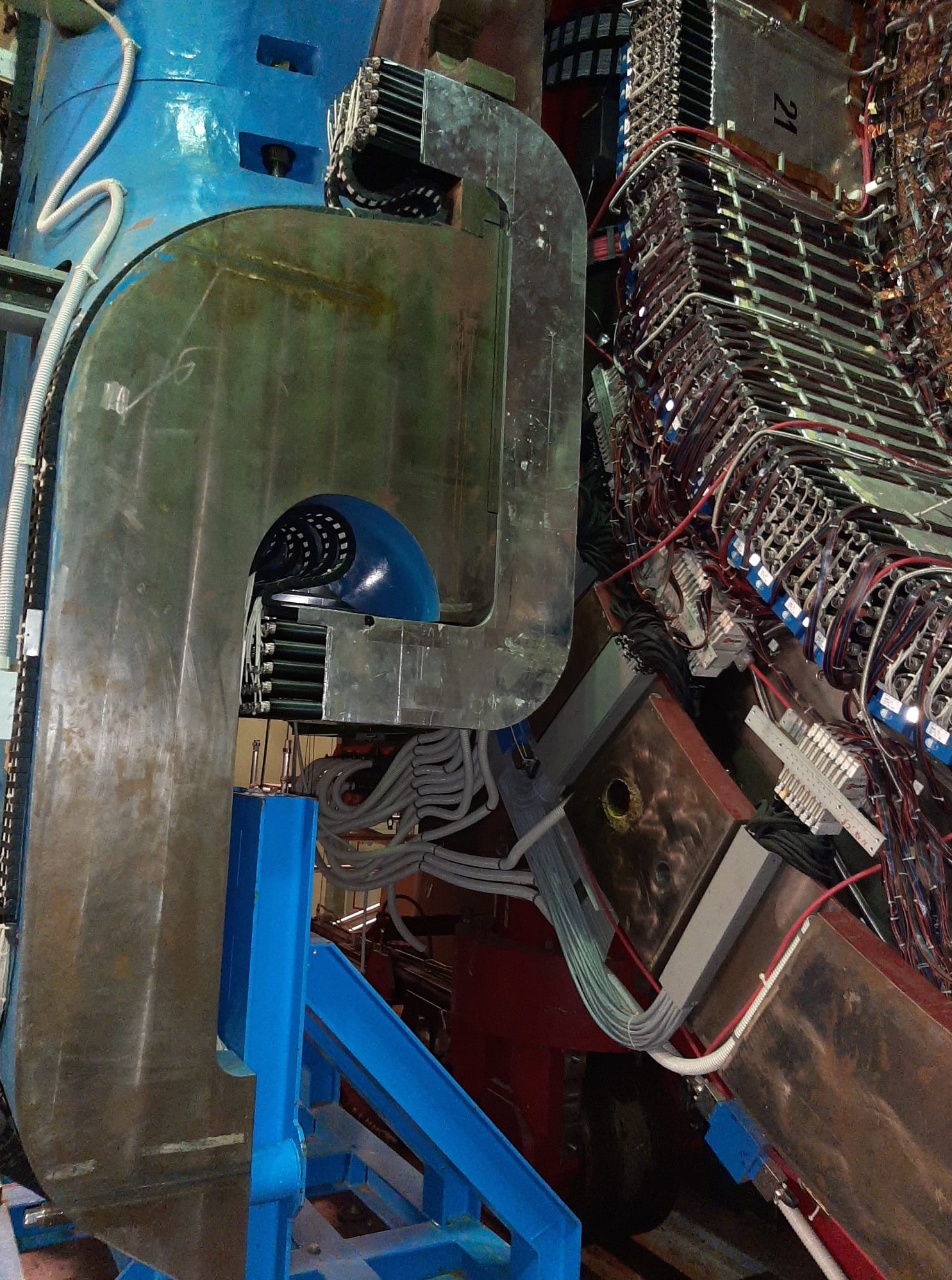}
\includegraphics[width=0.6\textwidth]{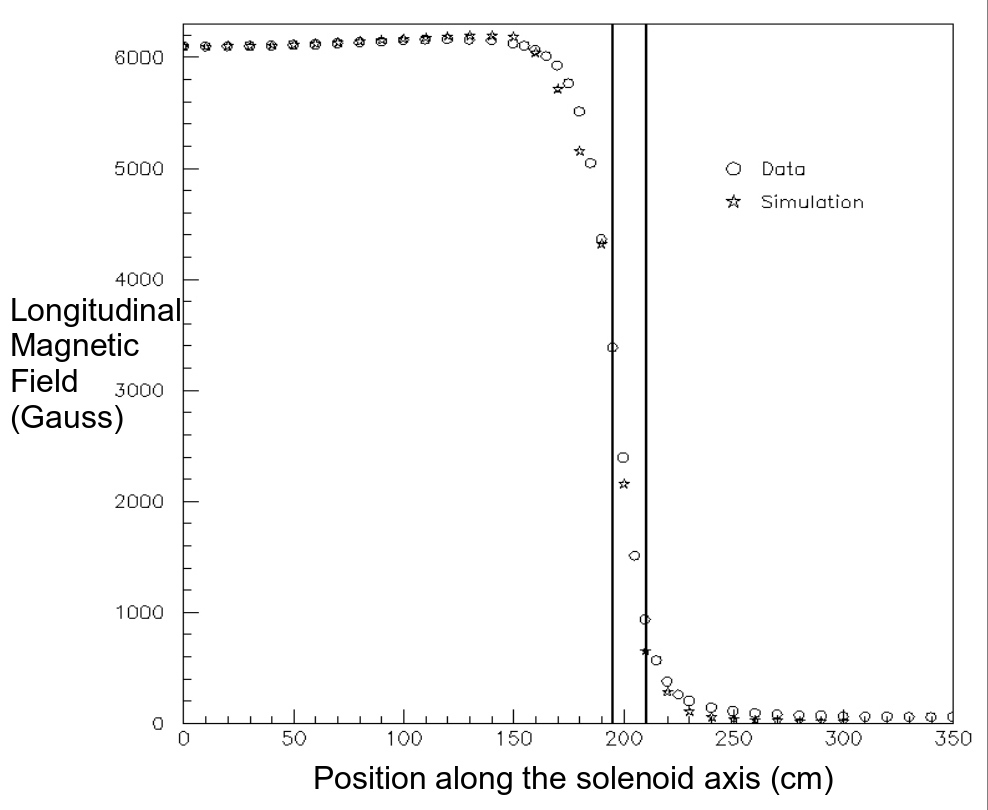}
\end{dunefigure}
The solenoidal longitudinal field component as measured in the KLOE installation phase is plotted on the right in Figure~\ref{fig:kloe_magnet} and compared with simulation (MAGNUS Monte Carlo program)~\cite{Ceccarelli:1997ua}.

\subsection{The KLOE Lead/Scintillating-Fiber Calorimeter}
\label{sec:kloe-calo}

\subsubsection{Detector Description}

The KLOE electromagnetic calorimeter~\cite{Adinolfi:2002zx} is a lead/scintillating-fiber sampling calorimeter. Scintillating fibers offer good light transmission over several meters,  sub-ns timing accuracy, and very good hermeticity.
The barrel calorimeter (see Figure~\ref{fig:barrel_picture_kloe}) is cylindrical and is located inside the KLOE magnet, close to the coil cryostat. Two additional calorimeters (endcaps) ensure hermeticity along the magnet endcaps. The  barrel (Figure~\ref{fig:barrel_picture_kloe}) consists of
24 modules, each of which is 4.3 m long, 23 cm thick and trapezoidal in
cross-section, with bases of 52 and 59 cm. Each
end-cap consists of 32 vertical modules that are 0.7--3.9 m long and 23 cm thick. Their cross-section is
rectangular, of variable width. Modules are bent
at the upper and lower ends to allow insertion into
the barrel calorimeter and also to place the phototube
axes parallel to the magnetic field. Due to the
large overlap of barrel and endcaps, the KLOE
calorimeter has no inactive gap at the interface
between those components.

\begin{dunefigure}[The KLOE calorimeter.]{fig:barrel_picture_kloe}
{A view of the  KLOE calorimeter. The far endcap is closed and ECAL modules can be seen as vertically oriented slabs. The barrel ecal modules are slabs that have a trapezoidal cross-section and that run along the barrel. The near yoke pole piece and end-cap calorimeter are open.}
   \includegraphics[width=0.6\textwidth]{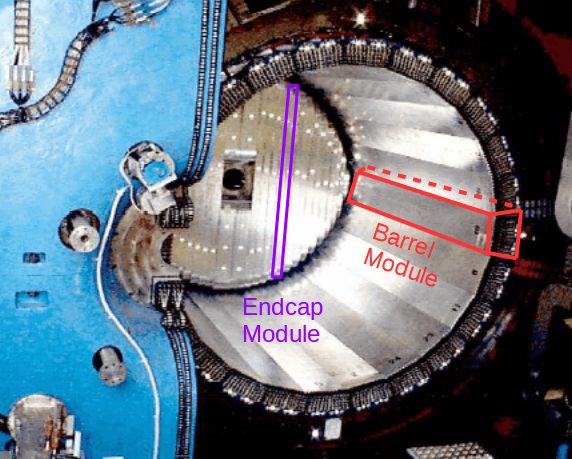}
\end{dunefigure}

All ECAL modules are stacks of approximately 200 grooved,
0.5 mm thick, lead foils alternating with 200 layers
of cladded 1~mm diameter scintillating fibers,
glued together with a special epoxy compatible with the fiber materials. The
average density is 5 g/cm$^3$; the radiation length is $\sim$~1.5~cm; and the overall thickness of the calorimeter is $\sim$~15 radiation lengths.
Light guides that match the almost square portions of the module end-faces to circular photo-cathodes are employed to read both ends of each module. The
readout subdivides the calorimeter into five planes
in depth. The first four planes are 4.4 cm deep and the last plane is 5.2 cm deep. Each plane is subdivided
in the transverse direction into 4.4 cm
wide elements, except at the edges of the trapezoidal modules.
Barrel modules are attached to the inner wall of
the coil cryostat.  Endcaps are divided into two halves allowing the opening for access to the chamber. 
The readout segmentation gives an $r-\phi$ (in the case of the barrel) or $x-z$ (in the case of the endcaps) readout resolution of 1.3 cm ($4.4/\sqrt{12}$~cm). The design is such that a particle crossing the calorimeter deposits energy in at least five readout regions or cells. The calorimeter weight is about 100 tons and the readout system includes 4880 phototubes.
The light readout system of one barrel module is shown in Figure~\ref{fig:barrel_guides_kloe}.
The light guides matching the module end-faces
to the photo-tube windows begin with a mixing
section and terminate with a Winston cone providing an area concentration factor of about 4.
Since light propagates at small angles with respect
to the fiber axis ($\sim$22$^\circ$ for the light traveling in a
plane containing the fiber axis), light losses in the large area reduction are very small. The overall efficiency of the guides is $\sim$80$\%$.

\begin{dunefigure}[The KLOE calorimeter light guides.]{fig:barrel_guides_kloe}
{Light guides at one end of a barrel module before the installation of the phototubes.}
    \includegraphics[width=0.5\textwidth]{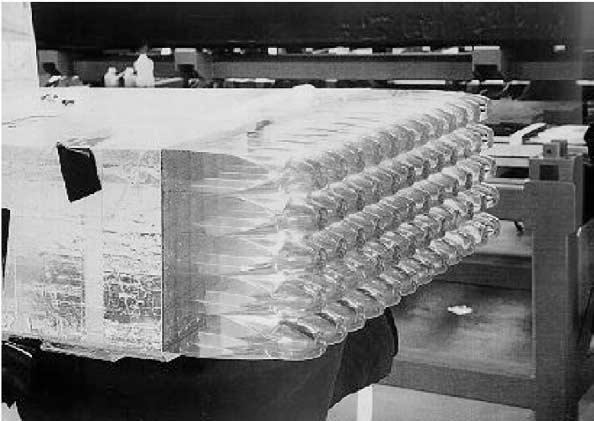}
\end{dunefigure}

The layout of the calorimeter inside the KLOE magnet is depicted in Figure~\ref{fig:layout_caloem}.
\begin{dunefigure}[Light guide location in the KLOE magnet.]{fig:layout_caloem}
{Front (top figure) and side (bottom figure) view of the calorimeters showing the light guides and their location inside the KLOE magnet. The units are in mm.}
    \includegraphics[width=0.7\textwidth]{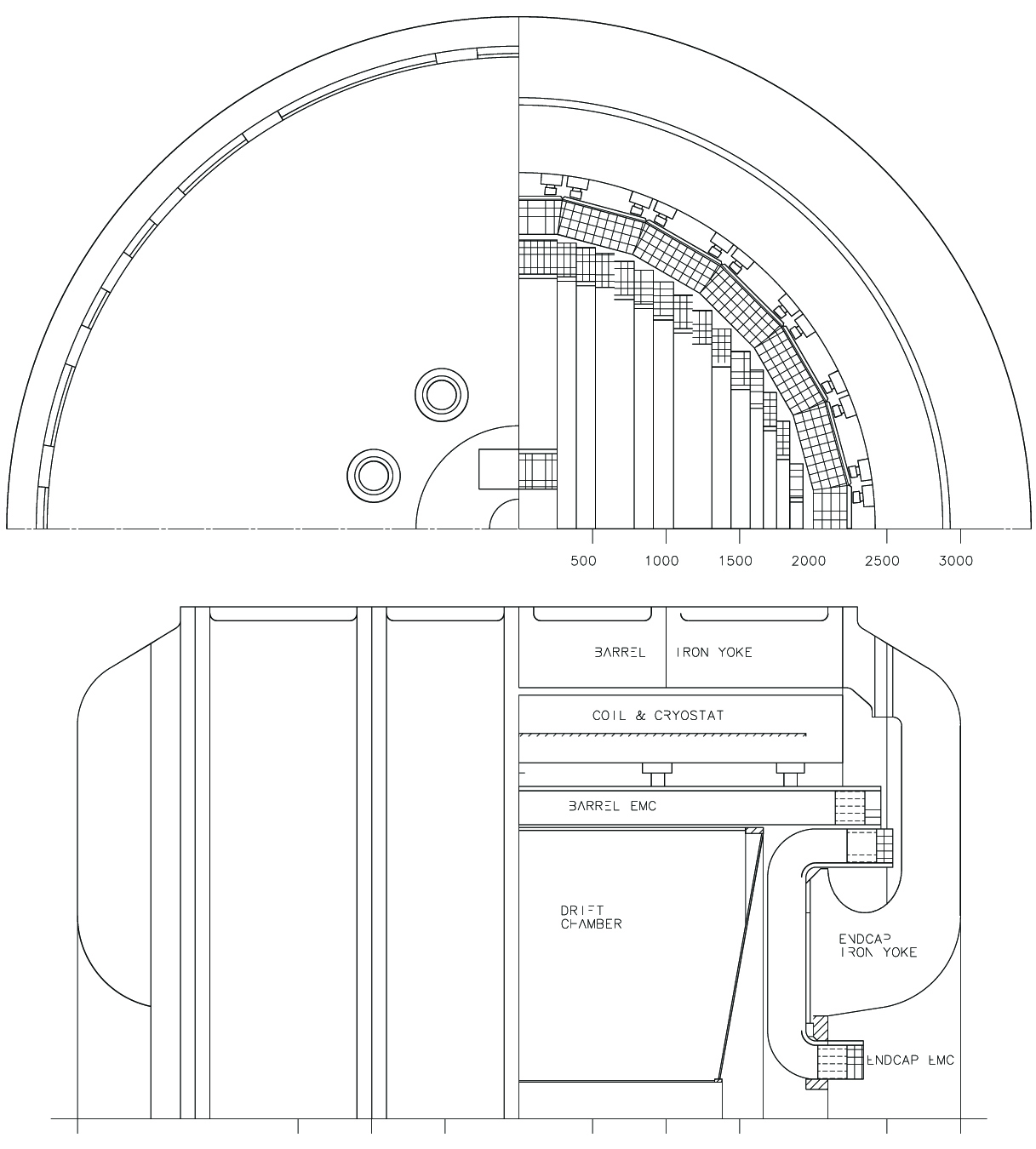}
\end{dunefigure}

\subsubsection{Reconstruction of Time, Position and Energy}
\label{sec:time_pos_ener_meas}

The calorimeter's readout granularity is defined
by the light collection segmentation. The fiber’s
direction is referred to in the following as longitudinal.
This segmentation provides the determination
of the position of energy deposits in $r-\phi$
for the barrel and in $x-z$ for the end-cap. A
calorimeter segment is called in the following a cell
and its two ends are labeled as A and B. For each
cell, two time signals, $t^{A,B}$ and two amplitude
signals $S^{A,B}$ are recorded from the corresponding
PM’s signals. The longitudinal position of
the energy deposit is obtained from the difference
$t^A - t^B$:
The particle arrival time $t$ and its coordinate $s$
along the fiber direction, the zero being taken at
the fiber center, are obtained from the times T$^{A,B}$ in TDC counts as
\begin{equation}
t({\rm ns})=\frac{t^A + t^B}{2}- \frac{t_{0}^A + t_{o}^B}{2}-\frac{L}{2v}
\end{equation}
\begin{equation}
s({\rm cm})=\frac{v}{2}\left(t^A-t^B -t_{o}^A+t_{o}^B\right)
\end{equation}
with $t^{A,B}$ = $c^{A,B}\times T^{A,B}$, where $c^{A,B}$ (in ns/TDC counts) are the TDC calibration constants and $t_0^{A,B}$ are the overall time offsets. L and $v$ are the cell length (cm) and the light velocity (cm/ns) in the fiber, respectively.

The energy signal, E, on each side of a cell $i$ is obtained from S as
\begin{equation}
E_i^{A,B}({\rm MeV})=\frac{S_i^{A,B}-S_{0,i}^{A,B
}}{S_{M,i}}\times k_E
\end{equation}
All signals $S$ above are in ADC counts. $S_{0,i}$ are the offsets of the amplitude scale. $S_{M,i}$ is the response for a minimum ionizing particle crossing the calorimeter center. Dividing the  equation above by $S_{M,i}$ accounts for PM response, fiber light yield and electronics gain. The $k_E$ term gives the energy scale in MeV, and it is determined from showering particles of known energy.
In order to obtain a calorimeter response independent of the position, a correction factor $A_i^{A,B}(s)$, due to the attenuation along the fiber length, is applied. The cell energy, $E_i$, is taken as the mean of the measurements at both ends,
\begin{equation}
E_i({\rm MeV}) = \left(
E_i^{A}A_i^{A}+E_i^{B}A_i^{B}
\right)/2
\end{equation}

The energy and time resolution of the calorimeter were evaluated in the commissioning and running phases of KLOE and were found to be 
\begin{itemize}
    \item Energy resolution: $\sigma/E = 5\%/ \sqrt{E(\mathrm{GeV})}$,
    \item Time resolution: $\sigma = 54/ \sqrt{E(\mathrm{GeV}})$~ps.
\end{itemize}
\subsubsection{Calibration and Performance}
As described in the literature \cite{Adinolfi:2002zx}, the ECAL has been extensively calibrated with cosmic muons and photons. 
 The response to neutrons has been studied on a small scale prototype
in a low energy neutron beam \cite{Anelli:2007zz}.
After installation in the \dword{nd} hall at Fermilab, further checks will be performed in situ using cosmic muons, stopping particles, neutral pions, etc.

Preliminary studies were done to characterize the \dword{ecal} performance using a configuration of \dword{sand} with the entire magnetized inner volume filled with an STT system \cite{bib:docdb13262}(described in Sec.~\ref{sec:sand:STT}).\footnote{This version of the tracker is presented as an option in Sec.~\ref{sec:sand:STTonly}. Here, the study illustrates the \dword{ecal} performance and is not strongly dependent on the tracker details.}  A sample of $\nu_{\mu}$ CC interactions with one $\pi^0$ and an interaction vertex located inside the active volume of the straw tubes  was used to look at $\pi^0$ reconstruction using \dword{ecal} information.  As shown in (Figure~\ref{fig:ecal_perfor}, left) a resolution  of 15\% was achieved.  The neutrino energy for $\nu_\mu$ CC events was reconstructed as well.  In this study, neutrons, neutral pions, and photons were reconstructed mainly from the information provided by the electromagnetic calorimeter and a circular fit approximation was used to reconstruct the charged particle momentum component in the bending plane.   An energy resolution for the core of the distribution of better than 7\% was obtained (Figure~\ref{fig:ecal_perfor}, right).


\begin{dunefigure}[\dword{sand} \dword{ecal} performance measures.]{fig:ecal_perfor}
{Left: $\pi^0$'s invariant mass computed  using the ECAL information for $\nu_{\mu}$ CC interactions with one $\pi^0$.
 \hspace{3mm} Right: Reconstructed neutrino energy in $\nu_\mu$ CC events using ECAL information. }
    \includegraphics[width=3.2in]{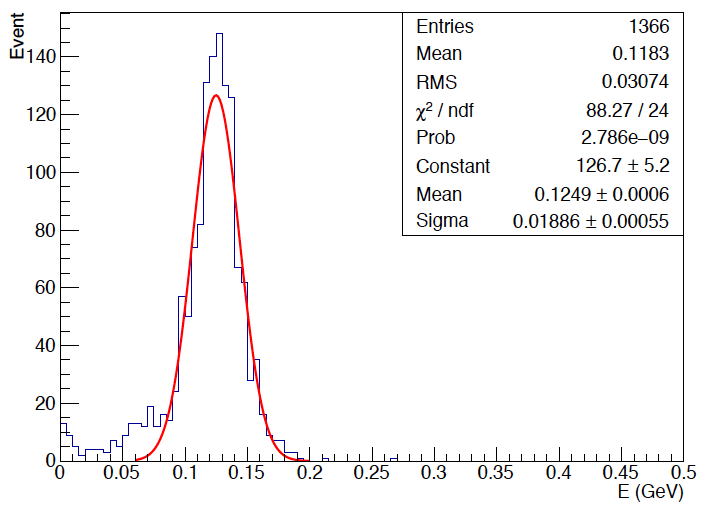}
    \includegraphics[width=3.2in]{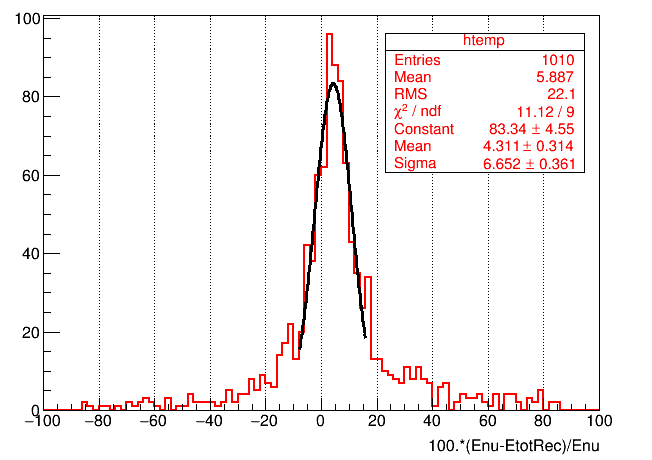}
\end{dunefigure}

\subsection{Inner Target/Tracker}
\label{sec:trackers}


In the reference design, the \dshort{3dst} acts as an active target for neutrino interactions. Low density trackers located between the 3DST and the ECAL provide a high resolution momentum measurement of muons and other charged particles that exit the \dshort{3dst}. In addition, the trackers can provide useful PID information. Two technology options are being considered: atmospheric pressure time projection chambers and straw tube trackers. The reference design with these two options will be referred to as 3DST+TPCs and 3DST+STT, respectively. The alternative design being considered for SAND has most of the volume inside the ECAL filled with STT modules. A thin liquid argon active target located inside the electromagnetic calorimeter and upstream of the tracking system is foreseen for both options.

\section{Technologies for the Inner Target Tracker}
\label{ch:sand-opt}

 \subsection{Three-Dimensional Projection Scintillator Tracker}
\label{sec:3dst}

The active target of the \dword{sand} reference design is a \dword{3dst}, made of many $1 \times 1 \times 1~\text{cm}^3$ plastic scintillator cubes, each optically isolated and read out by three orthogonal \dword{wls} fibers. The scintillator is composed of polystyrene doped with 1.5\% of paraterphenyl (PTP) and 0.01\% of POPOP. After fabrication the cubes are covered by a reflecting layer made by etching the scintillator surface with a chemical agent, resulting in the formation of a white polystyrene micropore deposit over the scintillator. Three orthogonal through-holes of \SI{1.5}{mm} diameter are drilled in each cube to accommodate the \SI{1.0}{mm} diameter \dword{wls} fibers. 

This novel geometry can provide a full angular coverage to any particle produced by neutrino interactions and reduce the momentum threshold for protons down to about 300 MeV/c (if at least three hits per view are required) \cite{Sgalaberna:2017khy}. Being a fully active detector, \dword{3dst} can also provide a calorimetric measurement of the energy deposited by low-momentum hadrons that are untracked due to short range.

The 3DST detector concept is shown in Figure~\ref{fig:cube}. The design of \dword{3dst} is fairly advanced since it is very similar to that for the \dword{sfgd} detector, which is under construction for the T2K ND280 (near detector) upgrade.  
\begin{dunefigure}[A few plastic scintillator cubes with \dword{wls} fibers.]{fig:cube}
{An assembly of 8 plastic scintillator cubes with \dword{wls} fibers.}
  \includegraphics[width=2.5in]{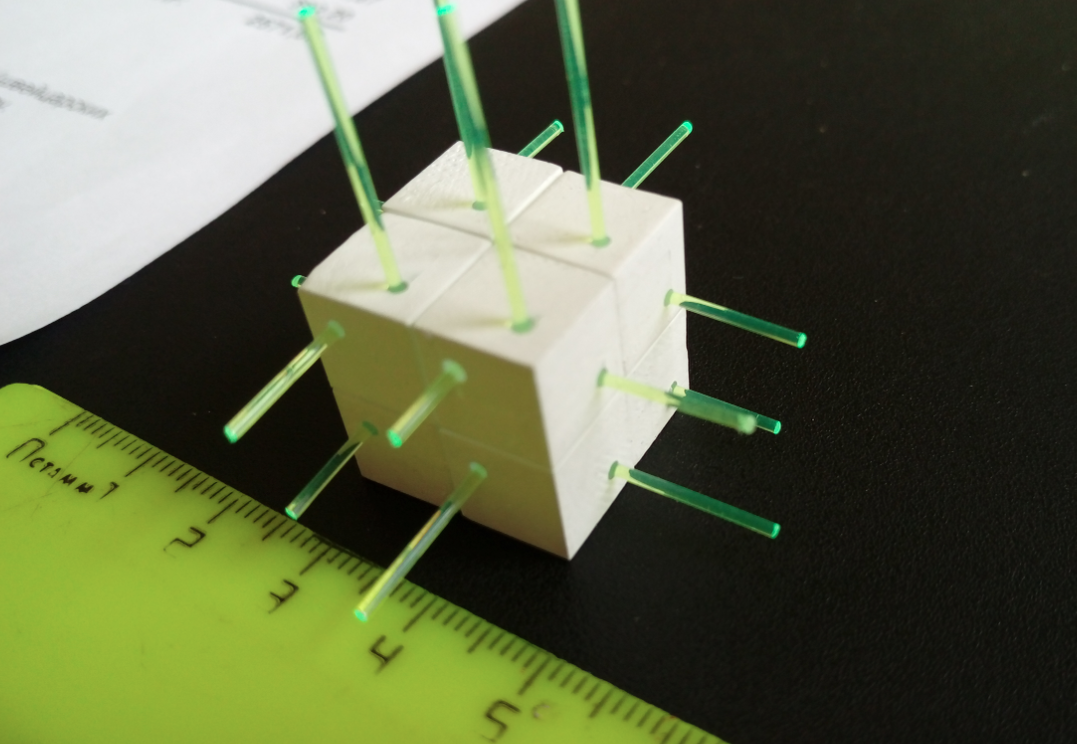}
\end{dunefigure}


The total size of the \dword{3dst} detector is $2.53~\text{(width)}~ \times 2.36~\text{(height)}~\times 2.04~\text{(depth)}~ ~\text{m}^3$ and takes into account the mechanical box that supports the load of the detector, the clearance between detectors, and the light readout interface between the \dword{wls} fibers and the \dwords{sipm}. The active detector has  $240 \times 224 \times 192 \approx \num{10.3e6}$ scintillator cubes. The total active mass is \SI{11.8}{ton}.  Table~\ref{tab:3dststats} shows the expected total number of events for different neutrino interaction topologies.

The angular resolution has been studied with simulations for electrons and muons~\cite{bib:docdb19972}. In the simulation, electrons and muons were generated at a point \SI{40}{cm} inside the front face of the 3DST with random angles $0^0< \theta < +30^\circ$ and $0^\circ < \phi < 360^\circ$. Hits were weighted by the number of photoelectrons and fit with a 3D fitting algorithm.  For \SI{1.5}{GeV/c} muons, a representative momentum, the resolution was \SI{15}{mrad}. For electrons with the same momentum, the resolution was \SI{30}{mrad}\footnote{For comparison, MINERvA's angular resolution for electrons of the same energy was about \SI{11}{mrad}, but that was with a better developed reconstruction~\cite{Park:2013dax}.}. In this study some staggering was introduced into the matrix of cubes to mitigate possible aliasing effects. A small improvement was found for the muons but not for the electrons. The momentum resolution for muons stopping in the fiducial volume is better than 3\% \cite{Abe:2019whr}. 


\begin{dunetable}[\dword{3dst} event rates]{|l|c|l|c|}{tab:3dststats}{Projected event rates per year for the \dword{3dst} detector, assuming the \SI{120}{GeV}, three horn, optimized \dword{lbnf} beam. The rates correspond to a fiducial volume of \SI{11.0}{tons}. }
\dword{fhc} Beam\span\omit & \textbf{\dword{rhc} Beam}\span\omit \\    
\rowtitlestyle %
Process & \textbf{Rate} & \textbf{Process} &  \textbf{Rate} \\ \toprowrule
All \numu-CC & \num{1.5e7}               & All \anumu-CC & \num{5.5e+06} \\
\hline
\hspace{1em} CC $0\pi$ & \num{4.4e+06}      & \hspace{1em}  CC $0\pi$ & \num{2.4e+06}    \\
\hspace{1em} CC $1\pi^{\pm}$& \num{4.3e+06}  & \hspace{1em}  CC $1\pi^{\pm}$ & \num{1.6e+06} \\
\hspace{1em} CC $1\pi^0$ & \num{1.3e6}    & \hspace{1em}  CC $1\pi^0$ & \num{5.4e+05}   \\
\hspace{1em} CC $2\pi$ & \num{1.9e6}      & \hspace{1em}  CC $2\pi$ & \num{5.1e+05}    \\
\hspace{1em} CC $3\pi$ & \num{8.3e5}      & \hspace{1em}  CC $3\pi$ & \num{1.6e+05}    \\
\hspace{1em} CC other & \num{1.9e6}       & \hspace{1em}  CC other & \num{3.0e+05}  \\
\hline
\hspace{1em} \numu-CC COH $\pi^{+}$ & \num{1.3e5}       & \hspace{1em}  \anumu-CC COH $\pi^{-}$ & \num{1.1e5}  \\
\hspace{1em} \anumu-CC COH $\pi^{-}$ & \num{1.2e4}       & \hspace{1em} \numu-CC COH $\pi^{+}$ & \num{1.6e4}  \\
\hline
\hspace{1em} \numu-CC ($E_\text{had}<\SI{250}{MeV}$) & \num{2.4e6} & \hspace{1em} \anumu-CC ($E_\text{had}<\SI{250}{MeV}$) &  \num{1.9e6} \\
\hline
All \anumu-CC & \num{7.1e5}              & All \numu-CC & \num{2.3e+06} \\
All NC & \num{5.3e+06}                     & All NC & \num{2.9e+06} \\
All \nue+\anue-CC & \num{2.6e+05}                & All \anue+\nue-CC & \num{1.7e+05}   \\
$\nu\ e \to \nu\ e$ & \num{2.0e3}               &  $\nu\ e \to \nu\ e$ & \num{1.1e3}          \\
\end{dunetable}

Care has been taken to minimize the material budget between the active volume of the \dword{3dst} and the surrounding gas trackers to a few \% of $X_{\text{rad}}$. This maximizes the efficiency for low momentum charged particles to pass through the passive material and into the active gas tracker volume where it is possible to do precise momentum reconstruction and complementary \dword{pid}.


\subsubsection{Characterization of the 3D plastic scintillator concept with beam tests}
\label{sec:design-cube-characterization}

The response of the plastic scintillator cubes, the active part of 3DST, have been tested in several beam tests at the \dword{cern}~\cite{Mineev:2018ekk}.  A small prototype of $5\times5\times5$ cubes collected data in the T10 test-beam area at \dword{cern} in 2017, with the goal of characterizing the response of the plastic scintillator cubes.  The detector was instrumented with 75 \dword{wls} fibers, 1 mm diameter Y11(200) Kuraray S-type of 1.3 m length. One end of the fiber was attached to a photosensor, the other end was covered by a reflective aluminum-based paint (Silvershine). The photosensors in the beam test were \dword{mppc} 12571-025C with a $1\times1~\text{mm}^2$ active area and 1600 pixels. The data were collected with a 16-channel CAEN digitizer DT5742 with 5 GHz sampling rate and 12-bit resolution. The average light yield was about 40 p.e./MIP in a single fiber, and the total light yield from two fibers in the same cube was measured on an event-by-event basis to be about 80 p.e., as expected. The light cross-talk probability between a cube fired by a charged particle and a neighboring cube was studied. The light measured in the neighbor cube was about 3.7\% of the light collected from the fired cube. The timing resolution for a single fiber was about \SI{0.95}{ns}. When a cube was read out by two \dword{wls} fibers, the timing resolution was \SI{0.7}{ns}.  This would further improve if the light collected by all the three \dword{wls} fibers (approximately as $\sqrt{\text{number~of~fibers}}$) were used. More details can be found in \cite{Mineev:2018ekk}.

In summer 2018, a new prototype made of 9,216 cubes with a size of $8 ~(\text{height}) \times 24~(\text{width}) \times 48~(\text{length})~\text{cm}^3$ collected additional data in the CERN T9 test-beam line.  A different electronic readout, based on the CITIROC chip used in the Baby MIND detector \cite{Basille:2019mcp}, was adopted.  The analysis results confirmed the performance of the 2017 beam tests.  Some event displays are shown in Figure~\ref{fig:testbeam2018}.  These data are useful for validation of the reconstruction tools currently under development.

\begin{dunefigure}[Event displays from \dword{3dst}/SuperFGD testbeam at CERN.]{fig:testbeam2018}
{Event displays showing a photon conversion (top) and a stopping proton (bottom) from data collected at the 2018 test beams at the CERN T9 area.}
\includegraphics[width=6.0in]{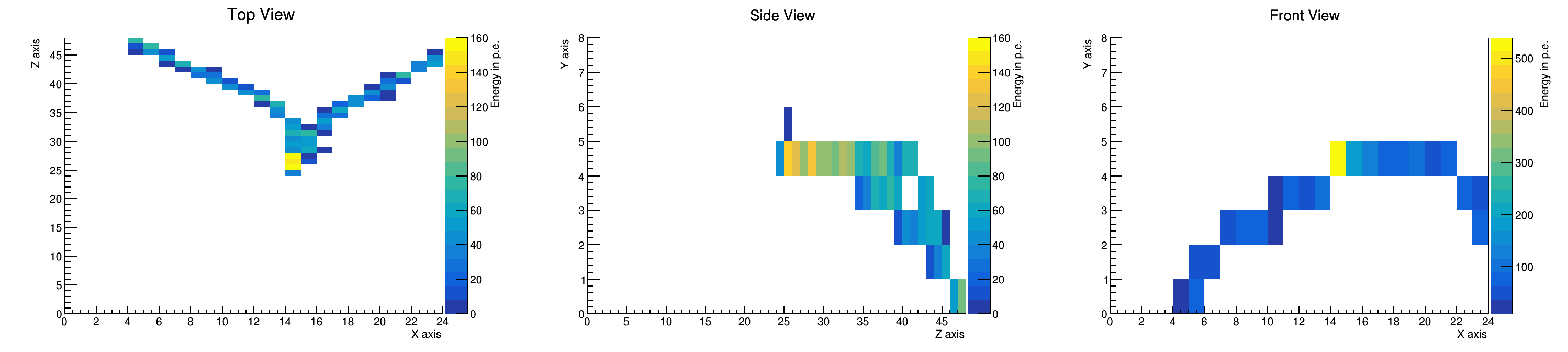}
\includegraphics[width=6.0in]{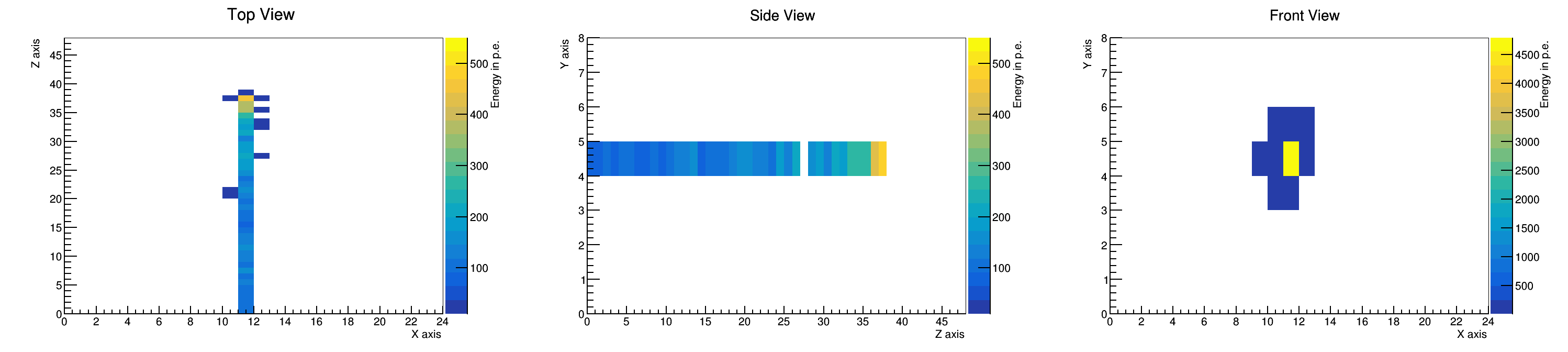}
\end{dunefigure}

\subsubsection{The mechanical box}
\label{sec:design-mechanics}

A mechanical box contains the plastic scintillator cubes and supports the whole \dword{3dst} structure.  It is made of a sandwich of two carbon-fiber (CF) skins, a few mm in thickness, with an AIREX (rigid foam with a density $\sim 0.2~\text{g}/\text{cm}^3$) core, that is a few cm in thickness. The exact dimensions of each component will be optimized with finite element analysis and stress tests in the laboratory.  The CF-based sandwich has 3~mm diameter holes placed with a pitch of about 10~mm through which the fibers penetrate to guide the scintillation light to the \dword{sipm}s.

\subsubsection{The light readout system}
\label{sec:design-light-readout}

The \dword{3dst} light readout system and elements of the mechanical box are shown in Figure~\ref{fig:box-readout}.  The \dword{wls} fibers exit the CF-box through holes and bring the scintillation light outside the mechanical box. A plastic optical connector is glued to the end of each fiber.  The fiber end and connector are polished with a diamond-cutting machine to reduce the internal light reflection.  The scintillation photons are measured with \dword{sipm}s which are coupled to the optical connectors and fibers.
In order to maximize the light detection efficiency, the alignment between the \dword{wls}-fiber end and the \dword{sipm} is extremely important. A glass-resin epoxy layer (readout interface) that is glued on top of the CF-sandwich accomplishes this task.  It has many 3~mm diameter holes that host the plastic optical connectors and precisely align them where they couple to the \dword{sipm}.
The \dword{sipm}s use \dword{mppc} technology. 64 \dword{mppc}s ($8\times8$) are surface mounted on a printed-circuit board (\dword{mppc}-PCB). The \dword{mppc}-PCB is screwed on the readout interface and the alignment between the \dword{wls} fiber and the \dword{mppc} is provided by the precise positioning of the \dword{mppc}-PCB.

 \begin{dunefigure}[Elements of the light readout system integrated on the mechanical box.]{fig:box-readout}
{Elements of the light readout system integrated on the CF-based mechanical box are shown.}
   \includegraphics[width=6.in]{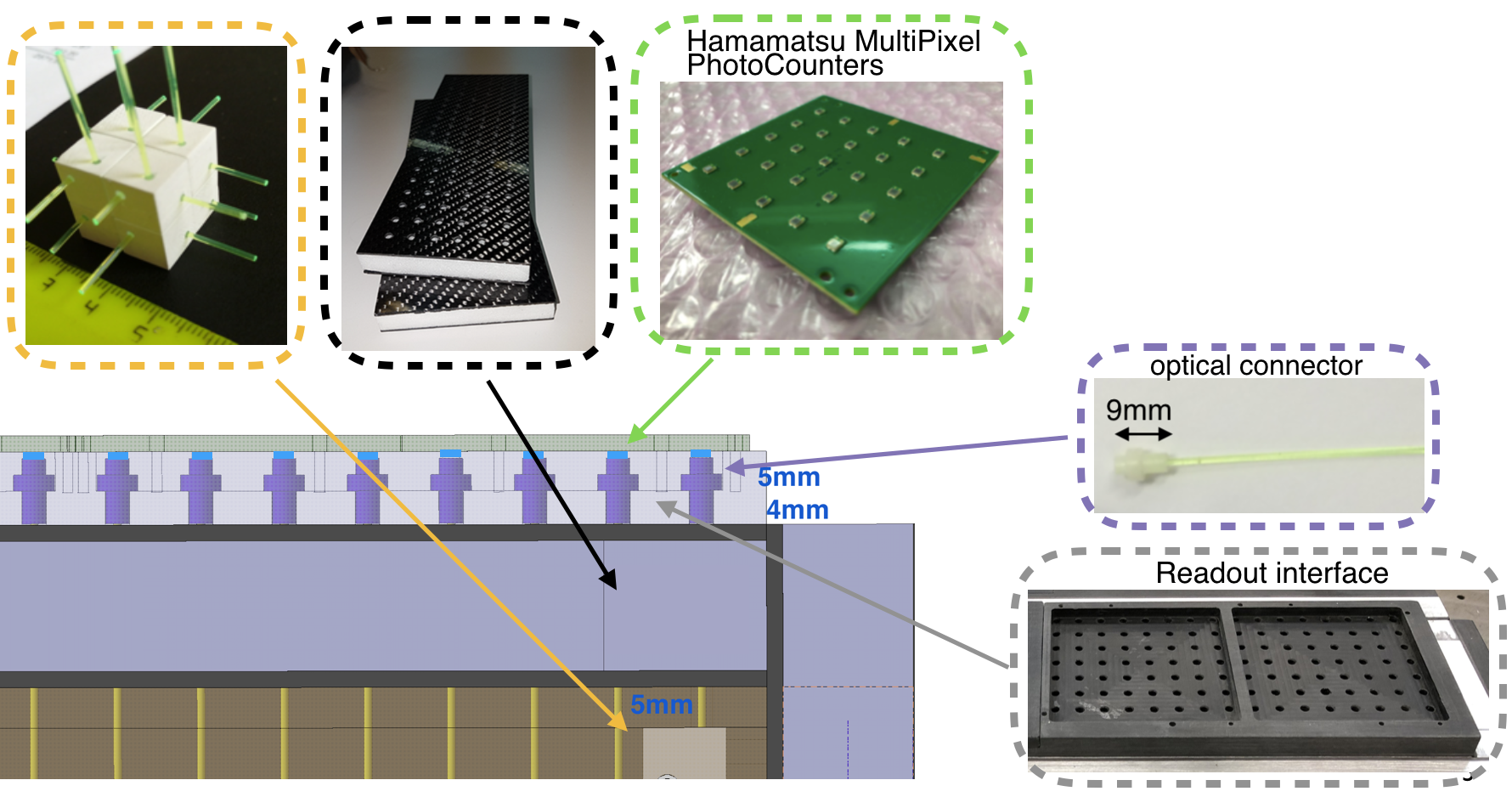}
\end{dunefigure}

\subsubsection{The front-end electronics}
\label{sec:design-fee}

The current design from the T2K \dword{sfgd} detector provides an excellent starting point for the design of the 3DST front-end electronics (FEE).  It is based on the CITIROC chip that can measure the \dword{mppc} signal by providing both the highest charge signal peak value and the time-over-threshold (ToT). An FPGA provides the time stamp.  For the \dword{3dst},  a custom ASIC is being pursued that allows for a more flexible board design, improved timing resolution, reduced power consumption, and significantly lower production costs.  

\subsubsection{The light readout calibration system}
\label{sec:design-calibration}

Since the \dword{sipm} response is sensitive to variations in temperature and humidity, it is important to provide continuous monitoring and calibration. This will be done by measuring the  \dword{sipm} gain, i.e. the number of \dword{adc} counts corresponding to a single photoelectron.  The calibration system is currently under development and is based on the concept developed for the CALICE project \cite{Kvasnicka:2012hwa}. The R\&D aims to develop a very compact system that can be integrated in the \dword{3dst} readout interface.  The idea behind the current design is shown in Figure~\ref{fig:led-principle}. It consists of injecting LED light into clear fibers that are laid along the far end (i.e., on the side away from the \dword{sipm} ) of the \dword{wls} fibers as they emerge from the mechanical box.  The clear fiber is notched with high precision.  The injected LED light travels along the fiber and, when it encounters a notch, some of the light is scattered at almost $90^{\circ}$ toward the \dword{wls} fiber edge opposite the notch.  Some of the light that escapes the clear fiber is captured by the \dword{wls} fibers and is used for calibration.

 \begin{dunefigure}[Elements of the LED calibration scheme for \dword{wls} fibers.]{fig:led-principle}
{Left: the notches made on a fiber are shown as well as its working principle. This figure is taken from \cite{Kvasnicka:2012hwa}.
Right: a notched fiber illuminated with LED light is shown. The light exiting from each single notch is visible.}
\includegraphics[width=4.in]{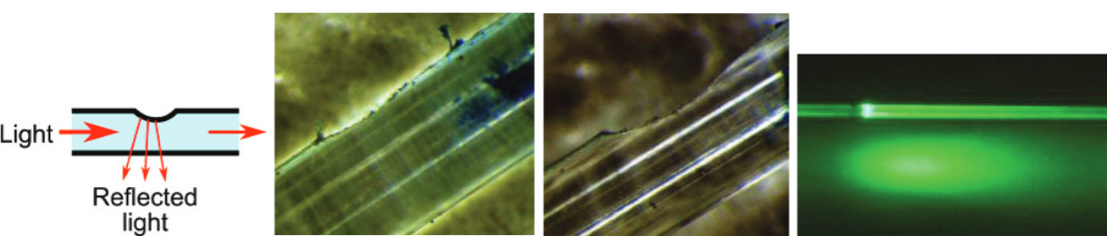}
\includegraphics[width=2.in]{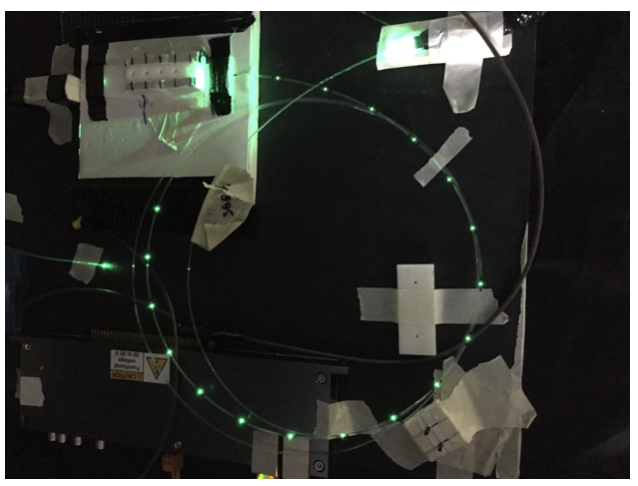}
\end{dunefigure}

\subsubsection{Current prototypes and future R\&D}
\label{sec:design-rd}

The \dword{t2k} \dword{sfgd} design is almost finalized and will be installed in the \dword{t2k} \dword{nd280} complex in the end of 2021 with the aim of collecting neutrino beam data in 2022.  This design is a good launching point for the \dword{3dst} design.  Because the \dword{3dst} is substantially larger than the \dword{sfgd}, additional R\&D is planned to test different cube production techniques in order to improve the light yield and the production speed. Additionally, the \dword{fee} will be further developed and customized for DUNE's needs.

In addition to the \dword{sfgd} prototype mentioned in~\ref{sec:design-cube-characterization}, a smaller prototype of 8 x 8 x 32 cubes, called the US-Japan prototype, was constructed in 2019. The cubes, fiber and the front-end electronics of the US-Japan prototype are the same as those in the \dword{sfgd} prototype, while the MPPC’s and mechanical box have been updated as mentioned in~\ref{sec:design-calibration}.  

In December 2019, both the \dword{sfgd} prototype and the US-Japan prototype were exposed to a neutron test beam at \dword{lanl} \cite{bib:docdb15763}.  The focus of this test was to examine the neutron response of the detector, validating studies indicating the \dword{3dst} can reconstruct neutrons produced in neutrino interactions via \dword{tof} on an event-by-event basis.  During these tests, the \dword{sfgd} was exposed to a beam with neutron energy ranging from 1-800 MeV for over 60 hours at a rate of more than 3kHz.  The US-Japan prototype was exposed to the neutron beam for a few hours. The neutron detection ability in both prototypes was demonstrated clearly and detailed data analyses are being done. Figure~\ref{fig:LANL_photo} shows the prototypes in the Los Alamos beamline (top) and neutron-induced energy deposit hit candidates in the US-Japan prototype (bottom).

 \begin{dunefigure}[Prototypes in the Los Alamos neutron beam test facilty and event displays.]{fig:LANL_photo}
{Photos of prototypes exposed in the Los Alamos National Lab neutron beam test facility and neutron-induced hit candidates in US-Japan prototype. Top Left: \dword{sfgd} prototype. Top Right: US-Japan prototype. Bottom: Neutron-induced hit candidates in the US-Japan prototype. Each candidate has three 2D views and only YZ views are shown here.}
\includegraphics[width=5.8in]{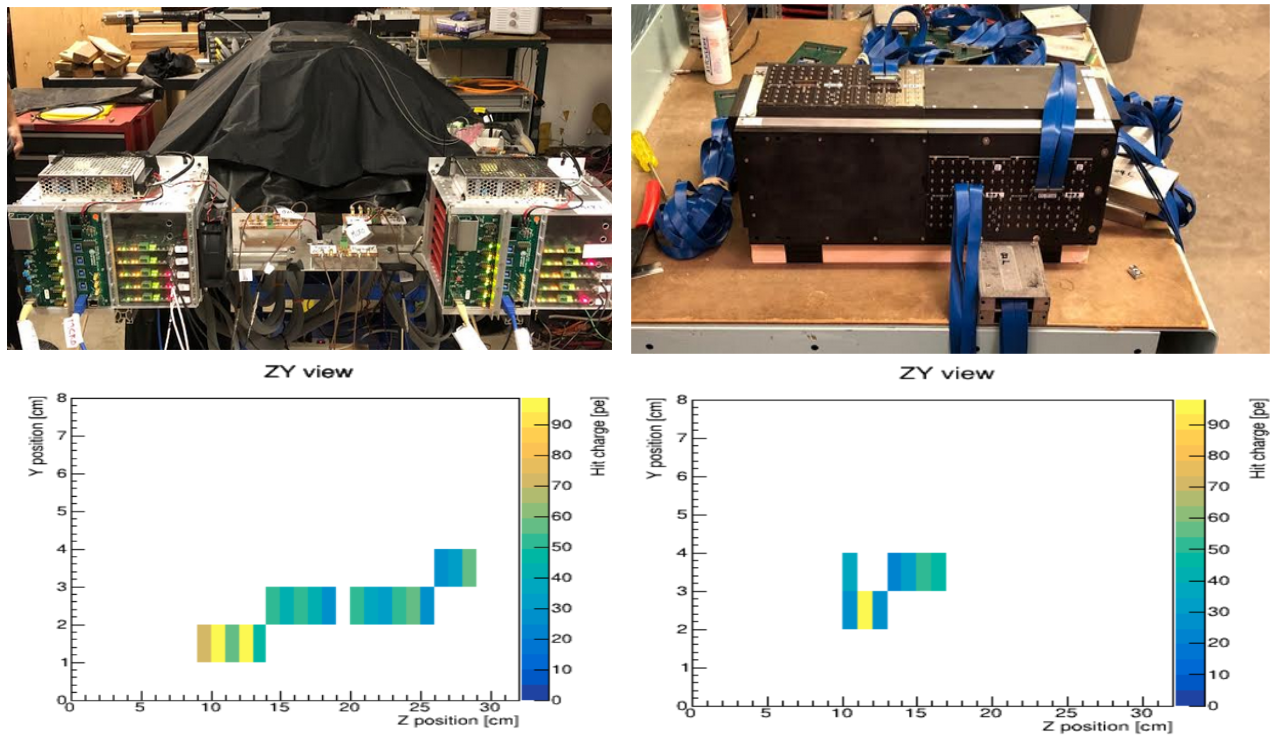}
\end{dunefigure}



\subsection{Straw Tube Tracker Technology and Design} 
\label{sec:sand:STTdesign} 

\subsubsection{A Compact Modular Design} 

The Straw Tube Tracker (STT) is designed to offer a control of the configuration, chemical 
composition and mass of the neutrino targets similar to the one achieved in 
electron scattering experiments~\cite{Petti:2019asx,bib:docdb13262}. 
The base tracker technology for a  \dword{stt} is provided by low-mass straws (5 mm diameter, 12 $\mu m$ walls, 20 micron gold-plated tungsten wire, operated with Xe/CO$_2$ 70/30 gas at 1.9 atm) similar to the ones used in many modern experiments for precision physics or the search for rare processes~\cite{Sergi:2012twa,Anelli:2015pba,Nishiguchi:2017gei,Lee:2016sdb,Gianotti:2013ipa}. The single hit space resolution for the straws is projected to be $< 200~\mu m$.

Thin target layers (typically 1-2\% $X_0$) of various passive materials with high chemical purity can be dispersed 
between the layers of straws distributing the target mass throughout the volume and separating the target 
mass from the low mass tracking system. In the current design, these targets will be polypropylene foils (also acting as a radiator for the transition radiation detector) and carbon.  The average density is kept low enough to obtain a total detector 
length comparable to the the radiation length, for an accurate measurement of the four-momenta of the 
final state particles. The passive targets account for up to 97\% of the total detector mass. 
This feature, combined with the excellent vertex, angular, momentum, and timing 
resolutions are key factors to correctly associate neutrino interactions to each target material. 

Figure~\ref{fig:STT-CompactModule} shows the design of one default STT module, equipped with 
neutrino target layers, providing optimized tracking and particle identification. The straw layers are shown on the right. Shown in the middle is a radiator made of a series of thin, polypropylene foils (119 foils 18 $\mu$m thick with 117 $\mu$m air gaps) for $e/\pi$ separation via transition radiation. To the left is a target layer (of polypropylene in this case). The target and radiator layers act as the main targets in the detector. They can be dismounted/replaced if necessary, though this may be difficult in SAND and is not anticipated to be done often. The average density of the detector can be tuned between a maximum 
of $\SI{0.18}{g/cm^3}$ -- corresponding to the thickness of the radiator and target in Fig.~\ref{fig:STT-CompactModule} --   
and a minimum of $\SI{0.005}{g/cm^3}$ if only the straw layers are present. 
A broad range of target materials like C, Ca, Fe, Pb, etc. can be installed in place of the target radiator, provided 
that they can be manufactured in the form of thin planes. The tracker under discussion here will only have the C target layers.
The STT modules equipped with 
target materials are interleaved to guarantee the same average acceptance.

 \begin{dunefigure}[Compact STT module.]{fig:STT-CompactModule}
{Drawing of a compact STT module including three main elements (left to right): 
(a) a tunable polypropylene CH$_2$ target; (b) a radiator with 119 polypropylene foils for $e^\pm$ ID; (b) four straw layers XXYY (beam along $z$ axis and B field along $x$ axis). Some of the radiator foils are replaced with thin carbon targets.
}
      \includegraphics[width=0.55\textwidth]{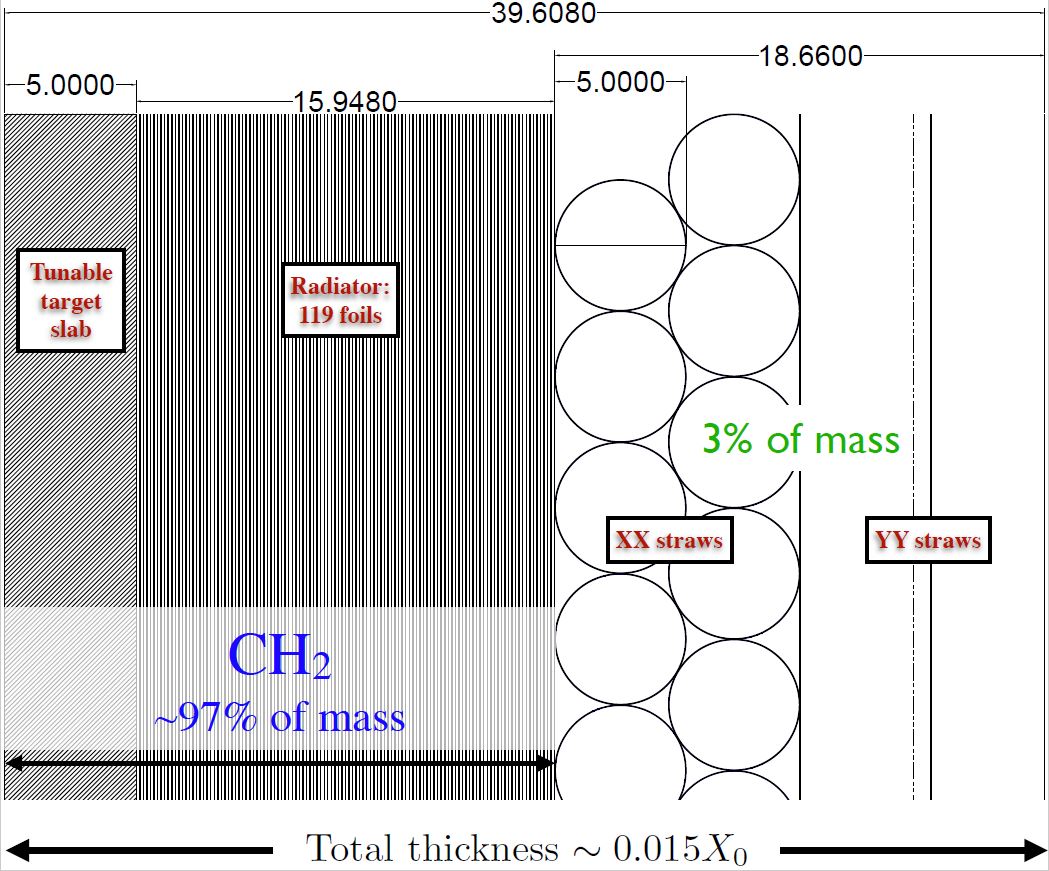}
\end{dunefigure}

A detailed 3D CAD engineering model of the STT modules and assembly was created in order to study the various constraints on the support frames with a realistic detector. Figure~\ref{fig:STT-CompactModule-CAD3D} shows the corresponding design of two STT modules equipped with CH$_2$ and graphite targets.  
The material chosen for the support frames is C-composite, resulting in an average amount of material crossed by the particles of $\sim 0.1 X_0$.
An analysis of the deformations was done using finite element analysis. The maximal deformations in the center of each frame beam are typically a few mm and decrease rapidly with the size of the modules. These results indicate that the design implemented is realistic and adequate to be installed within the \dword{sand} magnet.

 \begin{dunefigure}[CAD model of STT module with radiator and target slab.]{fig:STT-CompactModule-CAD3D}
{Left picture: Detailed 3D engineering CAD model of one STT module equipped with CH$_2$ target slab (in brown color) 
and radiator (in blue color). The average density of the detector can be fine tuned between 0.005 g/cm$^3$ 
(without target slab and radiator) and 0.18 g/cm$^3$ (configuration in the drawing). Both the target slab and the 
radiator can be unmounted during data taking by removing four corner screws.
Right picture: Engineering CAD model of of one STT module equipped with graphite target (in black color). 
The tracking part is composed of four straw layers XXYY and is the same as in the CH$_2$ module on the left.}
      \includegraphics[width=1.00\textwidth]{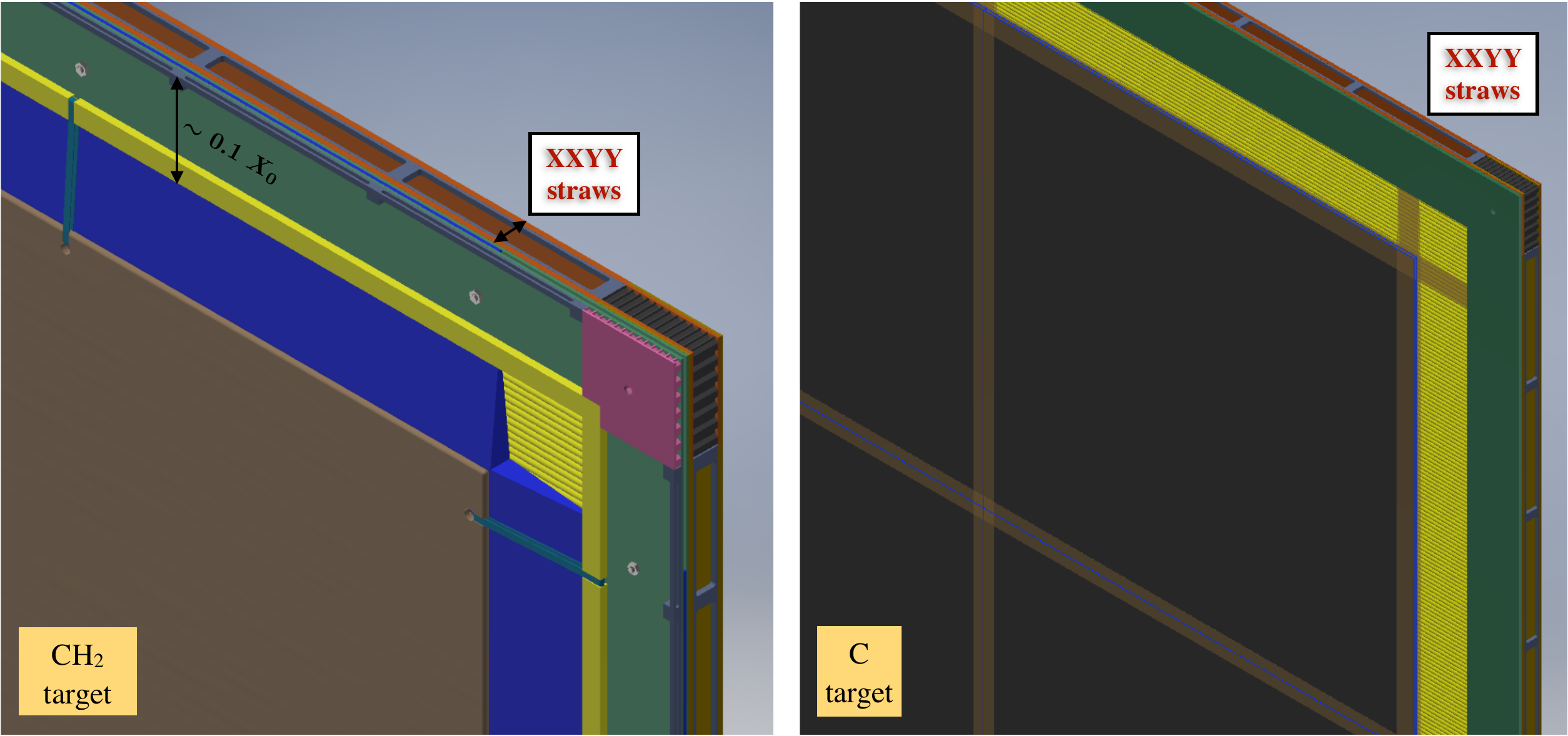}
\end{dunefigure}

The front-end electronics readout is based upon the VMM3a custom \dword{asic} \cite{DeGeronimo:2013noa}, developed at Brookhaven National Laboratory and CERN, which is a component of the ATLAS upgrade, as well as several other detectors. Each ASIC chip provides the readout for 64 individual straws, allowing the use of compact FE electronic boards fully integrated within the frames of the STT modules (Figure~\ref{fig:STT-CompactModule-CAD3D}). The low-power consumption and per-channel cost are similarly useful for a compact detector readout. The ASIC chip provides precise detector hit charge and time measurements for the straws. Preliminary studies indicate that the range of available 
gain options, the low electronics noise, a 12-bit ADC, 10 bits global DAC and 5-bit channel specific trim DACs are compatible with the STT requirements. Each FE board includes up to 8 ASIC chips and is controlled by an FPGA. The FE readout board FPGAs transfer VMM3a data over gigabit links to the Front-End LInk eXchange (FELIX) PCIe cards, similar to the implementation in Proto-DUNE and in the DUNE FD.

\subsubsection{Concept of ``Solid" Hydrogen Target}
\label{sec:sand:STTsolidH} 

The control of the configuration, chemical composition and mass of the targets provided by the STT allows the implementation of a ``solid" hydrogen target by subtracting measurements on dedicated graphite (pure C) targets from those on the CH$_2$ plastic targets described above~\cite{Duyang:2018lpe}. Each graphite target is 4 mm thick and is composed of a stack of 61 cm x 61 cm tiles mounted in front of a four layer XXYY straw assembly (the same as in Fig.~\ref{fig:STT-CompactModule}). Figure~\ref{fig:STT-CompactModule-CAD3D} shows the complete assembly of one STT module equipped with graphite target. The thickness of the graphite plates is tuned to match the same fraction of radiation length (1-2\% $X_0$) as the combined CH$_2$ radiator and target slab it is replacing. The gas mixture used for the STT modules equipped with nuclear targets (without radiator) is Ar/CO$_2$ 70/30 with an internal pressure of about 1.9 atm. Modules equipped with graphite plates are interleaved with CH$_2$ modules throughout the  STT volume in order to guarantee the same detector acceptance for CH$_2$  and C targets. The graphite targets are an essential element of this concept: they automatically provide all types of interactions, as well as reconstruction effects, relevant to achieve a model-independent subtraction of the C background in selecting the $\nu(\bar\nu)$-H CC interactions (Sec. 5.6.3.2).

\subsubsection{Prototyping and Tests}
\label{sec:sand:STTproto} 

The core technology required to build the STT is well established and 
the need of major detector R\&D is not anticipated.  The STT design described in Sec.~\ref{sec:sand:STTdesign} 
combines an off-the-shelf VMM3a readout with the same straws currently being produced for COMET~\cite{Nishiguchi:2017gei} and other experiments. This design benefits from the 
extensive expertise and R\&D activities performed for those projects. 
All the components required to build the \dword{stt} can be manufactured industrially by vendors, 
and will be assembled into the final \dword{stt} modules at selected production centers. 

A small \dword{stt} prototype is being tested at JINR and CERN using FE boards with VMM3a readout 
from the Mu2e experiment provided by Brookhaven National Laboratory. This prototype is used to 
validate the straw performance with the VMM3a readout in a configuration similar to the 
one planned for DUNE. Extensive tests of the straw properties, operational conditions, and 
possible aging effects are performed by GTU, JINR, and IIT Guwahati groups with the same 
straws as in the \dword{stt} design. Three straw production lines equipped with the ultrasonic 
welding technology foreseen for STT are currently operational and two additional lines dedicated 
to the STT production are planned. A prototype of the graphite target has been tested at USC. 
The VMM3 readout is supported by a continuous 
R\&D activity aimed at reducing the power and size used by the chip. It is planned to expose 
complete \dword{stt} prototypes to a test beam with various low-energy particles ($\mu,e,\pi,p,n$) at CERN in the near future.

\subsection{Time Projection Chambers}
\label{sec:tpc}

TPCs are a well established technology to enable precise tracking and particle identification covering a large volume with a relatively low number of electronic channels. The proposal of using TPCs for tracking in SAND is based on the successful experience of the T2K near detector ND280~\cite{Abe:2011ks} and its upgrade~\cite{Abe:2019whr}.  In T2K, the TPCs have played a crucial role enabling the near detector to constrain uncertainties in the measurement of neutrino oscillation parameters. Aside from the precise tracking and good particle identification characteristics, the low density of the \dword{tpc} volume reduces backgrounds due to neutrino interactions happening outside the main target fiducial volume. 
The design of the ND280 TPCs was the result of a dedicated R\&D program which reached an highly optimized solution, including custom electronics. The ND280 TPCs have demonstrated very good reliability and good, stable performance over ten years of running. 

It is planned for \dword{sand} to use the improved TPC design being deployed for the ND280 upgrade which relies on resistive Micromegas detectors. This technology results from detector development R\&D for the International Linear Collider~\cite{Dixit:2003qg}.  This design is demonstrating very good performance in beam and cosmic tests of prototypes for the ND280 upgrade~\cite{Attie:2019hua}. The main advantage of the design relative to past designs is that it gives an improvement in the spatial resolution with a lower number of channels due to the spreading of the charge onto multiple anode pads. It also has very robust protection of the electronics against possible sparks. 

\subsubsection{TPC general design}
The design of the TPCs for \dword{sand} is based on three rectangular chambers in the downstream and upper/lower side of the \dword{3dst}, filling the low-density gas tracker volume shown in Figure~\ref{fig:sand-geometry}. The design is based on the ND280 experience~\cite{Abgrall:2010hi}.

The dimensions of the three TPCs will be at least \SI{240}{cm} along the magnetic field direction to provide good coverage of the \dword{3dst}. The downstream chamber will have a height of \SI{300}{cm} and a thickness of \SI{77}{cm}. The upper and lower chambers will be \SI{57}{cm} thick and extend for \SI{141}{cm} along the upper and lower edges of the \dword{3dst}.

\begin{dunefigure}[TPCs in \dword{sand} and muon reconstruction efficiency.]{fig:TPCsForSAND}
{Left: A cutaway view of the \dword{sand} reference design with the \dword{3dst} and the three surrounding \dword{tpc}s .
Right: The efficiency to reconstruct muons generated by neutrino interactions in the 3DST that escape and enter into the TPCs. Note, the acceptance is low for backward going muons but it is not zero since tracks can be bent into the TPCs by curvature in the magnetic field.}
    \includegraphics[width=3.2in]{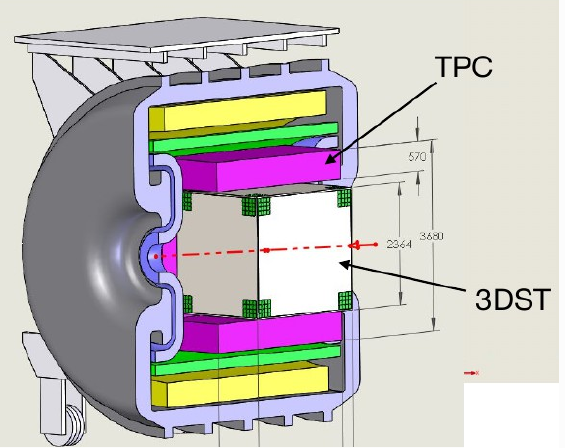}
    \includegraphics[width=3.2in]{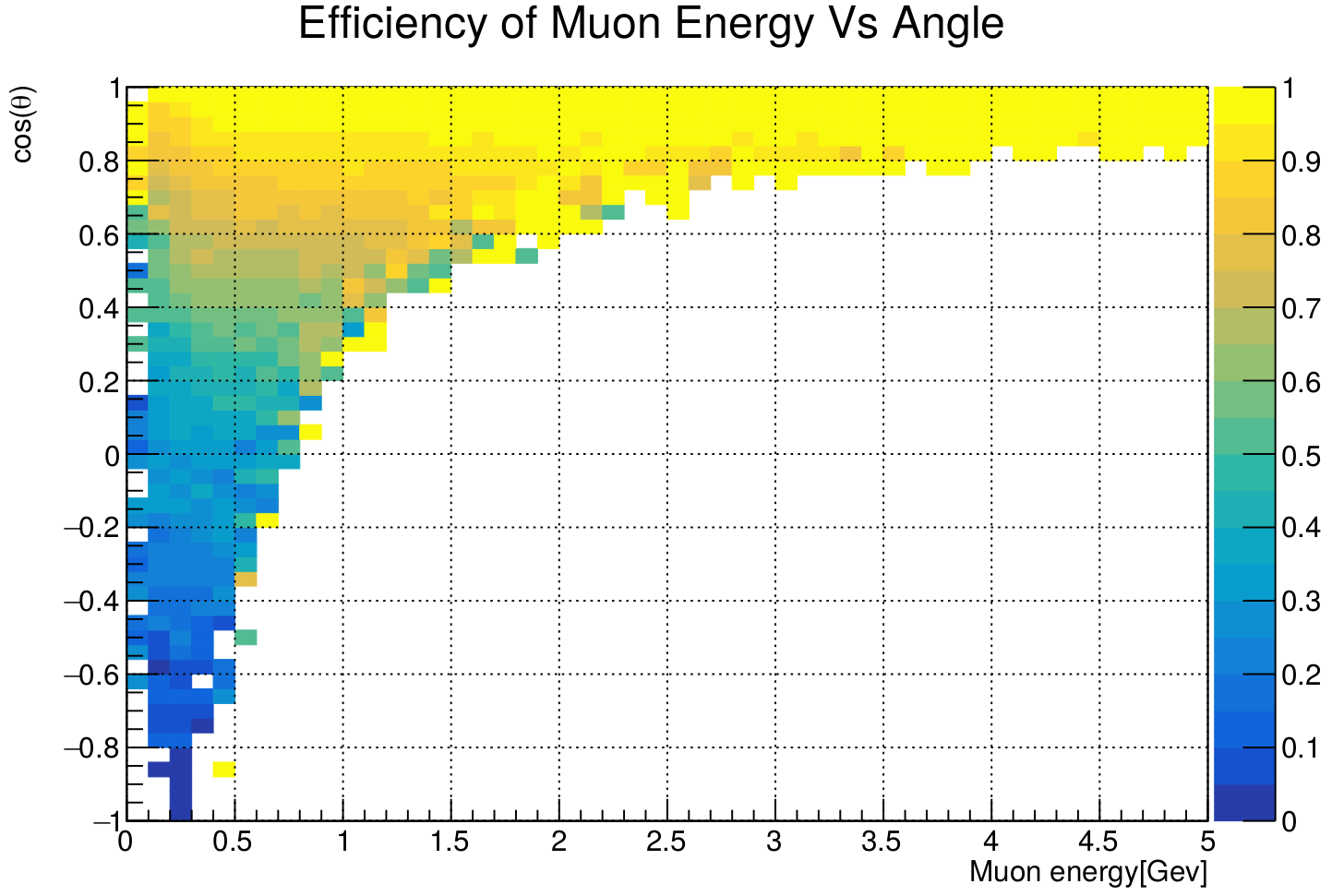}
\end{dunefigure}


For the ND280 TPCs (and for the proposed \dword{sand}  design), the volume is filled with a dedicated non-flammable mixture of gas, which has been optimized for low transverse diffusion, large drift velocity (7.5 cm/$\mu$sec) and to minimize the effect of impurities (30~m attenuation length). \footnote{The TPC gas mixture used by T2K is argon $+$ 2\% isobutane $+$ 3\% carbon tetraflouride.}
The purity of the gas (O$_2<$10~ppm, H$_2$O$<$10~ppm, CO$_2<$100~ppm) is monitored by dedicated small gas-monitoring chambers. The general stability of the performance is monitored by the photo-electron signals induced by a laser calibration system.
The chambers can be operated at 200-300~V/cm drift field. The field cage is designed to minimize the non-planarity of cathode/anodes ($<$0.2~mm) inducing electric distortions ($\Delta E/E < 10^{-4}$),  and to minimize the amount of material to avoid large multiple Coulomb scattering. 
The readout planes are instrumented with bulk Micromegas~\cite{Giomataris:2004aa} modules, providing a signal-over-noise ratio of 100, with pixeled readout anode with pads of $7\times10$~mm$^2$ for the vertical TPCs.  The horizontal TPCs are instrumented with resistive Micromegas modules~\cite{Dixit:2003qg}, where the pads are covered with a layer of insulating material and a layer of resistive material. The avalanche signal is induced in the pads through capacitive coupling and is thus spread over multiple pads. By studying the distribution of the charge over multiple pads, a better resolution can be obtained than from the direct signal deposited in a single pad. This technology enables good resolution with larger pads and thus a lower number of channels. The pad size of the horizontal TPCs is $10\times 11$~mm$^2$.

\subsubsection{TPC performances and specifications}
The preliminary design of the \dword{sand} \dword{tpc}s is very similar to those used in ND280 and the ND280 upgrade.  So, the ND280 experience and prototype tests provide realistic benchmarks for projecting performance in \dword{sand}.
The performance of the ND280 TPCs is described in Ref.~\cite{Abgrall:2010hi}.  The spatial resolution is about $700\mu$m, driven by the pad size ($7\times10$~mm$^2$). As shown in Figure~\ref{fig:TPCSpatialReso}, even better performance has been demonstrated by using resistive Micromegas detectors. A first TPC prototype was tested in CERN test beam~\cite{Attie:2019hua} showing spatial resolution as good as 200-500$\mu$m with pad size of $9.8\times7$~mm$^2$. Preliminary results from a more optimized Micromegas module, tested in the DESY beam in presence of magnetic field, show further improvements in the resolution. In general, the amount of charge spread between different pads can be tuned by changing the resistivity of the resistive foil and the glue thickness (thus the capacitance). Such parameters, together with the pad size, can be adapted to the needed specifications in terms of spatial resolution.

 \begin{dunefigure}[Resolutions of TPCs.]{fig:TPCSpatialReso}
{Left: spatial resolution of the vertical TPCs in ND280 with beam events. Right: spatial resolution obtained with a prototype of resistive Micromegas module on a test beam of 0.8 GeV positrons, pions and protons for different drift distances. Figures taken from ref.~\cite{Abgrall:2010hi} and ref.~\cite{Attie:2019hua}. }
  \includegraphics[width=7cm]{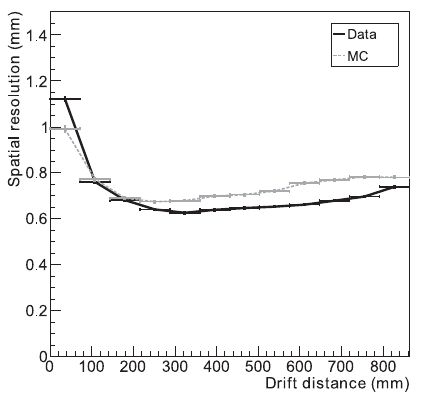}
  \includegraphics[width=8cm]{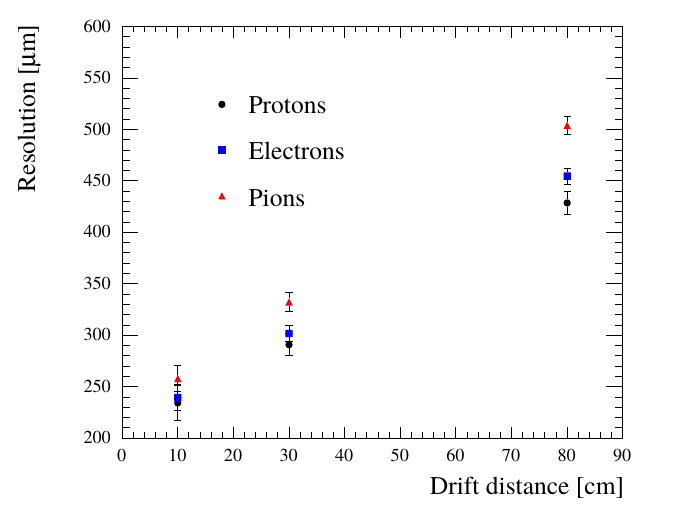}
\end{dunefigure}

In the ND280 TPCs, the spatial resolution of $700\mu$m in the 0.2T magnetic field corresponds to a momentum resolution of 10\% at 1~GeV. This performance level was sought by T2K because it matches roughly the  smearing of the reconstructed neutrino energy due to the Fermi momentum of the nucleons inside the nucleus, which is around 100-200~MeV/c. 
\dword{sand} will sit in a neutrino flux of higher energy but will profit from a stronger magnetic field. On the other hand, the sample of interactions on hydrogen, which can be selected in the 3DST as shown in Sec.~\ref{sec:sand:Hselection}, will not be limited by the Fermi momentum but by the precision in the measurement of neutron momentum through \dword{tof}, which is on the order of 50-100 MeV~\cite{Munteanu:2019llq}.  This calls for a resolution of about 5\%, matching with the one obtained by using momentum via range for muons stopping in 3DST.

For a resolution of about 200~$\mu$m and pad size ($1\times 1.1$~cm$^2$) as used in the \dword{nd280} upgrade \dword{tpc}, the \dword{sand} \dword{tpc} can feature a momentum resolution of a few percent at 3~GeV and better than 2\% at 1 GeV with about 70k channels.  This is a preliminary estimate, since the resistivity can be increased and the number of pads further decreased. ND280 achieved a 2\% uncertainty on the overall momentum scale.

The TPCs will also allow for particle identification by dE/dx. For example, the ND280 TPCs were able to use dE/dx to reduce the  misidentification rate of muons as electrons to 0.2\% below \SI{1}{GeV/c} and about 0.5\% between 1-\SI{2}{GeV/c} \cite{Abe:2014usb}. The latter range is typical for muons in \dword{dune} and a similar performance is expected. 


\subsubsection{TPC Micromegas modules and electronics}


The shape of the Micromegas readout modules can be easily adapted to different geometries and the readout electronics used in ND280 can be directly deployed in \dword{sand}. The Micromegas modules of the ND280 vertical-TPCs are rectangular with $34\times36$~cm$^2$ size, hosting 1728 pads each. Each module is read by 6 Front-End Cards (FECs), each of which is instrumented with 4 AFTER ASICs~\cite{Baron:2008zza}. The data from each module are further processed by a Front-End Mezzanine (FEM) card which sends them through optical fibers to 18 Data-Concentrator cards (DCC) placed outside the magnet (2 for each readout plane). A similar architecture can be envisaged for SAND: in the baseline hypothesis of 70,000 channels, this would correspond to a production of about 1000 ASICs, 250 FECs, 40 FEM and about 10 DCC.

The experience in terms of production and qualification of such electronics for ND280 has been very positive. The 72 Micromegas modules have been produced and tested in about 15 months with 92\% of the produced modules satisfying the required specifications. The quality control was positive for 90\% (97\%) of the FECs (FEMs). The total fraction of dead channels is 0.01\% and, in 10 years of operation, the electronics experienced only 1 FEM failure and only 2 HV filters had to be repaired. 

The Micromegas are operated with a gain of 1500 with less than one spark per hour. The TPC performance reported above relies on the very good performance of the Micromegas modules which feature a gain uniformity of 2\% and an energy resolution of better than 9\% for each module (with a uniformity better than 8\%) with a stability of 3\% between modules.

\subsection{LAr Active Target}
\label{sec:meniscus} 

An active liquid argon target of about 1 ton is foreseen in the upstream part of the magnetic volume for all the design options being considered. The main motivation is to constrain nuclear effects on Ar and to have a complementary Ar target permanently located on-axis for cross-calibration. The thickness of the LAr volume is small enough ($\sim\number1 X_0$) to reduce energy loss, showering and multiple scattering, as the outgoing particles will be analyzed by the downstream detector elements. Figure~\ref{fig:SAND-LAr-target} shows a conceptual drawing of the 
cryostat which will host the active LAr target inside the inner vessel. The cryostat walls are made of C-composite material reinforced with an internal thin
aluminium foil, resulting in an overall thickness of a small fraction of radiation length. The exact positioning, size, and shape of this LAr ``meniscus" will be the object of optimization. The LAr meniscus will be instrumented with an optical system which will collect UV scintillation light on fine segmented focal planes.
The cryogenic system can be reduced to essentials, with the LAr circulation going through the aperture of \dword{ecal} endcaps. 
Detailed studies were performed with the STT-only design option to evaluate the reconstruction quality and the acceptance 
for various particles produced in the LAr target. The momentum and angular resolutions 
obtained were consistent with the ones for events originating within the STT fiducial volume~\cite{bib:docdb13262}.

 \begin{dunefigure}[Active LAr target for SAND.]{fig:SAND-LAr-target}
{Conceptual design of the active LAr target to be located in the upstream part of the volume inside the \dword{ecal}. The detailed structure of the cryostat is visible in the picture on the right.}
      \includegraphics[width=0.70\textwidth]{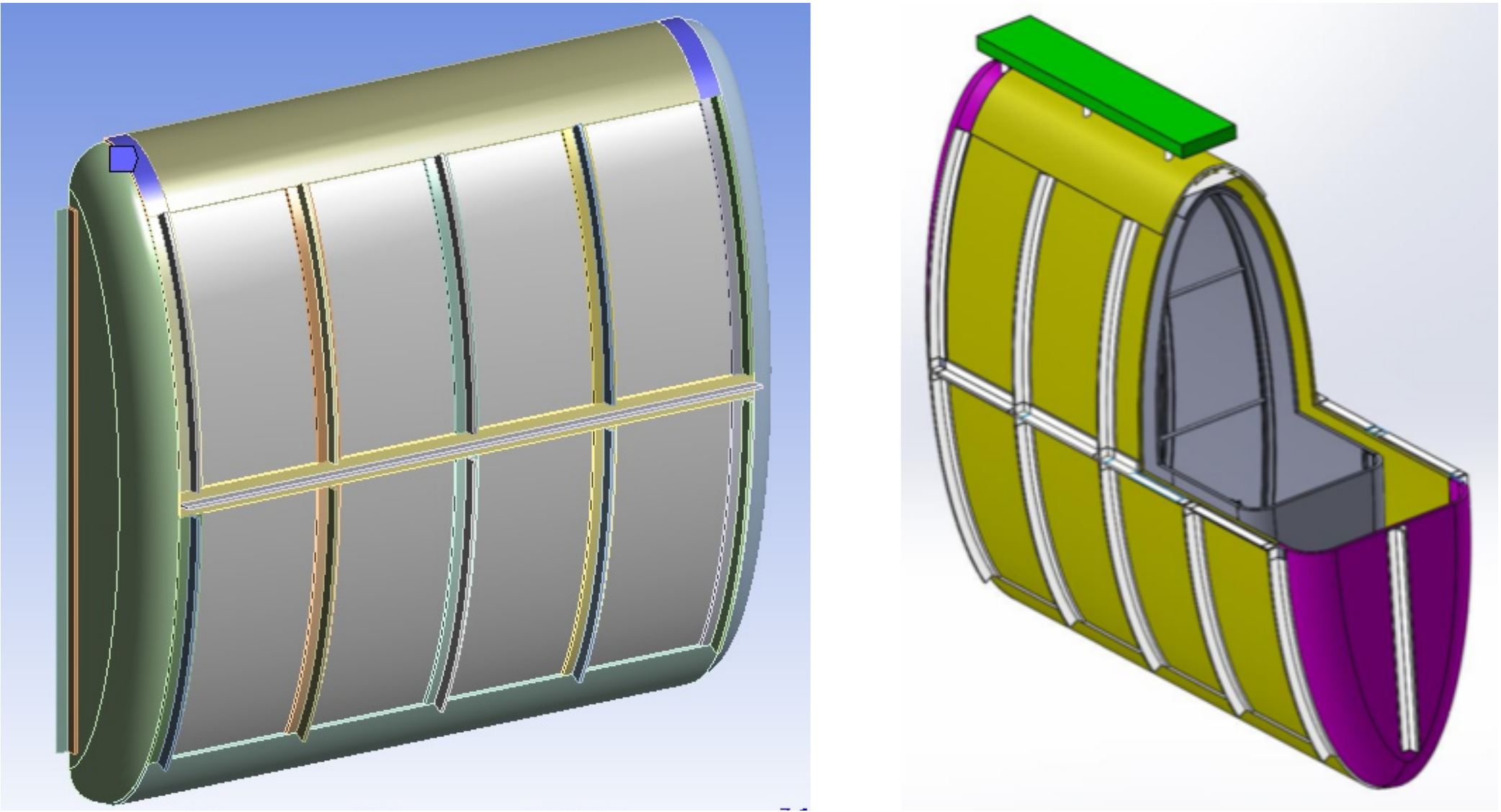}
 \end{dunefigure}
 

\section{Design Options for the Inner Target Tracker}

\subsection{The 3DST+TPCs design option}

In the 3DST+TPCs design option the 3DST target tracker is surrounded with the TPCs described in Sec.~\ref{sec:tpc}. This option is illustrated in Fig.~\ref{fig:sand-geometry}. The physics performance studies featuring the 3DST were done using this option. For the purposes of tracking particles leaving the 3DST the TPCs and STT perform similarly. In particular, the beam monitoring performance does not depend on the choice of TPCs vs STT. The 3DST+TPCs option is, in some ways, the least intricate option and therefore the most straightforward option to understand. The 3DST+STT and STT-only options have more nuances as described below.

\subsection{The 3DST+STT design option}
\label{sec:sand:STT}
A tracker based on well-established straw tube technology is another option for a low density tracker surrounding the \dword{3dst}. In this design option, known as 3DST+STT, modules containing layers of straws are used as the tracking elements for particles leaving the 3DST. Some modules will also contain transition radiator layers made of a large number of thin polypropylene foils to aid in particle identification.  Also, some additional mass is added to some of the modules in the form of solid polypropylene\footnote{Also referred to as CH$_2$ in the text.} target layers and graphite\footnote{Also referred to as carbon (C) in the text.} target layers. The additional mass will yield some neutrino interactions within the high resolution STT tracking regions.


\label{sec:sand:3dst_stt} 

The flexibility and modularity of the design described in Sec.~\ref{sec:sand:STTdesign} allows the use of STT as a low density tracker in combination with the 3DST for the 3DST+STT reference design variant.  In this variant the 3DST is surrounded on each of the four sides (top, bottom, left and right) by STT modules without targets and radiators (pure tracking modules), in which the four straw layers (Figure~\ref{fig:STT-CompactModule}) are located in the XZ or YZ planes.  Two tracking modules with 8 straw layers each are located in the upstream region to track backward going particles exiting from the 3DST volume. 

Downstream of the 3DST are the following STT modules:
\begin{itemize}
\item Three special tracking modules with 8 straw layers each (these modules are a variant of the standard STT modules).
\item Followed by 25 regular modules, consisting of:
  \begin{itemize}
  \item Twenty three modules equipped with polypropylene targets and radiator foils
  \item The 23 modules are interleaved with 2 modules equipped with only graphite targets and no radiators.
  \end{itemize}
\item Finally there are 5 modules with radiators and no targets.
\end{itemize}

Simulations indicate that this geometry can provide a good acceptance and reconstruction for particles emitted from each side of the 3DST. The polypropylene and graphite targets will be optimized and can be removed/modified during data taking. The maximal fiducial target mass corresponds to about 1.44 tons of polypropylene and 160 kg of graphite, for an average density of the downstream STT section of about 0.15 g/cm$^3$. The STT modules equipped with polypropylene and graphite targets enable various physics measurements with interactions occurring within the STT targets, as described in Sec.~\ref{sec:sand:nu-nucleus}.


\subsection{The STT-only design option} 
\label{sec:sand:STTonly} 




Another option being considered for SAND has most of the volume inside
the ECAL filled with STT modules, with the exception of a small upstream
region instrumented with a thin LAr target (Sec.~\ref{sec:meniscus}). Figure~\ref{fig:SAND-GeometryFullSTT} illustrates the layout of STT-only configuration. 
In the default configuration the total STT mass is about 7.4 tons, with a fiducial target mass of about 4.7 tons of CH2 (78 modules) and 504~kg of graphite (7 modules). In this configuration the targets represent about 97\% of the total detector mass - the mass of the straws being 3\% - for an average density of 0.18 g/cm$^3$ ($X_0 \sim 2.8$~m). The first upstream STT module following the LAr target, as well as the last four downstream modules, have neither target nor radiator, and are composed of six straw XXYYXX layers. The total thickness of the STT in this configurations is 1.33 $X_0$. The total number of straws in the entire STT is 234,272 which also is the number of readout channels. A detailed description of this option, along with the results of detector simulations, event reconstruction and physics sensitivity studies is available in reference~\cite{bib:docdb13262}.

\begin{dunefigure}[STT-only configuration.]{fig:SAND-GeometryFullSTT}
{Geometry of the STT-only configuration for the SAND inner target/tracker.
The different STT modules are visible from the YZ view on the left:
the blue modules are equipped with graphite targets and
are interleaved with standard CH$_2$ modules shown in green.}
\includegraphics[width=0.85\textwidth]{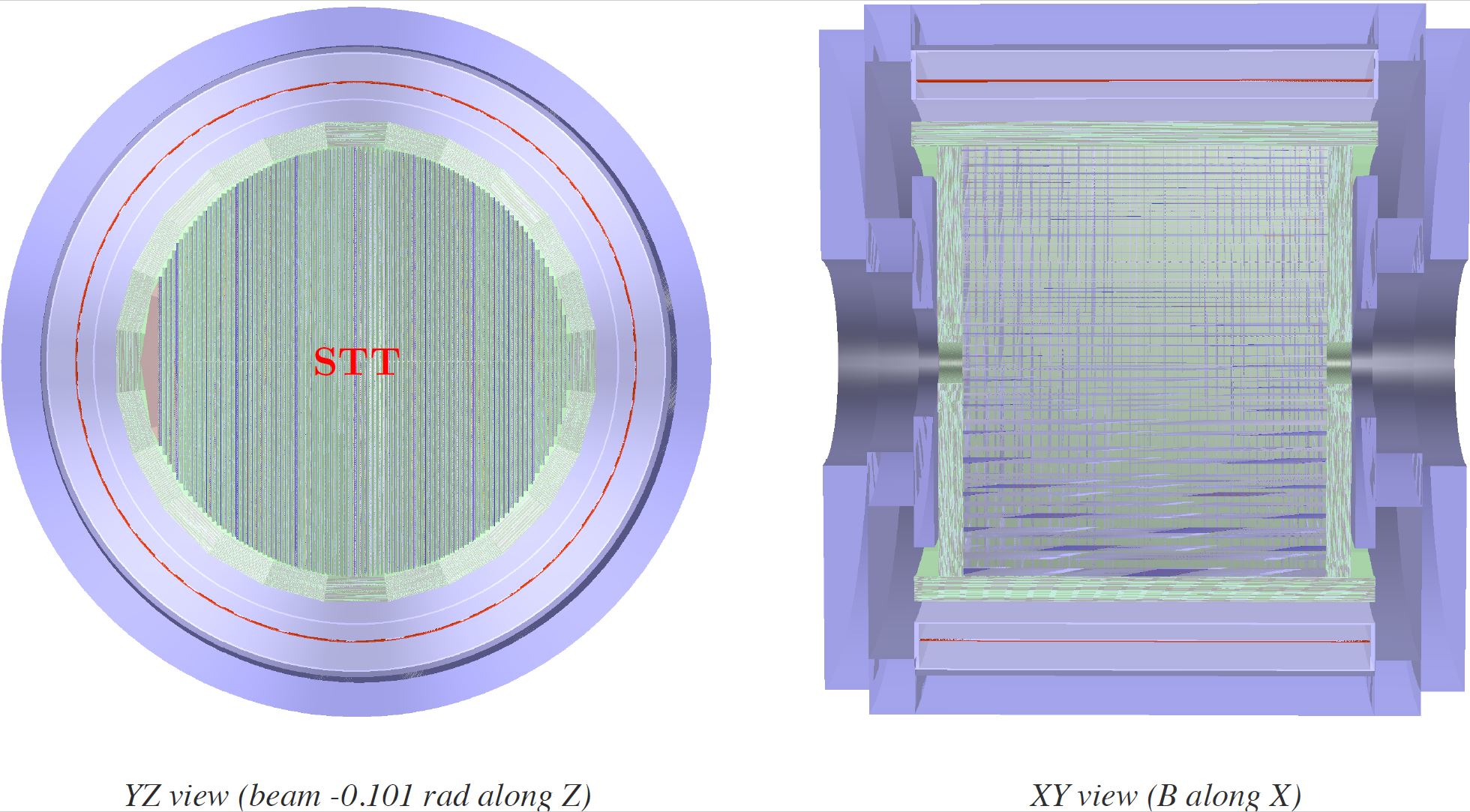}
\end{dunefigure}


The performance was studied using different simulation packages and the nominal DUNE flux. The default simulation was performed with the complete GENIE and GEANT4 chain. In addition the FLUKA package was used to cross-check the neutrino event generator and the detector simulation. A preliminary track reconstruction using the STT digitized hits and circular fit approximation was implemented. The results showed a momentum resolution of 3.1 \% (core, 5.9 \% RMS) averaged over the entire momentum spectrum and track lengths within the STT volume. The corresponding angular resolution was 1.7~mrad (core, 8.0~mrad RMS).


The momentum scale uncertainty can be calibrated with the mass peak
obtained from the $K_s^0 \to \pi^+ \pi^-$ decays reconstructed within the
STT tracking volume. Similarly, the angular scale can be calibrated with
both $K_s^0 $ and $\Lambda \to p \pi^-$ decays. With the default FHC
neutrino flux, an average of about 142,000 $K_s^0 \to \pi^+ \pi^-$ and
280,000 $\Lambda \to p \pi^-$ decays per year can be reconstructed within
the STT volume in SAND. A similar technique was used by the NOMAD
experiment which achieved a momentum scale uncertainty < 0.2\% with only
$\sim$30,000 $K_s^0 $ decays~\cite{Altegoer:1997gv,Wu:2007ab}.



Particle identification is available throughout the STT volume exploiting the dE/dx ionization signals in the straw gas, range, and the Transition Radiation (TR) produced by e$^\pm$ with $\gamma>$1000 in the radiator foils. The electron identification is particularly relevant in DUNE since the most critical measurements involve e$^\pm$ , e.g. $\nu$-$e^-$ elastic scattering, $\nu_e(\bar \nu_e)$ CC, $\pi^0 / \gamma$, etc. The NOMAD experiment obtained a $\pi^\pm$ rejection factor of 10$^3$ with an electron identification efficiency > 90\% for E$_e$ > 1GeV, exploiting the TR effect. The design of the radiators in the STT module (Fig.~\ref{fig:STT-CompactModule}) was optimized at energies < 2 GeV with dedicated simulations of the production and detection of TR photons. The results indicate that a similar performance as in NOMAD is obtained with tracks in STT at least 75 cm long.

\section{Detector and Physics Performance} 
\label{sec:sand:physics}


The addition of the KLOE magnet and \dword{ecal} to the \dword{dune} \dword{nd} is a relatively recent event.  For this reason, the evaluation of the performance of \dword{sand} is largely limited to individual sub-systems, which are discussed in the previous Sections. Dedicated studies are needed to finalize the design of \dword{sand} as an integrated on-axis detector, as well as to evaluate its overall  performance. The following summarizes existing and ongoing studies and provides some insight into the topics feeding into the design optimization \cite{bib:docdb13262,bib:docdb16103}.


\subsection{On-axis Beam Monitoring} 
\label{sec:beam-monitoring}

The \dword{dune} \dword{fd} is on the beam axis in a wideband neutrino beam.  The extraction of neutrino parameters relies on measuring the on-axis oscillation-induced spectral distortion of the beam.  A primary function of the \dword{nd} is to measure the beam prior to oscillations in order to tune the beam model.  Both deliberate and unanticipated changes in the beam spectrum must be tracked.  \dword{numi} experience shows the value of this constant monitoring.  At \dword{numi}, the near detector data was used not only to track changes in the beam necessitating changes in the beam model, but also to help diagnose the reasons for unanticipated changes in the beam spectra (such as target degradation and horn tilt).  This experience informs \dword{dune}, and the constant on-axis monitoring of the beam flux and spectral stability is deemed crucial to achieving the long-term goals of the experiment.

An important element of the \dword{dune} \dword{nd} conceptual design is \dshort{duneprism} where the \dword{ndlar} and \dword{ndgar} take data in off-axis positions, as discussed in Chapter~\ref{ch:prism}. \dword{sand} is the only component of the DUNE ND permanently located on-axis, and can continuously monitor the beam spectrum. The on-axis monitoring is critical, in part, due to the fact that the spectrum on the beam axis is more sensitive to some changes in the beam parameters than that off-axis  \cite{numi-cenf-nd}.  The other reason is that \dshort{duneprism} is dynamic and the constant on-axis monitoring helps ensure the changes in the off-axis flux are due to the movement of the detectors and not changes in the beam itself. 

To be useful as a beam monitor and diagnostic tool, \dword{sand} must be able to monitor the beam spectrum, profile, and event rate in a statistically significant way over a relatively short time frame\footnote{The timescale and significance are defined by the requirements \underline{ND-M8}, \underline{ND-M9} and their derived requirements.}. \dword{sand} can fulfill this role
by using CC neutrino interactions in the upstream barrel ECAL (which provides most of the beam monitoring sensitivity in SAND) as well as in the \dword{3dst} and \dword{stt}. In all cases, charged tracks are momentum analyzed in the low density trackers.

\subsubsection{Impact of Beam Monitoring on Oscillation Results}



The impact of unobserved beam distortions on DUNE's oscillation measurements was gauged by studying the effect of a 3-sigma shift in the current through the LBNF focusing horns\footnote{To be precise, the variation was equivalent to a shift in the horn current that is three times the tolerance.}.  This causes a shift in the mean and the normalization of the neutrino energy spectrum at the on-axis location that is similar to what would be expected from a variety of beam distortions. The distortion has a much smaller effect on the energy spectrum at off-axis locations.

In the study we considered a dataset with a 15 year nominal DUNE exposure that was comprised of 50\% undistorted data collected with the ND-LAr and ND-GAr detectors on-axis and 50\% distorted data collected with the detectors off-axis. The ND data taken off-axis is not included in the analysis. The study assumed that there is no on-axis beam monitor so the distorted data are present in the far detector spectrum but are not corrected for. Three bias conditions were considered, where the beam distortion was applied to FHC only, RHC only and both FHC and RHC fluxes. For each case, an Asimov study was performed, including all nuisance systematic parameters and oscillation parameters of interest, where the true values of the oscillation parameters are set to NuFit4 nominal values~\cite{Nufit:2018}, and a constraint on the value of $\theta_{13}$ (with uncertainties taken from the NuFit4 result) was applied. Normal ordering was assumed. Figure~\ref{fig:no-beam-monitor} shows the best fit oscillation parameter values, and 90\% confidence intervals in the $sin^{2}\theta_{23}$--$\Delta m^{2}_{32}$, $\delta_{\mathrm{CP}}$--$sin^{2}\theta_{23}$ and $\delta_{\mathrm{CP}}$--$\theta_{13}$ planes for the unbiased case, and the three biases considered.

\begin{dunefigure}[Oscillation Parameter Biases Caused by A Lack of Beam Monitoring]{fig:no-beam-monitor}
{The oscillation parameter postfit 90\% confidence contours with true and fake data best fit values. FHC, RHC and both flux changes were assumed and shown in different colors.  The best fit $\chi^{2}$ values are given in the legend.}
    \centering
    \includegraphics[width=0.45\textwidth]{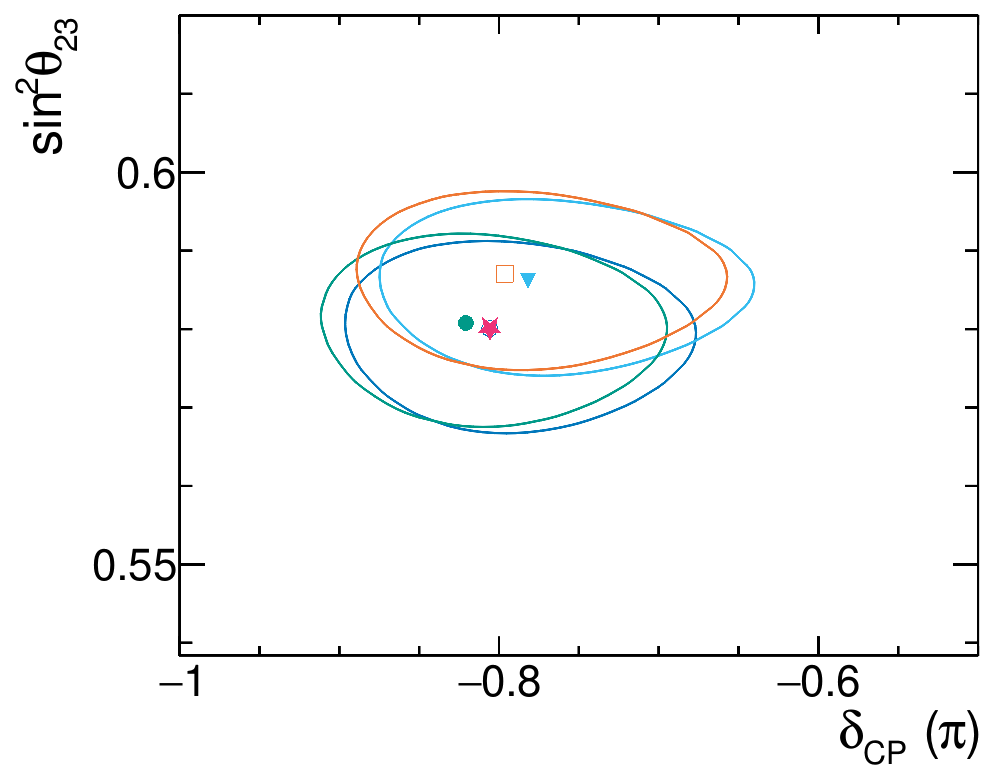}
    \includegraphics[width=0.45\textwidth]{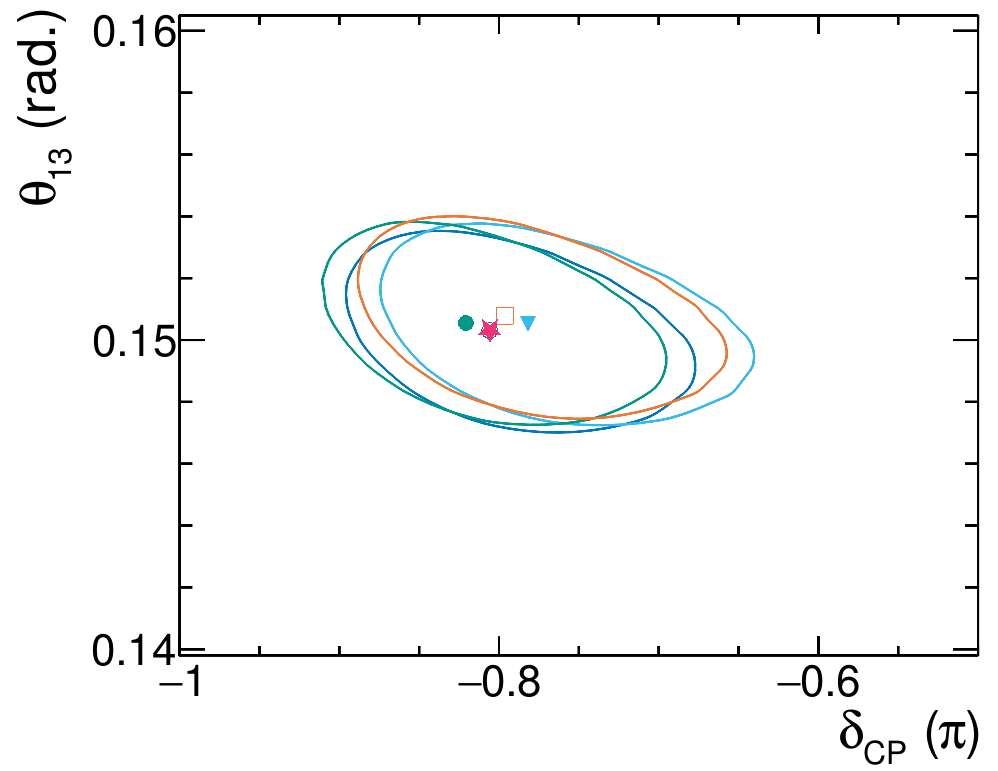}\\
    \includegraphics[width=0.45\textwidth]{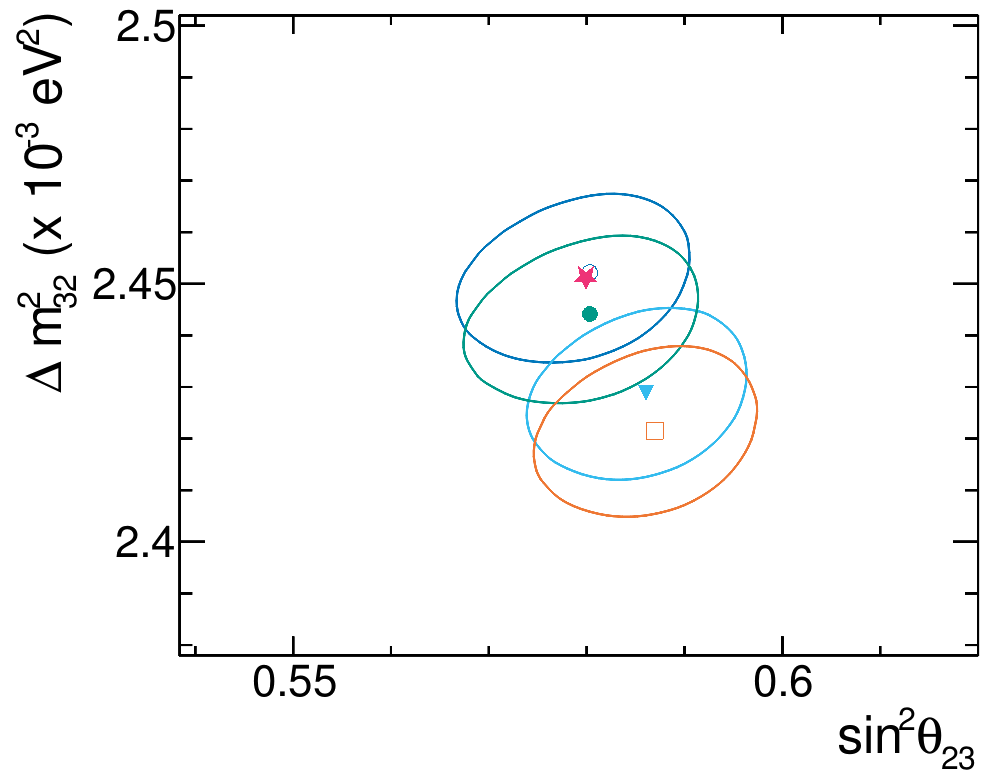}
    \includegraphics[width=0.45\textwidth]{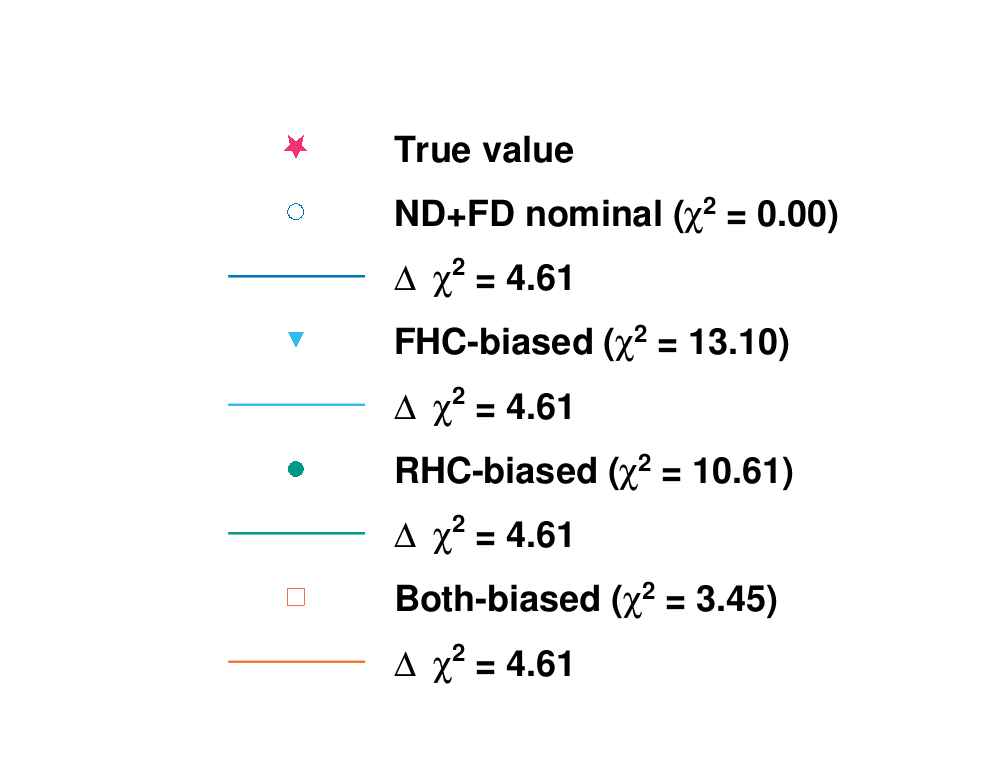}
    
\end{dunefigure}

Large biases are seen in the $\Delta m^{2}_{32}$ and sin$^{2}(2\theta_{23})$ space and this is true for various true oscillation parameter sets. There are noticeable biases in $\theta_{13}$ and $\delta_{CP}$ as well. Without an on-axis beam monitor, the postfit $\chi^{2}$ won’t provide us any obvious sign of a problem in the flux. Therefore, an on-axis beam monitor is necessary to mitigate the similar issues.



\subsubsection{Beam Monitoring with the 3DST+STT/TPC option} 
\label{sec:beam-monitoring-ecal3dst}

The 3DST has sufficient mass that statistically significant samples of \numu CC interactions can be collected over short times.  Although it is a tracking detector, it benefits from being surrounded by low density tracking detectors. 
For neutrinos interacting in the fiducial volume of the 3DST, the charged particles are measured in the 3DST and the low density tracker surrounding the 3DST.  In particular, except for the small fraction that stop in the \dword{3dst}, muons produced in \numu CC interactions can be momentum analyzed with high precision in the tracker.  In the case of the TPCs, for example,  the muon detection acceptance and reconstruction performance is shown in Sec.~\ref{sec:time_pos_ener_meas}. The \dword{stt} option for the low density tracker will have similar acceptance.

Better statistics and improved beam monitoring sensitivity is achieved by including the barrel \dword{ecal} as part of the target fiducial mass. The outermost layer of the ECAL is used as as a veto for external activity. The inclusion of the four inner layers of the ECAL increases the rate of neutrino interactions significantly. 


A study was done in the 3DST+TPCs variant of the reference design to estimate beam monitoring performance. Events were simulated in the 3DST and ECAL using DUNE's standard flux, GENIE, and GEANT versions. In the study, \numu-CC interactions in the ECAL were selected with a requirement that the muon needed to travel for at least \SI{20}{cm} in the \dword{3dst} or in the low-density tracker. There was no specific requirement placed on the hadronic part of the event. The muon energy reconstruction included the smearing in the ECAL, TPC and 3DST regions based on their tracking resolution. The smearing was derived from the Gluckstern formula~\cite{Gluckstern:1963ng} and anchored, in the case of the TPCs, to the performance of the ND280. The hadronic energy was obtained by collecting all the non-muon deposited energy in the sensitive volumes, i.e. 3DST, TPC and ECAL scintillator. The total reconstructed neutrino energy is the sum of the muon reconstructed energy and the collected hadron energy.  Further details of the simulated detector smearing and reconstruction may be found in~\cite{bib:docdb16103}.

Interactions in the 3DST were used to provide an independent sample to the ECAL. The reconstructed neutrino energy was obtained in the same way as described for the ECAL events. Aside from two layers of cubes on each side, all inner cubes are considered as fiducial targets and all interactions in this volume is considered as signals. The sum of the 3DST sample and the ECAL sample was used to study the sensitivity of \dword{sand} to beam variations.


Two input beam spectra were compared for each variation of the beam parameters considered. We simulate the statistics equivalent to one week of data taking ($3.78 \times 10^{19}$ \dshort{pot}) with the nominal beam setting. For each variation of the beam parameters listed in Tab.~\ref{tab:beam-spectrum-sensi-3DST} (which correspond to one standard deviation of the systematics from the beam modeling) we obtain a varied sample by re-weighting the nominal sample using the ratio of the corresponding spectra with respect to the nominal one. The sensitivity to the variations of the beam settings considered is quantified with the corresponding $\Delta \chi^2$ between the two samples and a significance is calculated as $\sqrt{\Delta \chi^2}$. Since the samples are not statistically independent we consider only the statistical uncertainty of the nominal sample.

Table~\ref{tab:beam-spectrum-sensi-3DST} summarizes the beam monitoring capability of this scheme for a number of shifts in the beam parameters. Tab.~\ref{tab:beam-spectrum-sensi-3DST} also shows the performance of a beam monitoring scheme based on four non-magnetized, \SI{7}{ton} modules that measure the neutrino interaction rate, but not the spectrum, at locations 0, 1, 2 and \SI{3}{m} from the beam axis position in the ND hall. This non-magnetized design was investigated as a potentially cost effective option for the beam monitoring role. The table shows that the spectral measurements made by \dword{sand} are more significant.  In addition, the spectra will contain useful diagnostic information on the changes in the beam. This can be seen, for example, in Figure~\ref{fig:bm-significance}. That figure shows the significance of the spectral shift in the reconstructed muon energy for neutrino interactions in the fiducial volume of the \dword{3dst} for representative horn shifts.  Both statistical and detector effects are included.

\begin{dunetable}[Sensitivity to beam changes: 3DST+TPCs option]{c||c|c||c|c|c}{tab:beam-spectrum-sensi-3DST}{The beam parameter description as well as the significance to the observation of a change in the beamline are shown for the 3DST+TPCs configuration. The significances are computed assuming 7-days data taking and considering 
neutrino interactions occurring in both the upstream barrel ECAL and within the 3DST. 
The GENIE 2.12 and edep-sim with GEANT4 v4.10 were used to simulate the neutrino interaction and final state particle energy deposit in the detector system. The neutrino energy reconstruction has been calculated from two contributions, the muon and the hadrons. The muons are required to travel at least 20 cm in either 3DST or TPC. The TPC momentum resolution described in Sec.~\ref{sec:tpc} has been applied. All the hadronic energy deposits are summed calorimetrically. There is a total 2\% rate uncertainty due to the proton number that is applied to the sensitivity calculation with the reconstructed neutrino energy. The sensitivity obtained that is based on the neutrino spectral information is compared to a ``Rate-only'' detector system consisting of four non-magnetized 7-ton modules that measure the beam rate and profile at 0, 1, 2, 3 meters from the on-axis position at the \dword{nd} site. This work is further described in~\cite{bib:docdb16103}.}
\textcolor{white}{ECAL+3DST option} & \multicolumn{2}{c}{Parameter description}    & \multicolumn{3}{c}{Significance, $\sqrt{\chi^2}$} \\\hline
 \rowtitlestyle
		Beam parameter &  Nominal & Changed & Rate-only & FHC & RHC \\\hline \hline
		proton target density & 1.71 g/cm$^{3}$ & 1.74 g/cm$^{3}$ & 0.02  & 8.51 & 5.65  \\\hline
		proton beam width & 2.7 mm & 2.8 mm & 0.02 &  4.67 & 2.93 \\\hline
		proton beam offset x & N/A & +0.45 mm & 0.09 &  2.84 & 1.70  \\\hline
		proton beam $\theta$ & N/A & 0.07 mrad & 0.03 & 0.50 & 0.42 \\\hline
		proton beam ($\theta, \phi)$& N/A & (0.07,1.571)~mrad & 0.00 & 0.51 & 0.35\\\hline
		horn current & 293 kA & 296 kA & 0.2 & 12.64 & 7.97 \\\hline
		water layer thickness & 1~mm & 1.5~mm & 0.5 & 5.30 & 3.20 \\\hline
		decay pipe radius & 2~m & 2.1~m & 0.5 & 7.45 & 4.20 \\\hline
		horn 1 along x & N/A & 0.5 mm & 0.5 & 4.77 & 2.94 \\\hline
		horn 1 along y & N/A & 0.5 mm & 0.1 & 3.53 & 2.27  \\\hline
		horn 2 along x & N/A & 0.5 mm & 0.02 & 0.85 & 0.62 \\\hline
		horn 2 along y & N/A & 0.5 mm & 0.00 & 0.24 &1.81 \\\hline
\end{dunetable}

\begin{dunetable}[Sensitivity to beam changes: STT-only option]{c||c|c||c|c}{tab:beam-spectrum-sensi-STT}{The beam parameter description as well as the significance for the observation of a change in the beamline are shown for the STT-only configuration. The significances are computed by using the neutrino spectral information, by assuming 7-days of data taking and by considering neutrino interactions occurring in both the upstream barrel ECAL and within the STT. 
A complete detector simulation taking into account the fine-grained structure of the electromagnetic calorimeter was implemented and the GENIE 2.12 and edep-sim with GEANT4 v4.10 were used to simulate the neutrino interaction and final state particle energy deposit in the detector system. In addition equivalent FLUKA simulations were performed as well, for a validation of the results at lower energy. A hit-based detector smearing was used for reconstructing the neutrino energy, taking into account the contributions from the muon and the hadrons.  The corresponding values for the ``Rate-only'' detector as in Tab.~\ref{tab:beam-spectrum-sensi-3DST} are also given for comparison. This work is further described in~\cite{bib:docdb13262}.}
\textcolor{white}{ECAL+STT option} & \multicolumn{2}{c}{Parameter description}    & \multicolumn{2}{c}{Significance, $\sqrt{\chi^2}$} \\\hline
\rowtitlestyle
Beam parameter &  Nominal & Changed & Rate-only & ECAL+STT  \\\hline \hline
proton target density & 1.71 g/cm$^{3}$ & 1.74 g/cm$^{3}$ & 0.02  & 4.4\\\hline
proton beam width & 2.7 mm & 2.8 mm & 0.02 & 6.1 \\\hline
proton beam offset x & N/A & +0.45 mm & 0.09 & 4.7 \\\hline
proton beam $\theta$ & N/A & 0.07 mrad & 0.03 & 0.5 \\\hline
proton beam ($\theta, \phi)$& N/A & (0.07,1.571)~mrad & 0.00 & 0.4\\\hline
horn current & 293 kA & 296 kA & 0.2 & 10.3\\\hline
water layer thickness & 1~mm & 1.5~mm & 0.5 & 4.7\\\hline
decay pipe radius & 2~m & 2.1~m & 0.5 & 6.9\\\hline
horn 1 along x & N/A & 0.5 mm & 0.5 & 3.8 \\\hline
horn 1 along y & N/A & 0.5 mm & 0.1 & 4.2 \\\hline
horn 2 along x & N/A & 0.5 mm & 0.02 & 0.5 \\\hline
horn 2 along y & N/A & 0.5 mm & 0.00 & 0.4 \\\hline
\end{dunetable}


Figure~\ref{fig:bm-prism} illustrates the result of a case study of a particularly insidious type of problem where a beam horn shifts while the \dword{ndlar} and \dword{ndgar} are at off-axis positions greater than 6~m.  In this case, the off-axis data would not show a significant beam change and the on-axis \dword{fd} spectrum generated with this data using \dshort{duneprism} analysis techniques (discussed in Chapter~\ref{ch:prism}) would generate a biased oscillation parameter measurement. In this study the first horn is shifted \SI{6}{mm} after \dword{ndlar} and \dword{ndgar} have moved to take data at locations more than \SI{6}{m} off-axis.  Figure~\ref{fig:bm-prism} shows fluxes at the FD corresponding to the oscillation parameters $\sin^2 \theta_{23} = 0.5$ and $\Delta m^2_{32} = 2.52 \times 10^{-3}~\text{eV}^2$. The \dword{fd} flux with shifted beam conditions is shown in blue. The \dword{nd}-to-\dword{fd} extrapolated flux when the shifted beam conditions are observed by SAND, and subsequently corrected, is shown in red. The \dword{nd}-to-\dword{fd} extrapolated flux where the shifted conditions were not observed in the ND is shown in green.  The sizable difference between the green and red curves indicates that the ND would improperly predict the FD if the beam shift was not observed.

 \begin{dunefigure}[Beam monitoring performance case study.]{fig:bm-significance}
{The significance of the spectral shift in the reconstructed neutrino energy on-axis for neutrino interactions in the fiducial volume of the \dword{3dst} for different horn shifts.}
  \includegraphics[width=0.6\textwidth]{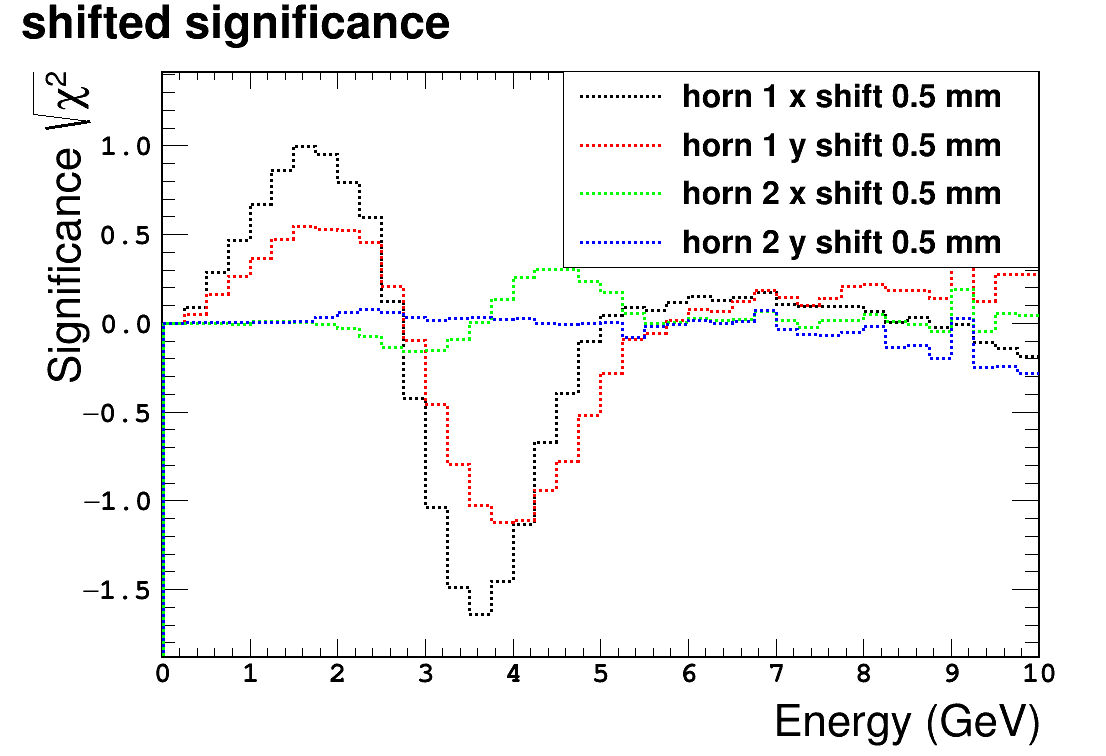}
\end{dunefigure}

 \begin{dunefigure}[Beam monitoring performance case study.]{fig:bm-prism}
{The muon neutrino flux obtained with the \dshort{duneprism} technique is shown after a horn shift when \dword{ndlar} and \dword{ndgar} have moved off-axis as described in the text.}
  \includegraphics[width=0.6\textwidth]{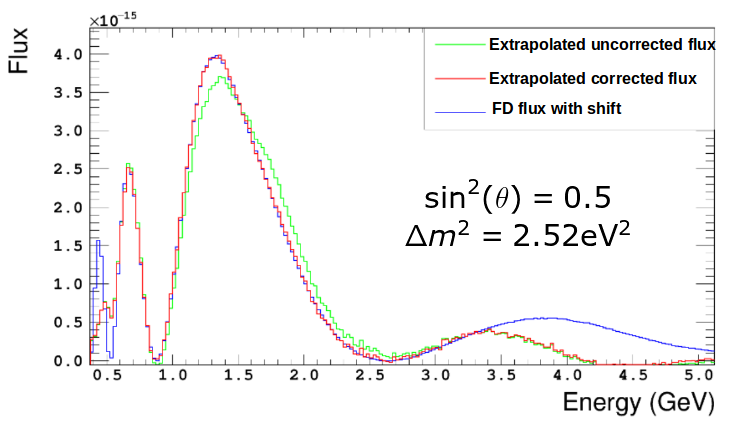}
\end{dunefigure}

The precision on the beam direction required by the DUNE experiment to keep the variation of the neutrino beam flux less than 1\% in all the energy bins is about 0.2~mrad~\cite{Acciarri:2016ooe}.  That corresponds to about 10~cm deviation of the beam center at the 3DST location. With the 3DST spanning more than 2~m width and height, a precision on the beam center position of $\sim 11$~cm can be achieved with one week of data taking.  With two weeks of data taking, the beam center is known to better than 8~cm. 

\subsubsection{Beam Monitoring with the STT-only option} 
\label{sec:beam-monitoring:ECAL} 


\begin{dunefigure}[STT-only configuration.]{fig:SAND-STTonly_BM-XshiftSens}
{Values of $\Delta \chi^2$ as a function of applied shift on the location of the beam axis along the X direction for the STT-only detector configuration. The independent sample from the upstream barrel ECAL and the STT fiducial volume are combined. The results obtained with three different methods are compared: X distribution (magenta curve), single $E_\nu$ distribution (blue curve), and two separate $E_\nu$ distributions for events with $X<0$ and $X>0$ (red curve). The horizontal line corresponds to a significance $\Delta \chi^2 = 9$. See text for details.}
\includegraphics[width=0.65\textwidth]{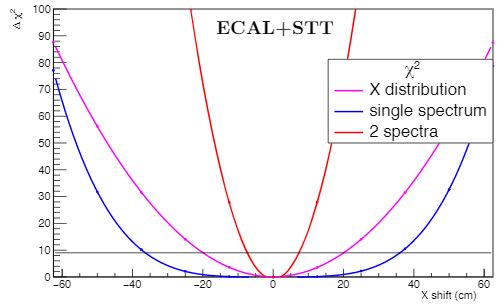}
\end{dunefigure}

The monitoring of variations in both the spectrum and profile of the beam
takes advantage of the large mass (22.8~t) and transverse dimension (up to
$\sim$ 4~m) of the upstream part of the barrel ECAL. Charged tracks exiting
from the ECAL are measured in the STT, resulting in an overall neutrino
energy resolution of about 7\%, to be compared to about 6\% for events from
the STT fiducial volume. Backgrounds from muons originating in the rocks of
the surrounding ND hall and from neutrino interactions in the magnet are
rejected at the level of $7\times 10^{-5}$ with about 95\% efficiency for
ECAL events, including all layers, using a combination of timing and
topological information. Detailed studies were performed to evaluate the sensitivity of the ECAL+STT configuration to the variations of the beam
settings: the values of the $\sqrt{\Delta \chi^2}$ obtained for the combined
ECAL+STT samples corresponding to one week of simulated data taking
($3.78 \times 10^{19}$ pot) are shown in Tab.~\ref{tab:beam-spectrum-sensi-STT}. A significance of $\geq 3$ ($\Delta \chi^2 \geq 9$) was achieved for the detection of most
variations~\cite{bib:docdb13262}.

\subsection{Neutron Detection} 
\label{sec:neutrons}

\subsubsection{Performance of the 3DST+TPCs option}

One goal of the \dword{dune} \dword{nd} is to improve measurements of neutrino interactions by observing particles and kinematic regions not studied at all (or very well) before.  The additional information may provide improved energy reconstruction, improved measurements of transverse kinematic variables, and a path to cross section model improvements.  One of the main advances in this direction for the \dword{nd} reference design is the ability to reconstruct neutrons via \dword{tof} using small energy depositions in plastic scintillator.  As seen in Chapter~\ref{ch:mpd}, neutrons with a kinetic energy above $\sim$50~MeV can be reconstructed in the \dword{ecal} of the \dword{ndgar} detector with a reasonable efficiency.  The target nucleus is argon in this case.  The \dword{3dst} can access neutrons down to a much lower kinetic energy, with an energy resolution that has a smaller tail than what is seen in the \dword{ndgar}.  The target nucleus in this case is usually carbon, but careful selection via transverse kinematic variables can enrich the sample in interactions on hydrogen.


Recently, MINERvA demonstrated the ability to tag neutrino-induced neutrons efficiently in scintillator~\cite{Elkins:2019vmy}. MINERvA's 2-dimensional strip geometry and O(4.5~ns) timing resolution allowed for basic \dword{tof} measurements. However, with fine 3-dimensional granularity and a very good time resolution (see sec.~\ref{sec:3dst}), the \dword{3dst} can detect neutrons with excellent purity and measure their kinetic energy via \dword{tof}.
The lowest threshold to detect a neutron is about \SI{50}{keV}. The analysis threshold will be larger due to the need to reject photon backgrounds. In the present analysis neutron candidates are formed out of clusters with an algorithm that combines adjacent hit cells that are above threshold and requires the cluster to be isolated from charged particle tracks. If at least 0.5~MeV reconstructed deposited energy is required for each neutron candidate cluster the resulting efficiency is 60\%.  If the threshold is lowered to 0.1~MeV, the efficiency goes up to ~75\%. This efficiency is relatively flat across most of the neutron kinetic energy range.

To be able to associate a neutron candidate to a particular event and do a clean reconstruction of its energy on an event-by-event basis, the analysis must handle three primary backgrounds.  The first background comes from neutrons and photons that were generated outside the fiducial volume (out-fiducial) as opposed to those that were generated inside of the fiducial volume.  These come from interactions primarily in the \dword{ecal}, magnet coil, and return yoke of \dword{sand}. This background is mitigated by using a relatively short time window to accept candidates.  Background neutrons that interact near the neutrino vertex typically travel a large distance and this takes time.  Since the \dword{3dst} is fully active, most of the true signal candidates are relatively close in space and in time to the neutrino interaction.  The purity of selecting in-fiducial clusters (mostly neutrons and gammas) is shown in Fig.~\ref{fig:sand-neutron-out-fv-purity} as a function of time and distance from the neutrino vertex.

The second background comes from neutrons generated by interactions of tracks emanating from the neutrino interaction itself, i.e. the secondary interaction background.  This background can be reduced, or eliminated, by removing candidates that fall within a conical volume of a certain size around the flight path of the particles (particularly pions) originating from  the parent neutrino interaction. The purity of the selection can be increased at the cost of efficiency. A third background comes from photons from the primary vertex or from secondary interactions of other particles in the event. Rejecting these backgrounds is under study using a multi-variate techniques.


The resolution for reconstructing neutron energies by TOF is shown for leading neutrons (first cluster in time) in Figure~\ref{fig:sand-neutron-eres}.


\begin{dunefigure}[In-fiducial purity for selecting neutron candidates in the \dword{3dst}.]{fig:sand-neutron-out-fv-purity}
  {The purity for selecting in-fiducial neutron candidate clusters (mostly due to neutrons and photons) as a function of the time and distance from the vertex. The time and vertex can be used in the analysis to tune the purity. The efficiency for detecting the clusters is 60\%, largely independent of time and space. The impurity in this plot is due to the out-of-fiducial neutron and gamma background. This plot was made with a \anumu-CC inclusive sample of events.}
\includegraphics[width=0.7\textwidth]{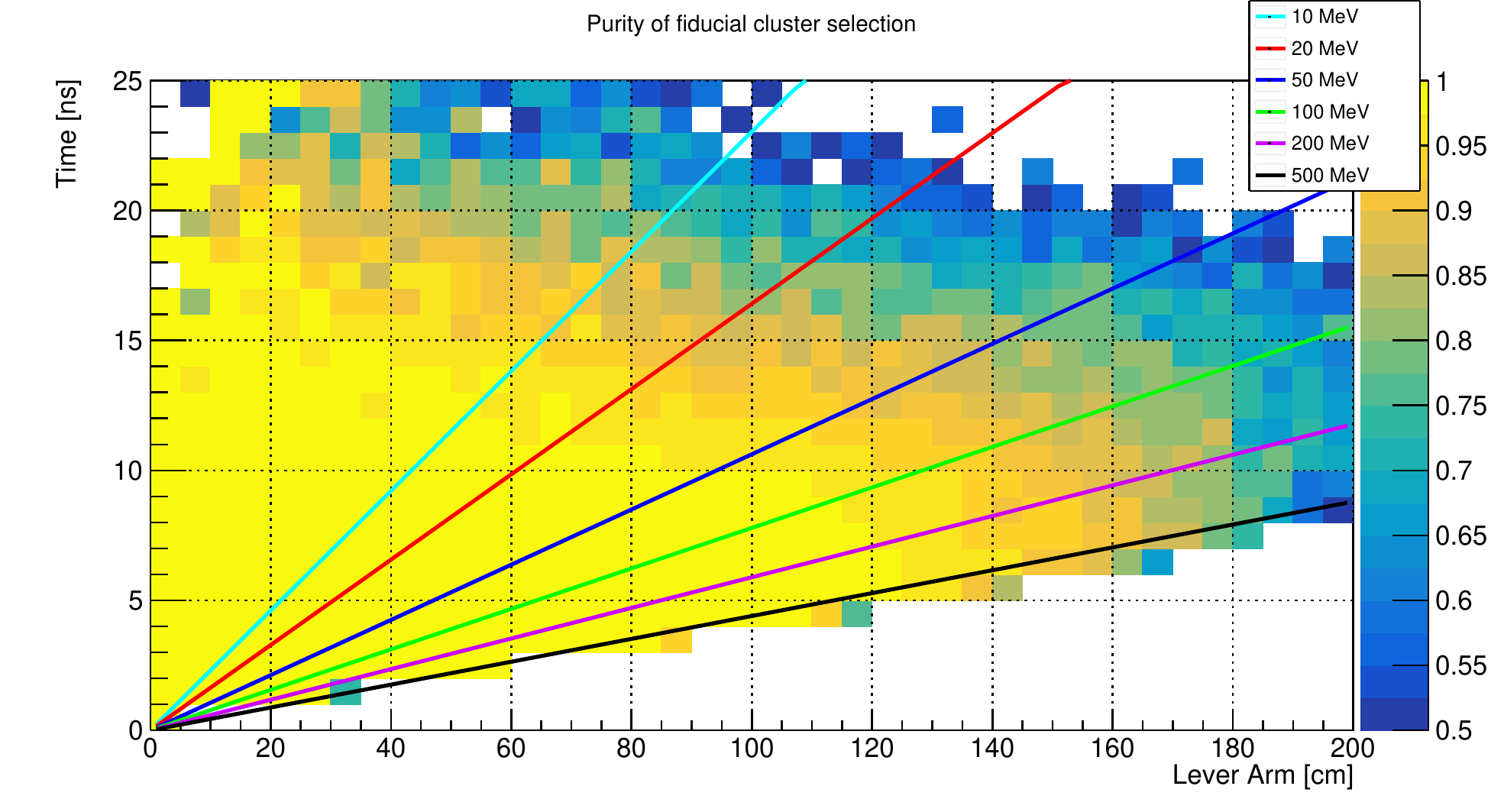}\\
\end{dunefigure}


 \begin{dunefigure}[Energy resolution for neutrons measured by the \dword{3dst}.]{fig:sand-neutron-eres}
{The fractional resolution of the reconstructed neutron kinetic energy for leading (first in time) neutrons created from neutrino events inside the fiducial volume of the 3DST. The axes are the time and distance of the neutron cluster from the neutrino vertex. The lines indicate specific neutron kinetic energies. This plot was made with a \anumu-CC inclusive sample of events.}
\includegraphics[width=0.7\textwidth]{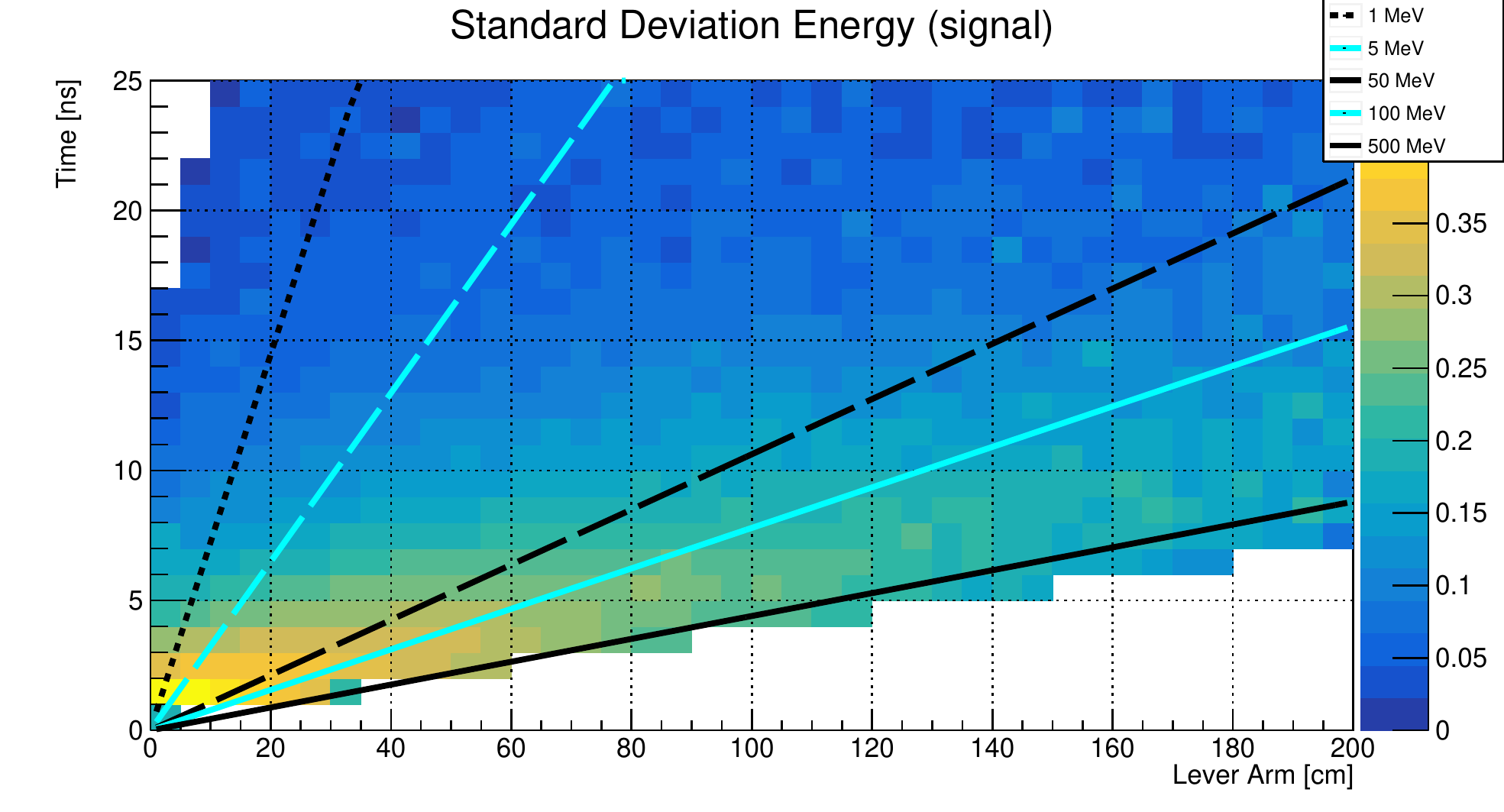}
\end{dunefigure}


\subsubsection{Performance of the STT-only option}
Studies of the neutron detection in \numu (\dword{fhc}) and \anumu (\dword{rhc}) CC interactions were also performed with the alternate design in which the entire magnetic volume is filled with STT. The ECAL has a large plastic fraction and excellent timing resolution (Sec.~\ref{sec:time_pos_ener_meas}) and has a good performance for neutrons. The large volume equipped with polypropylene targets in STT provides additional neutron detection capability, complementary to that of the ECAL. The average neutron reconstruction efficiency in the ECAL is about 55\% with a minimal energy deposited in a cell of 100~keV, rising above 70\% for neutron kinetic energies above 100~MeV\cite{bib:docdb13262}. The average neutron detection efficiency in STT is about 30\% with a minimal energy threshold of 250~eV in the straws. The corresponding combined ECAL+STT average neutron detection efficiency is about 75\%, excluding the double counting of neutrons detected in both ECAL and STT\cite{bib:docdb13262}.

The rejection of backgrounds from random neutrons and photons originating in the materials surrounding the STT fiducial volume as well as from secondary neutrons and photons produced in the STT has been studied. In the selection of $\nu(\bar \nu)$-H interactions (Sec.~\ref{sec:stthydrogen}) the constraints from energy-momentum conservation and from the kinematic analysis, in addition to timing and distance cuts, reduces the background contributions below
1\%.


\subsection{Measurement of \boldmath $\nu(\bar \nu)$-hydrogen Interactions}

\label{sec:sand:Hselection}

\subsubsection{Measurements with the 3DST+TPCs/STT option}


Neutrino interactions in hydrogen can also be reconstructed by using the large, active mass target of the \dword{3dst}. As described in \cite{Munteanu:2019llq,stv-neutron-dolan}, a sample of events enhanced by interactions in hydrogen can be obtained by requiring a small transverse momentum imbalance of the event.  To do this well, it is necessary to incorporate neutrons in the reconstruction, as well as all the charged particles produced in the neutrino interaction. As shown in Section~\ref{sec:neutrons}, the \dword{3dst}
is capable of doing this on an event-by-event basis with a nearly background-free sample.  Moreover the detection efficiency of the active plastic scintillator minimizes the number of neutrons that escape the detector without leaving any visible signal.

A hydrogen-selection purity of about 65\% can be obtained together with an efficiency of about 5\% with respect to all the true $\bar{\nu}_{\mu}$ \dword{cc} interactions without pions in the final state. Due to  the large mass  a large statistics sample can be obtained in the \dword{3dst}. In Figure~\ref{fig:3dst-hydrogen}, the separation between antineutrino interactions in carbon and hydrogen using a detector analogous to the \dword{3dst} is shown for a CC0$\pi$ sample.  The light density tracker downstream of the \dword{3dst} will allow for a good reconstruction of the muon exiting the \dword{3dst}.  For example, with the \dword{tpc} tracker, the typical momentum resolution for the muon is about 2\% (see sec.~\ref{sec:tpc}).

 \begin{dunefigure}[Missing transverse momentum in the\dword{3dst}.]{fig:3dst-hydrogen}
{The NEUT 5.4.0 predicted event rate of antineutrino \dword{cc} with no pions in the final state as a function of the missing transverse momentum in \dword{3dst}. The figure is taken from \cite{Munteanu:2019llq}.}
\includegraphics[width=0.50\textwidth]{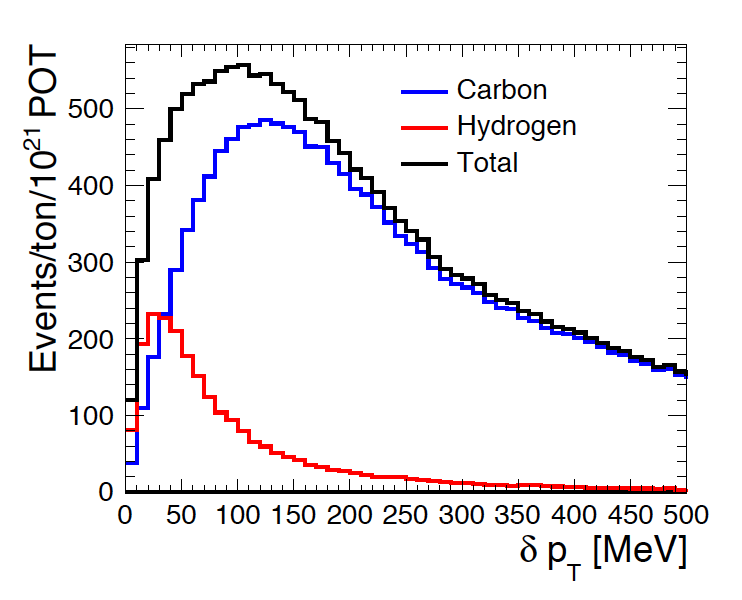}
\end{dunefigure}

\subsubsection{Measurements with the STT-only option}
\label{sec:stthydrogen}

The flexible design of the \dword{stt} offers the opportunity to extract measurements of $\nu(\bar \nu)$ CC interactions on a free proton by the subtraction of samples interacting on the polypropylene (CH$_2$) and graphite (C) targets of the \dword{stt}~\cite{Duyang:2018lpe}.~\footnote{This target and subtraction scheme is referred to in Section~\ref{sec:sand:STTsolidH} as a ''solid" hydrogen target.} One \dword{stt} configuration for this is discussed at the end of Section \ref{sec:sand:STT}. Another scheme under consideration would have the \dword{stt} with targets filling the volume inside the \dword{sand} \dword{ecal} and mentioned in Section~\ref{sec:sand:STTonly}.


The good angular and momentum resolution of the \dword{stt} allows the identification of the  interactions on hydrogen within the CH$_2$ target before subtracting the C background by using a kinematic analysis~\cite{Lu:2015hea,Duyang:2018lpe}. Since the H target is at rest,  CC events are expected to be balanced in a plane transverse to the beam direction (up to the tiny beam divergence) and the muon and hadron vectors are expected to be back-to-back in that plane. Events where the neutrino interacts off the carbon will be affected by both initial and final state nuclear effects, resulting in a significant missing transverse momentum and a smearing of the transverse plane kinematics.

%


The energy-momentum conservation allows the calculation of the energy of any individual particle produced in CC interactions on H using the measured four-momenta of the remaining detected particles. For charged particles the consistency between the calculated and measured energies offers additional discriminating power against interactions on nuclear targets. For neutrons, the energy calculated from energy-momentum conservation can be combined with the measured line of flight to reconstruct the neutron four-momentum vector. Using this technique for $\bar \nu_\mu p \to \mu^+ n$ interactions on H, the muon angular resolution of the STT allows a reconstruction of the energy of the neutrons (detected in either STT or in the ECAL) with a resolution of about 1\%

The kinematic differences described above have been exploited to separate H and C interactions using multi-dimensional likelihood functions. The functions leverage the kinematic differences between scattering off of hydrogen at rest versus a nucleon in a carbon nucleus (Figure~\ref{fig:STT-Hkine})~\cite{Duyang:2018lpe}.

Dedicated analyses allow the selection of all exclusive topologies in both $\nu$-H and $\bar \nu$-H interactions, as well as the corresponding inclusive samples~\cite{Duyang:2018lpe}. Results show that the typical purities of the selected H samples are in the range 80-95\%, with efficiencies of the kinematic selection in the range of 75-96\%, depending upon the specific process considered. The subtraction of the residual C background is entirely data-driven by using the corresponding measurements from the graphite target, automatically including all relevant processes and reconstruction effects.

 \begin{dunefigure}[C-H target separation in the STT.]{fig:STT-Hkine}
{Left plot: Example of kinematic identification of $\nu(\bar \nu)$H interactions 
for the $\nu_\mu p \to \mu^-p\pi^+$ CC topologies 
reconstructed in the \dword{stt}. 
The distributions of the logarithm of the 
likelihood ratio H/C for the H signal, the C background, and the CH$_2$ plastic (sum) are shown. 
Right plot: Efficiency (red color) and purity (blue color) as a function of the kinematic cut for the exclusive processes $\nu_\mu p \to \mu^-p\pi^+$ 
(solid lines) and $\bar \nu_\mu p \to \mu^+p\pi^-$ (dashed-dotted lines) on H~\cite{Duyang:2018lpe}. }
\includegraphics[width=\textwidth]{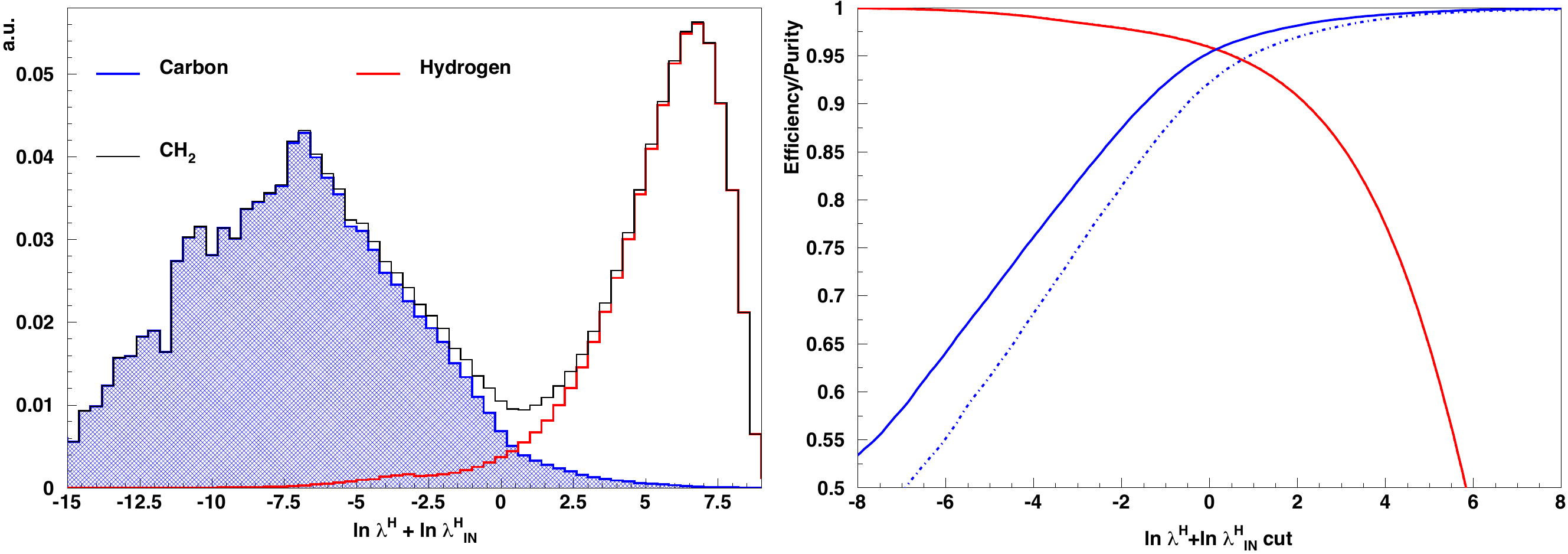}
\end{dunefigure}

Various studies were done to validate the kinematic selection, the impact of 
reconstruction effects and backgrounds using three different event generators,  
GENIE, NuWro, and GiBUU, and a complete detector simulation with both GEANT4 and FLUKA. 
Similar kinematic selections were also successfully demonstrated by NOMAD 
in more severe background conditions (rejections up to $10^5$) in various published 
analyses~\cite{Astier:2001yj,Naumov:2004wa}.

\subsection{Flux Measurements} 
\label{sec:sand:flux}

The \dword{sand} detector will provide measurements of the absolute and relative on-axis flux for the various components of the beam using different physics processes.  These measurements will be complementary to those performed using other components of the \dword{nd}. 
In this Section we briefly summarize the main 
measurements possible with \dword{sand}. 
More details on the techniques can be found in Chapter~\ref{ch:flux}. 

\subsubsection{Measurements made with the 3DST+TPCs/STT option}

\noindent 
\underline{\bf \boldmath $\nu e^- \to \nu e^-$ elastic scattering}.

This purely leptonic process is characterized by a well understood cross-section 
and can provide an accurate measurement of the absolute neutrino flux, as 
well as some limited spectral information. The signal process is not dependent on the target nucleus.  
While \dword{sand} will have substantially lower statistics than \dword{ndlar}, the systematic uncertainties in the measurement will be largely different. There will be some nuclear dependence in backgrounds but they are small
for this process. To the extent there is nuclear dependence in the background, the SAND measurement is a useful systematic crosscheck on the measurement made in \dword{ndlar}. For \dword{sand}, the bulk of the statistics available for this measurement will be in the \dword{3dst}.  Relative to MINERvA, which made this measurement\cite{Valencia:2019mkf}, the \dword{3dst} will have better statistics, and its 3D resolution to short hadron tracks gives it superior ability to reject backgrounds.


\mbox{  }

\noindent 
{\bf \underline{ $\bar \nu_\mu p \to \mu^+ n$ with low transverse momentum imbalance}}. 

In the \dword{3dst}, the $\bar \nu_\mu$ flux can be measured by requiring a low transverse momentum imbalance in $\bar \nu_\mu p \to \mu^+ n$ interactions combined with neutron detection with a precise \dword{tof} measurement. As shown in \cite{Munteanu:2019llq}, 
a resolution of about 5\% on the $\bar \nu_\mu$ energy can be achieved, being almost free of nuclear effects.

\noindent 
\underline{\bf \boldmath Low-$\nu$}.

The relative $\nu_\mu$ and $\bar \nu_\mu$ flux can also be measured by selecting 
inclusive CC interactions with small energy transfer, 
typically $\nu<0.25$ GeV or less, 
in the 3DST (CH). The large target mass available 
provides a sizable number of interactions (Tab.~\ref{tab:3dststats}) for this measurement. 
The uncertainties introduced by the nuclear effects and by the corresponding smearing on the 
reconstructed hadronic energy $\nu$ can be partially mitigated by the improved reconstruction 
of neutrons in \dword{sand} (Sec.~\ref{sec:neutrons}). 
An example is shown in Figure~\ref{fig:3dst-low-nu}.
As the largest fraction of the neutrino energy is taken by the final state muon, the low-density trackers would provide an accurate measurement of the momenta of the charged particles exiting the \dword{3dst}.  
Relative to \dword{ndlar} and \dword{ndgar} when they are on-axis, the different nucleus and detector technology will mean somewhat different systematic uncertainties and corrections that provide a useful crosscheck to the flux shape seen in the other detectors.  

 \begin{dunefigure}[Reconstructed versus true $\nu$ in the\dword{3dst}.]{fig:3dst-low-nu}
{The reconstructed versus true $\nu$ in \dword{3dst}. $\nu$ is defined as the energy carried by all the particles except the charged lepton. The left figure shows the case where neutrons are not detected, while the right figure corresponds to the case where all the neutron energy is reconstructed via \dword{tof}.}
\includegraphics[width=0.35\textwidth]{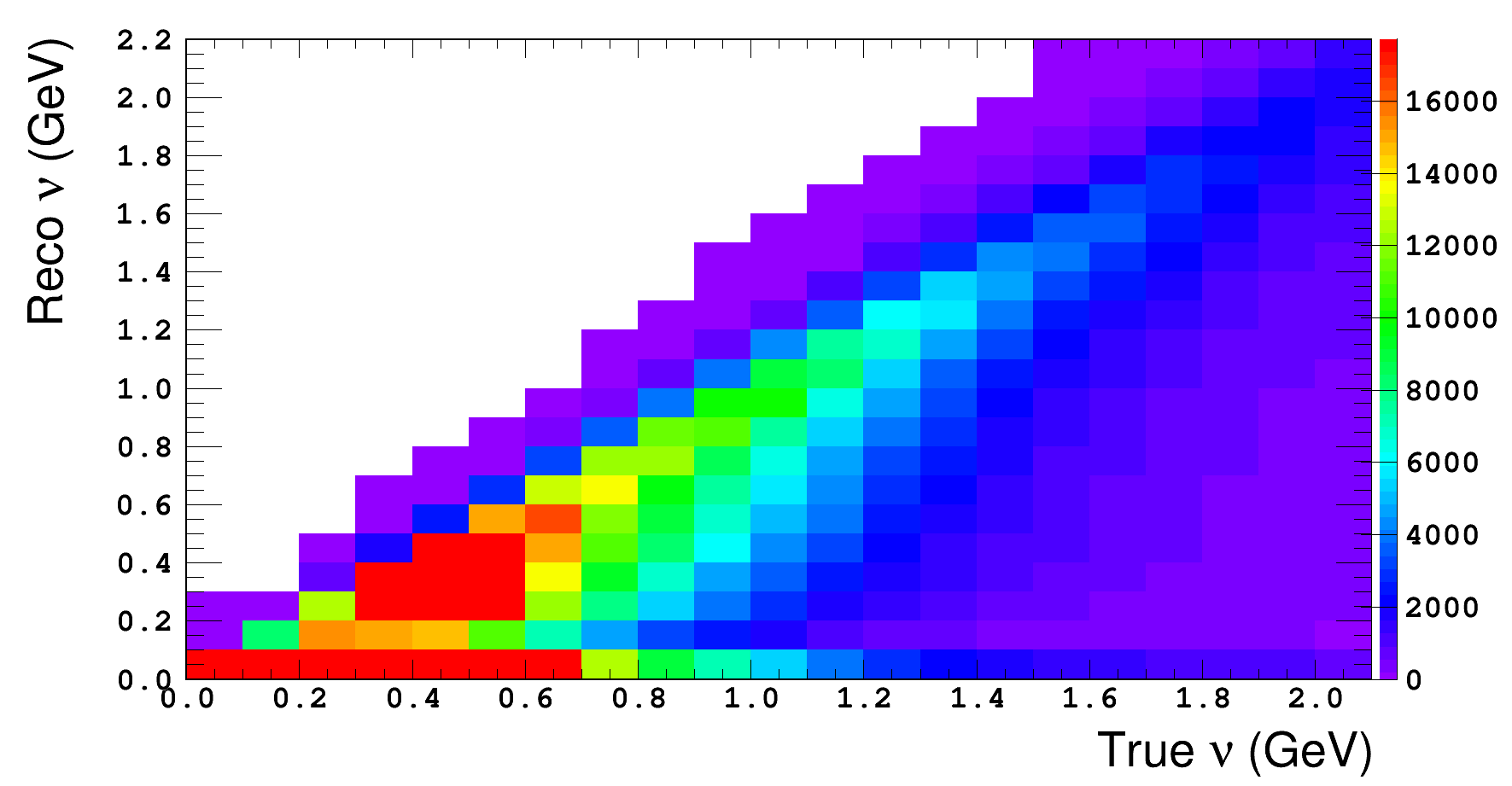}
\includegraphics[width=0.35\textwidth]{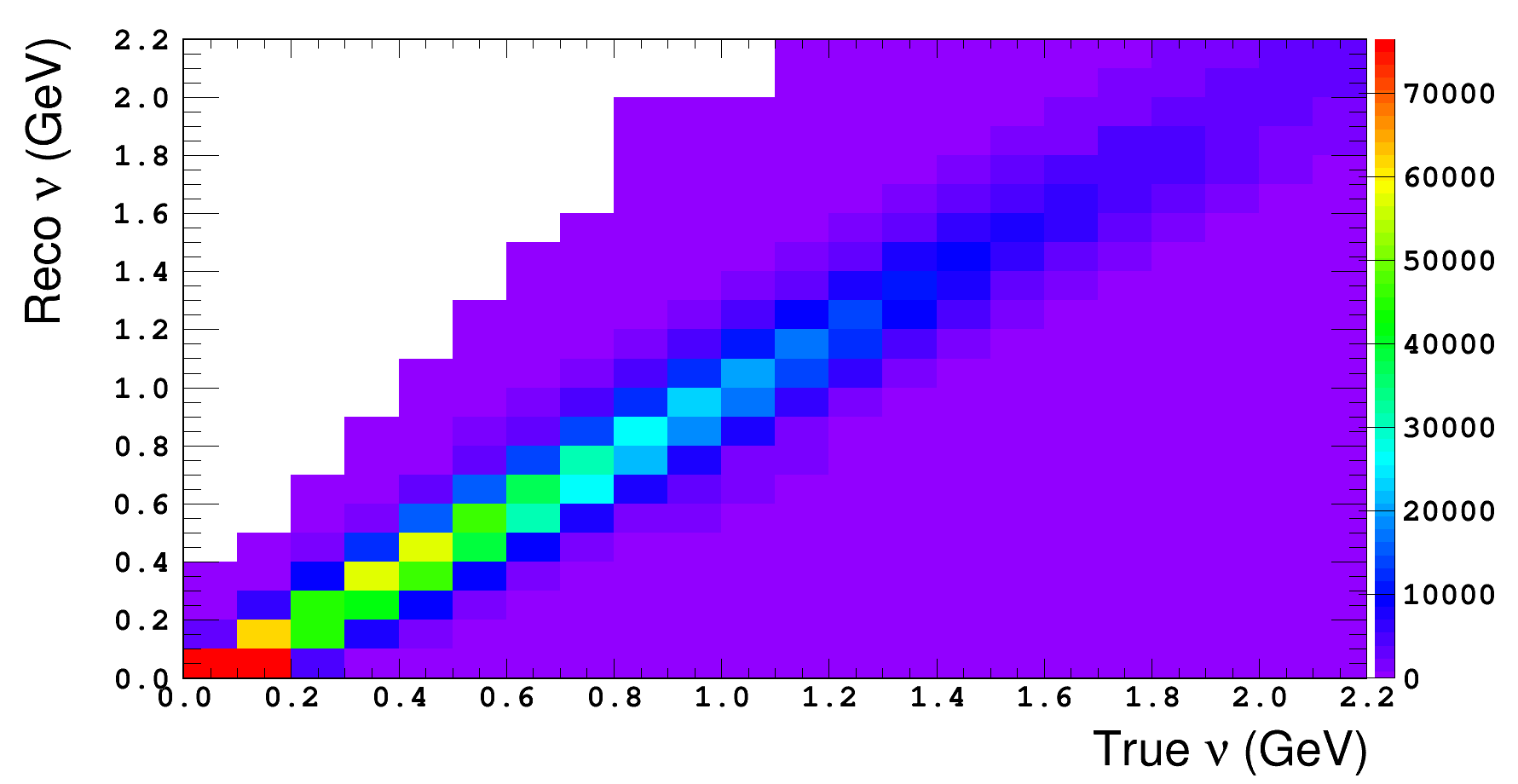}
\end{dunefigure}

\subsubsection{Measurements made with the STT-only option}

\noindent 
\underline{\bf \boldmath $\nu_\mu p \to \mu^- p \pi^+$ and $\bar \nu_\mu p \to \mu^+ n$ on H}. 


The selection of high statistics samples of $\nu(\bar \nu)$-H interactions described in Sec.5.5.3.2 allows an accurate determination of $\nu_\mu$ and $\bar \nu_\mu$ relative flux using exclusive $\nu_\mu p \to \mu^- p \pi^+$, $\bar \nu_\mu p \to \mu^+ p \pi^-$, and $\bar \nu_\mu p \to \mu^+n$ processes on hydrogen with small energy transfer $\nu$~\cite{Duyang:2019prb}. The systematic uncertainties relevant for the flux measurements can be directly constrained using data themselves. Figure~\ref{fig:STT-RelFluxH} shows the expected statistical and systematic uncertainties in the $\nu_\mu$ and $\bar \nu_\mu$ relative fluxes achievable in 5 years with the STT-only configuration.

\begin{dunefigure}[Relative $\nu_\mu$ (left) and $\bar \nu_\mu$ (right) flux
measurements in STT.]{fig:STT-RelFluxH}
{Left panel: expected statistical and systematic uncertainties in the
$\nu_\mu$ relative flux determination using $\nu_\mu p \to \mu^- p \pi^+$
exclusive processes on hydrogen assuming 5 year of data taking with the full
STT detector option in SAND~\cite{Duyang:2019prb}.Right panel: same as
the previous for the $\bar \nu_\mu$ relative flux determination using $\bar
\nu_\mu p \to \mu^+ n$ exclusive processes on hydrogen in
STT~\cite{Duyang:2019prb}.}
\includegraphics[width=0.90\textwidth]{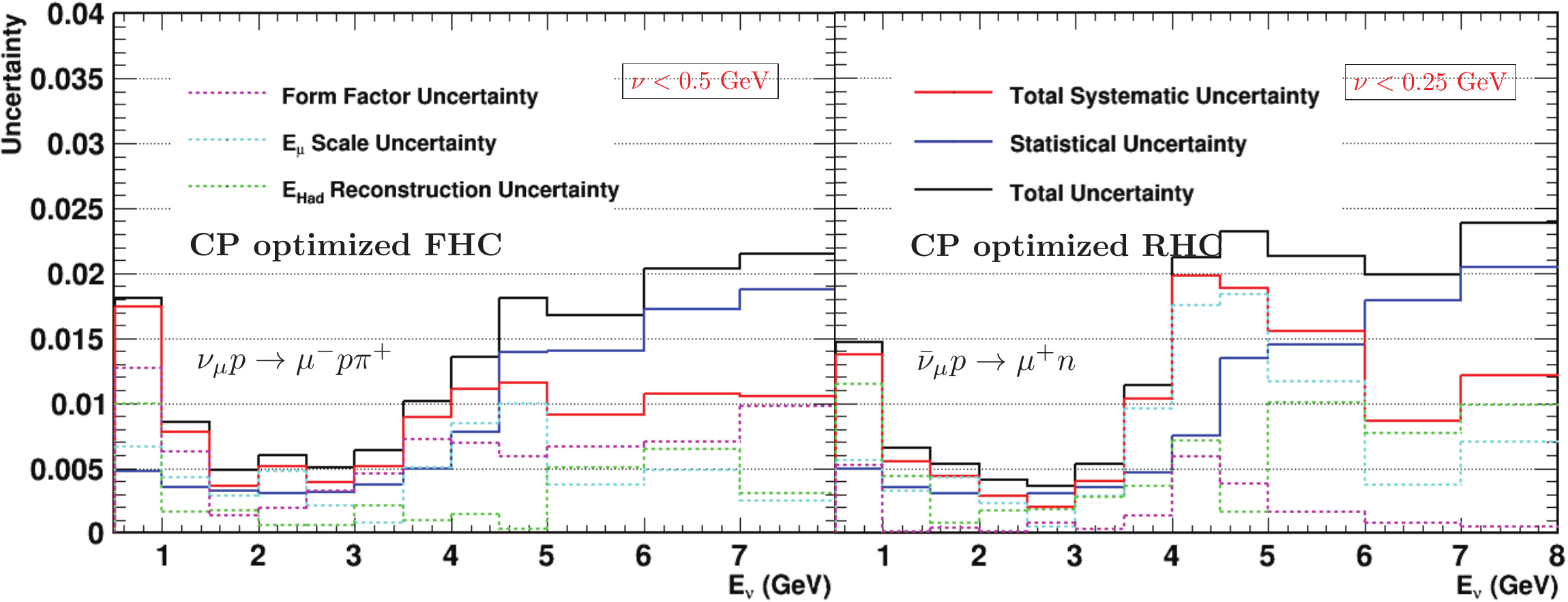}
\end{dunefigure}

Interactions on hydrogen solve the typical problems arising from  nuclear smearing and the small energy transfer reduces the systematic uncertainties on the energy dependence of the cross-sections. The measurement of the $\bar \nu_\mu p \to \mu^+ n$ interactions on H at small momentum transfer $Q^2\leq 0.05$ GeV$^2$ could also provide the absolute $\bar \nu_\mu$ flux, since the corresponding cross-section in the limit $Q\to 0$ is a constant known to high accuracy from neutron $\beta$ decay~\cite{Duyang:2019prb}. 

\underline{\bf \boldmath $\nu e^- \to \nu e^-$ elastic scattering}.

The \dword{sand} configuration with the STT filling the entire magnetic volume offers an angular resolution $\sim$ 1.5 mrad combined with an excellent electron identification from transition radiation, resulting in the selection of 1200 $\nu e^-$ events/year with a total background of about 5\% composed of $\nu_e$ QE interactions without a reconstructed proton (3\%) and $\pi^0$ in NC interactions (2\%)~\cite{bib:docdb13262}. 



\underline{\bf \boldmath Low-$\nu$}.


The option of SAND filled with STT will be able to precisely measure the neutrino and antineutrino flux via the low-$\nu$ technique~\cite{bib:docdb13262} using the large statistics from the CH$_2$ targets. This measurement is complementary to the one with exclusive topologies on hydrogen described above and provides independent samples with different systematic uncertainties.

\subsection{Constraining \boldmath $\nu(\bar \nu)$-Nucleus Cross-sections and Nuclear Effects}  
\label{sec:sand:nu-nucleus}

Though a primary design feature of the \dword{dune} \dword{nd} is to mitigate the complications arising from the nuclear target on the final results by using the same target nucleus in the \dword{nd} and \dword{fd}, the use of an interaction model for various corrections is unavoidable.
The modeling of (anti)neutrino-nucleus 
scattering is particularly challenging 
at the \dword{dune} energies,
since it requires an understanding of complex nuclear 
effects affecting both the initial interaction with the bound nucleon and the re-scattering of the final state particles 
within the nucleus~\cite{Alvarez-Ruso:2017oui}. 
The smearing introduced by nuclear effects 
directly affects the reconstruction of the (anti)neutrino energy. In this context, 
the use of a relatively heavy 
argon target in \dword{dune} implies a larger nuclear smearing from a nucleus experimentally less known than the more commonly measured hydrocarbons.   Measurements from the \dword{sbn} program will be helpful to understand effects in argon, but are at the low end of the energy range relevant for \dword{dune}. Both the \dword{ndlar} and \dword{ndgar} components of the \dword{nd} will have active programs working on the argon cross section model in the \dword{dune} beam.

The study of neutrino interactions on the CH of the \dword{3dst} in \dword{sand} will be useful for a number of reasons.  The data on additional nuclei may help constrain models of nuclear effects.  Although various models describing carbon and argon interactions within the same physics framework exist~\cite{Amaro:2019zos}, the current neutrino generators still have limited predictive power. While disagreements between generators and data~\cite{Alvarez-Ruso:2017oui} can be accommodated in a number of different ways using a single nuclear target, the availability of a different nucleus, i.e. carbon, can help to resolve among them. An example is given in Ref.~\cite{bib:docdb16058}. 

The event-by-event addition of neutrons in the reconstruction provides information, not measured directly otherwise, that can improve the performance of transverse variable measurements which may prove useful for the evolution of nuclear models~\cite{Abe:2019whr,bib:docdb13262}.  Insights into the interaction model on carbon may be useful for the argon model as well.  As an example of this, multi-nucleon effects were observed and initially modeled using carbon data and they are an important component of the modeling of neutrino interactions on argon. 

Finally, at the start of the \dword{dune} program, the connection of \dword{dune} data to the large catalog of existing data on carbon from DUNE precursor experiments provides an early check for problems and surprises. Similar cross-checks may be particularly valuable when comparing \dword{dune} measurements to those coming from Hyper-K.

The \dword{sand} options with the STT can provide a pure C target (graphite). Kinematic analysis as described above
can  be used to select a relatively pure C sample from the large CH$_2$ targets in 
STT (Fig.~\ref{fig:STT-Hkine}). In addition, in the future, \dword{sand} options offer the opportunity 
to study a broad range of other nuclear targets, which can replace any of the default CH$_2$ and C 
targets, if desired. 

As described above in Section~\ref{sec:stthydrogen} the STT design enables the selection 
of high statistics samples of all the exclusive topologies, 
as well as of the corresponding inclusive samples, in 
$\nu(\bar \nu)$-H CC interactions. 
These CC interactions on hydrogen will provide insights on the structure 
of the free nucleon~\cite{bib:docdb13262,ESGprop}. 
In addition, a comparison of the measurements in H and in the graphite targets 
available within the same STT detector (with similar acceptance) may provide information useful for constraining nuclear effects and the corresponding systematic 
uncertainties~\cite{bib:docdb13262}. A similar measurement can be made with the proposed thin liquid argon target (Sec.~\ref{sec:meniscus}) to constrain 
the nuclear smearing introduced by the use of the argon nucleus in the FD.

\subsection{External Backgrounds} 
\label{sec:bkg}
The backgrounds at the \dword{nd} site arise mainly from cosmic radiation, ambient radioactivity, and beam-related neutrino interactions in the material surrounding the detector. The first two background sources can be  suppressed to negligible levels by requiring a time coincidence with the beam spill. The third one is more critical because of the smaller detector mass with respect to the external mechanical structure.
 
The expected rates of inclusive ($\nu_\mu$+$\bar \nu_\mu$+$\nu_e$+$\bar \nu_e$) CC+NC interactions are 0.135/ton/spill in the \dword{fhc} beam and 0.072/ton/spill in the \dword{rhc} beam, resulting in a total number of events in \dword{sand} of about 84 events/spill (\dword{fhc}) and 45 events/spill (\dword{rhc}), respectively. 

\subsubsection{Performance of the 3DST+TPCs/STT option}

A study of the beam-related neutrino induced background was made assuming the  \dword{sand} internal volume filled with a \dword{3dst} detector ($\sim 10.7$ t total mass) and an \dword{stt} tracking system. Inclusive $\nu_\mu$-CC interactions were simulated throughout the magnet, the ECAL and the trackers in \dword{sand}. A reduction of by a factor of $\sim 1.3\times 10^{-4}$  of the background from external interactions was achieved by a combination of timing information and topological cuts using both the calorimeter and the inner trackers. 

\subsubsection{Performance of the STT-only option}

Studies of the external backgrounds originating from interactions in the SAND magnet and ECAL were performed with the STT-only configuration. Different discriminant variables describing timing and topological information in both ECAL and STT are combined into a multivariate analysis. Overall a combined rejection factor of $3\times 10^{-5}$ against CC+NC external background was obtained, witn an efficiency of 92.7\% and a purity of 99.6\%~\cite{bib:docdb13262}. Similar results are obtained from a simple cut-based analysis. This analysis refers to a generic event selection without using reconstruction information. The selection of specific event topology and/or particle types will further enhance the rejection of external backgrounds.





\cleardoublepage

\chapter{Measurements of Flux and Cross Sections}
\label{ch:flux}

The \dword{dune} \dword{fd} will not measure the neutrino oscillation probability directly. Instead, it will measure the neutrino interaction rate for different neutrino flavors as a function of the reconstructed neutrino energy. It is useful to formalize the measurements that are performed in the near and  far detector modules in the following equations:

\begin{align}
\label{eq:fdrate}
\frac{dN^{FD}_{x}}{dE_{rec}}(E_{rec}) & = 
\int \Phi^{FD}_{\numu}(E_\nu)P_{\numu\rightarrow x}(E_\nu)\sigma^{Ar}_x(E_\nu)T^{FD,Ar}_x(E_\nu,E_{rec})dE_\nu\\
\frac{dN^{ND}_{x}}{dE_{rec}}(E_{rec}) & = 
\int \Phi^{ND}_{x}(E_\nu)\sigma^m_x(E_\nu)T^{d,m}_x(E_\nu,E_{rec})dE_\nu\
\end{align}
with
\begin{itemize}
    \item $x$  = \nue , \numu 
   \item $d$  = \mbox{detector index} (FD, ND\footnote{Generally, the ND index refers to ND-LAr, but could be ND-Gar or SAND, depending on the measurement.}) 
   \item $m$  = \mbox{interaction target/material, (e.g., H, C, or Ar)}
   \item $E_\nu$  = \mbox{true neutrino energy}
   \item $E_{rec}$  = \mbox{reconstructed neutrino energy} 
   \item $T^{d,m}_x(E_\nu,E_{rec})$  = \mbox{true to reconstruction transfer function} 
   \item $\sigma^m_x(E_\nu)$  = \mbox{neutrino interaction cross section} 
   \item $\Phi^{d}_x(E_\mu)$  = \mbox{un-oscillated neutrino flux} 
   \item $P_{\numu\rightarrow x}(E_\nu)$ = oscillation probability
   \item $\frac{dN^{d}_{x}}{dE_{rec}}(E_{rec})$  = \mbox{measured differential event rate per target (nucleus/electron)} 
\end{itemize}

There are equivalent formulae for antineutrinos. For simplicity, the instrumental backgrounds (wrongly selected events) and the intrinsic beam contaminations (\nue interactions in case of the appearance measurement) have been ignored. But an important function of the  \dword{nd} is also to quantify and characterize those backgrounds.

It is not possible to constrain well the \dword{fd} neutrino flux directly, but the near-to-far flux ratio is constrained by existing hadron production data and the beamline optics. As such, Equation~\ref{eq:fdrate} can be rewritten as 

\begin{align}
\frac{dN^{FD}_{x}}{dE_{rec}}(E_{rec}) & = 
\int \Phi^{ND}_{\numu}(E_\nu)R(E_\nu)P_{\numu\rightarrow x}(E_\nu)\sigma^{Ar}_x(E_\nu)T^{d,Ar}_x(E_\nu,E_{rec})dE_\nu\\
\end{align}
with
\begin{align}R(E_\nu) = \frac{\Phi^{FD}_{\numu}(E_\nu)}{\Phi^{ND}_{\numu}(E_\nu)}
\end{align}
taken from the beam simulation. 
It is not possible to measure only a near-to-far event ratio and extract the oscillation probability since many effects do not cancel trivially.  This is due to the non-diagonal true-to-reconstruction matrix, which not only depends on the underlying differential cross section, but also on the detector used to measure a specific reaction, expressed as 
\begin{align}
\frac{dN^{FD}_{x}}{dE_{rec}}(E_{rec})/{\frac{dN^{ND}_{\numu}}{dE_{rec}}(E_{rec})} & \neq  R(E_\nu)P_{\numu\rightarrow x}(E_\nu)\frac{\sigma^{Ar}_x(E_\nu)}{\sigma^{m}_{\numu}(E_\nu)}.
\end{align}
It is therefore important that the \dword{dune} \dword{nd} suite constrain as many components as possible. In other words, the DUNE near detector should provide data that allows, to the extent possible, the deconvolution of effects from the beam, interaction cross section, and detector response.  This  requirement drives much of the DUNE \dword{nd} design concept.

This chapter deals with two of the three convolved elements contributing to the event rate, the beam (flux) and the interaction cross section.
It begins in Section~\ref{sec:ch6-fluxbusiness} with a discussion of issues surrounding the determination of the flux from the beam simulation.   Section~\ref{sec:ch6-methods} provides an overview of many of the more important and common techniques for measuring the flux.  A reminder of the importance of cross section measurements for \dword{dune} is given in  Section~\ref{ch_xsec:sec_xsec}.   Section~\ref{ch:nu-xsec:interact} provides a survey of  the types of neutrino-nucleus interactions that dominate in the \dword{dune} energy range, and some discussion of how the \dword{nd} system's reconstruction will facilitate investigation of these interactions.  In Section~\ref{ch_xsec:sec_nuclei}, issues and challenges surrounding the scattering of neutrinos from nuclei are discussed.  Finally, a few case studies are presented in Section~\ref{ch:nu-xsec:nd-xsec} that illustrate aspects of the \dword{dune} \dword{nd}'s  power in this area.

\section{Flux prediction from beam simulation}
\label{sec:ch6-fluxbusiness}

It is not enough to measure the flux in the beam at the near detector.  The beam needs to be modeled as well as possible because the flux has to be predicted at different locations, including those where measurements cannot be made.  In particular, the flux model is the basis for the prediction of the event rate at the far detector.  It provides the expected spectrum of neutrinos that is modified by the oscillation model and then used in concert with the cross section and detector response models to predict the event rate.
The beam model is also used to evaluate the systematic uncertainties associated with the beam. To do this, elements of the beam model (positions of the horns and the currents in them, target geometry, etc.) can be shifted in the model to see the effect in the final results.  Finally, the beam model can be used as a helpful tool for diagnosing things that are not understood in the beam.  There have been instances in NuMI, for example, where a change in the beam spectrum over time was diagnosed and understood via the beam model before the offending beam elements could be pulled from the beamline and autopsied \cite{Bishai:2012kta}.

Neutrino beam fluxes are notoriously difficult to model well.  The state-of-the-art is described in references \cite{Aliaga:2016oaz} and \cite{Abe:2012av}, where the models and measurements are described for the NuMI  and T2K beams, respectively. While the near-to-far flux ratio is tightly constrained to the level of \SIrange{1}{2}{\%}, the same is not true for the absolute flux itself. \dword{t2k}, using hadron production data obtained from a replica target, can constrain the absolute flux at the  \dword{nd} to \SIrange{5}{6}{\%} in the peak region and to around 10\% in most of its energy range. The \dword{numi} beam has been constrained to 8\% using a suite of thin target hadron production data. 

The overall beam model must incorporate the following elements: proton beam, target geometry, a prediction for the hadron production (particle species, direction, energy, rate) from the target, geometries and electromagnetic structure of the focusing system, geometry of the decay and beam dump regions, and a model for the decay of the hadrons produced from the target.  Mistakes in any of these elements of the model can affect the predicted neutrino spectrum.  Uncertainties in the modeling show up, to varying degrees, as uncertainties in the predicted neutrino spectrum. 

A significant recent advance in the precision of beam models has come from improvements in the knowledge of the expected hadron spectrum and distribution produced from the target, which is used as input to the beam model.  These improvements have been enabled by systematic efforts to measure hadron production (in other experiments) using beams and targets that are similar to those used in the relevant neutrino beams \cite{Vladisavljevic:2018prd}\cite{Paley:2014rpb}.  

A separate and significant advance in the precision of beam models has come from tuning the output of the models to agree with the neutrino event spectrum as measured in a near detector in the given beam, i.e., a detector so close to the neutrino production source that (Standard Model) oscillations have no significant effect on the neutrino flux.  Since it is based on the events observed in the near detector, this tuning involves aspects of the beam model, cross section model, and the detector response model.

For DUNE to achieve the most precise and accurate beam model, the hadron data used as input to the beam model should be based on a  measurement of the hadron production from a target  system that is as similar as possible to the one implemented in the experiment.  If possible, it would be even better if the hadron production could be measured in a test beam including a set of replica horns, or in situ in the actual LBNF beam.  This would provide a measure of the hadrons after production off the target and propagation through the horns. It would constitute the most robust input possible for the model.  Such measurements are not in the scope of the DUNE project, but are important supporting experiments that might be considered.

DUNE makes use of the PRISM technique described in Chapter~\ref{ch:prism} to reduce the overall model dependence in the final results.  This technique involves making a linear combination of predicted \dword{nd} fluxes to mimic the expected oscillated \dword{fd} flux. If successful, this technique should reduce the dependence on the interaction model and some detector effects by effectively removing the spectral differences between the analyzed beam at the near and the far detectors.  It is important to note that PRISM depends on the beam model that predicts the fluxes at different off-axis angles and provides the basis set for the linear combination analysis. The beam model and the event rate and flux measurements used to tune and diagnose changes in the beam are critical to the success of PRISM.

\section{Flux measurements}
\label{sec:ch6-methods}
The process of extracting the incident neutrino flux from the data benefits from the use of multiple techniques, along with the variation of experimental and theoretical strengths and weaknesses that implies.  The flux measurement also benefits from constraints on the other things that affect the event rate, such as cross sections, detector effects, etc. 
Several of the most important techniques of constraining the flux are discussed below.


\subsection{Inclusive muon neutrino CC interactions}
\label{ssec:numu}
Reconstruction of the neutrino event spectrum from the high statistics inclusive CC \numu sample is among the first things most experiments do.  This sample is statistically rich compared to others being discussed in this section.  In fact, in modern long baseline oscillation experiments, the statistical error on such a sample is very small compared to systematic effects.  In T2K, this sample is not used to measure the flux as much as it is used to constrain the parameters in the flux along with the beam and cross section models \cite{Abe:2018wpn}.  It is through this constraining of parameters that the near detector information provides constraint to the FD oscillation analysis.  For NOvA, with similar near and far detectors, the CC \numu event sample in the \dword{nd} is used to predict the event rate in the \dword{fd} and extract the oscillation signal \cite{acero:2019ksn}.  This technique minimizes the uncertainty stemming from detector effects in the extrapolation because these effects cancel out, to some extent, between the near and far samples.  That said, the extrapolation to the \dword{fd} makes use of the flux, beam, and cross section models as constrained in the \dword{nd}.  

The weakness in using the CC \numu sample as it is used in these experiments, is that the event rate convolves flux, cross section, beam, and detector effects.  It works, but to achieve smaller uncertainties it is important to constrain individual elements of this complicated convolution to some extent otherwise.  For example, the input flux model has smaller uncertainties if the external hadron production model for the beam simulation is constrained with quality data; the flux shape and normalization is constrained rather cleanly using other flux measurement techniques (described below) that are largely independent of nuclear and cross section uncertainties;  and cross sections are constrained using more exclusive analyses. 

\subsection{Neutrino-Electron Elastic Scattering}
\label{ssec:e-nu-scatt}

Neutrino-electron scattering ($\nu \ e \rightarrow \nu \ e$) is a pure electroweak process with a calculable cross section at tree level. The cross section is flavor dependent since the $\nu_{e}$ scatters through both NC and CC processes.  This is well understood and the effect is small since the scattering signal is dominated by $\nu_{\mu}$ NC interactions.  The signal is independent of nuclear effects and uncertainties in the cross section.  The background does not share this simplicity, but it is small.  The final state consists of a single electron, subject to the kinematic constraint 

\begin{equation}
1 - \cos \theta = \frac{m_{e}(1-y)}{E_{e}},
\end{equation}

where $\theta$ is the angle between the electron and incoming neutrino, $m_{e}$ and $E_{e}$ are the electron mass and total energy, respectively, and $y$ 
is the fraction of the neutrino energy transferred to the electron. For \dword{dune} energies, $E_{e} \gg m_{e}$, and the angle $\theta$ is very small, such that $E_{e}\theta^{2} < 2m_{e}$. 

The overall flux normalization can be determined by counting $\nu \ e \rightarrow \nu \ e$ events. Events can be identified by searching for a single electromagnetic shower with no other visible particles. Backgrounds from $\nu_{e}$ \dword{cc} scattering can be rejected by looking for large energy deposits near the interaction vertex, which are evidence of nuclear breakup. Photon-induced showers from \dword{nc} $\pi^{0}$ events can be distinguished from electrons by the energy profile at the start of the track. The dominant background is expected to be $\nu_{e}$ \dword{cc} scattering at very low $Q^{2}$, where final-state hadrons are below threshold, and $E_{e}\theta^{2}$ happens to be small. The background rate can be constrained with a control sample at higher $E_{e}\theta^{2}$, but the shape extrapolation to $E_{e}\theta^{2} \rightarrow 0$ is uncertain at the \SIrange{10}{20}{\%} level.

For the \dword{dune} flux, approximately \num{100} events per year per ton of fiducial mass are expected with electron energy above \SI{0.5}{GeV}. For a \dword{lartpc} of fiducial mass of 60 tons (e.g. \dword{ndlar}), this corresponds to $\sim$\num{6000} events per year. The statistical uncertainty on the flux normalization from this technique is expected to be $\sim$1\%. \dword{minerva} has achieved a systematic uncertainty just under 2\% and it seems plausible that \dune could do at least as well \cite{Valencia:2019mkf}.  The performance of \dword{ndlar} for this measurement is discussed in Sec.~\ref{sec:lartpc-nu-e-scattering}. 

\dword{sand} will contain hydrocarbon targets that can also do this measurement with significant  statistics and with detector and reconstruction systematics largely uncorrelated with \dword{ndlar}.  
The signal is independent of the atomic number $A$ and the background is small; so, this sample can provide a good cross-check of the results seen in the \dword{ndlar}.  As an example, the performance of a MINERvA-like scintillator detector 
is shown in Figure~\ref{fig:nominal_det_constraint}.


\subsection{Scattering With Low Energy Transfer To The Hadronic System }
\label{ssec:intro-low-nu}
The inclusive cross section for \dword{cc} scattering $(\nu_l+N\rightarrow l^-+X)$ does not depend on the neutrino energy in the limit where the energy transfered to the nucleus $\nu = E_\nu-E_{l} $ is zero~\cite{bib:original_lownu}.  In that limit, the event rate is proportional to the flux, and by measuring the rate as a function of energy, one can get the flux ``shape.'' This measurement has been used in previous experiments and has the potential to provide a constraint in \dune with a statistical uncertainty $<1\%$.
In practice, one cannot measure the rate at $\nu=0$. Instead it is necessary to restrict $\nu$ to be less than a few \SI{100}{MeV}.  This introduces a relatively small $E_\nu$ dependence into the cross section that must be accounted for to obtain the flux shape. Thus the measurement technique depends on the cross section model but the uncertainty is manageable~\cite{bib:bodek_lownu}. This is particularly true if low-energy protons and neutrons produced in the neutrino interaction can be detected. 

\subsection{Measurements using neutrino-hydrogen interactions}
Studies have been done looking at the use of transverse momentum balance and exclusive state reconstruction to isolate samples of events enriched in interactions on hydrogen within a hydrocarbon target.  Exclusive states considered include $\nu_\mu p \to \mu^-p\pi^+$, $\bar \nu_\mu p \to \mu^+p\pi^-$ and $\bar \nu_\mu p \to \mu^+ n$ \cite{Munteanu:2019llq}\cite{Duyang:2019prb}.  These analyses are expected to yield samples relatively free from nuclear effects due to the enrichment in interactions on hydrogen, and exhibit an improved energy resolution due the reconstruction of simple exclusive states and the use of transverse momentum balance. This measurement in \dword{sand} is described in Chapter~\ref{ch:sand}.   


\subsection{Intrinsic Electron Neutrino Flux}
\label{ssec:beam-nue}
Electron neutrinos in a wideband beam come from two primary sources: kaon decays and muon decays. These ``beam'' \nue are an irreducible background in $\numu \to \nue$ oscillation searches. As such, the \dword{lbnf} beam was optimized to make the \nue flux as small as possible while maximizing the \numu flux. The production of $\pi^{\circ}$'s and, at low energy, charged pion-electron confusion, can lead to backgrounds that are difficult to remove completely. In the energy range relevant for oscillations (\SI{0.5}{GeV} - \SI{4.0}{GeV}) the predicted $\nue/\numu$ ratio varies between 0.5\% and 1.2\% as a function of energy. The beam \nue flux in the same energy range is strongly correlated with the \numu flux due to the decay chain $\pi^+\to\mu^+\numu$ followed by $\mu^+ \to \anumu{} e^+ \nue $ (and likewise for \anue). As a result, the \dword{lbnf}  beam simulation predicts that the uncertainty on the $\nue/\numu$ ratio varies from \SIrange{2.0}{4.5}{\%}. At the  \dword{fd}, in a 3.5 year run, the statistical uncertainty on the beam \nue component is expected to be 7\% for the $\nu$ mode beam and 10\% for the $\bar{\nu}$ mode beam. The systematic uncertainty on the beam \nue flux is therefore expected to be subdominant, but not negligible.


\section{The importance of cross section measurements}
\label{ch_xsec:sec_xsec}

As discussed at the start of this chapter, the measured event rates at the \dword{fd} are a product of convolved flux, cross sections, and detector effects.  Each of these convolved aspects is modeled en route to results. The measurements at the \dword{nd} are critical input to the models.  An important source of uncertainty that arises in the models in comparing the near and far event rates is that the neutrino spectrum is rather different between the two detectors.  \dword{dune} plans to use the PRISM technique to minimize the spectral difference in the fluxes analyzed between the near and far detectors.  In doing so, this will minimize uncertainties in the extracted oscillation parameters arising from the spectral differences as implemented in the  imperfect interaction (cross section) model.  That said, the spectral matching in the PRISM analysis will not be perfect and residual corrections that depend on the interaction model must be made.  The implementation of PRISM helps mitigate, but does not remove, the effects of an imperfect interaction model.  \dword{dune} will need an interaction model that is as accurate and well tuned to data as possible.  \dword{dune} will need a vibrant program of cross section measurements as input to that work.

Cross section studies also allow the investigation of  complex nuclei and their behavior under the weak interaction.  This nuclear physics is  crucial for neutrino physics since the properties of invisible neutrinos can only be inferred from their interactions with matter, i.e., those nuclei.

Cross section measurements to be made with the \dword{dune} \dword{nd} are needed.
While the short-baseline neutrino program \cite{Antonello:2015lea} will collect cross-section measurements on argon, the energy range seen in that program is lower than \dword{dune}'s; cross-sections measured at \dword{dune} energies by experiments like \dword{minerva} \cite{Aliaga:2013uqz} investigated different nuclei; \dword{argoneut}  \cite{Anderson:2012vc} took measurements on argon in the \dword{numi} beam, but was too small for good event containment and took data in a $\bar{\nu}_\mu$-dominated beam configuration, where the relatively small $\nu_\mu$ component was at a higher energy range. Thus, it is important for the \dword{dune} \dword{nd} to collect the data and quantify the relevant cross sections at the \dword{dune} energy scale.

Neutrino scattering from heavy nuclei such as DUNE's argon, which has an atomic number of 40, is complex. There is a large array of possible interaction mechanisms, many of which can produce identical final states. While neutrino-nucleon interactions are relatively well understood, the effects of the nucleus itself is poorly understood.  The motion of initial-state nucleons, nucleon-nucleon correlations, and \dword{fsi} between ejected nucleons and the rest of the nucleus are complicating factors. As our understanding of all these processes is insufficient to evaluate all of these possible effects from first principles, assumptions must be made.  The models used and the assumptions behind them can make big differences in the predicted cross sections and have  implications  for the extracted oscillation parameters. The \dword{dune} \dword{nd} needs to provide data for tuning and improving these models, and for evaluating the systematic uncertainties incurred through their use.

The need for the \dword{dune} \dword{nd} to make broad and systematic measurements of cross sections on argon in the appropriate energy range is underscored by the fact that quantitative assessments demonstrate the current models describe the data  rather poorly. Examples demonstrating this can be found in Ref.~\cite{Dolan:2018zye} where the models in GENIE, NEUT, and GiBUU cannot describe T2K and MINERvA measurements made using transverse kinematic imbalance; in~\cite{Abe:2018pwo}, where T2K makes a wide range of model comparisons; in Ref.~\cite{cai2019nuclear} which compares a number of models to projections of transverse kinematic imbalance; in~\cite{Tice:2014pgu}, which shows models comparing to the  MINERvA measurement of inclusive scattering on heavy nuclei; and MINERvA's attempt to tune GENIE to describe all MINERvA pion production data~\cite{Stowell:2019zsh}.

Though the need for quality cross section data from the \dword{dune} \dword{nd} is clear, it is important to note the data alone are not sufficient to produce good models.  An appropriate program of cross section physics within \dword{dune} should be coupled with support for the development of theoretical models and their implementation in event generators.  

\begin{dunefigure}[Effect of cross-section model on $\delta_{CP}$ sensitivity]{fig:xsec-cpv-bias}
{Effect on \dword{dune}'s sensitivity to the CP-violating phase if an incorrect cross section model is used in the reconstruction.  This illustrates the danger of not improving/tuning the cross section model using data taken with the \dword{nd}.}
\includegraphics[width=0.5\textwidth]{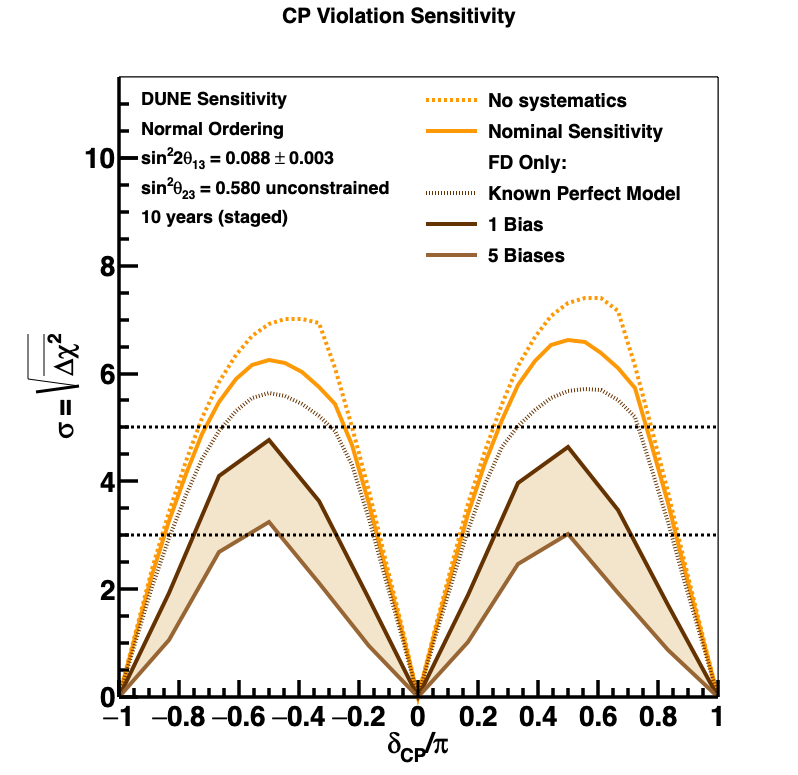}
\end{dunefigure}

Figure~\ref{fig:xsec-cpv-bias} demonstrates the importance of the cross section model when extracting oscillation parameters. The dotted brown line shows \dword{dune}'s sensitivity to the CP-violating phase $\delta_{CP}$, calculated using \dword{dune}'s nominal Monte Carlo event generator, \dword{genie}, assuming the interaction model is perfect, i.e., the same model is used for generation as in the \dword{fd} fit. To evaluate the effect of using an incorrect cross section model for our calculations, the simulated data are re-weighted using the cross section predictions of an alternative event generator, \dword{nuwro}. When these pseudo-data are fit using the nominal \dword{genie} model, the extracted values of $\delta_{CP}$ were found to be biased, due to differences in the energy spectra reconstructed in the different models. As it is unknown which model matches data, this bias must be translated to a systematic uncertainty, leading to a significant decrease in $\delta_{CP}$ sensitivity, as indicated by the dark brown lines. Given that these are not the only possible viable nuclear models, there is a strong possibility that more uncertainties due to model bias will need to be included, reducing sensitivity even further. The shaded brown area represents the loss in sensitivity that would come about if up to five biases of the size generated by the \dword{nuwro}-\dword{genie} model difference were in effect added in quadrature.

A better understanding of neutrino-nucleus interactions is the most effective way in which we can improve the precision of \dword{dune}'s oscillation measurements, as well as increasing the  understanding of nuclei and their weak interactions, which is key to other areas of physics such as double-beta decay. The \dword{dune} \dword{nd} 
gives us an ideal opportunity to explore this over a broad range of energies and make a strong contribution to the field.

\section{Interactions in the DUNE energy range}
\label{ch:nu-xsec:interact}
Depending on their energy, neutrinos have a probability of scattering from nucleons in various ways. We are typically interested in \dword{cc} 
interactions, which produce an identifiable lepton -  the neutrino's charged partner - in the final state, typically accompanied by other interaction products. Neutral current (NC) interactions are also of interest, though in the arena of oscillation physics that interest lies in how NC  events can create backgrounds to the important \dword{cc} processes.  At low energies, a neutrino undergoing a \dword{ccqe} with a neutron will convert it to a proton, which will be ejected along with the charged lepton partner. 
Higher-energy neutrinos are able to excite target nucleons to a resonant state, which decays to produce particles such as pions, or can scatter inelastically from the component partons, breaking up the nucleon and producing additional final-state hadrons.
These interactions are relatively well understood for isolated, stationary nucleons with distinctive final states that can be used to reconstruct the properties of the initial neutrino. However, the nuclear environment complicates the situation in two ways. The initial-state nucleons are subject to complex, isotope-specific momentum distributions and correlation effects, which can mean that neutrinos scatter from pairs or larger groups of nucleons, rather than from individual particles, invalidating our reconstruction equations. Additionally, final-state particles undergoing \dword{fsi} can be accelerated or decelerated due to interactions with the nucleus; pions may also be created or absorbed, and hadrons may undergo charge-exchange interactions with the nucleus as part of the \dword{fsi}. More work is needed to understand both these initial-state nuclear effects, and the \dword{fsi}, neither of which have yet been modeled in a way that matches experimental data well. 

There is an ongoing program of work to understand these nuclear effects, with neutrino event generators such as \dword{genie} and \dword{nuwro} generators incorporating various models. While there is a concerted effort  to streamline the way that we evaluate and combine new models for different parts of the interaction process (see, as an example, \cite{generator_wkshop}), a key part of this process will involve testing the models against physics data. The \dword{dune} \dword{nd} will form a vital part of this program.

\subsection{Quasi-Elastic Interactions}
\label{ch:nu-xsec:ccqe}
\dword{ccqe} interactions are typically considered the golden channel for oscillation experiments, due to their simple final state:
\begin{equation}
	\label{eq:nu-xsec:ccqe}
	\begin{aligned}
        \nu_{l} + n &\rightarrow l^{-} + p \\
        \bar{\nu}_{l} + p &\rightarrow l^{+} + n
     \end{aligned}
\end{equation}
where $l$~refers to the flavor (for \dword{dune}, typically $\mu$ or $e$) of the neutrino and its charged partner. For a pure \dword{ccqe} interaction on a stationary nucleon, the charged lepton kinematics can be used to reconstruct the incoming neutrino energy $E_\nu$ and squared four-momentum transfer $Q^2$ 
(as shown below for $\nu_\mu - n$ scattering):
\begin{equation}
	\label{eq:nu-xsec:enu}
E_\nu^{QE}=\frac{m_p^2 - (m_n-E_b)^2-m_\mu^2+2(m_n-E_b)E_\mu}{2(m_n-E_b-E_\mu+p_\mu\cos\theta_\mu)}
\end{equation}
\begin{equation}
	\label{eq:nu-xsec:qsq}
Q^2_{QE}=2E_\nu^{QE}(E_\mu-p_\mu\cos\theta_\mu)-m_\mu^2
\end{equation}
where $E_\nu$ and $E_\mu$ are the neutrino and muon energy; $p_\mu$ is muon momentum, and $\theta_\mu$ the angle between the muon and neutrino. The neutron, proton and muon masses are represented as $m_n$, $m_p$ and $m_\mu$, respectively, and $E_b$ is the nuclear binding energy. 
While the neutrino energy cannot be measured directly, the kinematics of the outgoing muon are typically straightforward to reconstruct, making \dword{ccqe} an attractive channel.

A theoretical cross-section expression for \dword{ccqe} on free nucleons, as a function of $Q^2$, calculated in 1972 by C. Llewellyn Smith, \cite{llewelyn-smith}, is still used today. This expression depends on the electromagnetic form factors of the nucleons. \dword{genie} provides several models for the vector form factors parametrized from electron scattering measurements; the default is BBBA05~\cite{Bradford:2006yz}.    The axial form factor contributes the larger uncertainty to the \dword{ccqe} cross section, and again \dword{genie} provides two models.   The first is a dipole form with the M$_{A}$ parameter set to 0.99 and an uncertainty the user can make larger or smaller.   Historically \dword{genie} assigned a very large uncertainty to this parameter to account for unmodeled multi-nucleon effects not present in the deuterium data.   The second is the Z-expansion form with parameters fit to the same deuterium data~\cite{Meyer:2016oeg}.   The single parameter dipole does not allow enough freedom to describe the uncertainty from the high Q$^{2}$ part of the spectrum; the Z expansion analysis overcomes this and also considers additional theoretical and experimental uncertainties in their analysis of the deuterium data, and can be combined with separate uncertainties on multi-nucleon effects.

The nuclear environment, however, complicates both the energy reconstruction and the distinctive signature of quasi-elastic events. This will dominate the discussion for the rest of this chapter.

\subsection{Resonant Pion Production}
\label{ch:nu-xsec:res}

The term \dword{res} refers to a class of neutrino interactions that proceed through an intermediate nucleon resonance, which typically decays into a nucleon and a pion. These interactions dominate the region of phase space where the \dword{hadw} is between 1 and \SI{2}{GeV}.
In these cases, the neutrino's interaction with a nucleon leaves it in an excited state ($N^*$ or $\Delta$ resonance), whose main decay mode involves the emission of one or more pions\footnote{Nucleons are also emitted in resonant decays, but the discussion here focuses on the pions.}.

Pion production starts at energies above \SI{200}{\MeV}. At low values of $W<$\SI{1.4}{\GeV}, \dword{res} decays typically produce a single pion, and are dominated by the weak excitation of the $\Delta(1232) P_{33}$ resonance. At higher \dword{hadw} values resonances in the second resonance region, $ P_{11}(1440)$, $S_{11}(1535)$ and $D_{13}(1520)$, become important. Their decays can emit multiple pions, kaons, and photons.  It is currently assumed that higher-energy resonances have small excitation cross sections, and have been assumed to have only small effects on existing cross section measurements. This assumption has not been tested for neutrino-argon scattering in the \dword{dune} energy range.

As resonance effects are  connected intrinsically to the hadronic products of the neutrino-nucleon interaction, both axial and vector form factors are relevant when modeling. As a probe of the nucleon axial vector response, the neutrino-nucleon interaction is quite useful in hadron physics. The description of the meson production mechanism is commonly modeled using the approach of Rein and Sehgal \cite{Rein:1980wg} and can be easily implemented in generators. 

For charged pion production, the available data sets are not well modeled; no current neutrino event generator agrees with data. \dword{minerva} \cite{Le:2019jfy, Eberly:2014mra, McGivern:2016bwh}, \dword{miniboone} \cite{AguilarArevalo:2010bm}, and T2K \cite{Abe:2016aoo, Abe:2019arf} each have published large datasets of pion production data.  Those results are difficult to reconcile in the context of models. Most generator models of resonant production are based around the Rein-Sehgal work, but exclude its modeling of interference. There are no models for non-resonant multi-pion production in the current neutrino event generators. The contributions of heavier resonances are  added explicitly to permit predictions over all kinematics. Most of the time this relies on updates to the outdated Rein-Sehgal parametrization, with educated guesses when it comes to the axial part.
A recent attempt \cite{Stowell:2019zsh} has been made to tune the strengths of various \dword{genie} pion production parameters to \dword{minerva} data, but studies are still needed to understand if the tune can be extrapolated successfully to describe the data from any other experiment. 

The transition region between \dword{res} and \dword{sis} (see Section~\ref{ch:nu-xsec:dis}) is also poorly understood. No model makes this transition smoothly, meaning that generators have to make inelegant assumptions to correct for this.

The \dword{dune} \dword{nd}, with the ability to study pion production with good PID, low thresholds, and high statistics, holds great promise in terms of providing data useful for understanding final states with pions.
 A good handle on resonant cross sections is crucial, as they constitute a significant fraction ($\sim$40\%) of the interactions seen at DUNE.  Also they are a background to quasi-elastic-like samples. Quasi-elastic-like event selection involves choosing events with no pions in final state.  Unfortunately, through \dword{fsi}, which can include pion absorption, and reconstruction limitations, \dword{res} events can mimic the \dword{ccqe} morphology.
A detailed discussion of \dword{fsi} effects can be found in Section~\ref{ch:nu-xsec:fsi}, while a breakdown of the types of interactions expected to generate different pion multiplicities in the \dword{nd} can be seen in \ref{ch:nu-xsec:pion-multiplicity}. A discussion of the effects of the nuclear environment on \dword{res} scattering can be found in Section~\ref{ch:nu-xsec:nucl-models}.

Current  \dword{res} models have been tested and implemented for target nuclei with $A < 20$.  Measurements in argon are necessary to provide data to tune against and to look for a proper understanding of how nuclear effects scale. 
In \dword{dune}, \dword{ndlar} will measure well the hadronic component of neutrino interactions with good liquid argon TPC resolution and will use the muon kinematics from the \dword{ndgar} measurements.  \dword{ndgar} will be able to measure charged particles with a very low energy threshold and unmatched PID.  These capabilities will yield spectacular data for constraining and understanding \dword{res} processes. 

With DUNE's goals in precision it is important to attain a greater understanding of the resonance channel.  The role of correlated nucleon pairs in resonance pion production is not fully understood.  There is tension between data sets that needs to be understood and clarified; more modern models need to be incorporated into generators; and liquid argon data in all ranges of energy is needed. The statistics, technology, energy range, and capabilities of the \dword{dune} \dword{nd} will facilitate the needed exploration of the resonance processes.

\subsection{Inelastic Scattering}
\label{ch:nu-xsec:dis}

Inelastic interactions present an interesting challenge for neutrino oscillation experiments in the few-GeV energy regime. Instead of being defined by a single final state, they are generally characterized by what they are not; they are not elastic, not resonant, and not coherent.  They are broadly divided into two major classifications; shallow- and deep- inelastic scattering.  \dword{sis} describes non-resonant meson (mainly pion) production with lower $Q^2$, typically < \SI{1}{\GeV}$^2$, and occupies the full W range $W > M_N + M_\pi$.  As $Q^2$ grows in these non-resonant interactions and the de Broglie wavelength of the neutrino allows for the resolution of the quarks inside the nucleus, the realm of \dword{dis} begins.  To aid in differentiating resonant produced pions from \dword{dis} quark-fragmented produced pions, a boundary at W = 2 GeV has been instituted.

The study of higher-energy neutrino-nucleus \dword{dis} interactions is advanced both theoretically and experimentally \cite{Tzanov:2005kr}\cite{Kulagin:2004ie}.  However, the study of lower-energy \dword{dis} (non-perturbative QCD), the transition region from \dword{sis} to \dword{dis}, and the complete realm of \dword{sis} interactions is largely unexplored.  Events in this region of kinematics will be a significant fraction of the events in \dword{dune}.   Although neutrino-nucleus single-pion resonant production has been studied (see Section~\ref{ch:nu-xsec:res}), there are additional multi-pion resonances, SIS/DIS non-resonant pion production and, significantly, the interference between all of these states that very much complicate the picture.  Furthermore, since the nuclear environment makes it difficult to disentangle resonance events from \dword{sis} events experimentally, it is mainly the measurement of inclusive cross section and theoretical investigation of the \dword{sis} region that are on-going.  

How this complicated picture is addressed is very MC generator dependent. \dshort{genie} simulates \dword{sis} and \dword{dis} in two stages. The first is the primary cross section model, which determines the kinematics of the outgoing lepton and hadronic systems.  Here \dshort{genie} uses the Bodek-Yang scaling formalism \cite{Bodek:2002ps}, which adapts a nucleon parton structure-function-based prediction to lower invariant masses.
The Bodek-Yang model predicts the entire inelastic cross section, not simply the nonresonant component. In a model such as \dshort{genie}, which contains explicit (modified Rein-Sehgal) calculations for lower-lying single-pion resonances, the portion that is ascribed to \dword{dis} is therefore the result of subtracting these resonances from the total prediction of Bodek-Yang.  Simply stated, \dshort{genie}-defined \dword{dis} is now not only the true kinematic \dword{dis} discussed above but rather a combination of \dword{sis} pion-, multi-pion resonant production and, finally, true \dword{dis} quark-fragmented pions.  Consequently, \dword{sis} does not exist as an independent production mode in \dshort{genie}.  Moreover, this approach does not correctly account for the impact of interference between the resonance and non-resonant production on outgoing hadron kinematics, introducing additional uncertainties into the prediction.

The second stage of simulation of non-resonant (\dword{sis} and \dword{dis}) inelastic scattering is hadronization, where a full final state with hadron identities, charges, and momenta is predicted using the four-momentum of the hadronic system given by the first stage.
Various theoretical models for hadronization, 
such as the Lund string model\cite{Andersson:1998tv}, can be applied to predict hadronization in neutrino-induced reactions.  The \dshort{pythia} program simulates interactions using the Lund model (in conjunction with others) and is used within \dshort{genie}.   However, because the assumptions in the string model lose predictive power as \dword{hadw} approaches the pion production threshold, \dshort{genie} contains a custom phenomenological model. This so-called \dword{agky} model is constructed from neutrino scattering data and is used below $\dshort{hadw}=\SI{3}{\GeV\per\square c}$ \cite{Yang:2009zx}.  Since, according to \dshort{genie}, about 80\% of non-resonant inelastic reactions in \dshort{dune} will have $\dshort{hadw} < \SI{3}{\GeV\per\square c}$, and approximately 50\% of reconstructed events will originate as non-resonant inelastic scattering \cite{tdr-vol-2}, any significant uncertainties in this model must be constrained by the \dword{nd}.  As an example of possible problems with the current \dshort{genie} model in this region, 
recent re-evaluations of the $\nu$-nucleon scattering bubble chamber results have suggested large deviations from values in \dshort{genie} \cite{Stowell:2019zsh}.   

A study demonstrating the impacts of changes in the \dshort{genie} model is presented in \ref{ch:nu-xsec:sis-dis}.  Together these uncertainties impact the oscillation sensitivities achievable by the experiment.  A more complete theoretical, experimental, and phenomenological understanding of this mix of resonant, \dword{sis} non-resonant, and \dword{dis} interactions is going to be necessary for precision physics in \dword{dune}. This is an important task for the \dword{nd}.

In higher-W and $Q^2$ (true) \dword{dis} interactions, there are contributions from the axial current in addition to the vector current in the weak force.   This means that different combinations of valence and sea quarks are sampled in neutrino interactions. This makes the precision measurement of the weak structure functions in neutrino scattering a significant and necessary complement to the electromagnetic (EM) structure functions of charged lepton scattering.  Highlighting this difference between the weak and EM interaction, both recent theoretical studies and experimental evidence now suggest that in the true \dword{dis} region nuclear effects for neutrino-nucleus interactions may be different as compared to the nuclear effects of e/$\mu$ nucleus interactions.  For both neutrino and electron/$\mu$ interactions, there are four distinct regions of nuclear media effects in increasing $x_\text{Bj}$.\footnote{  $x_\text{Bj} = \frac{Q^2}{2M\nu}$ with $\nu = E_{\nu} - E_\text{lepton}$ and M=mass of nucleon.}  The four regions are shadowing, anti-shadowing, EMC effect, and Fermi motion.  Although the shadowing and Fermi-motion regions have been addressed theoretically and phenomenologically, the explanations for anti-shadowing and the EMC effect are still under discussion.  The electromagnetic-weak differences in these nuclear effects  are of a similar size and in the same  $x_{Bj}$-range as the four effects.  This is a significant issue since the nuclear media modifications found in \dshort{genie} are based on those measured in charged lepton, not neutrino, scattering experiments \cite{Andreopoulos:2015wxa}, and the observed differences could impact the precision physics of DUNE.

\subsubsection{Nonresonant Inelastic Scattering and Long-baseline oscillations}
\label{ch:nu-xsec:sis-dis}

Neutrino interactions that produce hadrons not present in the initial state (thus not quasielastic, i.e., inelastic) and that do not proceed through explicit resonances are usually grouped together under the label \dword{dis}.
By extension, such reactions that occur at values of \dword{hadw} close to the pion production threshold (just below $\dshort{hadw}=\SI{1.1}{\GeV\per\square c}$) are sometimes known as \dword{sis}.

The \dword{agky} model begins from bubble chamber measurements of neutrino-induced hadron production.
The average multiplicity of charged hadrons \nch ~is fit to a two-parameter functional form observed by the original experimentalists to describe the data reasonably well at higher \dword{hadw}:
\begin{equation}
	\label{eq:nu-xsec:had-nch}
	\nch = a + b \cdot ln(\dshort{hadw}^2)
\end{equation}
The relationship between positive, negative, and neutral pion multiplicities is assumed to obey a rough average law: $\left\langle n_{\pi^0} \right\rangle \approx \frac{1}{2}\left( \left\langle n_{\pi^+} \right\rangle + \left\langle n_{\pi^-} \right\rangle \right)$.
An estimate for the impact of uncertainties in the model of Equation~\ref{eq:nu-xsec:had-nch} may be obtained by studying the effect of using a slightly more robust form fitted by a different group to a similar dataset \cite{Kuzmin:2013tza}.
Comparisons of the predicted \nch{} shape vs \dword{hadw} are given in Figure~\ref{fig:nd-xsec-nch-forms}.
The resulting effect on \dshort{genie}'s predictions for charged pion multiplicity and kinetic energy are illustrated in Figure~\ref{fig:nd-xsec-chgpi-distr}.

\begin{dunefigure}[Charged hadron multiplicity in different non-resonant scattering models]{fig:nd-xsec-nch-forms}
{
		Impact of various fitted forms for charged hadron multiplicity on the prediction (for neutrino-proton scattering) as a function of invariant hadronic mass.  GENIE's \dword{agky} model is blue; \dword{agky} with the value of $a_{ch}$ adjusted to fix a transcription error from its source is orange; an alternative fitted form by Kuzmin and Naumov is in green.
}	
			\includegraphics[width=0.8\textwidth]{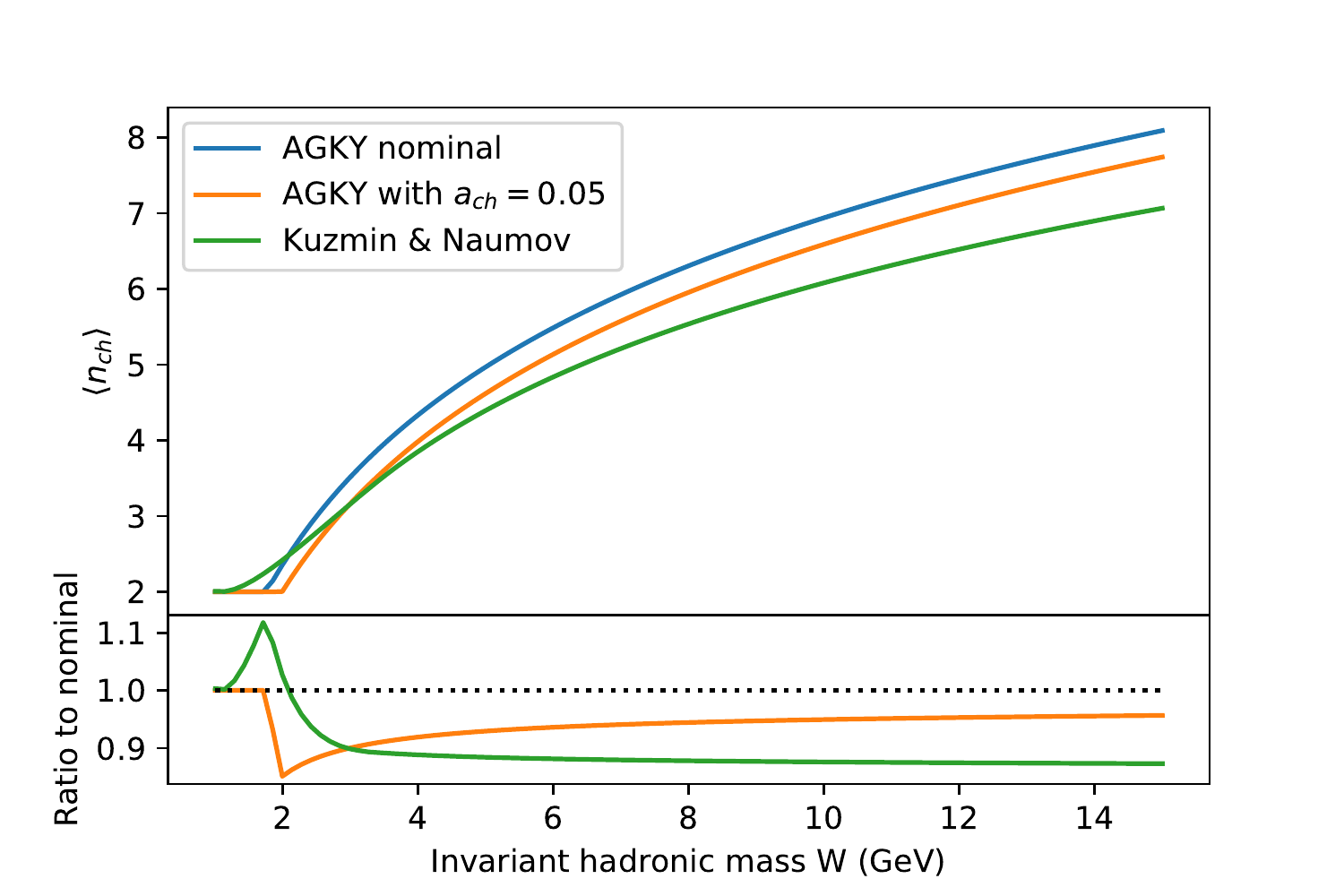}

\end{dunefigure}
\begin{dunefigure}[Charged pion multiplicity and energy in different non-resonant scattering models]{fig:nd-xsec-chgpi-distr}
	{
		\centering 
		\dword{genie} predictions for charged pion multiplicity (left) and kinetic energy (right) in the total \dword{cc} \numu inclusive sample using the variations on AGKY's \nch \ described in the text.
}		
		\includegraphics[width=0.49\textwidth]{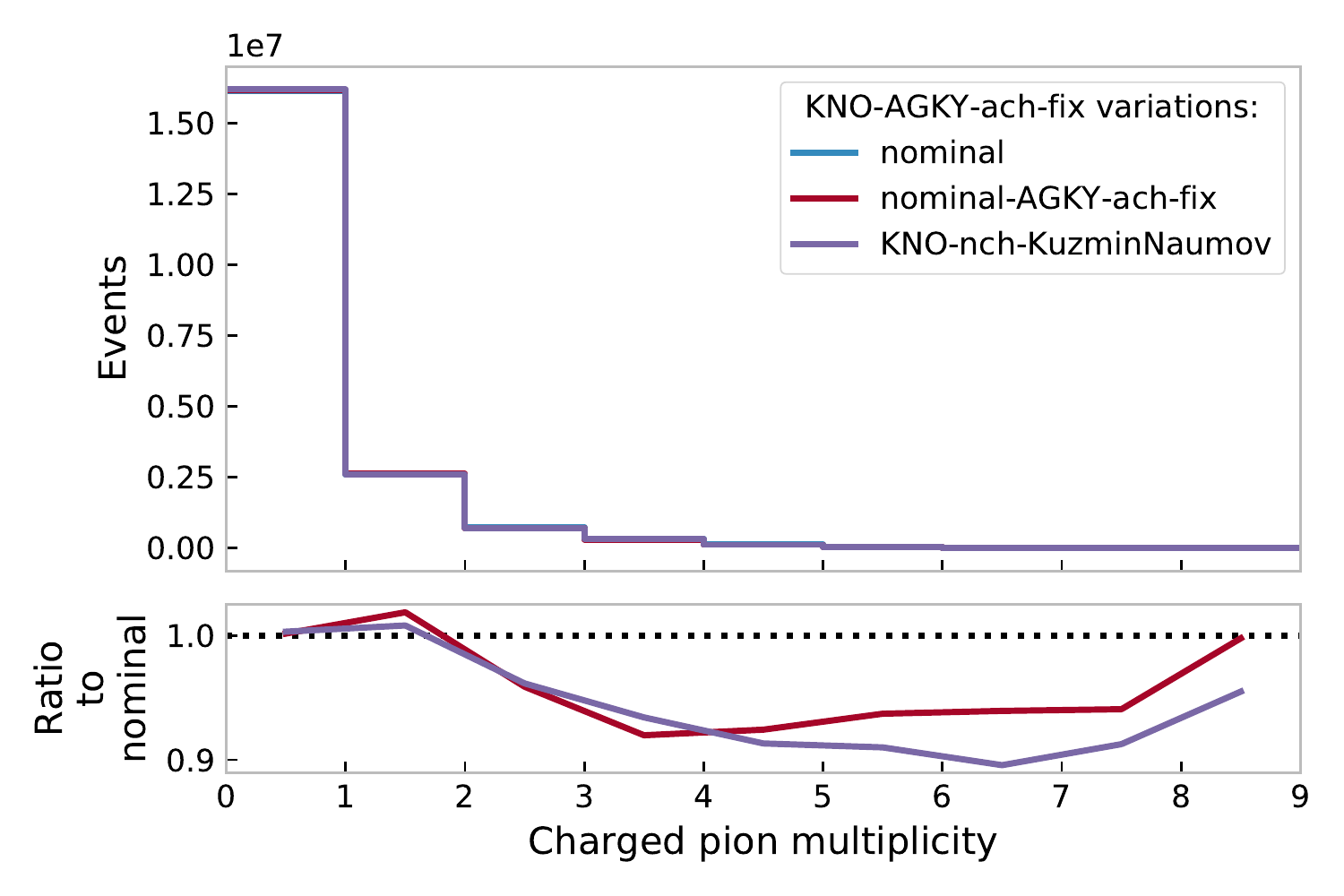}
		\includegraphics[width=0.49\textwidth]{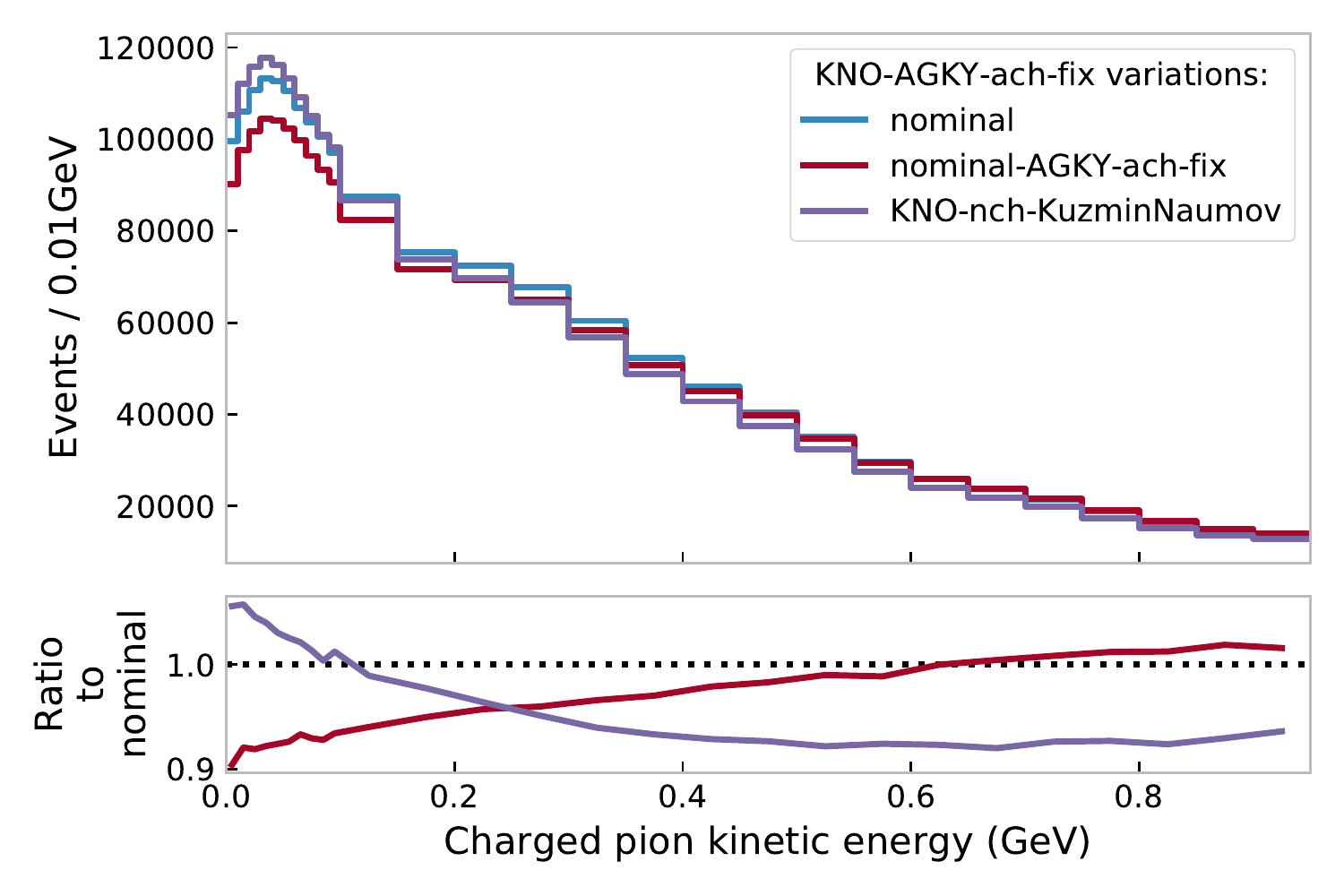}
	
\end{dunefigure}

These 5-10\% changes to the pion spectra result in few-percent differences in the predicted neutrino energy spectrum using the calorimetric approach taken for the FD TDR.
However, the resulting impact on oscillation sensitivities is relatively small, even without a constraint from the \dword{nd}, as demonstrated in Figure~\ref{fig:nd-xsec-disapp-contour-KNOnchKuzminNaumovFakeData}.
Therefore, constraining uncertainty in \nch ~is not regarded as a priority for the \dword{nd} but may become more important as the \dword{dune} measurements become more precise.
\begin{dunefigure}[Effect of non-resonant scattering model on oscillation sensitivity]{fig:nd-xsec-disapp-contour-KNOnchKuzminNaumovFakeData}
{
	Comparisons of oscillation sensitivity contours for \numu disappearance when fitting a mock dataset thrown with the nominal \dword{genie} cross section model (dotted contours) or the alternate Kuzmin-Naumov fit for the \nch ~distribution in the hadronization model.  Green are $3\sigma$ contours, pink are $5\sigma$ contours, and the gray contours correspond to much higher confidence at $\Delta \chi^2=50$ (to illustrate the effect near maximal disappearance).  The best fit point (red dot) moves less than 1\% in both \sinst{23} and \dm{32} from the true value (blue star).
}	
	\includegraphics[width=0.8\textwidth]{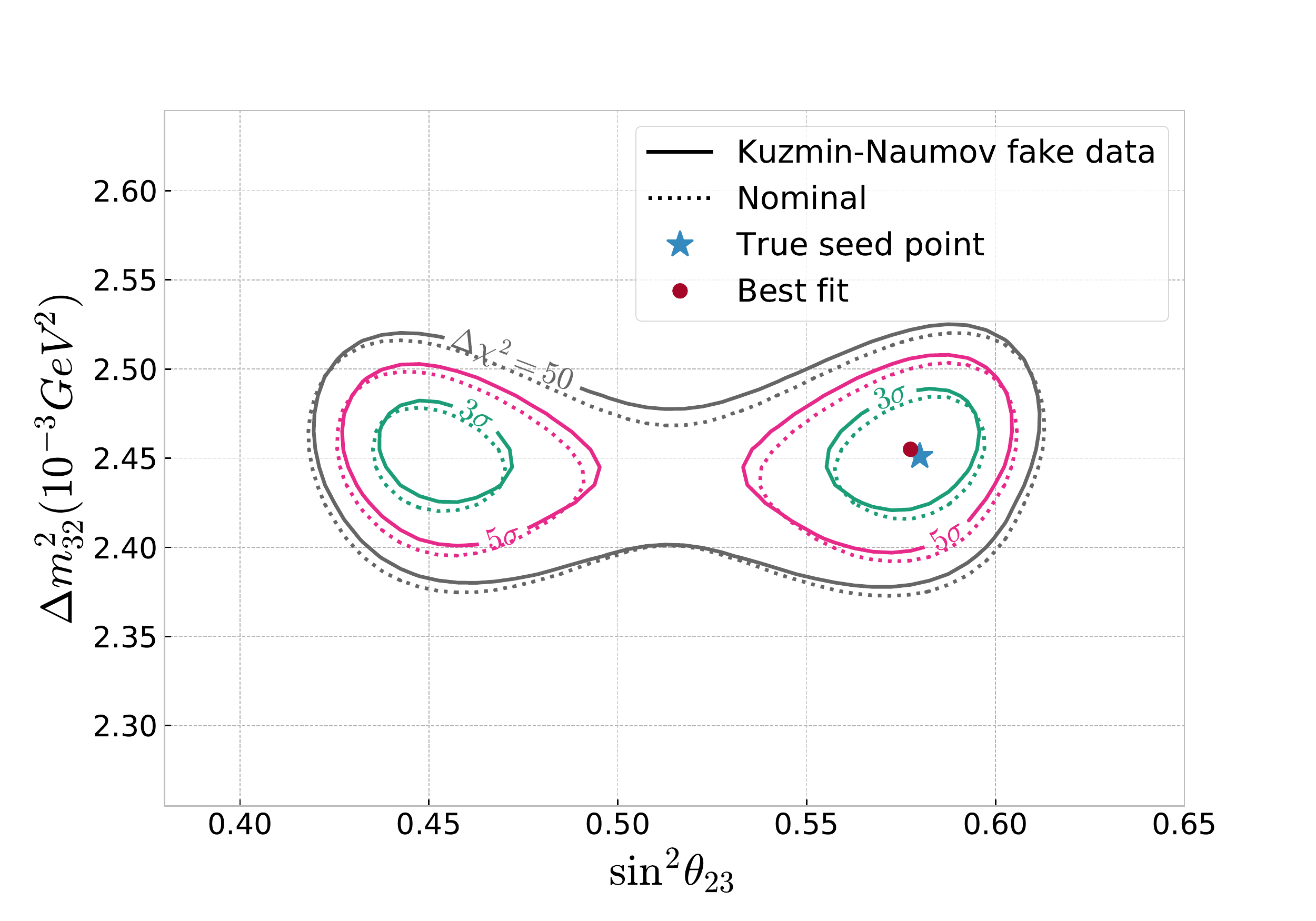}

\end{dunefigure}

The \dword{agky} model also prescribes the assignment of particle identities (nucleon, pion, kaon, or other strange particle) and momenta within the generated hadronic system.
The identities are determined based on fractions of each particle type produced at threshold for the available channels.
Constraining the fractions of particle types in neutrino reactions is one of the principal goals of the \dword{nd} complex.
The particle momenta in \dword{agky} are simulated by using a parametrization of the baryon momenta measured in bubble chamber data, and dividing the remaining phase space among the other particles in the event according to measurements from colliders.
A similar study to what is reported above for \nch{} was performed to investigate the impact of uncertainties in these parametrizations, but the effect on the predicted observables was negligible.

\subsection{Coherent Pion Production}
\label{ch:nu-xsec:coh}

Coherent scattering refers to interactions in which the nucleus remains in its ground state. Coherent elastic neutrino-nucleus scattering ($CE\nu NS$), $\nu A \rightarrow \nu A$, is important for non-standard neutrino interactions and dark matter searches, but does not impact oscillation analyses. However, the coherent production of photons and mesons can mimic signal events for neutrino oscillation searches, and needs to be well understood.  The coherent production of mesons on isoscalar nuclei also offers, in principle, a constraint on the ratio of neutrino to antineutrino fluxes.

While coherent production is relatively uncommon compared to scattering from nucleons, and is relevant only at the lower end of the \dword{dune} energy range, its ability to mimic the interactions used to measure \numu disappearance and \nue appearance means that it is important to evaluate this background. 

For \dword{cc} coherent scattering, charged mesons are produced:
\begin{equation}
	\label{eq:nu-xsec:coh-CC}
	\begin{aligned}
        \nu_{l} + A &\rightarrow l^{-} + m^{+} + A, \\
        \bar{\nu}_{l} + A &\rightarrow l^{+} + m^{-} + A,
     \end{aligned}
\end{equation}
where $m^{\pm} = \pi^{\pm}$ or $K^{\pm}$, $l$ can be any lepton flavor, and $A$ is the unaltered nucleus. For the nucleus to preserve its ground state, it is necessary that the kinematic impact on the nucleus in these processes is small.  Specifically, the magnitude of the kinematic variable $t$,
\begin{equation}
    \label{eq:nu-xsec:defineAbsT}
    \begin{aligned}
        |t| = |(p_\nu - p_l - p_m)^2|
    \end{aligned}
\end{equation}
will be near zero for coherent production of $m$.

These processes can mimic the final state topologies of other interaction types; in particular pion production through the $\Delta (1232)$ resonance, or through \dword{ccqe} interactions with \dword{fsi}, making it an important background to both \dword{ccqe} and resonant searches. 

The \dword{nc} coherent scattering process is
\begin{equation}
	\label{eq:nu-xsec:coh-NC}
	\begin{aligned}
        \nu_{l} + A &\rightarrow \nu_{l} + m^{0} + A' \\
        \bar{\nu}_{l} + A &\rightarrow \bar{\nu}_{l} + m^{0} + A,
     \end{aligned}
\end{equation}
where $m^{0} = \gamma$ or $\pi^{0}, \eta^{0}$, and so forth. The electromagnetic showers generated by the decays of these neutral particles can be misidentified as the event signature of a \nue interaction, a direct background to appearance searches. Though experimental results have been shown for liquid argon and scintillator~\cite{Abe:2013hdq} \cite{Wolcott:2015hda}, \dword{nc} coherent scattering is difficult to distinguish experimentally from other processes that have final state electrons, such as $\nu$-e$^{-}$ scattering and $\nu_{e}$ CCQE.



Diffractive pion production, $\nu_\mu p \to \mu^- \pi^+ p $, has similar dynamics to, and exists in a similar kinematic regime to, coherent scattering. While this complicates coherent measurements in materials containing both hydrogen and heavier nuclei, such as the scintillator in the \dword{3dst}, \dword{dune}'s argon-based detectors should not be affected. 



Models of coherent scattering have been implemented in neutrino event generators, and can be divided into two categories: those based on the partial conservation of the axial current (PCAC), and those based on microscopic models. The Rein and Sehgal (RS) model for coherent $\pi^{0}$ production is tuned with pion-nucleon elastic and inelastic scattering data \cite{REIN198329}. Still, the model has some issues with the prediction of pion angular distributions and gives a poor description of  pion-nucleus elastic scattering \cite{PhysRevD.80.013003}.
 Corrections that use directly experimental pion-nucleus elastic cross sections were proposed \cite{PhysRevD.79.053003,PhysRevD.80.033005} as an avenue of improving the model. The dependence of the coherent scattering cross section with the target is not well understood, although by definition it scales on approximately a per-nucleus rather than a per-nucleon basis.
 
  Microscopic models are constructed from particle production models on nucleons and perform a coherent sum over all nucleonic currents. These models of pion production are well described~\cite{PhysRevD.79.013002,PhysRevC.86.035504,PhysRevD.80.013003,PhysRevC.55.1964,PhysRevLett.96.241801,AlvarezRuso:2007tt,AlvarezRuso:2007it,PhysRevC.79.057601,PhysRevC.81.035502} and have recently incorporated photon emission~\cite{PhysRevC.89.015503,PhysRevC.86.035504}. The available models are restricted to low energy transfers, in the same region of phase space where weak particle production models and meson optical potentials are most applicable.  A version of the microscopic model of~\cite{AlvarezRuso:2007tt} has become available in GENIE and is used by T2K in a comparison with data~\cite{Abe:2016fic}. 

Diffractive scattering has not yet been implemented in most generators. Generators have the simple RS model implemented for coherent scattering. The model of Berger and Sehgal that uses pion-nucleus elastic scattering data as input~\cite{PhysRevD.79.053003} has been implemented as well in \dword{genie} and \dword{nuwro}.  

Table~\ref{tab:generatormodels} shows the current status of coherent scattering models in neutrino event generators.  The various implementations of the Rein-Sehgal PCAC model are insufficient for \dword{dune}'s precision measurements. The improved Berger-Sehgal model~\cite{Berger:2008xs} requires pion-nucleus elastic scattering data which is as yet unavailable for argon. Microscopic models are slowly being added to event generators but they need to be extended beyond the $\Delta(1232)$ region, as well as validated with other reactions such as coherent production of electrons and photons, and meson-nucleus scattering.  


\begin{dunetable}[Models included in the neutrino event generators]{c|c|c|c|c}{tab:generatormodels}{Each generator has a different approach to the transition region between RES and DIS. Also, the same model can be implemented differently in different generators. KNL-BRS: Kuzmin-Naumov-Lubushkin-Berger-Sehgal with axial form factor fit to MiniBooNE data. Graczyk-Sobczyk: relation between form factors and helicity amplitudes. }

Generator & SBN Experiments & Resonance model & Coherent model & FSI model \\ \toprowrule
NUANCE & MiniBooNE & Rein-Sehgal & Rein-Sehgal & Cascade \\
       & & & KNL-BRS & \\ \colhline

GENIE & MicroBooNE & Rein-Sehgal & Rein-Sehgal & INTRANUKE/hA \\ 
      & T2K & KNL-BRS & Bergel-Sehgal & \\
      & SBN (ND and FD) & & \\ \colhline

NEUT & T2K & Graczyk-Sobczyk & Rein-Sehgal & Hybrid Oset et al. \\ 
     & & & Bergel-Sehgal & + exp. Based tune  \\ \colhline

NuWro & T2K & Home-grown & Rein-Sehgal & Cascade \\
      & MicroBooNE & & Bergel-Sehgal &  \\
\end{dunetable}

The only measurement of coherent pion production in the \dword{dune} energy range was made on carbon-based scintillator at MINERvA, where it was found the coherent data does not agree well with predictions~\cite{Mislivec:2017qfz}. 

\section{Scattering From Heavy Nuclei}
\label{ch_xsec:sec_nuclei}

While basic neutrino-scattering models consider a stationary nucleon, particles in the nuclear environment undergo Fermi motion and are subject to complex effects. Furthermore, particles produced in neutrino interactions are subject to \dword{fsi} as they exit the nucleus. These effects can alter the final state dramatically.  These effects not only impact the energy and momenta of produced particles but also the composition. Because the initial state, hard scattering, and final-state interaction processes affect each other, and because several different processes can generate identical final states, it is extremely challenging to isolate and measure these effects. Furthermore, the complexity of heavy nuclei, such as argon, means that they cannot currently be simulated from first principles.  The models in use instead employ approximations. Validating and improving models of nuclear effects is the single most important task in cross section measurements.

\subsection{Base Nuclear Models}
\label{ch:nu-xsec:nucl-models}
 
Spectral functions describe the initial state of the nucleus in terms of a momentum distribution and removal energy of nucleons from the nucleus. The Relativistic Fermi Gas model \cite{Smith:1972xh}, where the nucleons move with a Fermi-Dirac momentum distribution and have a fixed removal energy, is a simple example of a nuclear spectral function. These models have been shown by experiments such as \dword{minerva} \cite{Aliaga:2013uqz} to be unable to reproduce \dword{ccqe}-like cross-section data 
\cite{Ruterbories:2018gub}. This is thought to be because they give an overly simplistic description of the nuclear initial state. More sophisticated spectral functions take the form of two dimensional distributions in momentum and removal energy space.  See, for example, \cite{Benhar:1997zz, Benhar:1999vv, Ankowski:2008sf, Benhar:1994hw}.


\dword{genie}'s implementation of the Relativistic Fermi Gas model includes Pauli blocking and a Bodek-Richie tail \cite{Bodek:1981wr} to model short-range correlations between nucleons. \dword{nuwro} also implements a Local Fermi Gas, with a position-dependent potential \cite{Negele:1970jv} and spectral function model \cite{Ankowski:2008sf} of the nucleus. \dword{gibuu} treats the nuclear ground state  within a local Thomas-Fermi approximation, with nuclear density profiles parametrized according to elastic electron-scattering data and Hartree-Fock nuclear-many-body calculations.

The nuclear medium also affects more complex scattering processes, including those that create pions. Medium effects pertaining to \dword{res} processes are implemented with different levels of sophistication in different generators. The \dword{gibuu} transport model includes a simulation of both nucleon and $\Delta$ spectral functions. NEUT \cite{Hayato:2009zz} assumes a fixed fraction of pionless $\Delta$ decays, using the results of Singh et al.~\cite{Singh:1971md}. \dword{nuwro} \cite{NuWro2012} takes the fraction to be neutrino-energy dependent. \dword{genie}  presently has none of these effects. 

\subsection{Multi-nucleon effects}
\label{ch:nu-xsec:mec}

It is now well known that nucleons within the nucleus do not behave as free particles; electron-proton scattering experiments on carbon showed that around 20\% of them formed correlated, mostly $np$, pairs \cite{Subedi:2008zz}. It is theorized that this is due to the presence of meson-exchange currents - a set of processes by which a pair or larger group of nucleons are bound by the exchange of a virtual pion or other meson.
If a neutrino scatters from one of these pairs, it is possible that both of the paired nucleons will be ejected from the nucleus; this is known as  \dword{2p2h} (two particle, two hole). At fixed three-momentum transfer, the cross section as a function of energy transfer has two discrete peaks, the first corresponding to quasielastic (or 1p1h) scattering, and the second due to $\Delta$ resonance production.  These 2p2h events fill the ``dip'' region between the peaks.
In the event that the second nucleon is not detected
\dword{2p2h} events can mimic the signature of a 1p1h event. However, as a second nucleon has carried away some of the incident neutrino's momentum, 
this can have serious repercussions for oscillation measurements because they rely on reconstructed energy spectra.

GENIE provides an implementation of \dword{2p2h} scattering from 
Ref. \cite{Nieves:2011pp, Gran:2013kda}. 
It also implements an effect of long-range correlations due to polarization in the nucleus that is modeled using the random phase approximation (RPA)~\cite{Morfin:2012kn}. Measurements of neutrino-carbon scattering at \dword{minerva} \cite{Rodrigues:2015hik} show that this model significantly underpredicts the cross section in the energy-momentum transfer space between the \dword{ccqe} and \dword{res} regions, where \dword{2p2h} processes contribute most. A tuned version that scales the \dword{2p2h} rate non-uniformly in that space, with an overall rate increase of 53\%, produced good agreement with quasi-elastic-like data in the low-energy \dword{numi} beam configuration \cite{Ruterbories:2018gub} over most phase space. Though this is an empirical tune, it is worth noting that some other models predict a significantly larger \dword{2p2h} contributions at MINERvA energies~\cite{dolan:2019bxf}.
A higher-energy \dword{minerva} study \cite{carneiro2019highstatistics} did not achieve good agreement with the tuned simulation, suggesting more work is needed to understand the nuclear effects.

\subsection{Final-state Interactions}
\label{ch:nu-xsec:fsi}

It is common to think of final-state interactions (FSI) as hadron re-interactions, occurring after a primary neutrino-nucleon interaction, as the hadrons produced in the primary interaction move through the nuclear medium. As such, the rate of final-state interaction is highly dependent on the structure of the residual nucleus. 
Multi-nucleon effects complicate FSI as they need to deal with not just the transport of the created hadron, but also the correlated pair. More realistic, quantum mechanical, treatments of FSI can show effects in both the outgoing lepton and hadron kinematics \cite{Nikolakopoulos:2018sbo, Ankowski:2017yvm}.

When it comes to modeling FSI, the theory is not specific to neutrino scattering. FSI need to describe the transport of hadrons through a nucleus. It is the theory of hadrons interacting with heavy nuclei, with the caveat that they ought to be simulated in a potential where they start in the middle of the nuclear medium. Also, note that the cross sections implemented in neutrino event generators are from hadron-nucleus scattering where the hadrons hit the outside of the nucleus.

\dword{fsi} can affect the nucleons knocked out in a quasi-elastic or resonant interaction, and can also affect pions, such as those produced in \dword{res} processes. Several types of \dword{fsi} can occur, including charge exchange, re-scattering, absorption and pion production. Thus, \dword{fsi} can affect not only the energy spectrum (due to accelerating or decelerating rescattering), but also the multiplicity of particles in the final state, as pions may be created or absorbed, or particles undergo charge-exchange interactions. For this reason, \dword{fsi} compromise the signature topologies of the primary interaction types. For example, a \dword{res} interaction such as $\nu_\mu + n \to \mu^- + p + \pi^0$ or $\nu_\mu + p \to \mu^- + p + \pi^+$, followed by \dword{fsi} pion absorption, will leave a \dword{ccqe}-like final state of a muon and a proton, as described in Section~\ref{ch:nu-xsec:ccqe}. However, while the morphology mimics CCQE, the quasi-elastic energy reconstruction in Equation \ref{eq:nu-xsec:enu} will generate an erroneous value of the neutrino energy. Conversely, a \dword{ccqe} interaction in which a pion is created in the \dword{fsi} can mimic the \dword{res} signature CC1$\pi$ topology explained in Section~\ref{ch:nu-xsec:res}.


From the experimental point of view, considering FSI means understanding that every morphology observed in the detector is the combination of many channels. 
Using an \dword{ndlar} topological selection as an example, Table~\ref{tab:ArgonCubeEvents} shows the fractional composition of the selection according to simulation using GENIE 2.12.6 (with an added version of Valencia's \dword{mec} model). The CC0$\pi$ selection is defined as charged current events with no pions observed in the final state, a topology that would usually be tied to CCQE events. However, the CC0$\pi$ topology contains contributions from CC quasi-elastic (CCQE) events (59\%), as well as resonance (RES) events with pion production where the produced pion has a momentum below detection threshold or is absorbed in the nuclear medium before it can exit the nucleus (18.8\%). \dword{mec} events are also present (15\%), and since we do not know the neutrino energy \textit{a priori}, low energy processes are also relevant. No matter what selection is being used, the full set of possible interactions must be considered. 

 
 Table \ref{tab:generatormodels} shows the neutrino event generators used by the short baseline neutrino detector program. Various models are available. Quantum mechanical models for hadron–nucleus experiments would, naively, be the most correct, but difficulties in tracking
multiple particles make such a calculation difficult.  Quantum Mechanical based hadron transport models, such as GiBUU \cite{Weil:2013gra}, and relativistic mean field models (see \cite{Amaro:2011qb,Megias:2014kia}, for example) are useful and in use. Semi-classical models have some success in describing pion–nucleus interaction data and are used widely in neutrino interaction generators. Unfortunately, even where generators use the exact same models, they may have different assumptions and/or implementations,  and this can lead to different event rate predictions as well as model dependence and a source of systematic uncertainty.

While the basic idea behind the models of FSI in the MC codes is always the same, numerical implementations are quite different reflecting the priorities of particular neutrino experiments (target, detection technique etc). Extending these predictions across target nuclei can be a problem when the A dependence of theses effects is not understood.  
New measurements in argon are necessary. That said, recent work has been fruitful in developing techniques to use carbon data to help argon target modeling \cite{bib:docdb16058}.

The best path toward understanding the effects of FSI is to measure the cross sections for as many final states as possible with neutrino beams.  But even if this is done well, there remains the complication of separating each of the effects that contribute to FSI. A new and promising technique is the use of \dword{tki}~\cite{Lu:2015tcr, Furmanski:2016wqo, Abe:2018pwo, Dolan:2018sbb, Lu:2018stk, Dolan:2018zye, Lu:2019nmf, Harewood:2019rzy, Cai:2019jzk, Cai:2019hpx, Coplowe:2020yea} as discussed 
in Section~\ref{ch:nu-xsec:trans-vars}.

\begin{dunetable}[Event rate prediction in near detectors]{c|c|c|c|c|c}{tab:ArgonCubeEvents}{ Events per year (1.1$\times$10$^{21}$ POT) in the forward horn-current ($\numu$-favoring) mode. The rates were computed with GENIE 2.12.10. 
The rates assume a \SI{50}{t} fiducial volume of liquid argon and a \SI{1}{t} fiducial volume of argon gas.}

&\multicolumn{3}{c|}{Interaction Channel} & \multicolumn{2}{c}{Event Rate}  \\ \colhline
&\multicolumn{3}{c|}{} & {ND-LAr} & ND-GAr   \\ \toprowrule
   CC &\numu & & &  $8.2 \times 10^7$ & $1.64 \times 10^6$ \\ \colhline
   &  & 0$\pi$ &  & $2.9 \times 10^7$ & $5.8 \times 10^5$ \\ \colhline
   &  & $1 \pi^{\pm}$ &    &$2.0 \times 10^7$ & $4.1 \times 10^5$  \\ \colhline
   &  & $1 \pi^{0}$ &     &$8.1 \times 10^6$ & $1.6 \times 10^5$ \\ \colhline
   &  & $2 \pi$ &     & $1.1 \times 10^7$ & $2.1 \times 10^5$\\ \colhline
   &  & $3 \pi$ &    & $4.6 \times 10^6$  & $9.3 \times 10^4$ \\ \colhline
   &  & other &   & $9.2 \times 10^6$  & $1.8 \times 10^5$  \\ \colhline
   & $\bar{\nu}_\mu$ &  &    & $3.6 \times 10^6$  & $7.1 \times 10^4$\\ \colhline
   & \nue &  &   & $1.45 \times 10^6$ & $2.8 \times 10^4$ \\ \colhline
NC & \multicolumn{3}{c|}{}    & $5.3 \times 10^5$  & $5.5 \times 10^5$\\ \colhline
$\nu+e$ &  \multicolumn{3}{c|}{} & $8.3 \times 10^3$  & $1.7 \times 10^2$  \\ \colhline

\end{dunetable}

\subsection{Electron-Nucleus Scattering}
\label{ch:nu-xsec:electron}
One of the hardest challenges to measuring neutrino-nucleus cross sections is the fact that neutrinos are invisible to detectors, meaning that the neutrino energy must be reconstructed from the final state kinematics. Furthermore, due to the multi-stage interactions needed to generate them, neutrino beams have broad energy spectra. For electrons, on the other hand, 
it is possible to generate mono-energetic beams and infer the full event kinematics from the four-vectors of the incoming and outgoing electron, without the need to measure the hadronic final state.
Charged lepton scattering from nucleons and nuclei is sensitive to the same underlying structure determined by QCD as  neutrino scattering from nuclei, i.e., the same initial nuclear state. As such, there are a number of ways that 
$e - A$ scattering data inform $\nu - A$ cross section modeling, as well as providing a benchmark for model testing and validation. Electron scattering data provide critical insights into the distributions of initial state momentum and energy of nucleons in nuclei and the importance of many-body currents and final state interaction effects. Electron scattering also provides fundamental experimental input on nucleon isovector elastic form factors and resonance transition form factors.

The \dword{clas} collaboration at Jefferson Lab \cite{Mecking:2003zu} has studied electron scattering from various nuclei, including argon. This has allowed them to observe nuclear effects that can be difficult to quantify in neutrino scattering, such as a direct measurement of  the make-up of \dword{src} nucleon pairs \cite{Carman:2019lkk} and their momentum distributions \cite{Cohen:2018gzh} (key to understanding \dword{2p2h} scattering), nuclear transparency to protons and neutrons (important for \dword{fsi} effects) \cite{Duer:2018sjb} and nucleon resonances\cite{Mokeev:2018zxt} (for \dword{res} pion production). 

The new CLAS12 campaign expands the original \dword{clas} energy range to include four beam energies between 1.1 and \SI{6.6}{\GeV}, giving excellent coverage of the \dword{dune} energy range. It also adds $^{40}$Ar as a target. 
The \dword{e4nu} collaboration is working on converting these electron-scattering measurements to neutrino cross section predictions, which can be compared with generator predictions, and with experimental data. 

While electron scattering 
(an electromagnetic process mediated by photons) 
is not exactly identical to neutrino scattering  (a weak process mediated by $W$ and $Z$ bosons), the processes are very similar and are affected by the same nuclear effects. 
If cross sections are scaled by a four-momentum transfer-dependent factor, $\frac{8\pi^2\alpha^2}{G_F^2Q^4}$, which accounts for the differences between electromagnetic and weak scattering, including the massive W and Z propogators, the vector components of electron- and neutrino-scattering cross sections become similar. This allows electron beams of known energy to be used to compare electron scattering results with the theories used by neutrino event generators.


For example,  $e\to e'p$ events - analogous to  \dword{ccqe} neutrino scattering interactions - can be studied to see how well the quasi-elastic neutrino energy reconstruction formula  (equation \ref{eq:nu-xsec:enu}) reproduces the known electron energy.

The \dword{dune} \dword{nd}  will be able to test the improved cross section predictions from \dword{clas}. Other experiments, such as \dword{ldmx} \cite{ankowski2019leptonnucleus}, are also exploring the possibility of making charged lepton scattering measurements to help \dword{dune}, making this a promising avenue to pursue. 

\section{Case studies of Cross Section Measurements at the Near Detector}
\label{ch:nu-xsec:nd-xsec}

The \dword{dune} near detectors provide us with a wealth of unique opportunities to understand neutrino-nucleus cross sections. This section highlights some case studies that indicate the near detectors' strengths in this area. 





\subsection{
Separating interaction channels by pion multiplicity}
\label{ch:nu-xsec:pion-multiplicity}

\begin{dunefigure}[CC $\nu_\mu$ Interaction rates on argon in the \dword{dune} dword{nd} by event morphology.]{fig:eventratehpgtpc}
{CC $\nu_\mu$ event rates on argon by interaction type expected in the \dword{dune} \dword{nd}. The simulation corresponds to $1.97\times10^{21}$ protons on target in the forward horn current mode (one spill per second for a year). }
\includegraphics[width=0.7\textwidth]{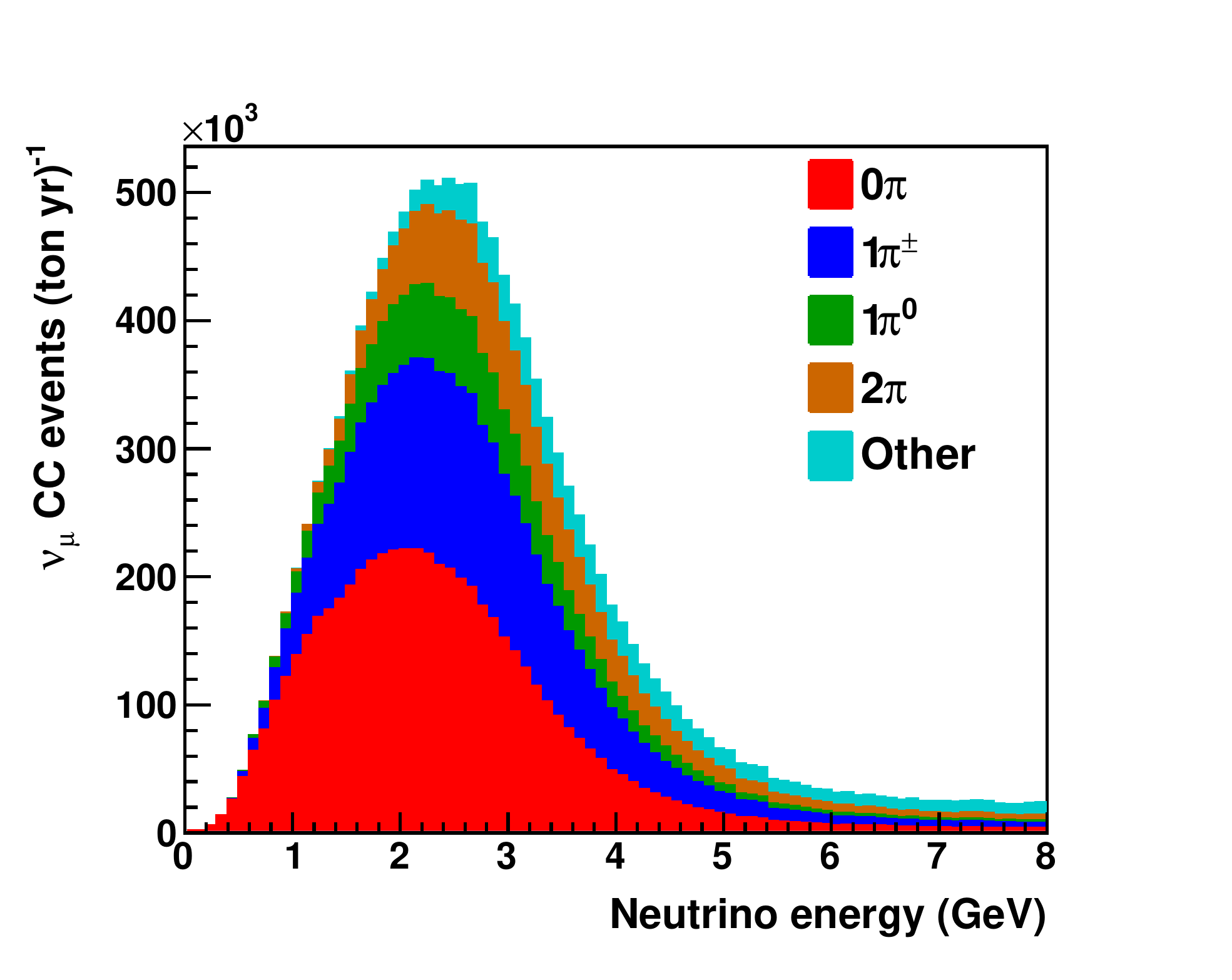}
\end{dunefigure}

The \dword{dune} flux profile means that a rich spectrum of interaction types will take place in the near detectors. Figure~\ref{fig:eventratehpgtpc} indicates the energy spectrum of different types of interactions expected in  \dword{ndgar}.
Liquid argon TPCs are sensitive to different final-state hadron topologies, and are excellent calorimeters. Also, argon provides excellent topological reconstruction, as shown in the event displays in Figure~\ref{fig:xsec-ubooneevents}, from the \microboone detector.

However, as LAr is a dense medium, hadron interactions in the detector volume can complicate the identification of exclusive final states. This is especially true for very low hadron energies, where particles may not travel far enough to reconstruct a track, and also at very high energies, where hadronic interactions are common. Exclusive cross section measurements can also be made in \dword{ndgar}, which has a much smaller target mass and thus reduced statistics, but a very low reconstruction threshold and a low density such that hadron scattering is rare.

\begin{dunefigure}[Zero- and one-pion neutrino-scattering events in liquid argon]{fig:xsec-ubooneevents}
{Event displays from the \microboone liquid argon detector, showing the excellent ability to distinguish $CC0\pi$ (left) and  $CC1\pi^{\pm}$ (right) charged-current \numu scattering events}
\includegraphics[width=0.45\textwidth]{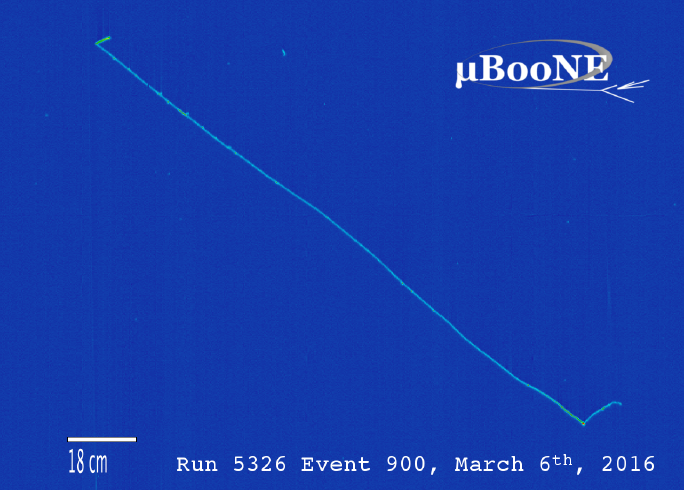}
\includegraphics[width=0.45\textwidth]{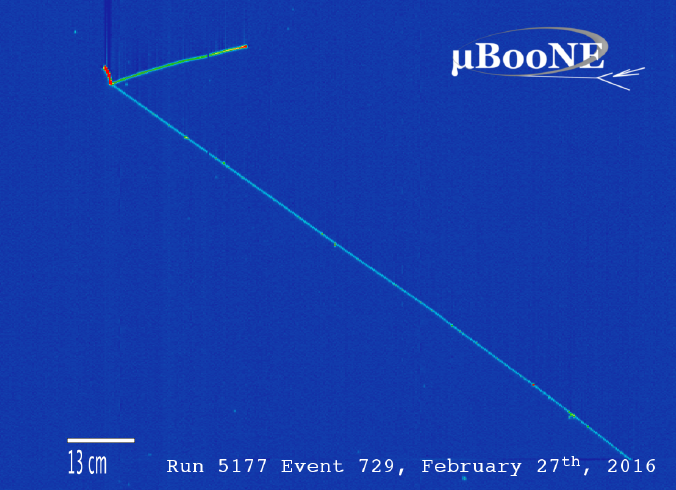}
\end{dunefigure}

\begin{dunefigure}[Pion multiplicity confusion matrices]{fig:xsec-confusion}
{Confusion matrices showing the ability to correctly reconstruct states based on the number of pions in their final state, in the \dword{ndlar} (left) and \dword{ndgar} (right). No confusion has been observed for the combinations filled in white; black cells have some, if minimal, confusion.}
\includegraphics[width=0.42\textwidth]{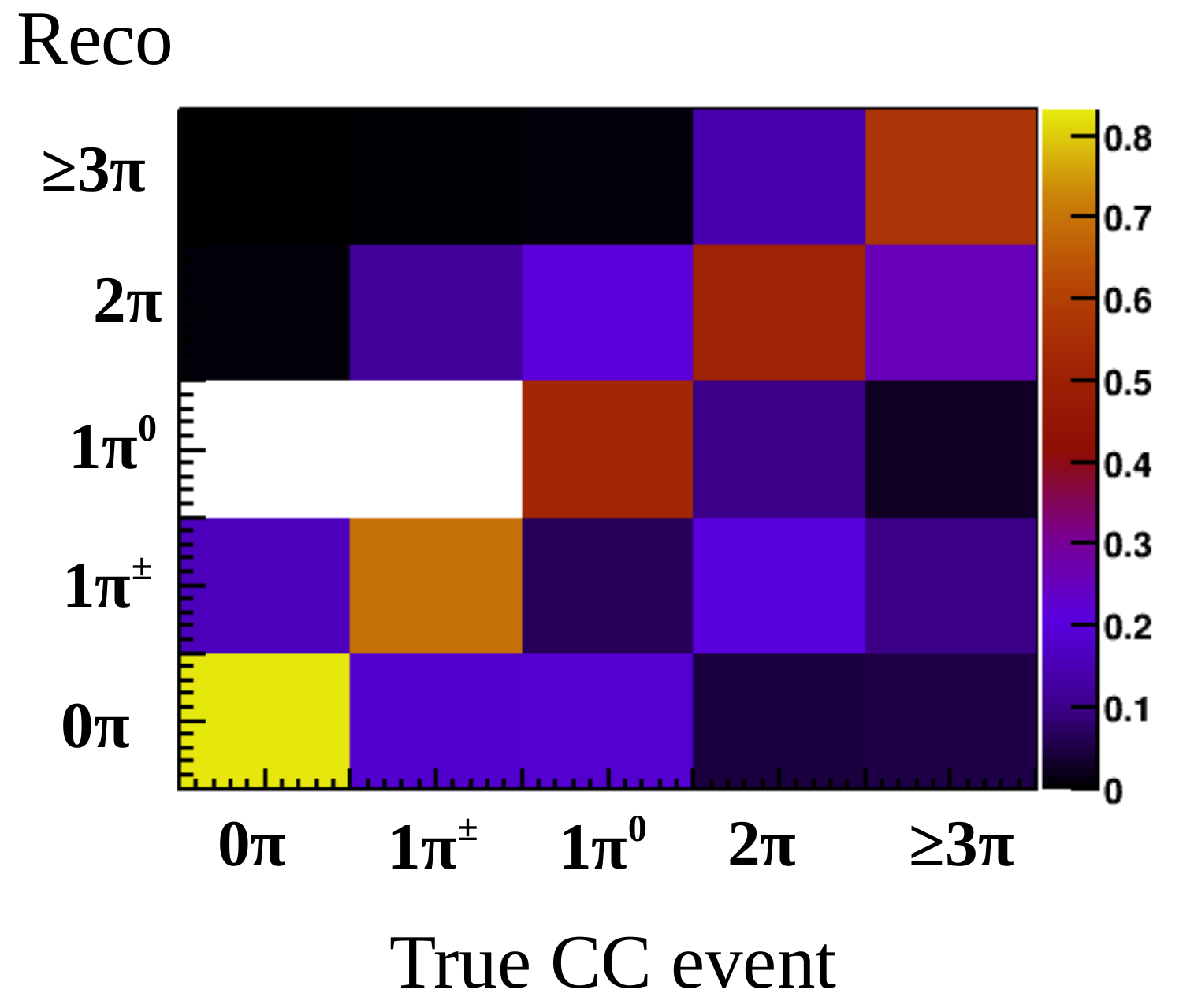}
\includegraphics[width=0.54\textwidth]{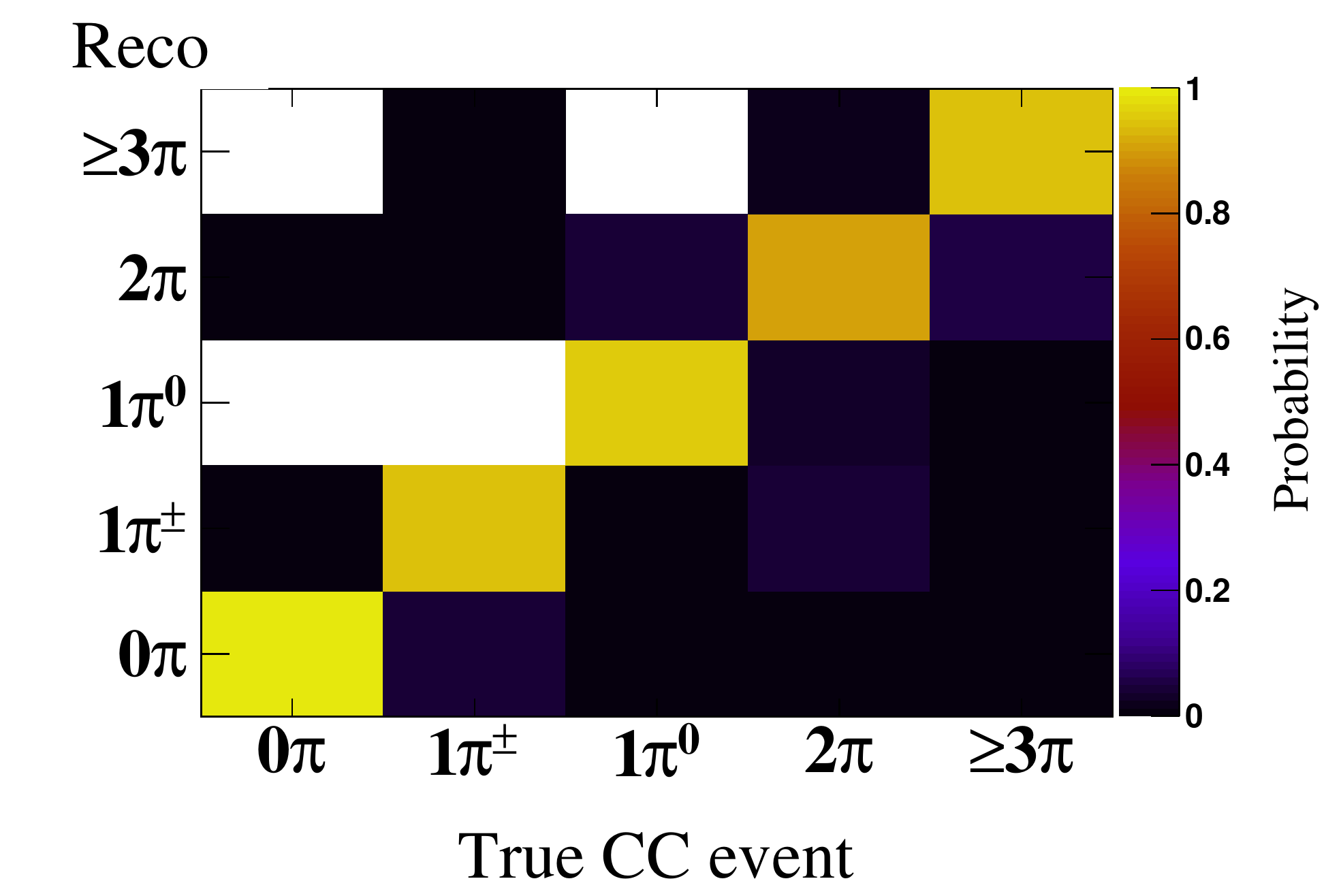}
\end{dunefigure}

This section presents a study of pion multiplicity.  
It uses a full simulation of both the \dword{ndlar} and \dword{ndgar} detectors.  For \dword{ndgar}, there is a simplified reconstruction in which true momenta are smeared using the Gluckstern formula \cite{Gluckstern:1963ng}, and pions and proton tracks are distinguished based on track length and energy deposition rate, dE/dx.
For \dword{ndlar}, events are simulated using Geant4. True energy deposits in active detector volumes are analyzed to look for hadron tracks. These tracks are followed until they either stop due to the hadron depositing all its energy, or interact inelastically. The dE/dx is segmented into chunks that are approximately the size of pixels. These dE/dx profiles are used to determine the PID. Neutral pions are considered reconstructable if both photons deposit at least \SI{20}{MeV} in the active volume.


Figure~\ref{fig:xsec-confusion} gives an indication of how well the near detectors can identify final states for charged-current \numu-Ar interactions, based on their pion composition. While the \dword{ndlar} (left) is excellent at counting the charged-hadron multiplicity, it must rely on dE/dx to distinguish between short pion and proton tracks, leading to some confusion with interacting protons. Nevertheless, the \dword{ndlar} will be able to reconstruct around 60\% of events into the correct category.
The \dword{ndgar} can identify low-energy protons, with near-perfect separation between protons and \dword{mip}s below \SI{1.5}{\GeV \per c}.  The ECAL  uses energy-momentum separation for muons and pions above \SI{1.5}{\GeV \per c} and the muon ID system is included.


The very high statistics of \dword{ndlar}, coupled with good performance in identifying exclusive final states, will enable multidimensional cross section measurements that probe correlations between kinematic variables in a way that has not previously been possible. The excellent sample purity and good resolution of \dword{ndgar} will facilitate precision measurements with small systematic uncertainties. In combination, the neutrino-argon cross section program at the DUNE near detector will go well beyond the current state-of-the-art on any nucleus.

For each of the exclusive samples, it is possible to study the dependence on various parameters, such as the squared four-momentum transfer $Q^2$, \dword{hadw} and energy and three-momentum transfer.
These distributions are heavily dependent on the interaction model used to generate the simulation, suggesting that the two detectors' data will be useful for distinguishing between models.

\begin{dunefigure}[$W$-dependence of generator models vs pion multiplicity]{fig:xsec-w-per-pion}
{The ratio of the reconstructed $Q^2$ distributions for \numu CC events for the \dword{nuwro} and \dword{genie} generators. Plots are shown for the \dword{ndlar} (left) and \dword{ndgar} (right), for final states including no pions (red), $1\pi^\pm$ (blue), $1\pi^0$ (green), 2 pions (purple) and 3 or more pions (orange).}
\includegraphics[width=0.45\textwidth]{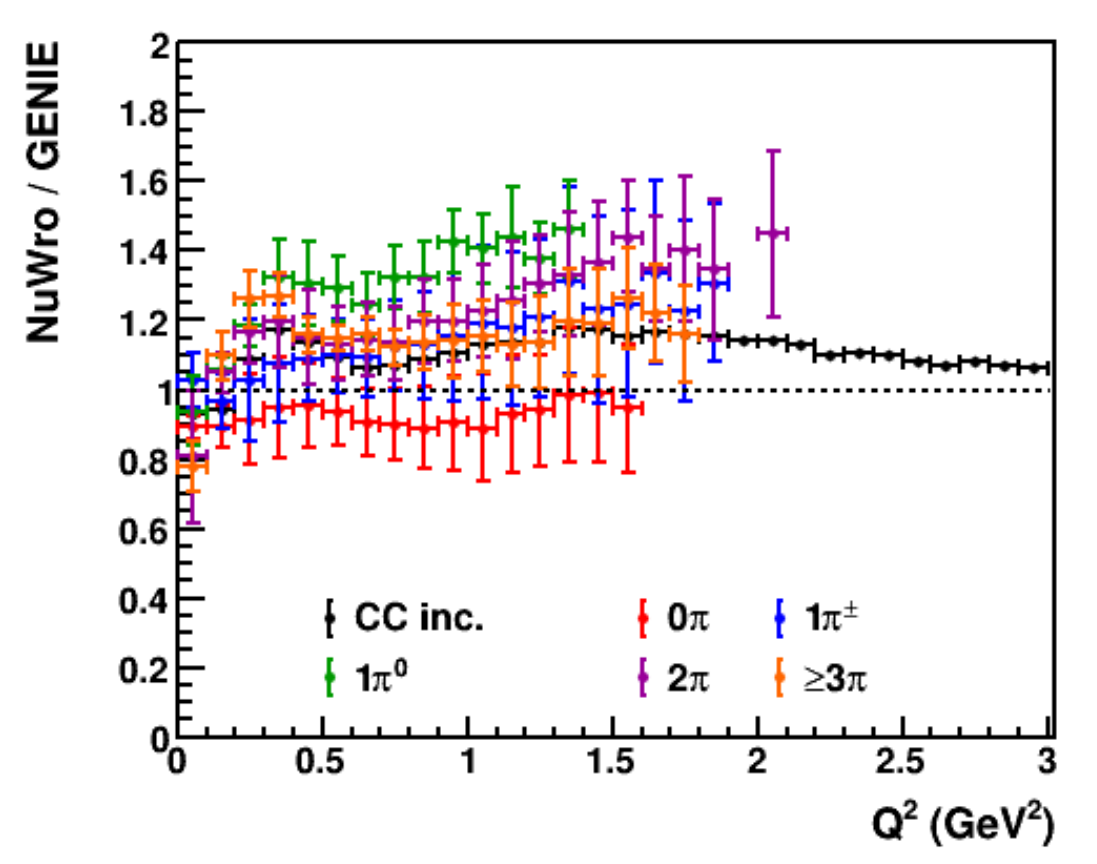}
\includegraphics[width=0.45\textwidth]{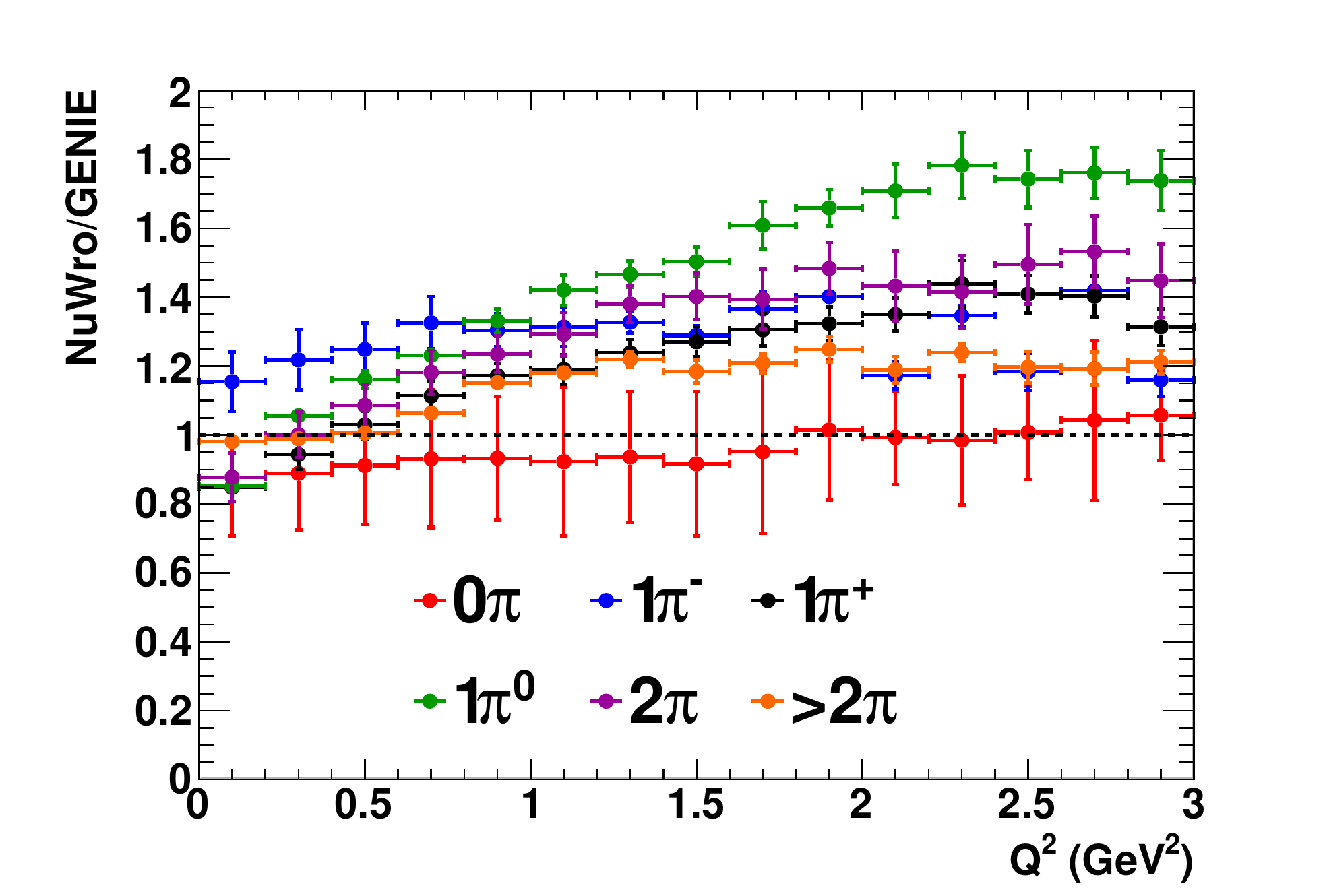}
\end{dunefigure}

Figure~\ref{fig:xsec-w-per-pion} shows the predicted $Q^2$ distributions for \numu CC interactions when the \dword{nuwro} event generator is used relative to that when the nominal \dword{genie} generator is used \footnote{In this study, for practical reasons the \dword{nuwro} sample was formed by reweighting the \dword{genie} sample in an 18-dimensional space using a boosted decision tree algorithm.}.  $Q^2$ is reconstructed from muon kinematics, using \ref{eq:nu-xsec:qsq}. (The neutrino energy is estimated by summing reconstructed final-state particle energies.)  It is plotted for the \dword{ndlar} (left) and \dword{ndgar} (right), for each of five different pion-multiplicity final states. The confusion matrices of Figure~\ref{fig:xsec-confusion} are used to estimate the uncertainty due to pion miscounting. Though the uncertainties are large, there is significant model spread between the \dword{nuwro} and \dword{genie} distributions, particularly where the final state includes pions. The \dword{nd} data are sensitive to these differences, and can be used both for model down-selection as well as tuning.


\subsection{Investigating Nuclear Effects Through Transverse Kinematic Imbalance}
\label{ch:nu-xsec:trans-vars}


Nuclear effects can make events arising from distinctly different interaction channels or processes indistinguishable experimentally.  This makes it very difficult to tease apart the effects for greater understanding. So, an important goal of neutrino interaction physics is to isolate observables that give a good separation between interaction channels or that isolate particular processes.  Measuring channels well where they can be separated provides confidence in the modeling of the confusion.


In recent work, some success has been achieved in learning about the nuclear effects in \dword{ccqe}-like samples by studying the \dword{tki} of the interaction products \cite{Abe:2018pwo, Lu:2018stk,Cai:2019hpx}. The transverse kinematic variables in use are shown in Figure~\ref{fig:xsec-trans-vars}.

\begin{dunefigure}[Schematic of transverse kinematic imbalance variables]{fig:xsec-trans-vars}
{Schematic definition of the transverse kinematic imbalance variables, from \cite{Lu:2018stk}}
\includegraphics[width=0.5\textwidth]{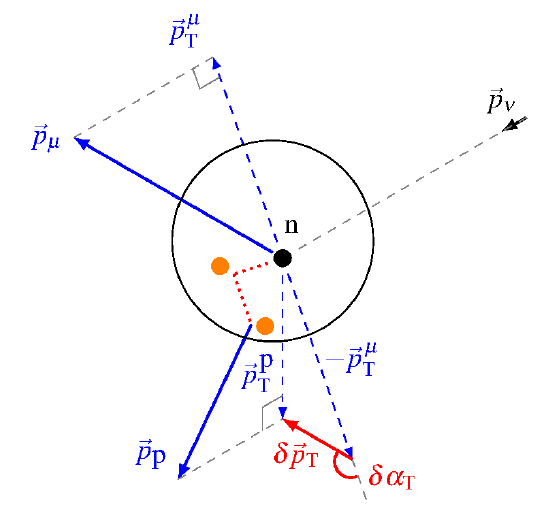}
\end{dunefigure}

For a neutrino scattering from a stationary target, the initial state momentum is along the neutrino's direction of travel.  Therefore, in this limit for \dword{ccqe} events, the transverse components of the final-state muon and proton should sum to zero.
Thus, for a \dword{ccqe} interaction on a non-stationary target neutron (where there is no intranuclear momentum transfer (\dword{imt}), which may arise from final-state interactions, or nucleon-nucleon correlations), the magnitude of the final-state transverse momentum imbalance should indicate the initial momentum of the neutron $\delta p = p_n$. This can be reconstructed from the final-state muon and proton kinematics, by the procedure outlined in \cite{Lu:2018stk}.
Furthermore, as the Fermi motion of a neutron in the nucleus is entirely independent of the neutrino interaction, any transverse momentum imbalance due to this Fermi motion should be isotropic. Thus, by looking at the direction of this imbalance, we can measure \dword{imt} effects. 

Consider the variable $\delta\alpha_\textrm{T}$ \cite{Lu:2015tcr}, defined by:
\begin{equation}
	\label{eq:nu-xsec:delta-alpha-t}
\delta\alpha_\textrm{T}\equiv\arccos\frac{-\vec{p}_\textrm{T}^{\,\mu}\cdot\delta\vec{p}_\textrm{T}}{p_\textrm{T}^{\,\mu}\delta p_\textrm{T}}
\end{equation}
In the absence of \dword{imt}, the distribution of $\delta\alpha_\textrm{T}$ should be flat. \dword{fsi} effects that accelerate the proton will lead to increased events at low values of $\delta\alpha_\textrm{T}$, whereas \dword{fsi} in which the proton has been decelerated will cause an increase in events at higher $\delta\alpha_\textrm{T}$. Thus, this variable can be used to investigate the strengths of the different \dword{fsi} components.


Analysis of \dword{minerva} and \dword{t2k} neutrino-carbon scattering data in these variables and projections of them \cite{Lu:2018stk,Abe:2018pwo,Cai:2019hpx} quantitatively show poor agreement with most generator models \cite{Dolan:2018zye,Dolan:2018sbb}. However, qualitatively there is particular difficulty in correctly modelling the transition between the Fermi motion dominated region and the region dominated by \dword{imt} in the \dword{minerva} measurement. With its excellent particle identification ability, the \dword{dune} \dword{nd}  will allow the use of transverse kinematic imbalance to investigate these effects in argon, perhaps shedding light on A-dependence.

Figures~\ref{fig:xsec-dalphat-minerva} and \ref{fig:xsec-pn-minerva} demonstrate the \dword{dune} detector's increased power in studying these kinematic imbalances, by comparing predicted distributions with those from \dword{minerva}. The predictions use the \dword{gibuu} model, which has been found to give good agreement with \dword{minerva} data. In each case, the $\nu_\mu$ flux is used to generate the simulation, and a final state consisting of a muon, proton, and no additional pions is considered. However, DUNE has a significantly larger phase space, with a full $4\pi$ angular acceptance, and a momentum acceptance of $p_\mu > $ \SI{0.0254}{\GeV\per c} and $p_p > $ \SI{0.0751}{\GeV\per c}. (\dword{minerva}, which relied on the MINOS near detector for muon charge identification and momentum determination, and whose technology led to challenges in proton/pion discrimination, was restricted to \SI{1.5}{\GeV\per c} $ < p_\mu < $ \SI{10}{\GeV\per c};  $\theta_\mu < 20^\circ$ and 
\SI{0.45}{\GeV\per c} $ < p_p < $ \SI{1.2}{\GeV\per c};  $\theta_p < 70^\circ$). Additionally, the \dword{minerva} detector is made of carbon-based scintillator (isospin $T=0$ for carbon), while the isospin $T=2$ for \dword{dune}'s argon.

\begin{dunefigure}[Predicted $\delta\alpha_\textrm{T}$  distribution for  MINERvA and DUNE near detector]{fig:xsec-dalphat-minerva}
{Differential cross section in the transverse boosting angle, $\delta\alpha_\textrm{T}$, as defined in Equation~\ref{eq:nu-xsec:delta-alpha-t}, for \dword{minerva} (left) and for argon (right) in  \dword{ndgar}.  The middle plot is for a carbon detector with the acceptance of \dword{ndgar} to separate the detector design from the nucleus for the comparison.}
\includegraphics[width=0.32\textwidth]{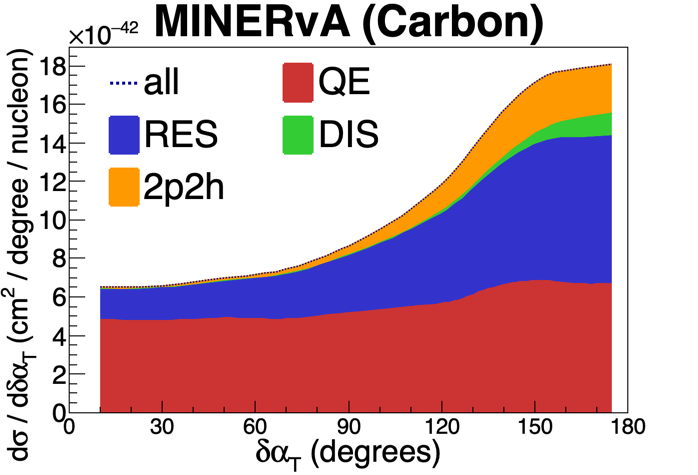}
\includegraphics[width=0.32\textwidth]{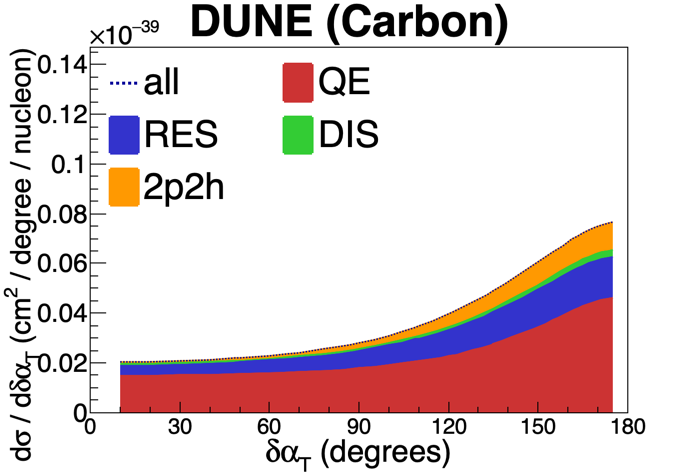}
\includegraphics[width=0.32\textwidth]{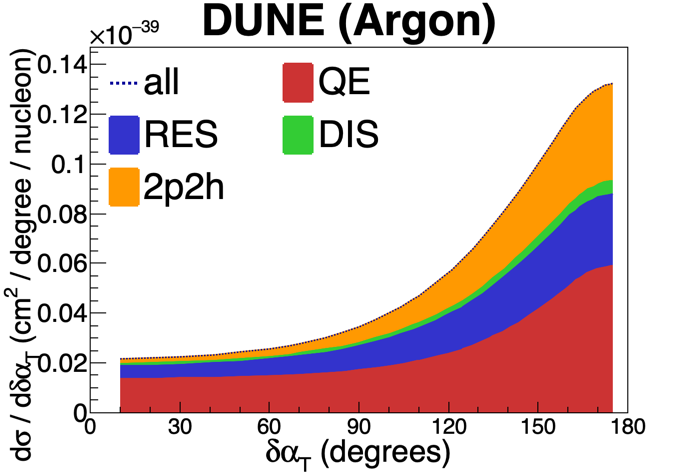}
\end{dunefigure}

Figure~\ref{fig:xsec-dalphat-minerva} shows the differential cross section as a function of the boosting angle 
$\delta\alpha_\textrm{T}$~\cite{Lu:2015tcr}. By comparing the prediction for neutrino-carbon scattering with the \dword{minerva} (left) and \dword{dune} (middle) phase spaces, it can be seen that \dword{dune}'s increased acceptance allows the detection of more low-energy protons, giving more quasi-elastic events with a higher $\delta\alpha_\textrm{T}$. When  the larger nuclei of \dword{dune}'s argon-based detector (right) are taken into account, there is an increased contribution from \dword{fsi} effects, leading to additional strength at high $\delta\alpha_\textrm{T}$ as compared with carbon (middle). This increased \dword{fsi} strength also leads to a larger $CC0\pi$ contribution from \dword{res} and \dword{dis} events followed by pion absorption. In the model, argon's higher isospin ($T=2$ vs. carbon's $T=0$) leads to an increase in high-$\delta\alpha_\textrm{T}$ \dword{2p2h} events.

\begin{dunefigure}[Predicted initial-state neutron momentum  distributions for  MINERvA and the DUNE near detector]{fig:xsec-pn-minerva}
{Differential cross section in the emulated nucleon momentum, $p_n$,  for \dword{minerva} (left), as studied in \cite{Lu:2018stk}, and for  argon (right) in \dword{ndgar}. The middle plot is for a carbon detector with the acceptance of \dword{ndgar} to separate the detector design from the nucleus for the comparison.}
\includegraphics[width=0.32\textwidth]{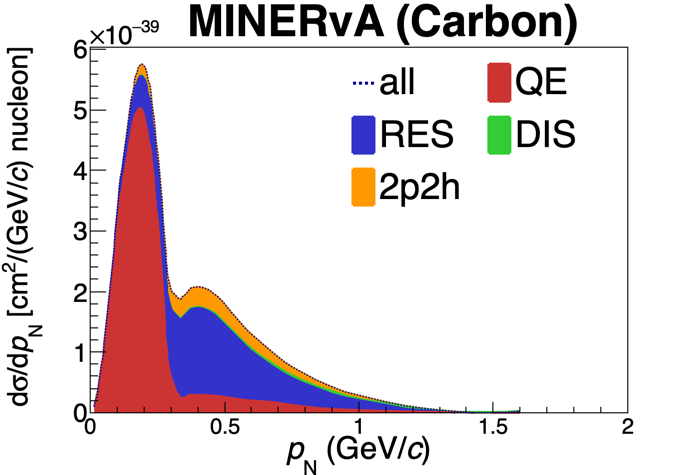}
\includegraphics[width=0.32\textwidth]{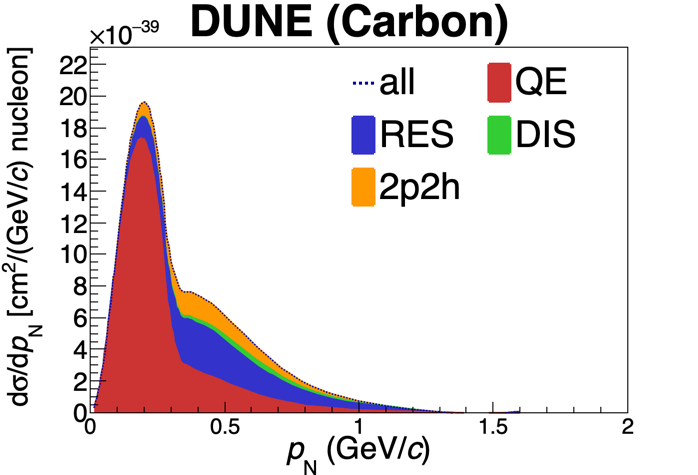}
\includegraphics[width=0.32\textwidth]{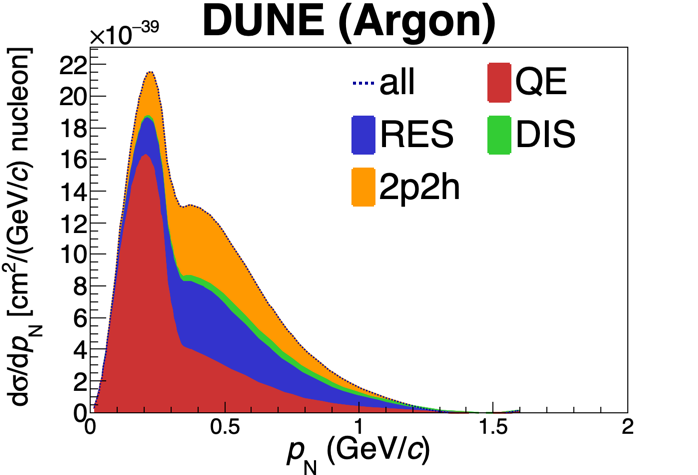}
\end{dunefigure}

Figure~\ref{fig:xsec-pn-minerva} shows the emulated nucleon momentum, $P_N$, which corresponds to the initial-state neutron momentum~\cite{Furmanski:2016wqo,Lu:2019nmf}. The \dword{dune} phase space in the neutrino-carbon scattering distribution (middle) shows a larger high $P_N$ tail than that for \dword{minerva} (left), due to \dword{dune}'s ability to identify low-momentum protons in the transition region which are below threshold for \dword{minerva}. In argon (right) a modest decrease is seen in \dword{ccqe} events at low momenta around \SI{0.2}{\GeV\per c} and increase above \SI{0.4}{\GeV\per c} due to increased \dword{fsi}, though this effect is small in this distribution that focuses on the initial state. However, an increase is seen, noticeable in both shape and magnitude in non-quasielastic events that experience \dword{fsi} in argon's large nucleus, and from additional \dword{2p2h} due to the isospin-2 nucleus.

In addition to these studies, by decomposing the momentum imbalance into its components along the Cartesian coordinate system defined by the neutrino and muon kinematics \cite{cai2019nuclear},  \dword{minerva} has been able to examine the base nuclear models used in generators, comparing the shape of their \dword{qe} peaks to data, observing discrepancies with \dword{genie}'s relativistic Fermi gas model. Other models, such as \dword{neut} and \dword{nuwro}'s spectral function models, as well as the \dword{gibuu} implementation, gave somewhat better agreement to the \dword{minerva} data. The comparisons underscore the need for generators that can better describe exclusive scattering data.  Again, no such study has yet been made on argon, leaving this to be done by experiments in the SBN progam and the \dword{dune} near detector.

\cleardoublepage

\chapter{Other Physics Opportunities with the ND}
\label{ch:bsm}

The physics driver for the design of the \dword{dune} \dword{nd} is the 3-flavor neutrino oscillation program.  As presented in this document, the \dword{nd} design optimal for this program is a very capable detector with a combination of technologies and targets.  When taken together with the high-intensity \dword{lbnf} proton and neutrino beams, the \dword{dune} \dword{nd} is a powerful laboratory for studying many \dword{sm} and \dword{bsm} physics topics.  \dword{dune} will take advantage of this and work to produce competitive and novel measurements in these areas, where possible. 

This chapter presents an incomplete survey of some of the \dword{sm} and \dword{bsm} topics of interest that might be explored by \dword{dune}. Much of the work here is in an early stage, as the design of the \dword{nd} has been evolving and the reconstruction software is not yet in place.  Also, particularly for the \dword{bsm} topics, it should be noted that the experimental and theoretical landscape may change before the \dword{nd} takes neutrino data.

In this chapter, Section~\ref{sec:bsm} presents a number of \dword{bsm} topics that illustrate the capabilities of the \dword{dune} \dword{nd} in this arena.  Where possible, estimates of potential sensitivity are given.  Following that, Section~\ref{sec:sm} discusses a number of interesting \dword{sm} physics measurements that might be done with the \dword{dune} \dword{nd}.

\section{Beyond the Standard Model Physics} 
\label{sec:bsm}

The role of the DUNE \dword{nd}  in most of the \dword{bsm} physics topics comes about by virtue of the intense LBNF beam and short baseline.
In the DUNE \dword{nd}, the unmagnetized liquid argon detector in \dword{ndlar} is followed downstream by \dword{ndgar}, which includes a highly capable, low-density tracker in a magnetized volume. This enables the DUNE \dword{nd} to achieve excellent momentum resolution for charged tracks produced in the \dword{ndlar} target volume that enter the  tracker region of \dword{ndgar}. Both \dword{ndlar} and \dword{ndgar} will take data off-axis in order to enable measurements of neutrino fluxes with different energies. Improved sensitivities to \dword{bsm} signatures will be possible in the off-axis locations due to the lower neutrino background. 

Relative to the Short-Baseline Neutrino program (SBN) program, the higher energy and the improved vertex resolution, charge measurement and momentum resolution available in the DUNE ND complex will extend the range of \dword{bsm} searches. 
The extent of the gain depends on the specific signature and analysis strategy employed.

The \dword{hpgtpc} in \dword{ndgar} may, in principle, give access to novel \dword{bsm} signatures beyond those accessible with liquid argon detectors, as the lower density enables a more precise measurement of vertex activity around the primary interaction.  It may also enable detection of electromagnetic \dword{bsm} signatures which would be difficult to detect in liquid argon, and allow for improved reconstruction of \dword{bsm} signatures such as tridents. The sign selection of charged particles via the magnetic field is also important. 

\subsection{Searches for light dark matter} 
\label{sec:bsm-dm}
A number of cosmological and astrophysical observations provide evidence for the existence of dark matter (DM) that constitutes $\sim$27\% of the mass-energy content of the universe but whose nature is still unknown~\cite{Kisslinger:2019ysx}. A compelling scenario is one where DM is made of particles that were in thermal equilibrium with the plasma  of the early universe due to their interactions with \dword{sm} particles. The production mechanism of DM in the early universe,  as well as the nature of DM interactions with \dword{sm} particles outside of gravity, are currently not understood. 

Recently, substantial attention has been paid to prospects for detecting light DM at neutrino  experiments with intense proton beams, such as DUNE. One possible scenario to ensure the correct thermal relic density involves DM states annihilating via light mediators. Consider a model in which the DM is a light weakly interacting massive particle (WIMP), below the GeV scale, that interacts with \dword{sm} particles via the exchange of a new light vector boson, allowing a coupling between the DM and the \dword{sm} at the renormalizable level. In the case of a gauge boson associated with a local $U(1)$ symmetry that mixes kinetically  with the photon as the mediator, light DM particles can be produced in the collision of protons on a target. For DUNE, these DM particles would travel to the DUNE \dword{nd}, where they could be detected through neutral-current-like interactions with the electrons or nucleons in the detector material via elastic scattering. Neutrinos constitute the main background for such a light DM searches. Interactions of DM with nuclei will have an experimental signature very similar to NC neutrino interactions on nuclei while DM-electron scattering would look like $\nu\,e^-\to\nu\,e^-$ or $\nu_e\,N\to e^- N'$ processes. These neutrino-scattering backgrounds can be suppressed using the timing and kinematics of the final-state electron or nucleons in the \dword{nd}. In addition, an effective way to reduce the neutrino-induced background in such a search is to look at events coming from an off-axis neutrino beam. A recent study considering the use of an off-axis beam for a DM search  shows a significant improvement in search sensitivity compared to on-axis data taking~\cite{DeRomeri:2019kic}. Neutrinos come from decays of charged mesons, which are focused by the magnetic horn system in the forward direction. Since DM is produced from the decay of neutral (unfocused) mesons,  the neutrino rate falls off faster than the DM rate when going off-axis.  This yields a substantial improvement in the signal to background ratio for the off-axis sample.

Consider a benchmark model in which the \dword{sm} gauge group is extended by an additional ``dark'' $U(1)_D$~\cite{DeRomeri:2019kic}. The \dword{sm} particle content is also extended to contain a new massive dark photon $V$ and the DM, which is a fermion $\chi$, charged under the (broken) $U(1)_D$ symmetry with a dark fine structure constant $\alpha_D \equiv \frac{g_D^2}{4 \pi}$. The relevant terms of the Lagrangian are
\begin{equation}
\mathcal{L} \supset -\frac{\varepsilon}{2} F^{\mu\nu} F^\prime_{\mu \nu} + \frac{M_{V}^2}{2} V_\mu V^{\mu} + \overline{\chi}i \gamma^\mu \left(\partial_\mu - ig_D  V_\mu\right)\chi - M_\chi \overline{\chi}\chi,
\label{eq:DMlagr}
\end{equation}
where $\varepsilon$ is the kinetic mixing parameter between the \dword{sm} $U(1)$ and the new $U(1)_D$, and $M_{V}$ and $M_\chi$ are the dark photon and DM masses, respectively. It is assumed that the DM is a thermal relic and that its initial abundance is isotropic in space. In this case, the DM/$V$ masses and couplings are such that the relic abundance matches the observed DM abundance in the universe. At DUNE, the DM flux will be dominantly produced in the decays of light pseudoscalar mesons -- mainly $\pi^0$ and $\eta$ -- that are produced in the DUNE target, and from proton bremsstrahlung processes $p + p \rightarrow p + p + V$. Assuming that the DM is lighter than half the mass of a pseudoscalar meson $\mathfrak{m}$ produced in the DUNE target, the DM is produced via two decays, those of on-shell $V$ and those of off-shell $V$, shown in Figure~\ref{fig:DMproduction}.

\begin{dunefigure}[DM production via pseudoscalar meson decays]{fig:DMproduction}
{Production of fermionic DM via two-body neutral pseudoscalar meson decay $\mathfrak{m} \to \gamma V$, when $M_{V} < m_\mathfrak{m}$ (left) or via three-body decay with off-shell $V$ $\mathfrak{m} \to \gamma \chi \overline{\chi}$ (right).}
\includegraphics[width=0.25\linewidth]{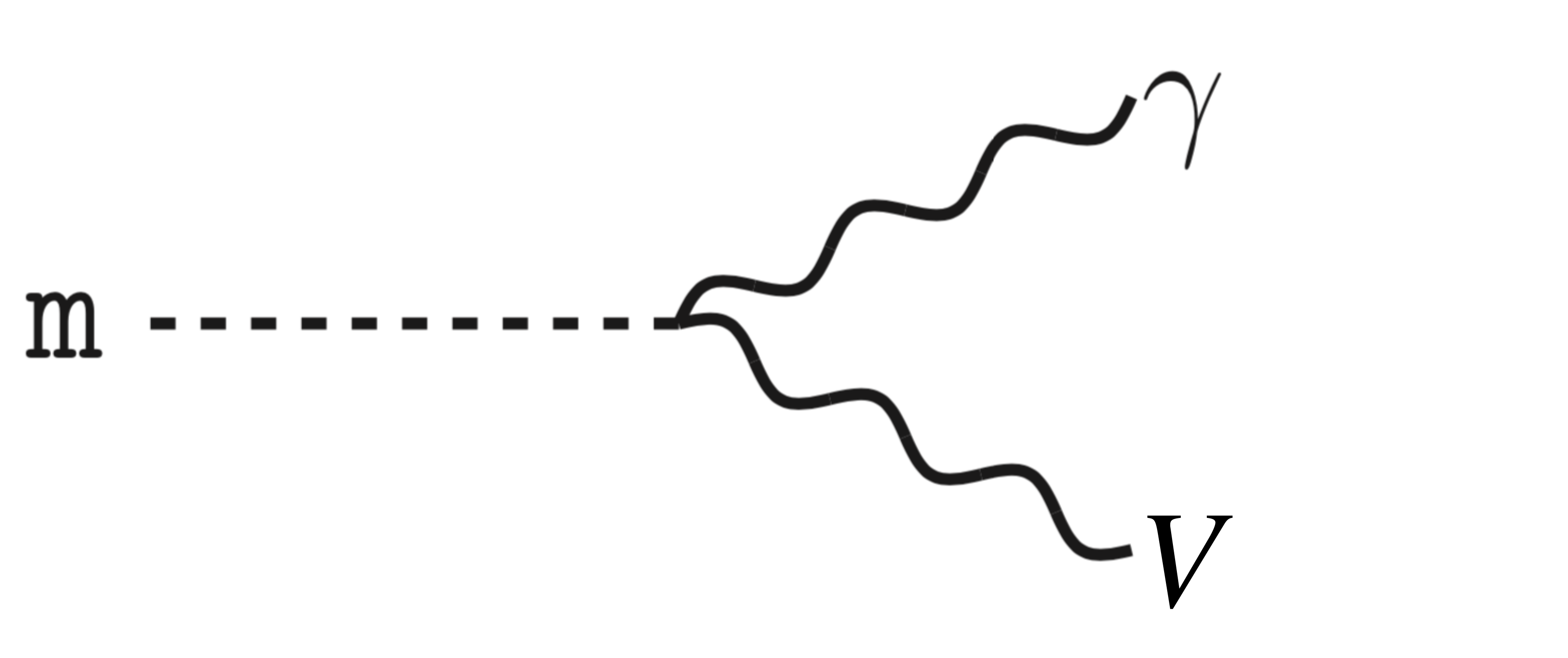}
\includegraphics[width=0.25\linewidth]{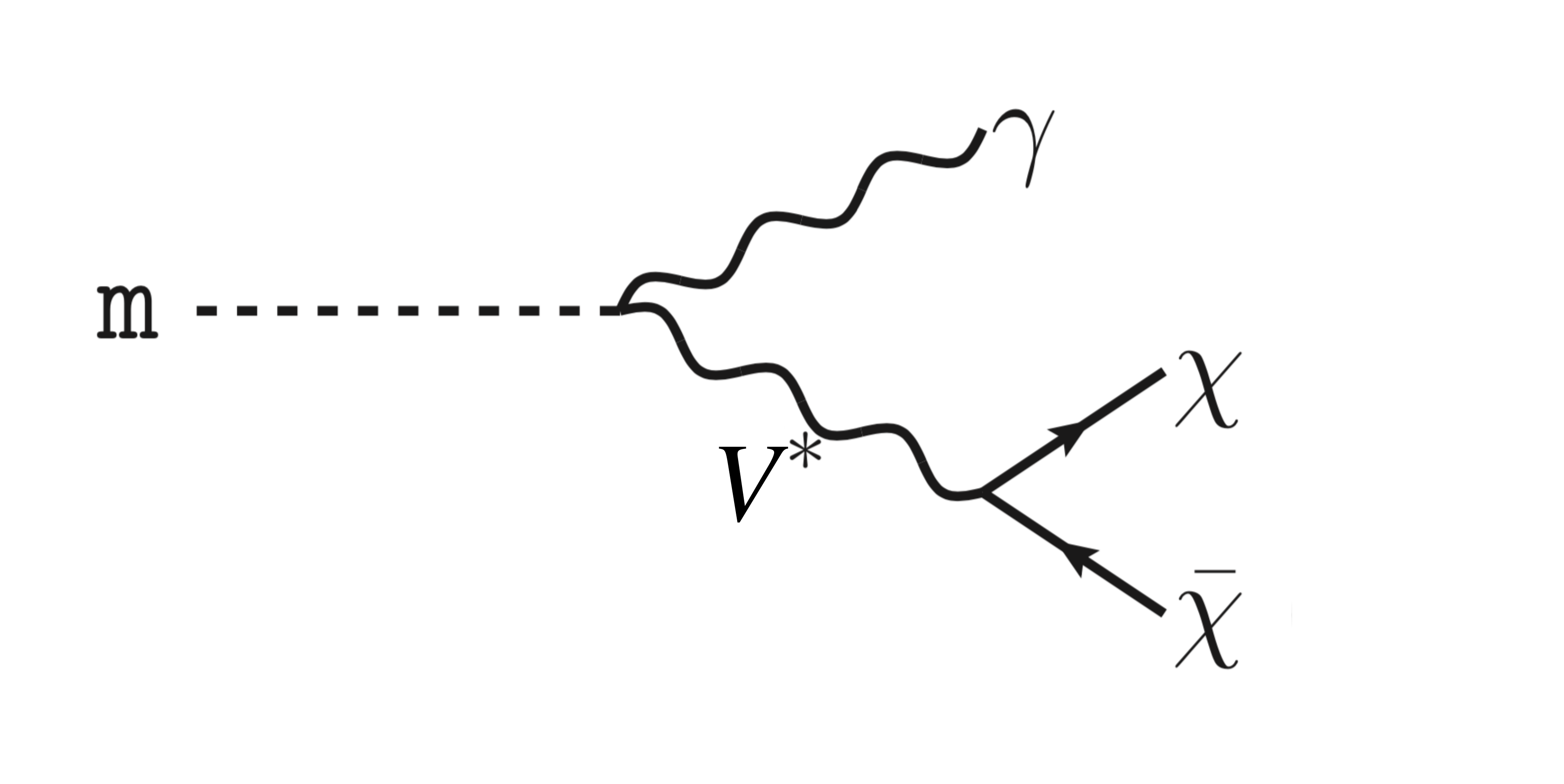}
\end{dunefigure}


For the range of dark photon and DM masses in which DUNE will set a competitive limit, the DM flux due to meson decays will dominate over the flux due to proton bremsstrahlung. 
If the DM reaches the \dword{nd}, it may scatter elastically off nucleons or electrons, via a $t$-channel dark photon. The focus here is on scattering off electrons because it has a smaller background. 
The differential cross section of this scattering, as a function of the recoil energy of the electron $E_e$, is given by
\begin{equation}
\frac{d\sigma_{{\chi}e}}{dE_{e}} 
= 4\pi \epsilon^{2}\alpha_D\alpha_{\mathrm{EM}} \frac{2m_{e}E_{\chi}^{2} - (2m_{e}E_{\chi} + m_{\chi}^{2})(E_e-m_{e})}{(E_e^{2}-m_{\chi}^{2})(m_{V}^{2}+2m_{e}E_{e}-2m_{e}^{2})^{2}}\,,
\end{equation}
where $E_{\chi}$ is the incoming DM energy and $\alpha_{\mathrm{EM}}$ is the electromagnetic fine-structure constant. 

The background to this scattering signal consists of any processes involving an electron recoil. As the \dword{nd} is located near the surface, some background events can be induced by cosmic rays but they will be vetoed by triggers and timing information. The dominant background will be from neutrinos coming in the DUNE beam, consisting of neutrinos scattering off electrons ($\nu_\mu e^- \to \nu_\mu e^-$ via a $Z$ boson) and electron (anti)neutrinos interacting with nucleons via charged-current processes ($\nu_e n \to e^- p$ or $\overline{\nu}_e p \to e^+ n$). The latter process has a much larger rate ($\sim$ 10 times higher) than the former and it does not look like the signal. It can be reduced by placing a cut on the outgoing electron kinematics, using the variable $E_e \theta_e^2$, where $\theta_e$ is the direction of the outgoing electron relative to the beam direction. Uncertainties in the $\nu_\mu$ flux could complicate such an analysis. Though studies are ongoing, a $10\%$
normalization uncertainty on the expected background rate is assumed here. Background events from $\pi^0$ mis-identification are expected to be subdominant thanks to the kinematic cut ($E_e\theta_e^2$) and the d$E/$d$x$-based particle identification of the ND-LAr.   

Assuming 3.5 years of data collection each in neutrino and antineutrino modes, results are shown for the case that all data are collected with the DUNE \dword{nd} on-axis and that where data collection is divided equally among four off-axis positions (0.7 years at each of \SI{6}{m}, \SI{12}{m}, \SI{18}{m}, and \SI{24}{m} off-axis). 
 Statistical, correlated systematic, and uncorrelated systematic errors are considered for each bin. For the correlated systematic uncertainty, a nuisance parameter A is included that modifies the number of (anti)neutrino-related background events in all bins (independently for neutrino and antineutrino beam modes) and it is assumed there is an overall flux-times-cross-section uncertainty which has a Gaussian probability with width $\sigma_A = 10\%$. The uncorrelated uncertainty in each bin is assumed to be parameterized by a Gaussian with a much narrower width, $\sigma_{f_i} = 1\%$. These uncertainties are included in the following test statistic as nuisance parameters and then marginalized over in producing a resulting sensitivity reach:
\begin{eqnarray}
\label{eq:teststat}
-2\Delta \mathcal{L} = \sum_i \frac{r_i^m\left( \left(\frac{\varepsilon}{\varepsilon_0}\right)^4 N_i^\chi + (A-1)N_i^\nu\right)^2}{A\left(N_i^\nu + (\sigma_{f_i} N_i^\nu)^2 \right)} + \frac{\left(A-1\right)^2}{\sigma_A^2}.
\end{eqnarray}
$N_i^\chi$ is the number of DM scattering events, calculated assuming $\varepsilon$ is equal to a reference value $\varepsilon_0 \ll 1$. $N_i^\nu$ is the number of irreducible background $\nu_\mu e^-$ scattering events expected in the detector at position $i$, and $r_i^m$ is the number of years of data collection in detector position $i$ during beam mode $m$ (neutrino or antineutrino mode).

The DUNE sensitivity assuming all on-axis data collection (DUNE On-axis) or equal times at each \dword{nd} off-axis position (DUNE-PRISM) are shown in Figure~\ref{fig:LDMsensitivity}. The 90\% CL sensitivity reach of the DUNE \dword{nd} is shown, assuming $\alpha_D = 0.5$ and $M_{V} = 3M_\chi$ (left panel) or $M_\chi = 20$ MeV (right panel).
\begin{dunefigure}[90$\%$ \dshort{cl} limit for Y as a function of $m_{\chi}$ at the ND]{fig:LDMsensitivity}
{Expected DUNE On-axis (solid red) and PRISM (dashed red) sensitivity using $\chi e^- \to \chi e^-$ scattering. We assume $\alpha_D = 0.5$ in both panels, and $M_V = 3M_\chi$ ($M_\chi = 20$ MeV) in the left (right) panel, respectively. Existing constraints are shown in grey, and the relic density target is shown as the black line. We also show for comparison the sensitivity curve expected for LDMX-Phase I (solid blue)~\cite{Akesson:2018vlm}.}
\includegraphics[width=\linewidth]{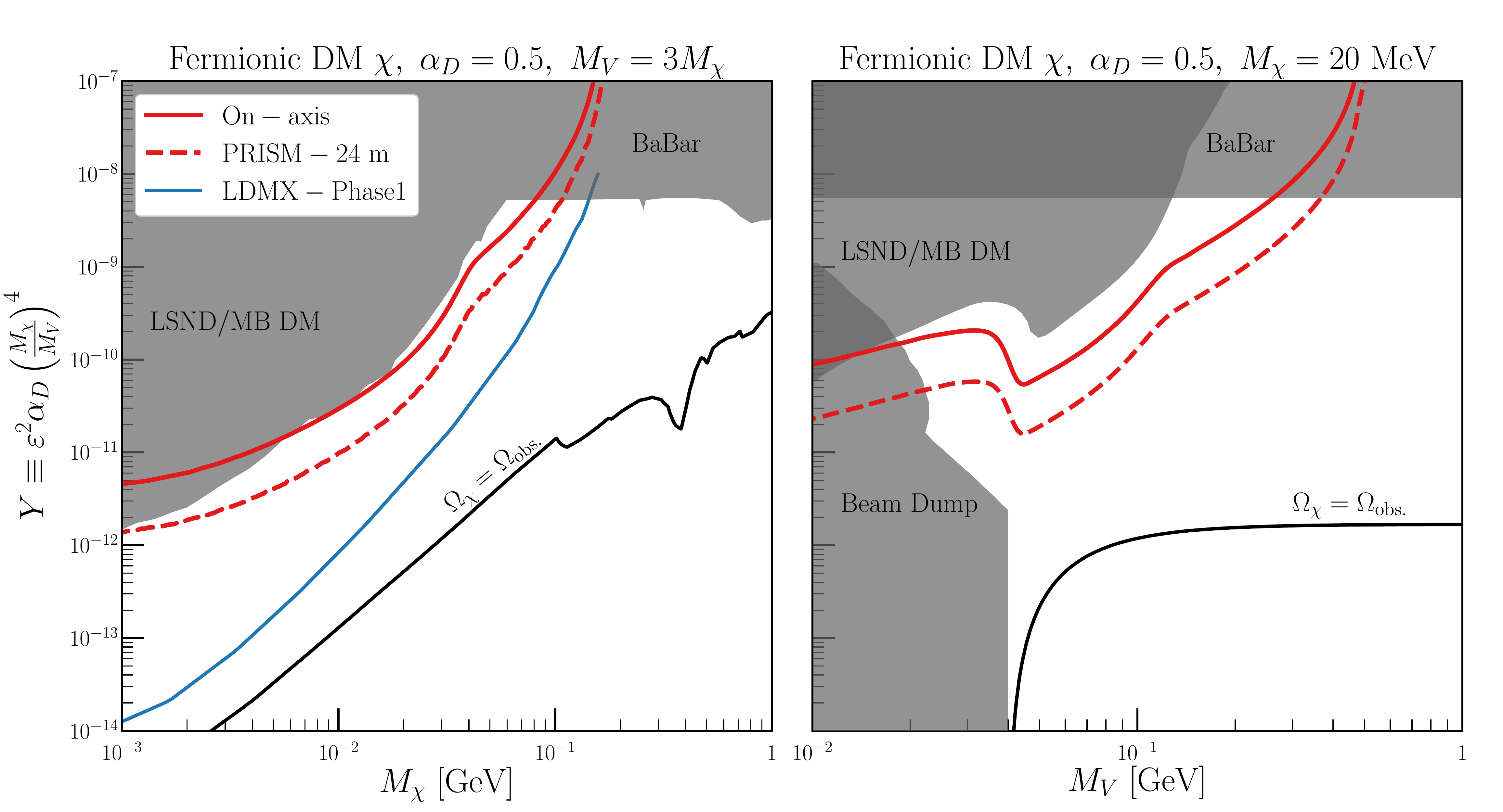}
\end{dunefigure}

Results are shown in terms of the DM or dark photon mass and the parameter $Y$, where
\begin{equation}
Y \equiv \varepsilon^2 \alpha_D \left(\frac{M_\chi}{M_V}\right)^4.    
\end{equation}
Assuming $M_V \gg M_\chi$, this parameter determines the relic abundance of DM in the universe today, and sets a theoretical goal in terms of sensitivity reach. In the same figure, the results of this study are compared to existing constraints, shown as grey shaded regions. The DUNE estimates significantly improve over those from LSND~\cite{deNiverville:2018dbu} and the MiniBooNE-DM search~\cite{Aguilar-Arevalo:2018wea}, as well as BaBar~\cite{Lees:2017lec} if $M_{V} \lesssim 200$ MeV. When $M_{V} < 2 M_\chi$, the limits from beam-dump experiments~\cite{Davier:1989wz,Batley:2015lha,Bjorken:1988as,Riordan:1987aw,Bjorken:2009mm,Bross:1989mp} are shown (right panel), as well as the lower bound obtained from matching the thermal relic abundance of $\chi$ with the observed one (black, dot-dashed). The sensitivity curves in the right panel show two interesting features related to the DM production mechanism. For a fixed $\chi$ mass, as $M_V$ grows, the DM production goes from off-shell to on-shell and back to off-shell. The first transition is responsible for the strong feature at $M_V=2M_\chi = 40$~MeV, while the second is the source for the slight kink around $M_V=m_{\pi^0}$. (The latter also appears in the left panel.)

\subsection{Neutrino tridents}
\label{sec:bsm-tridents}
Neutrino trident production is a rare weak process in which a neutrino, scattering off the Coulomb field of a heavy nucleus, generates a pair of charged leptons. The typical final state of a neutrino trident interaction contains two leptons of opposite charge (see Figure~\ref{fig:trident}). Table~\ref{tab:TridentEvents} shows the sizable number of trident events expected per year in the DUNE \dword{nd}. Both the excellent resolution and the magnetic field of \dword{ndgar}, which provides sign selection of the leptons in the final state, are likely to be very helpful in the reconstruction of neutrino trident events.


\begin{dunefigure}[Example diagrams for $\nu_\mu$-induced trident processes in the SM]
{fig:trident}
{Example diagrams for $\nu_\mu$-induced trident processes in the \dword{sm}. A second set of diagrams where the photon couples to the negatively charged leptons is not shown. Analogous diagrams exist for processes induced by different neutrino flavors and by antineutrinos. A diagram illustrating trident interactions mediated by a new $Z'$ gauge boson, discussed in the text, is shown on the top right.}
\centering
\includegraphics[width=0.3\textwidth]{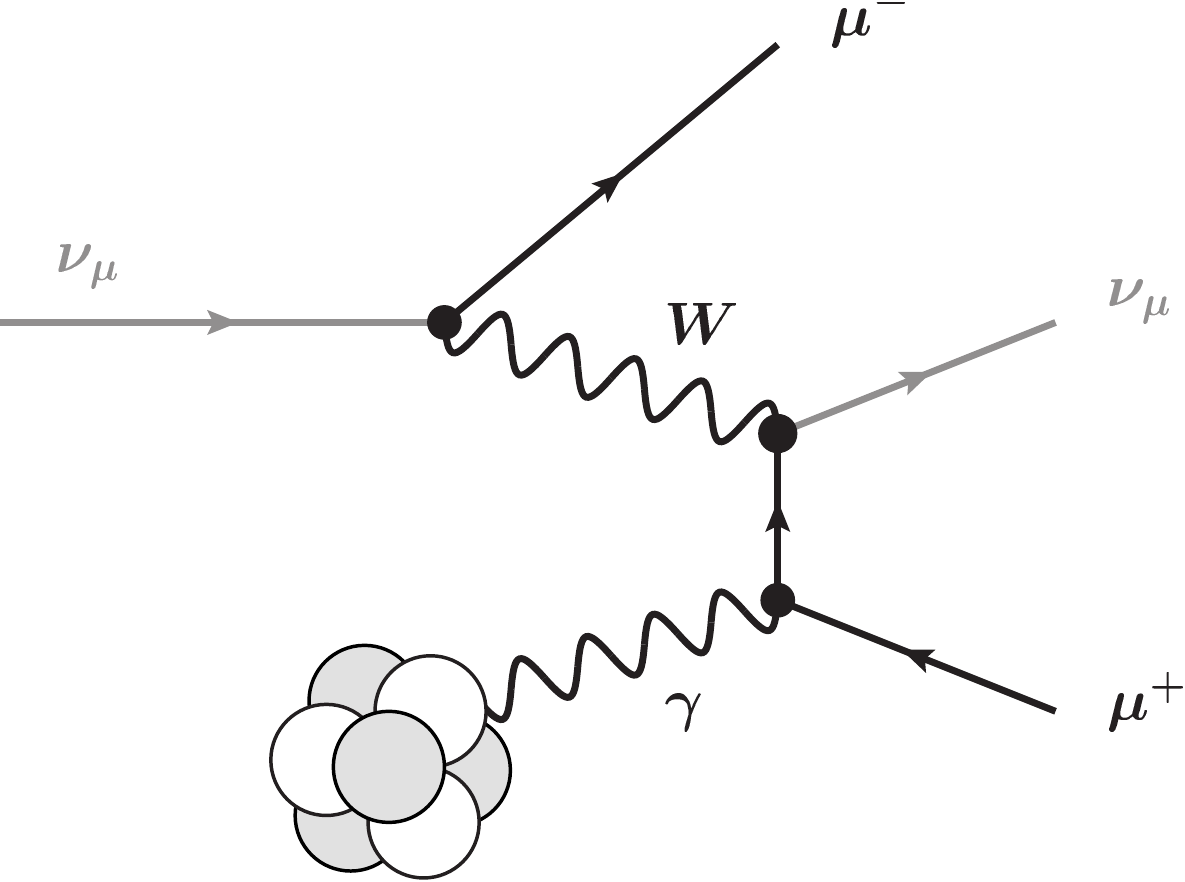} \qquad
\includegraphics[width=0.3\textwidth]{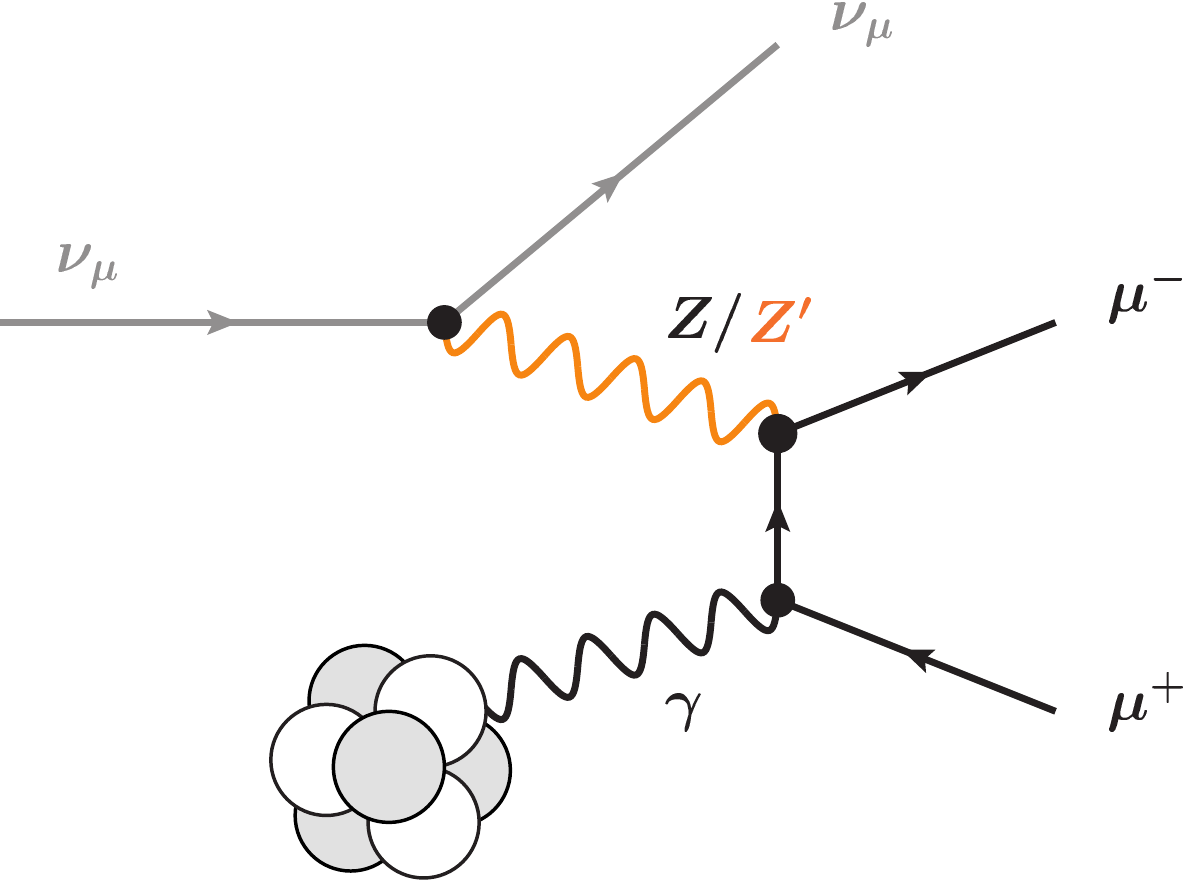} \\[\baselineskip]
\includegraphics[width=0.3\textwidth]{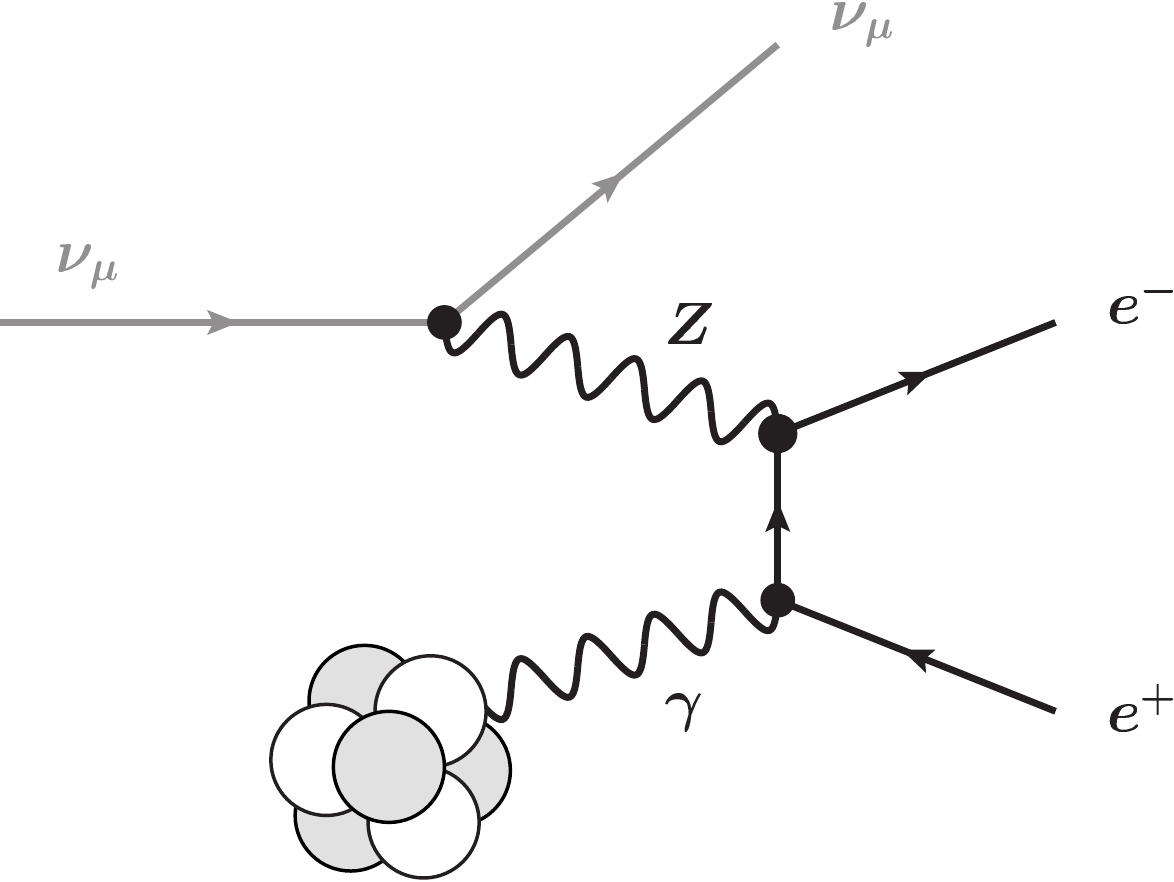} \qquad
\includegraphics[width=0.3\textwidth]{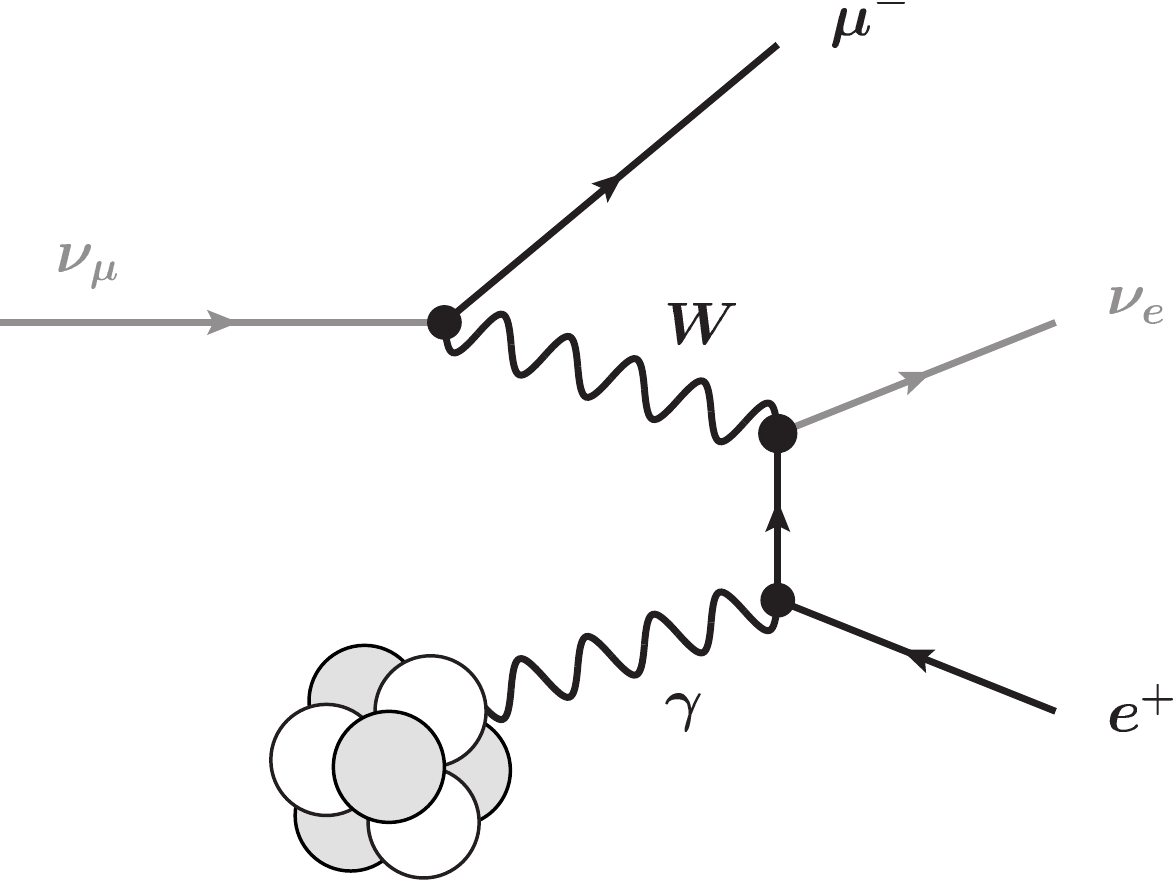} \\[\baselineskip]
\end{dunefigure}

Measurements of muonic neutrino tridents were carried out at the CHARM-II~\cite{Geiregat:1990gz}, CCFR~\cite{Mishra:1991bv}, and NuTeV~\cite{Adams:1999mn} experiments, and yielded results consistent with \dword{sm} predictions, but those measurements leave ample room for potential searches for new physics. As an example, a class of models that modify the trident cross section are those that contain an additional neutral gauge boson, $Z^{\prime}_0$, that couples to neutrinos and charged leptons. This $Z^{\prime}_0$ boson can be introduced by gauging an anomaly-free global symmetry of the \dword{sm}, with a particular interesting case realized by gauging L$_{\mu} -$L$_{\tau}$. Such a $Z^{\prime}_0$ is not very tightly constrained and could address the observed discrepancy between the \dword{sm} prediction and measurements of the anomalous magnetic moment of the muon, (g$-$2)$_{\mu}$.  DUNE can potentially discover or constrain the complete parameter space allowed for the $Z^{\prime}_0$ to explain the g-2 anomaly, as shown in Figure~\ref{fig:LmuLtau}. 

Another category of \dword{bsm} Physics models that can be probed through neutrino trident measurements are dark neutrino sectors. In these scenarios, \dword{sm} neutrinos mix with heavier 
singlet fermions (dark neutrinos) with novel interactions. Due to this mixing, neutrinos inherit or couple somewhat through the new interaction and may up-scatter to dark neutrinos. These heavy states in turn decay back to \dword{sm} fermions, giving rise to trident signatures. These scenarios can explain the smallness of neutrino masses and possibly the MiniBooNE low energy excess of events.  

\begin{dunetable}[Expected number of Standard Model trident events.]
{lcc}
{tab:TridentEvents}{Expected number of \numu{}(\anumu{})-induced Standard Model trident events at the DUNE near detector per ton of argon and year of operation in neutrino mode (first four rows) or antineutrino mode (last four rows). }
& ~~~Coherent~~~ & ~~~Incoherent~~~ \\ \toprowrule
$\nu_\mu \to \nu_\mu\ \mu^+\mu^-$& $1.17 \pm 0.07$ & $0.49 \pm 0.15$ \\ 
$\nu_\mu \to \nu_\mu\ e^+e^-$& $2.84 \pm 0.17$ & $0.18 \pm 0.06$ \\
$\nu_\mu \to \nu_e\ e^+\mu^-$ & $9.8 \pm 0.6$ & $1.2 \pm 0.4$ \\
$\nu_\mu \to \nu_e\ \mu^+e^-$ & $0$ & $0$ \\
$\bar\nu_\mu \to \bar\nu_\mu\ \mu^+\mu^-$& $0.72 \pm 0.04$ & $0.32 \pm 0.10$ \\
$\bar\nu_\mu \to \bar\nu_\mu\ e^+e^-$& $2.21 \pm 0.13$ & $0.13 \pm 0.04$ \\
$\bar\nu_\mu \to \bar\nu_e\ e^+\mu^-$ & $0$ & $0$ \\
$\bar\nu_\mu \to \bar\nu_e\ \mu^+e^-$& $7.0 \pm 0.4$ & $0.9 \pm 0.3$ \\
\end{dunetable}



\begin{dunefigure}[Existing constraints and DUNE sensitivity in the $L_\mu - L_\tau$ parameter space]{fig:LmuLtau}
{Existing constraints and projected DUNE sensitivity in the $L_\mu - L_\tau$ parameter space. Shown in green is the region where the $(g-2)_\mu$ anomaly can be explained at the $2\sigma$ level. The parameter regions already excluded by existing constraints are shaded in gray and correspond to a CMS search for $pp \to \mu^+\mu^- Z' \to \mu^+\mu^-\mu^+\mu^-$~\cite{Sirunyan:2018nnz} (``LHC''), a BaBar search for $e^+e^- \to \mu^+\mu^- Z' \to \mu^+\mu^-\mu^+\mu^-$~\cite{TheBABAR:2016rlg} (``BaBar''), a previous measurement of the trident cross section~\cite{Mishra:1991bv,Altmannshofer:2014pba} (``CCFR''), a measurement of the scattering rate of solar neutrinos on electrons~\cite{Bellini:2011rx,Harnik:2012ni,Agostini:2017ixy} (``Borexino''), and bounds from Big Bang Nucleosynthesis~\cite{Ahlgren:2013wba,Kamada:2015era} (``BBN''). The DUNE sensitivity shown by the solid blue line assumes 6.5 years of data running in neutrino mode, leading to a measurement of the trident cross section with 40\% precision.}
\includegraphics[width=0.9\textwidth]{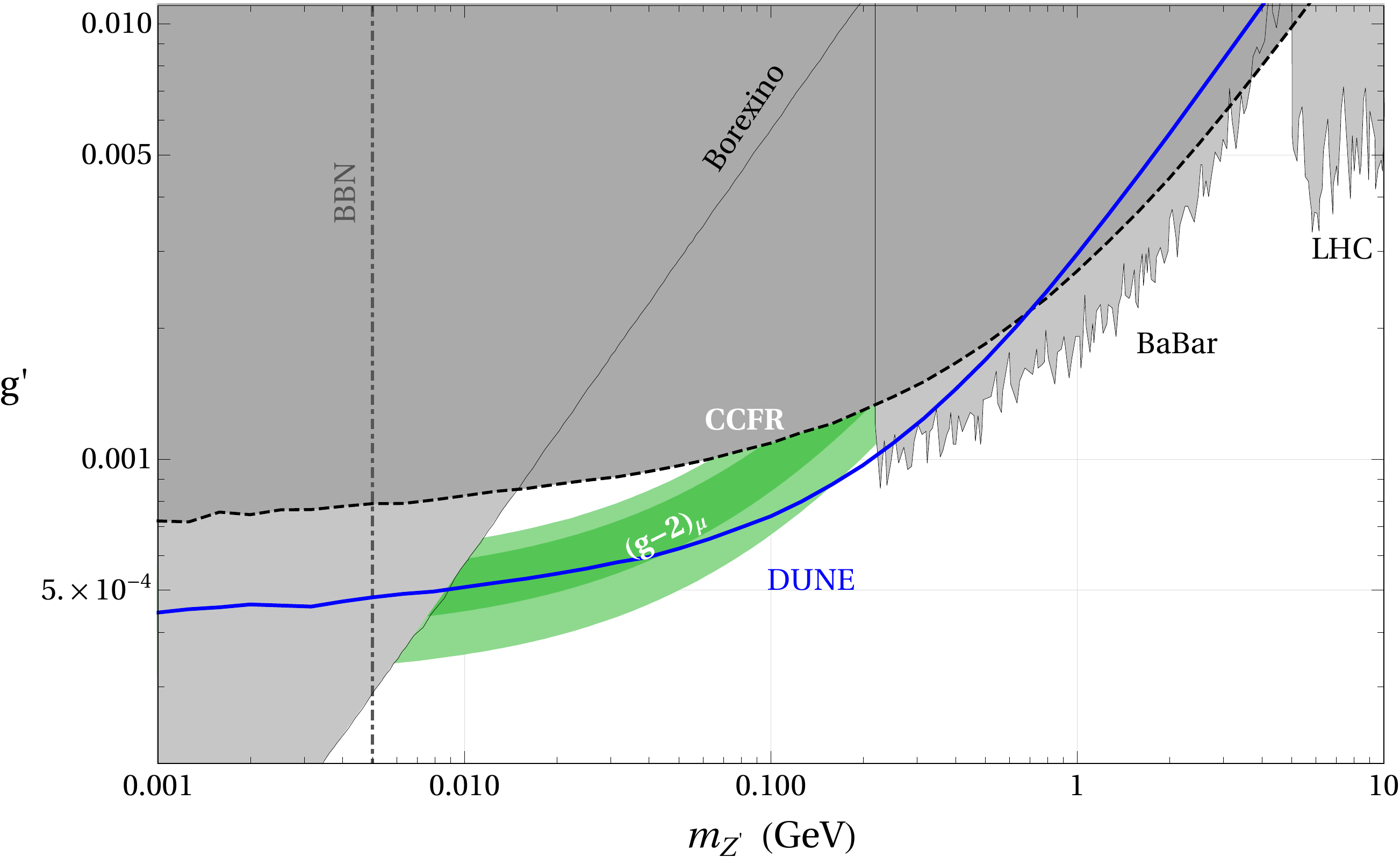}
\end{dunefigure}


\subsection{Search for Heavy Neutral Leptons} \label{sec:bsm-hnl}
The DUNE \dword{nd} can be used to search topologies of rare event interactions and decays that originate from very weakly-interacting long-lived particles, including heavy neutral leptons (HNL), right-handed partners of the active neutrinos, vector, scalar, or axion portals to the hidden sector, and light supersymmetric particles. Figure~\ref{fig:hnl_feynman_diagram} shows Feynman diagrams for some production processes. The high intensity of the LNBF source and the capability to  produce charmed mesons in the beam allow for accessing a wide variety of long-lived, exotic particles. Competitive sensitivity with possible future beam dump facilities, such as the one at CERN, is expected for the case of searches for decay-in-flight of sub-GeV particles that are also candidates for dark matter, and may provide an explanation for leptogenesis in the case of \dword{cpv} indications. DUNE would probe the lighter particles of the hidden sector.
The parameter space explored by the DUNE \dword{nd} extends to the cosmologically relevant region and is complementary to LHC heavy-mass dark-matter searches through missing energy and mono-jets.  It covers a similar range for HNL masses below 2 GeV as the one by the proposed SHiP experiment~\cite{SHiP:2018xqw}. Also, it can extend or confirm results from searches presently being carried out at NOvA or MicroBooNE, or in the near future with new SBN detectors.

Assuming these HNLs are the lightest particles of their hidden sector, they will only decay into \dword{sm} particles. 
Due to the expected small mixing angles, the particles can be stable enough to travel from the LBNF target to the \dword{nd} and decay inside the active fiducial region of the detector.
It is worth noting that, unlike a light neutrino beam, an HNL beam is not polarised due to the large HNL mass.
The correct description of the helicity components in the beam is important for predicting the angular distributions
of HNL decays, as they might depend on the initial helicity state.
In fact, there is a different phenomenology if the decaying HNL is a Majorana or a Dirac fermion~\cite{Balantekin:2018ukw, Ballett:2019bgd}.
Typical decay channels are two-body decays into a charged lepton and a pseudo-scalar meson, or a vector meson if
the mass allows it; two-body decays into neutral mesons; and three-body leptonic decays.

The results presented here are based on a recent study illustrating the potential sensitivity for  HNL searches with the \dword{dune} \dword{nd}~\cite{Ballett:2019bgd}, but are updated for the most recent LNBF neutrino flux predictions and include 
results for the antineutrino beam configuration. The sensitivity for HNL particles with masses in the range of 10 MeV to 2 GeV originating from decays of mesons produced in the neutrino beam target was studied.
The production of $D_s$ mesons (for both charges)  leads to high mass HNL production and also gives sensitivity to mixing in the tau sector. The dominant HNL decay modes to SM particles have been included, and basic detector constraints 
have been taken into account. 

\begin{dunefigure}[Diagrams for the production of HNLs and their decays.]{fig:hnl_feynman_diagram}
{Feynman diagrams (left) for the production of HNLs and (right) for their decays. The dashed line denotes the coupling to the Higgs vacuum expectation value, leading to the mixing of active neutrinos and HNLs via Yukawa couplings. Figure taken from \cite{Bonivento:2013jag}.}
\includegraphics[width=0.75\textwidth]{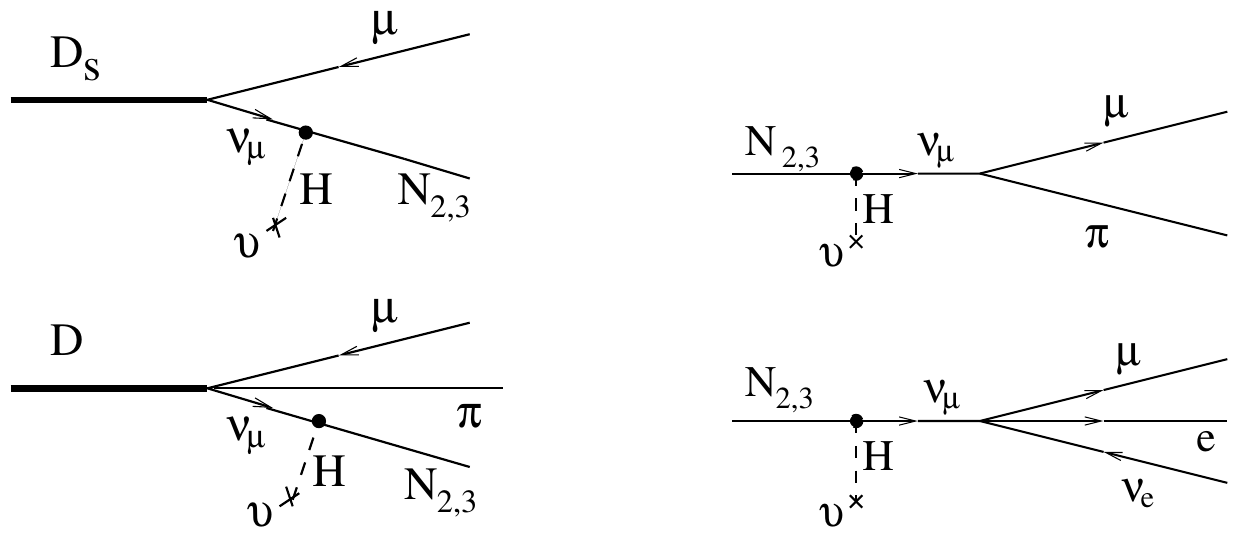}
\end{dunefigure}

The experimental signature for these decays is a decay-in-flight event with no interaction vertex, typical of
neutrino--nucleon scattering, and a rather forward direction with respect to the beam.
The main background to this search comes from \dword{sm} neutrino--nucleon scattering events in which the hadronic activity
at the vertex is below threshold.
Charged current quasi-elastic events with pion emission from resonances are background to the semi-leptonic decay channels,
whereas mis-identification of long pion tracks as muons can constitute a background to three-body leptonic decays.
Neutral pions are often emitted in neutrino scattering events and can be a challenging background for HNL decays that include a neutral meson or channels with electrons in the final state.

Figure~\ref{fig:sensa_hnl} shows the physics reach of the DUNE ND in its current configuration %
without backgrounds for both a Majorana and a Dirac HNL, after  six years
 and after 12 years of data taking, including the power upgrade of the LBNF facility.
The sensitivity was estimated assuming a total of \SI{6e21} 
POT and
\SI{2e22} 
POT, i.e., for a running scenario of six years with  a 120 GeV proton beam of \SI{1.2}{MW}, followed by six years of \SI{2.4}{MW} and using both the neutrino and antineutrino mode configurations.

For the \dword{nd}, both \dword{ndlar} and \dword{ndgar} 
are used to search for the HNL decays,  and the detectors were kept on axis for the whole data sample. \dword{ndlar} will suffer mostly from
background neutrino interactions. Neutrino-nucleus background could be additionally suppressed using the fast timing of the light detectors. Detailed simulation studies are 
required to quantify the background levels.
Decay channels included in the study are the dominant ones in the kinematic 
region under study, i.e., $e\pi, \nu e \mu, \nu ee, \nu \mu\mu, 
\nu \pi^0, \mu \pi$ channels. The most significant contributions to
the sensitivity come from the $e\pi$ and the $\mu \pi$ channels.
The mass range for HNLs up to \SI{2}{GeV} can be explored in all flavor-mixing channels. Figure~\ref{fig:sensa_hnl2} compares the sensitivity curves with 
those from SHiP and other constraints.

\begin{dunefigure}[$90\,\%$ \dshort{cl} sensitivity regions for dominant mixings $|U_{\alpha N}|^2$ (6+6 yrs, $\nu$ and $\overline{\nu}$)]{fig:sensa_hnl}
{The $90\,\%$ \dword{cl} sensitivity regions for dominant mixings $|U_{e N}|^2$, $|U_{\mu N}|^2$, and $|U_{\tau N}|^2$ are presented for DUNE ND. Sensitivity curves are shown that are reached after 6 and 12 years of data taking using an equal amount of beam time in the neutrino and anti-neutrino configuration. The regions are a combination of the sensitivity to HNL decay channels with good detection prospects. These are HNL$\to\nu e e$, $\nu e \mu$, $\nu \mu \mu$, $\nu \pi^0$, $e \pi$, and $\mu \pi$. The study is performed for Majorana neutrinos (solid) and Dirac neutrinos (dashed), assuming no background.}
\includegraphics[width=0.32\textwidth]{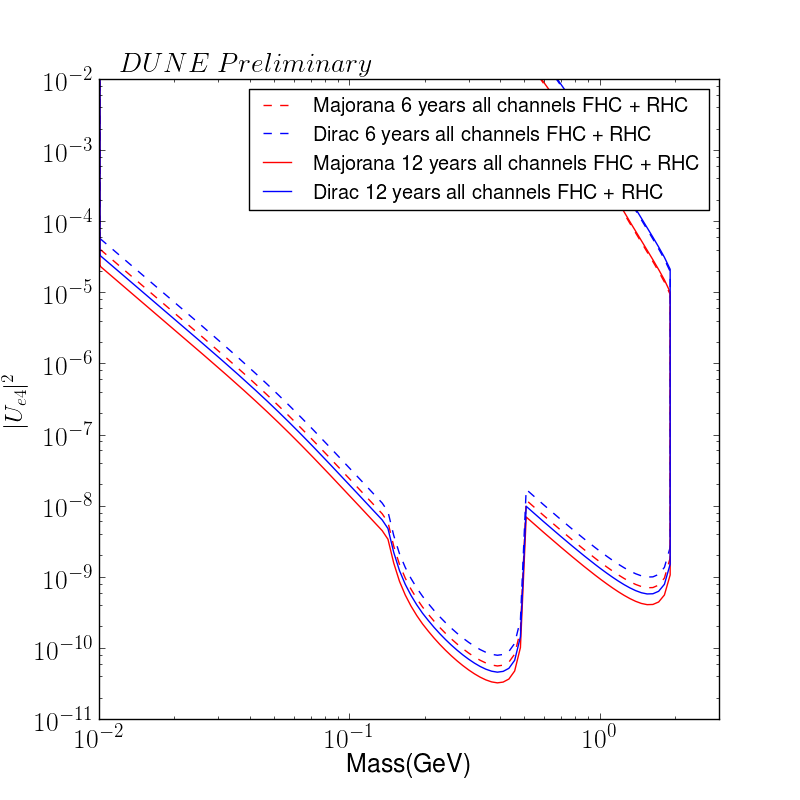}
\includegraphics[width=0.32\textwidth]{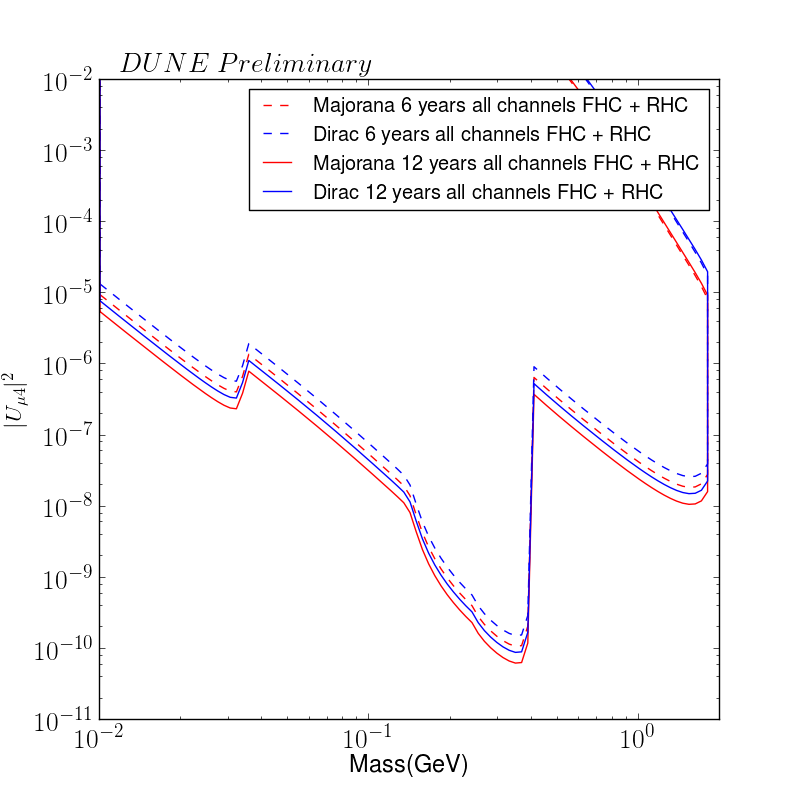}
\includegraphics[width=0.32\textwidth]{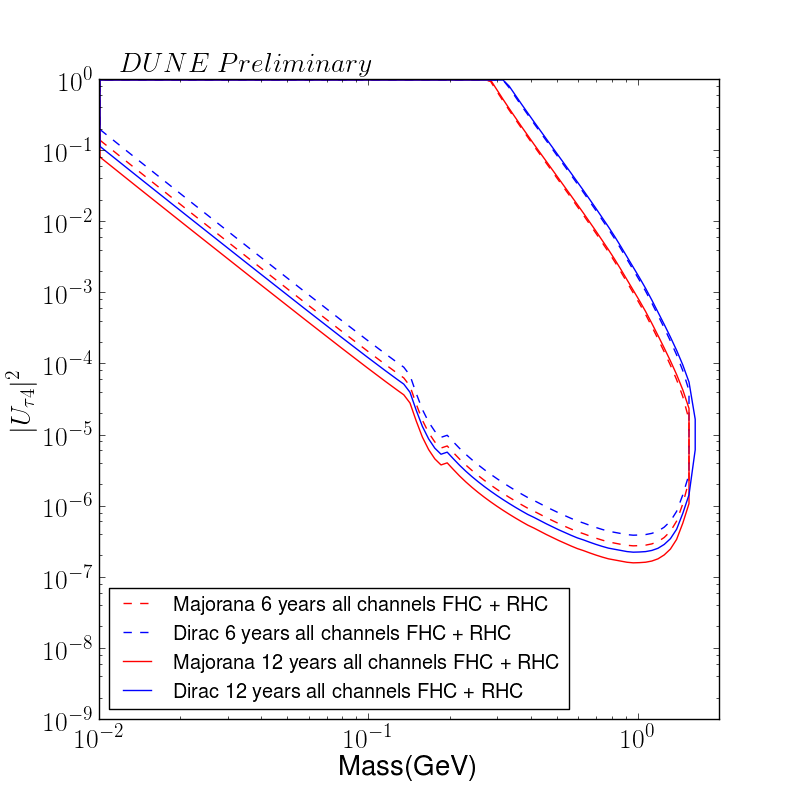}
\end{dunefigure}

\begin{dunefigure}[90\,\% \dshort{cl} sensitivity regions for dominant mixings $|U_{\alpha N}|^2$ (12 yrs, Majorana and Dirac)]{fig:sensa_hnl2}
{The 90\,\% \dword{cl} sensitivity regions for dominant mixings $|U_{e N}|^2$ and $|U_{\mu N}|^2$ are presented for DUNE ND (red). The study is performed for Majorana neutrinos (solid) and Dirac neutrinos (dashed), for 12 years of running, and assuming no background. The results are compared with predictions for SHiP and present data and theoretical constraints.}
\includegraphics[width=0.48\textwidth]{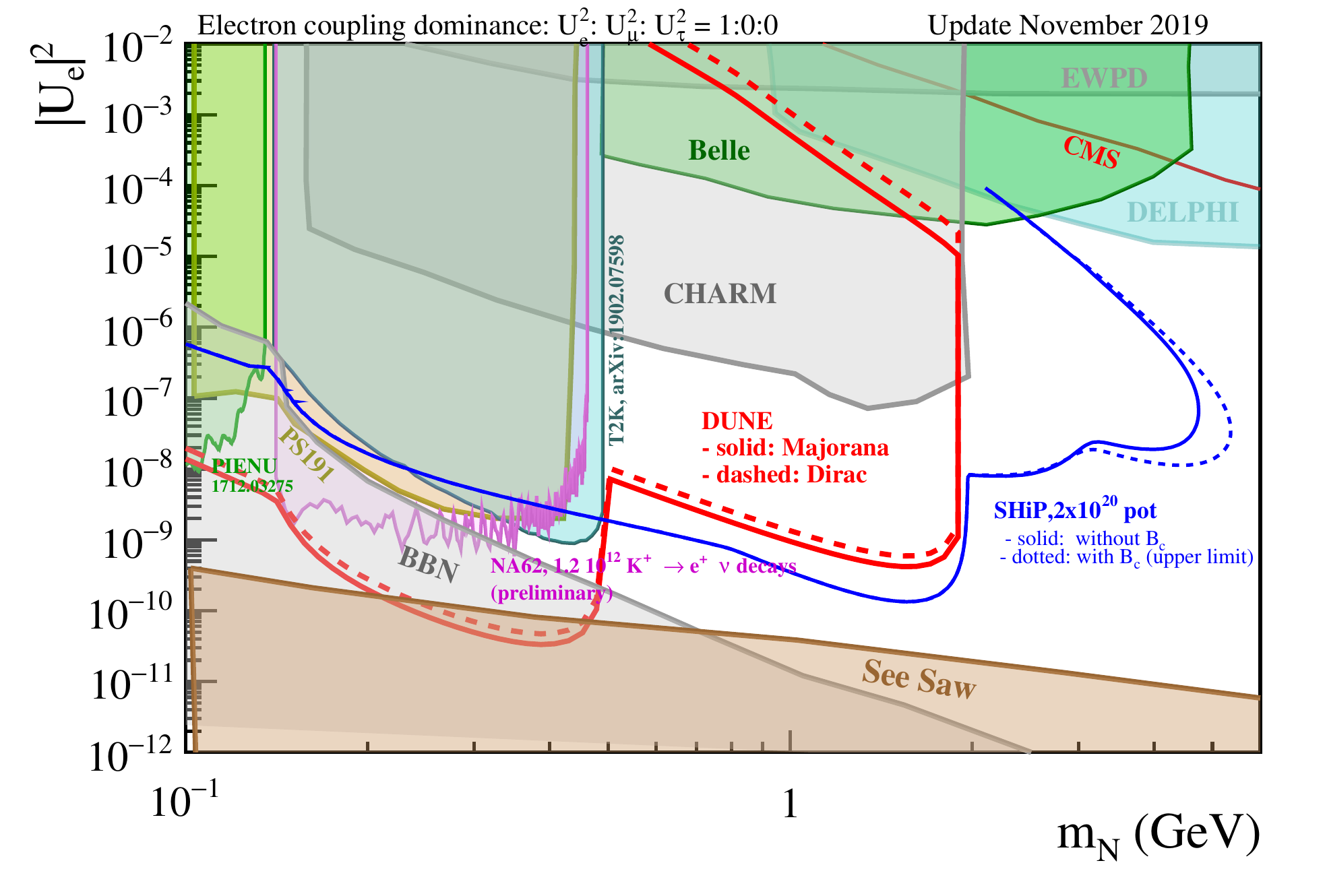}
\includegraphics[width=0.48\textwidth]{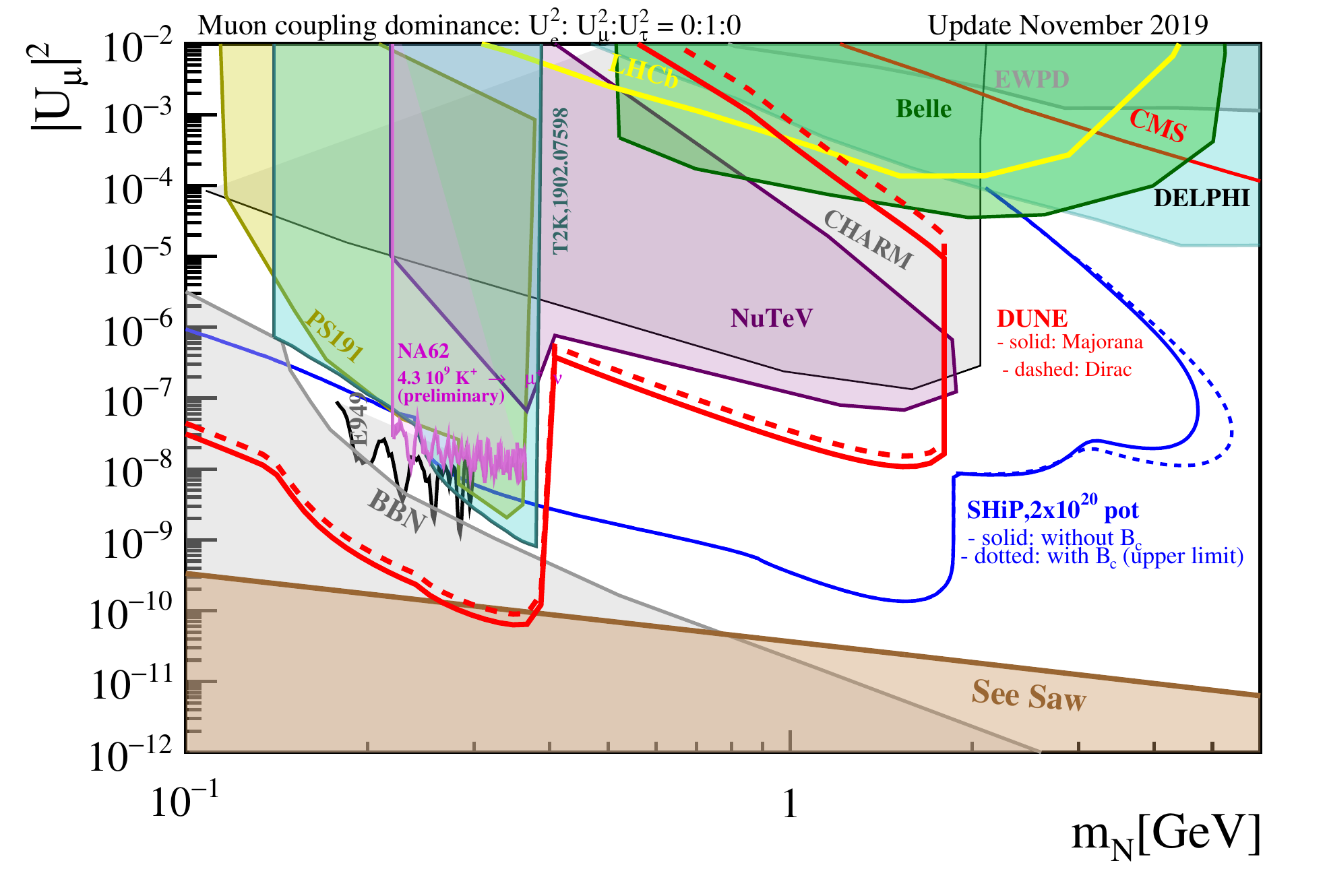}
\end{dunefigure}


The results show that DUNE will have an improved sensitivity at small values of the
mixing parameters $|U_{\alpha N}|^2$, where $\alpha=e,\,\mu,\,\tau$, compared to the presently available experimental
limits on mixing of HNLs with the three lepton flavors. At 90\% \dword{cl} sensitivity, DUNE can probe mixing parameters as low as 
$10^{-9}-10^{-10}$ in the mass range of \SIrange{300}{500}{MeV}, for  mixing with the electron or muon neutrino flavors. It is interesting 
to note that this would be the first such
search to go down to mixing angles favored by the seesaw mechanism. In the region above \SI{500}{MeV} the sensitivity
is reduced to $10^{-8}$ 
for $eN$ mixing and $10^{-7}$ for $\mu N$ mixing. The $\tau N$ mixing 
sensitivity is weaker but still covers an unexplored regime. A large fraction of the covered parameter space for all neutrino flavors falls in the region that is relevant for explaining the baryon asymmetry in the universe.
Detailed studies are in progress with the full detector simulations to validate these 
encouraging results and study the backgrounds.

In this study, the expected HNL flux is estimated from Ref.~\cite{Ballett:2019bgd}.  This rescales the standard neutrino fluxes with the ratio of the decay rates to HNL over standard neutrinos. This takes into account the different phase space available and possible enhancements due to the chirality flips required for the pseudoscalar meson decays. However, this procedure is not able to reproduce possible differences in the neutrino and HNL fluxes from differing kinematics. Indeed, for masses of the HNL close to that of the parent meson, the phase space is significantly reduced, resulting in small HNL velocities. This implies that the boost in the beam direction is more relevant than it is for neutrinos and can lead to an enhancement in the flux that reaches the detector. 
In ongoing work, a full simulation of the HNL decay from the parent mesons without reliance on the standard neutrino fluxes is being done so as to correctly account for these effects.

It is also of interest to consider searches for non standard decays of HNLs. They could be part of a new low energy sector which contains several new states (neutral fermions, gauge bosons, scalars, DM). Such a scenario could lead to interesting new decay channels that are being studied, including a process with   intermediate dark photons/Z' in the HNL decay.

\subsection{Sterile Neutrino Probes}
\label{sec:bsm-sterile}
Experimental results in tension with the three-neutrino-flavor paradigm, which may be interpreted as mixing between the known active neutrinos and one or more sterile states, have led to a rich and diverse program of searches for oscillations into sterile neutrinos. The combination of the DUNE Near Detector location and the LBNF beam energy spectrum will enable sensitive probes of sterile neutrino-driven oscillations in the L/E range of 0.01 to 1 eV$^2$, overlapping with the L/E range of the LSND signal. The large statistics provided by the LBNF beam and the highly-capable DUNE ND provide sensitivity to sterile mixing in various channels, specifically in probing short-baseline sterile-driven electron neutrino appearance and/or tau neutrino appearance, as well as stand-alone muon neutrino disappearance, or disappearance in association with the appearance measurements. These measurements will check and complement results from the  Short-Baseline Program underway at Fermilab. An example of the projected sensitivity for DUNE ND sterile probes is shown in Figure~\ref{fig:th24_nd}, showing the ND-only excluded parameter space for the angle $\theta_{24}$ as a function of the sterile mass splitting $\Delta m^2_{41}$ in a 3+1 model. The exclusion curves are obtained by looking for sterile-driven disappearance of $\nu_\mu$ \dword{cc} and \dword{nc} interactions. The red curve in Figure~\ref{fig:th24_nd} displays the sensitivity obtained when only normalization uncertainties are considered, while the blue curve shows how that sensitivity is reduced when a 1\% shift uncorrelated from bin-to-bin is added, exemplifying the systematic effects of uncertainties inducing spectral shape distortions.

The DUNE ND, in conjunction with the FD, will also enable the most precise Long-Baseline accelerator searches for sterile mixing, as described in the DUNE TDR. 
Further enhancements of the sterile neutrino mixing  sensitivity can be achieved by combining the DUNE \dword{nd}, capable of high-efficiency particle ID, with a precise muon monitor system for the LBNF beam, which would provide an independent constraint on the neutrino flux through measurements of the associated muon flux, not susceptible to mixing with sterile neutrinos.

\begin{dunefigure}[90~\% \dshort{cl} DUNE ND-only $\theta_{24}$ exclusion regions]
{fig:th24_nd}
{The 90~\% \dword{cl} DUNE ND-only $\theta_{24}$ exclusion regions when including normalization systematics, but no spectral shape systematics (red curve), and when including normalization and shape systematics. Both curves were computed using the GLoBES toolkit assuming a 3+1 model with one sterile neutrino. For reference, the projected DUNE limits significantly exceed the current limits from MINOS\cite{Adamson:2020jvo} and IceCube\cite{Aartsen:2020fwb} over much of the $\Delta m^2_{41}$ range.}
\includegraphics[width=0.7\textwidth]{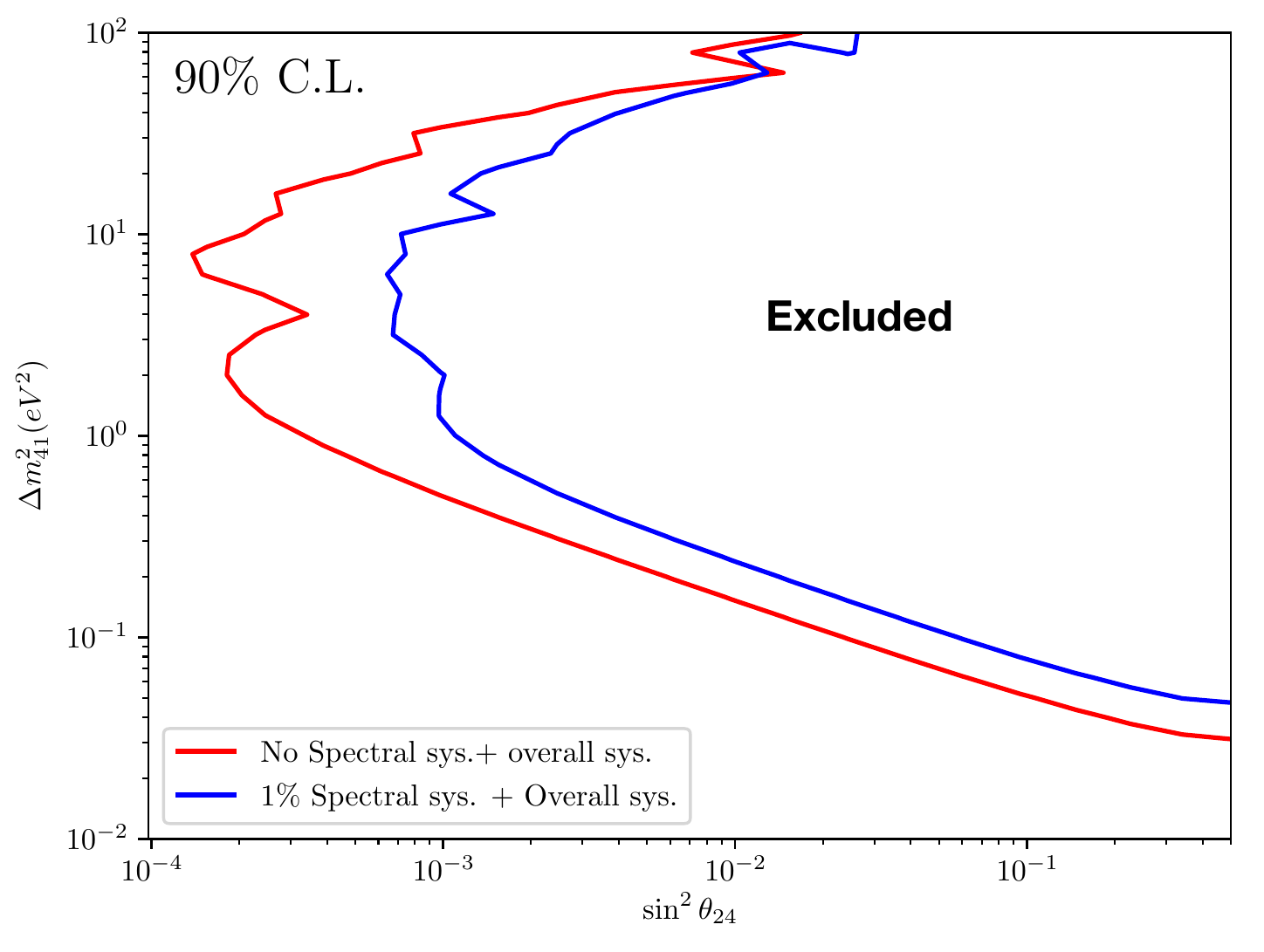}
\end{dunefigure}


\subsection{Searches for Large Extra Dimensions}
\label{sec:bsm-led}
The fact that neutrinos are massive and  their mass is much smaller than any other \dword{sm} fermion  is suggestive of the need for \dword{bsm} physics. One possibility to naturally explain the small size of neutrino masses is that there are large compactified extra dimensions, which were first proposed to solve the \dword{sm} hierarchy problem. In these large extra dimensions (LED) models, \dword{sm} gauge group singlets, such as right-handed neutrinos, are assumed to propagate in all dimensions, while the \dword{sm} particles can only propagate in the 4D brane. Assuming an asymmetry in the size of the extra dimensions, one can show the low-energy physics behavior can be described effectively with just the largest extra dimension. Therefore, the complete oscillation phenomenology can be described in terms of the size of the extra dimension $R$ and the mass of the lightest neutrino $m_0$, corresponding to $m_1$($m_3$) for normal (inverted) ordering.

Mixing between heavy Kaluza-Klein (KK) modes and active neutrinos will produce distortions in the oscillation pattern in the \dword{nd} which can be probed by looking for the disappearance of muon neutrinos. Particularly, the \dword{nd} can look for oscillations whenever $\Delta m^2\geq 0.1 \text{eV}^2$, which would be averaged out in the FD. Given that the mass splitting between the lightest neutrino and the KK modes is given by $\Delta m_{n 1}^2 = n^2/R^2+2m_0 n/R$, it is clear that being able to probe larger values of $\Delta m_{n1}^2$ corresponds to having access to smaller extra dimensions.

Thus, a \dword{nd} with a good energy resolution could greatly improve the reach of the \dword{lbnf} by probing the disappearance of muon neutrino events at short baselines. In Figure~\ref{fig:led_sens} the sensitivity to LED at the DUNE \dword{nd} at 90~\% \dword{cl} is shown for different cases depending on the information used in the fit and the level of the systematic uncertainties. The dark blue curve shows the sensitivity when the only source of systematic uncertainties is an overall normalization, such that the shape of the events is perfectly reconstructed, and would represent a best-case scenario. The muon neutrino disappearance and electron neutrino appearance channels, along with their antineutrino counterparts, are considered when computing the blue curve. The dashed red line depicts the sensitivity obtained using only the muon neutrino and antineutrino disappearance samples. The dark green lines show the sensitivities computed by including an energy-dependent systematic, labeled as `shape' in the plot, intended to represent how the sensitivities may be affected by cross section energy calibration uncertainties. 
In this particular case, we have introduced ``shape'' uncertainties as $1\%$, $2\%$, and $5\%$ shifts uncorrelated from energy bin to energy bin, to globally account for small spectral distortions. These ``shape'' shifts reduce considerably the sensitivity below $m_0\sim 5\times 10^{-2}\;\text{eV}$.

\begin{dunefigure}[90~\% \dshort{cl} exclusion regions for the \dshort{lbnf} with a \dshort{nd} with perfect spectral information]
{fig:led_sens}
{The 90~\% \dword{cl} exclusion regions for the \dword{lbnf} with a \dword{nd} with perfect spectral information (red and blue) and introducing an energy-dependent systematic (green) apart from the overall normalization uncertainties. Regions to the right of the blue and red curves, and above the green curves, are excluded.}
\includegraphics[width=0.9\textwidth]{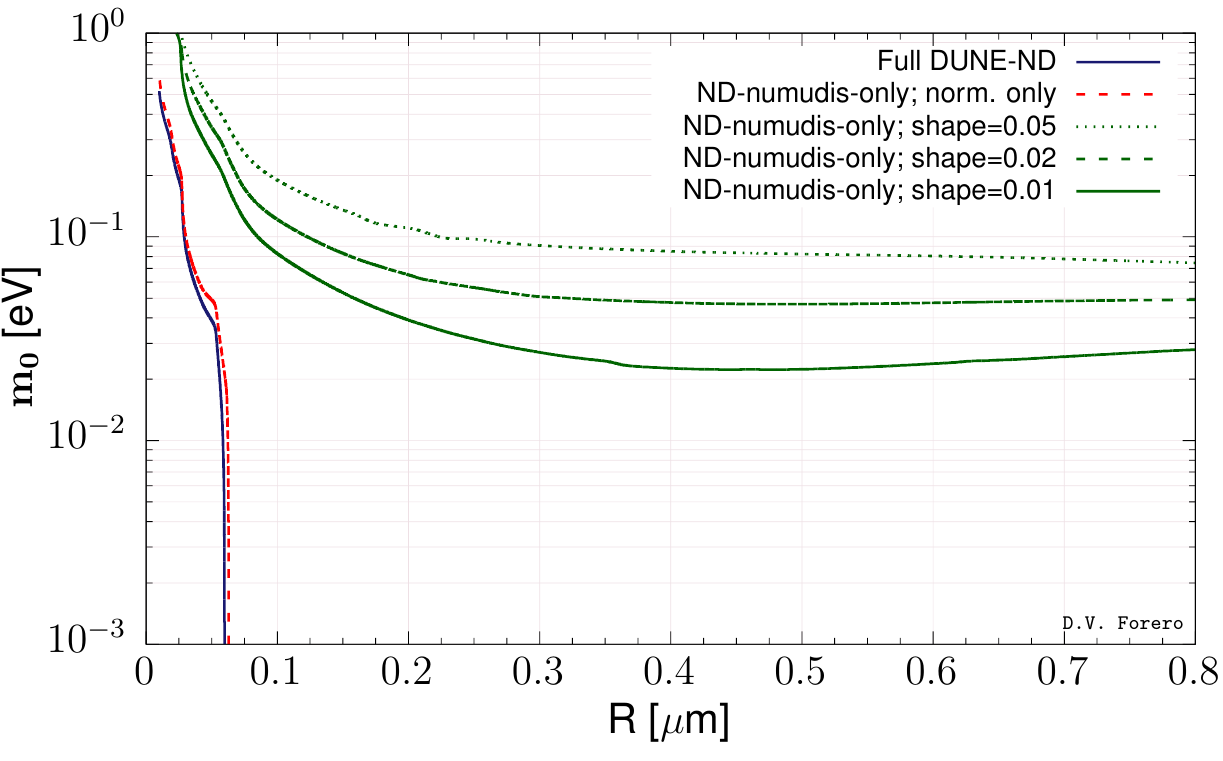}
\end{dunefigure}


\subsection{Non-Standard Neutrino Interactions}
\label{sec:bsm-nsi}
The DUNE ND may be sensitive to non-standard neutrino interactions, which can be probed  through non-standard deviations in the behavior of coherent elastic neutrino-nucleus scattering. The momentum transfer in these events is typically small ($<~100$~MeV), so the ability of the DUNE ND to measure low-energy neutrino interactions is essential for these searches. Sensitivity to these effects would require a very well characterized flux for it to be competitive with probes of the same phenomenon in coherent electron-neutrino scattering experiments.

Source and detector CC NSI can be studied in the ND well due to the expected high neutrino flux. General neutrino interactions in neutrino-electron scattering at the DUNE near detector can be associated with heavy new physics and their effect is to cause distortions in the electron recoil spectrum. The ND will give limits comparable but complementary to the ones from the analysis of neutrino oscillations in the FD.

A particularly intriguing probe arises from measuring scattering from dimension-7 Rayleigh operators~\cite{Altmannshofer:2018xyo}, which is presently poorly bounded, with the best bounds placed by the Borexino experiment. The expected signal is a \dword{nc} interaction with emission of a single hard gamma in the final state. The cross section for this process is enhanced for more energetic beam neutrinos and also by the atomic number of the nucleus. In both cases, DUNE presents advantages over Borexino and may yield stronger sensitivity to this process. 

Finally, the more common search for NSI affecting neutrino propagation through the Earth benefits from constraints on cross section and flux provided by a highly-capable \dword{nd} in the same way as the \dword{cpv} probe would. If the DUNE data are consistent with standard oscillations for three massive neutrinos, interaction effects of order 0.1 G$_{\rm{F}}$ can be ruled out at DUNE. DUNE could improve current constraints on $\epsilon_{\tau e}$ and $\epsilon_{\mu e}$ by a factor 2 to 5.

\subsection{Lorentz- and CPT-Symmetry Tests}
\label{sec:bsm-CPT}

The \dword{dune} \dword{nd} features excellent capabilities to perform competitive Lorentz and CPT tests. These closely intertwined spacetime symmetries form a cornerstone of present-day physics. Moreover, a fully consistent underlying theory incorporating both quantum and gravitational physics is widely believed to require adjustments to currently accepted fundamental principles at high energies, and in such a context, a breakdown of both Lorentz and CPT symmetry may occur in various approaches to underlying physics including string theory \cite{Kostelecky:1988zi,Kostelecky:1991ak}. The ensuing low-energy signals are amenable to experimental searches in a broad range of physical systems including numerous neutrino measurements \cite{Kostelecky:2008ts}.

To use data from the \dword{dune} \dword{nd} for systematic searches for imprints of Lorentz and CPT violation, a consistent and general test framework is needed. Mirroring other theoretical approaches to fundamental physics, effective field theory provides the standard theoretical tool kit for such purposes. This approach has yielded the Standard-Model Extension (SME), a framework that contains all Lorentz- and CPT-breaking corrections to the \dword{sm} and General Relativity that are compatible with realistic field theories. The SME predicts various modifications of ordinary neutrino propagation relevant for the \dword{dune} \dword{nd}. They include novel variations of oscillation patterns with energy, dependence on the beam direction, differences in flavor oscillations between neutrinos and antineutrinos, as well as oscillations between neutrinos and antineutrinos. The general relations governing these effects for physics set-ups such as the \dword{dune} \dword{nd} are given as Eq. (106) in Ref.~\cite{Kostelecky:2011gq}. 

The small baseline for the \dword{dune} \dword{nd} implies that conventional mass-oscillation effects can be disregarded. However, certain types of Lorentz and CPT violation can dominate at short distances leading to potentially observable signals \cite{Kostelecky:2011gq,Kostelecky:2004hg,Diaz:2009qk}. For example, the aforementioned dependence of oscillation patterns on the neutrino propagation direction would lead to sidereal changes in the flavor composition of (anti)neutrino beams. This represents a possible avenue for Lorentz and CPT tests in situations when absolute neutrino-flux calibration plays only a secondary role. This particular idea has already been exploited using the \dword{minos} \dword{nd} and the \dword{t2k} \dword{nd} to constrain minimal-SME coefficients \cite{Adamson:2008aa,Adamson:2012hp,Abe:2017eot}. The \dword{dune} \dword{nd} could similarly be employed to search for sidereal variations. The different alignment of the beam direction relative to \dword{minos} and \dword{t2k} would provide access to another set of minimal-SME coefficients, and previously unexplored nonminimal-SME coefficients could also be measured. The \dword{dune} \dword{nd} is therefore ideally positioned for substantial improvements of existing tests of Lorentz and CPT symmetry in the neutrino sector, and harbors the potential to yield various first-ever measurements of select types of Lorentz and CPT violation.

\section{Some Standard Model Physics opportunities} 
\label{sec:sm}

The powerful \dword{lbnf} beam and the capabilities of the \dword{dune} \dword{nd} enable an exciting program of \dword{sm} physics that goes beyond the cross-section measurements discussed in Chapter~\ref{ch_xsec:sec_xsec}.  
The list of topics presented here is intended to be illustrative rather than complete, and serious studies looking at the performance of the current \dword{dune} \dword{nd} design on these topics have not been done yet.    

\subsection{Electroweak mixing angle}

\dword{dune} can make a precise measurement of the electroweak mixing angle, sin$^{2}\theta_{W}$, using neutrino-nucleon or neutrino-electron scattering.  Such measurements probe a different range of momentum transfer than those done on the Z pole.  To date, the most precise measurements of sin$^{2}\theta_{W}$ using neutrino scattering are extracted from the neutrino DIS measurements of the ratio of the neutral-to-charged-current cross sections \cite{Allaby:1987vr,McFarland:1997wx} or the Paschos-Wolfenstein \cite{Zeller:2001hh} ratio
\begin{equation}
R^{-}=\frac{ \sigma^{NC}_{\nu N} - \sigma^{NC}_{\ensuremath{\bar\nu} N} }{\sigma^{CC}_{\nu N} -\sigma^{CC}_{\ensuremath{\bar\nu} N} }.
\end{equation}
Measurements of these ratios are dominated by theoretical uncertainties \cite{Kulagin:2004ie,Zyla:2020zbs}.  A sub-1\% measurement of sin$^{2}\theta_{W}$ in the \dword{dune} \dword{nd} seems plausible using a program of in-situ measurements to constrain some of these uncertainties along with some modest improvements in theory\cite{bib:docdb13262}.  The extraction of sin$^{2}\theta_{W}$ from leptonic scattering has lower theoretical uncertainties since it does not depend on knowledge of the structure of nuclei.  The value of sin$^{2}\theta_{W}$  comes from the measurement of the ratio 
\begin{equation}
R=\frac{ \sigma_{\numu e} } { \sigma_{\anumu e} } ,
\end{equation}
in which many uncertainties cancel.  The cross section for (anti)neutrino scattering from atomic electrons is small and statistics has been a limiting factor in previous measurements \cite{Dorenbosch:1988is,Ahrens:1990fp,Vilain:1994qy}.  The \dword{dune} \dword{nd} is well suited to do this measurement.  The number of events is large, relatively speaking, in ND-LAr and many of the systematics, such as the uncertainty in the \numu to \anumu flux ratio and the \nue – nucleus CC background, can be understood cleanly in the light trackers of ND-GAr and SAND.  Again, a sub-1\% measurement of sin$^{2}\theta_{W}$  seems plausible.

\subsection{Background to proton decay}

The proton decay mode p$\rightarrow$K$^{+}\ensuremath{\bar\nu}$ is favored in many supersymmetric GUT models.
The NC production of K$^{+}$ by atmospheric neutrinos is an important background to this process. 
For example, in water Cherenkov detectors, the atmospheric NC production  of a K$^{+}$ when no final-state particles are produced above Cherenkov threshold can produce a signal of a de-excitation photon followed by a $\mu^{+}$ and a Michel electron that is indistiguishable from the proton decay process.
The \dword{dune} \dword{nd} can measure the production of K$^{+}$ and K$^{0}$ by beam neutrinos and place constraints on the proton decay background.

SAND is a detector that should be able to make useful measurements of kaon production using either of the technologies under discussion for the inner tracker.  The \dword{stt} should be able to make precise measurements of K$^{0}$ production, as already demonstrated by the functionally similar NOMAD detector \cite{Astier:2001vi,Naumov:2004wa}.  K$^{0}$ production can be related to K$^{+}$ production.  The \dword{3dst} can do an analysis similar to what has been done by MINERvA to measure NC K$^{+}$ production \cite{Marshall:2016yho}.  The significantly better timing resolution and the use of a fine-grained three dimensional scintillator structure instead of strips should help substantially in the efficiency and cleanliness of the kaon tagging.  

\subsection{Strange particles and M$_{A}$ from hyperon decays}

 With the powerful \dword{lbnf} beam and the capabilities of the \dword{dune} \dword{nd}, there will be a rich program of physics involving strange particles.  Published work by NOMAD provides a sense of the richness of the topic \cite{Naumov:2004wa,Astier:2001vi}.  One item of interest is $\Lambda^{0}$ production in antineutrino \dword{ccqe} interactions, which can be studied in detail.  The polarization components of the decay along and transverse to the  $\Lambda^{0}$ momentum are sensitive to the axial form factor M$_{A}$ \cite{Akbar:2016awk}.  Though this has been looked at before \cite{Erriquez:1978pg}, the high statistics available in the \dword{nd} will make the measurement more interesting.

\subsection{QCD and nucleon structure}

Figure~\ref{fig:GeneratorW} shows comparisons of recent predictions of three widely used neutrino event generators for the W distribution for neutrinos and antineutrinos. Interactions at E$_{nu}$=2.5~GeV on argon are shown on top and interactions at E$_{nu}$=6.0~GeV on iron are on the bottom.  The disparity in the predictions illustrate that the \dword{sis} region and the transition into the higher-W \dword{dis} region, as well as non-resonant pion production, are all areas that need further experimental and theoretical effort \cite{SajjadAthar:2020nvy}. By virue of the low beam energy, the \dword{sis} and \dword{dis} regions will not be studied well by the SBN program.  \dword{minerva} and \dword{nova} will make some measurements in these regions on hydrocarbon targets.  However, there will be a need for the \dword{dune} \dword{nd} to make measurements on argon in these regions for model tuning and improvement, as a significant fraction of \dword{dune} data will fall in these kinematic regions.  

\begin{dunefigure}[Generator comparison of W distributions.]
{fig:GeneratorW}
{These plots show a comparison of the predictions for the W distibution for three neutrino event generators (NEUT 5.4.0, GENIE 2.12.10, and NuWro 18.021) for interactions on argon with E$_\nu$=2.5~GeV on the top and on iron for E$_\nu$=6.0~GeV on the bottom.  Neutrino interactions are shown on the left side and antineutrino interactions are shown on the right side.   Figures from Bronner in reference~\cite{Andreopoulos:2019gvw}.}
\includegraphics[width=0.9\textwidth]{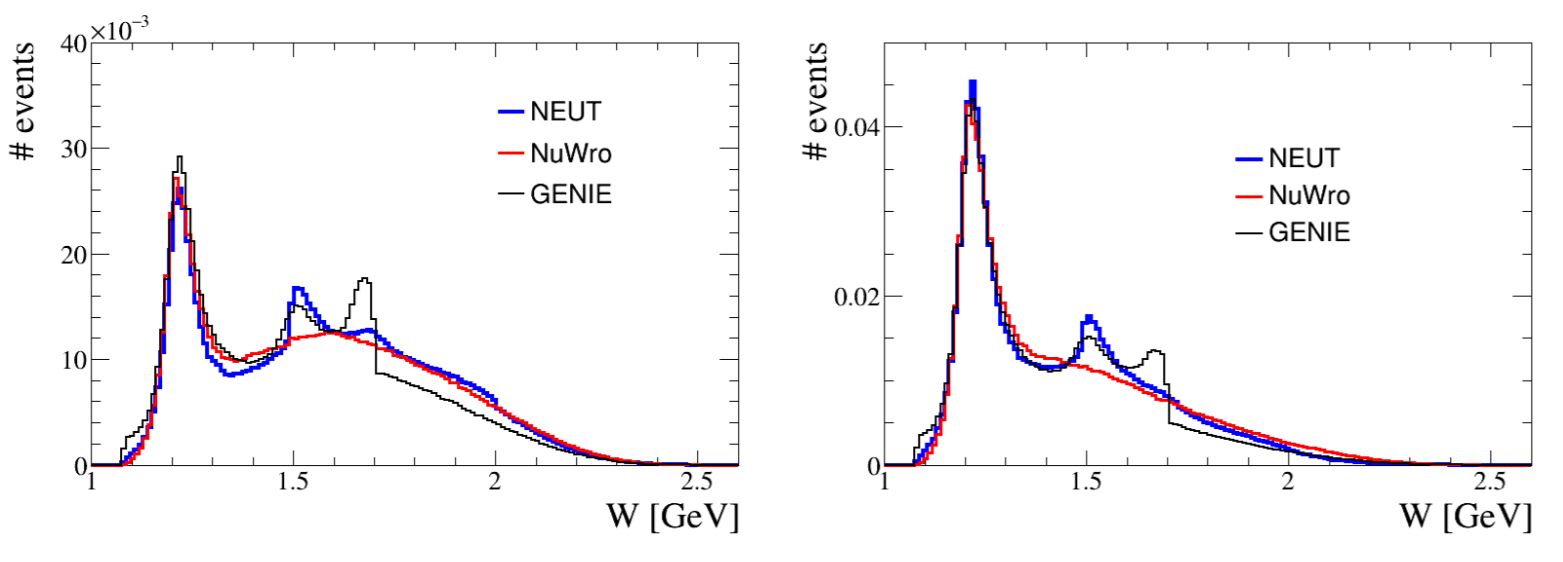}
\includegraphics[width=0.9\textwidth]{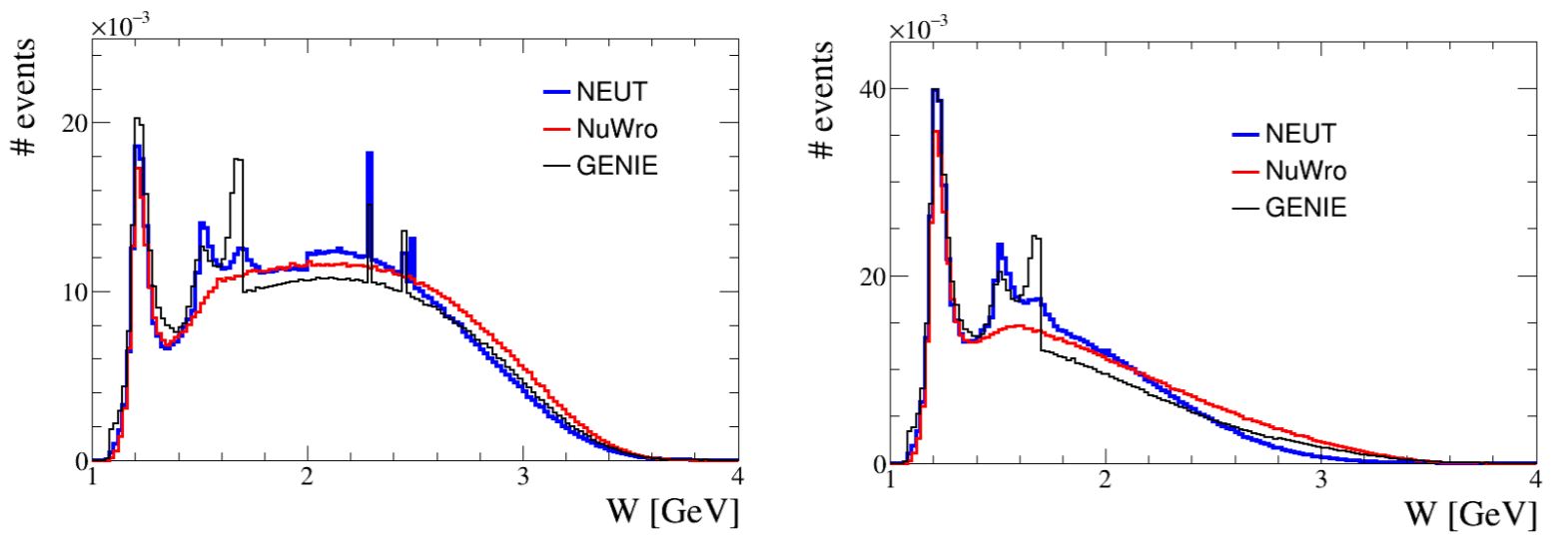}
\end{dunefigure}

There is some disagreement between current analyses as to the extent to which there is evidence for nuclear shadowing in $\nu$-A scattering, particularly at low-Q$^{2}$~\cite{Kopeliovich:2012kw}.  This is not the case for charged lepton scattering.   The source of this disagreement is not understood as yet, but might be a consequence of the flavor dependence of shadowing~\cite{SajjadAthar:2020nvy}. Measurements of di-muons and DIS events in the \dword{dune} \dword{nd} on argon and carbon, and perhaps other nuclear targets, would be a helpful in understanding this.

The measurement of inclusive $\nu$- and $\overline{\nu}$-induced charm production via opposite sign dilepton production would provide insight into the strangeness content of the nucleon~\cite{SajjadAthar:2020nvy,Alekhin:2018dbs}.  The statistics in the \dword{dune} \dword{nd} will be significantly greater than samples collected to date~\cite{Samoylov:2013xoa}.

Both the \dword{stt} and \dword{3dst} trackers under consideration for part of the SAND inner tracker contain considerable hydrogen bound in hydrocarbon.  Studies show that (anti)neutrino interactions on hydrogen can be selected with reasonable efficiency and purity.  This, along with the possibility of taking data on embedded targets and the carbon itself can lead to a rich program of nucleon structure and QCD studies~\cite{bib:docdb13262}.

\subsection{Isospin Physics and Sum Rules} 
\label{sec:isospin} 

Isospin physics is a compelling topic for \dword{dune}, which is  
looking for tiny differences between neutrino and antineutrino interactions. 
Accurate measurements of the $d/u$ content of
the nucleons can be obtained in STT~\cite{bib:docdb13262,ESGprop}  
using both $\nu$ and $\bar \nu$ interactions on hydrogen~\cite{Duyang:2018lpe,Duyang:2019prb}. 
In particular, the isospin symmetry allows a direct measurement of the free neutron 
structure functions $F_{2,3}^{\nu n} \equiv F_{2,3}^{\bar \nu p}$ 
and $F_{2,3}^{\bar \nu n} \equiv F_{2,3}^{\nu p}$. 
This measurement provides, in turn, a precise determination of 
the $d/u$ quark ratio up to values of Bjorken $x$ close to 1~\cite{Alekhin:2017fpf,Accardi:2016qay}.  

The Adler sum rule~\cite{Adler:1964yx,Allasia:1985hw}, 
$S_A=0.5 \int^1_0  (dx/x) ( F_2^{\bar \nu p} - F_2^{\nu p} ) = I_p$, 
gives the isospin of the target and can be measured as a function 
of the momentum transfer $Q^2$ using $\nu(\bar \nu)$ interactions 
on H and nuclear targets~\cite{Kulagin:2004ie,Kulagin:2007ju}. The value of $S_A$ is sensitive to possible 
violations of the isospin (charge) symmetry, heavy quark (charm) production, 
and strange sea asymmetries $s-\bar s$. 
The Gross-Llewellyn-Smith (GLS) sum rule~\cite{Gross:1969jf,Kim:1998kia}, 
$S_{GLS} = 0.5 \int^1_0  dx/x ( xF_3^{\bar \nu p} + xF_3^{\nu p} )$, 
can also be measured in $\nu$ and $\bar \nu$ interactions. The value 
of $S_{GLS}$ receives both perturbative and non-perturbative 
QCD corrections and its $Q^2$ dependence can be used to extract 
the strong coupling constant $\alpha_s$~\cite{Larin:1991tj,Kataev:1994rj}. Measurements with 
both H and various nuclear targets~\cite{Kulagin:2004ie,Kulagin:2007ju} would allow an investigation 
of the isovector and nuclear corrections. 

Isospin symmetry implies that $F_{2,3}^{\bar \nu p} = F_{2,3}^{\nu n}$ 
and that for an isoscalar target $F_{2,3}^{\bar \nu} = F_{2,3}^{\nu}$.
These relations as a function of $x$ and $Q^2$ can be used for precision 
tests of isospin (charge) symmetry using a combination of H and
isoscalar nuclear targets.

\cleardoublepage

\chapter{The ND Cavern and Facilities}
\label{ch:ndhall}


\section{Introduction}
\label{sec:chap-id:introduction}

\subsection{Near Detector Cavern Layout}
\label{sec:chap-id:introduction:layout}

The DUNE ND cavern, which accommodates the DUNE ND with its component subdetectors ND-LAr, ND-GAr, and SAND,  will be located on the western-most boundary of Fermi National Laboratory.   Figure~\ref{fig:nd_hall_location} shows a birds-eye view of the future LBNF/DUNE construction site with the FNAL main injector, the target hall, decay pipe, muon absorber, and the Near Detector hall locations indicated. A cross-sectional view of the near site beamline is shown in Figure~\ref{fig:near_site_xsec}. As illustrated the near detector hall will be located about \SI{570}{\m} from the proton beam target at an underground depth of approximately \SI{60}{\m}. This is the furthest possible separation from the target hall within the FNAL property boundary.

\begin{dunefigure}[Top view of the future LBNF/DUNE construction site]{fig:nd_hall_location}
{Birds-eye view of the future LBNF/DUNE construction site. The Near Detector cavern will be located at the west-most boundary of Fermi National Lab.}
\includegraphics[width=0.8\textwidth]{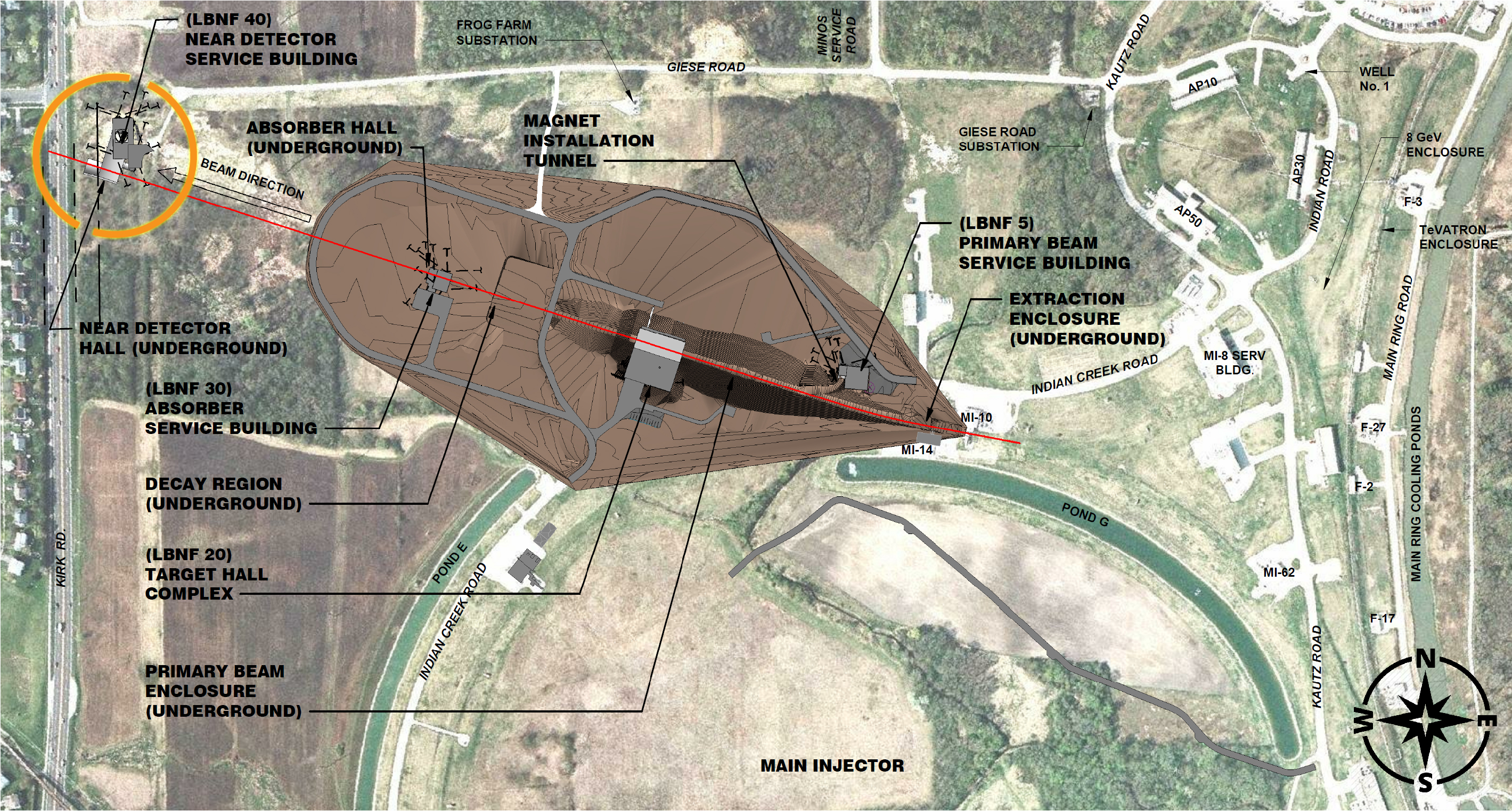}
\end{dunefigure}

\begin{dunefigure}[Cross-sectional view of the near site beamline]{fig:near_site_xsec}
{A cross-sectional view of the near site neutrino beamline. The near detector hall will be located about \SI{570}{\m} from the proton beam target at an underground depth of approximately \SI{60}{\m}.}
\includegraphics[width=0.8\textwidth]{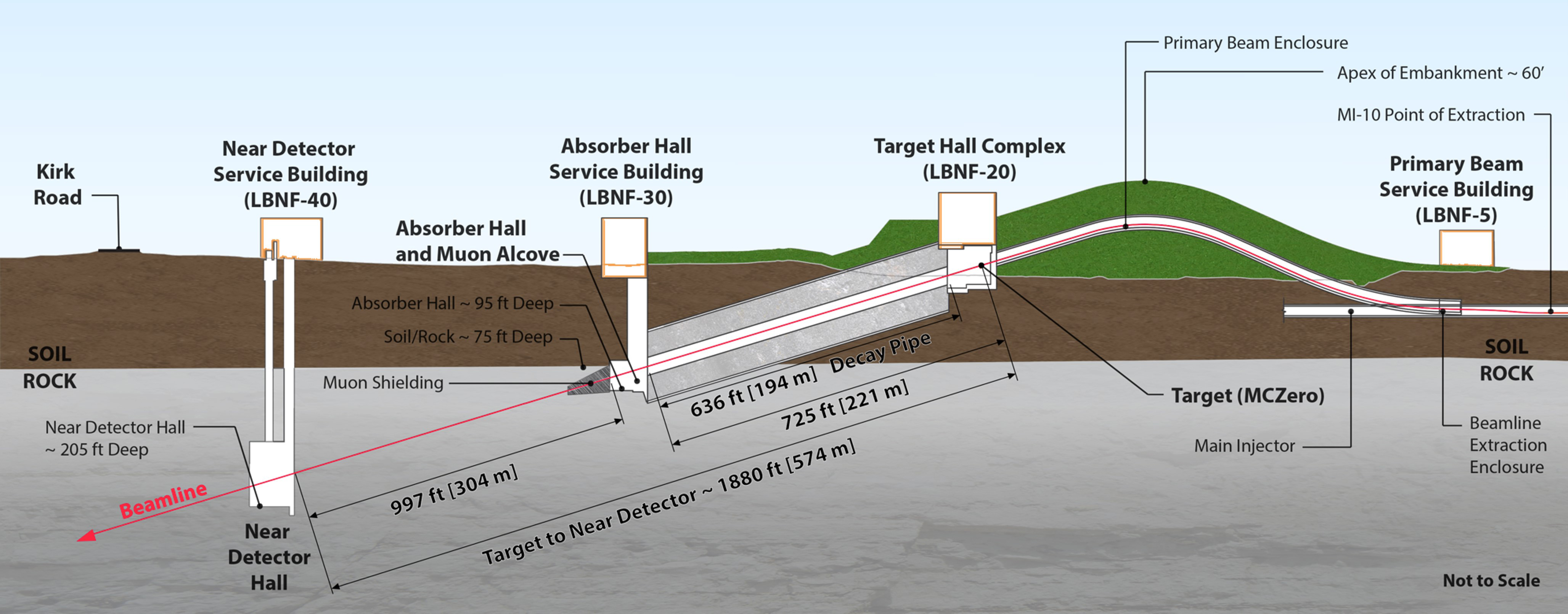}
\end{dunefigure}

The Near Detector hall complex consists of a surface building, the underground cavern, and a secondary egress shaft which has a separate air supply for fire safety. A large, \SI{38}{\ft} diameter primary shaft connects the surface building to the cavern and will be utilized for hoisting detector equipment underground. The shaft also accommodates utility and cryogenics lines plus an elevator. Figure~\ref{fig:nd_hall_3d} shows an architectural 3D model and detail drawing of the Near Detector hall complex. Main sizing parameters are summarized in Table~\ref{tab:cavern-sizing-param}.

\begin{dunefigure}[Cavern and surface building architectural drawings]{fig:nd_hall_3d}
{Near Detector cavern and surface building architectural drawings.}
\includegraphics[width=0.8\textwidth]{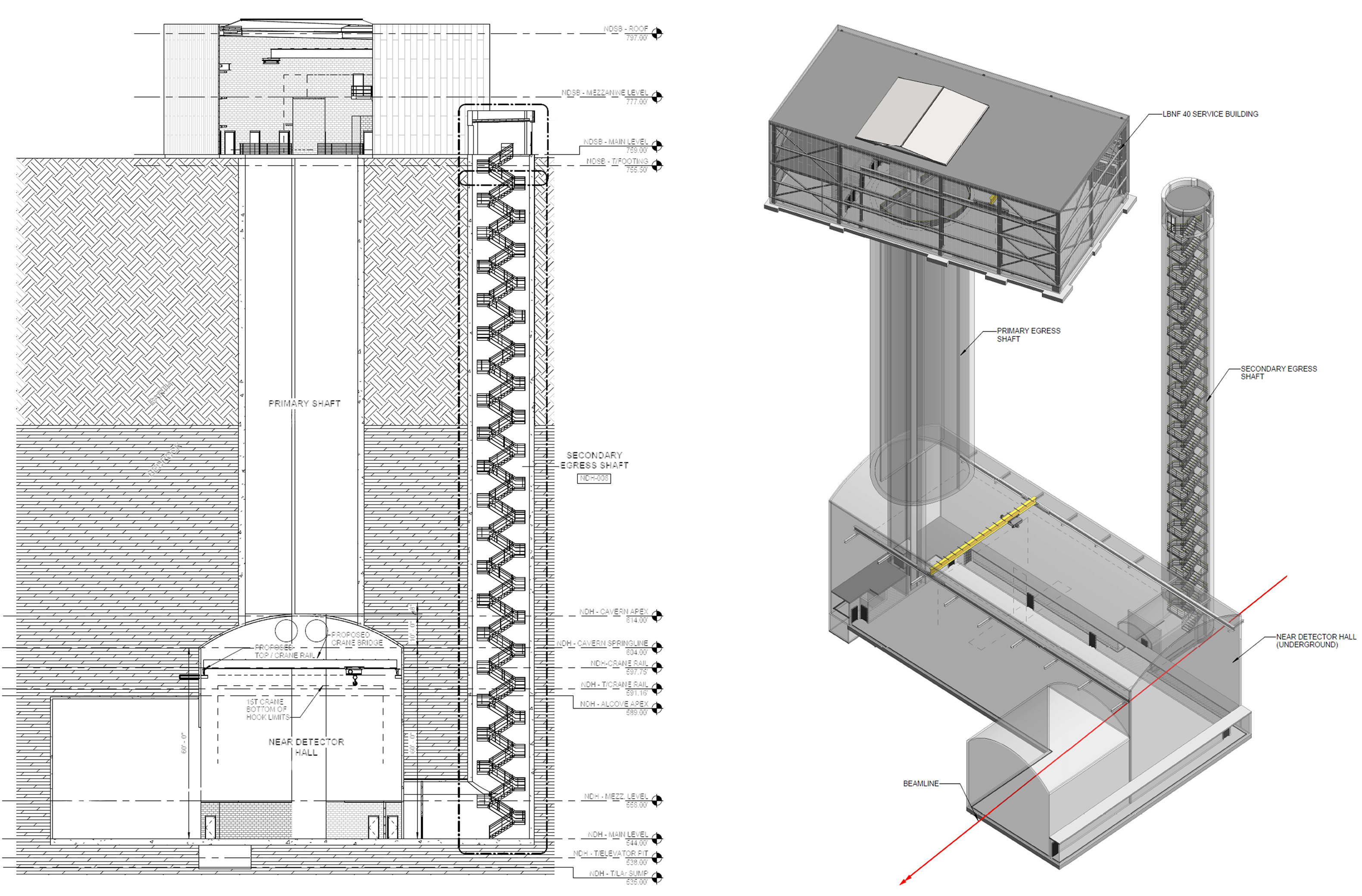}
\end{dunefigure}

\begin{dunetable}
[Underground cavern size parameters]
{cc}
{tab:cavern-sizing-param}
{Approximate underground cavern size parameters. Actual dimensions will be determined during the final design phase.}
Cavern Parameter & Dimension \\ \toprowrule
Main Cavern Length & \SI{166}{\ft} \\ \colhline
Main Cavern Width & \SI{63}{\ft} \\ \colhline
Main Cavern Height & \SI{50}{\ft} \\ \colhline
Alcove Width & \SI{40}{\ft} - \SI{2}{\in} \\ \colhline
Alcove Depth & \SI{50}{\ft} - \SI{6}{\in} \\ \colhline
Alcove Height & \SI{37}{\ft} \\ \colhline
Access Shaft Clear Diameter & \SI{38}{\ft} \\ 
\end{dunetable}

\subsection{Detector Arrangement and Neutrino Beamline}
\label{sec:chap-id:introduction:arrangement}

Figure~\ref{fig:nd_detector_arrangement} shows the three ND  subdetectors located at their nominal, on-axis beamline position. The ND-LAr and the ND-GAr subdetectors can move transverse to the neutrino beamline to permit off-axis flux measurements as part of the PRISM science program. PRISM requires an off-axis movement range of approximately \SI{30}{\m}. This distance determines the overall length of the main cavern. On the other hand, the SAND subdetector will act as a stationary, permanent beam monitor at a fixed position inside an alcove along the beam centerline.
 \dword{ndlar} and \dword{ndgar} will typically move together to facilitate the measurements of PRISM, but they can move separately for installation and maintenance as needed.  

\begin{dunefigure}[Cavern layout with ND-LAr, ND-GAr, and SAND detectors]{fig:nd_detector_arrangement}
{Near detector cavern arrangement of the ND-LAr, ND-GAr and SAND subdetectors plus the PRISM movement system.}
\includegraphics[width=0.9\textwidth]{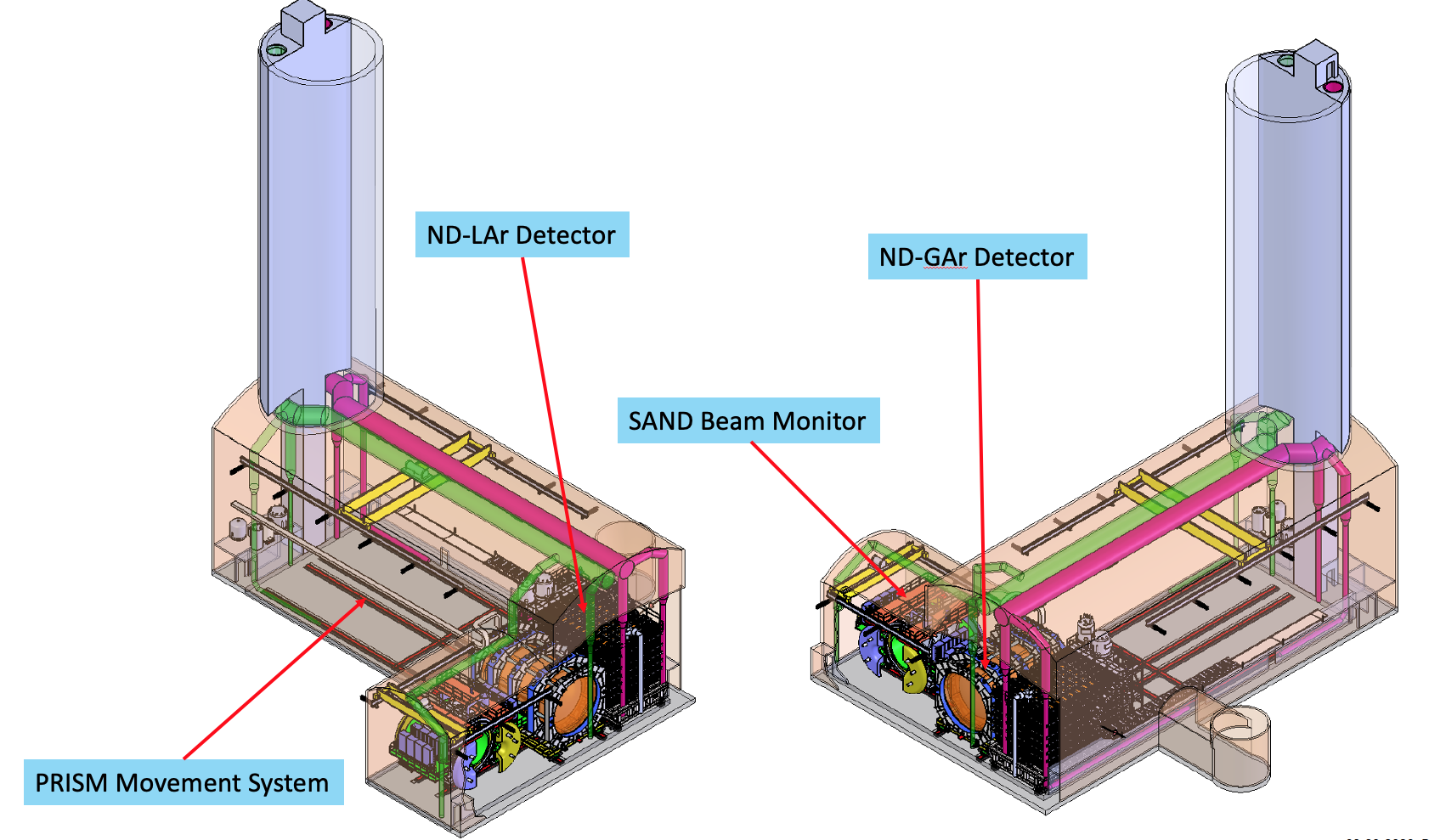}
\end{dunefigure}

The neutrino beamline which is directed towards the far site is angled by \SI{5.3}{\degree} with respect to the cavern horizontal plane. As shown in Figure~\ref{fig:beamline} the subdetectors are elevated such that the neutrino beamline passes through the center of the active volume of each subdetector. The figure also summarizes the main dimensions related to the positioning of the detectors inside the cavern.

\begin{dunefigure}[Neutrino beamline location inside the Near Detector cavern]{fig:beamline}
{Near detector positions relative to the cavern and the DUNE neutrino beamline which passes through the center of each detector active volume.}
\includegraphics[width=0.9\textwidth]{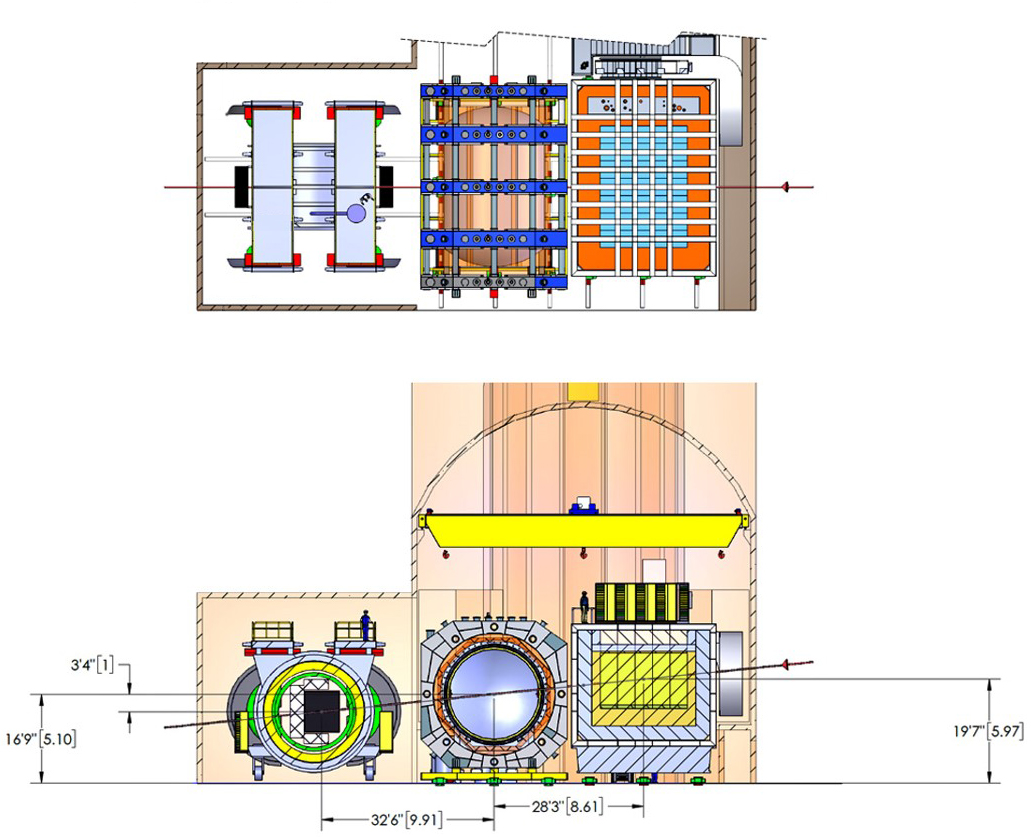}
\end{dunefigure}

The cavern floor consists of a \SI{24}{\in} thick concrete slab resting on the natural rock formation. The structure can support the significant subdetector weights which are summarized in Table~\ref{tab:detector-weights}.

\begin{dunetable}
[approximate detector weights]
{cc}
{tab:detector-weights}
{Approximate subdetector weight summaries.}
Detector & Approximate Weight \\ \toprowrule
ND-LAr Subdetector & \SI{880}{\metricton} \\ \colhline
ND-GAr Subdetector & \SI{710}{\metricton} \\ \colhline
SAND Beam Monitor & \SI{900}{\metricton} \\ \colhline
PRISM & included with detector weights \\ 
\end{dunetable}

\section{Near Detector Installation Details}
\label{sec:chap-id:details}

\subsection{ND-LAr Subdetector}
\label{sec:chap-id:details:lar}

A conceptual configuration for the ND-LAr subdetector setup is shown in Figure~\ref{fig:nd_lar_details}. Seven rows of five pixelated detector modules each (see Section~\ref{sec:lartpc-dimensions}) are suspended inside a large membrane cryostat. The conceptual design of the membrane cryostat is comparable to similarly sized cryostats built for previous or existing neutrino experiments (ProtoDUNE-SP, for example).

\begin{dunefigure}[Main ND-LAr detector components]{fig:nd_lar_details}
{Conceptual image of the ND-LAr subdetector. Seven rows of five pixelated detector modules each are suspended inside a large membrane cryostat. The figure shows prototype module design features which will be adapted to the final cryostat configuration.}
\includegraphics[width=0.8\textwidth]{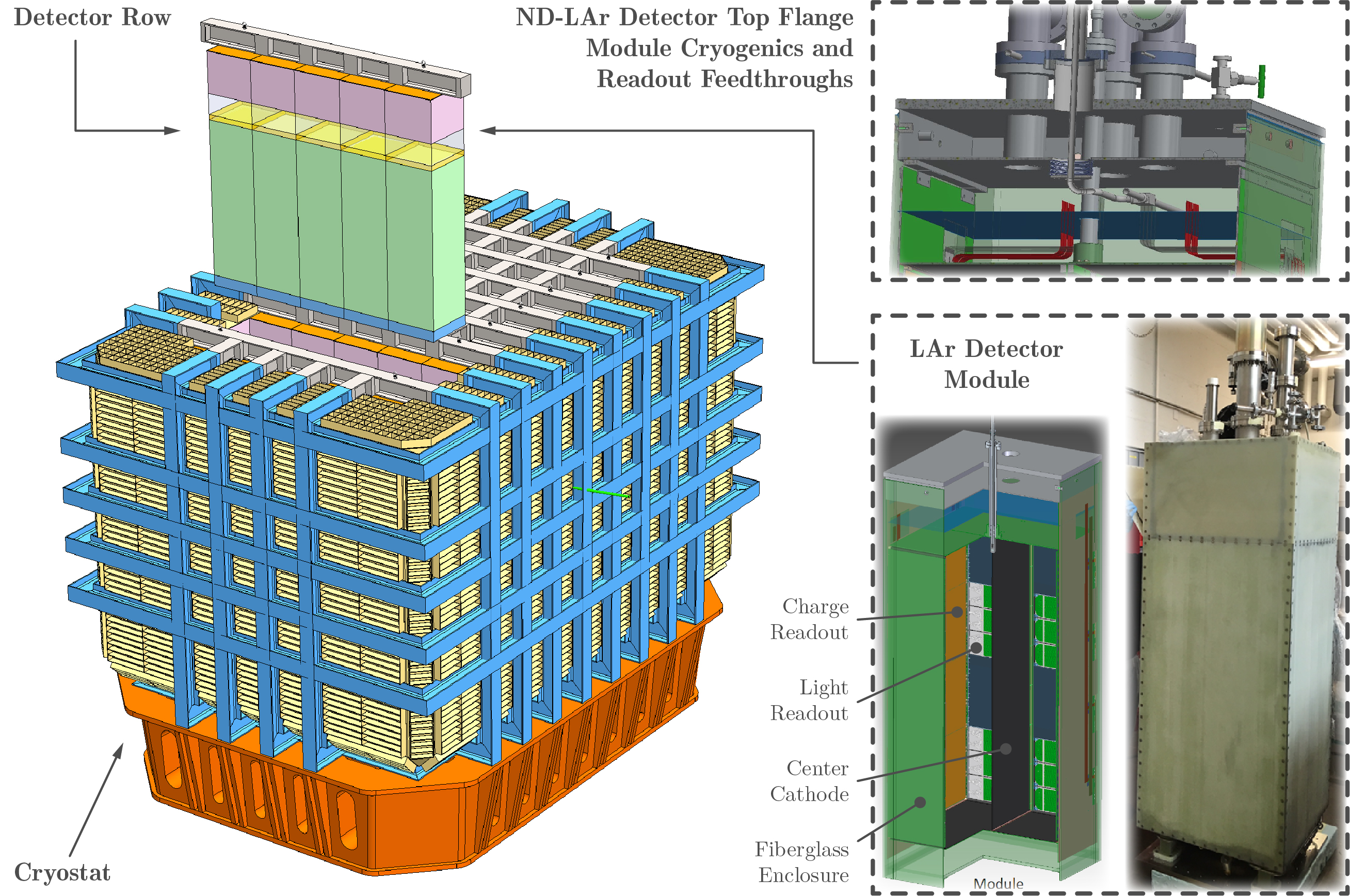}
\end{dunefigure}

The cryostat stores roughly \SI{300}{\metricton}s of liquid argon, and its outside dimensions are approximately \SI{11.4}{\m} wide (transverse to neutrino beam) x \SI{8.4}{\m} deep (along neutrino beam) x \SI{7.0}{\m} high. A cryogenic process flow diagram for the ND-LAr subdetector is shown in Figure~\ref{fig:nd_lar_flowdiagram}. A unique feature of the subdetector will be a large cable carrier (see Figure~\ref{fig:nd_lar_setup}) which houses flexible cryogenic pipes, power, and data acquisition cables. To minimize the quantity of flexible cryogenic piping in the cable carrier most of the argon purification system will be located on a two-level support system mounted on the moveable detector platform. The argon cryogenic system will be designed based on extensive experiences on ProtoDUNE enabling optimization of space needs on the detector platform.

\begin{dunefigure}[ND-LAr cryogenic process flow diagram]{fig:nd_lar_flowdiagram}
{Conceptual cryogenic process flow diagram for the ND-LAr subdetector. A unique feature of the ND-LAr detector will be the capability of moving the entire cryogenic purification system together with the detector for off-axis beam measurements.}
\includegraphics[width=0.8\textwidth]{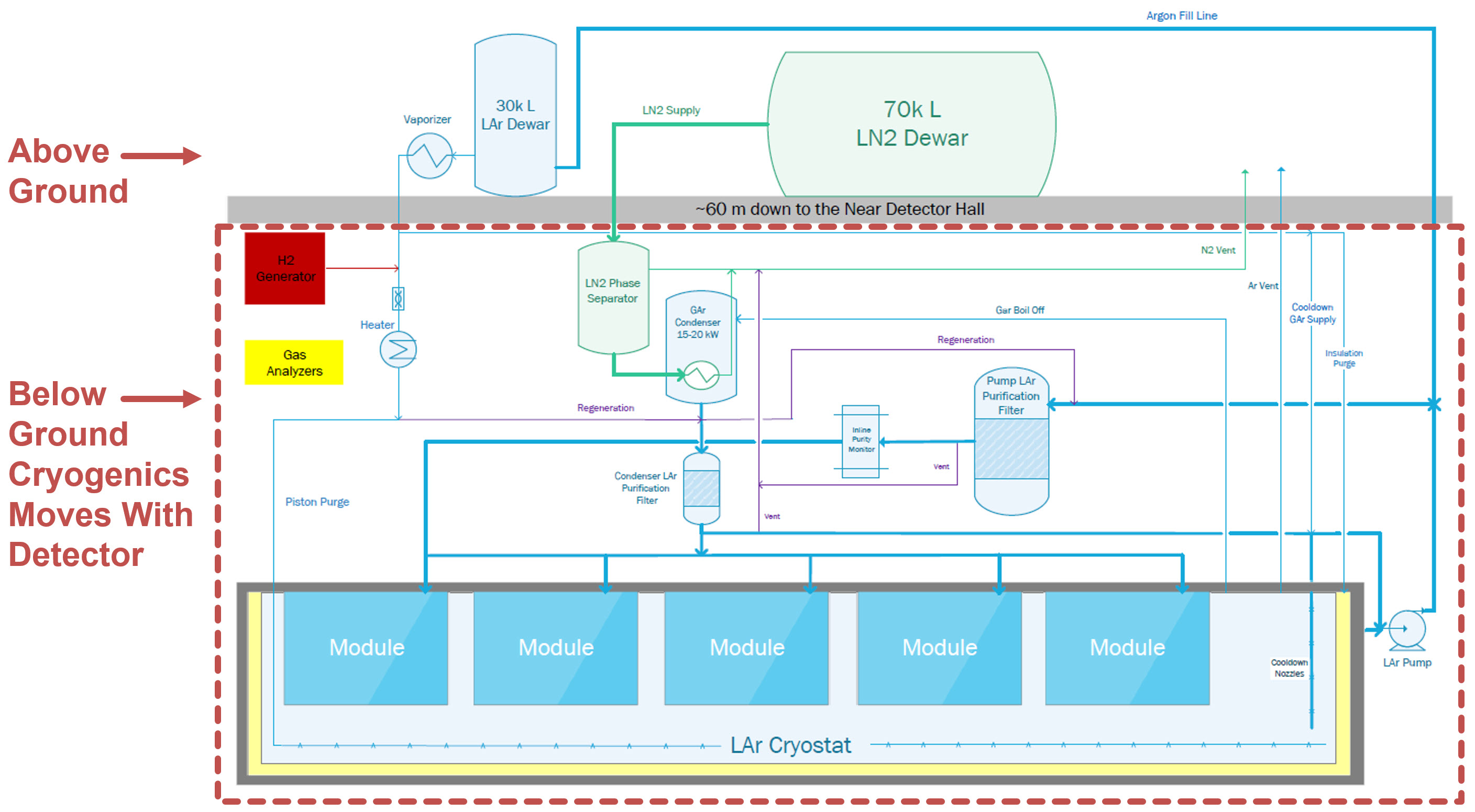}
\end{dunefigure}

\begin{dunefigure}[Detailed ND-LAr detector layout]{fig:nd_lar_setup}
{Illustration of the ND-LAr subdetector setup including the cryostat, the cryogenic purification system, and the PRISM movement system. A large cable carrier houses flexible cryogenic pipes, power, and data acquisition cables. \dword{ndgar} will have a similar, separate cable carrier.}
\includegraphics[width=0.8\textwidth]{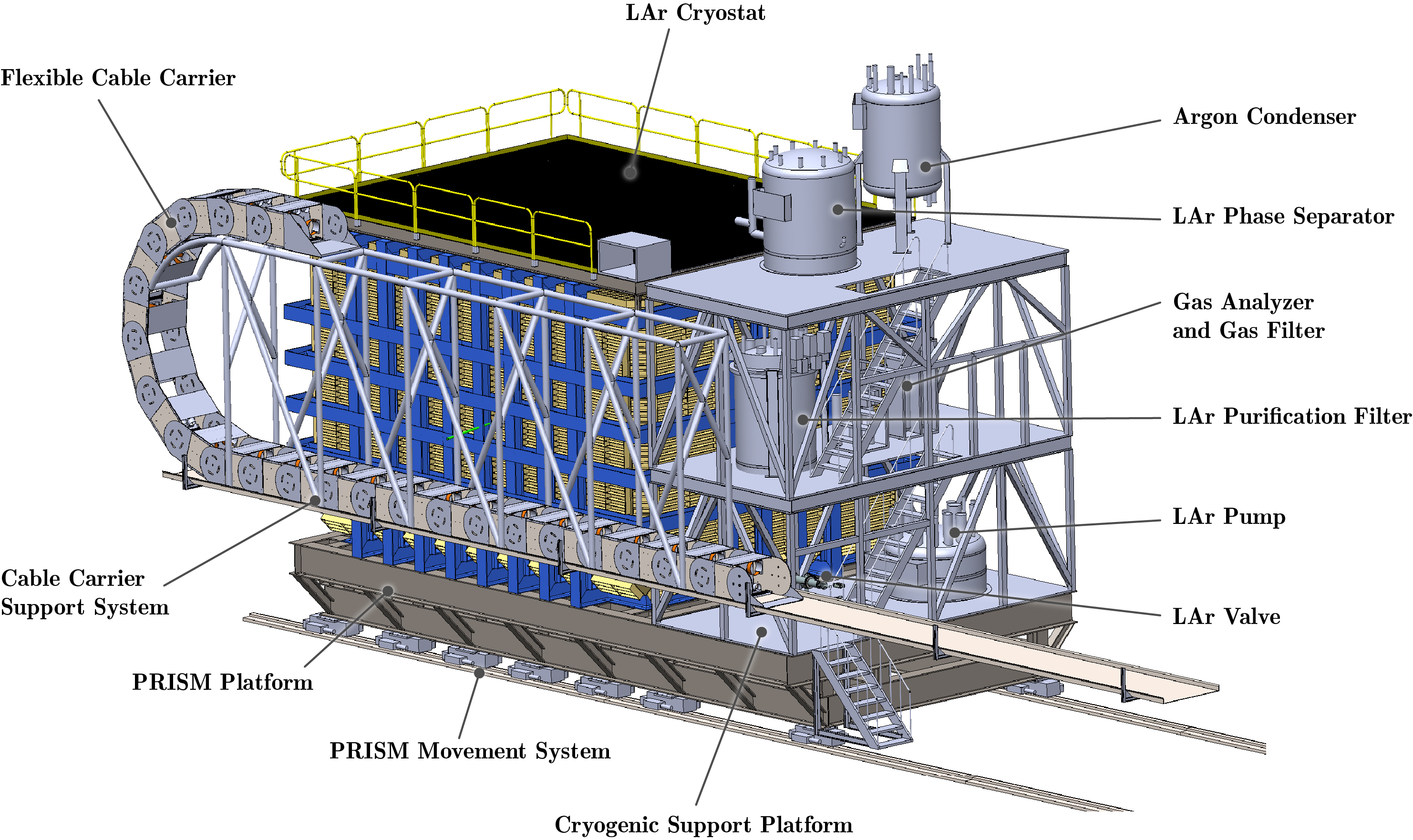}
\end{dunefigure}

The ND-LAr subdetector, which can only contain muons with energies up to approximately \SI{1}{\GeV}, needs to operate in conjunction with the ND-GAr subdetector (see Section~\ref{sec:chap-id:details:mpd}) to cover the full muon energy spectrum up to approximately \SI{5}{\GeV}. Passive material between the ND-LAr and the ND-GAr subdetectors must be minimized in order to maximize the muon detection efficiency. Therefore, the LAr cryostat is designed to include a large, low-mass fiberglass back-wall which is designed to interlock with the main steel beams of the cryostat.

\subsection{ND-GAr Subdetector}
\label{sec:chap-id:details:mpd}

A high-pressure argon gas detector is located directly downstream of the ND-LAr subdetector. This multipurpose subdetector will enable a broad physics program, including the measurement of muons that exit \dword{ndlar}. \dword{ndgar} requires a powerful and large superconducting magnet. The current reference design - as shown in Figure~\ref{fig:nd_gar_detector} - is built around a time projection chamber design based on the ALICE TPC. Consequently, the size of the pressure vessel tank and the electromagnetic calorimeter structure are determined by the TPC outer diameter. The ND-GAr outside dimensions are approximately \SI{12.8}{\m} wide (transverse to neutrino beam) x \SI{8.5}{\m} deep (along neutrino beam) x \SI{10}{\m} high.

\begin{dunefigure}[Conceptual ND-GAr detector configuration]{fig:nd_gar_detector}
{Schematic drawing and 3D view of the ND-GAr subdetector with 5-coil magnet configuration. A large, superconducting solenoid produces the magnetic field required for muon energy spectroscopy.}
\includegraphics[width=0.8\textwidth]{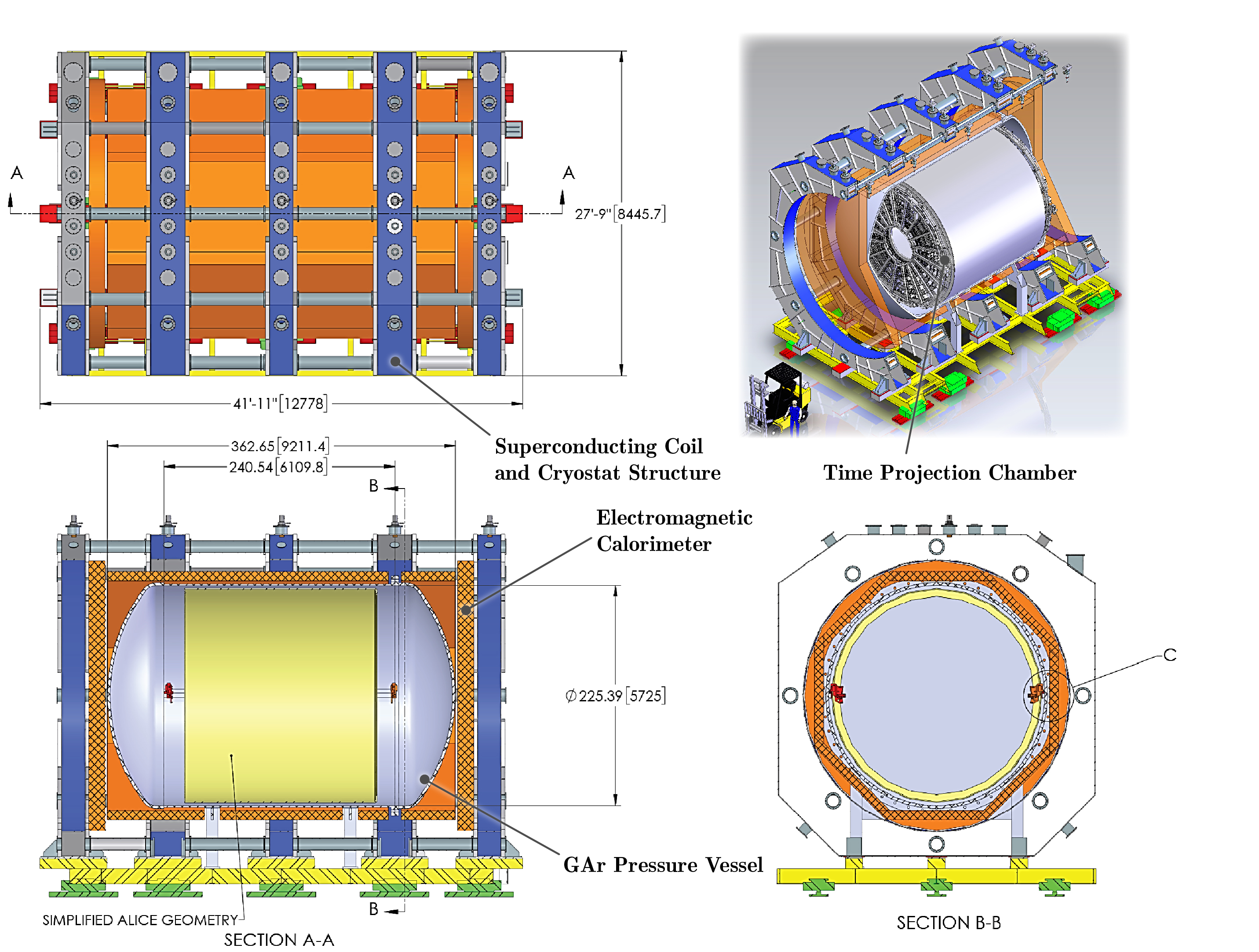}
\end{dunefigure}

The superconducting solenoid structure is optimized to minimize the size of the subdetector in the beam direction to reduce cavern excavation volume. In addition, the open magnet structure is designed to reduce the significant magnetic fringe field (100s of Gauss) extending towards the ND-LAr subdetector as well as the SAND beam monitor. The fringe field impact must be considered in the design of the detector support systems due to the added transverse forces. Further optimized solenoid coil configurations with or without iron yokes are under ongoing investigation in order to reduce the fringe field impact as well as the overall ND-GAr subdetector size.

The ND-GAr subdetector will require a helium refrigeration system either based on cryocoolers or a cryoplant. Two cryoplant design variations have been developed either utilizing a large cryoplant shared between the ND-GAr subdetector and the SAND beam monitor (see Section~\ref{sec:chap-id:details:sand}) or two smaller, separate cryoplants for each subdetector. The latter design has advantages with respect to staging of the two subdetectors. In addition, a separate and smaller coldbox dedicated to the ND-GAr subdetector could potentially be mounted adjacent to the subdetector limiting the need for long flexible cryogenic lines for PRISM operation.

Operation of the time projection system will require a dedicated gas circulation system. The detail design of the gas system has not yet been defined. Possible locations for installing components are at the shaft end of the cavern or on the surface building.

\subsection{SAND Beam Monitor}
\label{sec:chap-id:details:sand}

The \dword{nd} cavern includes an alcove to house a stationary beam monitor for continuous flux measurements, since ND-LAr  and ND-GAr are designed to be moved off-axis. An existing collider detector structure (KLOE, which is currently installed at INFN Frascati, Italy, see Figure~\ref{fig:sand_photos}) will provide the magnet system and ECAL for SAND. As shown in Figure~\ref{fig:sand_schematic}, SAND consists of a large superconducting solenoid which produces a $\sim$ \SI{0.6}{\T} magnetic field on-axis. The magnetic field is uniform over a large volume inside the subdetector due to a carefully designed and heavy steel yoke including large end plates which can be opened for access. The main magnet parameters are summarized in Table~\ref{tab:sand_coil_data}. The fully assembled subdetector has an approximate length of \SI{10}{\m} and height of \SI{11}{\m}.

\begin{dunefigure}[SAND detector installation photos]{fig:sand_photos}
{Parts of an existing detector (KLOE, which is currently installed at INFN Frascati, Italy) will be repurposed for use in the SAND beam monitor.}
\includegraphics[width=0.7\textwidth]{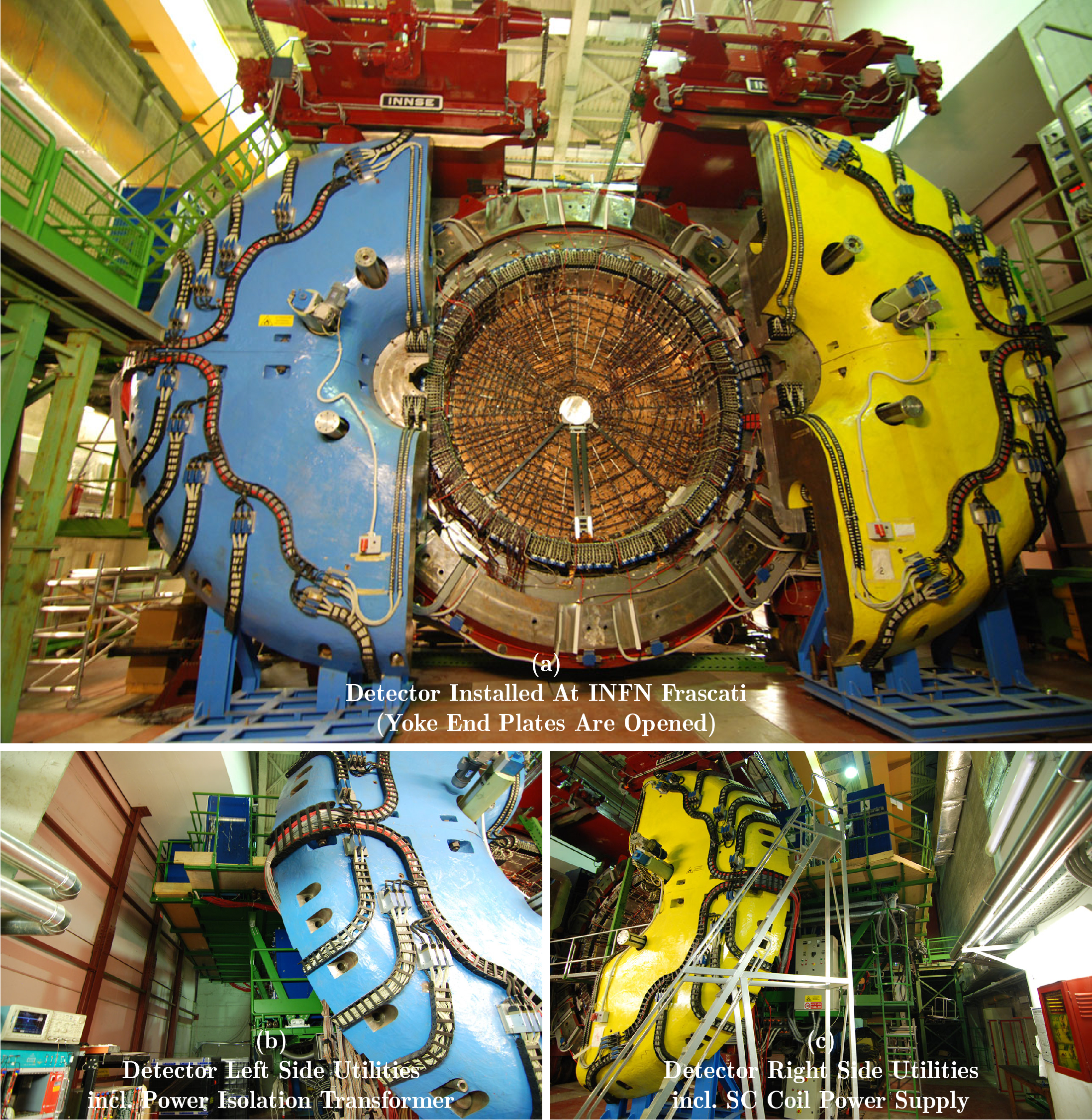}
\end{dunefigure}

The SAND subdetector comes with a lead-scintillating fiber calorimeter with photomultiplier readout. The central barrel has an inner diameter of \SI{4}{\m}, \SI{4.3}{\m} active length, and \SI{23}{\cm} thickness. Two calorimeter end caps as shown in Figure~\ref{fig:sand_schematic} close the barrel. The total weight of the calorimeter is approximately \SI{110}{\metricton}. The innermost diagnostic components, which must fit within the existing KLOE TPC volume constraints, are still in design development and will be based on 3DST, TPC, or straw tube technology.

\begin{dunefigure}[Detailed SAND beam monitor layout]{fig:sand_schematic}
{Schematic drawing of the primary SAND beam monitor components. The subdetector consists of a large superconducting coil surrounded by a thick iron yoke and end plates which produce a uniform magnetic field within the subdetector volume. The subdetector incorporates a lead-based electromagnetic calorimeter barrel. The detailed inner detector design is still ongoing and will be based on either 3DST, TPC, or straw tube technology.}
\includegraphics[width=0.8\textwidth]{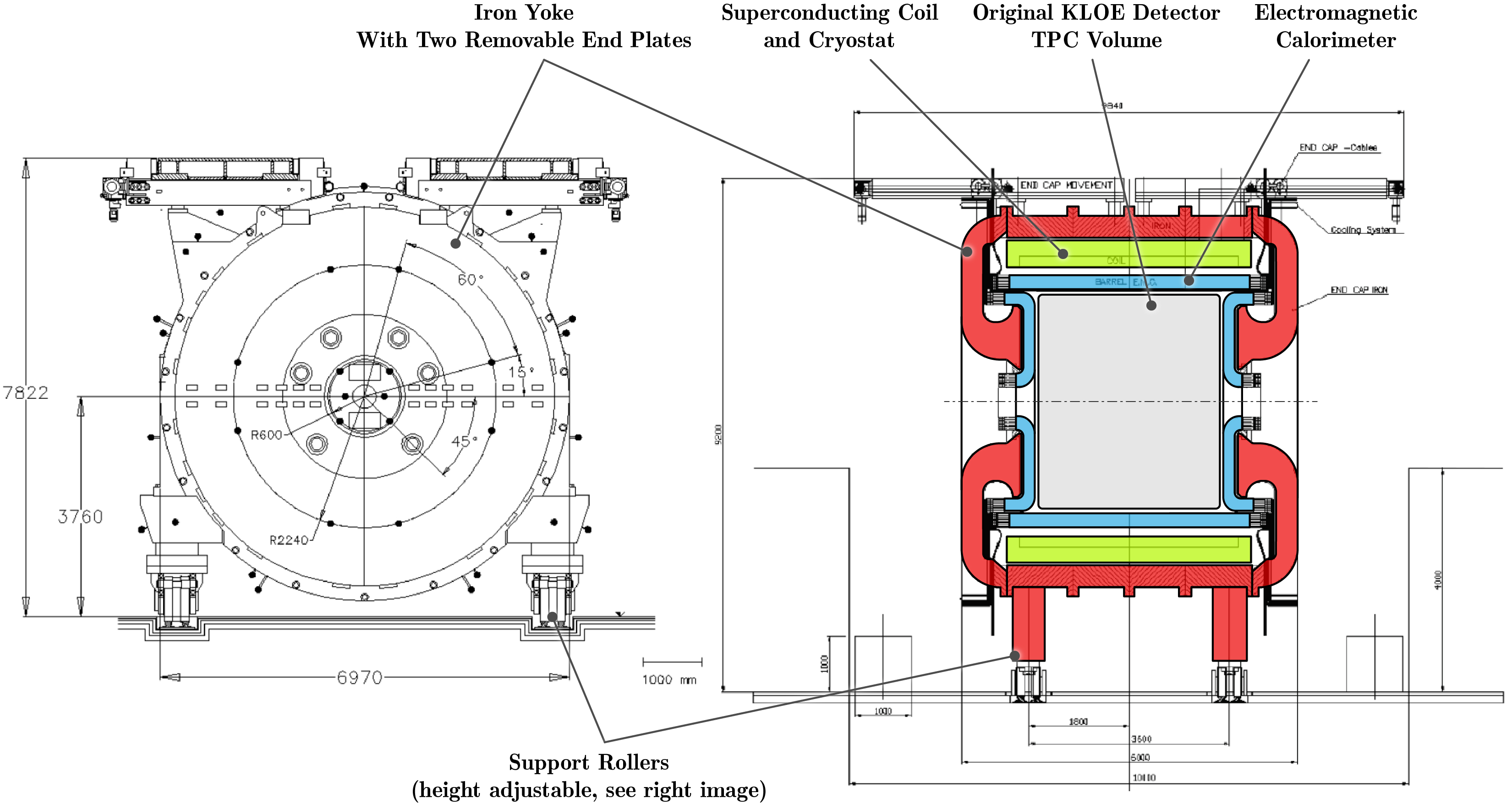}
\end{dunefigure}

\begin{dunetable}
[SAND magnet parameters]
{cc}
{tab:sand_coil_data}
{Important SAND magnet parameters.}
Parameter & Quantity \\ \toprowrule
Central Magnetic Field & \SI{0.6}{\T} \\ \colhline
Solenoid Coil Inner Diameter & \SI{5.19}{\m} \\ \colhline
Cryostat Inner Diameter & \SI{4.86}{\m} \\ \colhline
Cryostat Outer Diameter & \SI{5.76}{\m} \\ \colhline
Cryostat Length & \SI{4.4}{\m} \\ \colhline
Coldmass and Cryostat Weight & \SI{36}{\metricton} \\ \colhline
Iron Return Yoke Weight & \SI{475}{\metricton} \\ \colhline
Operating Current & \SI{2902}{\A} \\ \colhline
Stored Energy & \SI{14.3}{\MJ} \\ 
\end{dunetable}

The SAND beam monitor requires the installation of a cryoplant in the Near Detector facility. The superconducting coil is cooled by liquid helium supplied at 1.2~bar and 1.44~K, and the cryostat thermal shield is cooled by 70-80~K gaseous helium. The cooling requirements for SAND are modest, with a heat load to 4~K of less than 55~W, a heat load to the 70~K thermal intercepts of less than 530~W, and a helium flow along the current leads of 0.6~g/s. A small, commercially available cryoplant (approximately 200~W cooling capacity at 4.5~K) will be sufficient for SAND. Its coldbox can be located underground in the vicinity of the shaft as shown in Figure~\ref{fig:sand_cryo_location}. The plant compressor and oil removal skid will be located above ground in the surface support building (see Section~\ref{sec:chap-id:facility:surface}). The cryoplant requires liquid nitrogen for pre-cooling. The liquid nitrogen cryogenic system including a liquid nitrogen phase separator can be shared between the SAND liquid helium and the ND-LAr argon purification systems.

\begin{dunefigure}[SAND cryoplant location]{fig:sand_cryo_location}
{A small, commercially available cryoplant will be sufficient to cool the SAND superconducting solenoid. Its coldbox can be located underground in the vicinity of the shaft.}
\includegraphics[width=0.8\textwidth]{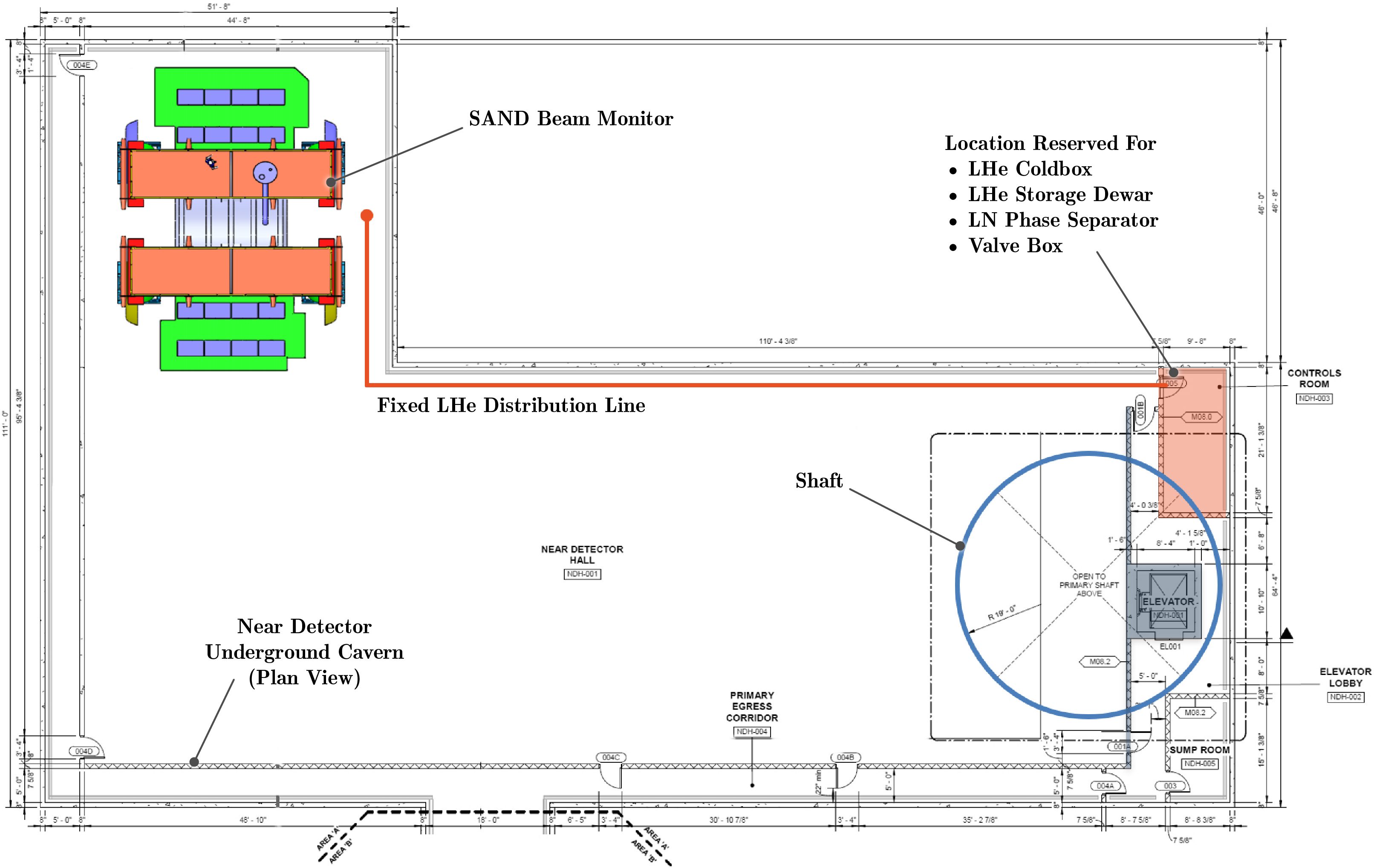}
\end{dunefigure}

\begin{dunefigure}[SAND support and movement system]{fig:sand_installation}
{SAND incorporates a versatile support system with large diameter steel rollers for smooth movement and a specialized hydraulic system to adjust the detector height. The steel rollers can be rotated by \SI{90}{\degree} which permits movement of the beam monitor from its assembly area to the final alcove location.}
\includegraphics[width=0.8\textwidth]{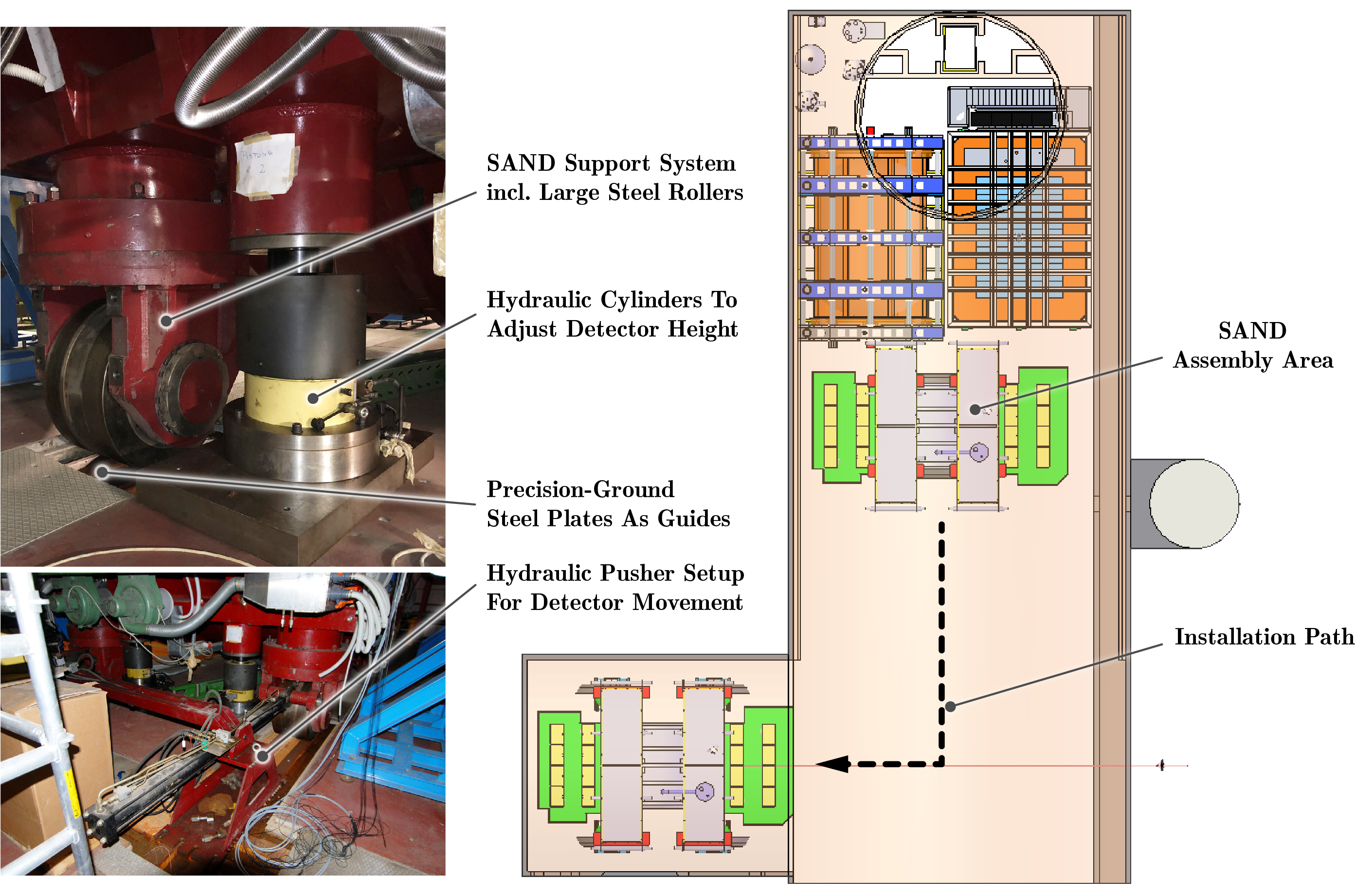}
\end{dunefigure}

SAND incorporates a versatile support system with large diameter steel rollers for smooth movement and a specialized hydraulic system to adjust the detector height, see Figure~\ref{fig:sand_installation}. A hydraulic pusher system can move the detector a couple of feet per stroke along the rail system. By repeatedly moving the pusher system the beam monitor can be positioned efficiently. The steel rollers can be rotated by \SI{90}{\degree} which permits changing installation direction. This process is highlighted in Figure~\ref{fig:sand_installation} which shows the planned SAND assembly area inside the underground cavern and the installation path towards the final alcove location. In addition, alternative plans are being developed which would permit the installation and removal of SAND while the other detectors are in place. Such a scenario would require the assembly of the SAND electronics racks and their mezzanine structures after moving the detector close to the alcove. 

\subsection{PRISM System}
\label{sec:chap-id:details:prism}

The PRISM system enables off-axis, energy-dependent neutrino beam measurements. A travel distance of \SI{30}{\m} from the nominal detector on-axis position is required to sample a wide enough energy spectrum. This requirement determines the main length of the Near Detector cavern excavation which consists of the PRISM movement range plus the ND-LAr detector width plus required conventional facility space needs on the walls. For PRISM measurements, the ND-LAr and ND-GAr move in tandem. The design also allows for individual movements that may be needed during installation and maintenance.

PRISM consists of two elements: (1) the individual detector movement platform, the motorized transport system, and rails embedded in the concrete floor; plus (2) fairly large and flexible cable carriers which support the required movement of cryogenic, power, and data lines. Figure~\ref{fig:nd_lar_setup} shows both elements for the ND-LAr subdetector.

Large and movable detectors have been built previously, primarily relying on hydraulic pushing systems (for instance, see Figure~\ref{fig:sand_installation}) which don’t permit continuous and automatic movement. Other setups incorporate rack and pinion drives which add complexity. For DUNE PRISM, an evolutionary next step incorporating  technology based on synchronized servo motor control systems will be developed. Figure~\ref{fig:prism_movement_system} highlights such a commercial (patented by Hilman Inc.), remotely operable transport system optimized for extremely heavy loads and smooth movement. It is based on a continuous track system running on a band of large-diameter, high-strength chain rollers connected to synchronized servo motors by gears and chains. Such a setup provides continuous back and forth movement without jolting.

A single transport system chain unit, as shown in Figure~\ref{fig:prism_movement_system}, can support up to \SI{200}{\metricton}. At least six to eight of such units are needed to support the >\SI{900}{\metricton} DUNE ND  detectors. The cost efficiency of using fewer chain units with higher load capacity will be investigated.
After the conceptual design phase, DUNE will work with the supplier to refine such system optimizations. Table~\ref{tab:prism_profile} summarizes a strawman PRISM movement profile and corresponding movement system requirements for an experimental off-axis run.

\begin{dunefigure}[PRISM movement system]{fig:prism_movement_system}
{Industrial transport systems for extremely heavy loads and smooth movement are commercially available. This figure shows a continuous track system running on a band of high-strength chain rollers driven by synchronized servo motors. The shown setup is patented by Hilman Inc. and would be well suited for the \dword{dune} \dword{nd} Prism movement system permitting smooth forward and back movement.}
\includegraphics[width=0.8\textwidth]{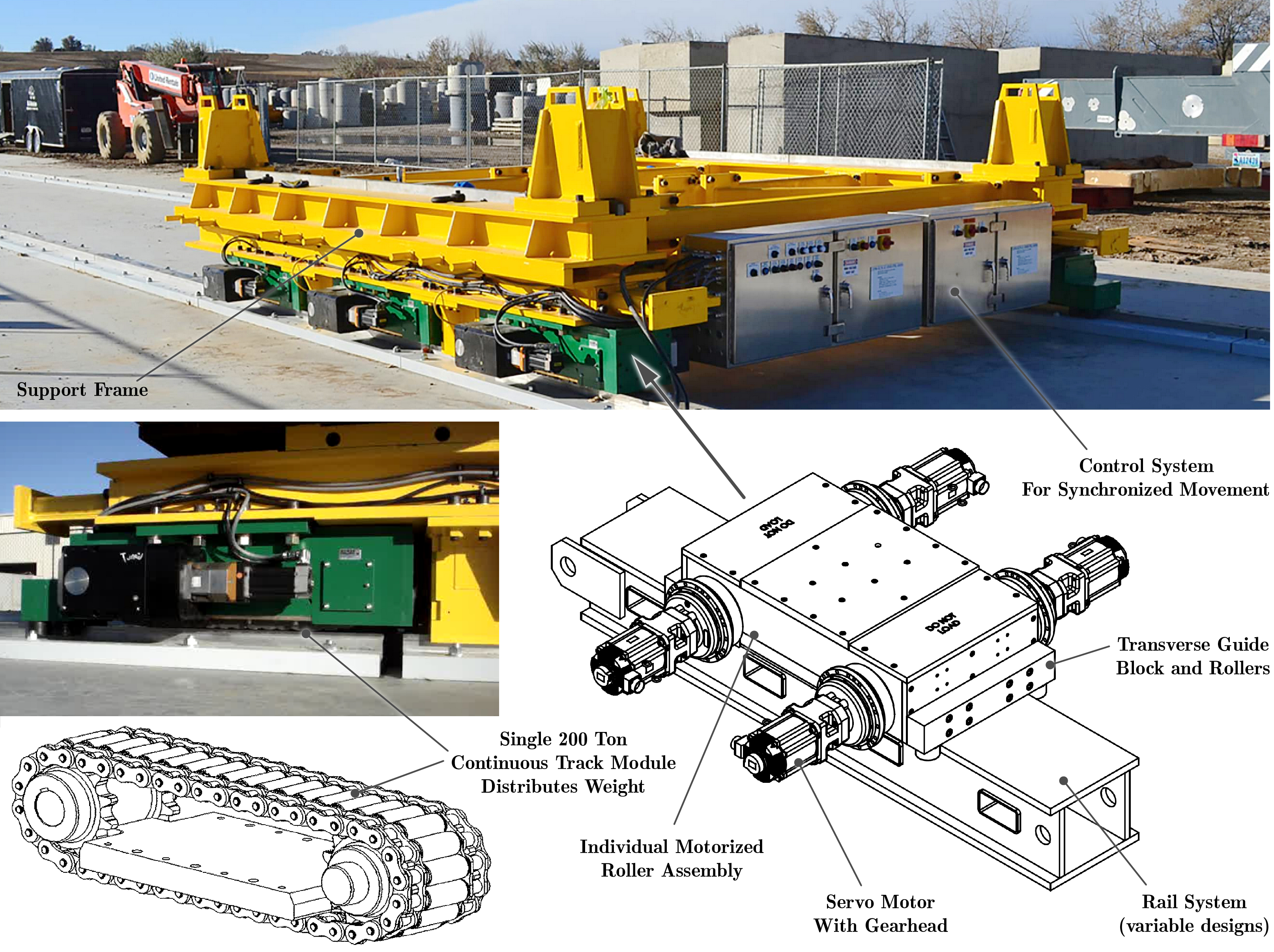}
\end{dunefigure}

\begin{dunetable}
[Strawman PRISM movement profile]
{cc}
{tab:prism_profile}
{Strawman PRISM movement profile for an experimental off-axis run.}
Parameter & Quantity \\ \toprowrule
Travel Distance & 30.5~m \\ \colhline
Average Movement Speed & 8.5~cm/min \\ \colhline
Top Movement Speed & 10.2~cm/min \\ \colhline
Time To Reach Top Movement Speed & 60~min \\ \colhline
Resulting Linear Acceleration & 0.17~cm/sec\textsuperscript{2} \\ \colhline
Time To Travel Entire Travel Distance Without Stopping & $\sim$6~hrs \\ \colhline
Planned Stops Per Experimental Run & $\sim$9 (locations may vary per run) \\ \colhline
Stop Location - Position Repeatability & <~1~cm \\ \colhline
Stop Location - Position Measurement Accuracy & 1~mm \\ \colhline
Experimental Run Time For A Full Round Trip & 2~weeks \\ 
\end{dunetable}

\begin{dunefigure}[PRISM cable chain configuration]{fig:prism_cable_chain}
{Heavy-duty cable chains are being developed for both movable DUNE ND subdetectors. The chains carry several flexible cryogenic lines of significant outside diameter, plus electrical power and data cables.}
\includegraphics[width=0.8\textwidth]{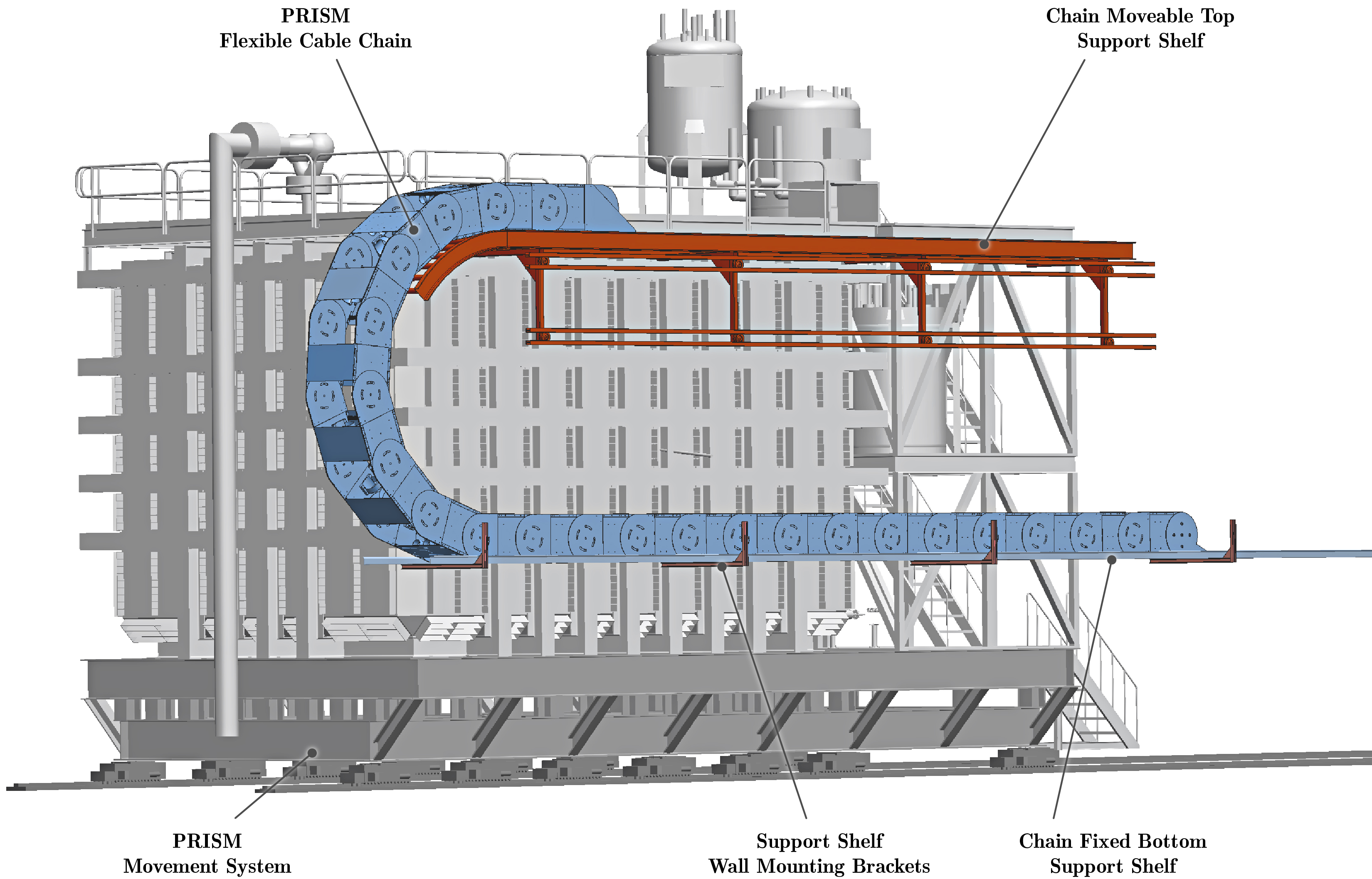}
\end{dunefigure}

To enable the movements of the \dword{ndlar} and \dword{ndgar} subdetectors, heavy-duty cable chains must be implemented, as  seen in Figure~\ref{fig:prism_cable_chain}. Depending on the subdetector needs, these cable chains will guide and support large and flexible cryogenic lines, electrical power as well as data cables. The cryogenic lines are made of corrugated, double-walled stainless steel pipes which are evacuated for thermal insulation.

The cable chain consists of custom-fabricated linkages made out of stamped aluminum sheets connected by low-friction bearings. Pipe holders inside the chain position and route the cryogenic lines and power cables. Due to the significant weight of the cables, the chain is not designed to be self-supporting. As shown in Figure~\ref{fig:prism_cable_chain} the bottom part of the cable chain is supported by a fixed platform, whereas the top part of the chain is carried by a moving shelf which follows at half the detector speed. The chain platform and the movable top shelf are supported from the cavern wall by conventional brackets.  The \dword{ndlar} cable chain can be seen well in the right plot of Figure~\ref{fig:nd_detector_arrangement}.

Due to the novel capabilities of the PRISM system  all core components will be prototyped, including the high-load  roller assemblies, the servo motor control system, as well as a representative length of flexible cable chain including evacuated cryogenic lines.

\section{Near Detector Facility and Installation Planning}
\label{sec:chap-id:facility}

\subsection{Surface Building and Rigging Access}
\label{sec:chap-id:facility:surface}

The \dword{nd} facility includes a steel-frame building located above ground on top of the primary shaft. An architectural model is shown in Figure~\ref{fig:nd_hall_3d}. Key dimensions are summarized in Table~\ref{tab:surface_building_dims}. The building is oriented parallel to the FNAL property boundary and includes architectural features to minimize impact on the neighbors. The front façade which is directed towards Wilson Hall incorporates transparent cladding which provides a well-lit highbay for equipment staging. See Figure~\ref{fig:surface_building_architectural} for a surface building conceptual architectural rendering plus floor plan.

\begin{dunefigure}[DUNE ND surface building architectural rendering]{fig:surface_building_architectural}
{The Near Detector facility includes a steel-frame building located above ground and on top of the primary shaft. The front façade incorporates transparent cladding which provides a well-lit highbay for equipment staging.}
\includegraphics[width=0.7\textwidth]{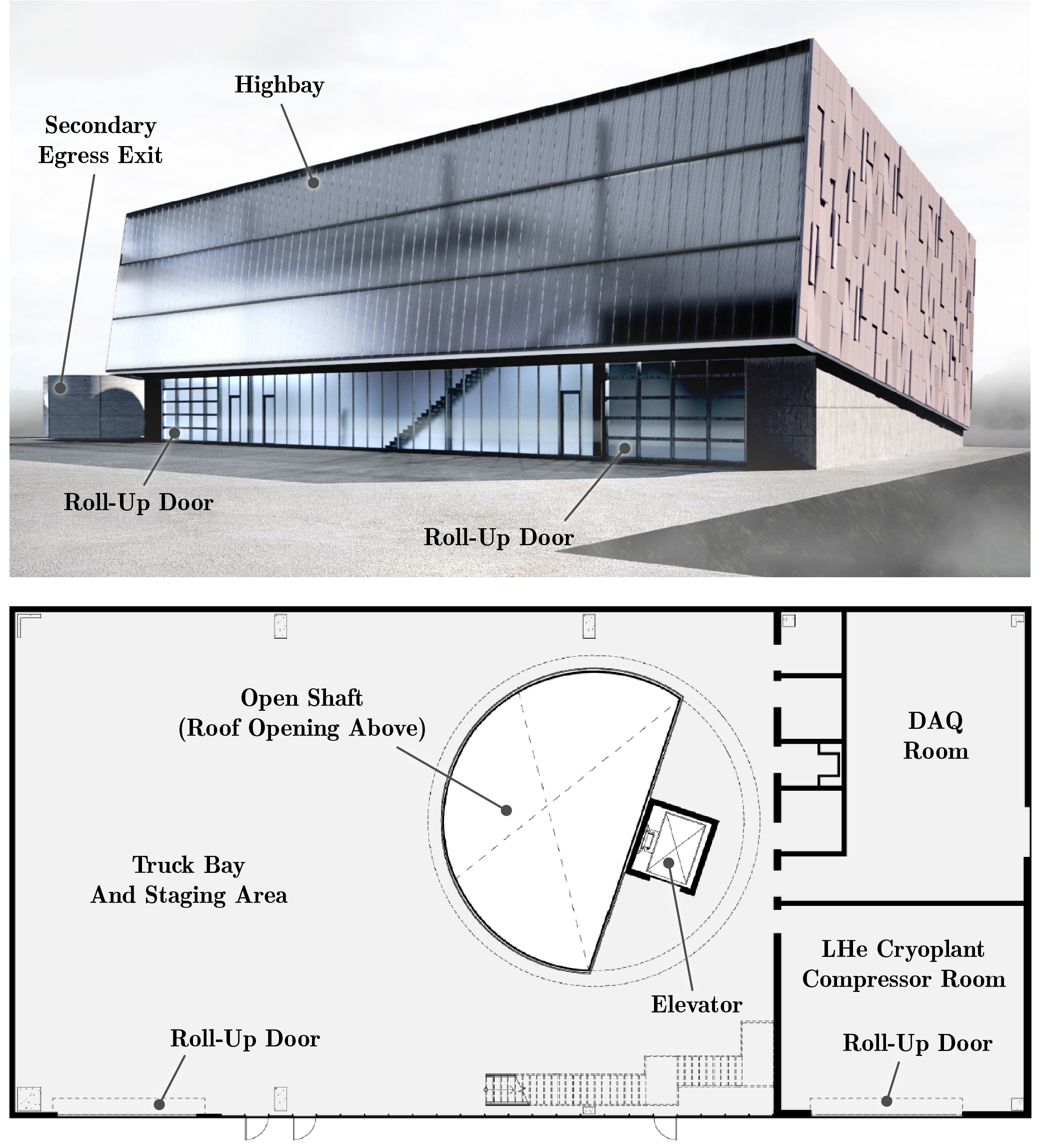}
\end{dunefigure}

\begin{dunetable}
[Surface building dimensions]
{cc}
{tab:surface_building_dims}
{Surface building dimensions.}
Parameter & Quantity \\ \toprowrule
Length & 128~ft \\ \colhline
Width & 64~ft \\ \colhline
Height & 38~ft \\ \colhline
Crane Capacity & 15~ton \\ \colhline
Primary Shaft Clear Diameter & 38~ft \\ 
\end{dunetable}

The surface building will house important support equipment for detector operation. As shown in Figure~\ref{fig:surface_building} a LHe cryoplant compressor room and a data acquisition (DAQ) server room will be located on the first floor. The compressor room will also accommodate a cryoplant oil purification system, an air compressor, and control racks. The DAQ room will contain a separate air conditioning unit and uninterruptible power supply.

All primary building mechanical systems will be located on the second floor, above the DAQ and compressor rooms. Most of that space will be occupied by the air handling system for the underground cavern which has to provide sufficient air flow (two redundant fan units providing 7,500~ft\textsuperscript{3}/min ventilation volume each) for oxygen deficiency hazard situations. The remaining rooms in the surface building accommodate building control systems, restrooms, storage, and fire suppression control systems.

The front side of the building will include a concrete driveway of sufficient size and strength to enable truck deliveries of detector equipment or cryogens. The LAr, LN, and GHe storage tanks will be located in the vicinity, see Figure~\ref{fig:surface_building}. The surface building highbay will permit the loading and unloading of standard tractor trailers. The highbay rollup door is dimensioned to allow backing in of a tractor trailer loaded with fully assembled LAr modules.

\begin{dunefigure}[Detailed surface building layout]{fig:surface_building}
{The surface building houses important support equipment for detector as well as building operation. In addition, the building provides rigging and staging capabilities to lower equipment into the underground cavern.}
\includegraphics[width=0.7\textwidth]{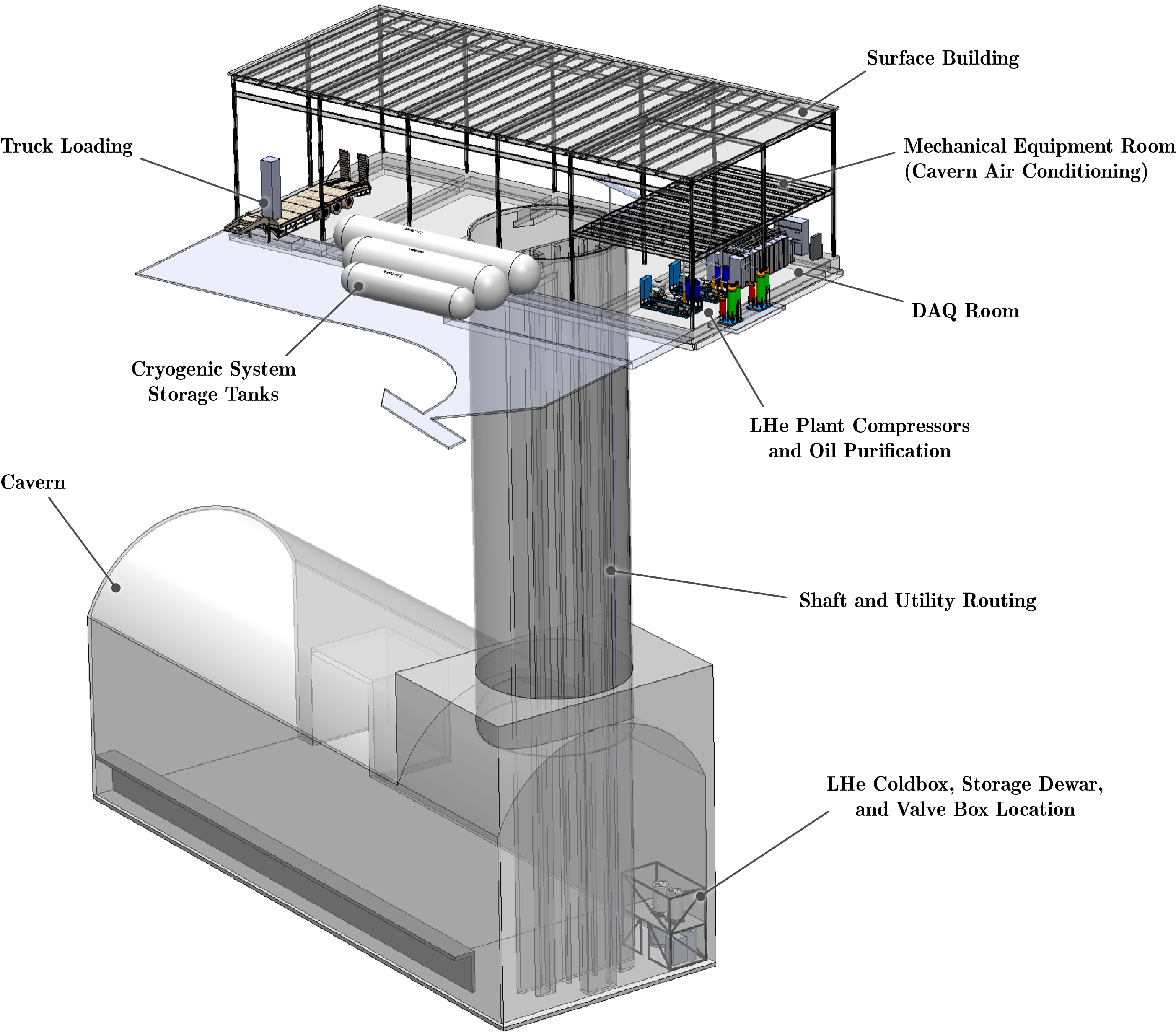}
\end{dunefigure}

The open shaft to the underground cavern provides detector rigging access. A significant portion of the shaft is also reserved for a personnel elevator and the large ventilation ducts. Electrical power and cryogenic distribution lines will also be routed through the shaft. Since the highbay has only limited building height and crane capacity, a large roof hatch located above the shaft will be utilized to lower heavy detector components underground. Figure~\ref{fig:surface_building_shaft_access} illustrates a few representative rigging setups utilizing rental cranes which can be positioned outside the surface building. The primary shaft diameter has been chosen to fit the ND-GAr pressure vessel and the SAND solenoid cryostat. Neither detector can be disassembled into smaller pieces or rotated. The large shaft size would also permit rigging fully welded and assembled ND-LAr cryostat warm structure panels underground which could significantly ease underground installation. 

\begin{dunefigure}[DUNE ND rigging access]{fig:surface_building_shaft_access}
{A roof hatch in the surface building and above the shaft will be utilized to lower large and heavy detector components underground utilizing rental cranes. The primary shaft diameter is large enough for the ND-GAr pressure vessel or the SAND solenoid cryostat. Such a shaft size would permit rigging fully welded ND-LAr cryostat warm structure panels underground which could significantly ease underground installation.}
\includegraphics[width=0.9\textwidth]{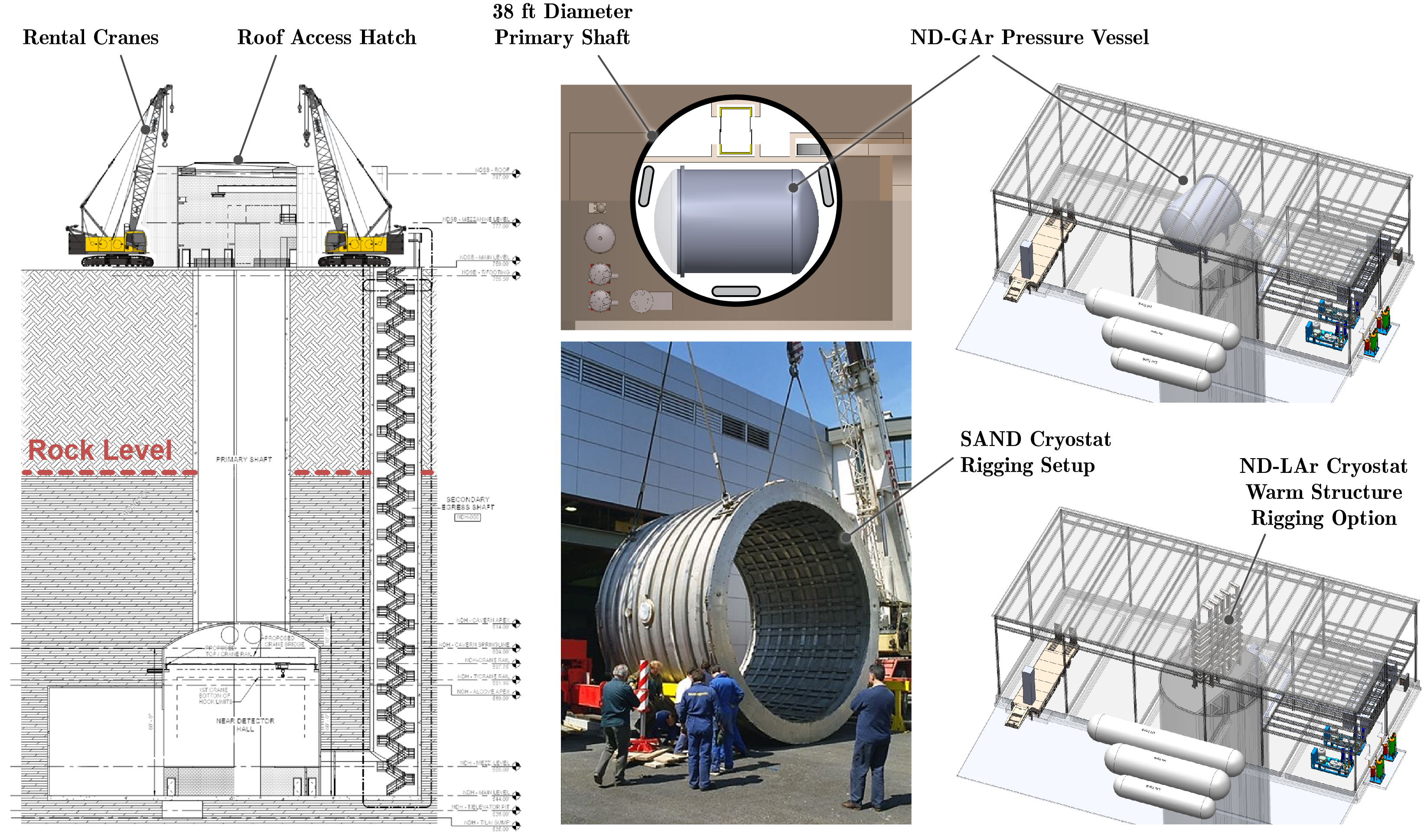}
\end{dunefigure}

\subsection{Auxiliary Building Systems}
\label{sec:chap-id:facility:additional}


A new electrical main power feed will be installed to service the Near Detector Facility. A preliminary estimate for detector power requirements is shown in Table~\ref{tab:power_needs}. A corresponding power distribution line diagram for the Near Detector scientific equipment has been developed, see Figure~\ref{fig:power_distribution}. Conventional Facilities will design and install the main circuit breaker panels and the conventional power transformers in the surface building as well as the cavern. All three subdetectors will be electrically isolated from building ground to reduce electrical noise and cross-talk. Therefore, the Near Detector Integration \& Installation electrical engineering group will design the respective subdetector isolation transformers and cable routing. A detailed grounding plan for each subdetector will be developed during the preliminary design phase. The ND-LAr subdetector will closely follow ProtoDUNE design approaches and lessons learned. The SAND subdetector is a self-contained detector with an isolated ground.

\begin{dunetable}
[Near Detector electrical power needs]
{cc}
{tab:power_needs}
{Near Detector Electrical Power Needs.}
System & Power \\ \toprowrule
LHe Cryoplant (surface building) & 500~kVA \\ \colhline
DAQ Room and UPS (surface building) & 75~kVA \\ \colhline
ND-LAr & 225~kVA \\ \colhline
ND-GAr & 225~kVA \\ \colhline
SAND & 500~kVA \\ \colhline
LHe Cold Box & 75~kVA \\ \colhline
PRISM Motors & 75~kVA \\ \colhline
LAr Cryogenics & 45~kVA \\ 
\end{dunetable}

\begin{dunefigure}[DUNE ND power distribution diagram]{fig:power_distribution}
{DUNE ND detector power distribution diagram.}
\includegraphics[width=1.0\textwidth]{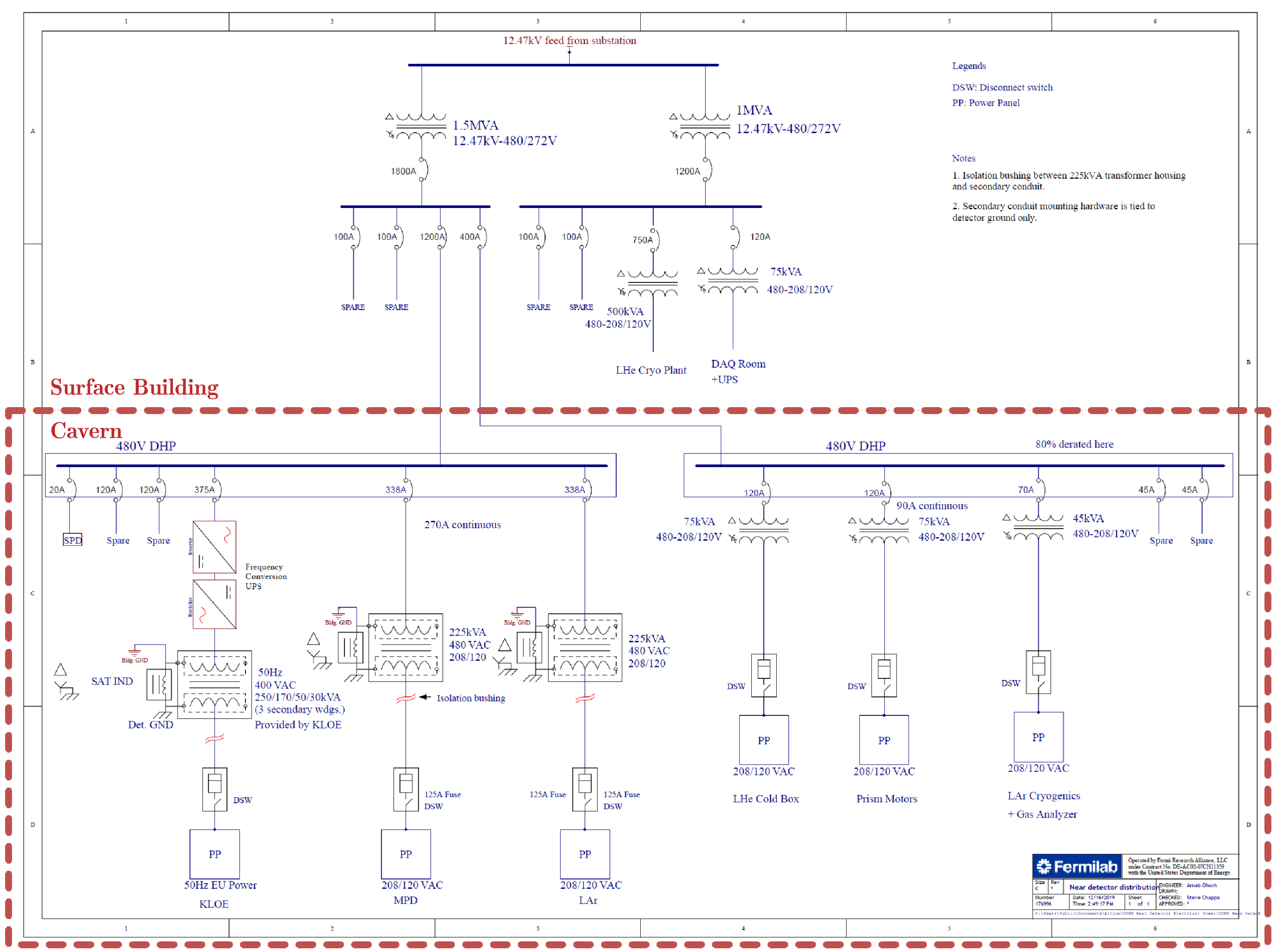}
\end{dunefigure}

The SAND subdetector is built based on European electricity standards. Therefore, the US power feed frequency must be converted from 60 to 50~Hz. As shown in Figure~\ref{fig:power_distribution}, a commercially available frequency conversion system similar to ones used in UPS units can be installed. This approach would be an elegant and efficient way of adapting the subdetector to the US line frequency and eliminating the need of modifying SAND components.

Additional building services will be required in addition to electrical power. These include:
\begin{itemize}[noitemsep,topsep=-10pt]
\item Air conditioning,
\item Industrial water cooling,
\item Compressed air,
\item Lighting,
\item Stand-by power,
\item Access control,
\item Fire safety systems, and
\item Oxygen deficiency hazard (ODH) safety systems.
\end{itemize}
Requirements for most of these systems are comparable to similar detector facilities and will be fully defined during the conventional facilities final design phase. More unique design features for these systems are described next.

The Near Detector cavern temperature should be controlled to \SI{21}{\celsius} $\pm$ \SI{5}{\celsius}. To limit condensation on cryogenic lines and the cryostat structures cavern humidity should ideally be maintained below 50$\%$~RH at \SI{21}{\celsius}. However, such an air conditioning system could become quite expensive since water may be continually present inside the cavern. A proposed, more cost-effective air conditioning approach would be to generate conditioned (dried) air at the surface building and blow it into the cavern. This approach can control temperature and would significantly reduce cavern humidity. However, such an approach cannot actively control or guarantee cavern humidity which will depend on the water present in the cavern or weather conditions on the surface. Localized heating or air flow may be required to minimize excessive ice forming on cryogenic systems or to reduce water formation on electronics and could be implemented after initial detector operational experiences.

\begin{dunefigure}[ODH Ventilation Duct Routing]{fig:odh_vents}
{Conceptual routing of the ODH ventilation ducts (the supply line is in green and the exhaust line is in dark pink) inside the Near Detector cavern. Only the LAr detector is shown for clarity. The exhaust line must be directed to the bottom of the cavern to be able to collect argon gas along the travel path of the ND-LAr detector. The large ventilation ducts will be routed through the main shaft to the surface building mechanical equipment room (see Figure~\ref{fig:surface_building}).}
\includegraphics[width=0.7\textwidth]{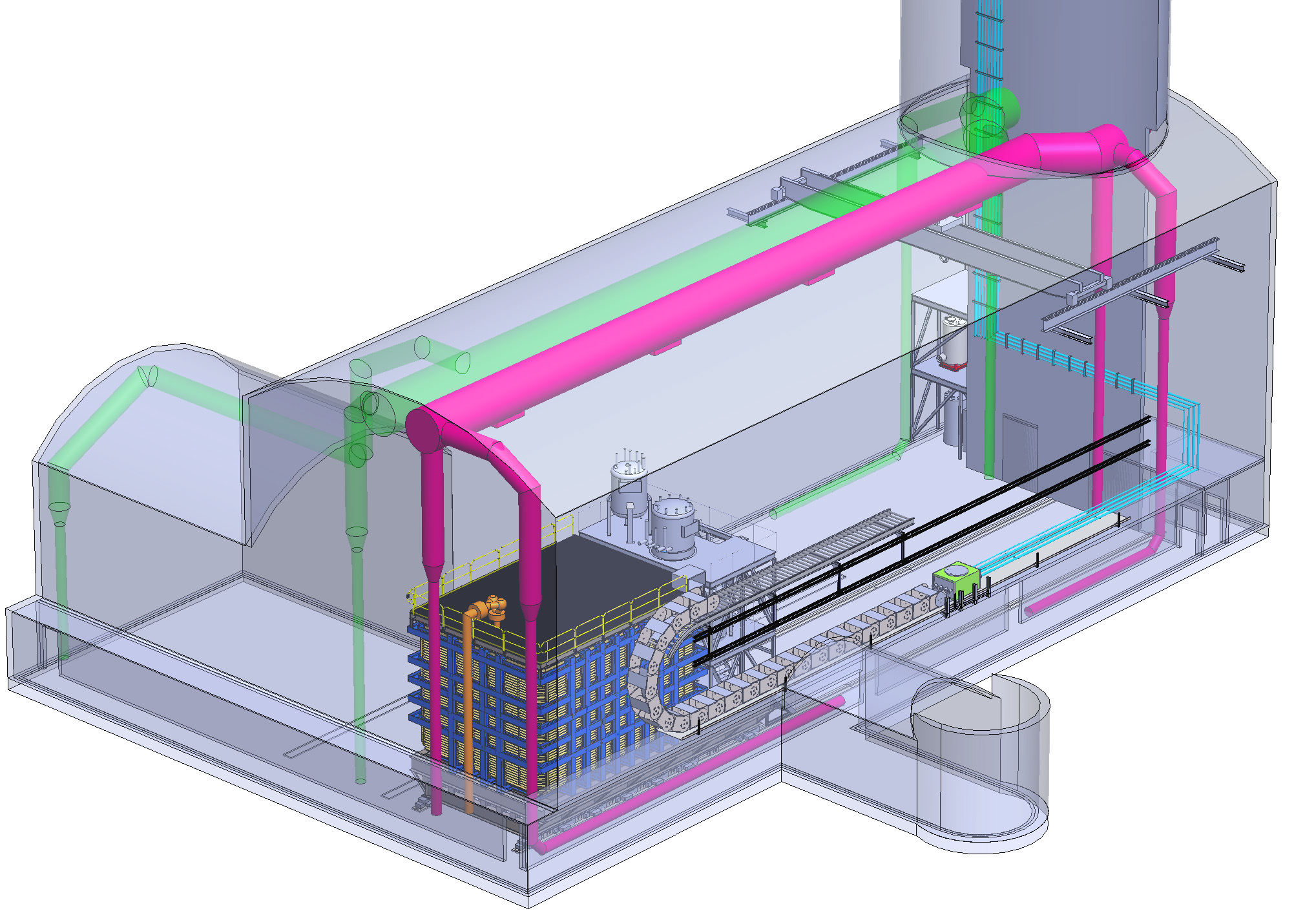}
\end{dunefigure}

An ODH scoping calculation has been performed for the Near Detector cavern. Based on this study the air ducts will be  initially designed to provide >7,500~ft\textsuperscript{3}/min ventilation volume. Two fan units will be implemented to provide redundancy. Since large quantities of liquid helium (lighter than air) as well as liquid argon (heavier than air) cryogens are present in the cavern the ventilation ducts will have adjustable louvers on the top as well as the bottom of the cavern. A suggested routing of the ventilation ducts is shown in Figure~\ref{fig:odh_vents}. Similar to ProtoDUNE the current ND-LAr detector design incorporates a cryostat emergency vent discharging to the bottom of the cavern floor in close vicinity to the ODH ventilation ducts. The cavern is expected to be designated class ODH-1. Such a hazard class would require the use of personal oxygen monitors and self-rescue oxygen packs. A detailed safety analysis according to the FNAL ES\&H manual will be performed once the subdetector designs have further matured.

Industrial water cooling will be required in the underground cavern for the LHe coldbox as well as detector power supplies and electronics. Compressed air will be needed for underground cryogenic valve operation. These systems plus standby power and lighting needs will be specified during the detector preliminary design phase. DUNE systems engineering as well as DUNE Integration \& Installation maintain integrated CAD models for clash detection between detectors and conventional facilities equipment.

\subsection{Installation Schedule}
\label{sec:chap-id:facility:installation}

An initial installation schedule, see Figure~\ref{fig:installation_schedule_ndgar}, has been developed based on the available conceptual design information for the cavern and the detectors (see previous sub-sections). Individual detector installations must be staggered in time due to movement constraints given by the size of the equipment and the limited space available underground. For instance, SAND should be moved in place before the other two detector installations can be finalized. The ND-LAr membrane cryostat requires final assembly and welding in situ. ND-GAr installation can only be initiated once the ND-LAr membrane cryostat has been completed.

\begin{dunefigure}[DUNE Near Detector Facility Installation Schedule]{fig:installation_schedule_ndgar}
{A conceptual \dword{dune} \dword{nd} Facility installation schedule. Installation will commence ("FY~x+1" in the displayed schedule) once cavern construction has been completed and authorization to use the \dword{nd} facility has been granted. Assembly of the full \dword{dune} \dword{nd} reference design including checkout and transition to operation will require approximately three years to complete.}
\includegraphics[width=1.0\textwidth]{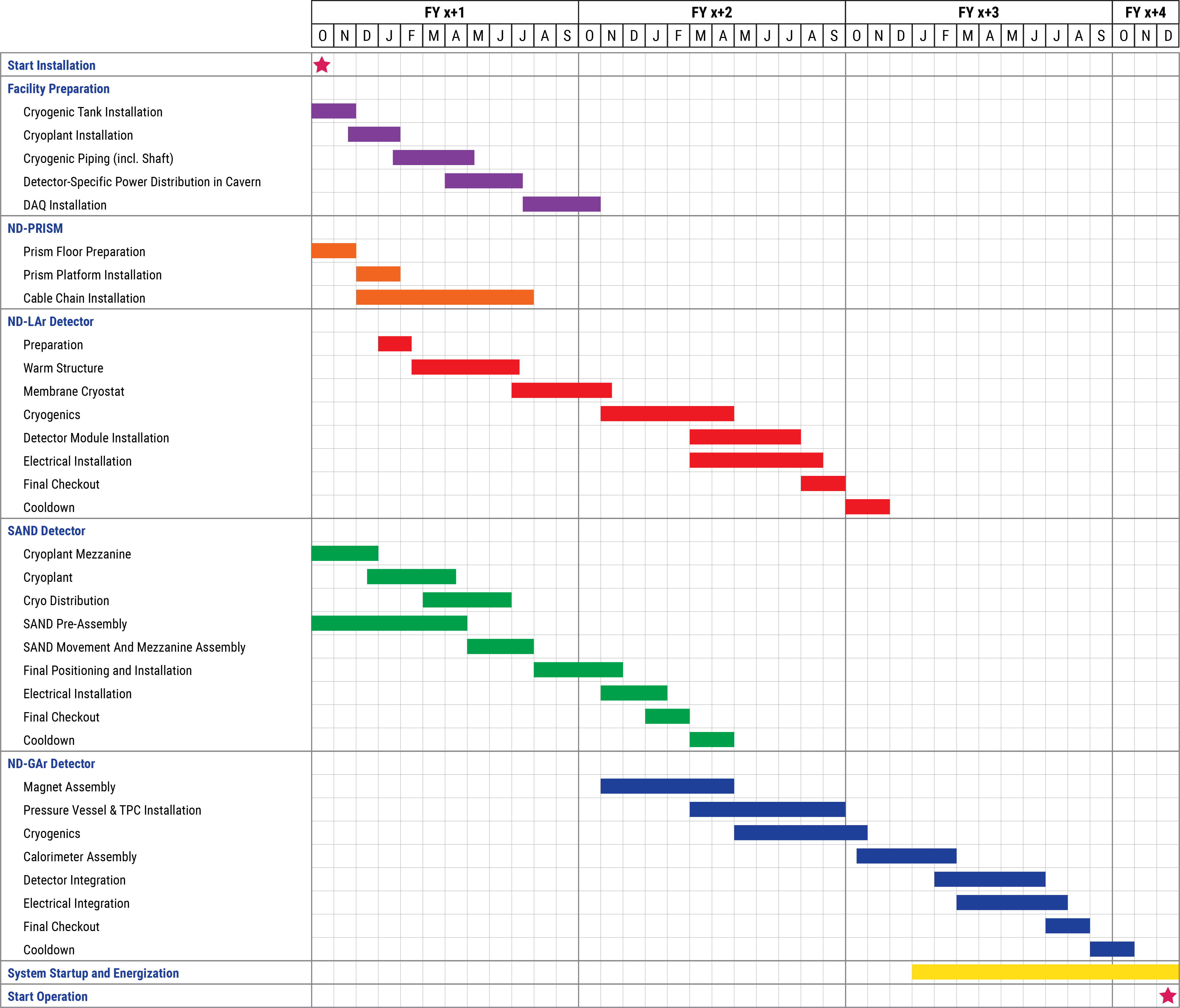}
\end{dunefigure}

Installation activities will commence once cavern construction has been completed and authorization to use the Near Detector facility has been granted. Assembly of the full DUNE Near Detector reference design including checkout and transition to operation will require approximately 3 years to complete. Near Detector installation will require close coordination between the detector consortia and the DUNE project. All on-site installation activities will be coordinated and scheduled by a DUNE~ND Integration \& Installation group to ensure safe occupancy of the installation areas and to optimize manpower utilization. In addition, the DUNE~ND Installation group will provide common rigging hardware, clean room structures, lifting platforms, and shared tooling and fixturing to perform typical installation tasks. To support the detector consortia during detector assembly, \dword{dune} \dword{nd} Integration \& Installation will employ a core group of four mechanical technicians, shift supervisors, an electrical technician and engineer, and two manufacturing engineers. At the same time, the individual detector consortia will provide additional personnel and specialized tooling needed to ensure that the three-year installation schedule can be achieved. This approach will require in-depth agreements between DUNE and the detector consortia. These agreements together with a detailed installation workflow will be developed during the final design phase once all detector component designs have matured.

Figure~\ref{fig:installation_schedule_ndgar} outlines a potential sequence of Near Detector installation activities. Once authorization to use the building structures has been granted essential detector infrastructure will be installed first. This work includes the installation of cryogenic tanks and cryoplant equipment in the surface building. Likewise, the ND-GAr and SAND LHe coldbox equipment and mezzanine structure (located underground, see Figure~\ref{fig:sand_cryo_location} and Figure~\ref{fig:surface_building}) plus the PRISM rail system and moving platforms will be installed during this first phase.

Next, major rigging activities will be initiated. SAND components will be lowered into the cavern to permit assembly of the main structural components including the coil cryostat. Assembly will occur underground in the vicinity of the shaft. Once the SAND cryostat has been inserted into the iron yoke and the yoke end plates have been installed the detector can be moved close to the beam monitor alcove. All subsequent SAND assembly steps will take place in this location which frees up space close to the shaft to complete parallel assembly of the ND-LAr detector warm structure and to initiate the installation of the membrane cryostat. The SAND cryogenic distribution line installation will be completed during that later time frame. All major rigging activities will require the rental and installation of gantry crane structures with sufficient (>50~ton) load capacity. The current LBNF preliminary cavern design only includes a 15~ton overhead crane which is not sufficient for any of the detector assembly steps.

The installation of the ND-LAr cryostat follows a process similar to ProtoDUNE. Once the PRISM movement frame has been installed the platform surface will be shimmed to bring it within the membrane cryostat flatness requirements. Next, the conventional, warm cryostat structure will be erected and the inner steel membrane welded leak tight. At this point the membrane cryostat company can proceed with assembling the insulating foam structure, the inner stainless-steel liner, and the bottom LAr valve feedthrough. A temporary top plate cover will be mounted to enable leak-checking of the full cryostat assembly.

An extended period of cryogenic installation activities follows the completion of the ND-LAr cryostat assembly. A prefabricated, large mezzanine structure will be erected to support the ND-LAr purification system, and all the cryogenic equipment and piping will be installed. That work also includes connecting the cryostat to the cable chain (see Figure~\ref{fig:prism_cable_chain}) which has been mounted to the cavern wall earlier.

Individual ND-LAr pixelated detector modules, which will be pre-assembled and tested at FNAL in a separate production facility (IREC, currently under construction), will be transported into the Near Detector surface building and assembled to a large flange combining an entire detector row (see Figure~\ref{fig:nd_lar_details}). These rows will be lowered into the cavern and immediately installed in the cryostat. Installation of all electrical utilities, racks, and control chassis will occur in parallel.

Once SAND has been moved into its final position and the ND-LAr membrane cryostat has been assembled the cavern space under the vicinity of the shaft becomes available for ND-GAr detector assembly. Major parts of the ND-GAr detector are still in an early conceptual design stage. Therefore, the proposed schedule is based on a preliminary detector configuration as described in Subsection~\ref{sec:chap-id:details:mpd}.

The ND-GAr detector magnet structure consists of five individual cryostats which will be lowered into the cavern in a pre-assembled state\footnote{This description is relevant for the 5-coil magnet design.  The procedures for constructing the SPY design are under study.}. Subsequently, these cryostats must be mounted on the PRISM support platform and connected to each other underground. Similarly, the TPC and its pressure tank will be pre-assembled above ground and lowered into the cavern as one unit. A special lifting fixture will permit the insertion of the tank into the magnet structure. Added time is allocated for effort to complete the complex assembly and installation of the TPC components underground. Cryogenic equipment installation and connections to the cryostat can proceed in parallel. Next, the ND-GAr electromagnetic calorimeter segments plus any additional tracking devices will be installed in between the cryostat structures and around the detector ends. At this point assembly of the main ND-GAr detector components will be complete and final detector and electrical integration can start followed by several months of checkout and magnet cooldown.

Finally, individual detector start-up can commence which is a multi-step process requiring subsystem sign-offs, safety approvals, and step-wise equipment energization. During that time frame coordination of and responsibility for Near Detector installation activities will gradually transition to the international detector consortia and their scientific operation teams. At this point, the DUNE Near Detector Facility is ready to start scientific operation.

%

\cleardoublepage

\chapter{Computing and DAQ for the ND}
\label{ch:comp}

\section{Introduction}

This chapter briefly introduces the 
computing model (Sections~\ref{sec:computingoverview}-\ref{sec:ndcomputingresourceusagescenario})  and
DAQ concept (Sections~\ref{sec:daq:intro}-\ref{sec:daq:refdes:datasel}) for the \dword{dune} \dword{nd}, with its three subdetectors: \dword{ndlar}, \dword{ndgar}, and \dword{sand},
as well as prototypes for these detectors.  It touches on the relationship of the \dword{nd} to the \dword{fd} and the physics program.  The \dword{dune} Collaboration will prepare a more complete design of the \dword{nd} computing model for the  Technical Design Report.

\section{Overview}
\label{sec:computingoverview}

\begin{figure}[H]
  \centering
    \includegraphics[width=7 in]{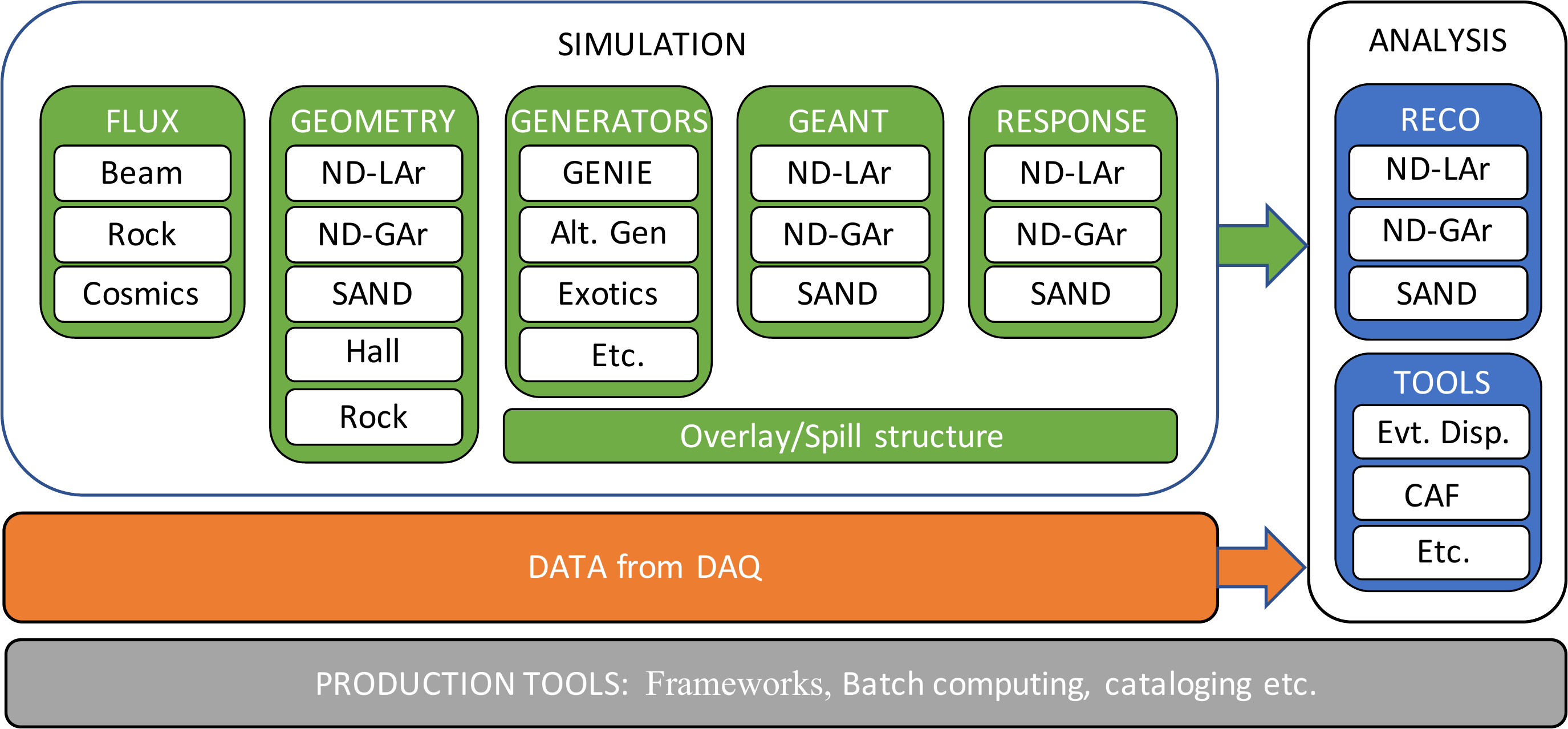} \caption[Overview DUNE ND Computing]{\label{fig:ndc_computing overview} Overview of DUNE \dword{nd} computing.}
  \end{figure}

\dword{dune} is planning on commissioning the first of four \dword{fd} modules between 2024 and 2026, and adding the remaining three modules in the time frame up to 2036.  The neutrino beam, to be provided by the LBNF, with 1.2 MW of protons on target is expected to be commissioned in 2028.  The \dword{dune} \dword{nd}, situated downstream from the decay pipe and absorber, is required to be ready to take and analyze data when the first beam is produced.  
The computing model for the \dword{dune} \dword{nd} shown in Figure~\ref{fig:ndc_computing overview} encompasses handling and processing of raw data produced by the detector subsystems, as well as the production, handling and processing of detector simulations through to common analysis files (CAF). Many of these steps will require the development of tools which integrate the output from the subdetectors to produce physics, and an analysis object which will meet the needs of the DUNE physics and data taking program. This integration work is currently in progress and a complete description of the integration requirements will be provided for the Near Detector Technical Design Report.  

The computing organization, personnel, infrastructure, and services are expected to have a significant overlap with those needed for the \dword{fd}~\cite{tdr-vol-1}.  Specifically, the Computing Consortium~\cite{tdr-vol-1} will also manage the near detector's computing services. We expect the data services, storage and distribution tools put to use for the FD will be substantially the same as those used by the ND, both to improve the efficiency of service provider support, and also convenience for collaborators who would like to analyze all of DUNE's data without learning how to use several redundant tools. The ND computing and software is being designed with an eye to integration across the ND subsystems and DUNE as a whole. The ND software groups meet regularly and are working on design towards a common data model that will be supported by standard DUNE computing tools and services. We expect to have this design more fully developed for the upcoming ND TDR, but don't anticipate major specialized ND specific computing needs.  It is also expected that the computing data volume and needs for the \dword{nd} will be small compared to that of the \dword{fd} in terms of raw data. However, simulations of the \dword{nd} may be comparable in size to the \dword{fd} data volume.  

In the following sections, data types, data volumes and CPU estimates are given for the three primary subsystems in steady-state operation.  Common assumptions are listed first. Descriptions of the data rate requirement for  \dword{nd} calibration running modes and large-scale prototyping efforts are also briefly discussed. 

\section{Steady-State Data Types and Volume Estimates}
\label{sec:comp-dataestimates}

In the following subsection we present estimates of the data volumes expected from each component of the \dword{dune} \dword{nd}. 

\subsection{Beam and Detector Downtime estimations}
\label{sec:comp-dataestimates-downtime}
In order to estimate the data volume from the \dword{nd}, a set of assumptions will be used about the standard operating conditions based on the current reference design and experience from other  neutrino experiments such as NOvA, T2K, MicroBooNE and MINERvA. The inter-spill time is expected to be 1.1~sec.  The beam is expected to run for 60\% of the possible time each year, leading to $\approx 1.7\times 10^7$ spills annually.  The estimated beam downtime includes both scheduled interruptions and random operational issues.  Included in the estimates are: summer shutdowns each year, typically lasting eight weeks; target and horn change-outs (about one week each); detector maintenance, repair, and calibration; and, in the case of \dword{ndlar} and the \dword{ndgar}, the time required to move to off-axis positions.  Effort will be made to align detector downtime with beam downtime to the maximum extent possible, but we assume a 5\% downtime for each of the primary subdetectors on top of the beam availability fraction.

\subsection{Detector Components}
\label{sec:comp-dataestimates-detctors}
With the annual spill count presented above, an initial estimate of the data volumes expected from each \dword{nd} detector subsystem using the current design and anticipated data collection modes is given in this section.  The results are summarized in Table~\ref{tab:nd_data_volume_estimates}.

\begin{dunetable}[Annual DUNE ND detector data volumes]{l|r}{tab:nd_data_volume_estimates}{Annual DUNE near detector data volume estimates, in Terabytes.  No compression is assumed. }
 {\bf Subdetector} & Terabytes \\ \hline \hline   
    {\bf \dword{ndlar}}     &  \\
    \quad\quad In-spill data & 144 \\
    \quad\quad Out-of-spill cosmics & 16 \\
    \quad\quad Calibration & 16 \\
    \quad\quad Total & 176 \\\hline
    {\bf \dword{ndgar}}           & \\
    \quad\quad In-spill data & 52 \\
    \quad\quad Out-of-spill cosmics & 10 \\
    \quad\quad Calibration & 6 \\
    \quad\quad Total & 68 \\\hline
    {\bf \dword{sand}}        & \\
        \quad\quad In-spill data & 4 \\
    \quad\quad Out-of-spill cosmics & 1 \\
    \quad\quad Calibration & 1 \\
    \quad\quad Total & 6 \\\hline
    {\bf Total ND} & {\bf 250}\\
\end{dunetable}

\subsubsection{\dword{ndlar}}
\label{sec:comp-dataestimates-argoncube}

The TPC readout comprises 12~million $3\times 3$~mm$^2$ pixel channels and $\sim$4200~photon detector channels.  The TPC pulses will be read out by electronics that, instead of digitizing waveforms, provide pulse times and integrals.  Each pulse produces 10~bytes of uncompressed data and there are expected to be 150,000 pulses per spill exceeding the readout threshold.  Neighboring pads falling below the readout threshold are assumed to also be read out.  A total of 3~MB of uncompressed data is anticipated per spill from the TPC.   The in-spill uncompressed data volume per year from the TPC is anticipated to be 54~TB.  Assuming a compression factor of 3 which was achieved using lossless compression in ProtoDUNE, 18~TB of compressed data will be written from the TPC in one year.  If the full waveforms are read out for the photon detectors, a larger amount, 5~MB/spill, is expected just from the photon detectors. Adding in the photon detectors, the number rises to 144 TB/year for uncompressed in-spill data, and 48 TB of compressed data per year, assuming that the photon detector data also compress by a factor of 3.

For calibrations, 300 runs are assumed to be taken per year, each generating 10~GB of data, for the TPC, and a similar set of runs for the photon detectors.  These runs include pulser runs, laser runs, radioactive source runs, or other special-condition runs that require taking data outside of the regular spills.  Since they are not tied to the spill timing structure, they can be collected at higher trigger rates and take less time.

In addition to the beam data, cosmic rays will contribute to the data volume.  For the \dword{ndlar} geometry in the ND hall, the anticipated rate
of cosmic rays is 100~Hz.  If all cosmic ray data were collected, the data volume would be approximately 1~MB/sec.  The scenario considered here is to
collect one spill's worth of cosmic ray data for every ten beam spills, for a data volume of 6.3 TB per year.  While the activity on the cosmic-ray triggers is expected to be much less than that on a beam spill, it is assumed that the cosmic-ray triggering will continue even when the beam is off.

The TPC-only out-of-spill and calibration numbers have been scaled by 8/3 to account for photon detector data, assuming full waveform readout in these samples, yielding the same ratio as in-spill data and the same compression. 

The CPU estimates for the \dword{ndlar} given in Table~\ref{tab:NDCPUPerEvent} are very rough estimates.  The data preparation and processing algorithms are still in development.  Machine-learning techniques, especially those that leverage GPU resources for training and inference, have been developed as the first solutions chosen for addressing the reconstruction challenges of \dword{ndlar}, and so it is anticipated that the computing model will include these components.  For the time being, though we simply estimate CPU hours.

\subsubsection{\dword{ndgar}}
\label{sec:comp-dataestimates-mpd}

\dword{ndgar} is composed of 678,136 readout pads in the TPC, and approximately 3~million channels in the \dword{ecal}.  In a typical spill, approximately 60 neutrino interactions will occur in \dword{ndgar}, primarily in the \dword{ecal}.  Approximately one in five spills will generate an interaction in the gas TPC, but particles entering the gas from interactions in the \dword{ecal} will provide the bulk of the data volume.  On average there are expected to be 130,000 pulses on TPC channels per event.  We assume 10 bytes per pulse and add additional overheads to arrive at a data volume of 2 MB of uncompressed data per spill from the TPC.  The calorimeter is expected to contribute approximately 1~MB per spill of uncompressed data.

For calibrations, 300 runs per year generating 10~GB of data per run are assumed for the TPC, and a similarly-sized set of calibration runs are assumed for the \dword{ecal}.  Cosmic rays are expected to be collected between spills and when the beam is off.

A third detector component, a muon tagger, designed to separate muons from charged pions, has not yet been designed and is not included in the totals.

The CPU estimates for the \dword{ndgar} given in Table~\ref{tab:NDCPUPerEvent} are for 60-interaction simulated events, where most interactions are in the \dword{ecal}, using the development version of the software dated October, 2019.  The use of GPUs has not yet been investigated for \dword{ndgar}, but machine-learning techniques are being explored to improve the reconstruction of tracks near crowded primary vertices.

\subsubsection{\dword{sand}}
\label{sec:comp-dataestimates-sand}

The \dword{sand} data volume estimate is computed assuming an inner tracker composed of the \dword{3dst} and gas TPCs, which is one of the options under consideration.  \dword{sand}'s \dword{3dst} component is composed of 11.5~million $1\times 1\times 1$~cm$^3$ scintillating cubes, read out by 153,600 fibers.  There are expected to be approximately 2160 hits per spill.   Each hit produces signals on three fibers to enable location of the hit in 3D, and each hit on a fiber produces 20 bytes of data without compression, for a total of 0.13 MB of data per spill.  The KLOE \dword{ecal}~\cite{Adinolfi:2002zx} uses 4850 PMTs, with an estimated 5500 total hits per spill.  Each hit produces 6 bytes of data, including time and energy measurements and the channel ID.  This corresponds to 0.033 MB of packed data per spill.  Choosing the TPC light tracking option for the estimate, 
the gas TPC channel counts are modeled after the ND280 gas TPCs~\cite{Abgrall:2010hi}.  This TPC contains 124,416 Micromegas readout channels.  The zero-suppressed, compressed data volume per spill in the ND280 gas TPC's is 0.12~MB, though it could be a factor of two higher in the high-intensity LBNF beam.  The data volume from SAND is expected to be 0.3~MB/spill, although the 0.13 MB/spill from the \dword{3dst} can be further compressed.  The data volume from SAND is 4.3 TB/year with these assumptions.  The \dword{ecal} is expected to be calibrated in situ with minimum-ionizing particles from the neutrino beam and from cosmic rays.  The amount of data from out-of-spill cosmic rays is estimated to be 20\% of that of the in-spill data, or approximately 1 TB.  The data volume from SAND is significantly smaller than that from the \dword{ndlar} and the \dword{ndgar} due to the relative sizes of the three-dimensional tracking volumes and the segmentation choices.

\begin{dunetable}[CPU estimates for the DUNE ND.]{l|r}{tab:NDCPUPerEvent}{CPU time to process one data event for the DUNE near detector components, in seconds, given mid-2020 reconstruction algorithms. These are rough estimates. Simulated events take more CPU time than data events due to the generation and simulation steps, overlay, and the handling of truth information.}
 {\bf Subdetector} & Seconds \\ \hline \hline 
   {\bf ND-LAr} &  \\
    \quad\quad Monte Carlo gen+sim & 100 \\
    \quad\quad Reconstruction & 60 \\\hline
   {\bf ND-GAr} &  \\
    \quad\quad Monte Carlo gen+sim & 100 \\
    \quad\quad Reconstruction & 12 \\\hline
    {\bf SAND} & \\
    \quad\quad Monte Carlo gen+sim & 100 \\
    \quad\quad Reconstruction & 10 \\\hline
\end{dunetable}

\section{Simulation}

As shown in Figure~\ref{fig:ndc_computing overview}, the simulations for the \dword{dune} \dword{nd} have a number of components: LBNF neutrino generator, cosmic ray generation, neutrino interactions in the detector volumes and the rock surrounding the hall, propagation of produced particles, and simulation of detector response. The beam flux simulation is generated by the LBNF Beam simulation group using the G4LBNF code. Cosmic-ray interactions will also be included in the future.  A complete GDML \dword{nd} geometry will be created using a framework called dunendggd, which contains the detector, hall, and rock volumes. Neutrino interactions in these volumes are simulated using the GENIE generator and the interaction products are propagated using GEANT4. Each detector group will maintain a simulation of response to particles which appear in the active regions of the detector. Simulated interactions are overlaid into the spill structure and the final result is expected to be a simulation which matches that of the DAQ output for data taking.  Given the extra computational steps needed for MC production, the CPU required for a Monte Carlo event is expected to be five times that of a data event.

It is planned that the entire chain will be demonstrated for the TDR, and solid estimates of simulations needed for data analysis will be available at that time. Based on the experience of other long-baseline neutrino experiments, such as NOvA and T2K, it is assumed that DUNE will generate run-matched simulations where the simulation is tuned to match data taking running conditions. In this way, the minimal simulation sample size will equal that produced from data taking plus additional Monte Carlo truth information. The level of truth information deemed necessary to retain for the simulated events can lead to an increase in data volumes by more than a factor of 10. It is assumed here that a simulated event will take five times as much storage as a data event, and that we will need four simulated events for every data event.  In NOvA, the base ND simulated data set is 15 times the size of the matched near detector data set. 

In addition, the generation of systematically shifted simulations for both detector response and interaction model uncertainties within the context of the central model (currently GENIE) can further multiply the required simulation volume by additional factors of 2-5 based on the experience of running neutrino experiments. On top of this, as part of the systematic studies, additional simulations are needed to accommodate additional neutrino interaction generators such as NEUT, GIBUU, and NuWro, as well as any  alternative physics generators deemed necessary by the physics working groups.   Special samples for BSM physics studies are expected to be much smaller than the Standard Model neutrino interaction samples due to the expected small rates of these signals.

In total, the simulation data set is expected to  be in excess of 20 times larger than the raw detector data from the \dword{nd}. This puts the total data size on the same scale as that expected for the raw data from the far detector. These estimates will be refined for the \dword{nd} TDR. 

\section{Analysis}

Physics analysis involves reading the processed data, selecting subsamples of interesting data, constructing smaller ntuples for further selection and calculation of measured observables, and repeatedly reading in these ntuples to optimize the sensitivity and evaluate systematic uncertainties.  The analysis CPU and storage needs will scale with the number of analyses and the numbers of collaborators interested in performing them.  Some ntuples will be shared by more than one analysis in order to save storage, CPU and analyzer time, and to improve the reliability of analyses.  The ntuple storage is expected to be much smaller than the Monte Carlo production output. The for reference, the reduced analysis ntuples on NOvA are a 3-6 percent the size of the related full datasets. These analysis ntuples are not given a separate column in the resource usage scenario shown in Table~\ref{tab:ndcomputingscenario1resourcesbyyear} and described further in Section~\ref{sec:ndcomputingresourceusagescenario}.  Some analyses will require access to the fully produced data and Monte Carlo samples however, as some machine-learning techniques are based on raw images rather than processed hits for example.  The CPU required to do analysis is estimated to be the sum of that needed to produce the data and the Monte Carlo samples.

\section{Large-Scale Prototypes - ProtoDUNE-ND}

In 2021, the $2 \times 2$ ArgonCube demonstrator~\cite{bib:docdb12571} will be moved into the MINOS-ND hall.  It has $2\times 2$ modules, and each module is 67~cm $\times$ 67~cm $\times$ 140~cm in size (LWH).  The total active mass is 2.4~t.  Existing detector components from MINERvA  will be repurposed as upstream and downstream trackers and calorimetry.  Twelve tracker modules, 10~ECAL modules, and 20~HCAL modules will be included.  Additionally, a small high pressure gas TPC test stand may be installed as part of ProtoDUNE-ND.  These prototypes will be operated in the NuMI neutrino beam.  The operational needs will require commissioning and testing of the detector functionality as well as reading out the detector during neutrino spills.  Monte Carlo samples will be required in order to compare with the observed data and extract measured quantities.  A rough estimate of the data volume needed by ProtoDUNE-ND is of order 100~TB per year, but it could be a few times larger.

\section{Resource Usage Scenario}
\label{sec:ndcomputingresourceusagescenario}

A resource usage scenario is presented in Table~\ref{tab:ndcomputingscenario1resourcesbyyear}. The numbers assume that the Monte Carlo simulation samples necessary for the TDR preparation will match the size of approximately one year's worth of near detector operations.  Monte Carlo samples will be simulated in the years before detector operations start in order to optimize physics analyses and staging options.  Once data-taking starts, it is assumed that the Monte Carlo samples needed will have four times as many events as the data.  The additional size of the Monte Carlo truth information is expected to make the Monte Carlo data volume twenty times the size of the raw data volume, although only a factor of two is assumed for the ProtoDUNE-ND simulation.  Early simulation samples are expected to be less detailed than later ones (starting in 2024), and so the size per MC event is expected to increase over time.  No speedups from GPU acceleration or compression of the data are assumed in this scenario.  As mentioned above, Monte Carlo is expected to take five times as much CPU per event as data to fully produce.

\begin{dunetable}[Near detector computing resource estimates for CY 2020 through 2030]{l|r|r|r|r|r|r|r|r|r|}{tab:ndcomputingscenario1resourcesbyyear}{Near detector computing resource estimates for CY 2020 through 2030}
         & Data & Raw & Test & Reco & Reco & Sim & Sim & Sim & Analysis \\ 
         \rowcolor{dunesky} & events & data & data & data & CPU & events & data & CPU & CPU \\
         \rowtitlestyle Year &  [M] &  [TB] &  [TB] &  [TB] &  [MHrs] &  [M] & [TB] & [MHrs] & [MHrs] \\ \toprowrule
        2020  & 0 & 0 & 0 & 0 & 0 & 10.0 & 200 & 1 & 1  \\
        2021  & 0 & 0 & 0 & 0 & 0 & 25.0 & 500 & 3 & 3  \\ 
        2022  & 10.0 & 100 & 300 & 200 & 0.3 & 10.0 & 200 & 1.2 & 2.2  \\ 
        2023  & 10.0 & 100 & 300 & 200 & 0.3 & 10.0 & 200 & 1.2 & 2.2 \\ 
        2024  & 0 & 0 & 0 & 0 & 0.3 & 10.0 & 400 & 1.2  & 2.2 \\ 
        2025  & 0 & 0 & 0 & 0 & 0.3 & 10.0 & 400 & 1.2 & 2.2 \\ 
        2026  & 0 & 0 & 0 & 0 & 0 & 50.0 & 2000 & 6 & 6 \\ 
        2027  & 0 & 0 & 0 & 0 & 0 & 50.0 & 2000 & 6 & 6 \\ 
        2028  & 25 & 250 & 500 & 250 & 2.4 & 100.0 & 5000 & 12 & 14.4 \\ 
        2029  & 25 & 250 & 500 & 250 & 2.4 & 100.0 & 5000 & 12 & 14.4 \\ 
        2030  & 25 & 250 & 500 & 250 & 2.4 & 100.0 & 5000 & 12 & 14.4  \\ \hline 
 \end{dunetable}

We have conservatively doubled the amount of CPU needed to reconstruct a data event relative to that in Table~\ref{tab:NDCPUPerEvent} in estimating the needs in Table~\ref{tab:ndcomputingscenario1resourcesbyyear}.
The data are expected to be processed on average twice in this scenario, as the first pass will be needed as inputs for calibration and the second pass will incorporate these calibrations and other improvements, and will be used as input for physics analyses.  Infrequent reprocessings of the entire ND data sample will be needed beyond 2030 to accommodate significant algorithmic improvements or needs for consistency with the \dword{fd} analyses.

In addition to the raw data stream, test data are expected to be generated by the DAQ during commissioning, debugging, and tests conducted between physics runs.  Generous accommodations for test data especially in the prototyping stage and early years of running of 2-3 times the physics data volume are assumed.  These data are not intended to be run through the production chain, although some may be in order to test the production chain's functionality or to stress the workload management system.  Analysis CPU needs are estimated to be twice that of the MC simulation CPU.

 The largest uncertainties in the resource estimates arise from the simulated data sample storage needs and the numbers of events required for simulation in order to address the full suite of systematic uncertainties needed for physics analyses.  The storage required for simulation is uncertain up to a factor of five, and the number of events in simulation is uncertain by a factor of two to five.  CPU requirements are highly uncertain as well.  As algorithms get more sophisticated and more alternative algorithms are explored, CPU usage is likely to go up.  Optimization work and the addition of compute accelerators such as GPUs will make event processing times go down.   An uncertainty of a factor of five in all CPU estimates is also reasonable. Again, we expect to have this design more fully developed and integrated with the complete DUNE Computing Model for the upcoming ND TDR.
 




\section{DAQ System Introduction}
\label{sec:daq:intro}

The \dword{nd} data acquisition (DAQ) system receives, processes, and records
data from the \dword{dune} \dword{nd}. It receives and synchronizes data from the subdetectors, and it buffers, reduces, and compresses the data for
processing.  Also, it builds event records for permanent storage. The DAQ receives triggering information from external systems (e.g. the accelerator complex) as well as performing
self-triggering decisions. The DAQ provides timing and synchronization for all subdetectors.
This section introduces the requirements and describes the reference design for the DAQ and outlines the further research needed to fully design the DAQ.

\section{DAQ System Requirements}
\label{sec:daq:req}

The DAQ
\begin{itemize}
    \item must be able to trigger and acquire data on indication of a beam spill signal received
from the accelerator complex;
    \item must be able to trigger and acquire data consistent with cosmic rays crossing the
detector(s);
    \item must provide the ability to distribute configurable time-synchronous commands to
the calibration systems, and capture the response of the detectors to calibration
signals;
    \item must be able to acquire data consistent with Ar-39 decay in the liquid argon subdetector;
    \item shall be able to trigger and acquire data without missing beam spills due to other
triggers;
    \item shall have an uptime that does not compromise the overall uptime of the \dword{nd};
    \item shall be able to run combinations of subdetectors independently;
    \item shall form a data record corresponding to every trigger to be transferred to offline
together with the metadata necessary for validation and processing;
    \item shall check integrity of data at every data transfer step. It shall only delete data
from the local storage after confirmation that data have been correctly recorded to
permanent storage;
    \item shall support storing triggered data with a variable size readout window, from few $\mu$s (calibration) to the full readout time of the drift detectors; and 
    \item shall be able to accept the continuous data stream from all subdetectors.

\end{itemize}

\section{Reference Design}
\label{sec:daq:refdes}

The \dword{dune} \dword{fd} DAQ is a system that is in an advanced state of technical readiness, and
which is detailed in the DUNE-SP~\cite{Abi:2020loh} TDR and the DUNE-DP Interim Design Report~\cite{Abi:2018rgm}. Many of the technical requirements
of the \dword{dune} \dword{fd} DAQ overlap strongly with the \dword{dune} \dword{nd} DAQ requirements, and
therefore the reference design for the \dword{dune} \dword{nd} DAQ is the \dword{dune} \dword{fd} DAQ, adapted to
the interfaces required for the \dword{nd}. Using the same solutions for \dword{nd} and \dword{fd} reduces the
effort required to design the DAQ system, and helps ensure long-term stability for the
overall DAQ systems at DUNE.
Relative to the \dword{dune} \dword{fd} DAQ, the \dword{dune} \dword{nd} DAQ must meet the following challenges:

\begin{itemize}
    \item A wider variety of interfaces to subdetectors, some of which may require interfaces
to legacy equipment.
    \item The use of the externally generated beam trigger and the generation of internally
generated triggers (cosmics) that may be propagated to other subdetectors.
\end{itemize}

\begin{dunefigure}[Reference ND DAQ design] {fig:nd-daqdesign}
{A simplified diagram of the reference \dword{nd} DAQ design. Subdetector systems are shown in blue, the upstream DAQ in pink, the data selection system in green, the backend in grey, the timing in orange, and the control system in lavender.}
\includegraphics[width=0.8\textwidth]{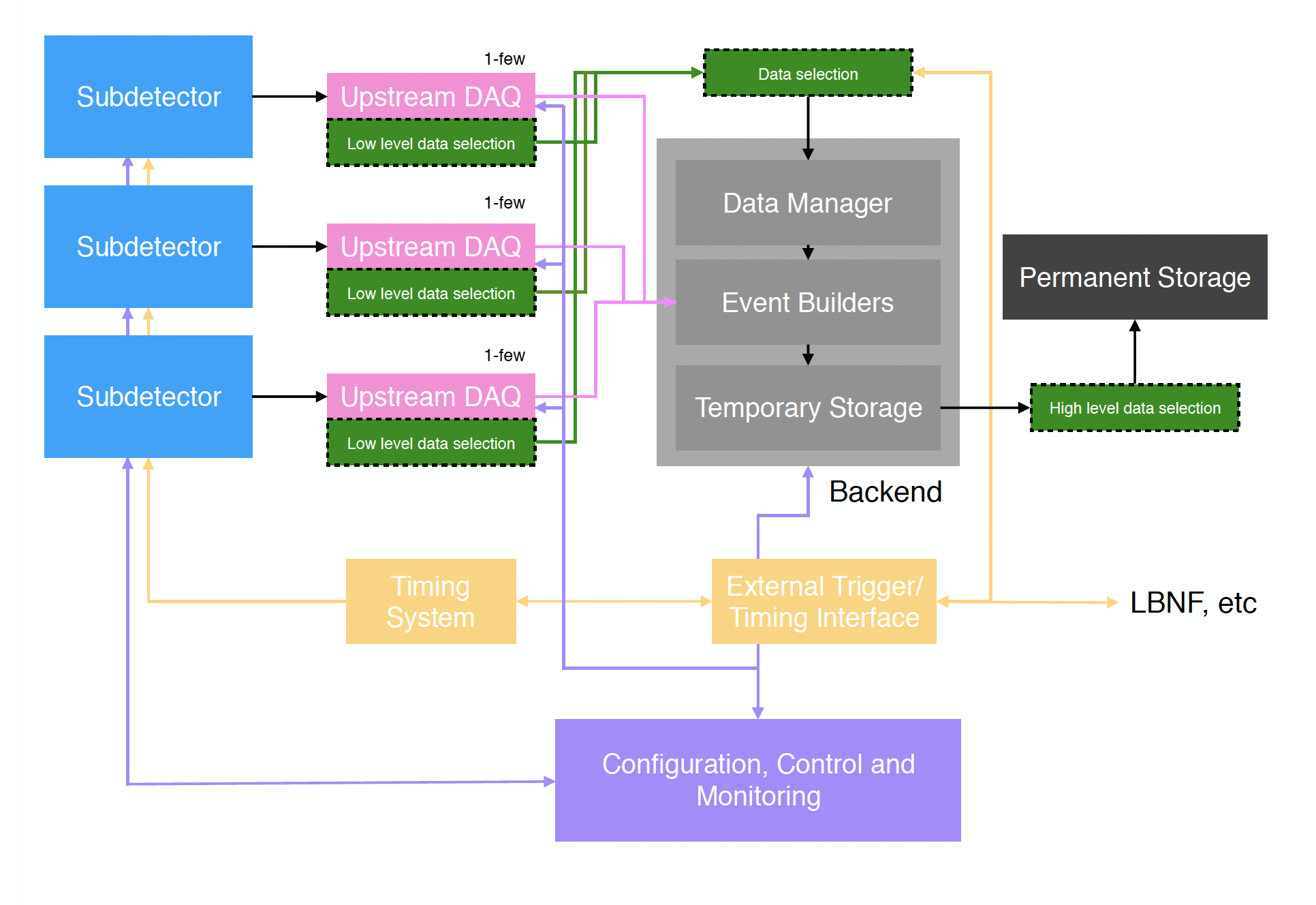}
\end{dunefigure}

Figure~\ref{fig:nd-daqdesign} shows a schematic of a \dword{nd} DAQ system based on the \dword{fd} DAQ system,
highlighting the essential subsystems of the DAQ: upstream DAQ, data selection, the backend, the timing system, and the control, configuration, and monitoring system. A
brief description of each essential subsystem is given in the remainder of this chapter, particularly highlighting where additional research must
be undertaken to define the requirements placed on the DAQ and the suitability of existing
systems for the \dword{nd}.

\subsection{Upstream DAQ}
\label{sec:daq:refdes:upstr}

The upstream DAQ forms the interface between the DAQ system and the front end electronics (FEE) of the various subdetectors. In the \dword{fd} DAQ, the hardware component of
this system is the DAQ readout unit (RU). The design of the RU is slightly different for SP
and DP, but essentially consists of a server with custom FELIX cards. The data buffering
requirements for the \dword{dune} \dword{fd} are higher than anticipated for the \dword{nd}, and so it is anticipated that RU of similar design will be suitable for the \dword{nd}. The number of RU necessary
is of order 10, pending the determination of the final configuration of, for example, the \dword{ndgar} \dword{ecal}.

The FEE design for the \dword{nd} subdetectors are not all in the same state of readiness.
The LArPIX concept for the liquid argon detector is in an advanced state, and work has
commenced to determine how the \dword{fd} upstream DAQ  units can interface with it. The nominal
design for the HPgTPC consists of a similar card, and so the two are likely tied together. Other
subdetectors are in a much less mature state, and as a result, the interface can be defined
for those subdetector FEE and they will conform to the standard.

\section{Data Selection}
\label{sec:daq:refdes:datasel}

The \dword{nd} DAQ will have to acquire data in a few modes.  Of primary importance
is the beam mode. The other modes are required for calibration, both controlled (e.g., light
injection) and uncontrolled (e.g., cosmics and Ar-39).
A major component of future work will be to determine what the needs of the data
selection system are. Dedicated studies are required to understand if
data selection can be done in software only (both low and high level) or if it requires some
hardware component.  Studies are also needed to determine if data selection information must be passed between detectors (e.g., a signal in the
\dword{ndgar} \dword{ecal} results in readout of the \dword{ndlar} photon detection system).

As the FEE for \dword{ndgar} are already mature, the data rate from the detector is
approximately known, and the zero-suppression contained within the FEE means that the
data rate is quite low, and that only high-level data selection (potentially within the DAQ
backend system) is necessary. 
The data rates from the FEE of the \dword{sand} \dword{ecal} are well understood.
The FEE data rates coming from other detectors are less well known at this point.

\subsection{Timing}
\label{sec:daq:refdes:timing}

The reference design for the timing system will be the DUNE SP timing system, which
has been tested in beam conditions at  ProtoDUNE, and contains both the ability to
provide a reference clock and synchronous external signals (e.g. the beam signal). The
nominal clock will be 62.5 MHz, which is compatible with the need for 1 ns relative timing
synchronization between subdetectors within the near detector.
\subsection{Backend DAQ}
\label{sec:daq:refdes:backend}

The DAQ backend (BE) is responsible for organizing the 
flow of data between various
components of the DAQ, building events, and sending them to permanent storage. This
functionality will be developed for the \dword{fd} DAQ, and the nominal design will be to adapt the
\dword{fd} DAQ BE to produce events in the data model for the \dword{nd}. The \dword{nd} data model (that
is, the precise format of events and their metadata grouped into runs) is to be determined
and forms a necessary part of future work. The computing needs for this subsystem for
the \dword{nd} are not yet known.

\subsection{Configuration, Control, and Monitoring}
\label{sec:daq:refdes:ccm}

The configuration, control, and monitoring system is essentially the human interface to the
DAQ system, allowing the user to coherently access  several components of the DAQ
system. It also handles automatic error handling, fault recovery, and resource management.
This functionality will be developed for the \dword{fd} DAQ, and the nominal design will be to
ensure that the development of the configuration, control, and monitoring system for DUNE includes the specific needs of
the \dword{nd}, including the ability to configure the various FEE of the subdetectors, execute the
run types necessary for the \dword{nd}, and monitor the performance of the DAQ. The computing
needs for this subsystem for the \dword{nd} are not yet known.

\cleardoublepage



\cleardoublepage
\printglossaries

\cleardoublepage
\cleardoublepage
\renewcommand{\bibname}{References}
\bibliographystyle{utphys} 
\bibliography{common/tdr-citedb}

\end{document}